\documentclass[a4paper,11pt,oneside,titlepage]{book}
\usepackage[T1]{fontenc}		
\usepackage[utf8]{inputenc}		
\usepackage[english]{babel}		
\usepackage{geometry}			
\usepackage{setspace}			
\usepackage[normalem]{ulem}  
\usepackage{fancyhdr}			
\usepackage{afterpage} 			
\usepackage{bbold}
\usepackage{adjustbox}
\usepackage{tikz-feynman}
\usepackage{tcolorbox}
\usepackage{xcolor}
\usepackage{tikz}
\usepackage[%
  colorlinks=true,
  urlcolor=blue,
  linkcolor=blue,
  citecolor=blue
]{hyperref}
\usepackage{etoolbox}
\usepackage{breqn}			
\usepackage{url}
\usepackage{color}			
\usepackage{xcolor}			
\usepackage{enumerate}			
\usepackage{enumitem}			
\usepackage{graphicx}			
\usepackage{epstopdf}			
\usepackage{pstool}			
\usepackage{psfrag}
\usepackage{subfig}		
\usepackage{array}			
\usepackage{tabularx}			
\usepackage[bottom]{footmisc}		
\usepackage{booktabs}			
\usepackage{longtable}			
\usepackage{caption}			
\usepackage{float}			
\usepackage{subfloat}			
\usepackage{rotating}			
\usepackage{rotfloat}			
\usepackage{amsmath} 			
\usepackage{amssymb}			
\usepackage{cancel}                     
\usepackage{mleftright}
\usepackage{listings}                   
\usepackage[swapnames,norules,nouppercase]{frontespizio}
\usepackage[intoc]{nomencl}
\usepackage[acronym,toc,nomain,nopostdot,nonumberlist]{glossaries}
\usepackage[square, numbers, comma, sort&compress]{natbib}
\usepackage{verbatim}                   
\usepackage{wrapfig}                 
\usepackage{blindtext}
\fancypagestyle{plain}{
                        \fancyhf{}                              
                        \fancyfoot[R]{\thepage}                 
                        \renewcommand{\headrulewidth}{0pt}      
                        \renewcommand{\footrulewidth}{0.4pt}    
                        }                                       
\hypersetup{colorlinks=true, linkcolor=black, citecolor=black, filecolor=black, urlcolor=red}
\definecolor{sapienza}{RGB}{130,36,51} 
\definecolor{cust1}{RGB}{85,85,85}
\definecolor{cust2}{RGB}{212,212,212}
\makenomenclature 
\makeglossaries 
\newenvironment{dedication}
{
  \phantom{.}
  \vspace{13cm}
  \begin{quote} \begin{flushright}}
{\end{flushright} \end{quote}}
\usepackage{calligra}

\begin{document}
\cleardoublepage
\frontmatter 	   
\pagestyle{empty}  

\begin{titlepage}
    \centering
    \vspace*{2cm}
    \includegraphics[width=12cm]{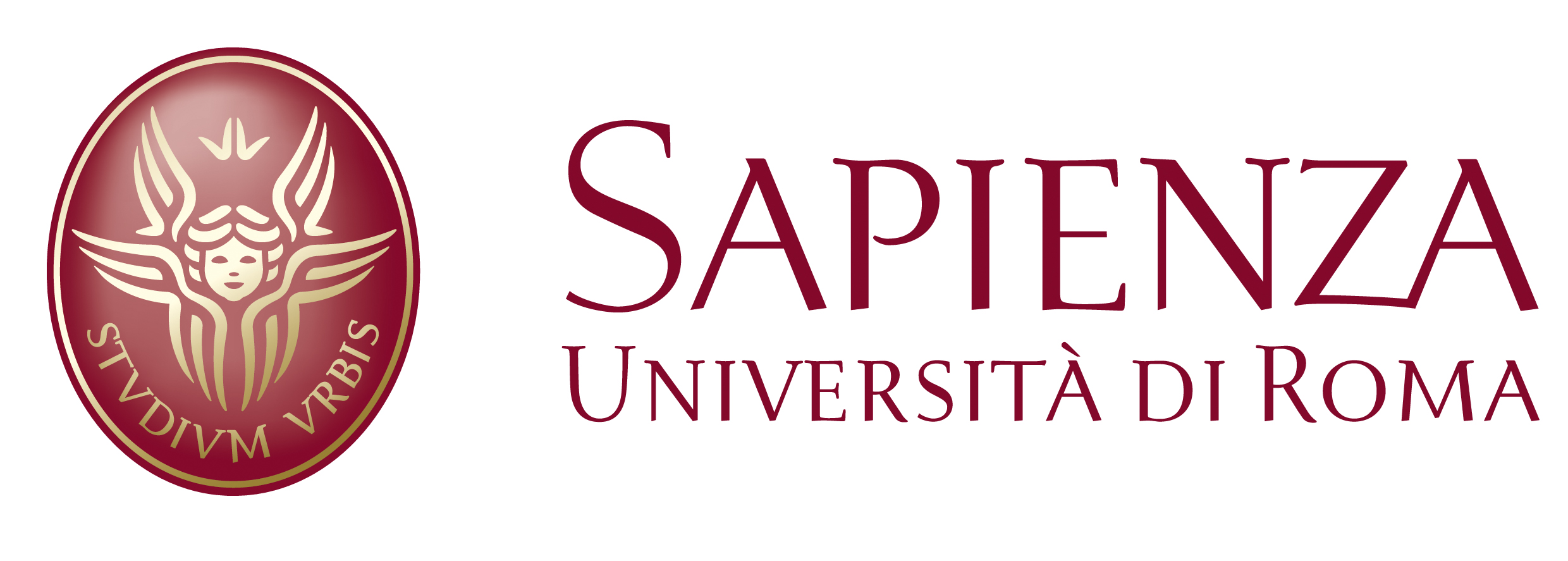}\par\vspace{1cm}
    {\scshape\LARGE Sapienza University of Rome \par}
    \vspace{0.5cm}
    {\scshape\Large Department of Physics \par}
    \vspace{2.5cm}
    {\huge\bfseries Cosmic Whispers of the Early Universe: \\[0.4cm]
    Gravitational Waves and Dark Matter \\[0.4cm]
    from Primordial Black Holes \par}
    \vspace{3cm}
    {\Large\itshape Antonio Junior Iovino \par}
    \vspace{1.5cm}
    \textbf{Thesis Advisor:} Prof. Alfredo Leonardo Urbano \par
    \vfill
    {\large Academic Year MMXXIII-MMXXIV (XXXVII Cycle)\par}
\end{titlepage}

\begin{dedication}

{\fontfamily{ptm}\selectfont
{\large A tutte le volte in cui non vi ho dato ascolto,
\\e a tutte quelle in cui non lo farò.

}
}

\end{dedication}

\newpage
\afterpage{\null\thispagestyle{empty}\clearpage} 
\thispagestyle{plain}			
\setlength{\parskip}{0pt plus 1.0pt}
\section*{Abstract}
Various mechanisms have been suggested for the formation of Primordial Black Holes (PBHs), and this thesis focuses on the standard mechanism based on the critical collapse of cosmological fluctuations. The underlying idea is that during inflation, a period of rapid expansion in the early Universe, large cosmological fluctuations could have been generated. After inflation, when the cosmic horizon reached a size comparable to these fluctuations, if the latter were high enough, they could collapse and form a PBH.

Beyond the fascinating possibility that these compact objects might make up all or part of the Dark Matter (DM) we observe today, their formation and existence is also associated with the generation of gravitational waves (GWs). These waves could contribute to the merger events observed by the LIGO/Virgo/KAGRA Collaboration (LVK) or account for signals detected by pulsar timing array experiments (PTA). In the first part of this thesis, we investigate the PBH scenario, examining the computation of the abundance beyond the Gaussian paradigm. In the second part, we discuss how PBH formation can produce a stochastic GW background and how observations, related to recent experiments such as LVK and PTA collaborations, can help in distinguishing between different PBH production models. Finally, in the last part, we investigate some aspects of the interplay between black holes and fundamental physics in the early Universe, describing some characteristics and challenges of various PBH production models.
\vfill
Keywords: Primordial Black Holes, Inflation, Cosmology, Gravity Waves, Dark Matter.

\thispagestyle{empty}
\mbox{} 
\begin{singlespace}
 \tableofcontents 	
\addcontentsline{toc}{chapter}{List of Publications}
 \chapter*{List of Publications}
\thispagestyle{plain}
The scientific work discussed in this thesis has been carried out at Sapienza University in Rome. Part
of this work was also performed at the National Institute of Chemical Physics and Biophysics in Tallinn and at the University of Geneva, which I kindly thank for the amazing hospitality.

\textbf{Papers appearing in this thesis}:
\begin{itemize}
    \item \cite{Ferrante:2022mui} \textit{Authors:}  G. Ferrante, G. Franciolini, \textbf{A. J. Iovino}, and A. Urbano.
    \\ \textit{Title:} Primordial non-Gaussianity up to all orders: 
    Theoretical aspects and implications for primordial black hole models.
    \\ \textit{Published in:} Phys. Rev. D 107, 043520 (2023).
    \item \cite{Franciolini:2023agm} \textit{Authors:}   G. Franciolini, \textbf{A. J. Iovino}, M.Taoso and A. Urbano.
    \\ \textit{Title:} Perturbativity in the presence of ultra slow-roll dynamics.
    \\ \textit{Published in:} Phys. Rev. D 109 (2024) 12, 123550.
    \item \cite{Ferrante:2023bgz} \textit{Authors:}  G. Ferrante, G. Franciolini, \textbf{A. J. Iovino}, and A. Urbano.
    \\ \textit{Title:} Primordial black holes in the curvaton model: possible connections to pulsar timing arrays and dark matter.
    \\ \textit{Published in:} JCAP 2023 (2023) 06, 057.
    \item \cite{Franciolini:2023pbf} \textit{Authors:}   G. Franciolini, \textbf{A. J. Iovino}, V.Vaskonen and H.Veermae.
    \\ \textit{Title:} Recent Gravitational Wave Observation by Pulsar Timing Arrays and Primordial Black Holes: The Importance of Non-Gaussianities.
    \\ \textit{Published in:} Phys. Rev. Lett. 131 (2023) 20, 201401.
    \item \cite{Ellis:2023oxs} \textit{Authors:} J. Ellis, M. Fairbairn, G. Franciolini, \textbf{A. J. Iovino}, M. Lewicki, M. Raidal, J. Urrutia, V.Vaskonen and H.Veermae.
    \\ \textit{Title:} What is the source of the PTA GW signal?
    \\ \textit{Published in:} Phys. Rev. D 109 (2024) 2, 023522
    \item \cite{Ianniccari:2024bkh} \textit{Authors:} A. Ianniccari, \textbf{A. J. Iovino}, A. Kehagias, D. Perrone and A. Riotto.
    \\ \textit{Title:} Primordial black hole abundance: The importance of broadness
    \\ \textit{Published in:}  Phys. Rev. D 109 (2024) 12, 123549 
    \item \cite{Andres-Carcasona:2024wqk} \textit{Authors:} M. Andrés-Carcasona, \textbf{A.J. Iovino}, V. Vaskonen, H. Veermae, M. Martínez and Ll.M. Mir
    \\ \textit{Title:} Constraints on primordial black holes from LIGO-Virgo-KAGRA O3 events
    \\ \textit{Published in:} Phys. Rev. D 110 (2024) 2, 023040.
    \item \cite{Iovino:2024tyg} \textit{Authors:}  \textbf{A.J. Iovino}, G. Perna, A. Riotto and H. Veermae
    \\ \textit{Title:} Curbing PBHs with PTAs
    \\ \textit{Published in:} JCAP 2024 (2024) 10, 050.
    \item \cite{Allegrini:2024ooy} \textit{Authors:} S. Allegrini, L. Del Grosso, \textbf{A.J. Iovino} and A.Urbano
    \\ \textit{Title:} Is the formation of  primordial black holes from single-field inflation compatible with standard cosmology?
    \\ \textit{Published in:} Peer review
\end{itemize}
\textbf{Other Papers:}
\begin{itemize}
    \item \cite{Ianniccari:2024eza} \textit{Authors:} A. Ianniccari, \textbf{A. J. Iovino}, A. Kehagias, D. Perrone and A. Riotto.
    \\ \textit{Title:} Black Hole Formation -- Null Geodesic Correspondence
    \\ \textit{Published in:}  Phys. Rev. Lett. 133 (2024) 8, 081401. 
    \item \cite{Ianniccari:2024ysv} \textit{Authors:} A. Ianniccari, \textbf{A. J. Iovino}, A. Kehagias, P.Pani, G. Perna, D. Perrone and A. Riotto.
    \\ \textit{Title:} Deciphering the Instability of the Black Hole Ringdown Quasinormal Spectrum
    \\ \textit{Published in:} Phys. Rev. Lett. 133 (2024), 211401 
    \item \cite{Iovino:2024sgs} \textit{Authors:} \textbf{A.J. Iovino}, S. Matarrese, G. Perna, A. Ricciardone and A. Riotto
    \\ \textit{Title:} How Well Do We Know the Scalar-Induced Gravitational Waves?
    \\ \textit{Published in:} Peer review
\end{itemize}

 \printglossaries
\end{singlespace}

\mainmatter	  
\clearpage
\pagestyle{fancy} 
\renewcommand{\chaptermark}[1]{\markright{\chaptername\ \thechapter.\ #1}{}}
\renewcommand{\sectionmark}[1]{\markright{\thesection.\ #1}}
\lhead{} 
\chead{}                   
\rhead{\slshape \rightmark} 
\lfoot{}
\cfoot{} 
\rfoot{\thepage}          
\renewcommand{\headrulewidth}{0.4pt} 
\renewcommand{\footrulewidth}{0.4pt}


\chapter{Prolog}
\thispagestyle{plain}
\vspace{-0.5cm}
\begin{minipage}[t]{0.5\textwidth}
\raggedright
\textit{"O frati", dissi "che per cento milia 
\\perigli siete giunti a l’occidente, 
\\a questa tanto picciola vigilia
\vspace{0.3cm}
\\d’i nostri sensi ch’è del rimanente, 
\\non vogliate negar l’esperienza, 
\\di retro al sol, del mondo sanza gente.     
\vspace{0.3cm}
\\Considerate la vostra semenza: 
\\fatti non foste a viver come bruti, 
\\ma per seguir virtute e canoscenza."}
\vspace{0.3cm}

Inferno Canto XXVI
\end{minipage}
\hfill
\begin{minipage}[t]{0.55\textwidth}
\raggedleft
\textit{"O brothers", I said, "who through a thousand
\\perils have reached the West, 
\\to this so brief vigil
\vspace{0.3cm}
\\of your senses which remains 
\\wish not to deny the experience, 
\\following the sun, of the world that has no people.     
\vspace{0.3cm}
\\Consider your origin;
\\you were not made to live as brutes, 
\\but to pursue virtue and knowledge."}
\vspace{0.3cm}

\end{minipage}
\vspace{0.5 cm}

Since the dawn of civilization, humanity strives to unravel the mysteries of the Universe's origin and evolution. Among these enigmas, the role of dark matter is particularly significant. Despite being invisible, its existence is inferred from the gravitational effects necessary to explain the observed structures of galaxies. Recent advancements in astronomical and cosmological observations have provided substantial evidence for dark matter, yet its true nature remains one of the most persistent puzzles in physics.

Among the various explanations proposed is the intriguing concept of primordial black holes (PBHs). Various mechanisms have been suggested for their formation, and this thesis focuses on the standard mechanism based on the critical collapse of cosmological fluctuations. The underlying idea is that during inflation, a period of rapid expansion in the early Universe, large cosmological fluctuations could have been generated. When the cosmic horizon reached a size comparable to these fluctuations, if they exceeded a certain threshold, they could collapse and form a PBH.

Beyond the fascinating possibility that these compact objects might make up all or part of the dark matter we observe, their formation is also associated with the generation of gravitational waves (GWs). These waves could contribute to the merger events observed by the LIGO/Virgo/KAGRA Collaboration (LVK) or account for signals detected by pulsar timing array experiments (PTA). In this thesis, we investigate the PBH scenario, examining the computation of abundance beyond the Gaussian paradigm. We also discuss how PBH formation can produce a stochastic GW background and how observations can help distinguish between different PBH production models. Finally, we describe some characteristics and challenges of various PBH production models.
\part{Introduction}
\chapter{Primordial black holes:
\\What, why, how and when?}\label{cap:intro}
\thispagestyle{plain}

\section{Inflationary Cosmology}
We start describing briefly the $\Lambda_{\rm CDM}$ model of cosmology. We introduce the inflationary paradigm, explaining how a period of rapid expansion in the early Universe shaped its uniformity. Additionally, we discuss the curvature perturbation and power spectrum, essential in discussing the formation of PBHs.
\subsection{$\Lambda_{\rm CDM}$ model}
The Standard Model of Cosmology $\Lambda_{\rm CDM}$, is the standard framework that provides a reasonably good description of the Universe's evolution on the largest scales.
It is based on two important observational facts: first, on large scales, the Universe is spatially homogeneous and isotropic. This observational fact is referred to as the Cosmological Principle. One of the main pieces of evidence is the nearly uniform temperature of Cosmic Microwave Background (CMB) radiation which explains the isotropy of our Universe. Second, the Universe is expanding.
It is important to note that the Cosmological Principle refers only to space. The Universe is neither homogeneous nor isotropic in time, a fact which allows us to introduce the inflation paradigm.
An isotropic and homogeneous Universe is described by the Friedman-Robertson-Walker (FRW) metric:
\begin{equation}
d s^2=-d t^2+a(t)^2\left[\frac{d r^2}{1-K r^2}+r^2 d \theta^2\right]
\end{equation}
where $a(t)$ is the cosmic scale factor. The constant $K$ is a curvature parameter and specifies the geometry of the Universe: $K=0$ is a flat Universe, $K=1$ is a closed Universe and $K=-1$ is an open Universe. From that, we can define the Hubble parameter, given by:
\begin{equation}
H(t)=\frac{\dot{a}(t)}{a(t)}.
\end{equation}
Current CMB observations indicate that $H_0 \approx 67.4 \pm 0.5(\mathrm{Km} \mathrm{s^{-1}}) \mathrm{Mpc^{-1}}$\,\cite{Planck:2018vyg}.
\\The energy-momentum tensor for a fluid with an energy density $\rho$ and pressure $p$ is
\begin{equation}\label{eq:Tmunu}
T_{\mu \nu}=(p+\rho) u_\mu u_\nu-p g_{\mu \nu}.
\end{equation}
Introducing Eq.\,\ref{eq:Tmunu} into the Einstein equations, we obtain two independent equations:
\begin{align}
H^2 & =\frac{8 \pi G}{3} \rho-\frac{K}{a^2}, \\
\frac{\ddot{a}}{a} & =-\frac{4 \pi G}{3}(\rho+3 p) .
\end{align}
These two equations are also called Friedmann equations. From these equations, the continuity equation can be derived
\begin{equation}
\dot{\rho}+3 H(\rho+p)=0 .
\end{equation}
In the cosmological context, the cosmic fluid can be described by a perfect fluid $(P=w \rho)$. The value of $w$ determines the kind of energy-matter content. For instance, $w=0$ corresponds to non-relativistic matter (dust). Instead, for relativistic particles, such as radiation, we have $w=1 / 3$. The case of vacuum-energy (cosmological constant) is $w=-1$.

Solving the continuity equation, with the equation of state, we get
\begin{equation}
\rho \propto a^{-3(1+w)}
\end{equation}
and then
\begin{align}
a(t) \propto t^{\frac{2}{3(1+w)}} & \text { if } w \neq-1 \\
a(t) \propto e^{H t} & \text { if } w=-1.
\end{align}
It is useful to define a conformal time $\eta$, given by,
\begin{equation}
\eta=\int \frac{d t}{a(t)},
\end{equation}
which can be used to define the particle horizon, that is, the maximum distance that light has traveled from $t_i$ up today,
\begin{equation}
\tau=\eta-\eta_i=\int_{t_i}^t \frac{d t}{a}.
\end{equation}
\subsection{The inflationary paradigm}
The inflationary paradigm\,\cite{Guth:1980zm} proposes that a small patch of space in the early Universe underwent a period of exponential growth, and this explains why the Universe appears nearly homogeneous, isotropic and flat. 

In order to preserve the concepts of locality and causality, we require that something changes the causal structure of the early Universe, giving time for different parts of space to influence each other. Moreover, the Universe must undergo a non-adiabatic period to account for the anomalously large entropy observed today. This could explain the anomalously high entropy values observed today. These modifications can be obtained postulating that the very early Universe underwent a period of accelerated expansion $\ddot{a}>0$ referred as inflation.

In order to describe the main features of cosmic inflation, cosmology must be combined with particle physics and therefore over the years many inflationary models were presented. 
The easiest way to implement an inflationary phase in the early Universe is to introduce a new scalar field $\phi$, whose energy density dominates the primordial Universe, called Inflaton.
The action for the inflaton field is:
\begin{equation}
S=\int \mathrm{d}^4 x \sqrt{-g} \mathcal{L}=\int d^4 x \sqrt{-g}\left(\frac{1}{2} g^{\alpha \beta} \partial_\alpha \phi \partial_\beta \phi-V(\phi)\right)
\end{equation}
where for the FLRW metric $\sqrt{-g}=a(t)^3$.
From the Euler-Lagrange equations, we obtain the following equation of motion:
\begin{equation}
\ddot{\phi}+3 H \dot{\phi}-\frac{\nabla^2 \phi}{a^2}+V^{\prime}(\phi)=0.
\end{equation}

This equation is similar to the harmonic oscillator, where the extra term $3 H \dot{\phi}$ is a friction term. The latter implies that a scalar field rolling down its potential suffers a friction due to the expansion of Universe.
At this point we want to determine under which condition the inflaton can lead to accelerated expansion.
The corresponding energy density $\rho_\phi$ and the Pressure density $P_\rho$ are:
\begin{align}
& \rho_\phi=\frac{\phi^2}{2}+V(\phi) \\
& P_\phi=\frac{\dot{\phi}^2}{2}-V(\phi)
\end{align}
In order to have an accelerated expansion, we need to require that:
\begin{equation}
V(\phi) \gg(\dot{\phi})^2
\end{equation}
Hence, in order to generate inflation, the inflaton, whose energy is dominant during inflation, has a potential energy that dominates over the kinetic term.
If the inflaton field rolls slowly, and at a nearly constant speed down it's potential, then it is said to be a slow-roll model of inflation. This distinction is quantified by the two slow-roll parameters
\begin{align}
& \epsilon=-\frac{\dot{H}}{H^2}=\frac{\dot{\phi}^2}{2 M_p^2 H^2} \\
& \eta=\frac{\dot{\epsilon}}{\epsilon H},\label{eq:eta}
\end{align}
The first slow-roll parameter, $\epsilon$, tracks how quickly the scalar field is moving down its potential, while the second slow-roll parameter, $\eta$, tracks the acceleration. This means that if $\epsilon$ is small and constant, then $\eta \approx 0$, and the model is classed as slow-roll. In this approximation
$\epsilon \ll 1$, then the Friedmann equation reads as
\begin{equation}
H^2 \approx \frac{V}{3 M_p^2} .
\end{equation}
In this approximation $\epsilon$ and $\eta$ can be recast in terms of the potential and its derivatives only:
\begin{align}
& \epsilon \approx \frac{M_p^2}{2}\left(\frac{V^{\prime}(\phi)}{V(\phi)}\right)^2 \\
& \eta \approx M_p^2 \frac{V^{\prime \prime}(\phi)}{V(\phi)}
\end{align}
As we will see in the Chapter\,\ref{cap:Models} in the context of single field inflationary scenarios a slow roll inflation is not enough to produce PBH and we need to introduce a particular phase called Ultra-slow roll or a spectator field.
\subsection{Curvature perturbation and Power spectrum}
During inflation quantum fluctuations of the inflaton field are produced and their wavelengths are stretched on large scales by the rapid expansion of the universe\,\cite{Riotto:2002yw}.
Hence when we include quantum fluctuations to the inflaton field
\begin{equation}
\phi(\boldsymbol{x}, t)=\phi(t)+\delta \phi(\boldsymbol{x}, t) ,
\end{equation}
we find that, at second order, the equation of motion for scalar field perturbations becomes
\begin{equation}
\ddot{\delta} \phi+3 H \dot{\delta} \phi-\frac{1}{a^2} \nabla^2 \delta \phi+\partial_\phi^2 V \delta \phi=0 .
\end{equation}
Moving to Fourier space it gives
\begin{equation}
\ddot{\delta} \phi+3 H \dot{\delta} \phi+\frac{k^2}{a^2} \delta \phi+\partial_\phi^2 V \delta \phi=0 .
\end{equation}
Generally these perturbations are studied using the correlation function at two different spatial positions, but at equal times. In Fourier space, this is identified with the power spectrum $\mathcal{P}_{\delta \phi}(k, t)$, defined by
\begin{equation}
\left\langle\delta \phi(\boldsymbol{k}, t) \delta \phi\left(\boldsymbol{k}^{\prime}, t\right)\right\rangle=\frac{2 \pi^2}{k^3} \mathcal{P}_{\delta \phi}(k, t) \delta\left(\boldsymbol{k}+\boldsymbol{k}^{\prime}\right),
\end{equation}
where $\delta\left(\boldsymbol{k}+\boldsymbol{k}^{\prime}\right)$ is a Dirac delta function.
After the inflation, the inflaton quickly decays into other fields in a stage called Reheating, as a consequence we need to relate the power spectrum to some other quantities. This can be achieved by considering each perturbation to be a separate FLRW Universe with a curvature $K$, more commonly parametrised in terms of the curvature perturbation $\zeta$ as
\begin{equation}
\mathrm{d} s^2=\mathrm{d} t^2-a^2(t) e^{2 \zeta} \mathrm{d} x^2 .
\end{equation}
The curvature perturbation is still problematic since it is a gauge dependent quantity. To overcome this, we introduce the gauge invariant comoving curvature perturbation defined as follows
\begin{equation}
\mathcal{R}=\zeta+\frac{\delta \rho}{3(\rho+P)} .
\end{equation}
In the comoving gauge, $\delta \rho=0$ and hence $\mathcal{R}=\zeta$. Additionally, on super-horizon scales, where the causal horizon at scale $k$ is larger than the Hubble horizon, $\delta \rho$ can be neglected and the two definitions are again equivalent. Therefore, it is common to see $\mathcal{R}$ and $\zeta$ used interchangeably.
The equation of motion for $\mathcal{R}$ reads
\begin{equation}
\ddot{\mathcal{R}}+\left(3+\eta_H\right) H \dot{\mathcal{R}}+\frac{k^2}{a^2} \mathcal{R}=0,
\end{equation}
known as the Mukhanov-Sasaki equation\,\cite{Sasaki:1986hm,Mukhanov:1988jd}. The comoving curvature perturbation power spectrum is then defined as for the inflaton perturbation,
\begin{equation}
\left\langle\mathcal{R}(\boldsymbol{k}, t) \mathcal{R}\left(\boldsymbol{k}^{\prime}, t\right)\right\rangle=\frac{2 \pi^2}{k^2} \mathcal{P}_{\mathcal{R}}(k, t) \delta\left(\boldsymbol{k}+\boldsymbol{k}^{\prime}\right) .
\end{equation}
where this quantity is evaluated for each mode on super-horizon scales where the value of the curvature perturbation is frozen
\begin{equation}
\mathcal{P}_{\mathcal{R}}(k)=\left.\mathcal{P}_{\mathcal{R}}(k, t)\right|_{k=a h} .
\end{equation}
The observation of CMB anisotropies provides an estimation of this power spectrum on large scales of $\mathcal{P}_{\mathcal{R}}(k) \approx 2 \times 10^{-9}$ with a mild dependence on $k$. However, there are currently no such measurements on smaller scales, leaving room for the power spectrum to potentially increase to significantly larger values, possibly leading to intriguing observable phenomena or the formation of objects like primordial black holes.
\section{Primordial black holes}
Having set the stage earlier, let's delve into the intriguing realm of PBHs.
\\The notion of PBHs potentially forming in the early Universe emerged in historical works by Y. Zel'dovich and I. Novikov\,\cite{Zeldovich:1967lct} alongside S. Hawking\,\cite{Hawking:1971ei}. Interestingly, it was during Hawking's focus on PBH studies that the famous black hole evaporation was discovered\,\cite{Hawking:1975vcx,Hawking:1974rv}. Subsequently, it became apparent that PBHs could contribute significantly to the dark matter puzzle, as outlined by B. Carr\,\cite{Carr:1974nx,Carr:1975qj} and G. Chapline\,\cite{Chapline:1975ojl}. These enigmatic entities might have also played a role in forming the large-scale structure through Poisson fluctuations\,\cite{Meszaros:1975ef}, potentially seeding the supermassive black holes we observe today.

PBHs, emerging from epochs far preceding matter-radiation equality, behave cosmologically like a cold, collisionless fluid. They stand as compelling dark matter candidates, especially if their masses exceed $\simeq 10^{-19} M_{\odot}$ to ensure lifetimes surpassing the age of the Universe\,\cite{MacGibbon:2007yq}. What is particularly captivating about PBH dark matter is its potential to be explained within the standard model of particle physics. In other words, PBHs can be, in principle, generated using the standard inflationary scenarios, without requiring the introduction of additional particles, but the inflaton. This explanation may require tweaks in the (largely unconstrained) description of the early Universe at small scales to account for the generation of substantial density perturbations that catalyze PBH formation.
Extensive investigations have been undertaken to probe and constrain the abundance of PBHs, leveraging their distinct characteristics at smaller scales. These studies might reveal observable phenomena such as lensing, electromagnetic emissions from accretion processes, and gravitational waves. We will discuss the constraints on the abundance in Chapter\,\ref{cap:NGS}.
\\Even if the PBHs are only a tiny fraction of the Dark matter, they can still explain some of the GW events we are seeing today, as we will describe in Chapter\,\ref{cap:GWs}, and could produce the primordial seeds for the formation of the supermassive BHs observed at high redshift\,\cite{Duechting:2004dk,Kawasaki:2012kn,Bernal:2017nec}.

In this section we briefly discuss several formation mechanisms of PBHs, their physical properties and eventually how to distinguish primordial from astrophysical black holes.
\subsection{Formation mechanisms}
After the hypothesis of their existence, many formation mechanisms for PBHs were devised in literature. In this thesis we will focus on the most studied scenario. It is based on the collapse of large overdensities generated during inflation\,\cite{Ivanov:1994pa,GarciaBellido:1996qt,Ivanov:1997ia,Blinnikov:2016bxu}. These overdensities are characterised by the density contrast $\delta=\delta \rho/\rho$, where $\delta \rho$ is the local change in energy density relative to $\rho$, the background density\footnote{Sometimes the same condition is expressed in terms of the compaction function $\mathcal{C}$. This quantity expresses the compactness or density of the matter inside a PBH. We will give more details in sec.\ref{sec:C1pre}.}. If the density contrast exceeds a critical value $\delta_c$, when the wavelength of the perturbations is of the same size of the Horizon after the inflation, then a PBH will be formed. This critical value is very large, so it is quite improbable to achieve a density contrast large enough to form a PBH. In general, as we will discuss in details in sec.\,\ref{sec:C1pre}, there is a non-linear relation between the density contrast $\delta$ and the comoving curvature perturbation $\mathcal{R}$. However, the important point to know is that we need an enormous boost of the amplitude of the curvature power spectrum (the amplitude must be around $\mathcal{O}(10^{-3})$) in order to produce a sizeable amount of PBHs.

In standard slow-roll inflation, the power spectrum for the comoving curvature perturbation $\mathcal{R}$ is approximately constant at $\mathcal{P}_{\mathcal{R}} \sim 10^{-9}$, and no PBHs form. To generate even a single PBH in a Hubble volume, the power spectrum must grow to $\mathcal{P}_{\mathcal{R}}>10^{-3}$\,\cite{Green:1997sz,Green:1999xm,Yokoyama:1998xd}. We will discuss several possibilities to get the required enhancement in Chapter\,\ref{cap:Models}.

There are a numerous mechanisms to form PBHs based on an early matter era \,\cite{Khlopov:1980mg,Polnarev:1985btg,Green:1997pr,Harada:2016mhb,Alabidi:2013lya}, modified gravity\,\cite{Barrow:1996jk,Papanikolaou:2021uhe,Mann:2021mnc}, scalar field instabilities\,\cite{Khlopov:1985fch,Cotner:2016cvr,Cotner:2019ykd}, collapse of cosmic strings\,\cite{Polnarev:1988dh,Hawking:1987bn,Garriga:1993gj,Caldwell:1995fu,MacGibbon:1997pu,Helfer:2018qgv,Ghoshal:2023fhh} and domain walls\,\cite{Rubin:2000dq,Rubin:2001yw,Garriga:2015fdk,Deng:2016vzb,Liu:2019lul,Kusenko:2020pcg,Gouttenoire:2023gbn}, phase transitions\,\cite{Crawford:1982yz,Kodama:1982sf,Jedamzik:1996mr,Kanemura:2024pae,Lewicki:2024ghw}, bubble collisions\,\cite{Hawking:1982ga,Moss:1994iq,Kitajima:2020kig,Kasai:2023qic,Escriva:2023uko}, standard model Higgs instability\,\cite{Espinosa:2017sgp,Espinosa:2018euj} and from unconfined quarks and gluons \cite{Alonso-Monsalve:2023brx}. Furthermore, the phenomenon of PBH formation started being investigated with the aid of dedicated numerical relativity simulations, which show that it follows the properties of critical collapse\,\cite{Choptuik:1992jv,Niemeyer:1997mt,Yokoyama:1998qw,Niemeyer:1999ak,Shibata:1999zs,Gundlach:1999cu,Musco:2004ak,Polnarev:2006aa,Musco:2008hv,Uehara:2024yyp,Ianniccari:2024ltb}.
\subsection{Physical properties: Primordial vs Astrophysical Black holes}
The no-hair theorem\,\cite{PhysRevLett.26.331,PhysRev.164.1776} states that in the late Universe long after their formation, the information about the formation of a black hole is lost, and the resulting black holes can be described in terms of only three parameters: its mass, spin and electric charge.
As we can describe below, these three parameters are enough to distinguish PBH from the ones formed by astrophysical processes. Moreover as we will discuss next, the gravitational waves can be an powerful tools to determine the nature of the observed black holes both alone or in binaries. 
Below we describe briefly the main features of some properties of BH, such as their masses, spin, the possibility to form binaries and their clustering. We stress that in principle the electric charge can also be important, but electromagnetic repulsion prevents BH from forming with significant charge, and any charge attained by accretion of charged particles would quickly neutralise by attraction of particles of the opposite charge.
\subsubsection{Mass}
Let's delve into the realm of astrophysical black holes, categorizing them based on their mass into three main types: stellar-mass, intermediate-mass, and supermassive.

At the upper end of the scale are supermassive black holes (SMBHs), tipically with masses $m \gtrsim 10^5 \mathrm{M}_{\odot}$. These giants are believed to occupy the cores of galaxies, forming through the accretion of material onto seeds from the earliest stars. However, challenges persist in explaining the rapid growth of these seeds to account for the presence of very distant SMBHs\,\cite{Banados:2017unc}.

Next, we encounter intermediate-mass black holes (IMBHs) with masses ranging from $\sim 10^2$ to $10^5 \mathrm{M}_{\odot}$. Candidates for IMBHs have been detected within our galaxy and neighboring galaxies\,\cite{Patruno:2006bw,Maccarone:2007dd}, although their existence remains debated. Lastly, we have the stellar-mass black holes, with masses typically falling in the range of $\sim 1-100 \mathrm{M}_{\odot}$.

Astrophysical black holes are the end products of stellar evolution. When a star exhausts its light element fuel, the radiation pressure wanes, and the stellar matter succumbs to the overwhelming pull of gravity. This collapse creates an area of huge density. In the case of massive stars, a supernova explosion may occur, expelling the outer stellar layers and leaving behind a stellar remnant. This remnant's mass is considerably less than the original star's mass before collapse. Below a critical threshold, known as the Tolman-Oppenheimer-Volkoff (TOV) limit\,\cite{PhysRev.55.374}, the degeneracy pressure among free neutrons prevents further collapse. However, if the remnant surpasses this limit, gravitational collapse ensues, birthing a black hole. Despite uncertainties surrounding these remnants, the TOV limit is estimated to be around $\simeq 1 \mathrm{M}_{\odot}$\,\cite{Gao:2015xle,Shibata:2019ctb}.

Observational and theoretical investigations suggest that the distribution of stellar-mass black holes exhibits two notable gaps: a lower gap between the heaviest neutron star and the lightest black hole ($\sim 2.5-5 \mathrm{M}_{\odot}$), and an upper gap around $\sim 50-150 \mathrm{M}_{\odot}$. While uncertainties persist regarding remnants near the TOV limit, evidence from X-ray binaries hints at the existence of the lower mass gap\,\cite{Bailyn:1997xt}\footnote{Some suggest that systematic errors might have affected this evidence or that objects within this gap might be beyond current observation techniques\,\cite{Thompson:2018ycv}}. The upper mass gap, on the other hand, stems from theoretical considerations. As a star collapses, efficient pair production occurs, destabilizing the star\,\cite{LIGOScientific:2018jsj}. This process leads to either a pulsational pair-instability supernova if the helium core is $\sim 30-64 \mathrm{M}_{\odot}$, or a pair-instability supernova if the core is $\sim 64-135 \mathrm{M}_{\odot}$. The former results in a reduced remnant and hence a smaller black hole, while the latter completely obliterates the star, leaving no remnant or black hole behind. These combined effects contribute to the upper gap in the black hole mass spectrum.

The Ligo-Virgo-Kagra (LVK) has detected objects falling within these mass gaps\,\cite{LIGOScientific:2018jsj,LIGOScientific:2020iuh,LIGOScientific:2020zkf,LIGOScientific:2024elc}, though their exact nature remains a topic of ongoing study\,\cite{Huang:2024wse,Dhani:2024jja,Fragione:2020han,Zevin:2020gbd,Franciolini:2021tla,Deluca:2020sae}.

PBHs present a distinct contrast in mass from astrophysical black holes. Theoretical considerations reveal no forbidden regions within their mass distribution. Furthermore, the absence of degeneracy pressure during PBH formation means that while an astrophysical black hole cannot be smaller than $\mathrm{M}_{\odot}$, a PBH could form with a mass $10^{-18} \mathrm{M}_{\odot}$ and endure to the present day. These disparities are crucial, as the presence of compact objects in the astrophysical mass gaps or strong evidence of a solitary compact object with $m<\mathrm{M}_{\odot}$ could serve as compelling "smoking gun" evidence for PBHs. The LVK has observed a few objects falling within these mass gaps, though these instances might be attributed to second-generation mergers. The potential future discovery of a sub-solar candidate would be a remarkable "smoking gun" for PBHs.

In Chapter\,\ref{cap:NGS}, we will discuss how to compute the abundance and mass distribution of a PBH population.
\subsubsection{Spin}
The spin of astrophysical black holes represents a crucial parameter that influences various astrophysical phenomena, ranging from accretion disk dynamics to gravitational wave signatures. 
In the context of accretion disks, the black hole's spin profoundly affects the disk's innermost stable circular orbit (ISCO), determining the location where material plunges into the black hole. Higher spins lead to smaller ISCOs, influencing the disk's luminosity, temperature profile, and emission properties \cite{McClintock:2013vwa}.
In the realm of gravitational wave astronomy, the spin of merging black holes leaves distinct features on the emitted gravitational waveforms. Spin-induced precession can lead to modulations in the waveform, providing valuable insights into the binary's orbital dynamics and individual black hole spins \cite{Pretorius:2007jn,Campanelli:2005dd}.
Much like the mass, the PBH spin is expected to be quite different to that of
astrophysical black holes. Unlike astrophysical BHs, PBH formation predicts a very small
spin\,\cite{Chiba:2017rvs,DeLuca:2019buf,Mirbabayi:2019uph}, where the potential effect of accretion will cause the PBH spins to increase significantly
away from zero\,\cite{DeLuca:2020bjf}. The LIGO instrument is sensitive to a combination of the spins called
the effective spin parameter $\chi_{\rm eff}$, and the LVK has found that the vast majority of
merger events have $\chi_{\rm eff}$ posteriors consistent with zero\,\cite{LIGOScientific:2018mvr}. 
Hence, this parameter can play a fundamental role in the determination of the nature of a black hole.
\subsubsection{Binaries}
Binary black hole (BBH) systems are fascinating cosmic phenomena that provide valuable insights into both astrophysical processes and primordial relics. Here we explore the distinctions between binary systems comprising solely astrophysical black holes and those where one or both components could originate from primordial sources.

Astrophysical BBHs can form through two primary mechanisms: dynamical formation and the common envelope scenario \cite{Rodriguez:2016kxx,OLeary:2016ayz,Belczynski:2001uc,Marchant:2016wow}. In the dynamical formation scenario, characterized by environments with high stellar densities such as globular clusters, black holes formed from the remnants of massive stars gradually migrate towards the cluster's core due to dynamical friction. Interactions between these black holes and other stellar binaries play a crucial role in BBH formation, with gravitational interactions and scattering events leading to the convergence of black holes into binary systems. This process often results in binaries with similar mass components due to the dynamical sorting process.

Alternatively, BBHs can arise from binary star systems undergoing the common envelope phase during their evolution. In this scenario, the binary system consists of two stars, one of which evolves into a black hole following stellar collapse. As the more massive star nears the end of its life and expands, it may exceed its Roche lobe, initiating mass transfer to its companion. This mass transfer process, driven by gravitational interactions and Roche lobe overflow, can lead to the formation of a common envelope surrounding both stars. Ultimately, the collapse of both stars into black holes within this shared envelope results in the formation of a BBH system.

Both dynamical formation and the common envelope scenario typically yield BBHs with similar mass components. In the common envelope scenario, the equilibrium in mass arises from the shared surrounding matter during the binary's evolution, leading to comparable black hole masses. Conversely, in dynamical formation processes within dense stellar environments, the most massive black holes tend to migrate towards the core, where they form binaries with similarly massive companions. Furthermore, frequent scattering events within these environments often exchange black holes between binaries, further promoting the formation of BBHs with similar mass components.

Regardless of the formation mechanism of the single PBH, a population of PBH binaries can be formed through two mechanisms that operates on two separate time scales (for a recent review on this topic see \cite{Sasaki:2018dmp,Franciolini:2021nvv}):
\begin{itemize}
    \item \textit{Early Universe formation:} a pair of PBHs can decouple from the Hubble expansion if their mutual gravitational attraction starts dominating over the Hubble flow. During the two PBHs falling towards each other, the gravitational field of the surrounding PBHs and other matter inhomogeneities exert torques on the bound system. As a result, the two PBHs avoid a head-on collision and the binary acquires angular momentum \cite{Nakamura:1997sm}.
    The characteristic formation redshift is of the order $z_{dec}\geq10^4$ for the masses relevant for current ground-based detectors \cite{Ali-Haimoud:2017rtz}.
    \item \textit{Late Universe formation:} PBH binaries can be formed due to dynamical captures in the present-day halos or clusters \cite{Bird:2016dcv,Korol:2019jud,Clesse:2016vqa,Bagui:2021dqi}. If a PBH traveling in space passes close to another PBH, the process may emit gravitational waves, reducing the kinetic energy of the first object which becomes bound to the latter.
\end{itemize}
If all binaries remain unperturbed until merger, the merger rate is so large that only a small fraction $\mathcal{O}(0.1\%)$ of dark matter (DM) is allowed in the mass range $1-100$ solar masses $M_{\odot}$ that LVK detectors probe \cite{Raidal:2017mfl,Ali-Haimoud:2017rtz,DeLuca:2020qqa}.  However, this statement holds as long as no perturbations are considered. The binaries can be perturbed by two main mechanisms:
\begin{enumerate}
    \item If the initial configuration contains a third PBH close to the PBH pair expected to form a binary, it is very likely to collide with the binary and hence to perturb the system (see for instance refs.\,\cite{Ioka:1998nz,Mouri:2002mc}).
    \item PBHs will form dense N-body systems relatively early, and binaries absorbed by these clusters become more likely to be perturbed (see for instance refs.\,\cite{Bird:2016dcv,Jedamzik:2020ypm}). 
\end{enumerate}
Here we do not enter into the details of how these perturbations can modify the merger rate of BBHs of primordial type. The important point is that binaries comprising at least one primordial black hole serve as unique laboratories. As we have already described, the contrast between PBHs, formed in the early Universe through gravitational collapse of high-density regions, and their astrophysical counterparts encompasses various aspects, including mass distribution, formation mechanism, and dynamical evolution. Analyzing the interplay between these two distinct populations can provide insights into the hierarchical structure formation process and the merger history of black holes throughout cosmic history \cite{Sasaki:2018dmp}. Additionally, tidal effects\,\cite{Binnington:2009bb,Damour:2009vw,Chia:2020yla,Crescimbeni:2024cwh} can be employed to differentiate sub-solar PBH candidates from neutron stars\,\cite{Cardoso:2019rvt} or exotic compact objects such as, Q-balls\,\cite{Coleman:1985ki}, fermion\,\cite{Lee:1986tr,DelGrosso:2023trq} and boson stars\,\cite{Liebling:2012fv}.
\subsubsection{Clustering}
Due to their immense gravitational attraction, black holes can cluster together. The study of astrophysical black hole clusters and their impact on the formation and evolution of supermassive black holes and large-scale structures, such as galaxies and galaxy clusters, remains a subject of extensive research in the astrophysical community. However, this field faces significant constraints imposed by the numerical simulations required\,\cite{Barack:2018yly,Volonteri:2010wz,Madau:2001sc,Namikawa:2016edr}. Therefore, we will focus here on briefly describing the essential requirements and characteristics for PBHs clustering.

After their formation, PBHs can be considered discrete objects distributed throughout the Universe. A key question is how their spatial distribution is defined. Specifically, in order to understand their possible cluster, we should determine if their spatial distribution follows a Poisson distribution (dictating random positioning in a given volume) or if they are more likely to be close to each other forming clusters of PBHs. When originating from Gaussian density perturbations, PBHs do not exhibit clustering at the formation\,\cite{Desjacques:2018wuu,Ballesteros:2018swv,MoradinezhadDizgah:2019wjf,DeLuca:2020jug,DeLuca:2022uvz}. This outcome arises from the fact that PBHs form at the scale of the horizon, making it challenging to establish correlations on scales larger than a few Hubble patches due to the equivalence principle. Therefore, substantial non-Gaussianities are required in principle to induce significant clustering of PBHs alongside the Poisson distribution.However, the presence of bias in these non-Gaussianities is constrained by observations of the CMB\,\cite{Tada:2015noa,Young:2015kda,DeLuca:2021hcf}.

Furthermore, if primordial black holes constitute only a small fraction of dark matter, their clustering becomes irrelevant. Conversely, for a substantial contribution to dark matter, clustering could elevate the merger rate in the late-time Universe to a level compatible with the detection rate of gravitational wave events by instruments like LIGO/Virgo\,\cite{Clesse:2016vqa,Raidal:2017mfl,Raidal:2018bbj,Ding:2019tjk}.
\vspace{0.5cm}
\section{The Gravitational waves revolution}
Gravitational waves (GWs) can be described as propagating deformations of the spacetime geometry. While recent years have marked a revolution in the search for GW signals, with recent discoveries opening a golden era and driving the international community to deduce the nature of these signals, their history has not always been smooth sailing. In fact, after their theoretical prediction in 1916 by Einstein\,\cite{1916SPAW.......688E} in his seminal work on general relativity, where he first predicted that accelerated matter would produce undulations of spacetime interpretable as waves, their existence remained debated until the late 1950s.

The first "indirect" confirmation of GWs came after the discovery of pulsars, which are rotating neutron stars emitting radio pulses\,\cite{Hewish:1968bj,Gold:1968zf}. It was soon realized that pulsars in binaries would emit GWs, causing the orbital period to evolve as predicted by Einstein's theory. This was observed in 1974 by R. Hulse and J. Taylor in the system known as PSR B1913+16\,\cite{Hulse:1974eb}, for which they received the 1993 Nobel Prize in Physics.

The first 'direct' detection of GWs from a binary black hole merger came in September 2015, after a long wait\,\cite{LIGOScientific:2016aoc}. This observation demonstrated the potential of GW signals to provide information on one of the most violent events in the Universe. R. Weiss, B. Barish, and K. Thorne were awarded the Nobel Prize in Physics in 2017 for their decisive contribution leading to this discovery.

In the summer of 2023, collaborations such as NANOGrav~\cite{NANOGrav:2023gor, NANOGrav:2023hde}, EPTA (in combination with InPTA)\,\cite{EPTA:2023fyk, EPTA:2023sfo, EPTA:2023xxk}, PPTA\,\cite{Reardon:2023gzh, Zic:2023gta, Reardon:2023zen} and CPTA\,\cite{Xu:2023wog} announced evidence of a signal compatible with a stochastic gravitational wave background (SGWB) explanation at frequencies in the nHz range, with characteristics similar to those expected from a population of supermassive black hole binary systems. This discovery opens new perspectives in the study of GW sources, complementing the mergers of stellar-mass black holes discovered by the LIGO-Virgo-KAGRA (LVK) collaborations.

In the following section, we will briefly describe the main sources of gravitational waves and provide an overview of current and ongoing experiments searching for GW signals.

\subsection{Gravitational wave sources}
GW signals can be more broadly distinguished in two main categories. First, individual detections of resolved sources.
Those events are produced from the final stages of the orbit and subsequent coalescence of two compact objects. 

The late evolution of a compact binary system in its later stages can result in a transient signal of GWs, which typically unfolds in three distinct phases: inspiral, merger, and ringdown.

During the inspiral phase, the components of the binary system orbit each other, gradually drawing closer as they emit gravitational waves and lose energy. This phase is characterized by a distinctive signal known as a "chirp," marked by an increasing frequency and amplitude as the components approach each other and accelerate. A key parameter associated with this phase is the chirp mass $\mathcal{M}$, which can be expressed in terms of the component masses $m_1$ and $m_2$ as:
\begin{equation}
\mathcal{M}=\frac{\left(m_1 m_2\right)^{\frac{3}{5}}}{\left(m_1+m_2\right)^{\frac{1}{5}}}.
\end{equation}
Additionally, the chirp mass can be approximated in relation to the gravitational wave frequency $f$ and its time derivative $\dot{f}$ as:
\begin{equation}
\mathcal{M} \approx\left(\frac{5}{96} \pi^{-\frac{8}{3}} f^{-\frac{11}{3}} \dot{f}\right)^{\frac{3}{5}}.
\end{equation}
This mass parameter, directly linked to the gravitational wave frequency, serves as the most well-constrained mass scale for a merger event. As the compact objects approach each other, touching surfaces mark the onset of the merger phase. During this phase, the two components merge into a single remnant object swiftly, generating the most intense gravitational waves. Subsequently, in the ringdown phase, the remnant object sheds excess energy and settles into its final state.

Binary merger events involve compact objects such as black holes, neutron stars, or white dwarfs.. While mergers involving white dwarfs fall outside the frequency band of current detectors, binary mergers typically fall into three categories: binary black holes (BBHs), binary neutron stars (BNSs), and mixed neutron star-black hole mergers (NSBHs). Distinguishing between these categories often relies on a combination of the chirp mass and tidal deformability. A chirp mass $\mathcal{M} \lesssim 2 \mathrm{M}{\odot}$ suggests a BNS, $\mathcal{M} \gtrsim 4.5 \mathrm{M}{\odot}$ suggests a $\mathrm{BBH}$, and other events are categorized as NSBHs. Tidal deformability measures how much the shape of the compact object changes due to gravitational forces exerted by its companion. Black holes exhibit zero tidal deformability, while neutron stars deform shortly before the merger phase. Moreover detection of an electromagnetic counterpart also indicates the presence of at least one neutron star.

On the other hand, individual mergers falling below the detection threshold will in any case contribute to a Stochastic Gravitational Wave Background (SGWB), potentially observable as a superposition of many undetected signals. A SGWB could also be generated in the early Universe by mechanisms different from binary coalescences, as described in details in sec.\ref{sec:GW3}.

\subsection{Gravitational waves detectors}
Gravitational wave detectors represent a pinnacle of human ingenuity, enabling us to probe the fabric of spacetime and uncover some of the Universe's most profound mysteries. Over the years, these detectors have undergone remarkable advancements, paving the way for groundbreaking discoveries and revolutionizing our understanding of the cosmos.
The detection of GWs is often categorized as either direct or indirect.
A direct detector searches for the effect of the local passing of a GW, while an indirect detection is found by examining the properties of astrophysical objects that emit gravitational waves.

Several experiments are ongoing or have been proposed in order to scan a huge frequencies range for the detectable gravitational waves. These experiments cover from the nano-Hz, such as the PTA collaborations (NANOGrav~\cite{NANOGrav:2023gor, NANOGrav:2023hde}, EPTA\,\cite{EPTA:2023fyk, EPTA:2023sfo, EPTA:2023xxk}, PPTA\,\cite{Reardon:2023gzh, Zic:2023gta, Reardon:2023zen} and CPTA\,\cite{Xu:2023wog}) and SKA\,\cite{Zhao:2013bba}, passing through higher frequencies in the Hz-kHz for the LVK\,\cite{KAGRA:2021kbb,LIGOScientific:2014pky,LIGOScientific:2016fpe,LIGOScientific:2019vic}, or in the range mHz-kHz  for the proposed future experiments, such as LISA\,\cite{LISA:2022kgy}, BBO/DECIGO\,\cite{Yagi:2011wg,Yagi:2013awa} and ET\,\cite{Branchesi:2023mws}, up to ultra high frequencies range in the MHz-GHz with future several possibility (for a review see \cite{Aggarwal:2020olq}).

Here we describe the main features of several categories of GWs detector such as interferometry, pulsar timing array and future detectors\footnote{In the recent years also the idea to search for GWs in the MHz-GHz is taken into consideration (see for a recent review ref.\cite{Aggarwal:2020olq}) with the so called ultra-high frequencies detectors.Indeed there are no known astrophysical objects which are small and dense enough to emit at frequencies beyond $10$ kHz, Then any discovery of Gws at higher frequencies would indicate new physics, linked for example to exotic astrophysical objects such as evaporating PBHs.}.

\subsubsection{Pulsar timing arrays}
\begin{figure}
  \includegraphics[width=0.499\textwidth]{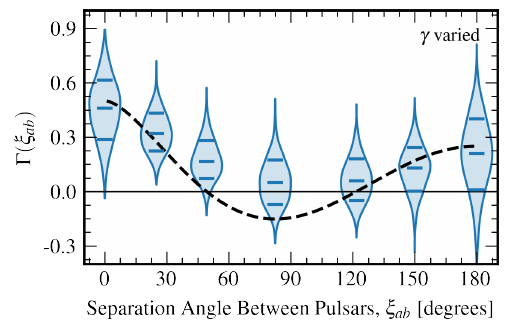}
  \includegraphics[width=0.499\textwidth]{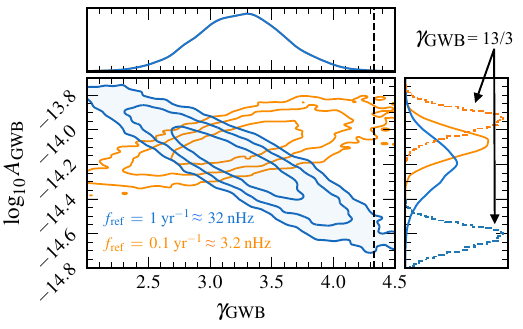}
  \includegraphics[width=0.999\textwidth]{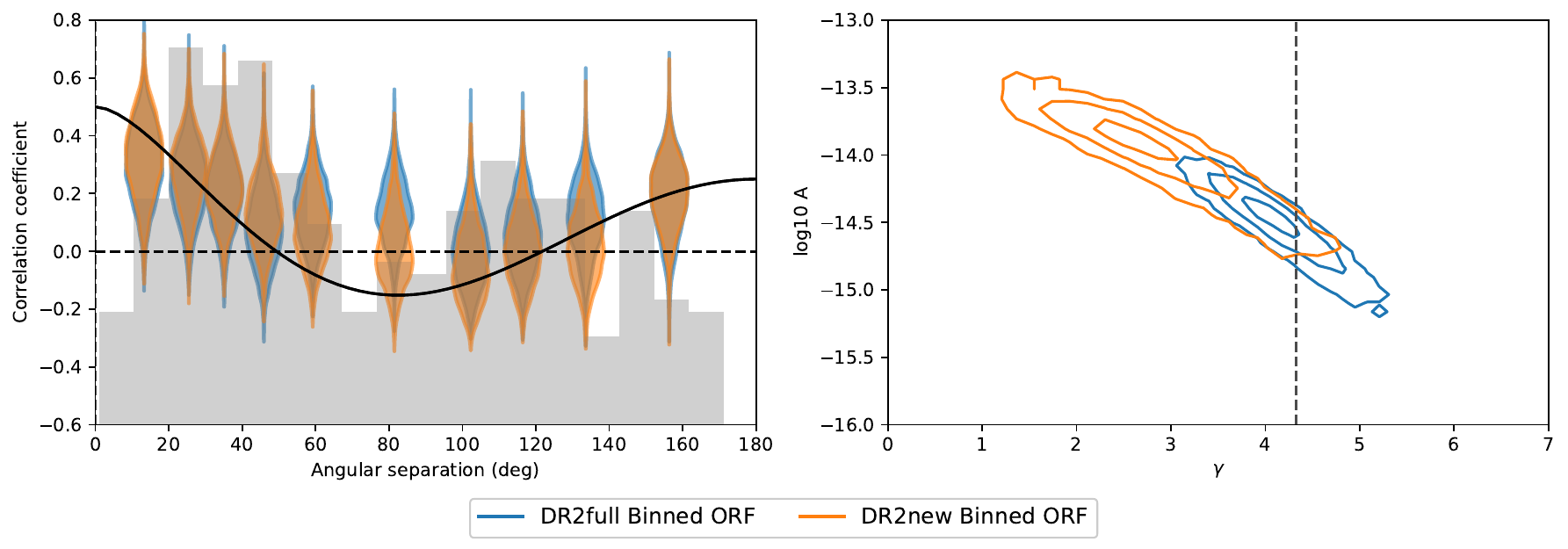}
  \caption{ \textit{\textbf{Top panel:}} Taken form ref.\cite{NANOGrav:2023gor}. The left panel is the angular-separation–binned inter-pulsar correlations, measured from 2,211 distinct pairings with the 67-pulsar array using the frequentist optimal statistic, assuming maximum-a-posteriori pulsar noise parameters and $\gamma=13/4$ common-process amplitude from a Bayesian inference analysis. The bin widths are chosen so that each includes approximately the same number of pulsar pairs, and central bin locations avoid zeros of the Hellings–Downs curve. The black line is the HD curve based on theoretical expectation of a GWB signal. The right panel is the corresponding 2D posterior for the amplitude and spectral index of the common correlated signal, showing 1/2/3 $\sigma$ contours.
  \textit{\textbf{Bottom Panel:}} Taken from ref.\,\cite{EPTA:2023fyk}. Binned overlap reduction function. Blue is for DR2full while orange is for DR2new. The left panel shows violins of the posterior of the correlation coefficients averaged at ten bins of angular separations with 30 pulsar pairs each.  The grey histogram is the arbitrarily normalised distribution of the number of pulsar pairs at different angular separations. The right plot is the same as for Nanograv but for the new dataset of EPTA.
  }
  \label{fig:HD}
\end{figure}

Pulsars are rapidly rotating compact objects (generally neutron stars) with strong magnetic fields that emits electromagnetic beams with a very precise period and hence can be used as precise clocks. If the Earth is in the path of these beams, pulsars are observable as a series of pulses of electromagnetic radiation at fixed time intervals. If a gravitational wave from another source passes between the pulsar and the Earth, the time between subsequent pulses will be modified slightly. Once the period of a pulsar is known, the passage of a GW in between the pulsar and the observer on the Earth can cause a variation of the time of arrival of a pulse\,\cite{Detweiler:1979wn}.This induces residuals in the pulse arrival time $t$ (with respect to the reference $t=0$ ) when compared to the pulsar timing model as
\begin{equation}
  R(t)=-\int_0^t \frac{\delta \nu}{\nu} d t  
\end{equation}
where $\nu$ is the pulse frequency, characteristic of each observed pulsar. In terms of the tensor perturbation $h_{i j}$ at the detector location $\vec{x}_{\mathrm{d}}$ and pulsar location $\vec{x}_{\mathrm{p}}$, the pulse frequency variation is given by
\begin{equation}
\frac{\delta \nu}{\nu}=-H^{i j}\left[h_{i j}\left(t, \vec{x}_{\mathrm{d}}\right)-h_{i j}\left(t-D, \vec{x}_{\mathrm{p}}\right)\right],
\end{equation}
where $H^{i j}$ is a geometrical factor depending on the propagation direction of the GWs relative to the direction of the pulsar at a distance $D$. It was shown by Hellings and Downs\,\cite{Hellings:1983fr} that timing residuals induced by GW signals for pulsars separated by an angular distance $\xi$ follows the correlation pattern
\begin{equation}
\Gamma\left(\xi_{a b}\right)  =\frac{3}{2} x \ln (x)-\frac{1}{4} x+\frac{1}{2}+\frac{1}{2} \delta_{a b},
\end{equation}
where
\begin{equation}
x  =\frac{1-\cos \xi_{a b}}{2} .
\end{equation}
This is known as the Hellings and Downs (HD) curve. The recent PTA data release by the NANOGrav~\cite{NANOGrav:2023gor, NANOGrav:2023hde} shows evidence of a Hellings-Downs pattern in the angular correlations as reported in the top panel of fig.\ref{fig:HD}. On the bottom panel of fig.\ref{fig:HD} there is the HD reported by the EPTA collaborations\,\cite{EPTA:2023fyk}, which in the case of the DR2F full dataset is much less evident than the one of NANOGrav.
Moreover the potential GW signal can be fitted by a power-law $\Omega_{\mathrm{GW}} \propto f^{\gamma-4}$, and as we can see the recent dataset disfavoured the SMBH case without environmental effect where the usual power law is $f^{2/3}$\,\cite{Sesana:2004sp}.
\subsubsection{Interferometry and the LIGO–Virgo-Kagra collaboration}
The concept behind an interferometer is to split light with a well-defined wavelength into two beams that traverse different paths, subsequently measuring the phase shift caused by their differing optical path lengths (OPLs) upon recombination. Typically, an interferometer comprises two arms perpendicular to each other, featuring a beam splitter at their intersection and mirrors at their ends. This setup is also sensitive to changes in arm length, such as those induced by a passing gravitational wave.

To generate an observable signal due to the change in arm length, the OPL must be approximately one quarter of the wavelength of the gravitational wave, necessitating extremely long arms (e.g., approximately 750 km for binary mergers of stellar-mass compact objects). However, constructing arms of such length is impractical. Therefore, the light is cycled up and down the arms multiple times to achieve the required OPL.

The most renowned interferometer detector is the Laser Interferometer Gravitational-wave Observatory (LIGO) in the USA, established in 1988, with construction starting in 1994\,\cite{Abramovici:1992ah}.It comprises two detectors, located in Hanford, Washington, and Livingston, Louisiana, each with $4$ km arms. With a separation of approximately $3000$ km, these detectors exhibit a significant time difference in gravitational wave arrival, aiding in distinguishing astrophysical signals from terrestrial noise\,\cite{LIGOScientific:2016aoc}. The localisation of the source can be improved with the addition of more detectors across a large area.
The first of these additional detectors is the Virgo detector\,\cite{Giazotto:1988gw}, located at Santo Stefano a Macerata, near the city of Pisa, in Italy, and consisting of $2 \mathrm{~km}$ arms. The second is the KAGRA detector\,\cite{KAGRA:2018plz}, located in the Kamioka mine in Japan, and has $3 \mathrm{~km}$ arms. A further detector known as LIGOIndia is planned to be set up in Aundha Nagnath, India, moved from the LIGO Hanford site where there are currently two identical detectors.
These collaborations goes under the name of LIGO-Virgo-Kagra (LVK).

The first detection of GWs by LIGO and Virgo in 2015~\cite{LIGOScientific:2016aoc} has allowed for a new independent way to study BHs and the growing number of confirmed events~\cite{LIGOScientific:2018mvr, LIGOScientific:2020ibl, KAGRA:2021vkt} is being used to infer their population~\cite{KAGRA:2021duu}. To explain the origin of the observed events, two conceptually different scenarios can be considered: astrophysical and primordial. Although the existence of astrophysical black holes is beyond doubt, the characteristics of their population are not well understood. These form binaries either in isolation or by dynamical capture in dense stellar environments~\cite{Mandel:2018hfr, Woosley:2002zz} and, depending on the formation channel, different astrophysical binary populations can be expected.
Several works have explored the use of GW data to find direct or indirect evidence of PBHs. Specifically targeted searches of subsolar mass compact objects, which would provide a smoking gun evidence of the existence of PBHs, have so far been unsuccessful~\cite{LIGOScientific:2018glc,LIGOScientific:2019kan,Nitz:2020bdb,Nitz:2021mzz,Nitz:2021vqh,Nitz:2022ltl,Miller:2020kmv,Miller:2021knj,Andres-Carcasona:2022prl}. Additionally, studies of the LIGO-Virgo GW data including both PBH and astrophysical populations have not found enough statistical significance to claim the existence of PBHs~\cite{Hutsi:2020sol, Hall:2020daa, Wong:2020yig, Franciolini:2021tla, DeLuca:2021wjr}. However, some of the component masses, in particular, in GW190521~\cite{LIGOScientific:2020iuh} and GW230529$\_$181500~\cite{LIGOScientific:2024elc}, fall in regions where astrophysical models do not predict them, potentially suggesting for a PBH population. Another method of studying the possible PBH population uses the stochastic GW background and the lack of its detection has also been recast as an upper limit to the PBH abundance~\cite{Raidal:2017mfl, Raidal:2018bbj, Vaskonen:2019jpv, Hutsi:2020sol, Romero-Rodriguez:2021aws, Franciolini:2022tfm}. 
\subsubsection{Future detectors}
If the current state of gravitational wave detectors is already exciting, the future appears even brighter, with a multitude of ground- and space-based detectors utilizing interferometric and PTA techniques proposed or planned for the next few decades.

Two forthcoming ground-based interferometer detectors are the Einstein Telescope (ET)\,\cite{Punturo:2010zz, Hild:2010id} and the LIGO Cosmic Explorer (CE)\,\cite{Reitze:2019iox}, both aiming to commence observations in the 2030s at frequencies similar to the current LIGO band. The ET will feature $10$ km arms arranged in a triangular configuration, while the CE will boast $40$ km arms in the traditional L-shape. These detectors will be sensitive to the same signals as current interferometer detectors like LIGO but with substantially improved sensitivity. A further ground-based detector is the Square Kilometer Array (SKA)\,\cite{Weltman:2018zrl}, a large radio telescope project with various scientific objectives, including gravitational wave detection using the PTA method.

Space-based detection offers several advantages for gravitational wave detection, notably the elimination of many terrestrial noise sources. Several space-based detectors have been proposed, all employing interferometry, with the Laser Interferometer Space Antenna (LISA) being the most prominent\,\cite{Bartolo:2016ami, Caprini:2019pxz, LISACosmologyWorkingGroup:2022jok}. LISA, funded and scheduled for launch within the next decade, comprises three arms arranged in an equilateral triangle, with an approximate arm length of $2.5$ million kilometers. LISA will primarily investigate the mergers of compact binaries at massive and supermassive scales, as well as the stochastic gravitational background. Other proposed interferometric space-based detectors include the Japanese DECIGO\,\cite{Kawamura:2019jqt} and the Chinese TianQin\,\cite{TianQin:2015yph,TianQin:2020hid}.

Another idea is to use the atomic interferometers based on ultracold atom technology, where ultracold atoms or molecules act as matter waves in order to test a similar frequencies range of LISA, both in space with AEDGE\,\cite{AEDGE:2019nxb,Badurina:2021rgt} than with a network of inteferometers on Earth with AION-km\,\cite{Badurina:2019hst,Badurina:2021rgt}.

These future detectors hold the potential to confirm or refute the existence of PBHs across  the entire un-evaporated mass range.

\part{Primordial Black Holes: 
\\Abundance and signatures}
\newcommand{\g}{{\rm G}}
\newcommand{\nn}{\nonumber}
\newcommand{\be}{\begin{equation}}  
\newcommand{\ee}{\end{equation}} 
\newcommand{\x}{{\bf x}} 
\newcommand{\dd}{{\rm d}}
\newcommand{\msun}{M_{\odot}}
\newcommand{\MPl}{\bar{M}_{\textrm{\tiny{Pl}}}}

\def\gw{gravitational wave\xspace}
\def\gwh{gravitational-wave\xspace}
\def\gws{gravitational waves\xspace}
\def\GW{Gravitational Wave\xspace}
\def\GWs{Gravitational Waves\xspace}
\def\cw{continuous wave\xspace}
\def\cws{continuous waves\xspace}
\def\cwh{continuous-wave\xspace}

\newcommand{\bea}{\begin{equation}\begin{aligned}} 
\newcommand{\eea}{\end{aligned}\end{equation}}

\newcommand{\ogw}{\tilde{\omega}}
\newcommand{\lgw}{\tilde{l}}
\newcommand{\mgw}{\tilde{m}}
\newcommand{\fgw}{f_{\rm gw}}
\newcommand{\mbh}{M_{\rm BH}}
\newcommand{\mb}{M_{b}}
\newcommand{\dotf}{\dot{f}_{\rm gw}}

\newcommand{\deff}{D_{\rm{eff}}}
\newcommand{\dlumi}{D_{\rm{L}}}
\newcommand{\cmass}{\mathcal{M}_{\rm{c}}}
\newcommand{\rr}{\mathcal{R}_{\rm{90}}}
\newcommand{\fpbh}{f_{\rm PBH}}
\newcommand{\meanm}{\langle m \rangle }
\newcommand{\varm}{\langle m^2 \rangle - \langle m \rangle^2 }
\newcommand{\td}{{\rm d}}
\newcommand{\Ra}{R_{\mathrm{ABH}}}
\newcommand{\mmin}{m_{\mathrm{min}}}
\newcommand{\mmax}{m_{\mathrm{max}}}

\newcommand{\Ndet}{N_{\mathrm{det}}}
\newcommand{\Nobs}{N_{\mathrm{obs}}}
\chapter{Computation of primordial black hole abundance}\label{cap:NGS}
\thispagestyle{plain}
The precise computation of the PBH abundance at formation is an extraordinary challenge. Not only does the methodology for determining their abundance undergo radical changes based on the formation mechanism, but even when restricting ourselves to one scenario, such as in our case the standard scenario for PBH formation. This scenario hypothesizes that PBHs emerge from the gravitational collapse of significant over-densities in the primordial density contrast field \cite{Ivanov:1994pa, GarciaBellido:1996qt, Ivanov:1997ia, Blinnikov:2016bxu}.
The primordial density contrast field being a random field, the computation of the abundance adopts a statistical nature, necessitating precise knowledge of the Probability Density Function (PDF) of density fluctuations. Since PBH formation constitutes a rare event, attention is drawn to the tail of the PDF where minor deviations from the Gaussian assumption generically yield exponential effects. Thus, understanding the role of Non-Gaussianities (NG) in the context of PBH formation proves crucial. 
Consequently, computing the PBH abundance becomes highly non-trivial.

In this chapter, after a detailed discussion of the NG nature of the curvature perturbation field in Section \ref{sec:C1pre}, we describe two techniques for computing the abundance in the presence of NGs: in Section \ref{sec:C1A}, building upon ref.\,\cite{Ferrante:2022mui}, we introduce a formalism based on threshold statistics applied to the compaction function. Concurrently, in Section \ref{sec:C1B}, we discuss the limitations of employing the average profile for compaction in abundance computations, as highly used in the literature and in ref.\cite{Ferrante:2022mui}. As a consequence we introduce a new formalism, based on ref.\,\cite{Ianniccari:2024bkh}, that goes beyond the use of average profiles of compaction.
We conclude describing some hints on how it possible to construct an observable whose critical threshold does not depend at all on the profiles of the peaks.

\section{The role of Non-gaussianities in the PBH abundance}\label{sec:C1pre}
During the recent years, it has become increasingly clear that limiting the analysis to the assumption that the PDF of density fluctuations follows the Gaussian statistics is theoretically flawed. 
The reason is twofold. 
First, density fluctuations in the primordial radiation field originate from curvature perturbations, previously stretched on super-horizon scales during inflation, after their horizon re-entry. 
In the long-wavelength approximation, the equation which relates curvature perturbations to density fluctuations is intrinsically non-linear\,\cite{Harada:2015yda}. For this reason, even under the \emph{assumption} that curvature perturbations follow exact Gaussian statistics (which is in general not true, as we shall discuss next), density fluctuations inherit an unavoidable amount of non-gaussianity from non-linear (NL) corrections\,\cite{DeLuca:2019qsy,Young:2019yug,Germani:2019zez}. In some points we refer to this kind of NGs as \textit{non-linearities}.
Second, as anticipated, curvature perturbations do not generically follow a Gaussian statistics.

Let us elaborate on the second point by considering two different scenarios\footnote{A deeper description of these kind of models is presented in Chapter\,\ref{cap:Models}.}. 
In single-field inflationary models, 
NG corrections are typically subdominant at scales relevant for CMB observations because suppressed by the tiny values of the slow-roll parameters; however, the formation of PBHs requires a strong violation of the slow-roll paradigm\,\cite{Motohashi:2017kbs}, usually achieved by the presence of an ultra slow-roll (USR) phase that boosts the amplitude of the power spectrum of curvature perturbations at scales relevant for PBH formation\,\cite{Inomata:2016rbd,Garcia-Bellido:2017mdw,Ballesteros:2017fsr,Hertzberg:2017dkh,Kannike:2017bxn,Dalianis:2018frf,Inomata:2018cht,Cheong:2019vzl,Ballesteros:2020qam,Iacconi:2021ltm,Kawai:2021edk}. 
This implies that at such scales, the suppression of NGs, due to the violation of the slow-roll condition, is no longer guaranteed.
Alternatively, one can envision a two-field scenario in which, in addition to the inflaton driving inflation and generating curvature perturbations at CMB scales, there exists a spectator field, dubbed curvaton\,\cite{Lyth:2002my}, which is responsible, after its decay into radiation, for the generation of curvature perturbations at scales relevant for PBH formation. 
In this case, it happens again that NG corrections are relevant. As a generic result, the smaller the fraction of total energy density in the curvaton field at the time of its decay, the larger NG becomes\,\cite{Malik:2002jb}.

The local NG of the curvature perturbation $\zeta$ is tipically parameterized by the power series expansion \cite{Bugaev:2013vba,Nakama:2016gzw,Byrnes:2012yx,Young:2013oia,Yoo:2018kvb,Kawasaki:2019mbl,Yoo2,
Riccardi:2021rlf,Taoso:2021uvl,Meng:2022ixx,Escriva:2022pnz}
\begin{align}\label{eq:FirstExpansion}
\zeta = \zeta_{\rm G} + \frac{3}{5}f_{\rm NL}\zeta_{\rm G}^2
 + \frac{9}{25}g_{\rm NL}\zeta_{\rm G}^3 + \dots\,,
\end{align}
where $\zeta_{\rm G}$ obeys the Gaussian statistics while the parameters $f_{\rm NL}$, $g_{\rm NL}$, $\dots$ (which,
in full generality, depend on the scale of the perturbation) encode deviations from the Gaussian limit. 
Interestingly, in both models discussed before (that is single-field inflation with an USR phase\,\cite{Atal:2019cdz,Tomberg:2023kli} and two-field curvaton models \cite{Sasaki:2006kq,Pi:2021dft}) it is possible to show, by means of the so-called $\delta N$ formalism, that in the previous expansion the parameters $f_{\rm NL}$, $g_{\rm NL}$, $\dots$ are not independent and actually give rise to a closed-form resummed expression for $\zeta$ of the schematic form  
\begin{align}
    \zeta(\vec{x}) = F(\zeta_{\rm G}(\vec{x}))\,,\label{eq:MainF}
\end{align}
for some appropriate function $F(\zeta_{\rm G})$ of the Gaussian component.
In general $F$ is a generic non-linear function of the Gaussian component $\zeta_{\rm G}$ and the attribute of locality refers to the fact that the value of $\zeta$ at the point $\vec{x}$ is fully determined by the value of $\zeta_{\rm G}$ at the same spatial point.\footnote{In other words, the function $F$ in eq.\,(\ref{eq:MainF}) must be independent on spatial derivatives of $\zeta_{\rm G}$ since the latter would make the value of $\zeta$ at $\vec{x}$ determined  by the values of  $\zeta_{\rm G}$ in a certain neighborhood of $\vec{x}$. 
} Whenever unnecessary, we will drop the explicit functional dependence from the spatial coordinates in $\zeta$ and $\zeta_{\rm G}$. 

In the course of this chapter we will describe the main results of our analysis using the generic functional form $\zeta = F(\zeta_{\rm G})$. 
However, in order to present and discuss explicit examples of phenomenological relevance, we will apply our formulas to the physics-case of the USR scenario and curvaton field.
Let us, therefore, explore in more detail these two specific theoretical set-up.

\subsection{Primordial NGs in USR and curvaton models}\label{subsec:NGs}
When presenting results inspired by the USR model, we will focus on primordial NGs with the following functional form\,\cite{Atal:2019cdz,Tomberg:2023kli}
\footnote{For $\zeta_{\rm G}> \mu_\star$, Eq.\,\ref{eq:USR} does not capture the possibility of PBHs formed by bubbles of trapped vacuum which requires a separate discussion \cite{Escriva:2023uko,Uehara:2024yyp}.}

\begin{align}
\label{eq:USR}
\zeta({\bf x})=-\mu_\star\ln\left(1-\frac{ \zeta_{\rm G}({\bf x})}{\mu_\star}\right),
\end{align}
with $\mu_\star$ a model-dependent parameter depending upon the transition between the ultra-slow-roll phase and the subsequent slow-roll phase. In general $\mu_\star$ is proportional to the slope at large $k$ of the power spectrum but in the analysis presented in this chapter we take $\mu_\star$ as a free parameter.
Moreover, one can also easily see that, by expanding Eq.\,\ref{eq:USR} at second order, $f_{\rm NL} = \frac{5}{6\mu_\star}$.
\\When presenting results inspired by the curvaton model, we will focus on primordial NG with the following functional form\,\cite{Sasaki:2006kq,Pi:2021dft}
\begin{align}\label{eq:MasterX}
\zeta = \log\big[X(r_{\rm dec},\zeta_{\rm G})\big]\,,
\end{align}
with
\begin{align}\label{eq:XFunction}
X(r_{\rm dec},\zeta_{\rm G}) \equiv& \frac{1}{\sqrt{2 (3+r_{\rm dec})^{1/3}}}
\Bigg\{
\sqrt{
\frac{
-3 + r_{\rm dec}(2+r_{\rm dec}) + [(3+r_{\rm dec})P(r_{\rm dec},\zeta_{\rm G})]^{2/3}
}{
(3+r_{\rm dec})P^{1/3}(r_{\rm dec},\zeta_{\rm G})
}
}  
\nn \\
+ &  
\sqrt{
\frac{(1-r_{\rm dec})}
{P^{1/3}(r_{\rm dec},\zeta_{\rm G})} -
\frac{P^{1/3}(r_{\rm dec},\zeta_{\rm G})}
{(3+r_{\rm dec})^{1/3}} 
+
\frac{
(2r_{\rm dec} + 3\zeta_G)^{2}
P^{1/6}(r_{\rm dec},\zeta_{\rm G})
}{r_{\rm dec}
\sqrt{-3 + r_{\rm dec}(2+r_{\rm dec}) +
[(3+r_{\rm dec})P(r_{\rm dec},\zeta_{\rm G})]^{2/3}
}}} 
\Bigg\}\,,
\end{align}
and 
\begin{align}
P(r_{\rm dec},\zeta_{\rm G}) \equiv 
\frac{(2r_{\rm dec} + 3\zeta_{\rm G})^{4}}{16 r_{\rm dec}^2} + \sqrt{
(1-r_{\rm dec})^3(3+r_{\rm dec}) + \frac{(2r_{\rm dec} + 3\zeta_{\rm G})^8}{256 r_{\rm dec}^4}
}\,.\label{eq:PFunction}
\end{align}
This functional form characterizes 
NG of the primordial curvature  perturbation in the curvaton model.
The parameter $r_{\rm dec}$ is the weighted fraction of the curvaton energy density  $\rho_{\phi}$ to the total energy density at the time of curvaton decay, defined by
\begin{equation}
r_{\rm dec} \equiv \left.\frac{3 \rho_{\phi}}{3 \rho_{\phi}+4 \rho_{\gamma}}\right|_{\rm curvaton\,\,decay}\,,
\end{equation}
where $\rho_{\gamma}$ is the energy density stored in radiation after reheating.
In eq.\,(\ref{eq:MasterX}) the gaussian random field $\zeta_{\rm G}$  
corresponds to $\zeta$ in linear approximation.
We remark that eq.\,(\ref{eq:MasterX}) is strictly valid under the approximation that the curvaton potential is quadratic (which implies the absence of any non-linear evolution of the curvaton field between the Hubble exit and the start of curvaton oscillation, see refs.\,\cite{Enqvist_2010,Fonseca:2011aa}).
\\The above expressions simplify in the limit case $r_{\rm dec}=1$ (that is the case in which the curvaton dominates the energy density of the Universe at the time of its decay) that gives
\begin{align}\label{eq:MasterX2}
\zeta_{\rm log} \equiv \log\big[X(1,\zeta_{\rm G})\big] = \frac{2}{3}\log\bigg(
1+ \frac{3}{2}\zeta_{\rm G}
\bigg)\,.
\end{align}
We note that, if we approximate eq.\,(\ref{eq:MasterX}) at the quadratic order in the gaussian field $\zeta_{\rm G}$, 
we find that
\begin{align}\label{eq:ZetaQ}
\zeta_{2} \equiv \zeta_{\rm G} + \frac{3}{5}f_{\rm NL}(r_{\rm dec})\zeta_{\rm G}^2\,,
~~{\rm with}~~
f_{\rm NL}(r_{\rm dec}) \equiv  \frac{5}{3}\bigg(
\frac{3}{4r_{\rm dec}} - 1 - \frac{r_{\rm dec}}{2}
\bigg)\,.
\end{align}
Similarly, at cubic order, we find
\begin{align}\label{eq:ZetaC}
\zeta_{3} = 
\zeta_{\rm G}  + \frac{3}{5}f_{\rm NL}(r_{\rm dec})\zeta_{\rm G}^2
 + \frac{9}{25}g_{\rm NL}(r_{\rm dec})\zeta_{\rm G}^3\,,
 ~~{\rm with}~~
g_{\rm NL}(r_{\rm dec}) \equiv \frac{25}{54}\left(
-\frac{9}{r_{\rm dec}} + \frac{1}{2} + 10r_{\rm dec} + 3r_{\rm dec}^2
\right)\,.
\end{align}
We refer to ref.\,\cite{Sasaki:2006kq} for a comprehensive derivation of eqs.\,(\ref{eq:MasterX},\,\ref{eq:XFunction},\,\ref{eq:ZetaQ},\,\ref{eq:ZetaC}).

\begin{figure}[!h!]
\begin{center}
\includegraphics[width=1.03\textwidth]{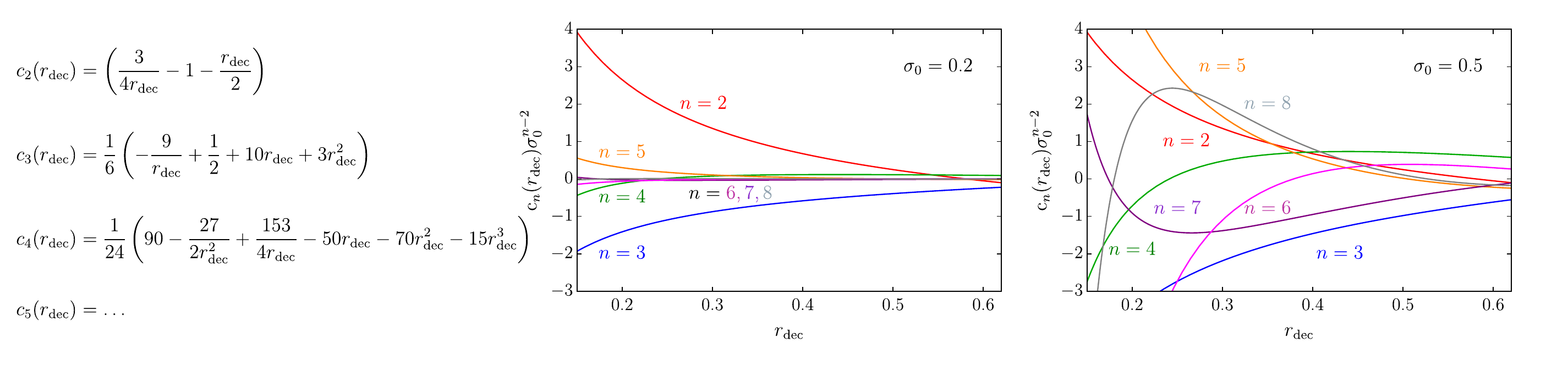}
\caption{
First few coefficients of the expansion in eq.\,(\ref{eq:ZetaSeries}) as function of $r_{\rm dec}$ (both analytically and numerically). 
In the figures we plot each coefficient $c_n(r_{\rm dec})$ rescaled by the appropriate power $\sigma_0^{n-2}$ in order to give a more realistic comparison of their relative size.
 }\label{fig:CnCoeff}
\end{center}
\end{figure}

The coefficients $f_{\rm NL}(r_{\rm dec})$, $g_{\rm NL}(r_{\rm dec})$, $\dots$ seem to diverge in the limit $r_{\rm dec}\to 0$ (meaning that the smaller the fraction of total energy density in the curvaton field at the time of its decay, 
the larger NG becomes). 
It is important for the rest of the analysis to have clear in mind the actual meaning of this apparent divergence.
Consider the simplest example of curvaton field, that is a massive real scalar field $\phi$ with quadratic potential $V(\phi) = m^2\phi^2/2$.
After the end of inflation the curvaton field  remains approximately constant until the Hubble rate becomes comparable to $m_{\phi}^2$; at this time, the curvaton starts oscillating around the minimum of its potential. The energy density of the oscillating curvaton is 
$\rho_{\phi} = m_{\phi}^2\phi^2$ and can be expanded into a background term $\bar{\rho}_{\phi}$ and 
a perturbation $\delta\rho_{\phi}$ according to
 $\rho_{\phi} = m_{\phi}^2(\bar{\phi}+\delta\phi)^2 = 
m_{\phi}^2\bar{\phi}^2 + m_{\phi}^2(2\bar{\phi}\delta\phi + \delta\phi^2) 
\equiv \bar{\rho}_{\phi} + \delta\rho_{\phi}$. 
Consequently, in the spatially flat gauge, the curvature perturbation associated with the curvaton field takes the form
\begin{align}
\zeta_{\phi} = \frac{2}{3}\frac{\delta\phi}{\bar{\phi}} + 
\frac{1}{3}\left(\frac{\delta\phi}{\bar{\phi}}\right)^2\,,
\end{align}
and it consists of a linear plus a quadratic (hence NG) term. 
The key point is that, after the end of inflation, the {\it total} curvature perturbation is the weighted sum (evaluated at the time of curvaton decay into radiation according to the so-called sudden decay approximation)
\begin{align}
\zeta = \frac{\dot{\rho}_{\gamma}}{\dot{\rho}}\zeta_{\gamma} + 
\frac{\dot{\rho}_{\phi}}{\dot{\rho}}\zeta_{\phi} = \underbrace{\frac{2r_{\rm dec}}{3}\frac{\delta\phi}{\bar{\phi}}}_{\equiv\,\,\zeta_{\rm G}}
+ \frac{3}{4r_{\rm dec}}\left(\frac{2r_{\rm dec}}{3}\frac{\delta\phi}{\bar{\phi}}\right)^2 = 
\zeta_{\rm G} + \frac{3}{4r_{\rm dec}}\zeta_{\rm G}^2\,,\label{eq:ZetaPhi}
\end{align}
where, importantly, we assume  that the curvature perturbation in radiation produced at the end
of inflation is negligible, $\zeta_{\gamma} \simeq 0$.\footnote{More in details, we assume $\zeta_{\gamma} \simeq 0$ at scales relevant for PBH formation. We tacitly work under the assumption that there exists 
some inflationary dynamics at the origin
of the curvature perturbation responsible, at much larger length-scales, for the generation of CMB
anisotropies.}  This means that the curvaton contribution to $\zeta$ is not just a  correction on the top of some leading (gaussian) term but it entirely defines $\zeta$.
Formally, in the limit $r_{\rm dec}\to 0$,  eq.\,(\ref{eq:ZetaPhi}) implies that 
$\zeta\to 0$ (so that we do not have any divergence). However, what eq.\,(\ref{eq:ZetaPhi}) is telling us is that, once we define the linear term to be 
$\zeta_{\rm G} \equiv 2/3\,r_{\rm dec}\delta\phi/\bar{\phi}$ and we fix it to some reference value, for decreasing $r_{\rm dec}$ the quadratic correction takes over the linear one, meaning that the level of NG of $\zeta$ increases. 
The above argument only captures the leading quadratic correction in the limit of small $r_{\rm dec}$ (cf. ref.\,\,\cite{Sasaki:2006kq} for the full non-linear relation between $\zeta_{\phi}$ and $\zeta$ that gives rise to the functional form in eq.\,(\ref{eq:MasterX})) but it is sufficient to clarify the meaning of the $r_{\rm dec}\to 0$ limit. More generally, one can setup the following power-series expansion 
\begin{align}\label{eq:ZetaSeries}
\zeta_N = \sum_{n = 1}^{N} c_n(r_{\rm dec}) \zeta_{\rm G}^{n}\,,~~~~~~~~
{\rm with}~~~~c_1(r_{\rm dec}) = 1\,.
\end{align}
We were not able to find a closed recursive expression for the generic coefficient $c_n(r_{\rm dec})$ but
it is not difficult to extract it (analytically or numerically) at any finite order $N$. 
In fig.\,\ref{fig:CnCoeff} we show the first few coefficients of this expansion as function of the parameter $r_{\rm dec}$. 
In order to give a fair idea of their relative size in the expansion $\zeta_N$, we plot the quantity $c_n(r_{\rm dec})\sigma_0^{n-2}$ (assuming $\zeta_{\rm G} = O(\sigma_0))$.\footnote{In other words, we rewrite the expansion in eq.\,(\ref{eq:ZetaSeries}) in the form
\begin{align}
\zeta_N = \sum_{n = 1}^{N} c_n(r_{\rm dec}) \zeta_{\rm G}^{n} 
= \zeta_{\rm G} + 
\sigma_0^2\left[
c_2(r_{\rm dec})\left(\frac{\zeta_{\rm G}}{\sigma_0}\right)^2 +
c_3(r_{\rm dec})\sigma_0\left(\frac{\zeta_{\rm G}}{\sigma_0}\right)^3 + 
\dots + c_n(r_{\rm dec})\sigma_0^{n-2}\left(\frac{\zeta_{\rm G}}{\sigma_0}\right)^n 
+\dots
\right]\,.
\end{align}
If we estimate $\zeta_{\rm G}$ with the variance of its PDF, $\zeta_{\rm G} = O(\sigma_0)$, the previous expansion is controlled by the coefficients $c_n(r)\sigma_0^{n-2}$.}
From this simple plot, we already get the idea that if $\sigma_0 \ll 1$ then the truncated expansion in $\zeta_N$  seems an appropriate approximation, and that only the first few orders are enough to capture the full result (cf. fig.\,\ref{fig:CnCoeff} with $\sigma_0 = 0.2$). Furthermore, we also note that the terms with $n=2$ and $n=3$ (that are, respectively, $f_{\rm NL}$ and $g_{\rm NL}$ in the commonly used language) contribute to $\zeta_N$ with opposite signs (and are comparable in absolute value); this suggests that truncating the expansion at the quadratic order without including also the cubic term may lead to incorrect results.
Finally, the last remark is that, as anticipated in the introduction, the condition $\sigma_0 \ll 1$ will crucially depend on the assumption about the power spectrum of the Gaussian variable $\zeta_{\rm G}$. 
We will come back shortly on this point. We anticipate that in the case of a broad power spectrum, the condition $\sigma_0 \ll 1$ will be badly violated. The previous discussion, therefore, suggests that in such a case the perturbative analysis based on $\zeta_{\rm G}$ will not  be applicable. 
The right-most panel of  fig.\,\ref{fig:CnCoeff} shows that if one takes $\sigma_0 = 0.5$ then any hierarchy between the various $c_n(r_{\rm dec})$ coefficients gets already lost.

The statistics of $\zeta$ are easy to get. 
We can obtain the PDF of the NG variable $\zeta$ starting from the gaussian PDF of $\zeta_{\rm G}$ (which we shall denote in the following as 
$\textrm{P}_{\rm G}$). Conservation of probability gives
\begin{align}
\textrm{P}(\zeta,r_{\rm dec}) = 
\textrm{P}_{\rm G}\big[
\zeta_{\rm G}^{(+)}(\zeta,r_{\rm dec})
\big]\left|
\frac{d\zeta_{\rm G}^{(+)}}{d\zeta}
\right| + 
\textrm{P}_{\rm G}\big[
\zeta_{\rm G}^{(-)}(\zeta,r_{\rm dec})
\big]\left|
\frac{d\zeta_{\rm G}^{(-)}}{d\zeta}
\right|\,,\label{eq:ExactPDF}
\end{align}
with 
\begin{align}\label{eq:ZetaRoots}
\zeta_{\rm G}^{(\pm)}(\zeta,r_{\rm dec}) & = 
\frac{2r_{\rm dec}}{3}\bigg[
-1 \pm \sqrt{
\bigg(\frac{3+r_{\rm dec}}{4r_{\rm dec}}\bigg)e^{3\zeta} +
\bigg(\frac{3r_{\rm dec}-3}{4r_{\rm dec}}\bigg)e^{-\zeta} 
}
\bigg]\,.
\end{align}
As a consistency check, integrating the PDF we find the total probability 
\begin{align}
\int_{\zeta_{\rm min}(r_{\rm dec})}^{\infty}\textrm{P}(\zeta,r_{\rm dec})d\zeta = 1,\,~~~~~~~~~~~
{\rm with}~~~ \zeta_{\rm min}(r_{\rm dec}) \equiv 
\frac{1}{4}\log\bigg(
\frac{3-3r_{\rm dec}}{3+r_{\rm dec}}
\bigg)\,.
\end{align}
Mathematically, the condition $\zeta \geqslant \zeta_{\rm min}(r_{\rm dec})$ follows from requiring eq.\,(\ref{eq:ZetaRoots}) to be real.

\subsection{The distribution of curvature perturbations}\label{sec:IntroStat}
The statistics of the NG variable $\zeta$ is, therefore,  completely determined by the statistics of $\zeta_{\rm G}$. Being the latter a gaussian random field, we only need to specify its variance $\sigma_0$ (assuming vanishing mean value) according to
\begin{align}
\textrm{P}_{\rm G}(\zeta_{\rm G}) = \frac{1}{\sqrt{2\pi}\sigma_0}\exp\bigg[
-\frac{\zeta_{\rm G}^2}{2\sigma_0^2}
\bigg]\,.
\end{align}  
The problem we face is that we are not interested in the statistics of $\zeta$ but rather in the statistics of 
the so-called density contrast field\,\cite{Harada:2015yda}
\begin{align}\label{eq:NonLinearDelta}
\delta(\vec{x},t) =  
-
\frac{2}{3}\Phi
\left(
\frac{1}{aH}
\right)^2 
e^{-2 \zeta(\vec{x})}
\bigg[
\nabla^2\zeta(\vec{x}) + \frac{1}{2} \partial_{i}\zeta(\vec{x})
 \partial_{i}\zeta(\vec{x})
\bigg]\,,~~~~{\rm with}~~~~\zeta(\vec{x}) = \log\big\{X[r_{\rm dec},\zeta_{\rm G}(\vec{x})]\big\}\,.
\end{align}
where we highlighted the spatial dependence of the random field $\zeta(\vec{x})$ and $\Phi$ captures the 
dependence of the relation between curvature and density contrast on the equation of state.
Written in terms of a constant equation of state parameter $\omega = p/\rho$ (with $\omega =1/3$ for a radiation-dominated Universe), one finds
\begin{align}
 \Phi \equiv \frac{3(1+\omega)}{5+3\omega}\,,
 \label{eq:DefPhi}
\end{align}
and  $\Phi = 2/3$ in a radiation fluid.

The time dependence in eq.\,(\ref{eq:NonLinearDelta}) comes from the scale factor $a=a(t)$ and the Hubble rate $H=H(t)$. 
The density contrast can be defined more precisely as $\delta(\vec{x},t) \equiv \delta\rho(\vec{x},t)/\rho_b(t)$ where 
$\rho_b(t)$ is the average background radiation energy density
$\rho_b(t) = 3H(t)^2/8\pi$
and $\delta\rho(\vec{x},t) \equiv \rho(\vec{x},t)- \rho_b(t)$ its perturbation. 
We are interested in the evaluation of eq.\,(\ref{eq:NonLinearDelta}) at $t = t_H$, that is the time when curvature perturbations re-enter the horizon. 

Eq.\,(\ref{eq:NonLinearDelta}) tells us two important things.  
First, information about first and second spatial derivatives of $\zeta(\vec{x})$ -- which, in turn, are also random variables -- is relevant to determine the spatial distribution of $\delta(\vec{x},t)$.
Second, additional NGs are present because of the non-linear relation between $\zeta$ and $\delta$\,\cite{DeLuca:2019qsy,Young:2019yug}.  

At the gaussian level, including information about the spatial derivatives of a random field is easy (see e.g.\,\cite{Bardeen:1985tr}). 
This is because the joint ten-dimensional probability density function (PDF) for the gaussian variable $\zeta_{\rm G}$ and its spatial derivatives takes the multi-normal form (under the assumption of spatial homogeneity and isotropy)
 \begin{align}\label{eq:GaussianPDF}
\textrm{P}_{\rm G}(\zeta_{\rm G},\zeta_{{\rm G},i},\zeta_{{\rm G},ij})
 d\zeta_{\rm G}d\zeta_{{\rm G},i}d\zeta_{{\rm G},ij} =& 
\bigg[\prod_{i=x,y,z}^{}\textrm{P}_{\rm G}(\zeta_{{\rm G},i})\bigg]
\bigg[\prod_{ij=xy,yz,xz}\textrm{P}_{\rm G}(\zeta_{{\rm G},ij})\bigg]\times
\\ \nn&\textrm{P}_{\rm G}(\zeta_{\rm G},\zeta_{{\rm G},xx},\zeta_{{\rm G},yy},\zeta_{{\rm G},zz})
d\zeta_{\rm G}d\zeta_{{\rm G},i}d\zeta_{{\rm G},ij}
\,,
 \end{align}
where we are using the short-hand notation $\zeta_{{\rm G},i} = \partial_{i}\zeta_{{\rm G}}$, 
$\zeta_{{\rm G},ij} = \partial_{ij}\zeta_{{\rm G}}$, and where
\begin{align}
\textrm{P}_{\rm G}(\zeta_{{\rm G},i}) = 
 \frac{1}{\sqrt{\pi \sigma_1^2}}\exp\left(-\frac{\zeta_{{\rm G},i}^2}{\sigma_1^2}\right)\,,~~{i=x,y,z}\,,~~
 \textrm{P}_{\rm G}(\zeta_{{\rm G},ij}) = 
  \frac{2}{\sqrt{\pi \sigma_2^2}}\exp\left(-\frac{4\zeta_{{\rm G},ij}^2}{\sigma_2^2}\right)\,,~~{ij=xy,yz,xz}\,,
\end{align}
with
   \begin{align}
  \textrm{P}_{\rm G}(\zeta_{\rm G},\zeta_{{\rm G},xx},\zeta_{{\rm G},yy},\zeta_{{\rm G},zz}) = 
  \frac{\exp\left(
 -\frac{1}{2}\tilde{\zeta}_{\rm G}^{\rm T} \tilde{C}^{-1}\tilde{\zeta}_{\rm G}
  \right)}{(2\pi)^{3/2}\sqrt{{\rm det}\tilde{C}}}\,,\hspace{0.3cm}
\tilde{C} \equiv   
\bordermatrix{     
  & {\scriptstyle \zeta_{\rm G}}         & {\scriptstyle \zeta_{{\rm G},xx}}     & {\scriptstyle \zeta_{{\rm G},yy}} & {\scriptstyle \zeta_{{\rm G},zz}} & \cr
    {\scriptstyle \zeta_{\rm G}}         & \sigma_0^2 & -\sigma_1^2/2 & -\sigma_1^2/2  &  -\sigma_1^2/2 \cr
    {\scriptstyle  \zeta_{{\rm G},xx} }  & -\sigma_1^2/2 & 3\sigma_2^2/8 & \sigma_2^2/8 & \sigma_2^2/8 \cr
    {\scriptstyle  \zeta_{{\rm G},yy}  } &  -\sigma_1^2/2 & \sigma_2^2/8 & 3\sigma_2^2/8 & \sigma_2^2/8 \cr
    {\scriptstyle  \zeta_{{\rm G},zz}  } &  -\sigma_1^2/2 & \sigma_2^2/8 & \sigma_2^2/8 & 3\sigma_2^2/8
}\,,
 \end{align}
with $\tilde{\zeta}_{\rm G} \equiv (\zeta_{{\rm G}},\zeta_{{\rm G},xx},\zeta_{{\rm G},yy},\zeta_{{\rm G},zz})^{\rm T}$.

The variances $\sigma_i^{2}$ can be computed integrating the power spectrum of curvature perturbations, $P_{\zeta}(k)$. 
The latter refers to the gaussian field $\zeta_{\rm G}$.
Furthermore, we include  the Fourier transform of the top-hat window
function in real space $W(k,R)$ (used to smooth the field over a finite volume of size set by the length scale $R$) and the linear transfer function $T(k,\tau)$ (which describes, being $\tau$ the conformal time, the linear evolution  of sub-horizon scales). In full generality, the variances are
\begin{align}
\sigma_j^{2} = \int 
\frac{dk}{k}
W^2(k,R) 
T^2(k,\tau) 
P_{\zeta}(k) 
k^{2j}\,,~~~~{\rm with}~~~~
\left\{
\begin{array}{ccc}
W(k,R)  & =  & 3\big[
\frac{\sin(kR) - kR\cos(kR)}{(kR)^3}\big]  \\
 &   &   \\
T(k,\tau)  & = & 3\big[
\frac{\sin(k\tau/\sqrt{3}) - (k\tau/\sqrt{3})\cos(k\tau/\sqrt{3})}{(k\tau/\sqrt{3})^3}\big]  
\end{array}.
\right.\label{eq:Variances}
\end{align} 
Notice that the expression of $T(k,\tau)$ in eq.\,(\ref{eq:Variances}) is strictly valid during radiation domination
with constant degrees of freedom.
We note that $\lim_{kR\gg 1}W(k,R) = 0$ and $\lim_{kR \ll 1}W(k,R) = 1$ meaning that, for fixed $R$, modes with comoving wavenumber $k\gg 1/R$ are smoothed away by $W(k,R)$; similarly, $\lim_{k\tau \gg 1}T(k,\tau) = 0$ and $\lim_{k\tau \ll 1}T(k,\tau) = 1$ meaning that $T(k,\tau)$ has the effect of smoothing out sub-horizon modes,
playing the role of the pressure gradients and dissipative effects.
It is important to stress that in the prescription presented in the Section \ref{sec:C1B} since all the quantities are evaluated superhorizon the radiation transfer function is simply the identity.
As an ansatz in this section we assume two representative case for the curvature power spectrum
 \begin{align}
{\rm log \mbox{-} normal\,power\,spectrum:}~~~~~~~& P_{\zeta}(k) = 
\frac{A}{\sqrt{2\pi}\sigma}\exp\left[
-\frac{\log^2(k/k_{\star})}{2\sigma^2}
\right]\,,\label{eq:LogNo}\\
{\rm broad\,power\,spectrum:}~~~~~~~& P_{\zeta}(k) = A\,\Theta(k - k_{\rm min})\,\Theta(k_{\rm max} - k)\,,
\label{eq:Bro}
 \end{align}
where $\Theta$ is the Heaviside step function. 
The log-normal functional form is a proxy for a power spectrum with a narrow peak, that is a 
peak characterized by a definite wavenumber $k_{\star}$ (with amplitude $A$ and width 
controlled by the parameter $\sigma$). In the broad case, on the contrary, 
the power spectrum does not possess a definite peak wavenumber but it extends 
over the large window in between $k_{\rm min}$ and $k_{\rm max}\gg k_{\rm min}$.
\begin{figure}[h]
\begin{center}
$$\includegraphics[width=.495\textwidth]{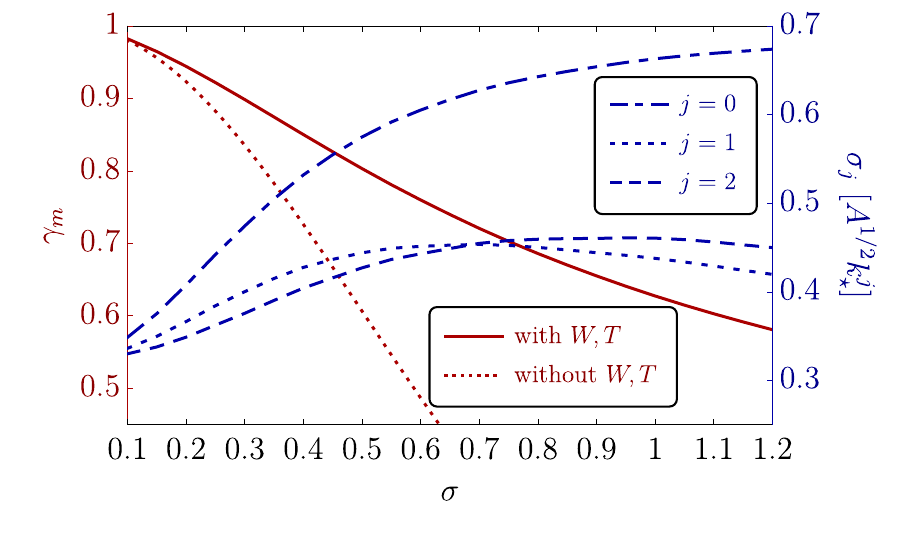}~
\includegraphics[width=.495\textwidth]{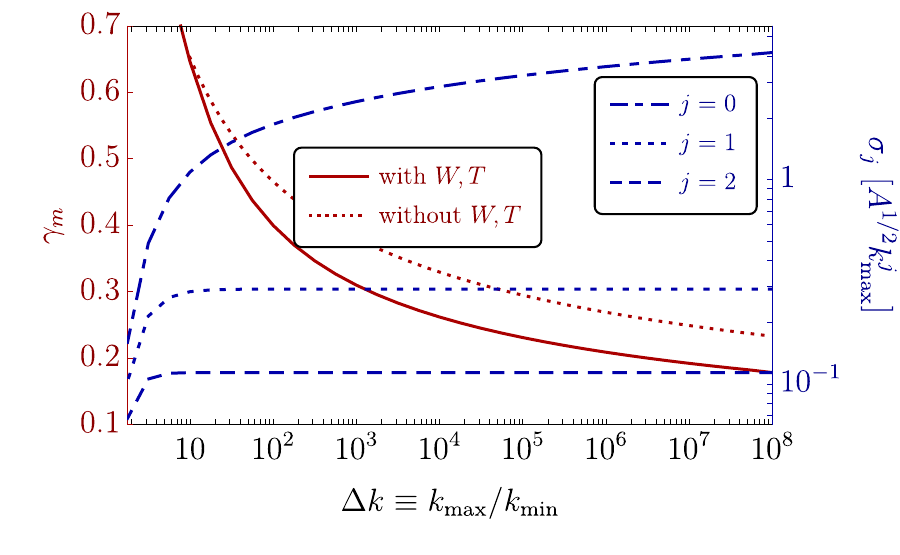}$$\vspace{-0.5cm}
\caption{
Variances (right-side axis) and $\gamma_m$ (left-side axis) computed according to, respectively, 
eq.\,(\ref{eq:Variances}) and eq.\,(\ref{eq:gammam}) in the case of the log-normal power spectrum (\textbf{\textit{left panel }}, as function of $\sigma$) and 
the broad power spectrum (\textbf{\textit{right panel }}, as function of $k_{\rm max}/k_{\rm min}$). 
As far as the variances are concerned, we show the cases with $j=0,1,2$. 
As far as $\gamma_m$ is concerned, the dotted red lines represent the analytic results given 
in eqs.\,(\ref{eq:anal1},\,\ref{eq:anal2}) in the absence of both the window and transfer function. 
We remark that in the computation of the variances window and transfer functions are included.
 }\label{fig:SpectralFeatures}  
\end{center}
\end{figure}
\\We compute numerically the variances in eq.\,(\ref{eq:Variances}) for the two power spectra in 
eq.\,(\ref{eq:LogNo}) and eq.\,(\ref{eq:Bro}). Furthermore, for future reference, we also compute the dimensionless parameter 
\begin{align}\label{eq:gammam}
\gamma_m = \frac{\sigma_1^2}{\sigma_0\sigma_2}\,.
\end{align}
We show our results in fig.\,\ref{fig:SpectralFeatures}. 
The parameter $\gamma_m$ is controlled by the width of the power spectrum. 
This point can be made more transparent if we temporarily neglect the presence of the window and transfer functions; in such a case, 
we get the following analytic 
expressions for the variances
\begin{align}
{\rm log\,normal\,power\,spectrum:}~~~&\sigma_j^2 = A k_{\star}^{2j} e^{2j^2\sigma^2}~~~~~~~~~~~~~\,\,
\Longrightarrow~~~ \gamma_m = e^{-2\sigma^2}\,,\label{eq:anal1}\\
{\rm broad\,power\,spectrum:}~~~~~~~& \sigma_j^2 = \frac{A k_{\rm max}^{2j}(1-\Delta k^{-2j})}{2j}
~~~
\Longrightarrow~~~\gamma_m = \frac{\Delta k^2 - 1}{\sqrt{(\Delta k^4 - 1)\log(\Delta k)}}\,,\label{eq:anal2}
\end{align}
where $\Delta k = k_{\rm max}/k_{\rm min} \gg 1$. 
In both cases, when the power spectrum has a large width (that is in the formal limits $\sigma,\Delta k \to \infty$) we have that 
$\gamma_m \to 0$. 
Furthermore, in both cases we have that $0<\gamma_m < 1$. In the log-normal case, $\gamma_m \to 1$ if $\sigma\to 0$; 
in the broad case, the relevant limit is $\lim_{\Delta k\to 1^{+}}\gamma_m = 1$.

 Qualitatively, these behaviors are respected in the presence of the window and transfer functions, that we include in  
 fig.\,\ref{fig:SpectralFeatures} 
 following ref.\,\cite{Musco:2020jjb}. 
 More precisely, with the window function $W(k,R)$ we smooth the curvature field over a region of size $R = r_m$ and we set the time-scale that enters in the transfer function $T(k,\tau)$ to be $\tau=r_m$. 
The scale $r_m$ measures the characteristic scale of
the density perturbation, and can be defined as the location where the
compaction function is maximized.
In fig.\,\ref{fig:SpectralFeatures} we use the numerical values for $k_{\star}r_m$ 
(as far as the log-normal power spectrum is concerned) and $k_{\rm max}r_m$ (in the broad case) derived as in 
ref.\,\cite{Musco:2020jjb}.

If we now combine fig.\,\ref{fig:SpectralFeatures} and fig.\,\ref{fig:CnCoeff}, we gain a very important intuition. If we consider a very narrow power spectrum (like the log-normal power spectrum in eq.\,(\ref{eq:LogNo}) 
with small $\sigma$ or the 
double-step power spectrum in eq.\,(\ref{eq:Bro}) with $1 <\Delta k \ll 10$) then we expect the expansion in eq.\,(\ref{eq:ZetaSeries}) to be a valid approximation already for small $N$. 
This is because we have in these cases $\sigma_0 \ll 1$. 
On the contrary, for  broader power spectra that violate the condition  $\sigma_0 \ll 1$ we expect a non perturbative analysis based on the full NG curvature perturbation variable $\zeta$ to be needed. 
Remarkably, this intuition---based on very simple consideration mostly involving the curvature perturbation instead of the density contrast---will turn out to be true. We shall elaborate more quantitatively the case of broad power spectra in section\,\ref{sec:TheBroad}.

Inferring the ten-dimensional joint PDF of $\zeta$ starting from the multi-normal distribution in 
eq.\,(\ref{eq:GaussianPDF}) is a very complicated task.
As anticipated, to make matters worse one is actually interested in computing the PBHs mass fraction (see e.g.\,\cite{Sasaki:2018dmp})
\begin{align}\label{eq:SimpleBeta}
\beta = \int_{\delta_c}^{\infty}
\mathcal{K}(\delta -\delta_c)^{\gamma}\,
\textrm{P}_{\delta}(\delta)d\delta\,,
\end{align}
where $\textrm{P}_{\delta}(\delta)$ is the PDF of the density contrast field that we integrate above a certain threshold above which over-densities are expected to collapse and form black holes. 
The scaling-law factor for critical collapse $\mathcal{K}(\delta -\delta_c)^{\gamma}$ accounts for the mass of the primordial black hole at formation written in units of the horizon mass at the time of horizon re-entry\,\cite{Choptuik:1992jv,Evans:1994pj,Musco:2012au}.
During radiation domination, the constants $\mathcal{K}$ and $\gamma$ have been numerically found to be given by $\mathcal{K} = O(1\div 10)$ (see e.g.\,\cite{Young:2019yug}) and $\gamma \simeq 0.36$.
For definiteness, we take $\mathcal{K} \simeq 3.3$ for a log-normal power spectrum and $\mathcal{K} \simeq 4.36$ for a broad power spectrum\,\cite{Musco:2023dak} .

Eq.\,(\ref{eq:SimpleBeta}) is valid within the Press–Schechter formalism.  
Computing $\beta$ in eq.\,(\ref{eq:SimpleBeta}) implies that one should {\it i)} extract the joint PDF of $\zeta$ from that of $\zeta_{\rm G}$ and 
{\it ii)} find a way to use it to compute the statistics of collapsing region characterized by extreme values of density contrast.

\subsection{A preliminary comparison between peaked and broad power spectrum}\label{sec:TheBroad}

Even at this point, without introducing any prescription to fully compute the abundance, we can describe the main differences expected when comparing the two cases of peaked and broad power spectra. Here, we focus on the curvaton case, but the considerations are generals.
\\We consider as formal analogue of eq.\,(\ref{eq:SimpleBeta}) the following integral
\begin{align}
\beta_{\rm NG} = \int_{\zeta_c}^{\infty}
\textrm{P}(\zeta,r_{\rm dec})d\zeta = 
\int_{-\infty}^{+\infty}
\textrm{P}_{\rm G}(\zeta_{\rm G})
\Theta\{\log\big[X(r_{\rm dec},\zeta_{\rm G})\big] - \zeta_c\}d\zeta_{\rm G}\,,\label{eq:ProxyBeta}
\end{align}
where our main focus is on the statistics of the $\zeta$ field (so that we did not include any scaling factor). 
Notice that, because of eq.\,(\ref{eq:ExactPDF}), this integral can be computed exactly once we fix the value of $r_{\rm dec}$ and the characteristic width $\sigma_0$ that controls the gaussian distribution of $\zeta_{\rm G}$. 
In the second equality in eq.\,(\ref{eq:ProxyBeta}) we recast the integration over the NG field $\zeta$ into an integration over the gaussian component $\zeta_{\rm G}$ subject to the constraint imposed by the Heaviside step function $\Theta$. 
In light of the power-series expansion proposed  in eq.\,(\ref{eq:ZetaSeries}), we introduce the approximated quantity 
\begin{align}
\beta_{\rm NG}^{(N)} = 
\int_{-\infty}^{+\infty}
\textrm{P}_{\rm G}(\zeta_{\rm G})
\,\Theta\left[
\sum_{n = 1}^{N} c_n(r_{\rm dec}) \zeta_{\rm G}^{n} - \zeta_c \right]d\zeta_{\rm G}\,.\label{eq:ProxyBetaApprrox}
\end{align}  
Our goal is to compare eq.\,(\ref{eq:ProxyBeta}) and 
eq.\,(\ref{eq:ProxyBetaApprrox}).
In the end, we aim to answer two reasonable questions: 
\begin{itemize}
    \item [{\it i)}]
at which order the truncated power series gives a $\beta_{\rm NG}^{(N)}$ that converges to the exact value given by eq.\,(\ref{eq:ProxyBeta})? 
\item [{\it ii)}]
Does this conclusion depend on the form of the power spectrum?
\end{itemize}
To answer these questions, the key point to bear in mind is that a power-series expansion of the form
\begin{align}
\sum_{n = 1}^{\infty} c_n(r_{\rm dec}) \zeta_{\rm G}^{n} = \log\big[X(r_{\rm dec},\zeta_{\rm G})\big]\,,\label{eq:ConvergenceTest}
\end{align} 
must be always accompanied, to make mathematical sense, by the information about its radius of convergence $R$, that is 
the region of values $-R<\zeta_{\rm G} < + R$ in which the above equality is strictly valid. 
This is a crucial information given that in eq.\,(\ref{eq:ProxyBetaApprrox}) we integrate, in principle, 
over all real values of $\zeta_{\rm G}$.
We address the computation of the radius of convergence $R$ for the above expansion in appendix\,\ref{app:Radius}, and 
we shall focus here on the implications.  
\begin{figure}[ht]
\begin{center}
$$\includegraphics[width=.52\textwidth]{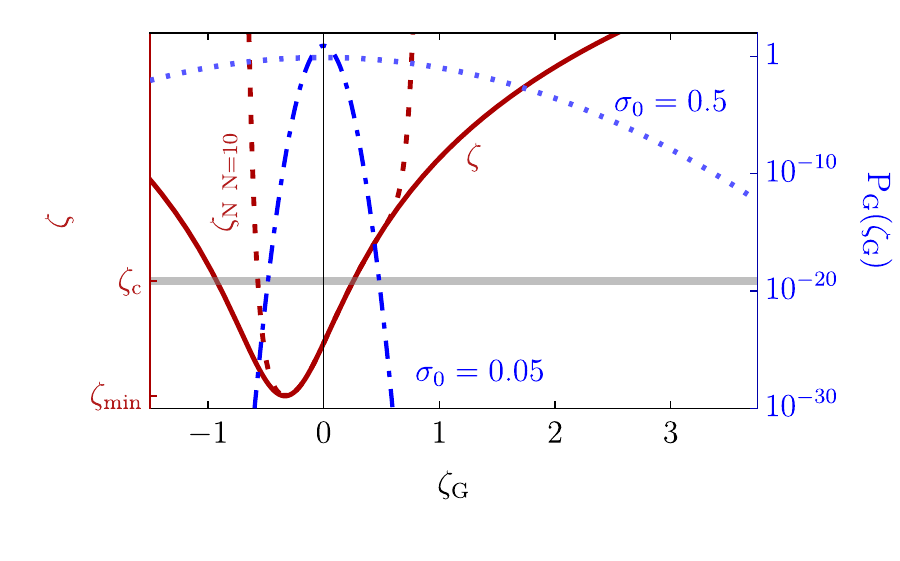}
\quad\quad\includegraphics[width=.52\textwidth]{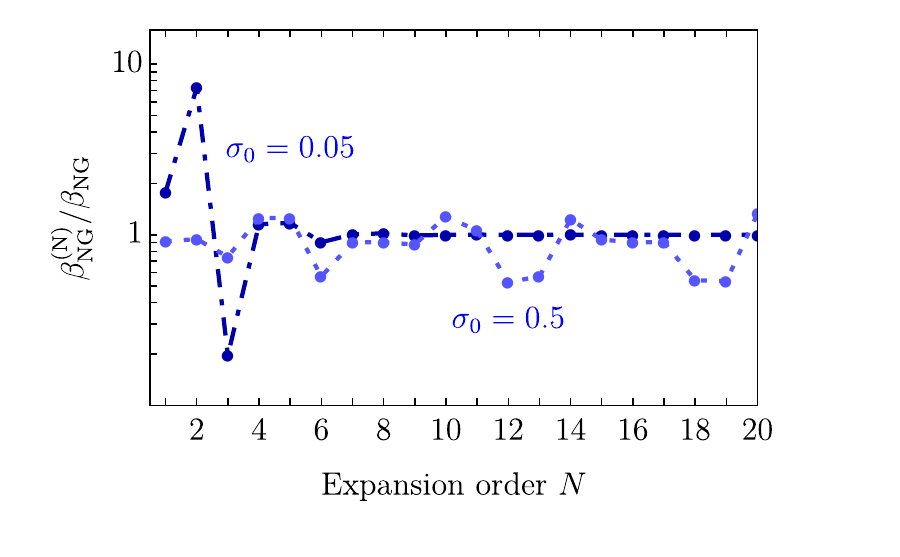}$$
\caption{
We fix $r_{\rm dec}=0.5$.  \textbf{\textit{Left panel. }}On the left-side $y$-axis we plot the function 
$\zeta = \log\big[X(r_{\rm dec},\zeta_{\rm G})\big]$ as function of the Gaussian field $\zeta_{\rm G}$ (solid red line) and its power-series expansion $\zeta_{10}$ (dashed red line). The two lines are superimposed and indistinguishable only within small range of values of $\zeta_{\rm G}$ centered around the origin. 
On the right-side $y$-axis we plot the Gaussian PDF (with zero mean) for two representative values of its variance $\sigma_0$ (blue dotted line, $\sigma_0 = 0.5$, and blue dot-dashed line, $\sigma_0 = 0.05$). 
\textbf{\textit{ Right panel. }}We plot the ratio 
$\beta_{\rm NG}/\beta_{\rm NG}^{(N)}$ (with $\beta_{\rm NG}$ and $\beta_{\rm NG}^{(N)}$ defined, respectively, as in eq.\,(\ref{eq:ProxyBeta}) and eq.\,(\ref{eq:ProxyBetaApprrox})) as function of the expansion order $N$. We plot both cases with 
$\sigma_0 = 0.5$ and $\sigma_0 = 0.05$.
 }\label{fig:PeakedBroad}  
\end{center}
\end{figure}
Consider the situation displayed in the left panel of fig.\,\ref{fig:PeakedBroad}. In this plot we show the graph of $\zeta = \log\big[X(r_{\rm dec},\zeta_{\rm G})\big]$ as function of the gaussian variable $\zeta_{\rm G}$ (solid red line). For definiteness, we take $r_{\rm dec}=0.5$. 
In addition, we show the graph of $\zeta_N$ in eq.\,(\ref{eq:ZetaSeries}) for some value of $N$ (dashed red line; we take $N= 10$ but the exact value of $N$ is not of crucial importance for the following argument).

From the comparison between $\zeta_N$ and $\zeta$, it is evident that the graph of $\zeta_N$ perfectly agrees with that of $\zeta$ for values of $\zeta_{\rm G}$ that lie into an interval centered in the origin and not too wide. This is because this interval lies within the radius of convergence of eq.\,(\ref{eq:ConvergenceTest}). However, as soon as we move too far from the origin the agreement between $\zeta_N$ and $\zeta$ gets unavoidably lost: in the language of eq.\,(\ref{eq:ConvergenceTest}) this happens because we go outside the interval of convergence of the power-series expansion. We refer to appendix\,\ref{app:Radius} for a careful discussion about the computation of the radius of convergence of eq.\,(\ref{eq:ConvergenceTest}) (see fig.\,\ref{fig:Radius} for a more detailed version of the left panel of fig.\,\ref{fig:PeakedBroad}).

The key point is the following. In eq.\,(\ref{eq:ProxyBetaApprrox}), the values of $\zeta_{\rm G}$ over which we integrate are weighted by the gaussian distribution $\textrm{P}_{\rm G}(\zeta_{\rm G})$ with width $\sigma_0$. 
We remind that, as illustrated in fig.\,\ref{fig:SpectralFeatures}, the broader the power spectrum the bigger the value of $\sigma_0$ compared to higher momenta $\sigma_i$. 
In the left panel of fig.\,\ref{fig:PeakedBroad} we superimpose in blue the profile of the gaussian distribution 
 $\textrm{P}_{\rm G}(\zeta_{\rm G})$ for two different benchmark values of $\sigma_0$, namely  $\sigma_0 = 0.05$ and $\sigma_0 = 0.5$. We argue the following points: 
\begin{itemize}
\item[$\circ$] If the power spectrum is very narrow -- thus $\sigma_0$ very small, like in the case $\sigma_0 = 0.05$ 
in our example -- the dominant contribution to the integral in eq.\,(\ref{eq:ProxyBetaApprrox}) comes from a region of $\zeta_{\rm G}$ that lies within the radius of convergence of eq.\,(\ref{eq:ConvergenceTest}). 
Consequently, we expect an excellent agreement between eq.\,(\ref{eq:ProxyBeta}) and eq.\,(\ref{eq:ProxyBetaApprrox}) already for relatively small values of $N$. \item[$\circ$] If the power spectrum gets broader -- thus for increasing $\sigma_0$, like in the case $\sigma_0 = 0.5$ in our example -- in the computation of the integral in eq.\,(\ref{eq:ProxyBetaApprrox}) the gaussian distribution will give more weight to values of $\zeta_{\rm G}$ that lie outside the region of convergence of eq.\,(\ref{eq:ConvergenceTest}). In this case, the agreement between eq.\,(\ref{eq:ProxyBeta}) and eq.\,(\ref{eq:ProxyBetaApprrox}) will get fatefully lost. 
Notice that, in this situation, Increasing $N$ arbitrarily will not fix the problem because, outside the region of convergence, the equality in eq.,(\ref{eq:ConvergenceTest}) does not hold.
\end{itemize}
To further corroborate these considerations, in the right panel of fig.\,\ref{fig:PeakedBroad} we show the value of $\beta_{\rm NG}^{(N)}/\beta_{\rm NG}$ for increasing $N$ and for the two values of $\sigma_0$ discussed before. 
In the case $\sigma_0 = 0.05$ we see that we have an excellent agreement between the full and the approximated computation already for $N\gtrsim 4$.  
On the contrary, if we consider $\sigma_0 = 0.5$ we see that the agreement between $\beta_{\rm NG}^{(N)}$ and $\beta_{\rm NG}$ starts deteriorating precisely because we are now including with some sizable probability the integration over a non-convergent region.

From this simplified exercise, we draw the following conclusions.
If we consider a very narrow power spectrum (like the log-normal power spectrum in eq.\,(\ref{eq:LogNo}) with small $\sigma$ or the double-step power spectrum in eq.\,(\ref{eq:Bro}) with $1 <\Delta k \ll 10$) we expect that the power-series expansion in eq.\,(\ref{eq:ZetaSeries}) will give reliable results. 
However, in the case of a broad power spectrum (like the log-normal power spectrum in eq.\,(\ref{eq:LogNo}) with large $\sigma$ or the double-step power spectrum in eq.\,(\ref{eq:Bro}) with $\Delta k \gg 1$) we expect that the power-series expansion in eq.\,(\ref{eq:ZetaSeries}) will give unreliable results. 	In the case of a broad power spectrum, therefore, understanding how to compute  eq.\,(\ref{eq:SimpleBeta}) without relying on the power-series expansion in eq.\,(\ref{eq:ZetaSeries}) is of pivotal importance.

\subsection*{Summary}
In this section, we discussed the fundamental role played by both kinds of non-Gaussianities in the computation of the PBH abundance. We also showed that the power series expansions for the primordial NGs (eq.\ref{eq:FirstExpansion}) lead to unreliable results when the shape of the curvature power spectrum becomes broader.
 
\section{Threshold statistics on the compaction function}\label{sec:C1A}

Many previous attempts in the literature have tried to address the problem of including NGs in the computation of PBHs abundance, with different levels of approximation. 
Various works have exploited the perturbative expansion in  eq.\,\eqref{eq:FirstExpansion} up to a finite number of orders\,\cite{Bugaev:2013vba,Nakama:2016gzw,Byrnes:2012yx,Young:2013oia,Yoo:2018kvb,Kawasaki:2019mbl,Yoo2,Yoo:2019pma,
Riccardi:2021rlf,Taoso:2021uvl,Meng:2022ixx,Escriva:2022pnz}, modelling either the distribution of curvature perturbation or density contrast (where the latter is the physical quantity that should be adopted\,\cite{Young:2014ana}).
Others have tried adopting a path-integral formulation\,\cite{Franciolini:2018vbk}, whose application requires the notion of the full list of n-th order point functions, or non-perturbative approaches\,\cite{Atal:2019cdz,Biagetti:2021eep,Kitajima:2021fpq} focused on specific USR models.
On the other hand, in refs.\,\cite{Shibata:1999zs,Musco:2018rwt,Young:2022phe} it was suggested that the so-called compaction function $\mathcal{C}$ — defined as twice the
local excess-mass over the co-moving areal radius — is the most suitable parameter to use to determine whether a perturbation collapses to form a PBH. 

In this section, we describe an exact formalism for the computation of the abundance of PBH in the presence of local NGs in the curvature perturbation field, going beyond the widely used quadratic and cubic approximations in the perturbative expansion of eq.\ref{eq:FirstExpansion}, and consider a completely generic functional form.
By adopting threshold statistics on the compaction function and using the average profile of the compaction, we address the computation of the abundance both for narrow and broad power spectra. This is important for matching the outcome of realistic inflationary dynamics. 
\subsection{The compaction function}
Starting from the metric in the comoving uniform-energy density gauge, where the curvature perturbation $\zeta(\bf x)$ on superhorizon scales appears 
\be
{\rm d}s^2=-{\rm d}t^2+a^2(t)e^{2\zeta(\bf x)}{\rm d}{\bf x}^2,
\ee
we define the compaction function as twice the local mass excess over the areal radius 
\begin{align}\label{eq:DefinitionCompaction}
\mathcal{C}(r,t) = \frac{2\left[M(r,t) - M_b(r,t)\right]}{R(r,t)}
\end{align}
where $M(\x,t)$ is the Misner-Sharp mass and   $M_{\rm b}(\x,t)$ its background value. The Misner-Sharp mass is the mass within a sphere of
areal radius 
\be
\label{R}
R(\x,t)=a(t)\tilde{r} e^{\zeta(\bf x)}
\ee
with spherical coordinate radius $\tilde{r}$, centered around position $\x$  and evaluated at time $t$. The compaction directly measures the overabundance of mass in a region and is therefore better suited than the curvature perturbation for determining when an overdensity collapses into a PBH. 
Furthermore, the compaction has the advantage of being  time-independent on superhorizon scales.  It  can be
written in terms of the density contrast as
\be\label{eq:Compa2}
\mathcal{C}(\x)=\frac{2\rho_{\rm b}}{R(\x,t)}\int {\rm d}^3\x  \,\delta(\x,t),
\ee
where $\rho_{\rm b}$ is the background energy density. On super-horizon scales, and adopting the 
 gradient expansion approximation,
 the density contrast is given by eq.\,(\ref{eq:NonLinearDelta}) that we now rewrite 
assuming spherical symmetry 
\begin{align}\label{eq:SphericalDelta}
\delta(r,t) = 
-\frac{2}{3}
\Phi
\left(\frac{1}{aH}\right)^2 
e^{-2\zeta(r)}\left[
\zeta^{\prime\prime}(r) + \frac{2}{r}\zeta^{\prime}(r) + \frac{1}{2}\zeta^{\prime}(r)^2
\right]\,.
\end{align}
Using the above expression and integrating over the radial coordinate, 
eq.\,(\ref{eq:Compa2}) takes the form
\begin{align}\label{eq:CompactionFull}
\mathcal{C}(r) = 
-2\Phi\,r\,\zeta^{\prime}(r)\left[
1 + \frac{r}{2}\zeta^{\prime}(r)
\right] = 
\mathcal{C}_1(r) - \frac{1}{4\Phi}\mathcal{C}_1(r)^2\,,~~~~~~~~~
\mathcal{C}_1(r) = -2\Phi\,r\,\zeta^{\prime}(r)\,,
\end{align} 
where $\mathcal{C}_1(r)$ defines the linear component of the compaction function.
The length scale $r_m$ is defined as the scale at which the compaction function is maximized, and, therefore, 
it verifies the condition 
\begin{equation}
\mathcal{C}^{\prime}(r_m) = 0
~~~~~~~~~~\text{that is }~~~~~~~~~~
\zeta^{\prime}(r_m) + r_m\zeta^{\prime\prime}(r_m) = 0    
\end{equation}
in terms of the comoving curvature perturbation. 
If we define $\mathcal{C}_{\rm max} = \mathcal{C}(r_m)$ as the value of the compaction at the position of the maximum, PBHs form if the maximum of the compaction function is above some threshold value, 
$\mathcal{C}_{\rm max} > \mathcal{C}_{\rm th}$.
If we consider the averaged mass excess within a spherical region of areal radius $R$, that is the ratio
$\delta M(r,t)/M_b(r,t)$, a direct computation shows that\,\cite{Musco:2018rwt} 
\begin{align}
\frac{\delta M(r,t)}{M_b(r,t)} = \frac{1}{V_b(r,t)}
\int_{S^2_R} d^{3}\vec{x}\,\delta\rho(\vec{x},t)
~~
\Longrightarrow~~
\delta_m = \frac{\delta M(r_m,t_H)}{M_b(r_m,t_H)} = \mathcal{C}(r_m) = 3\delta(r_m,t_H)\,,
\label{eq:MainCompa}
\end{align}
where the last equality follows from eq.\,(\ref{eq:SphericalDelta})  evaluated at horizon crossing
together with the condition $\zeta^{\prime}(r_m) + r_m\zeta^{\prime\prime}(r_m) = 0$ that defines $r_m$. 
Eq.\,(\ref{eq:MainCompa}) shows that the peak value of the compaction function $\mathcal{C}(r_m)$ equals 
$\delta_m$, that is 
the density contrast volume-averaged over a spherical region of areal radius set by the length scale $r_m$. 
This last point is crucial. 
As discussed in ref.\,\cite{Musco:2018rwt},
the gravitational collapse that triggers the formation of a PBH takes place
when the maximum of the compaction function $\mathcal{C}(r_m)$ is larger than a certain threshold
value. 
Therefore, the PBH formation can be investigated by directly studying statistics of the compaction function, and compute the probability of it overcoming the threshold $\mathcal{C}_{th}$, derived in numerical simulations of collapses of the volume-averaged density contrast $\delta_{c}$.
Indeed, consider now a family of compaction functions which have in common the same value of $r_m$, but a different curvature at the maximum parametrized by \cite{Escriva:2019phb}
\be
q=-\frac{1}{4}\frac{r_m^2 \mathcal{C}''(r_m)}{\mathcal{C}(r_m)},
\ee
Numerically, it has been noticed that the critical threshold depends on the curvature at the peak of the compaction function \cite{Escriva:2019phb,Musco:2018rwt}
\be\label{eq:Threshold}
\mathcal{C}_{th}(q)=\frac{4}{15}e^{-1/q}\frac{q^{1-5/2q}}{\Gamma(5/2q)-\Gamma(5/2q,1/q)},
\ee
such that $
\mathcal{C}_{th}(q\ll 1)\simeq 2/5$ and  $
\mathcal{C}_{th}(q\gg 1)\simeq 2/3$.
So, as discussed in\,\cite{Germani:2018jgr,Musco:2018rwt,Musco:2020jjb}, given a power spectrum we can compute the shape parameter and consequently the threshold $\mathcal{C}_{th}$ (or $\delta_c$). We have
 \begin{align}
{\rm log\,normal\,power\,spectrum:}~~~~~~~&
\mathcal{C}_{th}=\delta_c = \delta_c(\sigma)\,,\label{eq:deltacLog}
\\
{\rm broad\,power\,spectrum:}~~~~~~~&
\mathcal{C}_{th}=\delta_c = 0.56\,.\label{eq:deltacBroad}
\end{align}
In eqs.\,(\ref{eq:deltacLog},\,\ref{eq:deltacBroad}) we use the results of ref.\,\cite{Musco:2020jjb} that include the non-linear effects between curvature perturbations and density contrast.  
Furthermore, it should be noted that eqs.\,(\ref{eq:deltacLog},\,\ref{eq:deltacBroad}) are strictly valid during the radiation era. 
When phenomenologically relevant we will include the modifications to the threshold due to the quark-hadron QCD phase transition.
\\One final remark is in order. 
As discussed, we compute $\delta_c$ using the results of ref.\,\cite{Musco:2020jjb} that include the effect of the non-linear relation between curvature perturbations and density contrast. However, 
ref.\,\cite{Musco:2020jjb} assumes that curvature perturbations follow a Gaussian statistics. 
We expect that the presence of primordial NG will impact the value of $\delta_c$.\footnote{See Refs.\,\cite{Kehagias:2019eil,Escriva:2022pnz} for some efforts in this direction, showing the variation of threshold is around few percent for $|f_{\rm NL}| <\mathcal{O}(5)$.}
Also the position $r_m$ in which the compaction is maximized depends by the shape of the power spectrum \,\cite{Musco:2020jjb}.
Concretely, we have
 \begin{align}
{\rm log\,normal\,power\,spectrum:}~~~~~~~&
r_m k = \kappa(\sigma)\,,\label{eq:Logrk}
\\
{\rm broad\,power\,spectrum:}~~~~~~~&
r_m k = \kappa = 4.49\,,\label{eq:Broadrk}
\end{align}
where in the first equation $\kappa$ is a function of the log-normal width (that we take from ref.\,\cite{Musco:2020jjb}) while in the second one $\kappa$ takes the constant value valid for the broad power spectrum in eq.\,(\ref{eq:Bro}), cf. ref.\,\cite{Musco:2020jjb}.
\subsection{Computation of the mass fraction}
We now elaborate on the presence of primordial NGs, which are encoded in the relation 
$\zeta = F(\zeta_{\rm G})$ (see eq.\,\eqref{eq:MainF}).
The linear component of the compaction function takes the form
\begin{align}\label{eq:C1expl}
\mathcal{C}_1(r) = -2\Phi\,r\,\zeta_{\rm G}^{\prime}(r)\,
\frac{dF}{d\zeta_{\rm G}} = 
\mathcal{C}_{\rm G}(r)\,
\frac{dF}{d\zeta_{\rm G}}\,,~~~~{\rm with}~~~
\mathcal{C}_{\rm G}(r) =
-2\Phi\,r\,\zeta_{\rm G}^{\prime}(r)\,.
\end{align}
 Consequently, the compaction function reads
 \begin{align}\label{eq:CCgau}
\mathcal{C}(r) = 
\mathcal{C}_{\rm G}(r)\,
\frac{dF}{d\zeta_{\rm G}} 
 - \frac{1}{4\Phi}
 \mathcal{C}^2_{\rm G}(r)
 \left(\frac{dF}{d\zeta_{\rm G}}
 \right)^2\,.
 \end{align}
The compaction function depends on both the gaussian linear component $\mathcal{C}_{\rm G}$ and the gaussian  
 curvature perturbation $\zeta_{\rm G}$. 
 Both these random variables are gaussian; $\zeta_{\rm G}$ is gaussian by definition while  
 $\mathcal{C}_{\rm G}$ is defined by means of the derivative of the gaussian variable $\zeta_{\rm G}$. 

We start from the two-dimensional joint PDF of $\zeta_{\rm G}$ and $\mathcal{C}_{\rm G}$, which can be written as
 \begin{align}\label{eq:PDFCompa}
 \textrm{P}_{\rm G}(\mathcal{C}_{\rm G},\zeta_{\rm G}) 
 = \frac{1}{(2\pi)\sqrt{\det\Sigma}}
 \exp\left(
 -\frac{1}{2}Y^{\rm T}\Sigma^{-1}Y
 \right)\,,\end{align}
 with
 \begin{align}
 Y =\left(
\begin{array}{c}
 \mathcal{C}_{\rm G}  \\
  \zeta_{\rm G}    
\end{array}
\right)\,,~~~{\rm and}~~~
\Sigma =
\left(
\begin{array}{cc}
 \langle\mathcal{C}_{\rm G}\mathcal{C}_{\rm G}\rangle & \langle\mathcal{C}_{\rm G}\zeta_{\rm G}\rangle   \\
 \langle\mathcal{C}_{\rm G}\zeta_{\rm G}\rangle & \langle\zeta_{\rm G}\zeta_{\rm G}\rangle     
\end{array}
\right)\,,
 \end{align}
 where $\Sigma$ is the covariance matrix. The entries of $\Sigma$ are\,\cite{Young:2022phe}
 \begin{align}
 \langle\mathcal{C}_{\rm G}\mathcal{C}_{\rm G}\rangle & = \sigma_c^2 =
  \frac{4\Phi^2}{9}\int_0^{\infty}\frac{dk}{k}
  (kr_m)^4 W^2(k,r_m) T^2(k,r_m) P_{\zeta}(k)\,,\label{eq:Var1}
   \\
 \langle\mathcal{C}_{\rm G}\zeta_{\rm G}\rangle & = \sigma_{cr}^2 = 
 \frac{2\Phi}{3}\int_0^{\infty}\frac{dk}{k}(kr_m)^2
 W(k,r_m)
 W_s(k,r_m) T^2(k,r_m) P_{\zeta}(k)\,,
  \\
  \langle\zeta_{\rm G}\zeta_{\rm G}\rangle & = \sigma_r^2 =   \int_0^{\infty}\frac{dk}{k}
  W_s^2(k,r_m) T^2(k,r_m) P_{\zeta}(k)\,,\label{eq:Var3}
 \end{align}
where $W_s(k,r) = \sin(kr)/kr$ while $W(k,R)$ and $T(k,\tau)$ are given in eq.\,(\ref{eq:Variances}).  
As explained in the above preamble, the key point to build the variance of the compaction function, is to consider the density contrast volume-averaged over a spherical region of size set by $r_m$. This is done by means of the top-hat smoothing $W(k,R)$ in eq.\,(\ref{eq:Variances}). Concretely, we have set $R=r_m$ and all variances in this section are evaluated according to this choice.
Furthermore, we also set $\tau = r_m$ in the computation of the linear transfer function as required when computing the variance at the time of horizon crossing of the scale $r_m$\,\cite{Musco:2018rwt}.
Given a power spectrum, in this section we use the values for $r_m$ computed in ref.\,\cite{Musco:2020jjb}.

After computing the inverse of $\Sigma$ and its determinant, and completing the square in the argument of the exponential function,
 eq.\,(\ref{eq:PDFCompa}) can be recast in the form
\begin{align}\label{eq:PDFCompa2}
 \textrm{P}_{\rm G}(\mathcal{C}_{\rm G},\zeta_{\rm G}) =
 \frac{1}{(2\pi)\sigma_c\sigma_{r}\sqrt{1-\gamma_{cr}^2}}
 \exp\left(
 -\frac{\zeta_{\rm G}^2}{2\sigma_r^2}
 \right)
 \exp\left[
 -\frac{1}{2(1-\gamma_{cr}^2)}\left(
 \frac{\mathcal{C}_{\rm G}}{\sigma_c} - \frac{\gamma_{cr}\zeta_{\rm G}}{\sigma_r}
 \right)^2
 \right]\,,
 \end{align}
where 
\begin{align}
\gamma_{cr} = \frac{\sigma_{cr}^2}{\sigma_c \sigma_r}\,.\label{eq:GammacrDef}
\end{align} 
If we take the limit $\gamma_{cr}\to 1$ in eq.\,(\ref{eq:PDFCompa2}), we find the distribution
\begin{align}
\lim_{\gamma_{cr}\to 1}\textrm{P}_{\rm G}(\mathcal{C}_{\rm G},\zeta_{\rm G}) = 
\frac{1}{\sqrt{2\pi}\sigma_c\sigma_{r}} \exp\left(
 -\frac{\zeta_{\rm G}^2}{2\sigma_r^2}
 \right)\delta\left(\mathcal{C}_{\rm G}/\sigma_c -  \zeta_{\rm G}/\sigma_r\right)\,,
\end{align} 
where the last delta function forces the relation 
$\zeta_{\rm G} = \sigma_r\mathcal{C}_{\rm G}/\sigma_c$ once the integral over $\zeta_{\rm G}$ is performed. 
This is the essence of the so-called high-peak limit of ref.\,\cite{Young:2022phe}. 
This limit is strictly valid if one takes $\gamma_{cr}\to 1$ that is the limit in which one takes a monochromatic power spectrum of curvature perturbations (that gives precisely $\gamma_{cr} = 1$). 

The main goal of this prescription is to go beyond the high-peak limit. This is crucial as we are interested in the case of a broad power spectra for which the high-peak limit is not applicable. 
We approach this problem in a very simple way.
The prescription to compute the NG abundance (without taking the limit $\gamma_{cr}\to 1$)
is summarized in the following  master formula
\begin{align}
\beta_{\rm NG} & = \int_{\mathcal{D}}\mathcal{K}(\mathcal{C} - \mathcal{C}_{\rm th})^{\gamma}
\textrm{P}_{\rm G}(\mathcal{C}_{\rm G},\zeta_{\rm G})d\mathcal{C}_{\rm G} d\zeta_{\rm G}\,,\label{eq:CompactionIntegral}
\\
 \textrm{P}_{\rm G}(\mathcal{C}_{\rm G},\zeta_{\rm G}) & =
 \frac{1}{(2\pi)\sigma_c\sigma_{r}\sqrt{1-\gamma_{cr}^2}}
 \exp\left(
 -\frac{\zeta_{\rm G}^2}{2\sigma_r^2}
 \right)
 \exp\left[
 -\frac{1}{2(1-\gamma_{cr}^2)}\left(
 \frac{\mathcal{C}_{\rm G}}{\sigma_c} - \frac{\gamma_{cr}\zeta_{\rm G}}{\sigma_r}
 \right)^2
 \right]\,,
 \\
 \mathcal{D} & = 
\left\{
\mathcal{C}_{\rm G},\,\zeta_{\rm G} \in \mathbb{R}~:~~
\mathcal{C}(\mathcal{C}_{\rm G},\zeta_{\rm G}) > \mathcal{C}_{\rm th}  
~\land~\mathcal{C}_1(\mathcal{C}_{\rm G},\zeta_{\rm G}) < 2\Phi
\right\}\,,\label{eq:RegionD}
\end{align} 
where $\mathcal{C}$ is written implicitly in terms of $\mathcal{C}_{\rm G}$ and $\zeta_{\rm G}$ by means of eq.\,(\ref{eq:CCgau}); the two conditions in the integration domain $\mathcal{D}$ (with again $\mathcal{C}_1$ written in terms of $\mathcal{C}_{\rm G}$ and $\zeta_{\rm G}$ according to eq.\,(\ref{eq:C1expl})) select the so-called type-I perturbations.
We implement in eq.\,(\ref{eq:CompactionIntegral}) the two conditions that define $\mathcal{D}$ by means of Heaviside step functions. Eq.\,(\ref{eq:CompactionIntegral}) is the main result of this section: It describes the abundance of collapsing peaks in the context of threshold statistics and gives an exact expression in the presence of primordial NGs.

Instead of working with the full resummed expression $\zeta = \log\big[X(\zeta_{\rm G})\big]$, it is also possible to 
consider its power-series expansion in eq.\,(\ref{eq:ZetaSeries}) truncated at some given order $N$. 
The same integral in eq.\,(\ref{eq:CompactionIntegral}) will give an expression for $\beta_{\rm NG}^{(N)}$ at order $N$. This will be used to test the convergence of the series expansion in relation to the width of the spectrum.

In practice, at the prize of a slightly more complicated numerical integration (but at the end the computation of 
$\beta$ would have been numerical anyway) we can go beyond the high-peak limit and consider generic 
functional dependencies for local type NGs.
In the following, we are going to present 
worked examples by specializing our analysis 
to the expression $\zeta = \log\big[X(\zeta_{\rm G})\big]$.

\subsubsection{The log-normal power spectrum}

\begin{figure}[h]
\begin{center}
\includegraphics[width=1\textwidth]{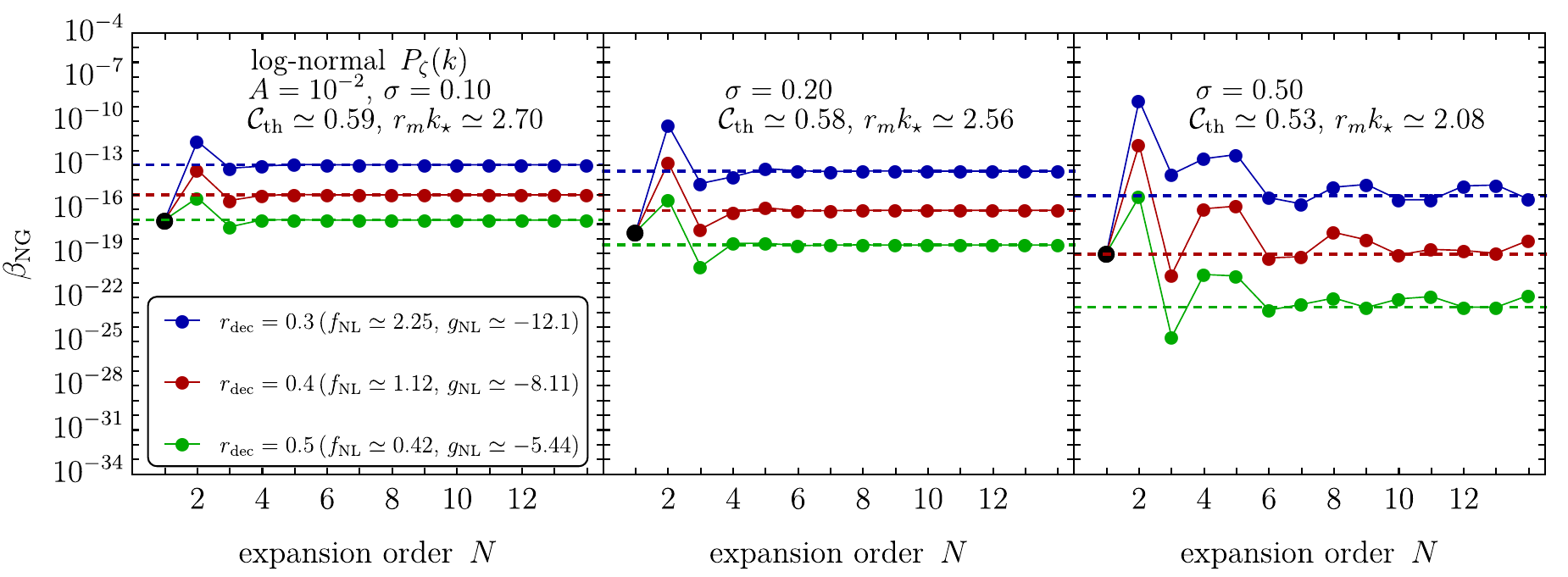}
\caption{
Computation of the primordial mass fraction $\beta$ of the Universe stored into PBHs at the formation time done using the prescription in eq.\,(\ref{eq:CompactionIntegral}).
We consider the log-normal power spectrum in eq.\,(\ref{eq:LogNo}) with fixed amplitude $A$ and 
three different values of $\sigma$; from left to right, we take $\sigma = 0.1$, $\sigma = 0.2$ and $\sigma = 0.5$. 
For each case, we investigate the impact of primordial NGs for three benchmark values of $r_{\rm dec}$.
This result includes both non-linearities and 
NGs of primordial origin motivated by the curvaton model, i.e. eq.\,(\ref{eq:MasterX}). 
We account for primordial NGs both perturbatively (by means of 
the power-series expansion in eq.\,(\ref{eq:ZetaSeries}) truncated at some finite 
order $N$ - filled dots)
and non-perturbatively (with the full resummed result in eq.\,(\ref{eq:MasterX}) - horizontal 
dashed lines). 
  }\label{fig:LogNoCompa}  
\end{center}
\end{figure}

First, we consider the log-normal power spectrum in eq.\,(\ref{eq:LogNo}).
In fig.\,\ref{fig:LogNoCompa} we plot the value of $\beta_{\rm NG}$, computed by means of eq.\,(\ref{eq:CompactionIntegral}), for different values of $\sigma$ and $r_{\rm dec}$ (see caption for details). 
We compare $\beta_{\rm NG}$ (dashed lines) with its perturbative expansion $\beta_{\rm NG}^{(N)}$ 
for different values of the expansion order $N$ (indicated in the $x$-axis in fig.\,\ref{fig:LogNoCompa}).

\begin{figure}[h]
\begin{center}
$$\includegraphics[width=.495\textwidth]{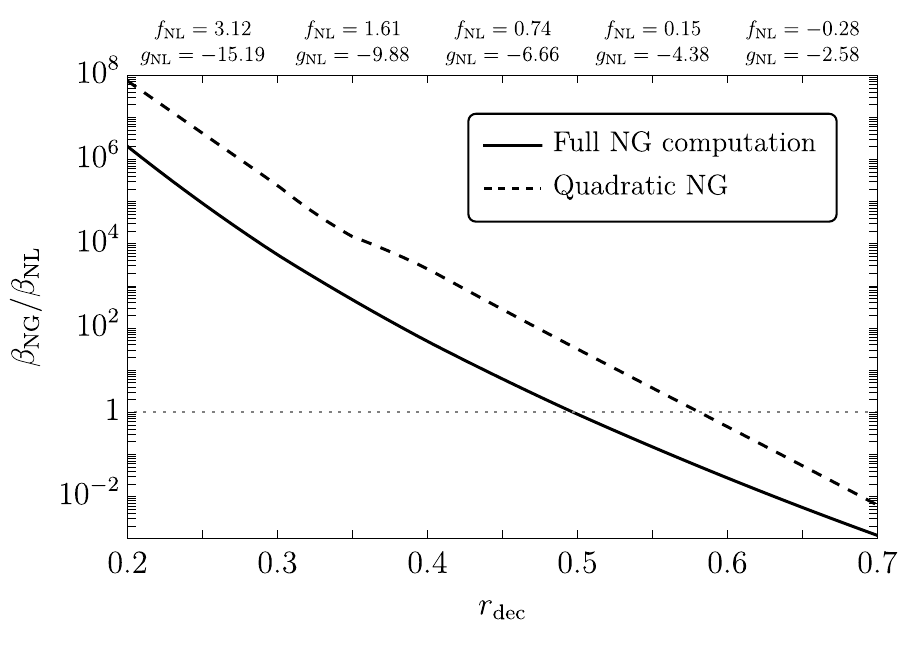}~
\includegraphics[width=.495\textwidth]{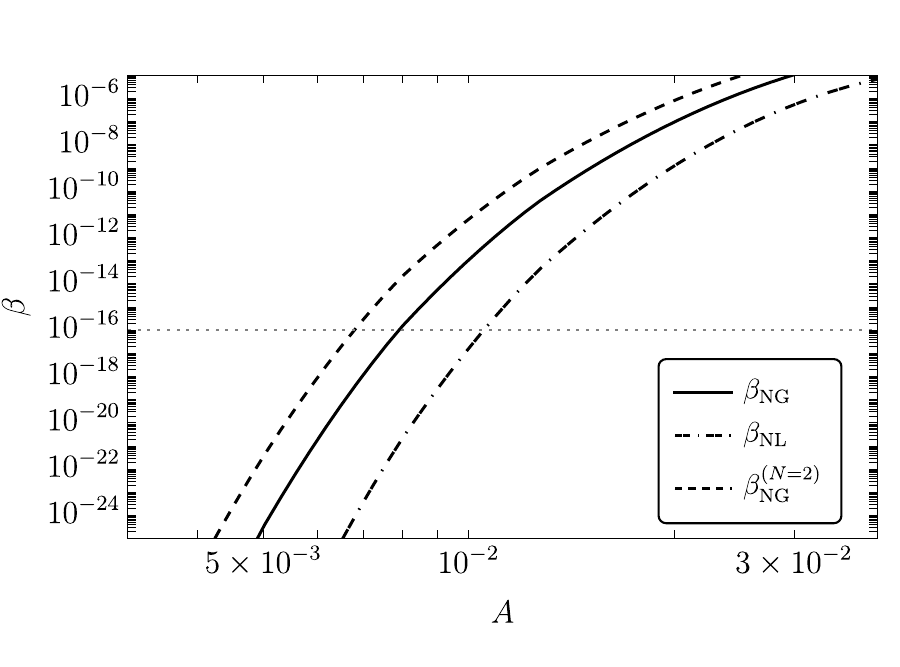}$$\vspace{-0.5cm}
\caption{
	\textbf{\textit{	Left panel. }} 
Ratio $\beta_{\rm NL}/\beta_{\rm NG}$ as function of the parameter $r_{\rm dec}$.
We evaluate $\beta_{\rm NG}$ at the quadratic order in the primordial NGs 
(see eq.\,(\ref{eq:ZetaSeries}) with $N=2$) and considering the full resummed expression for $\zeta$ 
(see eq.\,(\ref{eq:MasterX})). 
We focus on the log-normal power spectrum in eq.\,(\ref{eq:LogNo}) with fixed values of $\sigma$ and $A$. 
	\textbf{\textit{	Right panel. }} We take $r_{\rm dec} = 0.3$ and compute $\beta$ as function of the amplitude 
$A$ of the log-normal power spectrum (while $\sigma$ is kept fixed at the same reference value $\sigma = 0.1$ used in the left-side plot). 
We compare $\beta_{\rm NL}$ with 
$\beta_{\rm NG}$, computing the latter both at the quadratic order and with fully resummed primordial NGs. 
 }\label{fig:LogNoTest}  
\end{center}
\end{figure}

We find that quadratic NGs with positive $f_{\rm NL}$ enhance
the value of $\beta$, as expected.  
However, already the inclusion of the cubic term reverse 
this result. This is because the cubic correction derived from eq.\,(\ref{eq:ZetaC}) is characterized by a large and negative 
coefficient $g_{\rm NL}$. 
Given the analyzed values of $\sigma$ in fig.\,\ref{fig:LogNoCompa}, we are in the situation in which $\sigma_0 \ll 1$ and we indeed expected a good convergence of the NG expansion.
Therefore, the main conclusion we draw from this figure is the following: if we limit the analysis to narrow spectra (e.g. $\sigma = 0.10$, $\sigma = 0.20$) we find that the perturbative approach quickly converges to the exact result.

The net effect of primordial NGs displayed in fig.\,\ref{fig:LogNoCompa} depends on the specific value of $r_{\rm dec}$. To better illustrate this point, we show in the left panel of fig.\,\ref{fig:LogNoTest} the ratio between the results which only includes non-linearities $\beta_{\rm NL}$ (and Gaussian primordial curvature perturbation) and the NG result $\beta_{\rm NG}$ as function of $r_{\rm dec}$.  It is interesting to remark that, for $r_{\rm dec} \lesssim 0.6$, 
 the quadratic approximation for $\beta_{\rm NG}$ 
 systematically gives a result that is larger than $\beta_{\rm NL}$ (thus 
 $\beta_{\rm NG}/\beta_{\rm NL} > 1$). 
 This is because, as we can see from eq.\,\eqref{eq:ZetaQ}, the quadratic coefficient $f_{\rm NL}$ at $r_{\rm dec} = (\sqrt{10} - 2)/2 \approx 0.58$ changes from positive to negative values (for completeness, in the left panel of fig.\,\ref{fig:LogNoTest} we indicate on the top $x$-axis the values of the first two coefficients $f_{\rm NL}$ and $g_{\rm NL}$). From this perspective, the previous result is consistent with the expectation that local quadratic primordial NG with $f_{\rm NL} < 0$ ($f_{\rm NL} > 0$) tend to suppress (enhance) the PBH abundance compared to the case in which they are absent.
However, we find that including primordial NG at all-orders 
 modifies this conclusion: we find that
 $\beta_{\rm NG}/\beta_{\rm NL} > 1$ if 
 $r_{\rm dec} \lesssim 0.5$ 
 (instead of $r_{\rm dec} \lesssim 0.6$).

 In the right panel of fig.\,\ref{fig:LogNoTest} we set $r_{\rm dec} = 0.3$ and consider the computation 
 of $\beta$ as function of the  amplitude $A$. 
 As expected, we find that orders of magnitude differences in the computation of the mass fraction $\beta$ imply relatively small changes of $A$ because of the leading order scaling $\beta \sim e^{-1/P_{\zeta}}$. 
 For instance, we find that a relative change $\Delta A \simeq -25\%$ in the amplitude of the power spectrum (compared to the non-linear case) is needed in order to fit the reference value $\beta = 10^{-16}$. 

Let us summarize our findings in the case of the log-normal power spectrum.
\begin{itemize}
    \item[{\it i)}] Truncating the computation of the PBHs mass fraction $\beta_{\rm NG}$ at the quadratic order (that is, only including $f_{\rm NL}$) does not give reliable results, cf. fig.\,\ref{fig:LogNoCompa} and the left panel of fig.\,\ref{fig:LogNoTest}.
    \item[{\it ii)}] We find that the power-series expansion $\beta_{\rm NG}^{(N)}$ in eq.\,(\ref{eq:ZetaSeries}) does converge on the full result for appropriate $N$. 
    The optimal-truncation order $N$ depends on the width $\sigma$ of the log-normal, cf. fig.\,\ref{fig:LogNoCompa}. 
    For instance, if $\sigma = 0.1$ we find that $N=4$ gives the correct result while if 
    $\sigma = 0.5$ one should push the perturbative expansion up to $N\gtrsim 10$.
    \item[{\it iii)}] When translated in terms of 
    the amplitude $A$ of the power spectrum, 
    the various orders of magnitude changes in the PBH mass fraction are reabsorbed by re-tuning of $A$, which is typically less than a factor of 2 in the range of  $r_{\rm dec}$ that we explored.
\end{itemize}

\subsubsection{The broad power spectrum}

We now consider the broad power spectrum of curvature perturbations in eq.\,(\ref{eq:Bro}). 
This kind of power spectrum, widely used as a benchmark model in theoretical studies,  recently attracted some attention also on the phenomenological side. As shown in ref.\,\cite{DeLuca:2020agl}, PBHs obtained assuming a broad power spectrum of the type in eq.\,(\ref{eq:Bro})  may comprise the totality of dark matter observed in the present-day Universe and, at the same time, generate, as a second-order effect, a detectable signal  of gravitational waves that matches, both in frequency and amplitude, the tentative signal recently reported by the NANOGrav Collaboration. 
We also refer to ref.\,\cite{Franciolini:2022pav} for an explicit implementation of the idea in the context of USR inflationary models. We will see an implementation of such broad spectrum in the context of Curvaton models in Chapter\,\ref{cap:Models}.
The computation in refs.\,\cite{DeLuca:2020agl,Franciolini:2022pav} does assume the absence of NGs of primordial origin (while it fully does include NGs from non-linearities). 
However, primordial NGs may be expected in concrete models that feature a broad power spectrum.

\begin{figure}[t]
\begin{center}\vspace{-2cm}
\includegraphics[width=.98\textwidth]{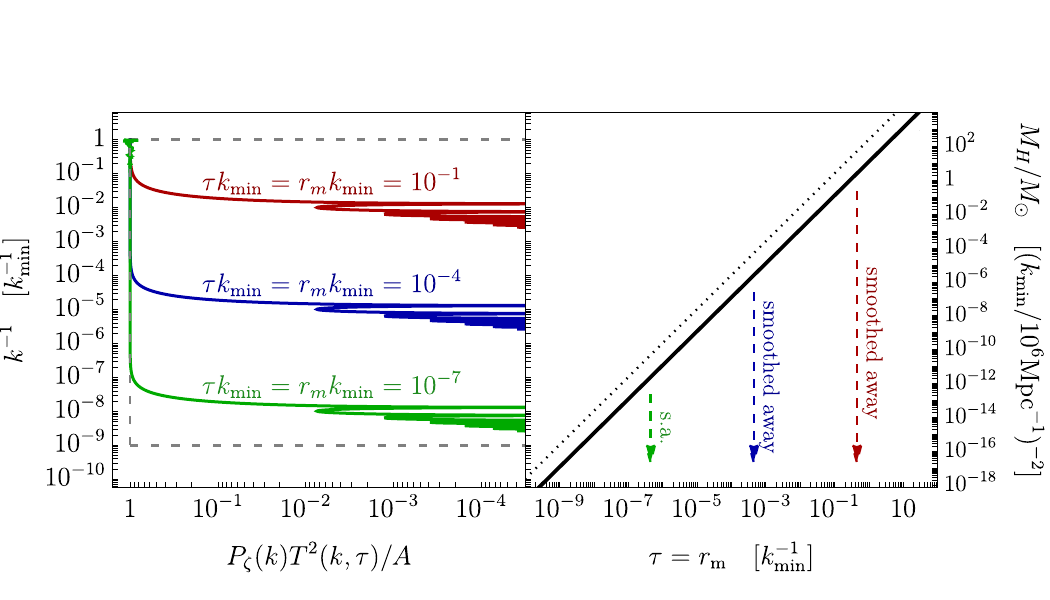}
\caption{
We plot (left-side $y$ axis, right panel) the relation $k^{-1} = r_m/\kappa$ in eq.\,(\ref{eq:Broadrk}) as function of $r_m$. Both length scales $k^{-1}$ and $r_m$ are written in units of $k_{\rm min}^{-1}$;
from eq.\,(\ref{eq:HorizonMass}), we read the corresponding horizon mass $M_H$ (rigth-side $y$ axis). 
We set $\tau = r_m$ and plot (left panel, rotated by 90$^{\circ}$ so to share the same $y$ axis with the right panel) the broad power spectrum in eq.\,(\ref{eq:Bro}) normalized to $1$, convoluted with  the 
the transfer function, and shown as function of
$k^{-1}$. 
The gray dashed line represents  the broad power spectrum without the transfer function.
We take $\Delta k = 10^9$, and consider three different choices of the comoving time-scale $\tau$.
As a whole, the figure shows that as time passes by one moves along the black line from left to right in the right panel and modes with 
decreasing $k$ (i.e. larger wavelengths and horizon mass $M_H$) enter the expanding cosmological horizon and are smoothed away by the transfer function. 
Consequently, at late times associated to the formation of heavier PBHs, the broad power spectrum effectively behaves as a peaked one (cf., for instance, the red curve in the left panel with $\tau = 10^{-1}k_{\rm min}^{-1}$) since the majority of modes, deep in the sub-horizon regime, have been smoothed away. 
 }\label{fig:SmoothedPS}  
\end{center}
\end{figure}

\begin{figure}[t]
\begin{center}
\includegraphics[width=1\textwidth]{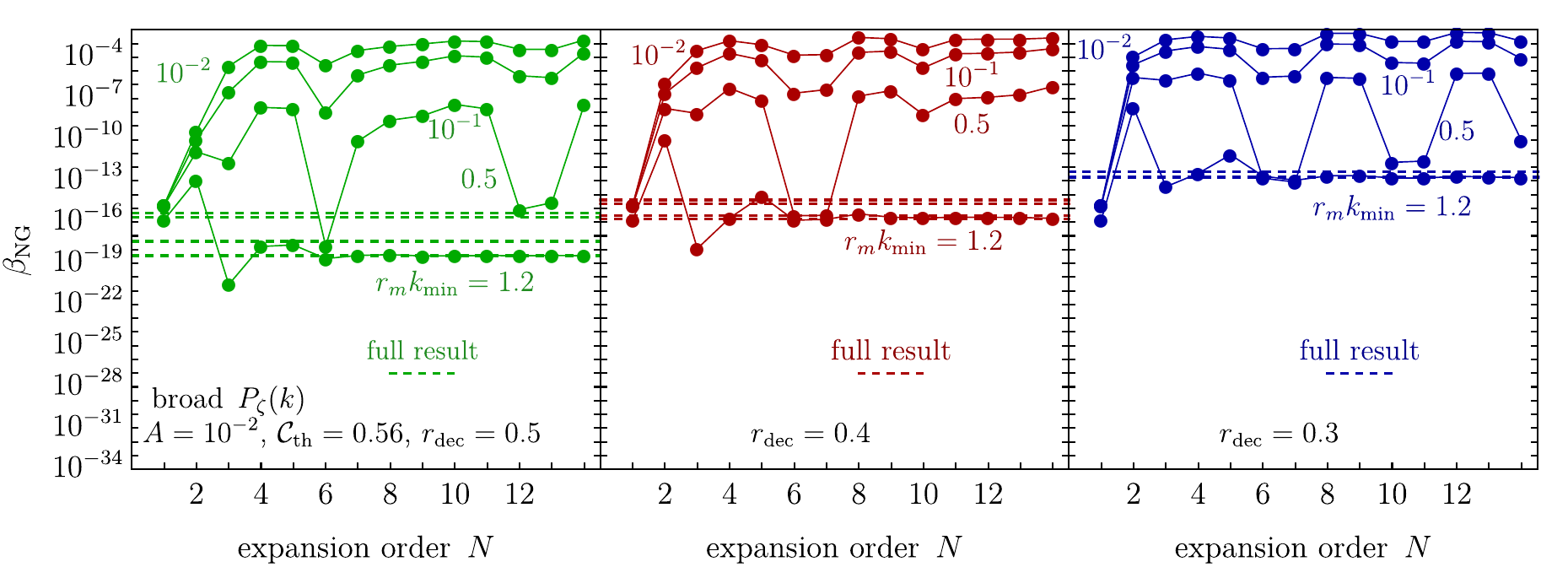}
\caption{
Left ($r_{\rm dec}=0.5$), central ($r_{\rm dec}=0.4$) and right ($r_{\rm dec}=0.3$) panels. 
We plot the NG mass fraction $\beta_{\rm NG}$ computed according to eq.\,(\ref{eq:CompactionIntegral}) and its  expansion $\beta_{\rm NG}^{(N)}$ in as function of the expansion order $N$. 
We consider different choices for $r_m k_{\rm min}$, with $\tau = R \equiv r_m$ in the linear transfer function and window function.
 }\label{fig:BroadCompa}  
\end{center}
\end{figure}

This is the case, for instance, of the curvaton model studied in ref.\,\cite{Inomata:2020xad,Ferrante:2023bgz} in which the curvature perturbation 
has the functional form in eq.\,(\ref{eq:MasterX}).
These considerations motivate the analysis proposed in this chapter.  
Motivated by the toy model for the power spectrum proposed in ref.\,\cite{DeLuca:2020agl}, we allow for a hierarchy 
 $k_{\rm max}\gg k_{\rm min}$ up to values $k_{\rm max}/k_{\rm min} = O(10^{9})$.
Let us illustrate our results, summarized in fig.\,\ref{fig:SmoothedPS} and in the three panels of fig.\,\ref{fig:BroadCompa}.

Furthermore, it is possible to relate the scale $r_m$ to the horizon mass $M_H$ 
via its relation with $k$  
\begin{align}
M_H \simeq 17 M_{\odot}
\left(
\frac{g_{\star}}{10.75}
\right)^{-1/6}\left(\frac{k/\kappa}{10^6 {\rm Mpc}^{-1}}\right)^{-2}
~~
\Longrightarrow
~~
r_m = \frac{\kappa}{k} \simeq 2.4\times 10^{-7} {\rm Mpc}
\left(
\frac{g_{\star}}{10.75}
\right)^{1/12}
\left(\frac{M_H}{M_{\odot}}\right)^{1/2},
\label{eq:HorizonMass}
\end{align}
that can be used to trade $r_m$ for $M_H$. 
In the right panel of fig.\,\ref{fig:SmoothedPS}, 
we show the relation $k^{-1} = r_m/\kappa$ in eq.\,(\ref{eq:Broadrk}) as function of $r_m$ (see caption). On the right-side $y$ axis, we show the horizon mass $M_H$ that corresponds to a given value of $k^{-1}$, cf. eq.\,(\ref{eq:HorizonMass}). We assume $g_{\star} = 106.75$ (radiation-dominated Universe).
We also show (left side of the plot, rotated by 90$^\circ$ to match the same $y$ axis with the right panel) the convolution $P_{\zeta}(k)T^2(k,\tau)$ for the broad power spectrum $P_{\zeta}(k)$ in eq.\,(\ref{eq:Bro}) as function of 
    $k_{\rm min}/k$ (and normalized to unit in amplitude). We plot the convoluted power spectrum for different values of 
    $\tau k_{\rm min}$. 
    This plot shows the effect of the linear transfer function on the broad power spectrum: sub-horizon modes with $\tau k \gg 1$ are suppressed. 
    Intuitively, what happens is that as time passes by (that is, for increasing $\tau$, going from the bottom green line with $\tau k_{\rm min} = 10^{-7}$ to the top red one with $\tau k_{\rm min} = 10^{-1}$) more and more modes with decreasing $k$ (thus larger comoving wavelengths) re-enter the horizon 
    and are subject to pressure gradients which effectively deplete their corresponding power. 
    A very similar effect is provided by the window function $P_{\zeta}(k)W^2(k,R)$: 
    modes with $kR\gg 1$ are smoothed away by the volume average. 
    
The above considerations are very interesting since they imply that, if we set $\tau = R \equiv r_m$ and consider the case $r_m k_{\rm min} = O(1)$, almost all modes of the power spectrum are suppressed, and 
the latter effectively looks like a very peaked one. 
This point is crucial since we argued that the broadness of the power spectrum controls the convergence of the power-series expansion. 

In the three panels of fig.\,\ref{fig:BroadCompa} we compare the NG PBHs mass fraction $\beta_{\rm NG}$ in eq.\,(\ref{eq:CompactionIntegral})
with its expansion $\beta_{\rm NG}^{(N)}$ as function of $N$. 
We consider different choices of $r_m k_{\rm min}$. 
If we take $r_mk_{\rm min} = O(1)$, we find that $\beta_{\rm NG}^{(N)}$ converges to the full result given by $\beta_{\rm NG}$ for $N$ large enough. 
This is expected because, as explained before, for $r_m k_{\rm min} = O(1)$ the broad power spectrum is effectively peaked since the majority of its modes have $r_m k \gg 1$ and are smoothed away. 
However, if we consider smaller and smaller values of $r_m k_{\rm min}$ then an increasing number of modes contributes to the variances, and the convergence of $\beta_{\rm NG}^{(N)}$ gets quickly lost. 
In this situation the computation of $\beta_{\rm NG}^{(N)}$ based on the power-series expansion is simply wrong (the series does not converge for whatever $N$) and one is forced to use the full result $\beta_{\rm NG}$. 
We are eventually interested in the computation of the PBHs mass fraction at different scales (that  is for different values of the PBH mass). 
In other words, if we combine eq.\,(\ref{eq:Broadrk}) and eq.\,(\ref{eq:HorizonMass}) we find
\begin{align}
    r_m k_{\rm min} \approx \left(\frac{M_H}{M_{\odot}}\right)^{1/2}
    \left(\frac{k_{\rm min}}{10^6\,{\rm Mpc}^{-1}}\right)\,,
\end{align}
and the computation of the abundance entails scanning over all the relevant horizon masses $M_H$. 
Only for values of $M_H$ such that above relation gives $r_m k_{\rm min} = O(1)$ we expect the perturbative computation based on $\beta_{\rm NG}^{(N)}$ to give reliable results; 
however, as soon as one takes smaller and smaller values for the horizon mass $M_H$ such that 
$r_m k_{\rm min} \ll 1$, we expect that the computation of $\beta_{\rm NG}^{(N)}$ based on the power-series expansion $\zeta_N$ is not reliable, for whatever expansion order $N$.  
The reason is precisely the one discussed in section\,\ref{sec:TheBroad}:
if the power spectrum is very broad (that is the case with $r_m k_{\rm min} \ll 1$), we go outside the radius of convergence of the series and beyond the range of validity of the power-series expansion $\zeta_N$.
For this reason, we strongly suggest to abandon the computation based on $\beta_{\rm NG}^{(N)}$ and focus on the full result described by $\beta_{\rm NG}$.

\begin{figure}[t]
\begin{center}
\includegraphics[width=0.495\textwidth]{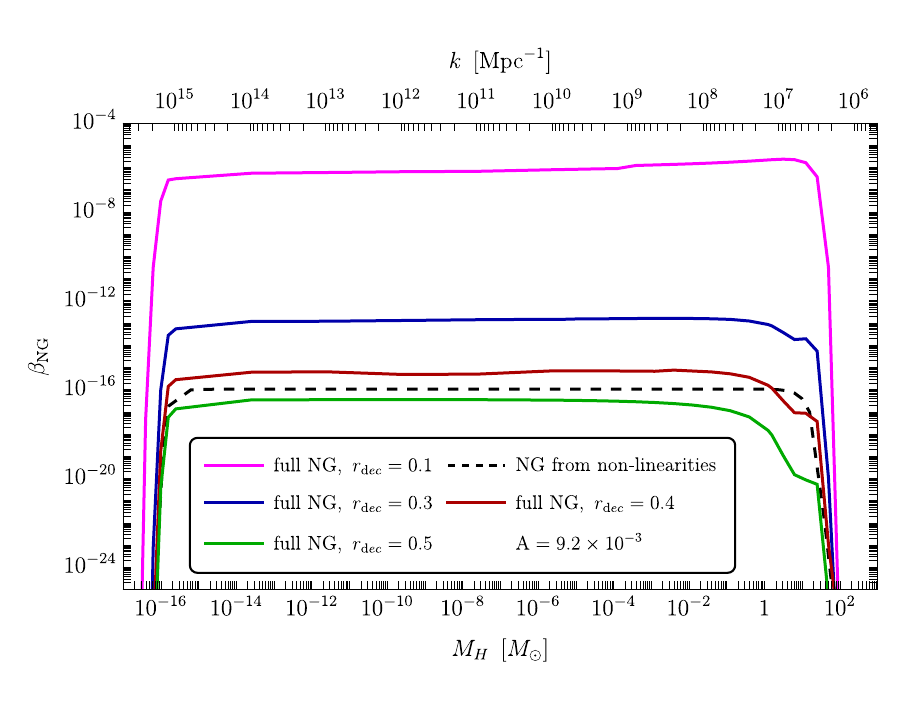}
\includegraphics[width=0.495\textwidth]{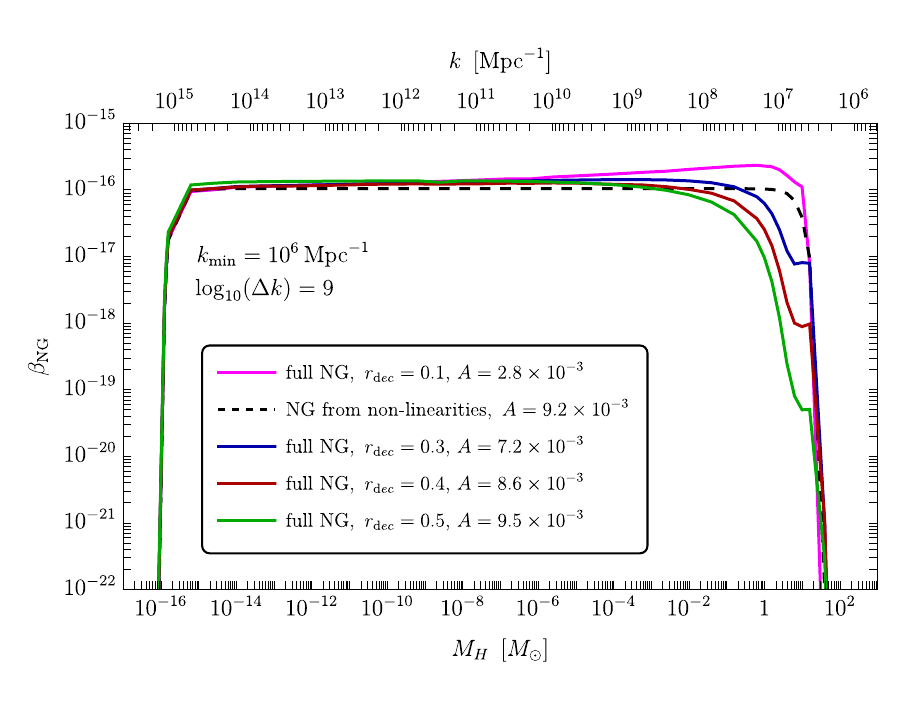}
\caption{
Computation of $\beta_{\rm NG}$ as a function of the horizon mass $M_H$ for different values of $r_{\rm dec}$. 
We fix $k_{\rm min} =10^6$ ${\rm Mpc}^{-1}$ and $\Delta k = 10^{9}$. 
The dashed black line corresponds to eq.\,(\ref{eq:CompactionIntegral}) in which we only include the effect non-linearities (i.e. assuming the absence of primordial NGs).
\textbf{\textit{Left panel.}}
The amplitude $A = 9.2 \times 10^{-3}$ is fixed in order to get $\beta_{\rm NG} \simeq 10^{-16}$ at $M_H = 10^{-15}\,\,M_{\odot}$ in the presence of only non-linearities.
Keeping the amplitude fixed, we show how several values of $r_{\rm dec}$ influence $\beta_{\rm NG}$. 
\textbf{\textit{Right panel.}}
For each $r_{\rm dec}$ we tune the amplitude $A$ in such a way that
$\beta_{\rm NG} \simeq 10^{-16}$ at $M_H \simeq 10^{-15}\,\,M_{\odot}$.
 }\label{fig:BroCompa2}  
\end{center}
\end{figure}

After having gained intuition on the convergence as a function of the smoothing scale when broad spectra are considered, let us analyze the impact of NGs from the curvaton model on the computation of $\beta_{\rm NG}$ as a function of the horizon mass $M_H$ in eq.\,(\ref{eq:HorizonMass}). 
This is shown in fig.\,\ref{fig:BroCompa2}. 
We explore the benchmark values $r_{\rm dec}=0.1,0.3,0.4,0.5$. 
As far as the power spectrum is concerned, 
we fix $k_{\rm min} = 10^6$ ${\rm Mpc^{-1}}$ and $\Delta k = 10^{9}$ while the amplitude $A$ is tuned, for each $r_{\rm dec}$
in such a way that we get $\beta_{\rm NG} \simeq 10^{-16}$ at $M_H = 10^{-15}\,\,M_{\odot}$. 
This choice is motivated by the fact that the above value of $\beta_{\rm NG}$ is what is typically needed in order to get a sizable abundance of asteroid-mass PBHs. On the top $x$-axis, we convert $M_H$ into $k$ according to eq.\,(\ref{eq:HorizonMass}) to highlight the relevant scales in $[{\rm Mpc^{-1}}]$.

We emphasize two important points:
\begin{itemize}
\item[$\circ$] 
First of all, 
we find that increasing the level of NG (that is decreasing the value of $r_{\rm dec}$ from $0.5$ to $0.1$) has the net effect of enhancing the overall mass fraction $\beta_{\rm NG}$. 
More explicitly, in fig.\,\ref{fig:BroCompa2}, in order to fit the same reference value $\beta_{\rm NG} \simeq 10^{-16}$, the amplitude $A$ decreases from 
$A\simeq 9.5\times 10^{-3}$ (for $r_{\rm dec} = 0.5$) to $A\simeq 2.8\times 10^{-3}$ (for $r_{\rm dec} = 0.1$). This trend is analogous to the one observed with a positive $f_{\rm{NL}}$ in a large PBH abundance.
\item[$\circ$] 
The second effect is more subtle, and regards the behavior of $\beta_{\rm NG}$ near the cut-off at large horizon mass. Looking at fig.\,\ref{fig:BroCompa2}, it is evident that the right-side of the distribution is altered by NG if compared to the part of $\beta_{\rm NG}$ at smaller $M_H$. 
In other words, the primordial NG breaks the $M_H$-independence of $\beta_{\rm NG}$ that would be expected on the basis of the scale invariance of eq.\,(\ref{eq:Bro}) (and that is verified, for instance, by the computation that only includes NG from non-linearities, cf. the black dashed line in fig.\,\ref{fig:BroCompa2} that remains approximately constant as function of $M_H$).
This is an interesting effect that is worth investigating.

\end{itemize}

\subsubsection{On the breaking of scale invariance}

\begin{figure}[t]
\begin{center}
\includegraphics[width=0.495\textwidth]{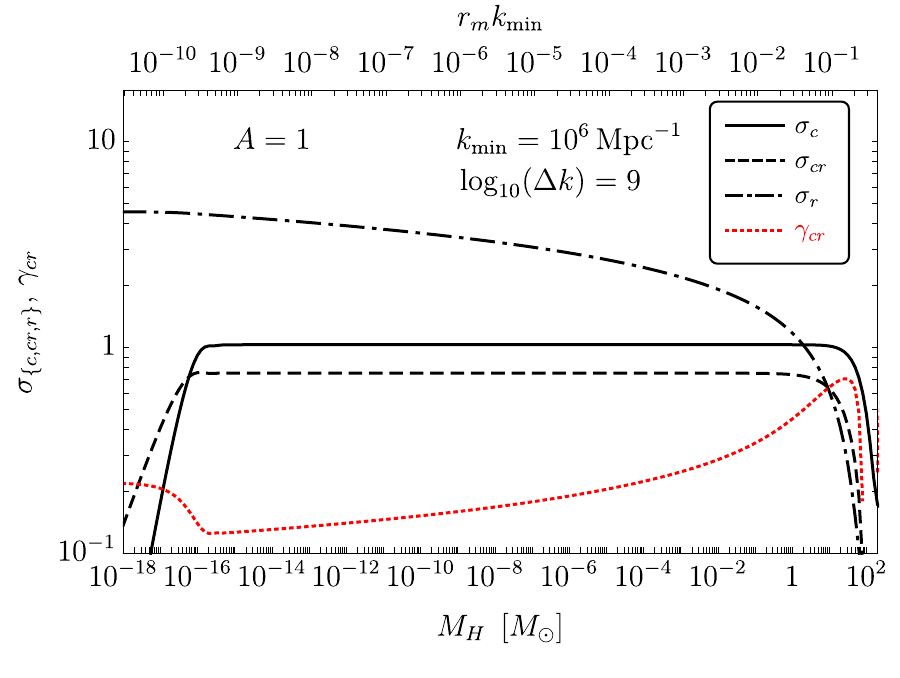}
\includegraphics[width=0.495\textwidth]{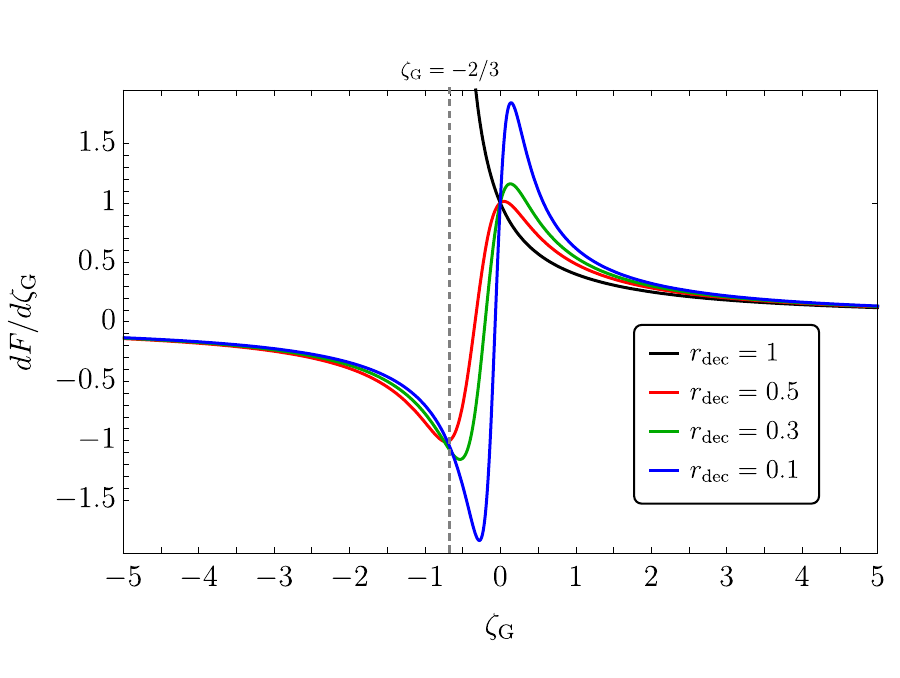}
\caption{
\textbf{\textit{Left panel.}} 
Variances $\sigma_{c,cr,r}$ in eqs.\,(\ref{eq:Var1}-\ref{eq:Var3}) as function of the horizon mass $M_H$. We take the parameters 
$k_{\rm min}= 10^{6}$ ${\rm Mpc^{-1}}$  and $\Delta k=10^{9}$, and normalize to 1 the amplitude of the power spectrum (we remind that the variances $\sigma_{c,cr,r}$ simply scale as $\sqrt{A}$). We also plot the correlation parameter $\gamma_{cr}$ (which, on the contrary, does not depend on $A$).
\textbf{\textit{Right panel.}} 
Functional dependence of $dF/d\zeta_{\rm G}$ on the Gaussian variable $\zeta_{\rm G}$ for different values of $r_{\rm dec}$. We consider the explicit case of the curvaton field.
 }\label{fig:discussion_broadinv_intro}  
\end{center}
\end{figure}

\begin{figure}[ht]\label{fig:discussion_broadinv}
\begin{center}
\includegraphics[width=1\textwidth]{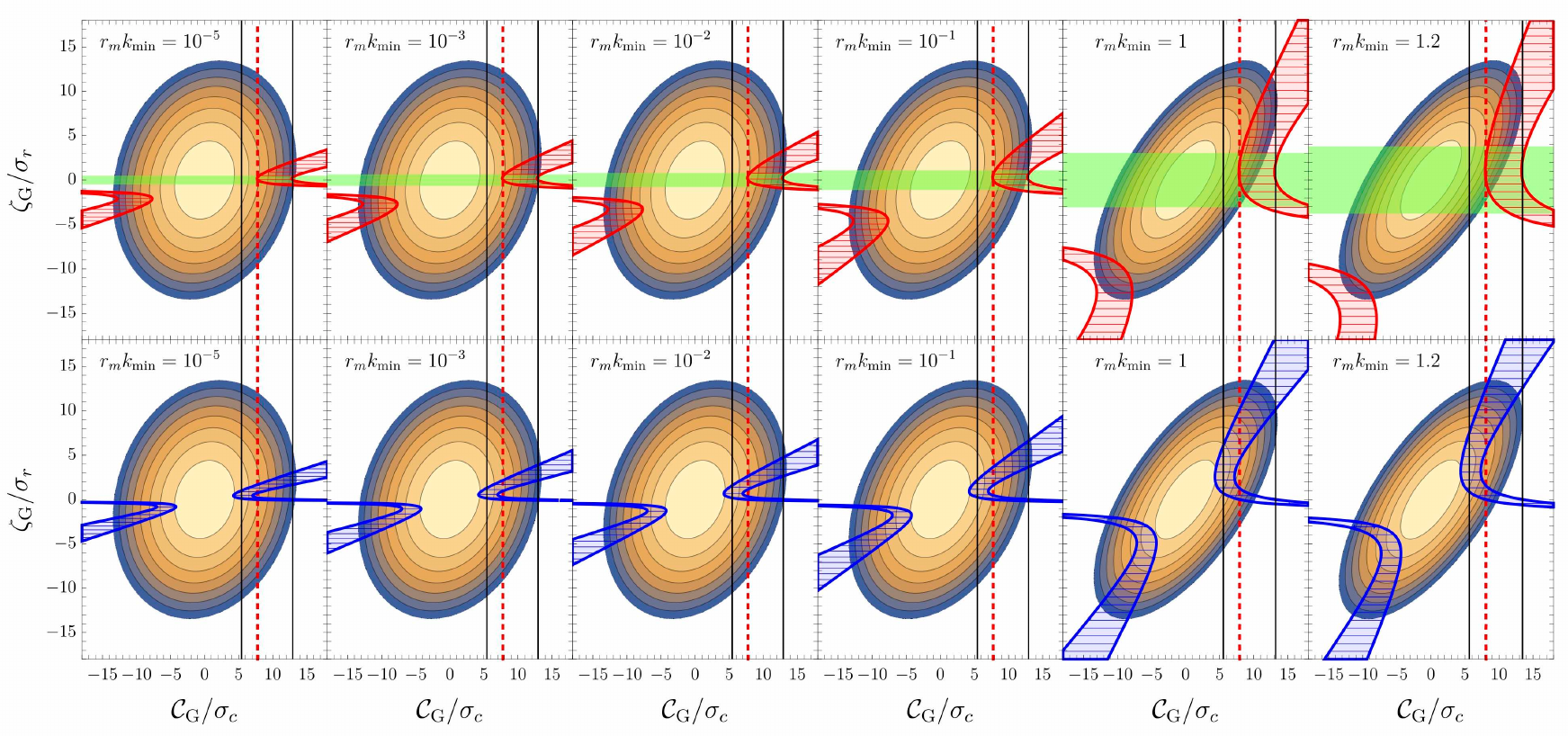}
\caption{
\textbf{\textit{Top row.}} Contour lines indicate values of the probability density $P_G\left(\mathcal{C}_G, \zeta_G\right)$. From the inner circle to the last purple one we locate values equally spaced in log scale in between $10^{-5}$ and $10^{-40}$. The region between the red solid lines is the parameter space defined as $\mathcal{D}$ in eq. (58). This is the region over which the $\operatorname{PDF} P_G\left(\mathcal{C}_G, \zeta_G\right)$ is integrated in the computation of the PBH mass fraction as it corresponds to perturbations above the threshold. Inside the horizontal region marked in green the power-series expansion in eq. (13) does converge to the full resummed result in eq. (4). The region limited by the two vertical black lines is defined by the condition $\mathcal{C}_{\mathrm{th}}<\mathcal{C}_{\mathrm{G}}<4 / 3$ (pure Gaussian case) while the region on the right side of the vertical dashed red line is limited by the condition $\mathcal{C}_{\mathrm{th}}<\mathcal{C}_{\mathrm{G}}-(3 / 8) \mathcal{C}_{\mathrm{G}}^2<2 / 3$ (thus including $N G$ from non-linearities). Notice that the right-edge of the above boundary falls outside the range of the $x$-axis. We consider the benchmark value $r_{\mathrm{dec}}=0.5$ while $k_{\min }=10^6 \mathrm{Mpc}^{-1}$ and $\Delta k=10^9$. The amplitude of the curvature power spectrum is fixed to the value $A=10^{-2}$. \textbf{\textit{Bottom row.}} We consider $r_{\mathrm{dec}}=0.1$.}
\end{center}
\end{figure}

In the computation of the PBH mass fraction, the $M_H$-dependence enters implicitly in eq.\,(\ref{eq:CompactionIntegral}) via the variances $\sigma_{c,cr,r}^2$ in eqs.\,(\ref{eq:Var1}-\ref{eq:Var3}). 
In the left panel of fig.\,\ref{fig:discussion_broadinv_intro}, we show the three variances $\sigma_{c,cr,r}^2$, together with their combination $\gamma_{cr}$ in 
eq.\,(\ref{eq:GammacrDef}), as function of the horizon mass $M_H$.  
The figure shows that 
while $\sigma_{c}$ and $\sigma_{cr}$ remain constant (apart from threshold effects) for a scale invariant spectrum, the value of $\sigma_{r}$ changes drastically within the first decade in $r_m k_{\textrm{min}}$ (looking at the plot starting from the right-side at $r_m k_{\textrm{min}} = O(1)$ values, see labels on the top $x$-axis) while only logarithmically in the remaining decades.
The $M_H$-dependence of $\sigma_{c}$, therefore, reflects the 
$M_H$-dependence observed in the right panel of fig.\,\ref{fig:BroCompa2} with the value of $\beta_{\rm NG}$ as function of $M_H$ becoming nearly $M_H$-invariant going towards smaller horizon mass. 
\\We remind that  $\sigma_{r}$ is the variance of the Gaussian variable $\zeta_{\rm G}$, cf. eq.\,(\ref{eq:Var3}). 
In light of the above result, it is, therefore, crucial to better understand how $\sigma_r$ enters in the computation of the PBHs mass fraction.
Consider again  eq.\,(\ref{eq:CompactionIntegral}). The change of variable 
$\zeta_{\rm G}\to \nu_{\rm G}\equiv \zeta_{\rm G}/\sigma_{r}$ seems to completely reabsorb the explicit dependence on $\sigma_{r}$. However, and crucially, 
in the presence of primordial NGs 
$\sigma_{r}$ appears back through the term $dF/d\zeta_{\rm G}$ in the non-linear relation in eq.\,(\ref{eq:CCgau}). Furthermore, 
 the coefficient $\gamma_{cr}$, which measures the correlation between $\zeta_{\rm G}$ and $\mathcal{C}_{\rm G}$, will also retain some $M_H$ dependence. 
 These simple observations suggest that a full understanding of the 
 effect of NGs requires to consider both the functional dependence of 
 $dF/d\zeta_{\rm G}$ on the Gaussian variable $\zeta_{\rm G}$ 
 and the two-dimensional probability distribution $\textrm{P}_{\rm G}(\mathcal{C}_{\rm G},\zeta_{\rm G})$.

First, consider $dF/d\zeta_{\rm G}$; we focus on the case of the curvaton, 
eq.\,(\ref{eq:MasterX}). 
In the right panel of fig.\,\ref{fig:discussion_broadinv_intro}, we show the behavior of $dF/d\zeta_{\rm G}$ as function of $\zeta_{\rm G}$ for different values of $r_{\rm dec}$. 
For $r_{\rm dec}<1$,
$dF/d\zeta_{\rm G}$ transits from small negative to small positive values through an heartbeat transition located at around $\zeta_{\rm G} \simeq 0$. 
The case $r_{\rm dec} = 1$ is somewhat special; in this case, 
$dF/d\zeta_{\rm G}$ is positive and 
grows for decreasing $\zeta_{\rm G}$ with the latter that is subject to the condition  $\zeta_{\rm G} > - 2/3$. 
This restriction is a consequence of 
the presence of a branch-point singularity that, for $r_{\rm dec} = 1$, is located on the real axis (while for 
$r_{\rm dec}<1$ it acquires an imaginary part and moves away from the real axis, cf. appendix\,\ref{app:Radius}).
\\In fig. 2.10 we show iso-contours of the two-dimensional probability distribution $\textrm{P}_{\rm G}(\mathcal{C}_{\rm G},\zeta_{\rm G})$. 
From left to right, different panels correspond to increasing  values of $r_m k_{\rm min}$. This is to capture the effect of the $M_H$-dependence on $\gamma_{cr}$. 
For $r_m k_{\rm min} = 1$ (right-most panel, large horizon mass) the power spectrum is very peaked and the correlation among 
$\mathcal{C}_{\rm G}$ and $\zeta_{\rm G}$ maximal ($\gamma_{cr} \to 1$, cf. the left panel of fig.\,\ref{fig:discussion_broadinv_intro}). For smaller $r_m k_{\rm min}$ (broader power spectrum and small horizon mass), the level of correlation decreases and the ellipses in fig.$2.10$ rotate towards a diagonal configuration ($\gamma_{cr} \to 0$).
The top (bottom) row refers to $r_{\rm dec} = 0.5$ ($r_{\rm dec} = 0.1$).
The regions shaded in red (top row) and blue (bottom row) in  fig. 2.10 are selected by the condition 
$ \mathcal{D} = 
\left\{
\mathcal{C}_{\rm G},\,\zeta_{\rm G} \in \mathbb{R}~:~~
\mathcal{C}(\mathcal{C}_{\rm G},\zeta_{\rm G}) > \mathcal{C}_{\rm th}  
~\land~\mathcal{C}_1(\mathcal{C}_{\rm G},\zeta_{\rm G}) < 2\Phi
\right\}$ (cf. eq.\,(\ref{eq:RegionD})) in the presence of both NGs and non-linearities. 
\\For comparison we also show: the region defined by the condition $\mathcal{C}_{\rm th} < \mathcal{C}_{\rm G} < 4/3$ and limited by the two vertical black lines (pure Gaussian case) and  the region defined by the condition $\mathcal{C}_{\rm th} < \mathcal{C}_{\rm G} - (3/8)\mathcal{C}_{\rm G}^2< 2/3$ that lies on the right-side of the dashed red line is (including NG from non-linearities; notice that the right-side of this constraint lies outside the range of the $x$-axis).

We are in the position to combine the two effects. For simplicity, we start from the benchmark case with  $r_{\rm dec} = 0.5$. 

Many important lessons can be drawn.
\begin{itemize}
    \item[{\it i)}] First of all we notice that, 
    going from the pure Gaussian case to the case in which only NGs from non-linearities 
    are included, the region over which the PDF 
    is integrated shifts towards larger values of  $\mathcal{C}_{\rm G}$ (where the PDF is smaller). Consequently,  
    including non-linearities decreases the PBH abundance in agreement with what found in ref.\,\cite{Young:2019yug,DeLuca:2019qsy}.
    \item[{\it ii)}] Second, we recover, from a different perspective, what we have already learned about the validity of the perturbative approach. 
  As discussed before, the perturbative approach is applicable in a very small range of values for $\zeta_{\rm G}$ centered around $\zeta_{\rm G} = 0$ and  within the radius of convergence of the power series expansion ($-0.173< \zeta_{\rm G} < +0.173$ for $r_{\rm dec} = 0.5$, cf. appendix\,\ref{app:Radius}). 
  If we consider 
  $r_m k_{\rm min} = O(1)$ (right-most panel in fig.2.10) we see that part of the region that is selected by the condition $(\mathcal{C}_{\rm G},\,\zeta_{\rm G})\in \mathcal{D}$ (and over which we integrate the PDF in eq.\,(\ref{eq:CompactionIntegral})) lies away from the convergence of the power series expansion but mostly  overlaps with negligible probability; on the contrary, the small part of the overlapping region in which the 
  probability is sizable (that is, 
  within the ellipses colored in yellow) lies within the convergence condition. 
  Consequently, as expected, 
  we find that for $r_m k_{\rm min} = O(1)$ the region inside the radius of convergence of the power-series expansion captures the relevant contribution to the PBH formation probability.
 
  On the contrary, if we move in fig. 2.10 towards smaller 
  $r_m k_{\rm min}$ we see that already for $r_m k_{\rm min} = 10^{-1}$ 
  the region selected by the condition $\mathcal{C}(\mathcal{C}_{\rm G},\zeta_{\rm G}) > \mathcal{C}_{\rm th}$ 
  has large probability 
  $\textrm{P}_{\rm G}(\mathcal{C}_{\rm G},\zeta_{\rm G})$ outside the radius of convergence thus invalidating, as expected, the applicability of the perturbative approach.
  \item[{\it iii)}] 
  In the presence of primordial NGs, we expect a smaller abundance of PBHs compared to the case in which only non-linearities are included. 
  This is indeed confirmed by the computation in the left panel of fig.\,\ref{fig:BroCompa2} (the line corresponding to the mass fraction with $r_{\rm dec} = 0.5$ lies below the dashed black line).   
  This is due to the effect of NGs in the non-linear relation,  eq.\,(\ref{eq:CCgau}). 
  There are two key points.
  
The first one is that the constraint $\mathcal{D}$ in eq.\,(\ref{eq:RegionD}) admits the solution (with $\Phi =2/3$)
\begin{align}
\frac{4(1-\sqrt{1-3\mathcal{C}_{\rm th}/2})}{3} 
< \mathcal{C}_{\rm G}\frac{dF}{d\zeta_{\rm G}} <
\frac{4}{3}\,,\label{eq:NGConstraint}
\end{align} 
that is a trivial generalization of the one we get in the case in which only non-linearities are included, 
$4(1-\sqrt{1-3\mathcal{C}_{\rm th}}/2)/3 
< \mathcal{C}_{\rm G} <
4/3$.  
However, eq.\,(\ref{eq:NGConstraint}) now has two branches depending on the sign of $dF/d\zeta_{\rm G}$
 since, with the only exception of the special case with $r_{\rm dec} = 1$, $dF/d\zeta_{\rm G}$ takes both positive and negative values (cf. fig.\,\ref{fig:discussion_broadinv_intro}, right panel). 
 As shown in fig.$2.10$ (the two regions in red and blue) both these possibilities are realized with negative 
 $dF/d\zeta_{\rm G}$ (thus negative $\zeta_{\rm G}$, cf. fig.\,\ref{fig:discussion_broadinv_intro}, right panel) that requires negative $\mathcal{C}_{\rm G}$ and viceversa.
 
 The second key point is the following.
 Looking at the right panel of fig.\,\ref{fig:discussion_broadinv_intro}, we notice that 
 $|dF/d\zeta_{\rm G}| \leqslant 1$ for $r_{\rm dec} = 0.5$. 
 More in detail, we find that 
 $dF/d\zeta_{\rm G} \simeq 1$ for $\zeta_{\rm G} \simeq 0.05$, 
 $dF/d\zeta_{\rm G} \simeq -1$ for $\zeta_{\rm G} \simeq -0.7$
 with 
 $|dF/d\zeta_{\rm G}|$ that becomes smaller as $|\zeta_{\rm G}|$ increases. 
 Consequently, as we move towards larger $|\zeta_{\rm G}|$, the variable $\mathcal{C}_{\rm G}$ is forced -- in order to compensate in eq.\,(\ref{eq:NGConstraint})  the change of $dF/d\zeta_{\rm G}$  -- to take larger values compared with the non-linear case. As a direct consequence, the 
 part of the parameter space that is selected by eq.\,(\ref{eq:NGConstraint}) falls in a region with smaller PDF thus smaller PBH mass fraction.
 
 \item[{\it iv)}] The discussion at point {\it iii)} immediately explains also why we get a larger PBH mass fraction for some other values of $r_{\rm dec}$. 
 Consider, for instance, the case $r_{\rm dec} = 0.1$. 
 From the left panel of fig.\,\ref{fig:BroCompa2} we know that in this case the mass fraction of PBHs is much larger compared to the pure non-linear case. 
 This is because, as evident from the right panel of fig.\,\ref{fig:discussion_broadinv_intro}, in this case it is possible to have $|dF/d\zeta_{\rm G}| > 1$ in a small range of values of $\zeta_{\rm G}$ corresponding to the heartbeat transition. Consequently, when $|dF/d\zeta_{\rm G}| > 1$ the variable $\mathcal{C}_{\rm G}$ takes smaller values thus falling in a region in which the PDF is larger. This is evident by inspecting in fig.2.10 the interplay between the ellipses of constant PDF with the region selected by the integration constraint $\mathcal{D}$.
\end{itemize}

To sum up, there are two important take-home messages we learn from the above discussion. 
\\{\it a)} In the presence of primordial NG, $\beta_{\rm NG}$ depends on $M_H$ (that is, the
formation of PBHs of various masses does not happen with equal probability as na\"{\i}vely expected on the basis of a scale-invariant power spectrum). 
This is a model-independent statement.
\\{\it b)} The size and impact of the effect, on the contrary, is model-dependent, and controlled by $dF/d\zeta_{\rm G}$. We explored the physics-case of the curvaton field but we stress that our analysis can be easily applied, step-by-step, to other physical models.

\subsection{Computation of the PBH mass distribution}
The mass distribution of PBHs at the end of the formation era is directly derived from the mass fraction $\beta_{\rm NG}$ as
\begin{align}
f_{\rm PBH}(M_{\rm PBH}) 
\equiv
\frac{1}{\Omega_{\rm  DM}} 
\frac{d\Omega_{\rm PBH}}
{d\log M_{\rm PBH}}\,,
~~
{\rm with}
~~
\Omega_{\rm PBH} = 
\int
d \log M_{H} \left(\frac{M_{\rm eq}}{M_{H}} \right)^{1/2}\beta_{\rm NG}(M_{H})\,,
\label{eq:diffmassfraction}
\end{align}
where $M_{\rm eq} \approx 2.8\times 10^{17}\,\,M_{\odot}$ is the horizon mass at the time of matter-radiation equality and 
$\Omega_{\rm  DM}$ is the cold dark matter density of the Universe ($\Omega_{\rm  DM} \simeq 0.12\,h^{-2}$ with $h = 0.674$ for the Hubble parameter). 
Our task is the generalization of the computation carried out in ref.\,\cite{Young:2019yug,DeLuca:2019qsy,Franciolini:2022tfm}, that only includes non-linearities and assumes gaussian primordial curvature perturbations, to the case in which local primordial NG are also present with the generic functional form $\zeta = F(\zeta_{\rm G})$. 

The differential fraction of PBHs with mass $M_{\rm PBH}$ over the totality of dark matter at present day can be computed by rearranging the integration \eqref{eq:CompactionIntegral}.
Requiring over-threshold perturbations ${\cal C} \geq {\cal C}_{\rm th}$, and using eq.\,\eqref{eq:CCgau}, one finds that the critical values of the linear component ${\cal C}_{\mathrm G}$ are
\begin{equation}
\mathcal{C}_{\mathrm{G}, \mathrm{th}, \pm}=2 \Phi\left(\frac{d F}{d \zeta_{\mathrm{G}}}\right)^{-1}\left(1 \pm \sqrt{1-\frac{\mathcal{C}_{\mathrm{th}}}{\Phi}}\right)
\end{equation}
so we limit our integration to values in the range
\begin{equation}
\mathcal{C}_{\mathrm{G}, \mathrm{th},-} \leq \mathcal{C}_{\mathrm{G}} \leq 2 \Phi\left(\frac{d F}{d \zeta_{\mathrm{G}}}\right)^{-1},
\end{equation}
where the minus sign was chosen as we focus only on the type-I branch~\cite{Musco:2020jjb} of solutions.
Also, by inverting the definition of mass during the critical collapse,
we get the explicit relation between ${\cal C} $ and the horizon mass $M_H$ as
\begin{equation}
M_{\mathrm{PBH}}=\mathcal{K} M_H\left[\left(\mathcal{C}-\frac{1}{4 \Phi} \mathcal{C}^2\right)-\mathcal{C}_{\mathrm{th}}\right]^\gamma \quad \text { or } \quad \mathcal{C}_{\mathrm{G}}=2 \Phi\left(\frac{d F}{d \zeta_{\mathrm{G}}}\right)^{-1}\left[1-\sqrt{1-\frac{\mathcal{C}_{\mathrm{th}}}{\Phi}-\frac{1}{\Phi}\left(\frac{M_{\mathrm{PBH}}}{\mathcal{K} M_H}\right)^{1 / \gamma}}\right]
\end{equation}
Requiring the argument of the square root to be real coincides with the requirement for threshold perturbations.
Also, the condition ${\cal C} \leq 2 \Phi$ indirectly selects a maximum mass $M_{\rm PBH}$ that can be formed at each epoch, where the epoch of formation is parametrized by $M_H$.
With this change of variable, we can express the integration over $ d{\cal C}_{\rm G} d \zeta_{\rm G}$ in eq.\,\eqref{eq:CompactionIntegral} in terms of $d{M}_{\rm PBH}d \zeta_{\rm G}$, and then consider the differential mass fraction \eqref{eq:diffmassfraction}. 
Finally, the abundance of a given PBH mass $M_{\rm PBH}$ comes out of the integration across the possible epochs of formation, parametrized by $M_H$. Therefore, we obtain the following master formula
\begin{equation}\label{eq:MasterAbundance}
\begin{array}{r}
f_{\mathrm{PBH}}\left(M_{\mathrm{PBH}}\right)=\frac{1}{\Omega_{\mathrm{DM}}} \int_{M_{\mathrm{H}}^{\min }\left(M_{\mathrm{PBH}}\right)} d \log M_{\mathrm{H}}\left(\frac{M_{\mathrm{eq}}}{M_{\mathrm{H}}}\right)^{1 / 2}\left[1-\frac{\mathcal{C}_{\mathrm{th}}}{\Phi}-\frac{1}{\Phi}\left(\frac{M_{\mathrm{PBH}}}{\mathcal{K} M_H}\right)^{1 / \gamma}\right]^{-1 / 2} \\
\quad \times \frac{\mathcal{K}}{\gamma}\left(\frac{M_{\mathrm{PBH}}}{\mathcal{K} M_{\mathrm{H}}}\right)^{\frac{1+\gamma}{\gamma}} \int d \zeta_{\mathrm{G}} P_{\mathrm{G}}\left(\mathcal{C}_{\mathrm{G}}\left(M_{\mathrm{PBH}}, \zeta_{\mathrm{G}}\right), \zeta_{\mathrm{G}} \mid M_{\mathrm{H}}\right)\left(\frac{d F}{d \zeta_{\mathrm{G}}}\right)^{-1},
\end{array}
\end{equation}
where the integrand also includes the determinant of the Jacobian and the horizon mass dependence of 
the PDF $P_{\rm G}(M_{\rm H})$ 
is inherited by the smoothing scale $r_m(M_H)$ controlling the variances 
$(\sigma_{c},\sigma_{cr}, \sigma_{r})$, fixing the horizon crossing epoch.
As shown in\,\cite{Franciolini:2022tfm,Musco:2023dak}, $\gamma(M_H)$, $\mathcal{K}(M_H)$, $\delta_c(M_H)$ and $\Phi(M_H)$ are functions of the mass of the horizon for epochs close to the QCD phase transitions (around the formation of solar mass PBHs). 
At that epoch, we also account for the variation of $g_{\star}(M_H)$ relating spectral modes to the horizon mass $M_H$ in eq.\,\eqref{eq:HorizonMass} \cite{PhysRevD.81.104019}.
The total fraction of PBHs is given by the integral 
\begin{equation}
f_{\rm PBH}=\int f_{\rm PBH}(M_{\rm PBH})d\log M_{\rm PBH}.
\end{equation}

\begin{figure}[t]
	\begin{center}
\includegraphics[width=.99\textwidth]{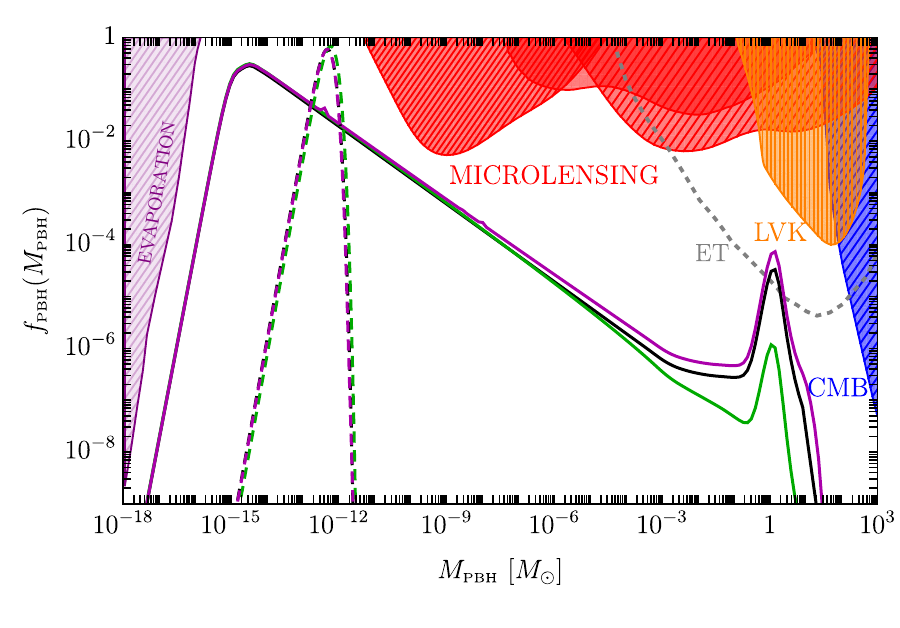}
		\caption{
Mass function resulting from a broad power spectrum (solid lines, with $k_{\rm min}=10^{6}$ $ {\rm Mpc}^{-1}$ and $ \Delta k =10^{8.5}$ $ {\rm Mpc}^{-1}$) and from a Log-Normal power spectrum (dashed lines, with $\sigma=0.5$ and $k_{\rm peak}=6 \times 10^{12}$ $ {\rm Mpc}^{-1}$) with different amount of NGs, i.e. values of $r_{\rm dec}$. 
The amplitude of the power spectrum has been re-scaled for each value of $r_{\rm dec}$ such that PBHs comprise the totality of dark matter. 
The most stringent experimental constraints are shown (for a deeper description of the constraints see Sec.\,\ref{sec:Clas}). Minimum testable abundance from PBH mergers with the Einstein Telescope as derived in \cite{DeLuca:2021hde} (see also \cite{Ng:2022agi,Martinelli:2022elq}).  
}\label{fig:fPBH1} 
	\end{center}
\end{figure}

In fig.\,\ref{fig:fPBH1} we show mass function for different power spectrum ansatz. As we can see from this plot with a fine-tuning of the amplitude of the power spectrum is always possible to have the same amount of DM under the form of primordial black holes. For this reason, in order to discriminate models for PBH production we need another observables. As we will discuss in Chapter \ref{cap:GWs}, we can use the SIGWs.

\subsubsection{Implications for PBHs in the stellar mass range}
\begin{figure}[t]
	\begin{center}
		$$\includegraphics[width=.46\textwidth]{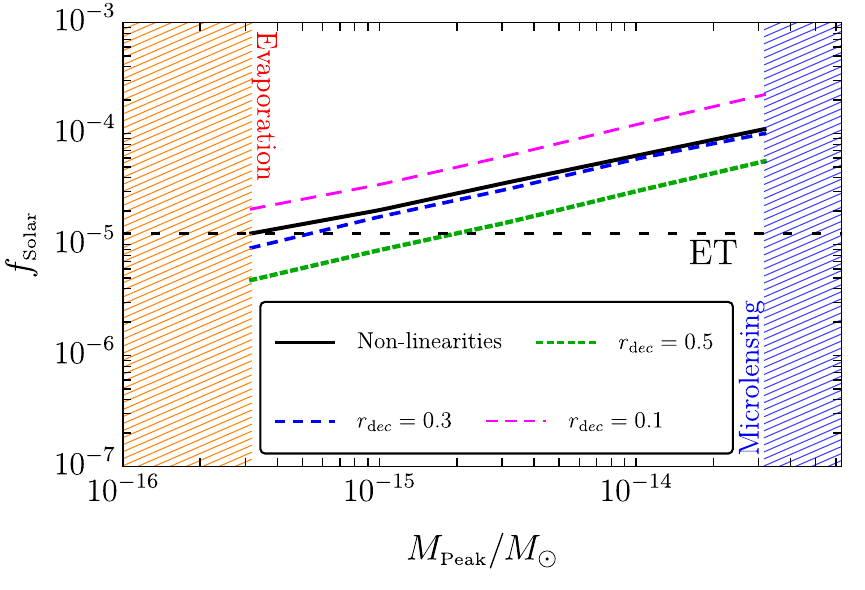}
		\qquad\qquad\includegraphics[width=.46\textwidth]{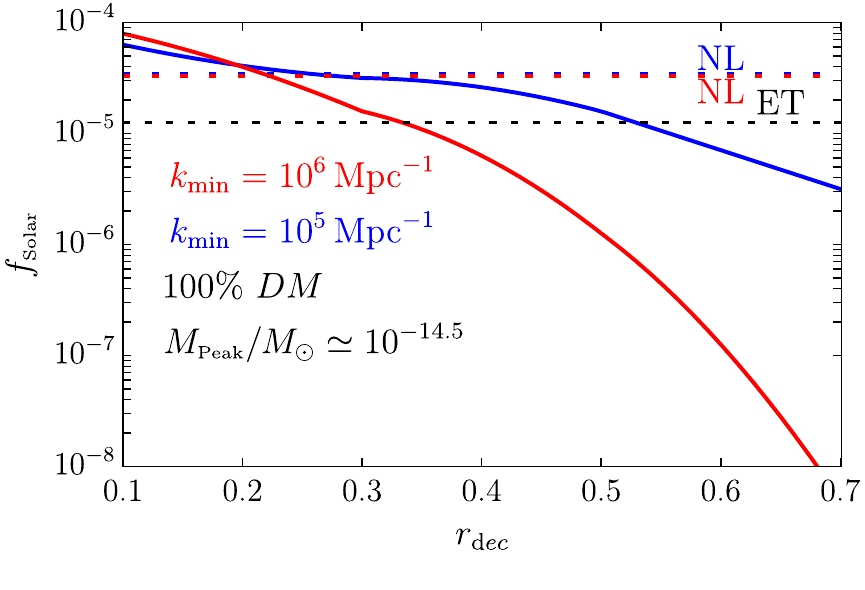}$$
		\caption{
	\textbf{\textit{	Left panel. }} 
Height of the peak at the QCD Mass scale $f_{\rm Solar}$,
assuming a broad power spectrum with $k_{\rm min}=10^{5}$ $ {\rm Mpc}^{-1}$ as a function of the location of the asteroidal mass peak dominating the contribution to the dark matter. 
We compare the result obtained considering only the effect of non-linearities with three different values of $r_{\rm dec}$. 
The dashed regions at the sides represent positions of the main peak which are ruled out by the constraints given by evaporation or microlensing. 
\textbf{\textit{	Right panel. }} 
Height of the peak at the QCD scale for different values of $r_{\rm dec}$. 
We choose as benchmark values 
$k_{{ \rm min}}=10^{6}$ ${\rm  Mpc}^{-1}$ (red) 
and $k_{{ \rm min}}=10^{5}$ ${\rm  Mpc}^{-1}$ (blue) respectively.
The plot shows a decreasing trend of $f_{\rm Solar}$ when $r_{\rm dec}$ increases, resulting from the effect of NG reducing the abundance of high masses. 
The dashed lines represent the height of the peak at the QCD scale computed by taking into account only non-linearities while the dashed black one shows the lower bound of the sensitivity of the Einstein Telescope (ET) experiment to PBH mergers as derived in \cite{DeLuca:2021hde}.
}
\label{fig:parScan}  
\end{center}
\end{figure}

The plot in fig.\,\ref{fig:fPBH1} shows the presence, around solar masses, of a second peak in $f_{\rm PBH}(M_{\rm PBH})$ which, as widely known in literature, is typically caused by the QCD phase transition \cite{Jedamzik:1996mr,Byrnes:2018clq, Musco:2023dak}.  
With the inclusion of NGs in a scenario characterized by a broad, and nearly scale invariant, power spectrum, the PBH abundance is further modulated and can be enhanced/suppressed compared to the case of gaussian primordial curvature perturbations. 
We highlight few main trends:
\begin{itemize}
    \item[$\circ$] 
Moving the position of the main peak of asteroidal mass PBHs in the allowed window between evaporation and microlensing constraints while keeping $f_{\rm PBH} =  1$,  one finds that the height of the QCD peak, dubbed $f_{\rm Solar} = f_{\rm PBH}(1.25 M_\odot)$, changes approximately with a power law $f_{\rm Solar} \propto M_{\rm PBH}^{1/2}$.
This is shown in the left panel of fig.\,\ref{fig:parScan}
and it is caused by the leading order behaviour of the mass distribution $f_{\rm PBH} \propto M_{\rm PBH}^{1/2}$ obtained for scale invariant mass fraction $\beta(M_H)$.
Indeed, choosing $k_{{\rm min}} = 10^5$ ${\rm Mpc}^{-1}$, 
$\beta_{\rm NG}$ still remains approximately scale invariant close to the QCD epoch. Hence, imposing PBHs account for the totality of dark matter, a shift to the right of the mass characterising 
the main peak $M_{\rm Peak}$ 
cause an increase of $f_{\rm Solar}$. 
\item[$\circ$]
When the large scale cut-off $k_{\rm min}$ is moved closer to the QCD mass scale, 
one obtains larger violations of the scale invariance around $M_{\rm PBH} \approx M_\odot$ (see fig.\,\ref{fig:BroCompa2}). 
Therefore, larger values of $k_{\rm min}$ leads to deviations from the overall trend highlighted in the previous point and $f_{\rm Solar}$ inherits a stronger dependence to $r_{\rm dec}$. This is highlighted in fig.\,\ref{fig:parScan} where we show results for 
$k_{\rm min} =10^{5}$  ${\rm Mpc}^{-1}$ in blue while 
$k_{\rm min} =10^{6}$  ${\rm Mpc}^{-1}$ in red.
\item[$\circ$]
As we can see from the right plot in fig.\,\ref{fig:BroCompa2}, the value of $r_{{\rm dec}}$ controls the deviation from a scale independent $\beta_{\rm NG}$.
This translates into modified secondary peak amplitude $f_{\rm Solar}$. 
The right-hand plot in fig.\,\ref{fig:parScan} shows how $f_{\rm Solar}$ varies for different values of $r_{\rm dec}$. Interestingly enough, the height of the peak grows as $r_{\rm dec}$ decreases (that is, for larger primordial NG).
This is nothing but the effect already discussed in .\,\ref{fig:BroadCompa},\,\ref{fig:BroCompa2} (at the level of the mass fraction $\beta_{\rm NG}$) but now reflected on the computation of $f_{\rm PBH}$.
\end{itemize}
\noindent
It is important to note that an enhanced effect of NGs on the secondary peak would allow for a larger merger rate of stellar mass mergers that may be potentially constrained at current and future ground base GW detectors \,\cite{Pujolas:2021yaw,DeLuca:2021hde,Franciolini:2022tfm}.
Let us summarize the results of this section. We devised a general formula in Eq.,\ref{eq:MasterAbundance} giving the PBH abundance when the density contrast field is non-Gaussian, assuming threshold statistics on the compaction and an average profile for the latter\footnote{Similar results were independently obtained in ref.\cite{Gow:2022jfb}, which appeared later with respect to ref.\cite{Ferrante:2022mui}.}.
We carefully assess under which conditions the conventional perturbative approach can be trusted. In the case of a narrow power spectrum, this happens only if the perturbative expansion is pushed beyond the quadratic order (with the optimal order of truncation depending on the width of the spectrum). Most importantly, we demonstrate that the perturbative approach is intrinsically flawed when considering broad spectra, in which case only the non-perturbative computation captures the correct result.
 
\section{Going beyond the average profile of the compaction}\label{sec:C1B}
One intrinsic and therefore unavoidable source of uncertainty in calculating the PBHs abundance arises from the inability to predict the value of a given observable with zero uncertainty, e.g. the compaction function or its  curvature at its peak, in a given point or region. This is due to the fact that the theory delivers only stochastic quantities, e.g. the curvature perturbation, of which we know only the power spectrum and the  higher-order correlators. Therefore, we are allowed to calculate only ensemble averages and  typical values, which come with  intrinsic uncertainties quantified by, for example,  root mean square deviations. 

Since the critical PBHs abundance depends crucially on the curvature of the compaction function at its peak, the natural question which arises is the following: in order to calculate the PBHs abundance, which value of the critical threshold should we use? In other words, which value of the curvature should one adopt to derive the formation threshold?

A natural answer to this question might be to use the average profile of the compaction function, and this is done routinely in the literature and in the previous section. After all, most of the Hubble volumes are populated by peaks with such average profile at horizon re-entry.

In this section, based on ref.\cite{Ianniccari:2024bkh} we wish to make a  simple, but  relevant observation:
only if the power spectrum of the curvature perturbation is very peaked, the critical threshold for formation is determined by the average value of the curvature of the compaction function at the peak; in the realistic cases in which  the power spectrum of the curvature perturbation is not peaked,  the critical threshold for formation is determined by the broadest possible compaction function. This is because the  abundance is dominated by the smallest critical threshold, which corresponds to the broadest profile. In such a case, the threshold for the compaction function is fixed to be 2/5.

\subsection{The average profile}
One question to pose is the following: which profile should one make use of to calculate the critical value for PBH abundance, given that it depends on the peak profile? The natural answer, routinely adopted in the literature,  would be the average profile of the compaction function with the constraint that  there is a peak of the curvature perturbation at the center of the coordinates with value $\zeta(0)$. This is the most obvious answer  as the average profile is the most frequent, statistically speaking. Supposing for the moment that $\zeta(r)$ is Gaussian, such an average profile would be
\be
\langle C_{\rm G}(r)\rangle_{\zeta(0)}=-\frac{4}{3}r
\langle \zeta'(r)\rangle_{\zeta(0)}=-\frac{4}{3}r\langle \zeta(r)\rangle_{\zeta(0)}'=-\frac{4}{3} r\frac{\xi'(r)}{\xi(0)}\zeta_0,
\ee
where 
\be
\xi(r)=\int\frac{{\rm d} k}{k}{\cal P}_\zeta(k) \frac{\sin k r}{k r}
\ee
is the two-point correlation of the curvature perturbation. In such a case the value of $r_m$ where the most likely compaction function has its maximum would then be fixed by the equation
\be
\xi'(r_m)+r_m\xi''(r_m)=0.
\ee
A standard choice is therefore to calculate the curvature of the peak of the compaction function as\footnote{This  is clearly not  correct as, for instance,  the average of the ratio of two stochastic variables is not the ratio of their averages.}

\be
q=-\frac{1}{4}\frac{r_m^2 \langle \mathcal{C}''_{\rm G}(r_m)\rangle_{\zeta(0)}}{\langle C_{\rm G}(r_m)\rangle_{\zeta(0)}}\left[1-\frac{3}{8}\langle C_{\rm G}(r_m)\rangle_{\zeta(0)}\right].
\ee
The crucial point is  that, the smaller the value of the curvature, the smaller the value of the threshold. Since the PBHs abundance has an exponentially strong dependence on the threshold, one expects  that broad compaction functions should be very relevant in the determination of the abundance of PBHs even though they are more rare than the average profiles. This is what we discuss next.
\subsection{The importance of being broad: the Gaussian case}
Here we  assume $C_{\rm G}(r_m)$ and $C_{\rm G}''(r_m)$ to be  Gaussian (and correlated) variables. This will allow us to gain some analytical intuition.   We define 
\be
 \sigma^2_0=\langle  C_{\rm G}^2(r_m)\rangle, \,\,\,\sigma^2_1= -\frac{1}{4}r_m^2\langle  C_{\rm G}''(r_m) C_{\rm G}(r_m)\rangle,\,\,\,{\rm and}\,\,\,\sigma^2_2=\frac{1}{16}r_m^4\langle {C_{\rm G}''(r_m)}^2\rangle.
\ee
Such correlations are easily computed knowing that the Fourier transform of the linear compaction function reads
\be
C_{\rm G}({\bf k},r)=\frac{4}{9}k^2 r^2 W(kr)\zeta({\bf k}),\,\,\,\,W(x)=3\frac{\sin x-x\cos x}{x^3},
\ee
where $W(x)$ is the Fourier transform of the Heaviside window function in real space.
We will  use the conservation of the probabilities 
 \begin{eqnarray}
     P\left[\mathcal{C}(r_m),\mathcal{C}''(r_m)\right]{\rm d}\mathcal{C}(r_m){\rm d}\mathcal{C}''(r_m)&=&
{\cal P}\left[C_{\rm G}(r_m),C_{\rm G}''(r_m)\right]{\rm d}C_{\rm G}(r_m){\rm d}C_{\rm G}''(r_m)\nonumber\\&=&\widetilde{{\cal P}}\left[C_{\rm G}(r_m),q_\zeta\right]{\rm d}C_{\rm G}(r_m){\rm d}q_\zeta
 \end{eqnarray}
where 
\begin{eqnarray}
{\cal P}\left[-\frac{1}{4}r_m^2  C_{\rm G}''(r_m),  C_{\rm G}(r_m)\right]&=&\frac{1}{2\pi\sqrt{{\rm det}\,\Sigma}}
\,{\rm exp}\left(-\vec V^T \Sigma^{-1}\vec V/2\right),\nonumber\\
\vec{V}^T&=&\left[-\frac{1}{4}r_m^2  C_{\rm G}''(r_m),  C_{\rm G}(r_m)\right],\nonumber\\
\Sigma&=&\left(\begin{array}{cc}
\sigma^2_2 & \sigma^2_1\\
\sigma^2_1 & \sigma^2_0\end{array}\right).
\end{eqnarray}
Where again we define the parameter
\be
\gamma=\frac{\sigma_1^2}{\sigma_2\sigma_0},
\ee
which plays an important role in the following and indicates the broadness of a given power spectrum of the curvature perturbation. The 
closer $\gamma$ is to unity, the more spiky is the peak of the curvature perturbation.

\subsubsection{The average of the curvature}
The average  curvature of the linear compaction function $C_{\rm G}$ can be computed by using the conditional probability to have a peak at $r_m$\footnote{In fact we use threshold statistics rather than peak statistics to elaborate our point. However, regions  well above the corresponding square root of the variance are very likely local maxima\,\cite{Hoffman:1985pu}. }
\be
\langle q_{\rm G}\rangle=\int_0^\infty{\rm d}q_\zeta\,q_\zeta\,P[q|C_{\rm G}(r_m)>C_{\rm G,\mathcal{C}}(q_{\rm G})],
\ee
with 
\be
P[q|C_{\rm G}(r_m)>C_{\rm G,\mathcal{C}}(q_{\rm G})]=\frac{\widetilde{{\cal P}}[q_{\rm G},C_{\rm G}(r_m)>C_{\rm G,\mathcal{C}}(q_{\rm G})]}{\widetilde{{\cal P}}[C_{\rm G}(r_m)>C_{\rm G,\mathcal{C}}(q_{\rm G})]}.
\ee
The conditional probability, in the limit of large thresholds, becomes
\begin{eqnarray}
    P[q_{\rm G}|C_{\rm G}(r_m)>C_{\rm G,\mathcal{C}}(q_{\rm G})]&\simeq &\frac{\left(1-\gamma^2\right)^{1/2}\sigma_2 C_{\rm G,\mathcal{C}}(q_{\rm G})}{\sqrt{2\pi}\sigma_0\left[(q_{\rm G}-\gamma\sigma_2/\sigma_0)^2+(1-\gamma^2)\sigma_2^2/\sigma_0^2\right]^{1/2}}\cdot
\nonumber\\
&\cdot& \exp\left[-\frac{(q-\gamma\sigma_2/\sigma_0)^2C^2_{\rm G,\mathcal{C}}(q_{\rm G})}{2(1-\gamma^2)\sigma_2^2}\right].
\end{eqnarray}
For a monochromatic power spectrum of the curvature perturbation, that is $\gamma\simeq 1$, we recognize the Dirac delta and the value of $q_{\rm G}$, which minimizes the exponent and maximizes the PBHs abundance, is the average value $\langle q_{\rm G}\rangle=\sigma_2/\sigma_0$. 

Departing from $\gamma\simeq 1$, and integrating numerically, one discovers departures from the value $\gamma\sigma_2/\sigma_0$ for the average of $q_{\rm G}$, but not dramatically, and one has 
\be
\label{XY}
\langle q_{\rm G}\rangle\simeq \gamma\frac{\sigma_2}{\sigma_0}.
\ee
Hence for very broad spectrum, $\gamma \rightarrow 0$, one has $\langle q_{\rm G}\rangle \rightarrow 0$.
\subsubsection{Recomputing the PBHs abundance}
\begin{figure}[htb]
\begin{minipage}[t]{0.45\linewidth}
\includegraphics[width=\linewidth]{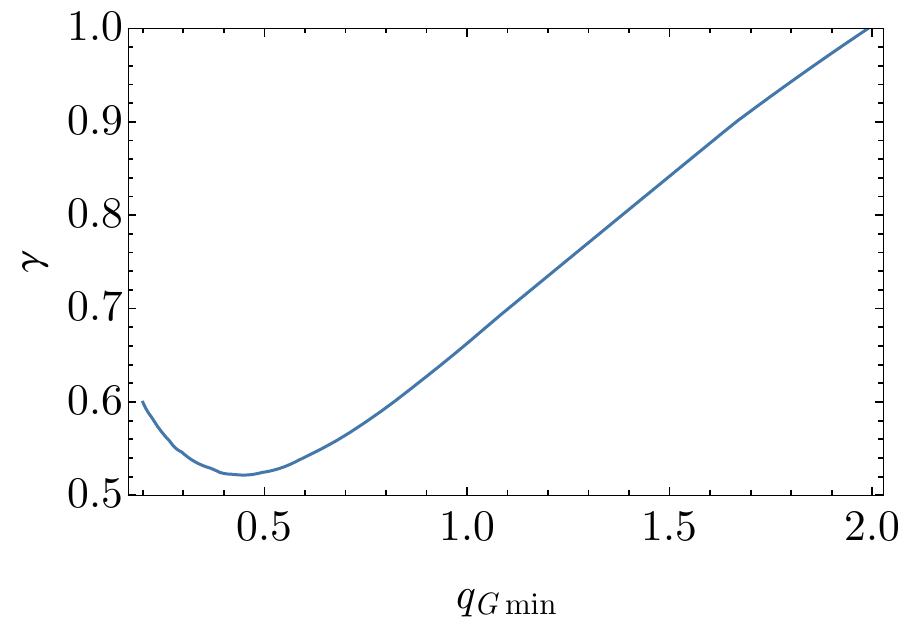}
\caption{Plot of $q_{\rm G \; \rm{\min}}$ as a function of $\gamma$ for $\sigma_2/\sigma_0=2$.}
\label{fig:gamma_qmax}
\end{minipage}
\hfill
\begin{minipage}[t]{0.45\linewidth}
\includegraphics[width=\linewidth]{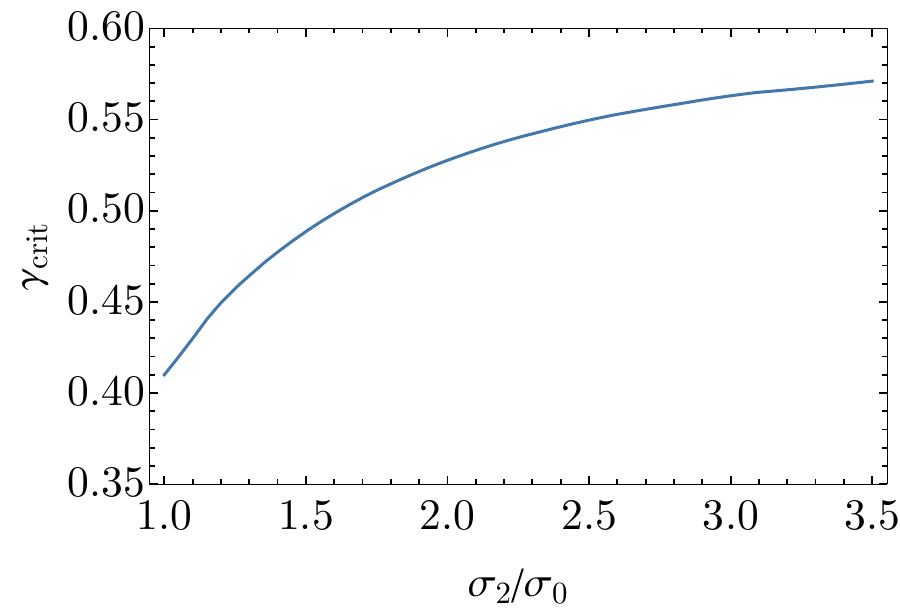}
\caption{Critical value of $\gamma$ as a function of $\sigma_2/\sigma_0$}
\label{fig:gamma_crit}
\end{minipage}%
\end{figure}

\begin{figure}[hbt]
\centering
  \includegraphics[width=14cm]{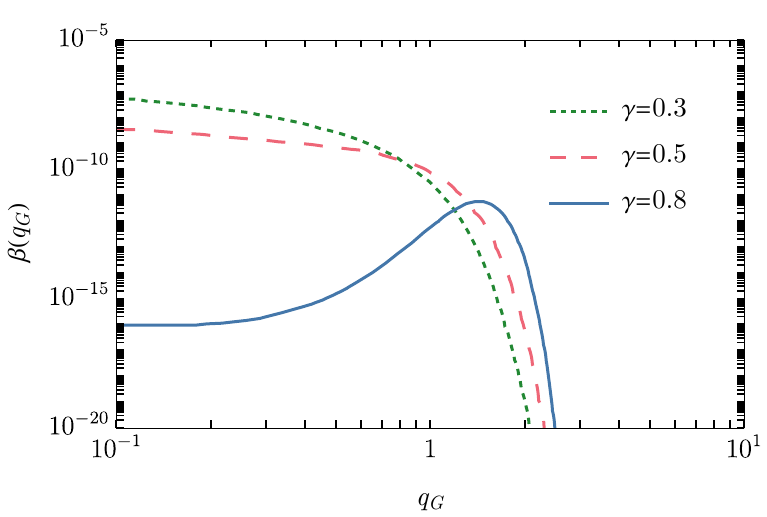}
  \caption{The PBHs formation probability as a function of $q_{\rm G}$ for $\sigma_0=\sigma_2/2=0.05$ and three different values of $\gamma=(0.3,0.5,0.8)$ for which $\langle q_{\rm G}\rangle =(0.6,1,1.6)$ for the Gaussian case.}
  \label{fig:three_gammas}
\end{figure}

The PBHs abundance is given by\footnote{We do not account for the extra factor counting the mass of the PBH with respect to the mass contained in  the horizon volume at re-entry as we give priority to getting analytical results. We will reintegrate it at the end.}
\be
\beta=\int_{C_c(q)}^\infty{\rm d}\mathcal{C}(r_m)\int_{-\infty}^0{\rm d}\mathcal{C}''(r_m) P\left[\mathcal{C}(r_m),\mathcal{C}''(r_m)\right]=\int_{0}^\infty {\rm d}q_{\rm G}\int_{C_{\rm G,\mathcal{C}}(q_{\rm G})}^\infty{\rm d}C_{\rm G}(r_m) \widetilde{{\cal P}}\left[C_{\rm G}(r_m),q_{\rm G}\right].
\ee
Going back to the initial probability, it can be written as 
\begin{align}
  {\cal P}\left[-\frac{1}{4}r_m^2  C_{\rm G}''(r_m),  C_{\rm G}(r_m)\right]&=\frac{1}{2\pi}\frac{1}{\sigma_2\sigma_0\sqrt{1-\gamma^2}}{\rm exp}\left[-\frac{r_m^4}{16} \frac{C_{\rm G}''(r_m)^2}{2\sigma_2^2}\right]\times\nonumber\\  &{\rm exp}\left[-\frac{1}{2(1-\gamma^2)}\left(\frac{C_{\rm G}(r_m)}{\sigma_0}+\gamma\frac{ r_m^2 C_{\rm G}''(r_m)}{4\sigma_2}\right)^2\right].
\end{align}
For a monochromatic, very peaked,  power spectrum of the curvature perturbation, where $\gamma\simeq 1$, the probability reduces to 
\be
\lim_{\gamma\rightarrow 1}{\cal P}\left[-\frac{1}{4}r_m^2  C_{\rm G}''(r_m),  C_{\rm G}(r_m)\right]=\frac{1}{\sqrt{2\pi}}\frac{1}{\sigma_2\sigma_0}{\rm exp}\left(-\frac{r_m^4 {C_{\rm G}''(r_m)}^2}{16\cdot 2\sigma_2^2}\right)\delta_D\left(\frac{C_{\rm G}(r_m)}{\sigma_0}+\frac{ r_m^2 C_{\rm G}''(r_m)}{4\sigma_2}\right),
\ee
which fixes 
\be
q_{\rm G}=\frac{\sigma_2}{\sigma_0}=\frac{\sigma_2\sigma_0}{\sigma_0^2}=\frac{\sigma^2_1}{\sigma^2_0}=\langle q_{\rm G}\rangle,
\ee
and
\begin{eqnarray}
\beta&=&\int_{0}^\infty{\rm d}q_{\rm G}\int_{C_{\rm G,\mathcal{C}}(q_{\rm G})}^{4/3}{\rm d}C_{\rm G}(r_m){\cal P}\left[-\frac{1}{4}r_m^2  C_{\rm G}''(r_m),  C_{\rm G}(r_m)\right]\nonumber\\
&=&\int_{C_{\rm G,\mathcal{C}}(\langle q_{\rm G}\rangle)}^{4/3}{\rm d}C_{\rm G}(r_m)\frac{1}{\sqrt{2\pi}\sigma_0}{\rm exp}\left(-\frac{\mathcal{C}^2_\zeta(r_m)}{2\sigma_0^2}\right)
=\frac{1}{2}{\rm Erfc}\left[\frac{C_{\rm G,\mathcal{C}}(\langle q_{\rm G}\rangle)}{\sqrt{2}\sigma_0}\right],
\end{eqnarray}
where
\be
C_{\rm G,\mathcal{C}}(q_{\rm G})\simeq\frac{4}{3}\left(1-\sqrt{\frac{2-3C_{\mathcal{C}}(q_{\rm G}) }{2}}\right).
\ee
Therefore for monochromatic spectra of the curvature perturbation the PBHs abundance is fixed by the value of the threshold corresponding to the average value of the curvature of the compaction function at its peak
\be
\mathcal{C}^{\rm peaked}_{\rm G,\mathcal{C}}=C_{\rm G,\mathcal{C}}(\langle q_{\rm G}\rangle).
\ee
For a generic power spectrum, we change the variables from $(-r_m^2 C_{\rm G}''(r_m),C_{\rm G}(r_m))$ to $(q_{\rm G},C_{\rm G}(r_m))$ and making use of the conservation of the probability we obtain
\begin{eqnarray}
\beta(q_{\rm G})&=&\int_{C_{\rm G,\mathcal{C}}(q_{\rm G})}^{4/3}{\rm d}C_{\rm G}(r_m)|C_{\rm G}(r_m)|\,{\cal P}\left[q C_{\rm G}(r_m),  C_{\rm G}(r_m)\right]\nonumber\\
&\simeq& \frac{\sqrt{1-\gamma^2}\sigma_2}{2\pi\sigma_0\left[(q_{\rm G}-\gamma \sigma_2/\sigma_0)^2+(1-\gamma^2)\sigma^2_2/\sigma^2_0\right]}\cdot\nonumber\\
&\cdot&\exp\left[-\frac{\left[(q_{\rm G}-\gamma \sigma_2/\sigma_0)^2+(1-\gamma^2)\sigma^2_2/\sigma^2_0\right]\mathcal{C}^2_{\zeta,\mathcal{C}}(q_{\rm G})}{2(1-\gamma^2)\sigma^2_2}\right].\nonumber\\
&&
\end{eqnarray}
We see that the square of the critical threshold is replaced by an effective squared critical threshold
\be
\mathcal{C}^2_{\rm G,\mathcal{C}}(q_{\rm G})\Big|_{\rm eff}
=\left[(q_{\rm G}-\gamma \sigma_2/\sigma_0)^2+(1-\gamma^2)\sigma_2^2/\sigma_0^2\right] \mathcal{C}^2_{\rm G,\mathcal{C}}(q_{\rm G}).
\ee
This equation is fundamental and its minimum is determined by the equation 
\be
\label{min}
(q_{\rm G}-\gamma\sigma_2/\sigma_0)C_{\rm G,\mathcal{C}}(q_{\rm G})+\left[(q_{\rm G}-\gamma\sigma_2/\sigma_0)^2+(1-\gamma^2)\sigma_2^2/\sigma_0^2\right]\frac{{\rm d}C_{\rm G,\mathcal{C}}(q_{\rm G})}{{\rm d}q_\zeta}=0.
\ee
For peaked profiles where $\gamma\simeq 1$ we have
\be
(q_{\rm G}-\sigma_2/\sigma_0)\left[C_{\rm G,\mathcal{C}}(q_{\rm G})+(q_{\rm G}-\sigma_2/\sigma_0)\frac{{\rm d}C_{\rm G,\mathcal{C}}(q_{\rm G})}{{\rm d}q_{\rm G}}\right]=0
\ee
and the mimimum lies at the value of the average $q_{\rm G}=\langle q_{\rm G}\rangle=\sigma_2/\sigma_0$. For broad spectra $\gamma\ll 1$, the effective threshold is minimized for $q_{rm G}\simeq 0$ as it reduces to 
\be
\label{eq:large_q0}
\mathcal{C}^2_{\rm G,\mathcal{C}}(q_{\rm G})\Big|_{\rm eff}
=\left[q_{\rm G}^2+\sigma_2^2/\sigma_0^2\right]\mathcal{C}^2_{\rm G,\mathcal{C}}(q_{\rm G}).
\ee
and the threshold $C_{\rm G,\mathcal{C}}(q_{\rm G})$ is also minimized for small $q_{\rm G}$. There is in general a critical value of $q_{\rm G}$ for which the abundance is always dominated by the broad spectra. We can see this behavior by plotting the curve $(q_{\rm G\;\rm{min}},\gamma)$ obtain from the Eq.\,(\ref{min}), as shown in Fig.\,\ref{fig:gamma_qmax}. As we start decreasing from $\gamma=1$ where the minimum is in $\sigma_2/\sigma_0$, also the value of $q_{\rm G,\;\rm{min}}$ decreases, up until a critical value $\gamma_{\rm crit}$.
For values of $\gamma$ below this point the function $C_{\rm G,\mathcal{C}}(q_{\rm G})|_{\rm{eff}}$ does not have a minimum, but is monotonically increasing with $q_{\rm G}$, hence the minimum lies at the boundary of the interval, i.e. $q_{\rm G}=0$. The transition is therefore very sharp after  the critical value.\\
It is also possible to evaluate the position of this minimum for different values of the parameter $\sigma_2/\sigma_0$, as shown in Fig. \ref{fig:gamma_crit}. We can understand the behavior because having larger values of this parameter the transition happens for larger values of $\gamma$, being easier to enter in the regime of Eq. (\ref{eq:large_q0}), where $\sigma_2/\sigma_0$ dominates.
\\To show this explicitly, in Fig. \ref{fig:three_gammas} we plot the formation probability for  three different values of $\gamma$. It demonstrates that the abundance is dominated by the broadest profiles when the curvature perturbation is not very spiky and not by the average value of $q_{\rm G}$.
The corresponding critical value needed to be used is therefore
\be
\label{cc}
C_{\zeta,\mathcal{C}}(q_{\rm G}\simeq 0)\simeq\frac{4}{3}\left(1-\sqrt{\frac{2-3\cdot 2/5 }{2}}\right)\simeq 0.49.
\ee
\noindent

\subsection{The importance of being broad: the non-Gaussian case }
As a matter of fact, the curvature perturbation generated in models producing large overdensities is typically non-Gaussian as discussed in sec.\ref{sec:C1pre}.
\\We assume again that the initial curvature perturbation is non-Gaussian, but a function of a Gaussian component 
\be
\zeta(r)=F[\zeta_{\rm G}(r)].
\ee
and indicating the derivatives of $F$ with respect to $\zeta_{\rm G}$ by $F_n =\dd F(\zeta_{\rm G})/ \dd \zeta_{\rm G}$.
 The maximum of the compaction function can be found solving the equation
\begin{equation}
    C_{\rm G}'(r_m) = F_1(\zeta_{\rm G}) \mathcal{C}'_{\g}(r_m) + C_{\g}(r_m) \zeta_{\rm G}'(r_m)  F_2(\zeta_{\rm G}) = 0,
\end{equation}
as long as $C_{\rm G}(r_m)<4/3$.
The next  step is to define  the  following  Gaussian and correlated variables  
\begin{equation}
    x_0 = \zeta_{\rm G}, \quad x_1 = r\zeta'_\g, \quad x_2 = r^2\zeta''_\g, \quad x_3 =r^3\zeta'''_\g,
\end{equation}
for which the condition of the maximum becomes
\begin{equation}
    x_2 = -x_1\left(1 + x_1 \frac{F_2(x_0)}{F_1(x_0)}\right).
\end{equation}
One can construct the corresponding   probability distribution as 
\begin{equation}
    P(x_0,x_1,x_2,x_3) = \frac{1}{(2\pi)^2\sqrt{{\rm det}\,\Sigma}}
\,{\rm exp}\left(-\vec V^T \Sigma^{-1}\vec V/2\right), 
\end{equation}
where
\begin{equation*}
    \vec{V}^T= \left[x_0,x_1,x_2,x_3\right],
\end{equation*}
and
\begin{eqnarray}
\Sigma&=&\left(\begin{array}{cccc}
\sigma^2_0 & \gamma_{01}\sigma_1\sigma_0 &\gamma_{02}\sigma_2\sigma_0 &\gamma_{03}\sigma_3\sigma_0\\
\gamma_{01}\sigma_1\sigma_0 & \sigma_1^2 &\gamma_{12}\sigma_2\sigma_1 &\gamma_{13}\sigma_1\sigma_3\\
 \gamma_{02}\sigma_2\sigma_0 & \gamma_{12}\sigma_2\sigma_1 &\sigma^2_2&\gamma_{23}\sigma_2\sigma_3\\
\gamma_{03}\sigma_3\sigma_0& \gamma_{13}\sigma_1\sigma_3 &\gamma_{23}\sigma_2\sigma_3 &\sigma_3^2\end{array}\right)
\end{eqnarray}
is constructed from the different correlators with\footnote{Notice that here for clarity the index for the various $\sigma_i$ is related to the number of derivatives of $\zeta_{\rm G}$, differently from the definition in the previous subsection.}
\begin{equation}
    \sigma_i^2 = \langle x_i^2\rangle, \quad \gamma_{ij}=\frac{\langle x_i x_j\rangle}{\langle x_i^2 \rangle^{1/2}\;\langle x_j^2 \rangle^{1/2}}.
\end{equation}
Next, we need to  convert all the relevant variables in terms  of the Gaussian ones   $x_i$ ($i=0,\cdots,3)$. 
First we have
\begin{equation}
    C_{\g} = -\frac{4}{3} \, x_1,
\end{equation}
and the  derivatives of $C_{\rm G}$ can be written in terms of $x_1$ and $x_2$ as
\begin{eqnarray}
    C_{\rm G} &=& -\frac{4}{3}\; x_1 F_1(x_0), \nonumber\\
    rC_{\rm G}' &=& -\frac{4}{3} \left(F_1(x_0) (x_1 + x_2) + x_1^2 F_2(x_0) \right), \nonumber\\
    r^2C_{\rm G}'' &=& -\frac{4}{3} \left[ F_1(x_0) (2x_2+ x_3) + 2 x_1^2 F_2(x_0) + 3 x_1x_2 F_2(x_0) + x_1^3 F_3(x_0)\right].
\end{eqnarray}
The  PBHs  abundance for a given value of the curvature $q$ will read 

\begin{equation}\label{eq:Master2}
    \beta(q) = \int_{D}\mathcal{K}\left(\mathcal{C} - C_c(q)\right)^{\gamma}\; p(x_0, C_{\rm G}, x_2, q)\;\delta\left( F_1(x_0) (x_1
    + x_2) + x_1^2 F_2(x_0) \right),
\end{equation}
where the domain of integration is 

\begin{equation}
    D = \left\{x_2\in \mathbb{R}, C_{\rm G,\mathcal{C}}(q)<C_{\rm G}<\frac{4}{3}\right\},
\end{equation}
with 
\be
\label{c1}
C_{\rm G,\mathcal{C}}(q)\simeq\frac{4}{3}\left(1-\sqrt{\frac{2-3C_{\mathcal{C}}(q) }{2}}\right).
\ee
We have reintroduced the  scaling-law factor for critical collapse $\mathcal{K}(\mathcal{C}-C_c(q))^\gamma$  which  accounts for the mass of the PBHs at formation written in units of the horizon mass at the time of horizon re-entry, with  $\mathcal{K}\simeq  3.3$  for a log-normal power spectrum and $\gamma\simeq 0.36$ \cite{Choptuik:1992jv,Musco:2012au,Musco:2023dak}. By using the conservation of probabilities we can finally write 
 \be
p\left[\zeta_{\rm G}, C_{\rm G},x_2,q\right]=P\left[x_0,x_1,x_2,x_3\right]|{\rm Det}\,  J|,
\ee
where 
\begin{equation}
    {\rm Det}\, J  = \frac{3}{4}\left(\frac{4x_1 +2F_1(x_0)x_1^2}{1+F_1(x_0)x_1}\right),
\end{equation}
and at the maximum
\begin{equation}
    x_3= \frac{-4q(1+\frac{1}{2}x_1F_1(x_0))x_1}{1+x_1F_1(x_0)}-2x_2-2x_1^2\frac{F_2(x_0)}{F_1(x_0)}-3x_1x_2\frac{F_2(x_0)}{F_1(x_0)}-x_1^3\frac{F_3(x_0)}{F_1(x_0)}.
\end{equation}
We rewrite the Gaussian probability in the following form 
\be
P(x_0, x_1, x_2, x_3) = \frac{1}{4\pi^2 \sqrt{\det{\Sigma}}} \exp{\left(-\frac{(\sigma_0 \sigma_1 \sigma_2 \sigma_3)^2}{2\det{\Sigma}} \sum_{i,j=0}^3 \frac{\kappa_{ij} x_i x_j}{\sigma_i\sigma_j} \right)},
\ee
where the $\kappa_{ij}$'s will depend on all the $\gamma_{lm}$, and they can be computed by performing the inverse of the matrix $\Sigma$, matching with the definition. 
Performing the change of variables we get 
\begin{eqnarray}
    p(\zeta_{\rm G}, C_{\rm G}, x_2, q) &=&  \left|\frac{9 C_\g \left(3 C_\g F_1-8\right)}{8 \
\left(3 C_\g F_1-4\right)} \right| \frac{1}{4\pi^2 \sqrt{\det{\Sigma}}}\cdot\nonumber\\
&\cdot& \exp{\left(-\frac{(\sigma_0 \sigma_1 \sigma_2 \sigma_3)^2}{2\det{\Sigma}} \frac{1}{4096 \
F_1^4}\left[ A(\zeta_{\rm G}, C_\g) q^2 + B(\zeta_{\rm G}, C_\g) q + \mathcal{C}(\zeta_{\rm G}, C_\g) \right] \right)},\nonumber\\
&&
\end{eqnarray}
where we have defined the following functions of $\zeta_{\rm G}$ and $C_\g$
\be
    A(\zeta_{\rm G}, C_\g)=\frac{9216 \kappa_{33} C_\g^2 F_1^4 
\left(8-3 C_\g F_1\right){}^2}{\sigma _3^2 
\left(4-3 C_\g F_1\right){}^2},
\ee
\begin{eqnarray}
B(\zeta_{\rm G}, C_\g)&=&\frac{192 C_\g F_1^2 \left(3 C_\g 
F_1-8\right)}{\sigma _0 \sigma _1 \sigma _2 
\sigma _3^2 \left(3 C_\g F_1-4\right)} \left[3 \kappa _{33} \sigma _0 
\sigma_1 \sigma _2 C_\g \left\{9 C_\g F_1 
\left(C_\g F_3+4 F_2\right)-27 C_\g^2 F_2^2-32 
F_1^2\right\}\right. \nonumber\\
&&\left. +4 \sigma _3 
F_1 \left\{3 \kappa _{23} \sigma_0 \sigma_1 C_\g \left(4 F_1-3 C_\g 
F_2\right)+4 \sigma_2 F_1 \left(4 \kappa _{30} \sigma_1 \zeta_{\rm G}-3 \kappa _{13} 
\sigma_0 C_\g\right)\right\}\right],
\end{eqnarray}
\begin{eqnarray}
\mathcal{C}(\zeta_{\rm G}, C_\g)&=&-\frac{\kappa_{33}}{\sigma_3^2}\left\{ 4374  C_\g^6 F_1 
F_2^2 F_3 - 5832  C_\g^5 
F_1^2 F_2 
F_3+ 17496  C_\g^5 
F_1 F_2^3- 27216  C_\g^4 
F_1^2 F_2^2+\right.\nonumber\\
&&\left.20736  C_\g^3 
F_1^3 F_2 - 729  
C_\g^6 F_1^2 F_3^2+ 5184  C_\g^4 
F_1^3 F_3-9216  C_\g^2 
F_1^4-6561  C_\g^6 F_2^4 \right\}+\nonumber\\
&& -\frac{1}{\sigma _0 \sigma _1 
\sigma _2 \sigma _3}24 C_\g \
F_1 \left(-9 C_\g F_1 \left(C_\g F_3+4 
F_2\right)+27 C_\g^2 F_2{}^2+32 F_1^2\right)\nonumber\\
&&\left(3 
\kappa _{23} \sigma_0 \sigma _1 C_\g \left(4 F_1-3 C_\g F_2\right)+4 \sigma _2 
F_1 \left(4 \kappa _{30} \sigma _1 \zeta_{\rm G}-3 \kappa _{13} \sigma _0 C_\g\right)\right)+\nonumber\\
&&+16 F_1^2 
\left(\frac{24 C_\g F_1 \left(3 C_\g 
F_2-4 F_1\right) 
\left(3 \kappa _{12} \sigma _0 C_\g-4 \kappa _{20} \sigma _1 \zeta_{\rm G}\right)}{\sigma _0 \sigma _1 \sigma _2}+\right.\nonumber\\
&&\left. +\frac{9 \kappa _{22} C_\g^2 \
\left(4 F_1-3 C_\g F_2\right){}^2}{\sigma _2^2}+\frac{16 F_1^2 \left(-24 \kappa _{10} \sigma _1 \sigma _0 C_\g \zeta_{\rm G}+9 \kappa _{11} \sigma_0^2 C_\g^2+16 \kappa_0 \sigma _1^2 \zeta_{\rm G}^2\right)}{\sigma _0^2 \sigma _1^2}\right),\nonumber\\
&&
\end{eqnarray}
and  each  function $F_n$ is intended to be $F_n(\zeta_{\rm G})$.

\subsubsection{An illustrative example}
\begin{figure}
    \centering
    \includegraphics[width=14cm]{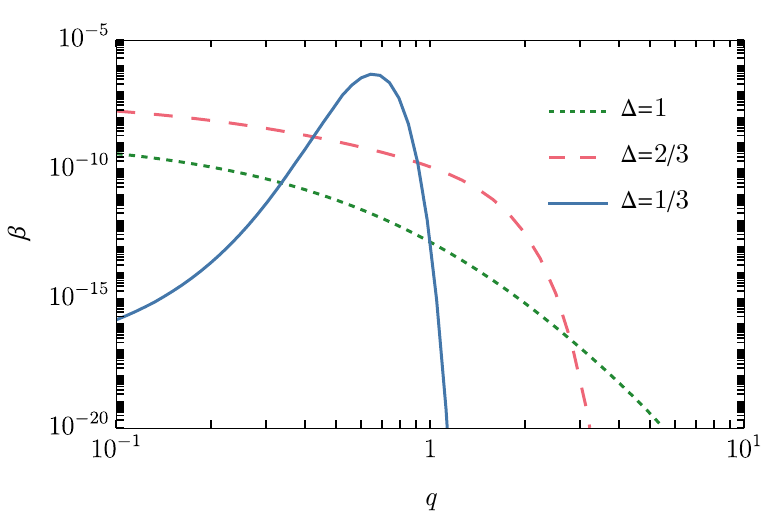}
    \caption{Mass fraction $\beta$ for the non-Gaussian scenario computed with several values of $\Delta$, where we fix $\mu_* = 5/2$, $k_*r_m=2$ and the amplitude of the power spectrum $A=10^{-2}$.} 
    \label{fig:nonlin_gamma_}
\end{figure}

\noindent
We consider the following illustrative example which typically arises in models in which the curvature perturbation is generated during a period of ultra-slow-roll\,\cite{Atal:2018neu, Biagetti:2018pjj, Karam:2022nym,Atal:2019cdz,Tomberg:2023kli}\footnote{For $\zeta_{\rm G}> \mu_\star$, Eq. (\ref{n}) does not capture the possibility of PBHs formed by bubbles of trapped vacuum which requires a separate discussion \cite{Escriva:2023uko,Uehara:2024yyp}.}

\be
\label{n}
\zeta({\bf x})=-\mu_\star\ln\left(1-\frac{ \zeta_{\rm G}({\bf x})}{\mu_\star}\right),
\ee
with $\mu_\star$ a model-dependent parameter depending upon the transition between the ultra-slow-roll phase and the subsequent slow-roll phase. To focus only on the impact of primordial non gaussianity, in this analysis we take $\mu_\star$ as a free parameter.
We take the power spectrum of the Gaussian component to be  a log-normal power spectrum

\be\label{eq:PS}
{\cal P}_\g(k)=\frac{A}{\sqrt{2\pi}\Delta}{\rm exp}\left[-\ln^2(k/k_\star)/2\Delta^2\right].
\ee
Our results are summarized in Fig.\,\ref{fig:nonlin_gamma_} where we have assumed  $k_\star r_m=2$ for the plot\footnote{This is not strictly correct as the value of $r_m$ is a function of the shape of the power spectrum and can slightly change in presence of large primordial non-guassianities\,\cite{Kehagias:2019eil}.
Nevertheless our choice is the ballpark of typical values and we explicitly checked that changing the value of $r_m$ within this range does not change the result qualitatively.}.
The  broadness of the power spectrum is controlled by the parameter  $\Delta$. We observe that by increasing the value of $\Delta$, enlarging the power spectra, again the PBHs formation probability is dominated by the broadest profiles. We have checked that for very peaked power spectrum, as in the case for $\Delta=1/3$, the abundance is peaked again around the average of $q$.

\subsection{Comparison between the two prescriptions}
\noindent
As final goal we compare the calculation presented above, accounting for the curvature of the compaction function at its peak, with the prescription based on threshold statistics on the compaction function, reported in sec.\ref{sec:C1A}, where the only explicit dependence on $q$ is encoded in $C_c(q)$.
In order to make this comparison we consider two cases: $\beta_0$, in which we do not adopt any transfer function ($T=1$) since everything is determined on superhorizon scales, and $\beta_T$, in which we consider the radiation transfer function assuming a perfect radiation fluid, as adopted in Ref. \cite{Ferrante:2022mui}.

\begin{figure}[h!]
    \centering
    \includegraphics[width=14cm]{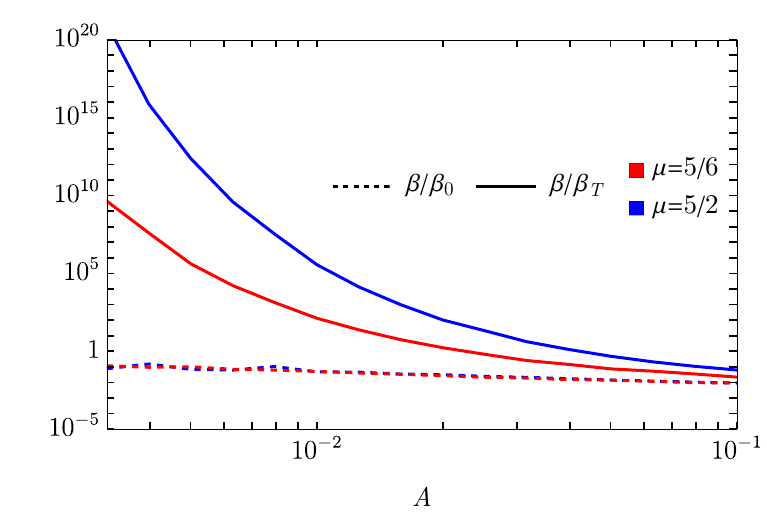}
    \caption{Ratio between mass fraction $\beta$ for the non-Gaussian case between the two prescriptions presented in this chapter. We fix the shape parameter $q=0.5$ (as a consequence also the threshold using Eq. \ref{eq:Threshold}) and the shape of power spectrum $\Delta=1$ while we vary the amplitude.} 
    \label{fig:compa}
\end{figure}

In Fig. \ref{fig:compa}, we show a comparison between the two prescriptions using the typical non-Gaussian relation in the ultra-slow-roll scenario (see Eq.(\ref{n})) with a log-normal power spectrum (see Eq.\ref{eq:PS}) with several benchmark values for $\mu_\star$. We fix $\Delta=1$ in the plots, but we have found analogous results also varying this parameter.
As we can understand from Fig. \ref{fig:compa}, evaluating the quantities on superhorizon scales, i.e. the ratio $\beta/\beta_0$, the prescription based on ref.\cite{Ferrante:2022mui} marginally overestimates the abundance of PBHs. This discrepancy arises because, unlike the prescription used in the literature, where an average profile is employed, the effective threshold is slightly larger than  the averaged case, as evident from Eq.\ref{eq:large_q0}. Nevertheless an equivalent amount of PBHs requires a slightly larger amplitude of the curvature perturbation power spectrum.

The situation is different when we include the radiation transfer function, i.e. the ratio $\beta/\beta_T$. The presence of the transfer function decreases the values of the variances and, as a consequence, it reduces the amount of PBHs.
We discuss phenomenological implications of these differences in sec.\ref{sec:subGWS}.

All the usual statistical approaches concerning the computation of the PBH abundance are exponentially dependent on the threshold. As a consequence we expect that our claim can be extended to other approaches.

\subsection*{Summary}
In this section, we have shown that the abundance of PBHs is dominated by the broadest profiles of the compaction function, even though they are not the typical ones, unless the power spectrum of the curvature perturbation is very peaked. The corresponding threshold is, therefore, always 2/5. We present an extension of the master formula reported in Eq.\ref{eq:Master2} to compute the abundance of PBHs.
We stress that one may be tempted to think that one can always reabsorb NG corrections in the computation of the PBH abundance by means of small retuning of model parameters. While generically true, this is not harmless, as it would spoil the model predictions that depend on the power spectrum (such as the amplitude of the induced gravitational-wave signal) and are directly sensitive to the inclusion of NG corrections.
Furthermore, with the aim of constraining PBHs with future GW detections, missing a precise description of NG corrections may prevent us from setting reliable bounds on their abundance and the fraction of dark matter they can account for. We will elaborate on this point in Chapter \ref{cap:GWs}.
\section{A small step forward: beyond the peaks profiles }
On more general grounds, given the dependence of the critical threshold on the profile of the compaction function, the natural question is if it possible to construct an observable whose critical threshold does not depend at all on the profiles of the peaks. 
In Ref.\,\cite{Escriva:2019phb} it has been proven numerically  that the volume average of the compaction function, calculated in a volume of sphere of radius $R_m$
\be
\overline{\mathcal{C}}(R_m)=\frac{3}{R_m^3}\int_0^{R_m}{\rm d} x\, x^2 \, \mathcal{C}(x)
\ee
has a critical threshold equal to 2/5 independently from the profile. In the case of a broad compaction function, whose critical threshold is 2/5, and since
$\overline{\mathcal{C}}(R_m)\simeq \mathcal{C}(R_m)$, it is trivial that the volume average has the same critical value 2/5. The case of a very spiky compaction function corresponds to a flat Universe with in it a sphere of radius $R_m$ and constant curvature $K(R)=\mathcal{C}(R)/R^2$, that is $\mathcal{C}(R)$ scales like $R^2$. One then obtains 
     
     \be
\overline{\mathcal{C}}(R_m)=3\frac{\mathcal{C}(R_m)}{R_m^5}\int_0^{R_m}{\rm d} x\, x^4=\frac{3}{5}
    \mathcal{C}(R_m)=\frac{3}{5}\cdot\frac{2}{3}=\frac{2}{5},
    \ee
     where it is used that for very spiky compaction functions the critical value is 2/3.
 In the case of non-Gaussian perturbations the universal threshold remains 2/5 \cite{Escriva:2022pnz} for the realistic cases in which the non-Gaussian parameter is positive \cite{Firouzjahi:2023xke}.
Notice that one can construct easily another observable whose threshold is independent from the profile. 
Indeed the compaction function is related to the local curvature of the Universe by the relation $\mathcal{C}(R)=K(R)R^2$. Given a curvature perturbation $\zeta(r)$, a compaction function $\mathcal{C}(R)$ with maximum in $R_m$ and the corresponding curvature $K(R)$,  one  consider a new perturbation with curvature
$$
\overline{K}=\Theta_H(R_m-R) \int_0^{R_m}{\rm d} x\, x^2 \, K(x), $$  
that is a spherical local closed Universe with curvature $\overline{K}$ with radius $R_m$ surrounded by a flat Universe. This corresponds to a new  infinitely peaked compaction function equal to $\Theta_H(R_m-R)R^2\overline{K}$ whose threshold will be  always 2/3 \cite{Harada:2013epa}, independently from the profile of the initial compaction function.
Assuming a universal threshold, one can then write  the probability that the  volume average compaction function is larger than 2/5  even for the non-Gaussian case as (we use here threshold statistics to make the point, one could similarly use peak theory).

\begin{eqnarray}
    \label{ppp}
    P\left[\overline{\mathcal{C}}(R_m)>2/5\right]&=&\Big<\Theta_H\left[\overline{\mathcal{C}}(R_m)-2/5\right]\Big>\nonumber\\
    &=&\frac{1}{2\pi}
\int\left[D \mathcal{C}(r)\right]P\left[ \mathcal{C}(r)\right]
\int_{2/5}^\infty{\rm d}\alpha
\int_{-\infty}^\infty {\rm d}\phi\,e^{i\phi(\overline{\mathcal{C}}(R_m)-\alpha)}
\end{eqnarray}
which can be written as
\be
P\left[\overline{\mathcal{C}}(R_m)>2/5\right]=
\int_{2/5}^\infty{\rm d}\alpha
\int_{-\infty}^\infty {\rm d}\phi\,e^{-i\phi\alpha}\cdot Z[J],
\ee
with
\begin{eqnarray}
 Z[J]=\int\left[D \mathcal{C}({\bf x})\right]P\left[ \mathcal{C}({\bf x})\right]e^{i\int {\rm d}^3 x\, J({\bf x}) \mathcal{C}({\bf x})},\quad J({\bf x})=\,V^{-1}_{R_m}\,\phi\,\Theta_H (\overline{r}_m-r)\, ,
\end{eqnarray}
and the measure $\left[D \mathcal{C}(r)\right]$  is such that 
\begin{eqnarray}
\int\left[D \mathcal{C}({\bf x})\right]P\left[ \mathcal{C}({\bf x})\right]=\int\left[D \mathcal{C}(r)\right]P\left[ \mathcal{C}(r)\right]=1. 
\end{eqnarray}
The correlators are determined by the  expansion of the  partition function $Z[J]$  in terms of the source $J$,  while the corresponding expansion of $W[J]=\ln Z[J]$ generates the connected correlation functions. We will denote the latter as 
\begin{eqnarray}
    \xi^{(n)}({\bf x}_1,\cdots,{\bf x}_n)=
    \frac{\delta}{\delta J({\bf y}_1)}\cdots\frac{\delta}{\delta J({\bf y}_n)}\ln Z[J],
\end{eqnarray}
and the connected cumulants of the volume average linear compaction function as 
\begin{eqnarray}
\langle\overline{\mathcal{C}}^n(R_m)\rangle&=&\frac{1}{V^{n}_{R_m}}
\int {\rm d}^3 x_1\cdots {\rm d} ^3 x_n\prod_{i=1}^n \xi^{(n)}({\bf x}_1,\cdots,{\bf x}_n)
\, \Theta_H(R_m-x_i)\nonumber\\
&=&\prod_{i=1}^n\int\frac{{\rm d}^3 k_i}{(2\pi)^3}P_N({\bf k}_1,\cdots,{\bf k}_n)W(k_1R_m)\cdots W(k_n R_m)\,\delta_D^{(n)}({\bf k}_1+\cdots+{\bf k}_n),\nonumber\\
\langle C_{\rm G}({\bf k}_1),\cdots,C_{\rm G}({\bf k}_n)\rangle &=& P_N({\bf k}_1,\cdots,{\bf k}_n) \delta_D^{(n)}({\bf k}_1+\cdots+{\bf k}_n).
\end{eqnarray}
Then, we may write 
\begin{eqnarray}
    \ln Z[J]&=& \sum_{n=2}^\infty\frac{(-1)^n}{n!}\int {\rm d}^3 {\bf y}_1\cdots \int {\rm d}^3 {\bf y}_n \,
 J_{i_1}({\bf y}_1)\cdots J_{i_n}({\bf y}_n)
 \xi^{(n)}({\bf x}_{i_1},\cdots,{\bf x}_{i_n})\nonumber\\
 &=& \sum_{n=2}^\infty\frac{(-1)^n}{n!}
\phi^n \langle\overline{\mathcal{C}}^n\rangle .
\end{eqnarray}
	Using the above expression for the connected partition function, we find that the one-point statistics of Eq. (\ref{ppp}) can be written as 
	\begin{equation}
\begin{aligned}
\label{11}
P\left[\overline{\mathcal{C}}(R_m)>2/5\right]
 =&
(2\pi)^{-1/2}\int_{2/5}^\infty {\rm d} a
\,
\exp\left\{\sum_{n=3}^\infty \frac{(-1)^n}{n!} 
\langle\overline{\mathcal{C}}^n\rangle
\frac{\partial^n}{\partial a^n}\right\}
\exp{\left(-\frac{a^2}{2\sigma_{\overline{\mathcal{C}}}^2}\right)}\\
=&
(2\pi)^{-1/2}\int_{2/5}^\infty {\rm d} a
\,\left(1-\frac{1}{3!}\langle\overline{\mathcal{C}}_\zeta^3\rangle\frac{{\rm d}^3}{{\rm d} a^3}+
\frac{1}{4!}\langle\overline{\mathcal{C}}^4\rangle\frac{{\rm d}^4}{{\rm d}a^4}+\cdots\right)\exp{\left(-\frac{a^2}{2\sigma_{\overline{\mathcal{C}}}^2}\right)}\\
=& h_0(2/5)+\frac{1}{\sqrt{2\pi}}\sum_{n\geq 3}\frac{1}{2^{\frac{n}{2}}n!}\frac{c_n}{\sigma_{\overline{\mathcal{C}}}^{n-1}}e^{-4/50\sigma_{\overline{\mathcal{C}}}^2}H_{n-1}\left(\frac{2/5}{\sqrt{2}\sigma_{\overline{\mathcal{C}}}}\right),
\end{aligned}
	\end{equation}
where 
\begin{eqnarray}
h_0(2/5)=\frac{1}{2}
{\rm Erfc}\left(\frac{1}{\sqrt{2}}\frac{2/5}{\sigma_{\overline{\mathcal{C}}}}\right),
\end{eqnarray}
$\sigma_{\overline{\mathcal{C}}}$ is the variance,  $H_n$ are Hermite polynomials and we  have defined in Eq. (\ref{11}) the parameters $c_n$ as 
\begin{eqnarray}
    c_n=
  \sum_{\hat{p}[n]}\,\,\,\,\prod_{\substack{p_1m_1+\cdots p_rm_r=n\\p_i\geq 0,m_i\geq 3}}
\frac{n!}{m_1!\cdots m_r! \, p_1!\cdots p_r!}\,\langle\overline{\mathcal{C}}^{m_1}\rangle^{p_1}\cdots \langle\overline{\mathcal{C}}^{m_r}\rangle^{p_r}, \label{110}
\end{eqnarray}
where  $\hat{p}[n]$ denotes the partitions of the integer $n$ into numbers $m_i\geq 3$. 

Given the statistics of the curvature perturbation, one can calculate the abundance of PBHs using the volume average of the linear compaction function, relying solely on superhorizon quantities. Generally, determining the statistics of the curvature perturbation can be challenging  and computing the connected cumulants is highly non-trivial. We are planning to tackle this task in the next future.
\chapter{Constraints on the PBH abundance}\label{sec:Clas}
In this chapter we provide a brief overview of the categories of constraints on the PBH abundance in Section\,\ref{sec:Clas} and subsequently we delve into a detailed description, following ref.\,\cite{Andres-Carcasona:2024wqk,Iovino:2024tyg}, of how we update the constraints utilizing both LIGO-Virgo-KAGRA O3 data and the PTA dataset.

\begin{figure}[h!]
    \centering
    \includegraphics[width=14cm]{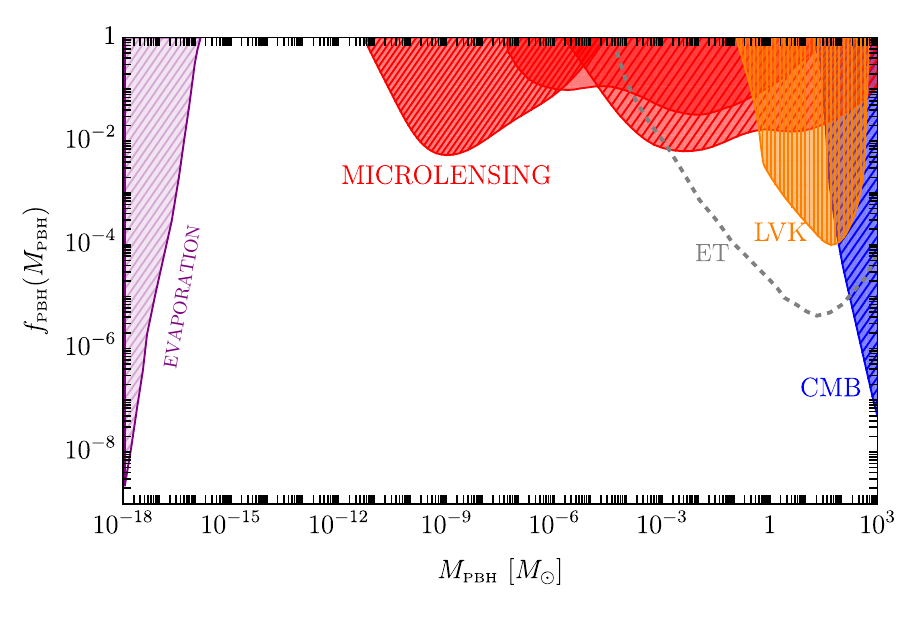}
    \caption{All constraints on the fraction of $\mathrm{DM}$ in the form of PBHs, $f_{\mathrm{PBH}}$, with mass $M_{\mathrm{PBH}}$, coming from PBH evaporation, microlensing, gravitational waves (LVK) and PBH accretion (CMB).} 
    \label{fig:ConTOT}
\end{figure}

The abundance of PBHs relative to the observed dark matter content today is constrained by a multitude of experiments. In Fig. \ref{fig:ConTOT}, we present the most stringent constraints as a function of PBH mass in the region of interest for this thesis. We can identify three broad windows: the asteroidal mass range between $10^{-12}-10^{-17}$ $\mathrm{M}_{\odot}$, where the entirety of dark matter can be accounted for by the existence of PBHs; the intermediate region between $10^{-12}-10^{-3}$ $\mathrm{M}_{\odot}$, where the abundance is particularly constrained by microlensing experiments; and finally, the subsolar-solar mass range, $10^{-3}-10^{3}$ $\mathrm{M}_{\odot}$, where PBHs can explain some observations related to gravitational wave experiments such as PTA and LVK, as also discussed in the subsequent chapter. In this section, we first provide a general overview of the categories of constraints and then explain how we have updated the abundance constraints based on LVK O3 data.

We caution the reader that some constraints reported in the literature were derived under specific assumptions (e.g., all are reported for a monochromatic mass function and unclustered PBHs) and may be relaxed or strengthened with model-dependent modifications of the formation mechanism. Indeed, for constraints in the late Universe, we typically assume that PBHs are not clustered, i.e., they follow a Poissonian spatial distribution. Whether any clustering would alleviate or tighten constraints has been a matter of debate, but it is now universally agreed that PBHs formed from inflaton perturbations should be unclustered if the initial perturbations are Gaussian in nature \cite{Ali-Haimoud:2018dau,Desjacques:2018wuu,Bringmann:2018mxj,Suyama:2019cst,MoradinezhadDizgah:2019wjf}.
\section{Classification of the costraints}
\subsubsection{Evaporated PBHs}
PBHs with $M_{\mathrm{PBH}} \lesssim 10^{-17} M_{\odot}$ evaporate in a time-scale comparable with the age of the Universe. As the energy emitted through the Hawking evaporation becomes increasingly larger for lighter BHs, this process may lead to the production of detectable signatures in the case of ultralight PBHs\,\cite{Balaji:2024hpu}. The emitted radiation has a black-body spectrum with a temperature $T \propto 1 / M_{\mathrm{PBH}}$. The purple region in fig.\,\ref{fig:ConTOT} comes  the most constraining parts of several listed analysis coming from extra-galactic gamma-rays\,\cite{Arbey:2019vqx}, $e^{ \pm}$observations by Voyager 1\,\cite{Boudaud:2018hqb}, positron annihilations in the Galactic Center\,\cite{DeRocco:2019fjq} and gamma-ray observations by INTEGRAL\,\cite{Laha:2019ssq,Laha:2020ivk}. Similar constraints in this mass window were derived in Refs.\cite{Ballesteros:2019exr,Dasgupta:2019cae,Inomata:2020lmk}.
\subsubsection{Gravitational microlensing}
PBHs, being very compact objects, can lead to significant lensing signatures on the electromagnetic radiation reaching the detectors from background sources. Then, the most stringent constraints in the mass range $10^{-10} \lesssim M_{\mathrm{PBH}} \lesssim 10^{3} M_{\odot}$ come from microlensing searches by Subaru HSC\,\cite{Niikura:2017zjd,Smyth:2019whb}, lensing searches of massive compact halo objects towards the Large Magellanic Clouds (MACHO) \,\cite{Macho:2000nvd}, fast transient events near-critical curves of massive galaxy cluster Icarus\,\cite{Oguri:2017ock,Kawai:2024bni}, and observations of stars in the Galactic bulge by the Optical Gravitational Lensing Experiment (OGLE)\,\cite{Mroz:2024wag}. They are shown in red in Fig.\,\ref{fig:ConTOT}.
\subsubsection{Gravitational Wave}
PBHs can assemble in binaries both in the early- and late-time Universe, leading to observable signals at current LIGO and Virgo detectors. 
Several works have explored the use of GW data to find direct or indirect evidence of PBHs. Specifically targeted searches of subsolar mass compact objects, which would provide a smoking gun evidence of the existence of PBHs, have so far been unsuccessful~\cite{LIGOScientific:2018glc,LIGOScientific:2019kan,Nitz:2020bdb,Nitz:2021mzz,Nitz:2021vqh,Nitz:2022ltl,Miller:2020kmv,Miller:2021knj,Andres-Carcasona:2022prl}. Additionally, studies of the LIGO-Virgo GW data including both PBH and ABH populations have not found enough statistical significance to claim the existence of PBHs~\cite{Hutsi:2020sol, Hall:2020daa, Wong:2020yig, Franciolini:2021tla, DeLuca:2021wjr}. However, some of the component masses, in particular, in GW190521~\cite{LIGOScientific:2020iuh} and GW230529$\_$181500~\cite{LIGOScientific:2024elc}, fall in regions where astrophysical models do not predict them, potentially suggesting for a PBH population. Another method of studying the possible PBH population uses the stochastic GW background and its possible detection has also been recast as an upper limit to the PBH abundance~\cite{Raidal:2017mfl, Raidal:2018bbj, Vaskonen:2019jpv, Hutsi:2020sol, Romero-Rodriguez:2021aws, Franciolini:2022tfm}. 

Moreover, for large values of $f_{\mathrm{PBH}}$ close to unity, PBH clustering evolution can reduce the merger rate by enhancing binary interactions in dense environments. However, this effect is not sufficient to reduce the rate to a level that would be compatible with LVK observations in the standard scenario.

The constraints in orange in fig.\,\ref{fig:ConTOT} are taken  from LVK and we will discuss them in detail in the second part of this section.
In gray we indicate the future bound that next generation of ground based detectors will set on the PBH abundance \cite{DeLuca:2021hde,Pujolas:2021yaw,Franciolini:2022htd,Martinelli:2022elq,Franciolini:2023opt,Branchesi:2023mws}.
\subsubsection{PBH accretion and CMB}
Accretion of baryonic matter on PBHs leads to various effects both in the early and late time Universe. As PBH accretion of gas generates the emission of ionizing radiation in the early-time Universe, a PBH population would have an impact on the CMB temperature and polarization anisotropies\,\cite{Ali-Haimoud:2016mbv,Poulin:2017bwe,Serpico:2020ehh}. In particular, Ref.\,\cite{Serpico:2020ehh}  the catalysing effect of the early DM halo forming around individual PBHs is considered. Since the relevant emission takes place at high redshift, its physics is independent of uncertainties in the accretion model due to the onset of structure formation and so, in the standard scenario, PBH clustering is not expected to play a major role\,\cite{Ballesteros:2018swv,Inman:2019wvr,Hutsi:2019hlw}.
Other constraints come from comparing the late-time emission of electromagnetic signals from interstellar gas accretion onto PBHs with observations of galactic radio and X-ray isolated sources\,\cite{Manshanden:2018tze,Hektor:2018qqw} and X-ray binaries\,\cite{Inoue:2017csr}, and the effect of PBH interaction with the interstellar medium compared to data from Leo $\mathrm{T}$ dwarf galaxy observations\,\cite{Lu:2020bmd,Takhistov:2021aqx}.
These constraints are shown in blue in Fig.\,\ref{fig:ConTOT}.
\subsection{Updated constraints using LVK O3}\label{sec:LVK}
In this section, following closely ref.\cite{Andres-Carcasona:2024wqk} we update the constraints on the PBH abundance using the GW data from LIGO-Virgo up to the third observational run (O3) studying the population of the observed events. One of our aims is to derive constraints which do not depend significantly on the underlying formation scenario. Thus, we consider a variety of different PBH mass functions. We take a closer look at mass functions arising in scenarios in which PBHs are produced via critical collapse of large overdensities~\cite{Musco:2008hv, Niemeyer:1999ak, Yokoyama:1998xd}. In such scenarios, PBH formation is affected by the cosmological background. In particular, the QCD phase transition, during which the cosmological equation of state is softened, catalyzes the formation of PBHs in the solar mass range~\cite{Jedamzik:1996mr, Byrnes:2018clq, Musco:2023dak}. We also account for the sizeable effect primordial non-Gaussianities can have on the mass function and show that their effect on the constraints is, however, quite marginal.
\subsubsection{Methods}

We mainly adopt a hierarchical Bayesian approach. We define the population parameters that describe the production rate of binary mergers as $\Lambda$ and the data measured from $\Nobs$ observed events as $\{d\}$. The associated likelihood is~\cite{LIGOScientific:2020kqk, Thrane:2018qnx,Mandel:2018mve, Mastrogiovanni:2023zbw}
\be \label{eq:hierarchicalBayesian}
    \mathcal{L}(\{d\}|\Lambda) \propto e^{-N(\Lambda)}\prod_{i=1}^{\Nobs}\int \mathcal{L}(d_i|\theta)\pi(\theta|\Lambda)\td \theta\, ,
\ee
where $N$ is the expected number of observed events. On the other hand, $\mathcal{L}(d_i|\theta)$ is the event likelihood, given some parameters $\theta$, and $\pi(\theta|\Lambda)$ is called the hyperprior and defines the distribution of mass, spin, redshift and PBH merger rate.

The single event likelihood is usually not available. Instead, posterior samples are provided. Using Bayes' theorem, the $N_p$ posterior samples and the likelihood are related as $p(\theta|d_i,\Lambda)\propto\mathcal{L}(d_i|\theta)\pi_{\varnothing}(\theta|\Lambda)$, where $\pi_{\varnothing}(\theta|\Lambda)$ is the prior used for the initial parameter estimation. Therefore, the integral appearing in Eq.~\eqref{eq:hierarchicalBayesian} for each event can be approximated using Monte-Carlo integration as~\cite{Mastrogiovanni:2023zbw,LIGOScientific:2020kqk}
\be
    \int\!\mathcal{L}(d_i|\theta)\pi(\theta|\Lambda)\td \theta 
    \!\approx\!\frac{1}{N_p}\!\sum_{j=1}^{N_p}\!\frac{\pi(\theta_{ij}|\Lambda)}{\pi_{\varnothing}(\theta_{ij}|\Lambda)}
    \!\equiv\!\frac{1}{N_p}\!\sum_{j=1}^{N_p}w_{ij} .
\ee
To ensure numerical stability during this integration, the effective number of posterior samples per each event, defined as~\cite{Mastrogiovanni:2023zbw,Talbot:2023pex,LIGOScientific:2021aug}
\be
    N_{\mathrm{eff},i} = \bigg[  \displaystyle\sum_j^{N_p} w_{ij} \bigg]^2 \!\bigg/\, \displaystyle\sum_j^{N_p} w_{ij}^2 \,,
\ee
is set to $20$. 

The number of expected events is~\cite{Mastrogiovanni:2023zbw,LIGOScientific:2021aug,LIGOScientific:2020kqk,Vaskonen:2019jpv,Hutsi:2020sol}
\be \label{eq:Nexpected}
    N(\Lambda) = T\int p_{\mathrm{det}}(\theta)\pi(\theta|\Lambda)\td \theta\, ,
\ee
where $p_{\mathrm{det}}(\theta)$ denotes the detection probability. It depends primarily on the masses and redshift of the system~\cite{LIGOScientific:2020kqk}. For this reason, the effect of the spin on the detection probability will be ignored in this analysis. We use a semianalytical estimate of $p_{\mathrm{det}}$, following the approach of Ref.~\cite{Vaskonen:2019jpv},
\be
    p_{\mathrm{det}}(\theta)=\int_{\frac{\mathrm{SNR}_c}{\mathrm{SNR}(\theta)}}^1 p(\omega)\td \omega \,,
\ee
where SNR denotes the signal-to-noise ratio and $p(\omega)$ the probability density function of the projection parameter $\omega\in[0,1]$ defined as (see e.g.~\cite{Hutsi:2020sol})
\be
    \omega^2 = \frac{(1+\iota^2)^2}{4}F_+(\alpha,\delta,\psi)^2+\iota^2F_\times(\alpha,\delta,\psi)^2 ~,
\ee
where $F_+(\alpha,\delta,\psi)$ and $F_\times(\alpha,\delta,\psi)$ represent the antenna patterns of an L-shaped detector. Assuming a uniform distribution of the inclination $\iota\in(-1,1)$, right ascension $\alpha\in(0,2\pi)$, cosine of the declination $\cos(\delta)\in(-1,1)$ and polarization $\psi \in (0,2\pi)$; the PDF of the projection parameter can be obtained and integrated to give the detection probability. Here, the critical SNR for detection is set to $\mathrm{SNR_c}=8$. 
The hyperprior used is
\be
    \pi(\theta|\Lambda) = \frac{1}{1+z}\frac{\td V_c}{\td z} \frac{\td R}{\td m_1 \td m_2}(\theta|\Lambda) \, ,
\ee
where $V_c$ is the comoving volume and $R$ the merger rate. 

Following the same prescription as for the posterior samples, we evaluate Eq.~\eqref{eq:Nexpected} using Monte-Carlo integration over $N_g$ generated samples as~\cite{Mastrogiovanni:2023zbw}
\be
    N(\Lambda)\approx \frac{T}{N_g}\sum_{j= 1}^{N_{\mathrm{det}}}\frac{\pi(\theta_j|\Lambda)}{\pi_{\mathrm{inj}}(\theta_j)}\equiv \frac{T}{N_g}\sum_{j=1}^{N_{\mathrm{det}}}s_j\, ,
\ee
where $\pi_{\mathrm{inj}}(\theta_j)$ is the prior probability of the $j$th event. Similarly, a numerical stability estimator can be defined for the injections as \cite{Farr:2019rap,Mastrogiovanni:2023zbw}
\be
    N_{\mathrm{eff,inj}} = \frac{\bigg[\displaystyle \sum_j^{N_{\mathrm{det}}} s_j \bigg]^2}{\displaystyle\sum_j^{N_{\mathrm{det}}} s_j^2-\displaystyle\frac{1}{N_{g}}\bigg[\displaystyle\sum_j^{N_{\mathrm{det}}} s_j\bigg]^2}\, ,
\ee
which we set to $N_{\mathrm{eff,inj}}>4\Nobs$. Finally, the log-likelihood is evaluated as~\cite{Mastrogiovanni:2023zbw}
\begin{equation}
\ln \mathcal{L}(\{d\} \mid \Lambda) \approx-\frac{T_{\text {obs }}}{N_g} \sum_j^{N_{\text {det }}} s_j+\sum_i^{N_{\text {obs }}} \ln \left[\frac{T_{\text {obs }}}{N_p} \sum_j^{N_p} w_{i j}\right]
\end{equation}
We implement this computation on top of the ICAROGW code, as it already includes these functionalities~\cite{Mastrogiovanni:2023zbw}.

In the case without any detected events, the likelihood~\eqref{eq:hierarchicalBayesian} reduces to~\cite{Vaskonen:2019jpv,Hutsi:2020sol}
\be
    \mathcal{L}(\{d\}|\Lambda) \propto e^{-N(\Lambda)} \, ,
\ee
corresponding to a simple Poisson process. Therefore, in the case where all the events detected by LVK are considered to be of astrophysical origin, the 2$\sigma$ confidence level upper limits can be placed by discarding the region of the parameter space where $N>3$~\cite{Vaskonen:2020lbd,Hutsi:2020sol}.

To analyze the scenarios with multiple BH binary populations, two approaches will be followed: an agnostic and a model-dependent one.  To perform the agnostic analysis, random subsamples of the events are selected and assumed to be primordial~\cite{Hutsi:2020sol}. Then, the Bayesian inference using the likelihood of Eq.~\eqref{eq:hierarchicalBayesian} is performed. Combining the results for the different subsamples, the contour with the $2\sigma$ CL upper limits can be obtained. While it is practically impossible to test all combinations, as stated in Ref.~\cite{Hutsi:2020sol}, the upper limits converge fast with a relatively small set of subsamples, especially if these are chosen by considering observed events in small mass intervals.

For the model-dependent case, the merger rate and population of ABHs is assumed to follow a given distribution and the fit using the likelihood of Eq.~\eqref{eq:hierarchicalBayesian} with the full set of parameters of both the PBH and ABH models is performed. In any case, all limits arising from this method should lie between the constraints using the agnostic approach and the one assuming that all events are astrophysical.

\subsubsection{BH binary populations: PBH population}

PBH binaries can arise from various channels. Characterized by the time of formation, one can distinguish binaries formed in the early Universe before matter domination as the first gravitationally bound non-linear structures in the Universe, and binaries formed later in DM haloes (for a review see~\cite{Raidal:2024bmm}). Both processes can be divided by the number of progenitors of the binary, i.e., whether the binary formation involves two or more PBHs. 

The early Universe formation channels include a two-body channel in which close PBHs decouple from expansion and form a highly eccentric binary, with the initial angular momentum provided by tidal forces from surrounding PBHs and matter fluctuations~\cite{Nakamura:1997sm, Ioka:1998nz, Ali-Haimoud:2017rtz, Raidal:2018bbj, Vaskonen:2019jpv, Hutsi:2020sol}, and a three-body channel in which a compact configuration of three PBHs forms a three-body system after decoupling from expansion and a binary is formed after one of the PBHs is ejected from this system~\cite{Vaskonen:2019jpv}. The latter process produces generally much harder and less eccentric binaries than the former. However, as compact three-body systems are less likely than two-body ones, the three-body channel is subdominant to the two-body one, unless PBHs make up a considerable fraction, above $\mathcal{O}(10\%)$, of DM. In the late Universe, PBHs can form binaries during close encounters in the DM haloes if they emit enough of their energy into GWs to become bound~\cite{Mouri:2002mc, Bird:2016dcv}, and, if the PBH abundance is sufficiently large, also three-body encounters in DM haloes can become relevant~\cite{Franciolini:2022ewd}. However, both late Universe formation channels are subdominant to the early Universe channels, and we will not consider their contribution in this study.

The merger rate from the early Universe two-body channel is~\cite{Raidal:2018bbj, Vaskonen:2019jpv, Hutsi:2020sol}:
\begin{multline}\label{eq:R2}
    \frac{\td R_{\mathrm{PBH},2}}{\td m_1 \td m_2}
    = \frac{1.6\times 10^6}{\text{Gpc}^3\text{yr}^1}
    \fpbh^{\frac{53}{37}}\left[ \frac{t}{t_0}\right]^{-\frac{34}{37}}\left[ \frac{M}{\msun}\right]^{-\frac{32}{37}}   \times 
    \eta^{-\frac{34}{37}} S  \left[ \psi,\fpbh,M \right]\frac{\psi(m_1)\psi(m_2)}{m_1 m_2}\, ,
\end{multline}
where $\fpbh\equiv \rho_{\mathrm{PBH}}/\rho_{\mathrm{DM}}$ denotes the total fraction of DM in PBHs, $t_0$ the current age of the Universe, $M=m_1+m_2$ the total mass of the binary, $\eta=m_1m_2/M^2$ its symmetric mass ratio, $S$ the suppression factor, $\psi(m) \equiv \rho_{\rm PBH}^{-1}\td \rho_{\rm PBH}/\td \ln m$ the PBH mass function and $\meanm$ its mean. Note that $\int \td \ln m\, \psi(m) = 1$ and the average over any mass-dependent quantity $X$ is defined via the number density as\footnote{We remark that different conventions for the mass function have been used in Refs.~\cite{Raidal:2018bbj,Vaskonen:2019jpv,Hutsi:2020sol}.}
\be
    \langle X \rangle  
    \equiv \int \frac{\td n_{\rm PBH}}{n_{\rm PBH}} X
    = \frac{\int \td \ln m \, m^{-1}\, \psi(m) X }{\int \td \ln m \, m^{-1} \,\psi(m)}  \,,
\ee
where $\td n_{\rm PBH} = m^{-1} \td \rho_{\rm PBH}$ is the differential number density of PBHs.

We compute the suppression factor $S$ as the product of two contributions $S=S_1S_2(z)$. The first factor accounts for the initial configuration and perturbations in the surrounding matter and discards all those initial configurations that contain a third PBH within a distance smaller than $y$. It can be approximated as~\cite{Hutsi:2020sol}
\be\label{eq:S1}
    S_1\approx 1.42\left[ \frac{\langle m^2\rangle/\meanm }{\tilde{N}(y)+C}+\frac{\sigma_M^2}{\fpbh^2} \right]^{-\frac{21}{74}}e^{-\tilde{N}(y)}~,
\ee
where $\sigma_M\simeq 0.004$ is the rescaled variance of matter density perturbations,  $\tilde{N}(y)$ is the number of expected PBH to be formed inside a sphere of comoving radius $y$ and can be approximated as~\cite{Raidal:2018bbj}
\be
    \tilde{N}(y)\approx \frac{M}{\meanm}\frac{\fpbh}{\fpbh+\sigma_M}~,
\ee
and $C$ is a fitting factor
\be
    C= \fpbh^2\frac{\langle m^2\rangle/\meanm^2}{\sigma_M^2}
    \times 
    \left\{\left[ \frac{\Gamma(29/37)}{\sqrt{\pi}}U\left(\frac{21}{74},\frac{1}{2},\frac{5}{6}\frac{\fpbh^2}{\sigma_M^2}\right)\right]^{-\frac{74}{21}}-1\right\}^{-1}\, ,
\ee
$U$ denotes the confluent hypergeometric function and $\Gamma$ is the gamma function. The second term of the suppression factor, $S_2(z)$, accounts for the disruption of PBH binaries due to encounters with other PBHs in the late Universe~\cite{Vaskonen:2019jpv}. It can be approximated by~\cite{Hutsi:2020sol}
\be\label{eq:S2}
    S_2(z)\approx\min\left[ 1,\,0.01\chi^{-0.65}e^{-0.03\ln^2\chi}\right]~,
\ee
where $\chi = (t(z)/t_0)^{0.44}\fpbh$. This approximation is accurate if $z \lesssim 100$, which is within our region of interest.

The merger rate of binaries formed from compact three-body configurations in the early Universe is~\cite{Vaskonen:2019jpv, Raidal:2024bmm}
\begin{align}\label{eq:R3}
\frac{\mathrm{d} R_{\mathrm{PBH}, 3}}{\mathrm{~d} m_1 \mathrm{~d} m_2} \approx & \frac{7.9 \times 10^4}{\mathrm{Gpc}^3 \mathrm{yr}}\left[\frac{t}{t_0}\right]^{\frac{\gamma}{7}-1} f_{\mathrm{PBH}}^{\frac{144 \gamma}{259}+\frac{47}{37}}\left[\frac{\langle m\rangle}{M_{\odot}}\right]^{\frac{5 \gamma-32}{37}}\left(\frac{M}{2\langle m\rangle}\right)^{\frac{179 \gamma}{259}-\frac{2122}{3333}}\\&(4 \eta)^{-\frac{3 \gamma}{7}-1}  \mathcal{K} \frac{e^{-3.2(\gamma-1)} \gamma}{28 / 9-\gamma} \mathcal{F}\left(m_1, m_2\right) \frac{\psi\left(m_1\right) \psi\left(m_2\right)}{m_1 m_2}
\end{align}
where $\gamma \in [1,2]$ characterizes the dimensionless distribution of angular momenta $j$, which is assumed to take the form $P(j) = \gamma j^{\gamma-1}$ after the initial three-body system has ejected one of the PBHs. The value $\gamma=2$ corresponds to the equilibrium distribution while studies suggest a super-thermal distribution with $\gamma = 1$~\cite{Raidal:2018bbj,Stone:2019qvl}.The factor
$$
\overline{\mathcal{F}}\left(m_1, m_2\right) \equiv \int_{m \leq m_1, m_2} \mathrm{~d} \ln m \psi(m) \frac{\langle m\rangle}{m}\left[2 \mathcal{F}\left(m_1, m_2, m\right)+\mathcal{F}\left(m, m_1, m_2\right)\right],
$$
with
$$
\mathcal{F}\left(m_1, m_2, m_3\right) =m_1^{\frac{5}{3}} m_2^{\frac{5}{3}} m_3^{\frac{7}{9}}\left(\frac{m_1+m_2}{2}\right)^{\frac{4}{9}} 
\left(\frac{m_1+m_2+m_3}{3}\right)^{\frac{2}{9}}\langle m\rangle^{-\frac{43}{9}}
$$
accounts for the composition of masses in the initial 3-body system and assumes that the lightest PBH gets ejected. Thus, the 3-body channel tends to generate binaries from the heavier tail of the PBH mass spectrum. The factor $\mathcal{K}$ accounts for the hardening of the early binary in encounters with other PBHs. We will use $\gamma=1$ and $\mathcal{K} = 4$ as suggested by numerical simulations~\cite{Raidal:2018bbj}.\footnote{The most conservative choice corresponds to $\gamma=2$ and $\mathcal{K} = 1$.} 

The binaries formed from three-body systems are significantly less eccentric and much harder than those contributing to the merger rate of Eq.~\eqref{eq:R2}. This means that, unlike for the rate~\eqref{eq:R2}, the binary-single PBH encounters in DM (sub)structures will approximately preserve the angular momentum distribution and harden the binaries. As a result, these encounters will not diminish the merger rate.\footnote{We note that Eq.~\eqref{eq:R3} does not account for the potential enhancement of the merger rate due to the hardening of the binaries by binary-PBH encounters. However, the effect of this process was estimated to be relatively weak for binaries formed in PBH DM haloes~\cite{Franciolini:2022ewd} and we expect this estimate to be valid also for the early binaries.} Since the suppression factors of Eq.~\eqref{eq:S1} and Eq.~\eqref{eq:S2} only apply to Eq.~\eqref{eq:R2}, the three-body channel can become dominant for relatively large PBH abundances, when $f_{\rm PBH} \gtrsim 0.1$. 

Unlike the rate of Eq.\eqref{eq:R2}, the rate of Eq.~\eqref{eq:R3} is expected to be enhanced when PBHs are formed in clusters since the density of very compact three-PBH configurations would be larger. Thus, more binaries would be formed as such configurations decouple from expansion. For this reason, the constraints derived from Eq.~\eqref{eq:R3} are more robust in ruling out $f_{\rm PBH} \approx 1$ as well as initially clustered PBH scenarios. On a more speculative note, as a potential caveat for avoiding the merger rate constraints, one can imagine extreme cases of clustering, in which case the mean PBH separation is so small that all the early binaries would merge within much less than a Hubble time even if they were circular -- thus, they would not contribute to the current merger rate. However, extreme clustering of PBHs in the stellar mass range can introduce noticeable isocurvature perturbations at relatively large scales and will likely clash with the Lyman-$\alpha$ observations~\cite{DeLuca:2022uvz}.

\begin{figure}   
    \includegraphics[width=0.49\textwidth]{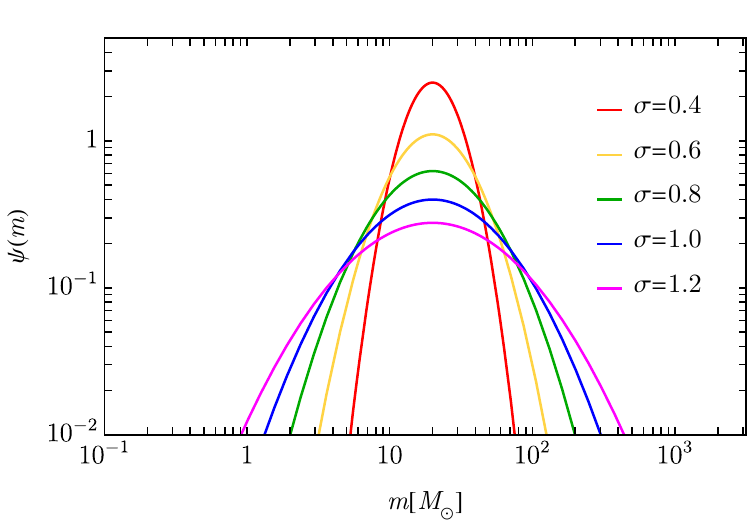}
    \quad
    \includegraphics[width=0.49\textwidth]{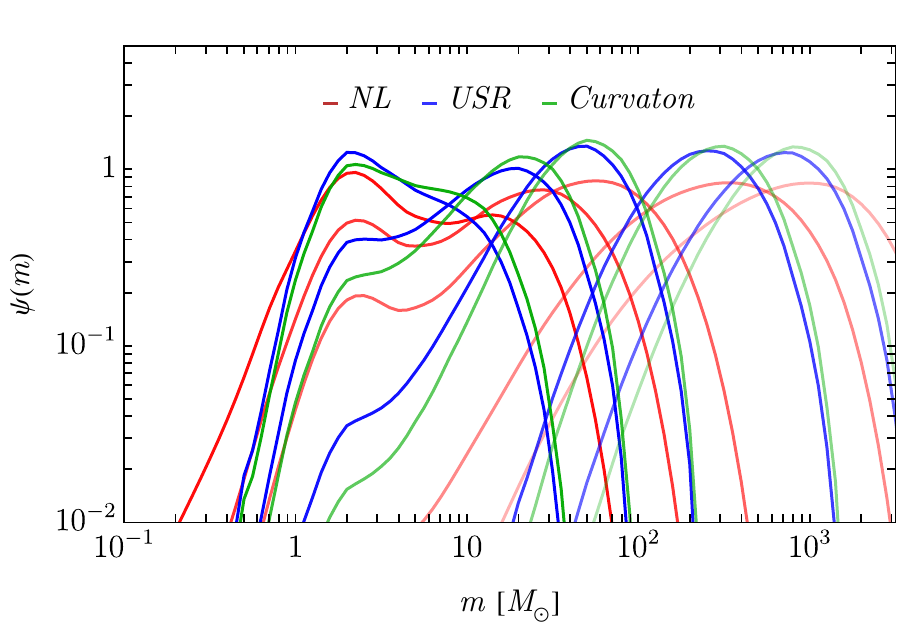}
    \caption{ Examples of extended mass functions considered in this analysis. \emph{Left panel:}  Log-normal mass function with $m_c=20 \msun$. \emph{Right panel:} Critical collapse mass function arising from a broken power-law curvature power spectrum with $\alpha=4$, $\beta=0.5$, different values of $k_*$, including only non-linearities (red), primordial non-Gaussianities from USR (blue), and curvaton (green) models. For the primordial non-Gaussianities, in the case of USR models, the choice of $\beta$ also determines the amount of primordial non-Gaussianities. For the curvaton model, we fix $r_{\rm dec}=0.1$.}
    \label{fig:mass_functions}
\end{figure}

In this analysis we consider the following PBH mass functions:

\begin{itemize}
    \item \textbf{Monochromatic:} The monochromatic mass function is
\be
    \psi(m) = m_c\delta(m-m_c)\,,
\ee
where $\delta$ is the Dirac delta function and $m_c$ is the mass where the distribution has the peak. It is likely the simplest possible and approximates models that predict a sharp peak at a particular mass. By construction, the mean of this distribution is $\meanm=m_c$ and the variance vanishes, $\varm=0$. If PBHs of different masses contribute independently to an observable that is constrained, the constraints obtained for a monochromatic mass function can be easily recast to other mass functions~\cite{Carr:2017jsz}. This is, however, not the case for the constraints arising from the PBH mergers and the constraint for each mass function must be derived separately.
\item \textbf{Log-normal:} The log-normal mass function
\be
    \psi(m)=\frac{1}{\sqrt{2\pi}\sigma}\exp\left\{ -\frac{\ln^2(m/m_c)}{2\sigma^2} \right\}\,,
\ee
where $m_c$ and $\sigma^2$ denote the mode and the width of the distribution. It is a good approximation of wider peaks in the mass spectrum and it is predicted by several PBH formation mechanisms. The mean of this distribution is $\meanm = m_c e^{-\sigma^2/2}$ and the variance $\varm= \meanm^2 (e^{\sigma^2}-1)$. The log-normal mass function is depicted in Fig.~\ref{fig:mass_functions} for different widths.

\item \textbf{Critical collapse mass function:}

The masses of PBHs formed from large curvature fluctuations follow the critical scaling law~\cite{Choptuik:1992jv, Niemeyer:1997mt, Niemeyer:1999ak} that gives rise to a PBH mass function with a power-law low-mass tail with an exponential high-mass cut-off~\cite{Vaskonen:2020lbd}. The exact shape of the mass function depends on the shape of the curvature power spectrum and the amount of non-Gaussianities~\cite{Bugaev:2013vba, Nakama:2016gzw, Byrnes:2012yx, Young:2013oia, Yoo:2018kvb, Kawasaki:2019mbl, Yoo:2019pma,
Riccardi:2021rlf, Taoso:2021uvl, Meng:2022ixx, Escriva:2022pnz, Young:2014ana, Shibata:1999zs, Musco:2018rwt, Young:2022phe}. Therefore, instead of using an ansatz directly on the mass function, we compute it numerically starting from the curvature power spectrum accounting for different types of non-Gaussianities. In this analysis, we assume a \emph{broken power-law} (BPL) shape for the curvature power spectrum:
\be \label{eq:PPL}
    \mathcal{P}_{\zeta}(k)
    = A \,\frac{\left(\alpha+\beta\right)^{\lambda}}{\left[\beta\left(k / k_*\right)^{-\alpha/\lambda}+\alpha\left(k / k_*\right)^{\beta/\lambda}\right]^{\lambda}},
\ee
where $\alpha, \beta>0$ describe, respectively, the growth and decay of the spectrum around the peak and $\lambda$ characterizes the width of the peak. Typically, $\alpha \approx 4$~\cite{Byrnes:2018txb,Karam:2022nym}.

To properly compute the abundance of PBHs, as described in this chapter, we include the NGs from non-linearities and primordial NGs. 

We consider several ansatz for primordial NGs: the USR models\footnote{This relation is derived assuming a constant-roll phase following the USR-like phase and was found to hold also in the stochastic formalism~\cite{Tomberg:2023kli}. Nevertheless, as quantum diffusion generally depends on the shape of the inflationary potential~\cite{Biagetti:2018pjj, Ezquiaga:2018gbw, Ezquiaga:2019ftu}, this relation may be modified if one accounts for the transition into the USR phase.}~\cite{Garcia-Bellido:2017mdw, Kannike:2017bxn,Ballesteros:2020qam,Inomata:2016rbd,Iacconi:2021ltm,Kawai:2021edk,Bhaumik:2019tvl,Cheong:2019vzl,Inomata:2018cht,Dalianis:2018frf,Motohashi:2019rhu,Hertzberg:2017dkh,Ballesteros:2017fsr,Karam:2022nym,Rasanen:2018fom,Balaji:2022rsy,Frolovsky:2023hqd,Dimopoulos:2017ged,Germani:2017bcs,Choudhury:2013woa,Ragavendra:2023ret,Cheng:2021lif,Franciolini:2023agm,Karam:2023haj,Mishra:2023lhe,Cole:2023wyx,Karam:2023haj,Frosina:2023nxu,Franciolini:2022pav,Choudhury:2024one,Wang:2024vfv}, and the the curvaton model~\cite{Enqvist:2001zp, Lyth:2001nq, Sloth:2002xn, Lyth:2002my, Dimopoulos:2003ii, Kohri:2012yw, Kawasaki:2012wr, Kawasaki:2013xsa, Carr:2017edp, Ando:2017veq, Ando:2018nge, Chen:2019zza, Liu:2020zzv, Pi:2021dft, Cai:2021wzd, Liu:2021rgq, Chen:2023lou, Torrado:2017qtr, Chen:2023lou, Cable:2023lca, Ferrante:2023bgz, Inomata:2023drn}.

To estimate the mass function we follow the prescription presented in sec.\ref{sec:C1A}. We note that corrections to the horizon crossing and from the non-linear radiation transfer function can affect the PBH abundance when compared to the current prescription, which relies on average compaction profiles (see sec.\ref{sec:C1B}). However, since the effect of the radiation transfer function involves uniform rescaling of all variances, we expect minimal alteration to the shape of the mass functions and thus also to our results.

The softening of the equation of state during the thermal evolution of the Universe on PBH formation, in particular, the QCD phase transition, can enhance the formation of PBHs~\cite{Jedamzik:1996mr, Byrnes:2018clq}. It is accounted for by considering that $\gamma\left(M_H\right), \mathcal{K}\left(M_H\right), \mathcal{C}_{\rm th}\left(M_H\right)$ and $\Phi\left(M_H\right)$ are functions of the horizon mass around $m=\mathcal{O}\left(M_\odot\right)$~\cite{Franciolini:2022tfm, Musco:2023dak}.

Examples of critical mass functions arising from curvature power spectra with $\beta = 0.5$ and $\alpha = 4$ are shown in the right panel of Fig.~\ref{fig:mass_functions} for different primordial non-Gaussianities. The enhancement around $1\,\msun$ arises due to the QCD phase transition. As a result, the QCD phase transition can generate multimodal mass distributions. However, as seen from Fig.~\ref{fig:mass_functions}, peaks around $1\,\msun$ are not universal. In particular, the effect of the QCD phase transition is negligible for mass functions peaked much above $1\,\msun$. 

For mass functions peaking close to $1\,\msun$, the QCD effect tends to be more pronounced in the non-linearities only (NL) scenario, indicating that primordial non-Gaussianities tend to curtail the impact of the QCD phase transition. In general, we find that primordial non-Gaussianities tend to reduce the width of the mass function in both the USR and curvaton models.
\end{itemize}

\subsubsection{BH binary populations: ABH population}

To describe the ABH population, we use the phenomenological POWER-LAW + PEAK model~\cite{KAGRA:2021duu}. This general parametrization allows us to cover a wide range of potential ABH scenarios that could serve as a foreground for PBH mergers. The differential merger rate for this model is 
\be\label{eq:R_ABH}
    \frac{\td \Ra}{\td m_1 \td m_2}
    \!=\! R_{\mathrm{ABH}}^0 p_{\mathrm{ABH}}^z(z)p_{\mathrm{ABH}}^{m_1}(m_1)p_{\mathrm{ABH}}^{m_2}(m_2|m_1) .
\ee
The probability density function of the primary mass is modeled as a combination of a power law and a Gaussian peak 
\be
    p_{\mathrm{ABH}}^{m_1}(m_1) = \left[ (1-\lambda)P_{\mathrm{ABH}}(m_1)+\lambda G_{\mathrm{ABH}}(m_1)\right]  S(m_1|\delta_m,\mmin) \, ,
\ee
where
\bea
    P_{\mathrm{ABH}}(&m_1|\alpha,\mmin,\mmax) \propto \Theta(m\!-\!\mmin) \Theta(\mmax\!-\!m) m_1^{-\alpha} \,,\\
    G_{\mathrm{ABH}}(&m_1|\mu_G,\sigma_G,\mmin,\mmax) \propto \Theta(m\!-\!\mmin) \Theta(\mmax\!-\!m) e^{-\frac{(m_1-\mu_G)^2}{2\sigma_G^2}}\,
\eea
are restricted to masses between $\mmin$ and $\mmax$ and normalized. The term $S(m_1|\delta_m,\mmin)$ is a smoothing function, which rises from 0 to 1 over the interval $(\mmin, \mmin + \delta_m)$:
\bea 
 \begin{cases} 
      0 & m< \mmin \\
      [f(m-\mmin,\delta_m)+1]^{-1} & \mmin\leq m< \mmin+\delta_m \\
      1 & m\geq \mmin+\delta_m\, .
   \end{cases}
\eea
with
\be
f(m',\delta_m) = \exp\left( \frac{\delta_m}{m'}+\frac{\delta_m}{m'-\delta_m}\right)\, .
\ee
The distribution of the secondary mass is modelled as a power law,
\be
    p_{\mathrm{ABH}}^{m_2}(m_2|m_1,\beta,\mmin) \propto \left(\frac{m_2}{m_1} \right)^\beta \,,
\ee
where the normalization ensures that the secondary mass is bounded by $\mmin \leq m_2 \leq m_1$. Finally, the redshift evolution considered is 
\be
    p_{\mathrm{ABH}}^z(z|\kappa) \propto (1+z)^\kappa\,,
\ee
where we leave $\kappa$ as a free parameter during inference, unlike in Ref.~\cite{Franciolini:2022tfm} where it was fixed to $\kappa = 2.9$ corresponding to the best fit value in~\cite{KAGRA:2021duu}.

\subsubsection{Results}

\begin{figure}
    \centering
    \includegraphics[width=0.8\textwidth]{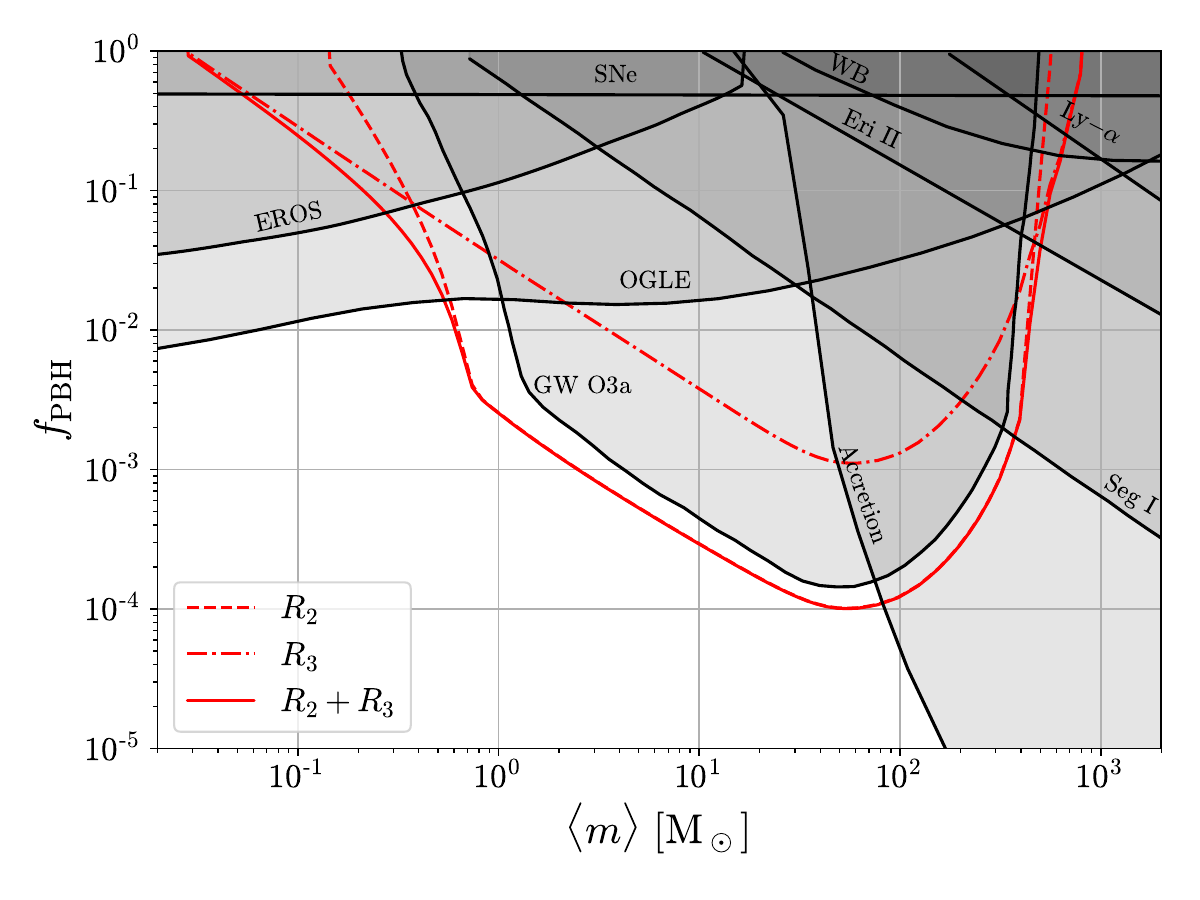}
    \caption{Constraints for the monochromatic mass function detailing the contribution from $R_{2}$ and $R_{3}$ assuming that none of the observed events has a primordial origin. The constraints shown in gray are described in the main text and assume a monochromatic mass function.}
    \label{fig:monofpbh}
\end{figure}

In the following, we will consider pessimistic and optimistic scenarios. In the pessimistic case, none of the observed events has a primordial origin. This scenario gives the most stringent constraints. In the optimistic case, we consider the possibility that a subset of the observed GW events were due to PBH mergers. We will approach this possibility in two ways: 1) fitting the data to a model that contains a mixed population of ABHs and PBHs and 2) fitting all different subsets of events with a PBH population. Both approaches have their advantages. The first approach makes it possible to impose specific characteristics of ABH binaries, which are, however, highly uncertain at present. The second method is completely agnostic about the ABH population and will thus result in the most conservative constraints.

\begin{figure*}
    \centering
    \includegraphics[width=0.45\textwidth]{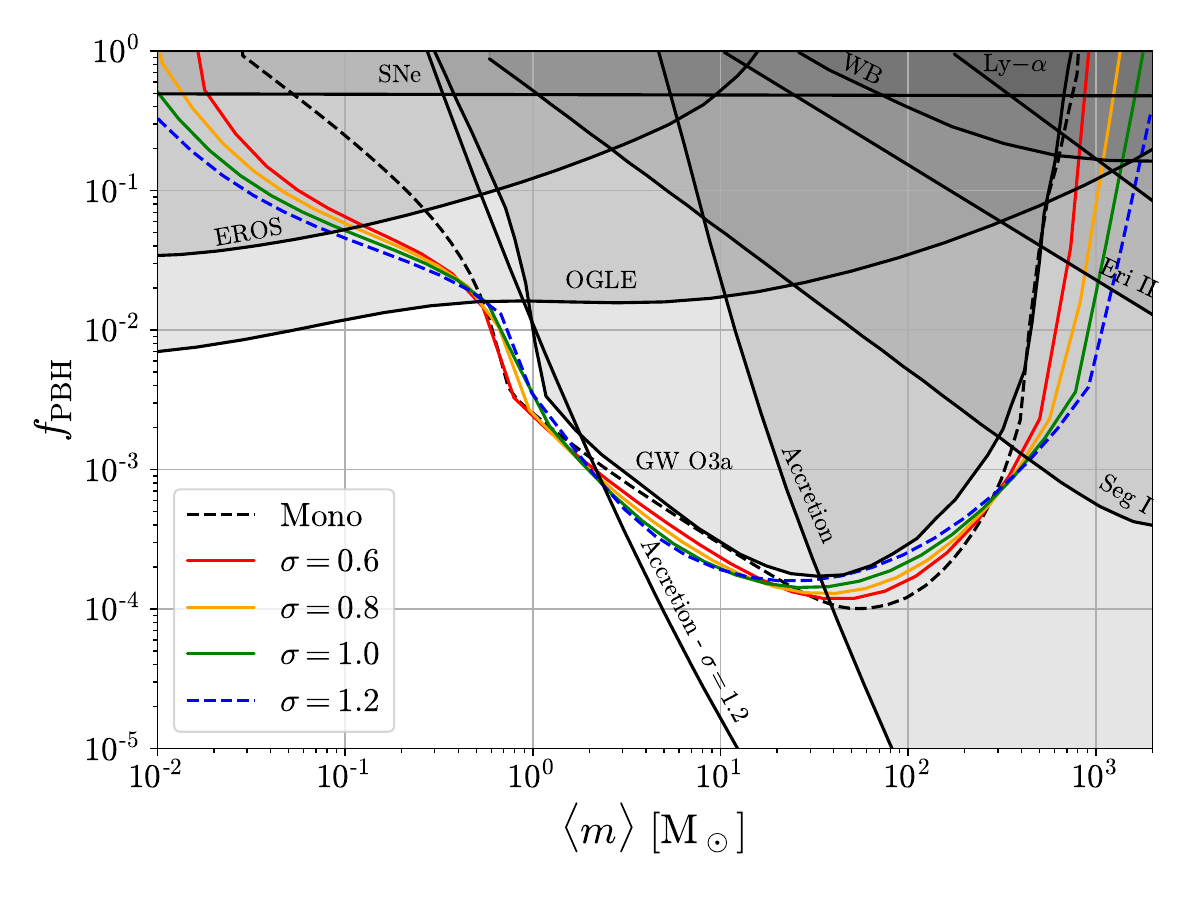}\quad
    \includegraphics[width=0.45\textwidth]{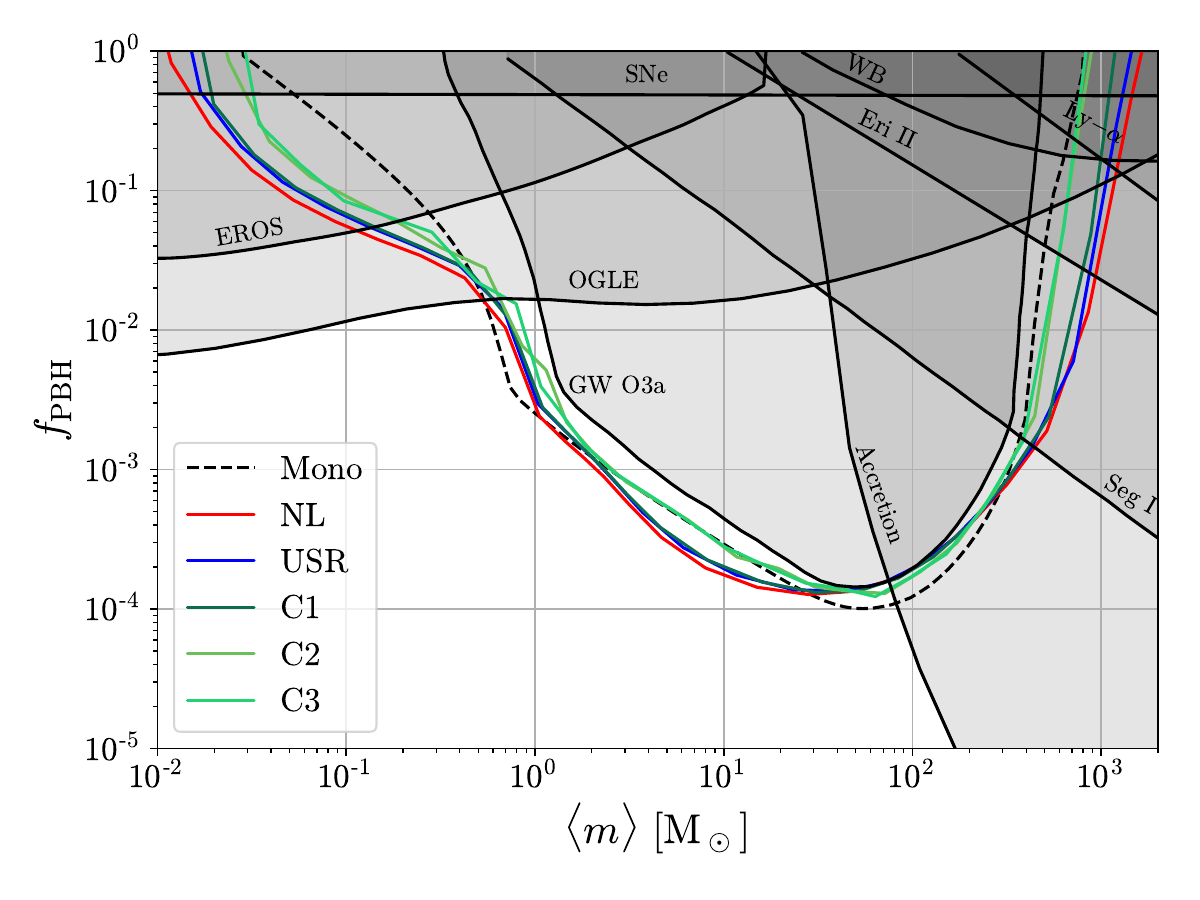}
    \caption{ Same as Fig.~\eqref{fig:monofpbh}, but for log-normal mass function \emph{(left panel)} and critical collapse mass functions \emph{(right panel)}. Parameters for the latter are reported in Table~\ref{tab:par}. The gray constraints are shown for a log-normal power spectrum with $\sigma=0.6$. The dashed line on the left panel shows, the accretion constraint for $\sigma=1.2$.}
    \label{fig:extfpbh}
\end{figure*}

Consider first the pessimistic case in which all events observed by LIGO-Virgo have an astrophysical origin. Although there is presently no evidence to indicate otherwise~\cite{KAGRA:2021duu,KAGRA:2021vkt}, this analysis assesses the current sensitivity of LVK to PBH populations in various PBH scenarios instead of constraining their abundance.

The limits that can be established in this way are shown in Figs.~\ref{fig:monofpbh} and~\ref{fig:extfpbh}. As expected, more observing time and a lower strain noise improve the existing limits obtained from the O3a data in the same region of the parameter space. In particular, for all of the considered mass functions, the new analysis reduces the window next to $\mathcal{O}(1)$ $\msun$ compared to O3a results~\cite{Hutsi:2020sol}. In Fig.~\ref{fig:monofpbh}, considering a monochromatic mass function, we show the contributions from the two- and the three-body channels. The latter begins to dominate when $f_{\rm PBH} > 0.1$ and strengthens the bounds at the edges of the constrained region. 

The other constrains reported here are GW O3a~\cite{Hutsi:2020sol}, EROS~\cite{EROS-2:2006ryy}, OGLE~\cite{Mroz:2024mse}, Seg1~\cite{Koushiappas:2017chw}, Planck~\cite{Serpico:2020ehh}, Eri II~\cite{Brandt:2016aco}, WB~\cite{Monroy-Rodriguez:2014ula}, Ly$-\alpha$~\cite{Murgia:2019duy} and SNe~\cite{Zumalacarregui:2017qqd}.

The limits for extended mass functions are shown in Fig.~\ref{fig:extfpbh}, assuming that all events have an astrophysical origin. The right panel shows the limits for the log-normal mass function with various widths. For small $\sigma$ the limit agrees with the monochromatic case. As illustrated in Fig.~\ref{fig:monofpbh}, the constrained region in the sub-solar mass range arises mostly from the three-body channel. In Fig.~\ref{fig:extfpbh}, we see that this tail is more sensitive to the width of the mass function than the constraint curves above the solar mass, which arise mostly from the two-body channel. 

The left panel of Fig.~\ref{fig:extfpbh}, shows the constraints obtained for the critical collapse mass functions in different models of non-Gaussianities listed in Table.~\ref{tab:par}. We have considered three scenarios: non-Gaussianities arising only due to the non-linear relation between the curvature perturbation and the density contrast, and models with additional primordial non-Gaussianities in the curvature perturbation arising in quasi-inflection-point models~\eqref{eq:zeta_IP} and curvaton models~\eqref{eq:zeta_cur}. Since the mass function depends mildly on the PBH abundance, we consider benchmark cases with $f_{\rm PBH}\simeq 10^{-3}$. 

Crucially, we find that the constraints are nearly identical in all cases considered and match approximately the constraints obtained in the monochromatic case. In particular, when comparing the constraints for log-normal and critical collapse mass functions, we do not find that the enhancement of mass spectra around $1 \msun$ due to the QCD phase transition has a noticeable effect on the constraints. We also find the impact of non-Gaussianities to be quite mild.

We remark that the mass functions considered here are relatively narrow as they span at most about 2 orders of magnitude. For much wider mass functions spanning several orders of magnitude, the bulk of the events would not fall into the LVK mass range and thus the constraints on $f_{\rm PBH}$ would be less stringent. On top of that, the available merger rate estimates have been derived assuming that the PBHs have masses of a similar order, thus the theoretical uncertainties are expected to increase for extremely wide mass functions. On the other hand, one must also consider that all constraints are affected by the shape of the mass function~\cite{Carr:2017jsz}. Compare, for example, the gray constraints in Figs.~\ref{fig:monofpbh} and~\ref{fig:extfpbh} corresponding to monochromatic and the relatively narrow log-normal mass function with $\sigma=0.6$. One can see that the constraint from accretion has moved towards lower mean masses. As shown by the dashed line in Fig.~\ref{fig:extfpbh}, for $\sigma=1.2$, the constraint from accretion will exclude most of the parameter space accessible by LVK. Thus, PBH scenarios with wider mass functions (for instance, with $\sigma > 1.2$) are less relevant for LVK as other observables such as accretion exclude the accessible PBH mass range.

\begin{table}
    \centering
    \begin{tabular}{p{3.cm}p{2cm}p{1.2cm}p{2cm}}
         \hline \hline
         &  $\log_{10} A$ &  $\beta$& $r_{\rm dec}$\\
         \hline
         only NL&  $[-1.9,-1.8]$& 3.0 & - \\
         USR&  $[-2.3,-2.2]$&  3.0& - \\
         curvaton (C1)&  $[-2.8,-2.7]$&  0.5& $0.1$\\
         curvaton (C2)&  $[-2.5,-2.4]$&  3.0& $0.1$\\
         curvaton (C3)&  $[-1.8,-1.7]$&  3.0& $0.9$\\
         \hline \hline
    \end{tabular}
    \caption{ Parameters used in the right panel of Fig.~\ref{fig:extfpbh}. We fix $\alpha=4$ and choose the amplitude $A$ (with one digit precision) in each case so that $f_{\rm PBH}\approx 10^{-3}$.}
    \label{tab:par}
\end{table}

\begin{figure*}
    \centering
    \includegraphics[width=\textwidth]{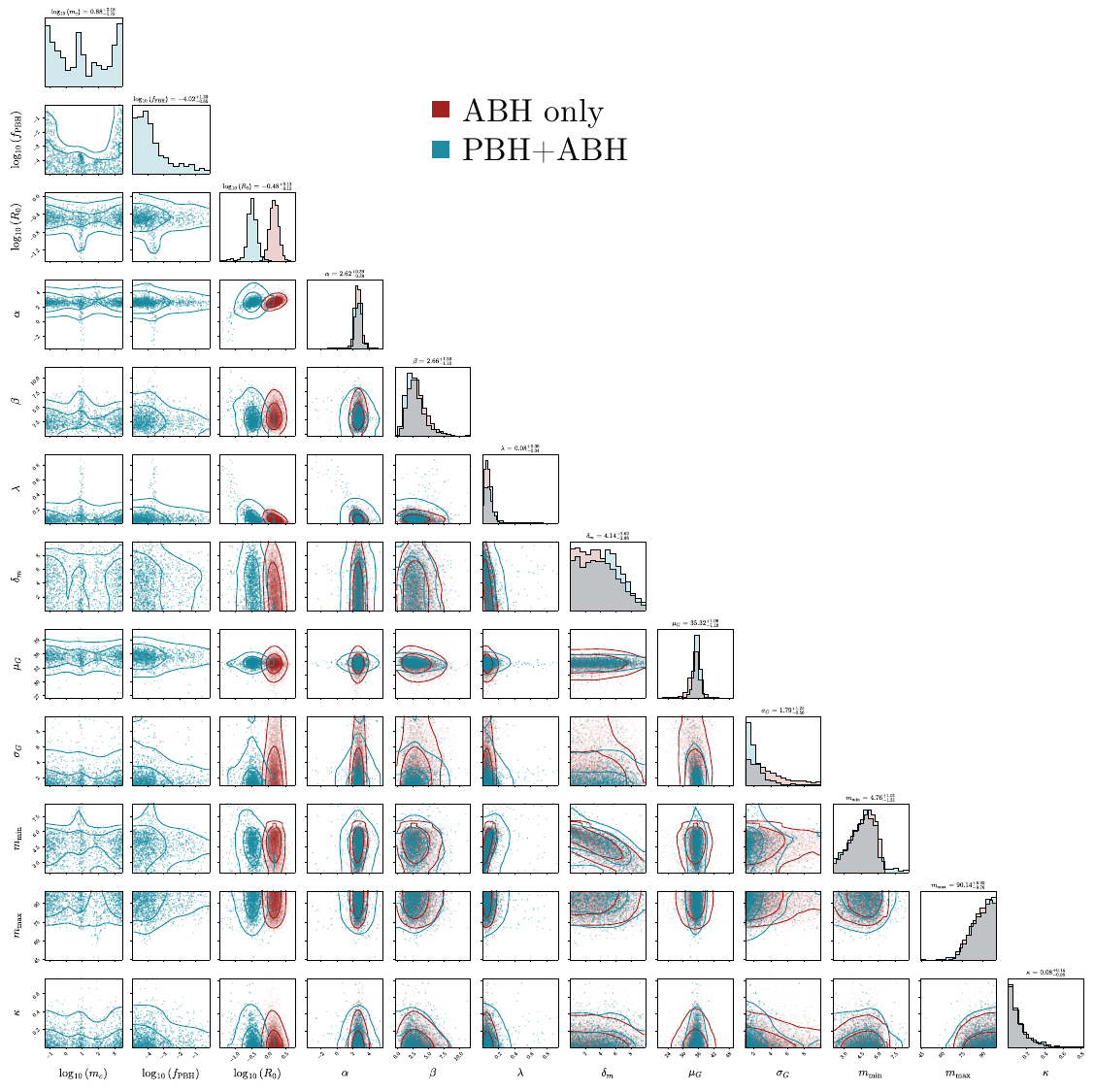}
    \caption{ Posteriors for the BH binary merger rate model combining PBH and ABH binaries (blue) and for the model containing only ABH binaries (red). The PBH binaries are assumed to have a log-normal mass function with width $\sigma=0.6$ and the ABH binaries are described by a phenomenological POWER-LAW + PEAK model. All masses are given in units of $\msun$ and $R_0$ in $\rm Gpc^{-3}yr^{-1}$.}
    \label{fig:PBHplusABH_LN_PP_sig06}
\end{figure*}

Allowing for the possibility, that some of the observed GW events have a primordial origin, does generally soften the constraints. In this case, two approaches are followed: a model-dependent and an agnostic one. 

For the agnostic approach, random subsamples of the events of the GWTC-3 catalogue are taken and assumed to be primordial. Then, the fit using Eq.~\eqref{eq:hierarchicalBayesian} is performed with the various subsamples and the posteriors are combined to obtain the 2$\sigma$ CL upper limits. For this analysis, we consider all the events that have a signal-to-noise ratio larger than $11$ and an inverse false alarm rate less than $4$ years. This ensures the usage of highly confident detections only.

\begin{figure}[h!]
    \centering
    \includegraphics[width=0.8\textwidth]{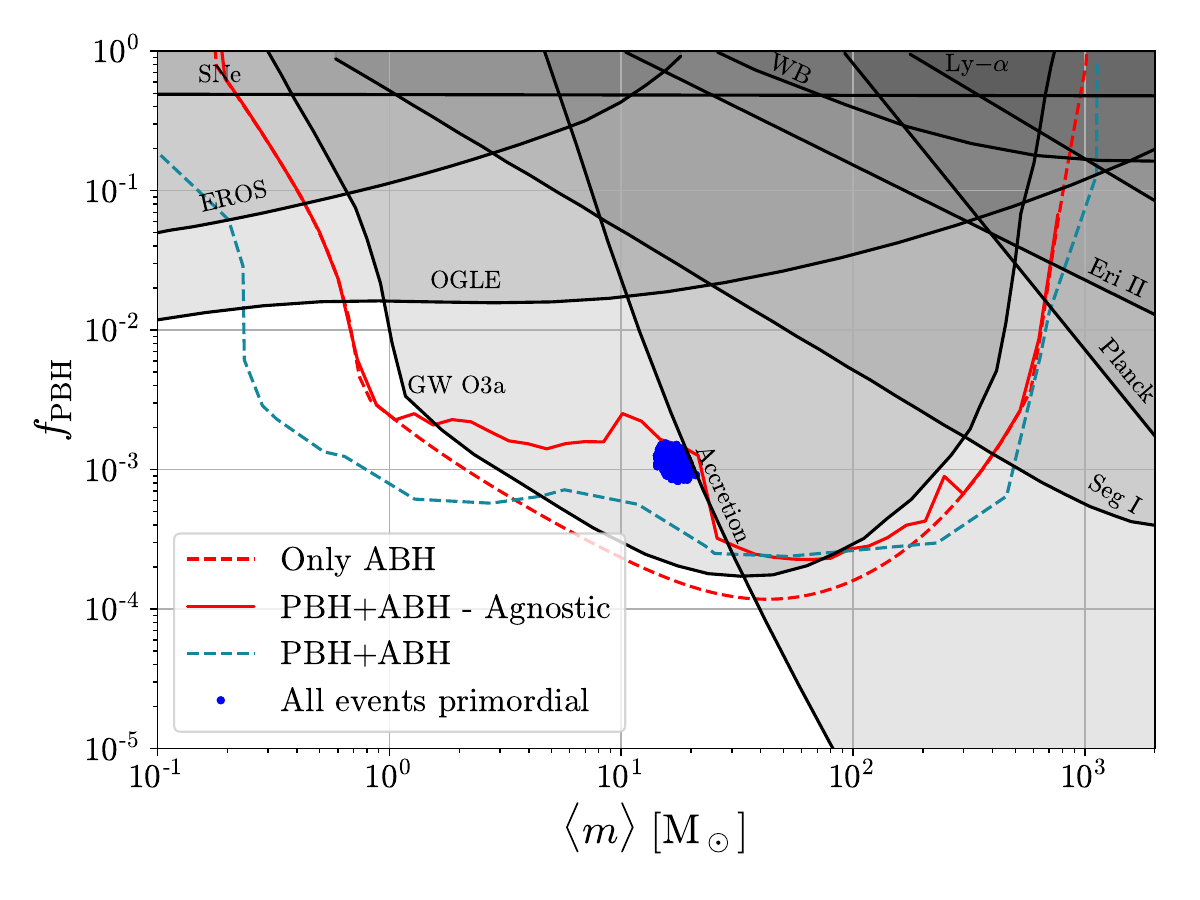}
    \caption{ Constraints for a log-normal mass function with $\sigma = 0.6$ assuming that an arbitrary fraction of events can be primordial (solid red line) and that only astrophysical events have been observed (dashed red line). The constraints on the PBHs obtained from fitting the mixed PBH+ABH model are shown by the blue line (see Fig.~\ref{fig:PBHplusABH_LN_PP_sig06}). The contribution from the three-body channel is omitted here.}
    \label{fig:LN06}
\end{figure}

Testing all possible combinations is computationally unfeasible, and therefore the strategy used in Ref.~\cite{Hutsi:2020sol} is followed. The mass range is divided into bands and for each one, all the events that have the primary mass falling inside it are taken to be primordial. In this way, the least restrictive case is considered and the most conservative constraints can be obtained. As shown in Ref.~\cite{Hutsi:2020sol}, the results converge fast with a relatively small set of subsamples, but following this strategy, the procedure is further optimized. The results are displayed in Fig.~\ref{fig:LN06} for a log-normal mass function with a fixed width of $\sigma=0.6$. As expected, the constraints are weakened when compared to scenarios assuming no primordial events. 

As shown before, the constraints are only mildly dependent on the shape of the mass function and, analogously to Figs.~\ref {fig:extfpbh} and~\ref{fig:monofpbh}, we expect that very similar constraints will hold for critical collapse mass functions. We neglected the contribution of the three-body formation channel~\eqref{eq:R3} in this analysis since, as was established in the previous subsection, it is mildly relevant only in a region of parameter space constrained by other observations.

In the case where an ABH population is modelled explicitly on top of a PBH one, the parameters defining both ABH and PBH binary populations must be fitted simultaneously. Such a fit is shown in Fig.~\ref{fig:PBHplusABH_LN_PP_sig06} using a nested sampling algorithm to scan the parameter space and the priors specified in Table~\ref{tab:priors}. It uses the phenomenological POWER LAW+PEAK model for the ABH population and the log-normal mass function with $\sigma=0.6$ for the primordial one. As a consistency check, we also performed the ABH population fit without PBHs (see Fig.~\ref{fig:PBHplusABH_LN_PP_sig06}) and found that it agrees with existing LVK results~\cite{KAGRA:2021duu}. The parameters of the ABH population in the combined ABH+PBH fit were relatively similar to the ABH only fit. One relevant difference was that the base rate of the ABH merger rate (see Eq.~\eqref{eq:R_ABH}) in the fit that included PBHs, $\log_{10} R_0 = -0.48(13)$, was reduced when compared to the ABH only fit $\log_{10} R_0 = -0.16(13)$, indicating that the combined fit prefers scenarios containing PBH subpopulations.

\begin{table}[h!]
	\centering
	\begin{tabular}{clc} 
		\hline \hline
		\textbf{Parameter} & \textbf{Prior} & \textbf{Units}\\
		\hline \hline
            $R_0$ & LogU$(10^{-2},10^3)$ & Gpc$^{-3}$yr$^{-1}$\\ 
		$\alpha$ & U$(-4,12)$ & - \\ 
            $\beta$ & U$(-4,12)$ & - \\ 
            $\mmin$ & U$(2,10)$ & $\msun$ \\ 
            $\mmax$ & U$(40,100)$ & $\msun$ \\ 
            $\delta_m$ & U$(0,10)$ & $\msun$ \\ 
            $\mu_G$ & U$(20,50)$ & $\msun$ \\ 
            $\sigma_G$ & U$(1,10)$ & $\msun$ \\ 
            $\lambda$ & U$(0,1)$ & - \\
            $\kappa$ & U$(0,5)$ & - \\
            $\fpbh$ & LogU$(10^{-5},1)$ & - \\
            $m_c$ & LogU$(10^{-1},10^3)$ & $\msun$ \\
            \hline\hline
	\end{tabular}
      \caption{Priors used for the PBH+ABH fit.}
      \label{tab:priors}
\end{table}

The $f_{\rm PBH}$ - $m_c$ projection of the posteriors on the top left corner in Fig.~\ref{fig:PBHplusABH_LN_PP_sig06} implies a constraint on the PBH abundance assuming an ABH binary population obeying the POWER LAW+PEAK model. It is consistent with the agnostic ones shown in Fig.~\ref{fig:LN06} except in the subsolar mass range, where it implies a stronger constraint. This is likely a result of insufficient resolution in the scan, because the most stringent constant is given by the red dashed line in Fig.~\ref{fig:PBHplusABH_LN_PP_sig06} corresponding to an expected $N=3$ primordial events. Below the dashed line, the expected number of observable PBH events is simply too small to yield a statistically significant constraint on the PBH abundance. These uncertainties in the fit to the ABH+PBH model illustrate a noticeable numerical drawback when compared to the agnostic approach, which was computationally much faster and more accurate than the scan of the ABH+PBH model -- we were not able to resolve the subsolar mass region with sufficient accuracy with the available computational resources.

\subsubsection{Summary}
We have updated the constraints on PBHs using data from the third observational run (O3) of the LIGO-Virgo experiment by adopting a hierarchical Bayesian approach. Overall, we found that in the mass range $1-300\msun$, the PBH abundance is constrained to $f_{\rm PBH} \leq 10^{-3}$ in all scenarios considered.

We have shown that the constraints are insensitive to the detailed shape of the mass function, and the order of magnitude of these constraints can already be captured by the monochromatic mass function. As a specific and theoretically well-motivated example, we considered mass functions generated by the critical collapse of primordial inhomogeneities. We included the effect of the QCD phase transition, which enhances the mass function around $1 \msun$, as well as the effect of non-Gaussianities. Although both phenomena have a non-negligible impact on the shape of the mass function, we found them to have a minor impact on the constraints.

To account for the foreground of ABH binaries, we followed two approaches. First, in the agnostic approach, which does not require an explicit model for ABH binaries, we allowed for the possibility that any subset of the observed BH-BH merger events could have a primordial origin. Second, we performed a fit in a mixed population of primordial and astrophysical BH binaries, with the latter described by the phenomenological POWER-LAW + PEAK model. Both approaches yielded a similar constraint on $f_{\rm PBH}$, indicating that the constraints on PBHs are relatively insensitive to the modelling of ABH binaries.

The analysis included the contribution from PBH binaries formed from three-body systems in the early Universe. In addition to the early Universe two-body channel, it generally gives the second strongest contribution and can dominate when $f_{\rm PBH} \gtrsim 0.1$. Consistent with this expectation, we found that including this channel mildly widens the constrained region. Overall, including the three-body PBH binary formation channel affects the LVK constraints in parameter regions already excluded by other observations and thus does not have a significant impact on the current PBH constraints.

We considered PBH scenarios with mass functions spanning at most a few orders of magnitude in mass and without significant initial clustering. However, we argued that even with these assumptions relaxed, combining LVK mergers with other observables, the abundance of stellar-mass PBHs remains strongly constrained. In particular, we demonstrated that CMB observations exclude stellar-mass PBHs with very wide mass functions, while Lyman-$\alpha$ observations are at odds with strong initial clustering~\cite{DeLuca:2022uvz}.
\subsection{Updated constraints on large scales}\label{sec:NANOCon}
In this section, following closely ref.\cite{Iovino:2024tyg} we update the constraints on the curvature power spectrum and the PBH abundance using the data from the PTA and CMB experiments.
Since PTAs are sensitive to frequencies of the order of the nHz, they are associated with the production of PBHs in the stellar mass range~\cite{Chen:2019xse,DeLuca:2020agl,Vaskonen:2020lbd,Kohri:2020qqd,Domenech:2020ers,Dandoy:2023jot,Franciolini:2023pbf}. Crucially, the SIGW background produced must not exceed the levels registered by the NANOGrav15. This condition strongly limits the maximum amplitude of the curvature power spectrum and the abundance of PBHs in the corresponding mass range. Additionally, if the power spectrum is enhanced at scales larger than accessible by PTAs, PBHs can serve as seeds of supermassive black holes~\cite{Duechting:2004dk, Bernal:2017nec, Serpico:2020ehh}. Nevertheless, in this case, the scales related to the formation process are strongly constrained by CMB spectral distortion analysis of the FIRAS collaboration~\cite{Fixsen:1996nj,Chluba:2012gq,Chluba:2012we,Chluba:2013dna,Bianchini:2022dqh} and, without NGs, the allowed PBH abundance is too small to provide a primordial origin for SMBHs seeds.
\subsubsection{PTA observations}
Consider the constraints on the SIGW signal from the current PTA observations. Using the NANOGrav 15 year dataset~\cite{NANOGrav:2023gor}, 
we compute the observational upper bound at 95\%  confidence level on $h^2\Omega^{\rm bound}_{\rm GW}$ in each of the first 14 frequency bins. The frequency of bin "$i$" is given by $f_i = i/T$, where $T =  16.03\,{\rm yr} = (1.98 \,{\rm nHz})^{-1}$ is the time of observation. We then compare the observation with theoretical prediction $h^2\Omega_{\rm GW}$ in each frequency bin. The constraint on $h^2\Omega_{\rm GW}$ in any given bin implies an upper bound on $A$ for a given model of NGs and a fixed shape of the curvature power spectrum. To combine the constraints of individual bins, we consider the strongest constraint on $A$. The upper bound on $A$ is then mapped to an upper bound on $f_{\rm PBH}$. More specifically, for every fixed benchmark model in Table \ref{tab:1}, we can map the pair $(A,k_*)$ uniquely into the pair $(f_{\rm PBH},\langle M_{\rm PBH}\rangle)$.
Due to the finite observational time, the spectral resolution of the detector is limited and can be estimated to be approximately $\Delta f = 1/T = 1.98 \,{\rm nHz}$. This will limit the PTA sensitivity to sharp features in the SIGW spectrum. To account for this, the contribution to the $i$-th bin can be estimated as
\be
    \Omega_{{\rm GW},i} = \int \frac{\td f}{f} \Omega_{\rm GW}^{\rm th}(f) W(f-f_i) \,
\ee
where $W$ denotes a window function. To understand the effect of limited spectral resolution, we will consider three different idealized cases
\begin{align}
    W_{\rm D}(f) &\propto \delta \left(f\right)\,,  \\
    W_{\rm S} (f) &\propto {\rm sinc}^2\left( \pi \frac{f}{\Delta f}\right) \,, \\
    W_{\rm TH} (f) &\propto \theta\left(f+\frac{\Delta f}{2}\right) \theta\left(\frac{\Delta f}{2} - f \right) \,,
\end{align}
which correspond to an infinitely accurate frequency resolution, a sharp binning of frequencies and a time-domain top-hat filter on $h_{\rm c}(t)$ with duration $T$, respectively. The window functions are normalized so that $\int \td \ln f \, W(f-f_i) = 1$. These window functions represent different limiting cases and can thus capture systematics related to the spectral resolution not otherwise accessible in our simplified analysis. As seen from Fig.~\ref{fig:GWfPBHMF}, the effect is expectedly larger at large scales and for heavy PBHs as these are mostly related to the signal at small frequencies. For instance, at $k\simeq10^{6}$ $ {\rm{Mpc}}^{-1}$, $W_{\rm D}(f)$ gives a mildly stronger constraint on $A$ when compared to the other choices, as it can resolve sharp peaks in the SIGW spectrum. This is reflected in the constraints on the PBH abundance, as shown on the right panel of Fig.~\ref{fig:GWfPBHMF}. This behavior is reverted at even smaller $k_*$, and $W_{\rm D}(f)$ is less constraining than the other choices. For the remainder of the analysis, we will consider the Dirac window function $W_{\rm D}(f)$.
\begin{table}[h!]
    \centering
    \begin{tabular}{p{1.2cm}p{5cm}p{1.5cm}}
         \hline \hline
         Case &  PS shape &   $f_{\rm NL}$  \\
         \hline\hline
         LN & LN $(\Delta=0.5)$ & $[-2,10]$ \\\\
         
         BPL1& BPL $(\alpha=4,\beta=3,\lambda=1)$&  $[-2,10]$ \\\\
         
         BPL2& BPL $(\alpha=4,\beta=0.5,\lambda=1)$&  $[-2,10]$ \\
         \hline \hline
    \end{tabular}
    \caption{ List of benchmark cases considered in this section.}
    \label{tab:1}
\end{table}
\begin{figure*}
    \centering
    \includegraphics[width=0.99\textwidth]{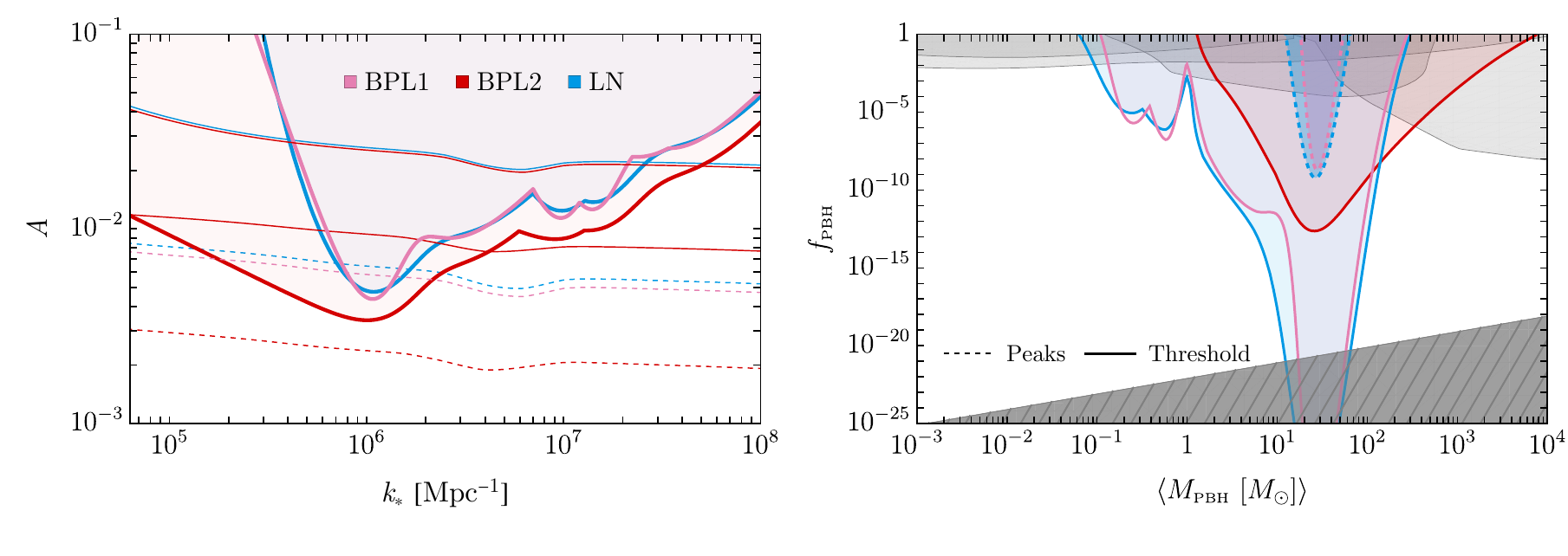}
    \caption{ \textit{Left panel:}
Constraints from NANOGrav15~\cite{NANOGrav:2023gor} on the amplitude of a log-normal curvature power spectrum, Eq.~\eqref{eq:LogNo}, with $\Delta=0.5$ (blue), and on a broken power-law curvature power spectrum, Eq.~\eqref{eq:PPL}, with $\alpha = 4$, $\gamma = 1$ for $\beta = 3$ (pink) and $\beta = 0.5$ (red),  assuming Gaussian fluctuations. The horizontal lines correspond to $f_{\rm PBH}=1$ using threshold statistics (solid) and theory of peaks (dashed), respectively, given the non-linearity-only scenario.
\textit{Right panel:} The implied NANOGrav15 constraints on PBH abundance.}
    \label{fig:GWfPBH}
\end{figure*}
\begin{figure*}
    \centering
    \includegraphics[width=0.99\textwidth]{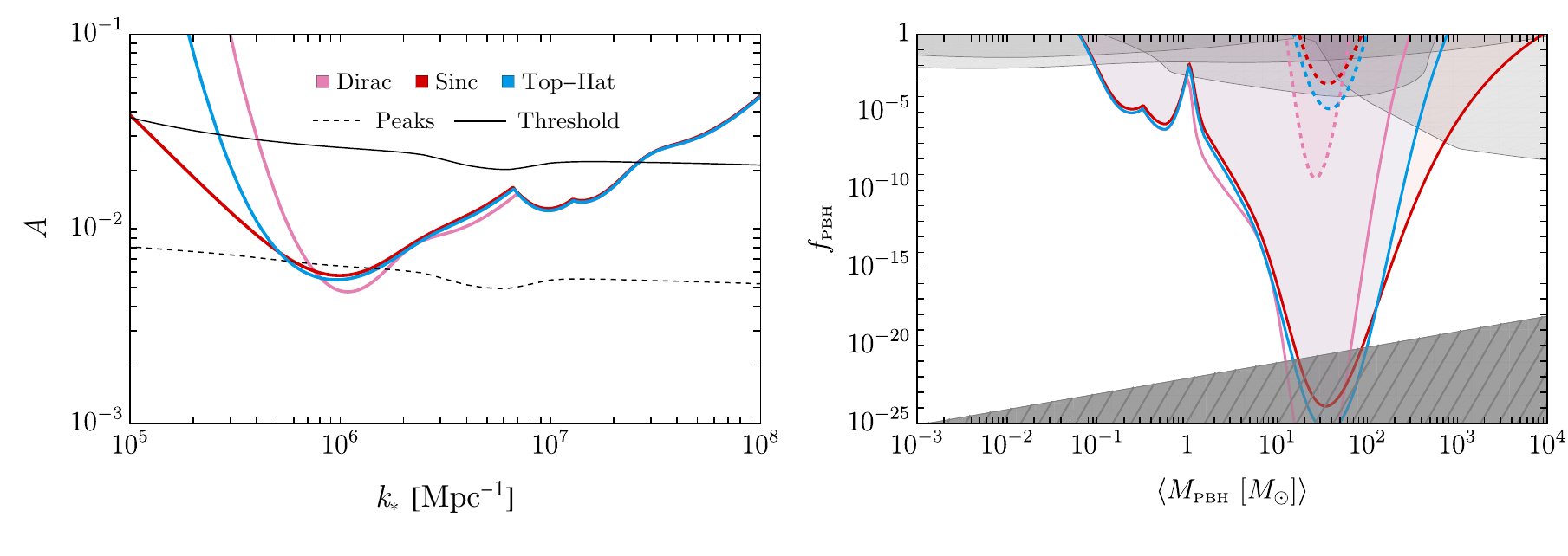}
    \caption{\textit{Left panel:}
Constraints from NANOGrav15~\cite{NANOGrav:2023gor} on the amplitude of a log-normal curvature power spectrum \eqref{eq:LogNo} with $\Delta=0.5$ assuming Gaussian fluctuations and changing the window function. The horizontal lines correspond to $f_{\rm PBH}=1$ using threshold statistics (solid) and theory of peaks (dashed), respectively, assuming NGs arising solely from the non-linear corrections.
\textit{Right panel:} Inferred constraints on PBH abundance.}
\label{fig:GWfPBHMF}
\end{figure*}
\begin{figure*}
    \centering
    \includegraphics[width=0.99\textwidth]{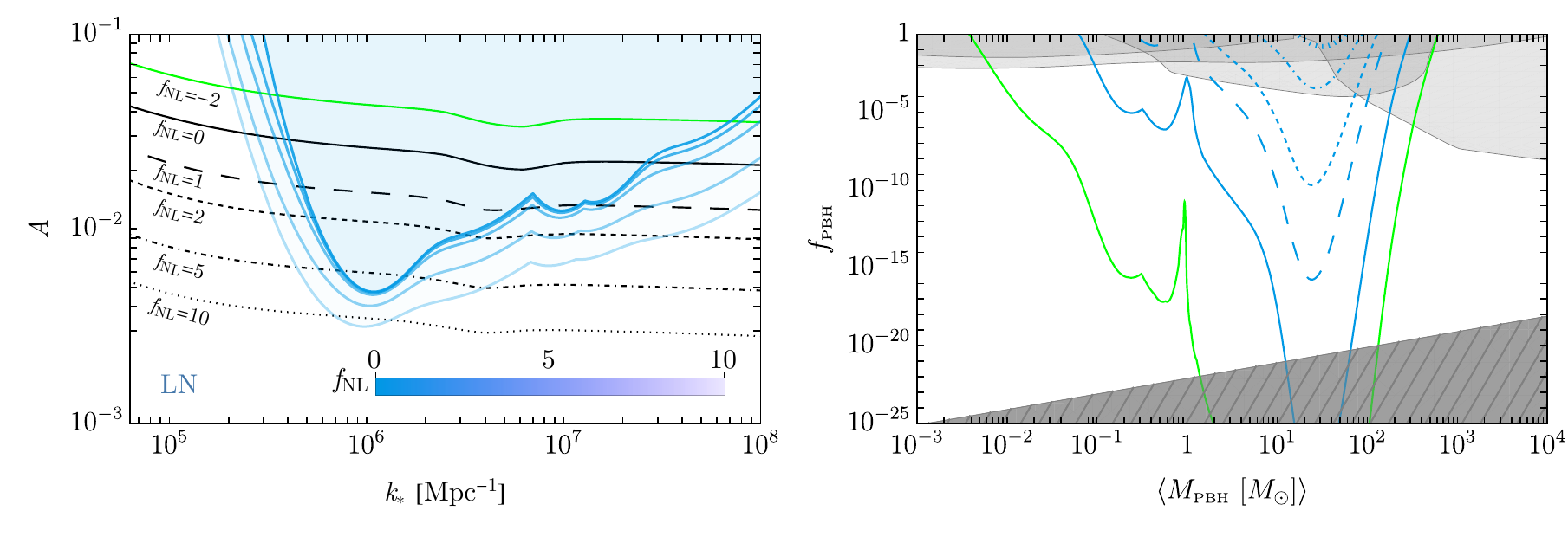}
    \vspace{5mm}
    \includegraphics[width=0.99\textwidth]{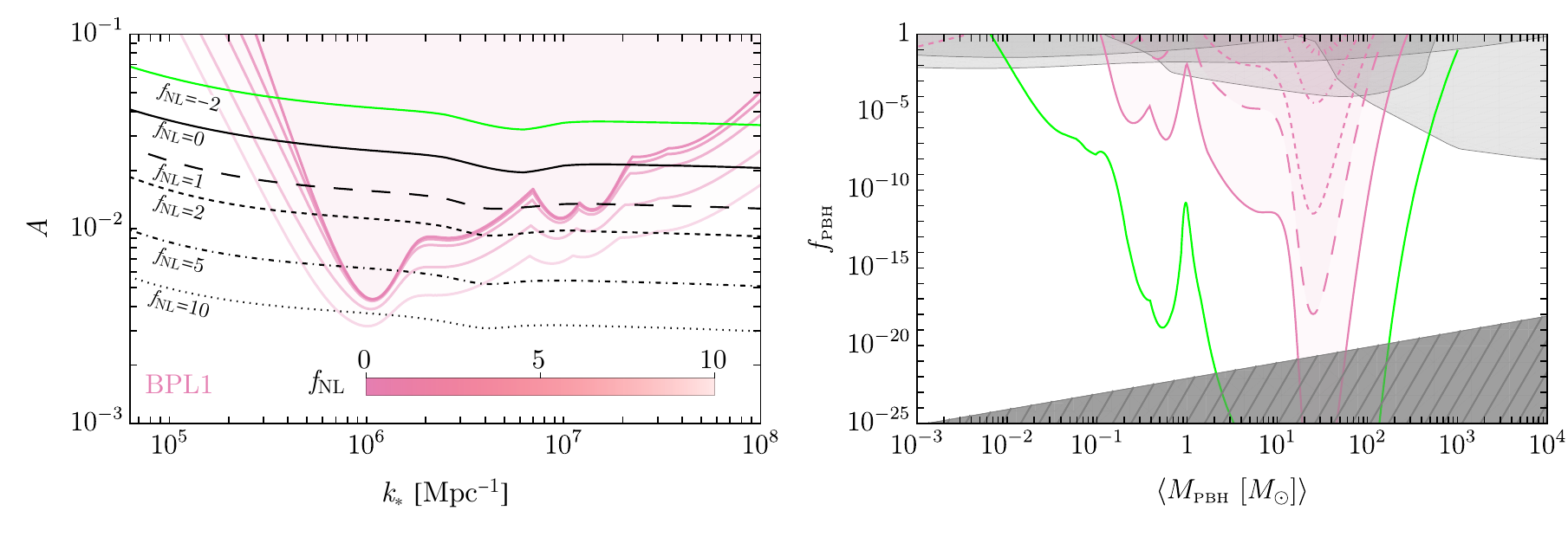}
    \vspace{1mm}
    \includegraphics[width=0.99\textwidth]{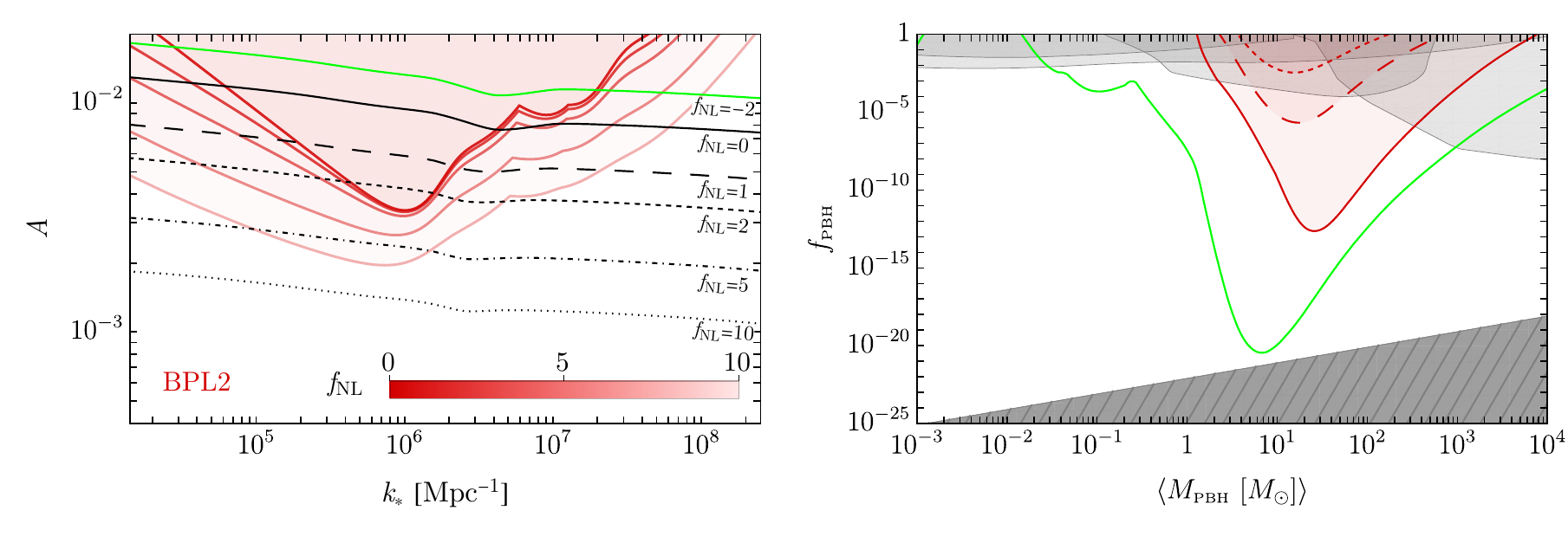}
    \caption{ 
\textit{Left panels:} Constraints from NANOGrav15~\cite{NANOGrav:2023gor} on the amplitude of a log-normal curvature power spectrum \eqref{eq:LogNo} with $\Delta=0.5$ (top panels) and on a broken power-law curvature power spectrum \eqref{eq:PPL} with $\alpha = 4$, $\gamma = 1$ for $\beta = 3$ (middle panels) and $\beta = 0.5$ (bottom panels), assuming quadratic primordial NGs only with $|f_{\rm NL}|\in [0,10]$. The nearly horizontal lines correspond to $f_{\rm PBH}=1$ using threshold statistics with several $f_{NL}$ values.
\textit{Right panels:} Inferred constraints on the PBH abundance using threshold statistics.}
\label{fig:GWfPBHNGs}
\end{figure*}

Another important theoretical uncertainty when estimating constraints on the PBH abundance arises due to the choice of the prescription when computing the PBH abundance. The right panels of Figs.~\ref{fig:GWfPBHMF} and~\ref{fig:GWfPBH} show the constraints on the PBH abundance arising when the abundance is computed using threshold statistics (solid lines) and peaks theory (dashed lines). In both figures, primordial NGs are neglected ($f_{\rm NL} = 0$) so the NGs arise only from the non-linear relation between density and curvature perturbations. As one can see, almost the entire parameter space accessible by LVK via GWs from PBH binary mergers~\cite{Hutsi:2020sol, Romero-Rodriguez:2021aws, Franciolini:2022tfm, Andres-Carcasona:2024wqk} is excluded when threshold statistics is assumed. On the other hand, with the peaks theory, only a small portion of the parameter space relevant to LVK is constrained. Despite this significant theoretical uncertainty, PTA observations can constrain the PBH abundance in both cases.

The dependence of the constraints on the shape of the power spectrum is illustrated in Fig.~\ref{fig:GWfPBH}. While relatively narrow spectra, such as the LN and the BPL1 cases, produce similar results, broader spectra, represented by the BPL2 case, are constrained also at larger scales due to the broadness of the spectrum. We remark that the constraints at large scales $k_* \lesssim 5 \times 10^6 {\rm Mpc}^{-1}$ arise mostly from the first bin by NANOGrav15. This is also the region in which the SIGW are most strongly constrained. In a similar vein, the constraints on the PBH abundance in the threshold statistics case are quite similar for all spectra except at large masses $\langle M_{\rm PBH} \rangle$. In particular, the strongest constraints touch the dashed region in Fig.~\ref{fig:GWfPBH} which corresponds to having less than a single PBH within the current Hubble volume. This line is given by $f_{\rm PBH} 4\pi \Omega_{\rm DM} M_{\rm pl}^2/H_0 < \langle M_{\rm PBH}\rangle$.

Fig.~\ref{fig:GWfPBHNGs} shows the effect of primordial NGs. Firstly, we underline that the NG corrections to SIGW depend on $f_{\rm NL}^2$ and are thus indifferent to the sign of $f_{\rm NL}$\,\cite{Perna:2024ehx}. On the other hand, this latter strongly impacts  $f_{\rm PBH}$. Indeed, negative primordial NGs suppress the tail of the PDF with respect to the Gaussian case, that is, they reduce the probability of large fluctuations that cross the threshold. So, a larger amplitude would be needed to obtain the same PBH abundance. Positive primordial NGs, on the other hand, have the opposite effect as they make large fluctuations more likely. 

As shown on the left panels of Fig.~\ref{fig:GWfPBHNGs}, for small $f_{\rm NL}$ we observe that the corrections to the SIGW affect only amplitudes at the tails, both at smaller and higher scales. Larger NGs ($f_{\rm NL} \geq 5$), on the other hand, shift the entire SIGW constraint curves. The effect on the $f_{\rm PBH}$ constraints, shown in the right panels of Fig.~\ref{fig:GWfPBHNGs}, can be qualitatively understood by comparing the shift of $f_{\rm PBH} = 1$ curves when compared to the SIGW constraint: Since the effect on $f_{\rm PBH} = 1$ is larger than the strengthening of the SIGW constraints, positive primordial NGs can completely remove constraints on the PBH abundance. We find that for $f_{\rm NL} = 5$, the PTA constraints on $f_{\rm PBH}$ are eliminated in the wide BPL2 case and become weaker than the constraints from other observables in the narrow LN and BPL1 cases.

The opposite is observed with negative NGs -- since they strengthen the constraints on SIGW in a similar way to positive NGs, but require a higher amplitude $A$ to achieve the same PBH abundance, the constraints on $f_{\rm PBH}$ become more stringent. This is indicated by the green line in Fig.~\ref{fig:GWfPBHNGs}. In particular, even a mild negative NG ($f_{\rm NL} = -2$) could exclude the existence of any PBHs in our present Hubble volume in the mass range $2-100 M_{\odot}$, given a narrow spectrum as in the LN and BPL1 cases.

However, one might mistakenly assume that by increasing the negative value of the coefficient $f_{\rm NL}$, the required power spectral amplitude for the same PBH abundance would always rise, tightening the constraints indefinitely. However, this is not the case, as the required amplitude reaches a maximum at $f_{\rm NL} \simeq -2$ and then decreases, eventually approaching the amplitude of the Gaussian case for very large negative values of $f_{\rm NL}$~\cite{Franciolini:2023pbf}. Consequently, the constraints for the case of $f_{\rm NL} = -2$ can be considered the most stringent.

Nevertheless, this claim should be taken \textit{cum grano salis}. Indeed, it is important to note that the prescription outlined in Ref.~\cite{Musco:2020jjb} to compute the threshold for PBH collapse only accounts for NGs arising from the non-linear relation between the density contrast and the curvature perturbations. In principle, also primordial NGs beyond the quadratic approximation should be taken into account when computing the threshold value. Following Refs.~\cite{Kehagias:2019eil, Escriva:2022pnz}, it appears that their effect on the threshold is small when the NGs are positive or mildly negative ($f_{\rm NL} \gtrsim -2$). In such cases, the threshold receives corrections of at most a few percent. For more negative NGs, the uncertainties on the exact value of the threshold can be sizeable, and could thus potentially modify the constraints introduced above\footnote{One alternative approach is to use the averaged value over a sphere of radius equal to the location of the maximum of the compaction function, $\Bar{C}_{\rm th} = 2/5$, since this value is fully non-perturbative and independent of NGs~\cite{Escriva:2019phb, Kehagias:2024kgk}. As discussed in Ref.~\cite{Ianniccari:2024bkh}, this requires determining the statistics of the curvature perturbation and computing the connected cumulants, which is a complex task and is left for future work.}.

For completeness, we analyze a common scenario for PBH formation. 

\begin{figure}[t]
    \centering
    \includegraphics[width=0.85\textwidth]{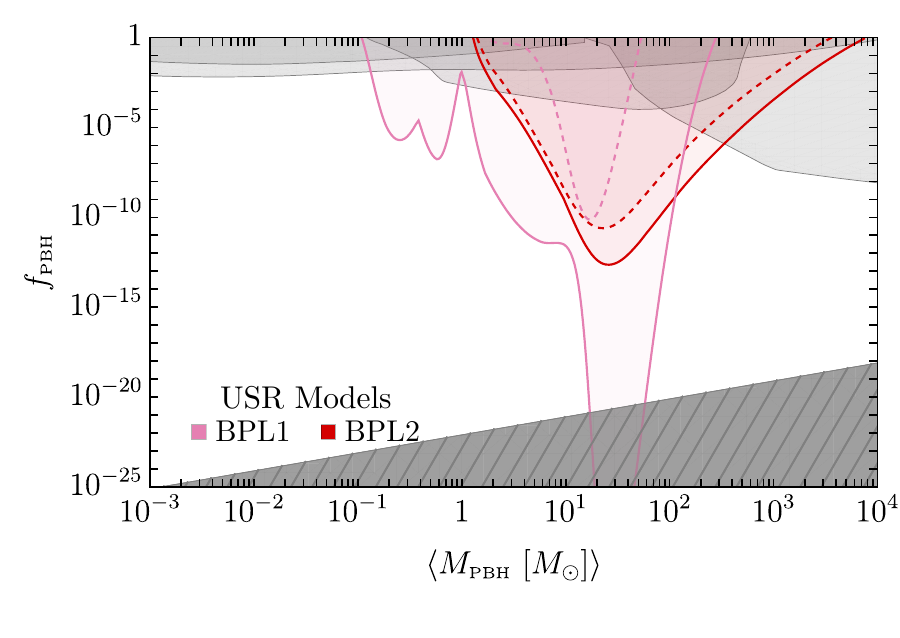}
    \caption{ 
The NANOGrav15 constraints on the PBH abundance using threshold statistics assuming the two cases for the broken power law reported in Tab.~\ref{tab:1} assuming the Gaussian approximation (solid) and with the full NG relation in Eq.~(\ref{eq:zeta_IP}).
}\label{fig:USR}
\end{figure}

In \emph{Ultra slow roll} (USR) models of single-field inflation, the peak in $\mathcal{P}_{\zeta}$ arises from a brief phase of ultra-slow-roll which is typically followed by slow-roll or constant-roll inflation dual to it~\cite{Atal:2018neu, Biagetti:2018pjj, Karam:2022nym}. In this case, the NGs can be related to the large $k$ spectral slope generated during the last inflationary phase~\cite{Atal:2019cdz, Tomberg:2023kli},
\be
    \zeta = -\frac{2}{\beta}\log\left(1-\frac{\beta}{2}\zeta_{\rm G}\right).
\ee
In Fig.~\ref{fig:USR}, we show the abundance of PBHs using the full NG relation Eq.~\eqref{eq:zeta_IP} for the two broken power law cases reported in Tab.~\ref{tab:1}.  The PBH abundance is computed using threshold statistics and compared with the Gaussian approximation. As seen from Fig.~\ref{fig:USR}, NGs can significantly affect the constraints on $f_{\rm PBH}$ when the spectrum is narrow (in the BPL1 case). For wider spectra (BPL2 case), the NG-induced suppression is milder and thus the final constraint is stronger. To make contact with the $f_{\rm NL}$ description of NGs, expanding Eq.~\eqref{eq:zeta_IP} yields $f_{\rm NL} = 5 \beta/12 $, so that $\beta = 3$ and $\beta=0.5$ correspond to $f_{\rm NL} = 5/4$ and $f_{\rm NL}= 5/24$, thus the NGs are relatively small and the NG corrections to SIGW have a subdominant effect. Consequently, for USR models, PTAs are more effective in constraining PBHs from wide curvature power spectra due to the weaker NG-induced suppression.

The other constraints reported here are GW O3~\cite{Andres-Carcasona:2024wqk}, EROS~\cite{EROS-2:2006ryy}, OGLE~\cite{Mroz:2024mse,Mroz:2024wag}, Seg1~\cite{Koushiappas:2017chw}, Planck~\cite{Serpico:2020ehh,Agius:2024ecw, Facchinetti:2022kbg}, Eri II~\cite{Brandt:2016aco}, WB~\cite{Monroy-Rodriguez:2014ula}, Ly$-\alpha$~\cite{Murgia:2019duy} and SNe~\cite{Zumalacarregui:2017qqd}. These constraints are shown for monochromatic mass functions.

\begin{figure*}
\centering
\includegraphics[width=0.85\textwidth]{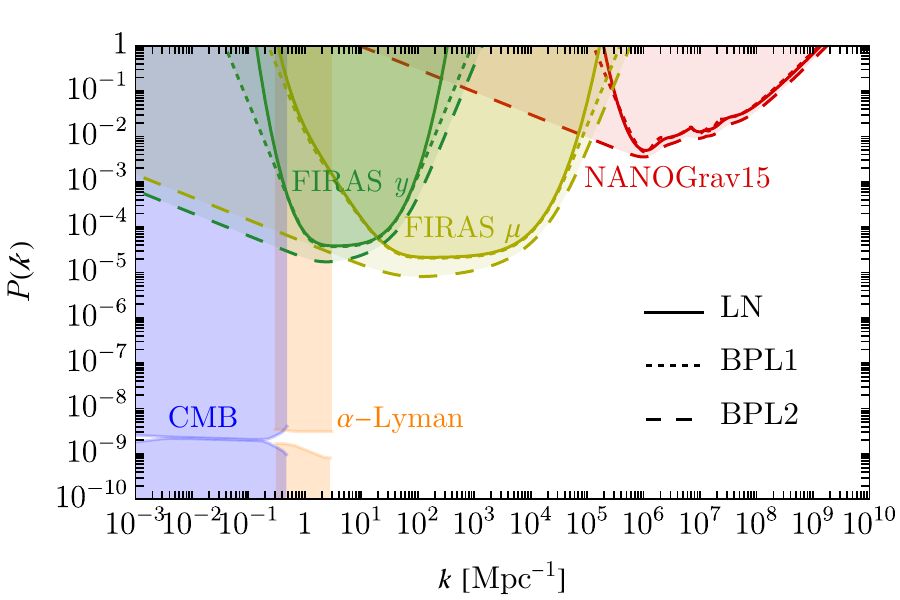}
\caption{ Constraints on the amplitude of the power spectrum, assuming negligible primordial NGs, from the FIRAS experiments, CMB, $\alpha$-Lyman and NANOGrav15 for the same cases reported in Tab.~\ref{tab:1}.}
\label{fig:muconstra}
\end{figure*}

\subsubsection{$\mu$ distortions and PBH seeds for SMBHs}

At the largest scales, the primordial power spectrum is strongly constrained by the CMB observations~\cite{Planck:2018vyg} and the Lyman-$\alpha$ forest data~\cite{Bird:2010mp}\footnote{For recent works on how future CMB experiments can probe the range of scales related to PTA, see Refs.~\cite{Cyr:2023pgw,Tagliazucchi:2023dai}}.
The CMB observations strongly constrain the curvature power spectrum at scales $10^{-4} \mathrm{Mpc}^{-1} \lesssim k \lesssim 1 \mathrm{Mpc}^{-1}$. 
At redshifts $z \lesssim 10^6$, energy injections into the primordial plasma cause persisting spectral distortions in the CMB. These distortions are divided into chemical potential $\mu$-type distortions created at early times and Compton $y$-type distortions created at $z \lesssim 5 \times 10^4$.
For a given curvature power spectrum $\mathcal{P}_{\zeta}(k)$ the spectral distortions are~\cite{Chluba:2012we,Chluba:2013dna}
\be
    X=\int_{k_{\min }}^{\infty} \frac{\mathrm{d} k}{k} \mathcal{P}_{\zeta}(k) W_X(k)
\ee
with $X=\mu, y$ where $k_{\min }=1$ $\mathrm{Mpc}^{-1}$ and the window functions can be approximated by
\be
    W_\mu(k)=2.2\left[e^{-\frac{(\hat{k} / 1360)^2}{1+(\hat{k} / 260)^{0.6}+\hat{k} / 340}}-e^{-(\hat{k} / 32)^2}\right], 
\ee
\be
W_y(k)=0.4 e^{-(\hat{k} / 32)^2}
\ee
with $\hat{k}=k /\left(1 \mathrm{Mpc}^{-1}\right)$.
The COBE/Firas observations constrain the $\mu$  distortions as $\mu \leq 4.7 \times 10^{-5}$~\cite{Bianchini:2022dqh} and $y\leq 1.5 \times 10^{-5}$ at $95\%$ confidence level~\cite{Fixsen:1996nj}.
The situation is summarised in Fig.~\ref{fig:muconstra}.
Including NGs modifies only the tails of the constraints and not the most constrained region~\cite{Sharma:2024img}.

Consequently, the amplitude of the power spectrum in the range of scales related to the FIRAS experiment is constrained to be at most $A=10^{-5}$.

We focus on the threshold statistics approach and we compute the mass fraction $\beta$ as a function of the parameters $f_{\rm NL}$ and $g_{\rm NL}$, assuming the power series ansatz. The power series expansion holds until we are in the perturbative regime, that is, the terms are ordered hierarchically with the higher orders being typically smaller than the lower ones. For a narrow power spectrum, this translates into a constraint for the amplitude of the power spectrum, $(3/5)\left|f_{\mathrm{NL}}\right| A^{1 / 2} \ll 1$, and $(9/25)\left|g_{\mathrm{NL}}\right| A \ll 1$.

Such large positive NGs can lead to the production of a non-negligible population of massive PBHs that can seed SMBHs.

\begin{figure*}
    \centering
\includegraphics[width=0.99\textwidth]{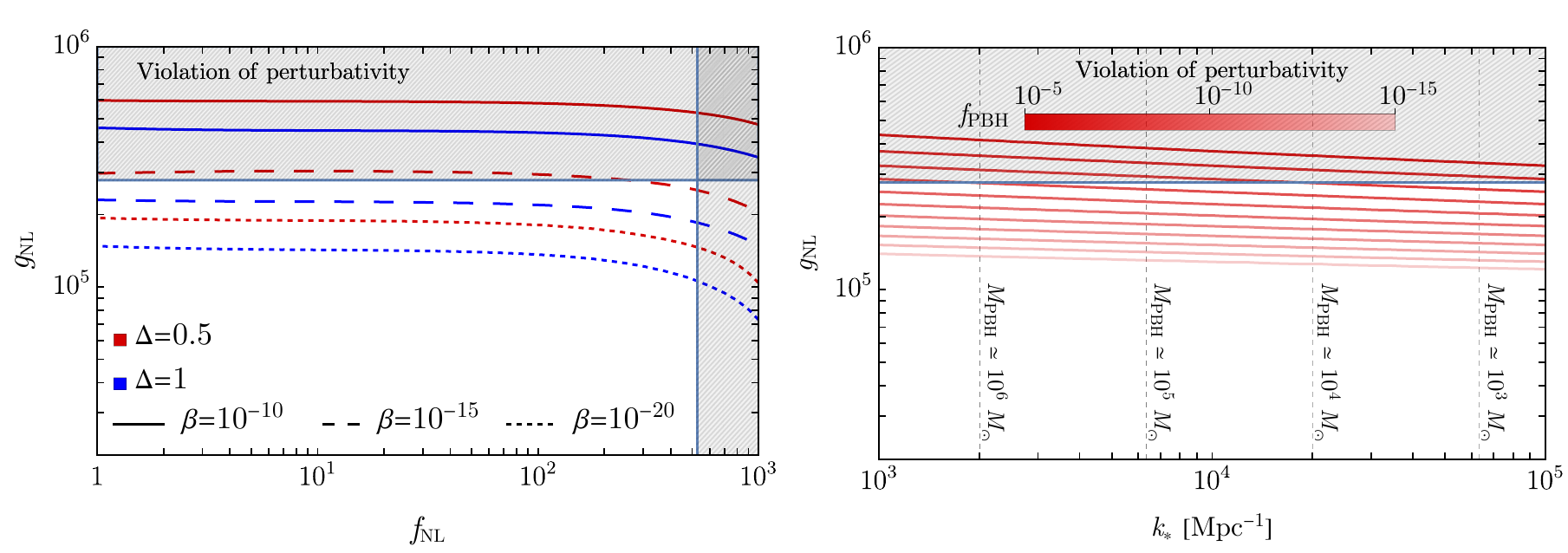}
\caption{ \textit{Left panel:} Computation of the mass fraction $\beta$ changing the NG parameters $f_{\rm NL}$ and $g_{\rm NL}$ with two benchmark cases for the log-normal power spectrum with an amplitude fixed to be $A=10^{-5}$. \textit{Right panel:} Computation of the PBH abundance $f_{\rm PBH}$ changing the scale of the peak $k_*$ and the NG parameter $g_{\rm NL}$. We fix $\Delta=1$, $A=10^{-5}$ and $f_{\rm NL} = 1$.}
\label{fig:fpbhgnl}
\end{figure*}
Indeed, the origin of SMBHs presents a significant challenge in the field of astrophysics. Although it is well known that they occupy the centres of most galaxies~\cite{Kormendy:1995er, Magorrian:1997hw, Richstone:1998ky}, the processes leading to their formation remain unclear.

In the context of PBHs, even a minimal presence of massive PBHs within the mass range $10^3-10^6 \, M_{\odot}$ could potentially seed for SMBHs~\cite{Duechting:2004dk,Kohri:2014lza,Bernal:2017nec}. As a naive computation to estimate the necessary abundance of these seeds, we follow Refs.~\cite{Vaskonen:2020lbd, Serpico:2020ehh}. Considering that SMBHs make up approximately $\mathcal{O}(10^{-4})$ of the stellar mass in their host galaxies, and that stars account for roughly $\mathcal{O}(10^{-2})$ of the total cosmic matter content~\cite{Fukugita:1997bi}, we deduce that the overall SMBH density is about $10^6$ times less than the dark matter density.

The primordial seeds' abundance can then be expressed as
\be\label{eq:SMBH}
    f_{\mathrm{PBH}} \sim \mathcal{O}(10^{-6}) \times \left\langle M_{\mathrm{PBH}}\right\rangle / M_{\rm SMBH}\, ,
\ee
with $\left\langle M_{\mathrm{PBH}}\right\rangle > 10^3 M_{\odot}$.
Nevertheless, the scales related to the formation of SMBHs are strongly constrained by the analysis of the CMB spectral distortion by the FIRAS collaboration~\cite{Fixsen:1996nj,Chluba:2012gq,Chluba:2012we,Chluba:2013dna,Bianchini:2022dqh}, and in the gaussian approximation, the corresponding abundance of PBHs is too small to furnish a primordial origin for Supermassive black holes seeds. 

As shown in the left panel of Fig.~\ref{fig:fpbhgnl}, in agreement with the literature~\cite{Unal:2020mts,Nakama:2017xvq,Hooper:2023nnl,Byrnes:2024vjt}, quadratic primordial NGs are not enough to produce a sizeable amount of PBHs to be the seeds of SMBHs. Consequently, differently from Ref.~\cite{Byrnes:2024vjt}, we find that already a sizeable cubic parameter $g_{\rm NL}$ is enough to overcome this hurdle. This discrepancy arises because their analysis is subject to a few simplifications: using the curvature perturbation instead of the density contrast to determine the collapse and neglecting the threshold's dependence on the shape of the curvature power spectrum.

According to Fig.~\ref{fig:fpbhgnl}, assuming a typical SMBH mass of approximately $10^{8}$ $M_{\odot}$, condition~\eqref{eq:SMBH}, without the violation of the perturbative criterium, holds across the range of PBH masses analyzed.
\subsubsection{Summary}
We have updated the constraints on the amplitude of the curvature power spectrum, ensuring that the amplitude of the produced SIGWs does not exceed the signal recently detected by the NANOGrav collaboration.

When PBHs are formed via the collapse of sizeable curvature perturbations, the constraints on SIGW will infer an upper bound on the PBH abundance. We derived these bounds in various scenarios and by estimating PBH abundance using threshold statistics as well as the theory of peaks. 

We found that, when using threshold statistics, only a small fraction of dark matter in the form of PBHs is available in the solar mass range regardless of the specific shape of the power spectrum. For narrow curvature power spectra, significant constraints $f_{\rm PBH}\lesssim 10^{-5}$ arise in the mass range $0.1-100 M_{\rm \odot}$, while for broader spectra, such constraints arise for heavier $5-500 M_{\rm \odot}$ PBHs.

These potential constraints can, however, be significantly relaxed in the presence of positive NGs, and are eliminated when $f_{\rm NL} \gtrsim 10$. However, we find that stellar mass PBHs tend to remain constrained when considering NGs present in typical USR models for PBH production. In these cases, wider spectra tend to infer strong constraints over a wider PBH mass range as they are associated with weaker NGs.

When the PBH abundance is estimated using the theory of peaks, however, constraints on $f_{\rm PBH}$ are strongly relaxed. Although they are not completely eliminated (in the absence of non-Gaussianities) they tend to be in the same order as existing constraints on the PBH abundance.

Finally, we discussed how large primordial NGs ($g_{\rm NL} \simeq 10^{5}$) permit the production of PBH seeds for SMBHs without violating perturbativity. Despite this possibility, the literature currently lacks explicit models capable of producing sufficiently large NGs. Indeed, as shown in this chapter and in refs.~\cite{Ferrante:2022mui, Ferrante:2023bgz}, in common scenarios, such as the curvaton and USR models, the general impact of the NGs is curtailed when one goes beyond the perturbative approach. In other words, in these models, using the full non-gaussian expression gives in general a smaller abundance compared to the power series expansion for not very narrow power spectra.

To improve future estimates, it is crucial to refine the computation of the PBH abundance, which is still subject to a series of theoretical uncertainties.
\newcommand{\cM}{\mathcal{M}}
\newcommand{\Msun}{M_\odot}
\newcommand{\vect}[1]{\boldsymbol{#1}}
\chapter{Gravitational wave signatures of primordial black holes}\label{cap:GWs}
\thispagestyle{plain}
PBHs formation occurs as large curvature perturbations re-enter the Hubble horizon after inflation and eventually collapse under the effect of gravity. The same enhanced scalar perturbations emit tensor modes thanks to second-order effects around the epoch of horizon crossing. This generates an observable stochastic gravitational wave background SGWB (see ref.\,\cite{Domenech:2021ztg} for a recent review). 
In this chapter we describe how to compute the SGWB produced in presence of large curvature perturbation (sec.\ref{sec:GW1}), then, in sec.\ref{sec:GW2}, based on\,\cite{Franciolini:2023pbf}, we consider the possibility that the recent PTA data can be explained by the SGWB associated with large curvature fluctuations generated during inflation. We conclude in sec.\,\ref{sec:GW3}, based on\,\cite{Ellis:2023oxs}, where we perform a Multi-Model Analysis (MMA), comparing the qualities of fits to the NANOGrav 15-year (NG15) data invoking SMBH binaries, driven either by GWs alone or together with environmental effects, with several proposed cosmological physics sources. 
\section{SGWB from PBHs}\label{sec:GW1}
In this section we compute the emission of GWs, neglecting initially higher order contributions to the SGWB. This is because, differently from what happens in the case of PBH formation which are extremely sensitive to the non-gaussian tail of the curvature distribution, 
the emission of GWs is dominated by the leading order in the cases we report below.
Therefore, as shown in refs.\,\cite{Cai:2018dig,Unal:2018yaa,Ragavendra:2021qdu,Adshead:2021hnm,Abe:2022xur,Garcia-Saenz:2022tzu}, corrections from higher orders terms only amount to a negligible contribution to the SGWB. Corrections due to NGs at quadratic order is reported in App.\ref{app:NGSIGW}.

The emission of GWs is dictated by the second order equation\,\cite{Tomita:1975kj,Matarrese:1993zf,Acquaviva:2002ud,Mollerach:2003nq,Ananda:2006af,Baumann:2007zm}
\begin{equation}
\left [ \frac{d^2}{d \eta^2} + k^2 - \left ( \frac{1-3w(\eta)}{2}  \right) 
{\cal H}^2 \right ]
 a(\eta)h _{\vec k}(\eta) 
= 4 a(\eta) {\cal S}_{\vec k}(\eta),
    \label{eq: EoM_gw}
\end{equation}
where $\eta$ is the conformal time, ${\cal H}\equiv a H$ is the conformal Hubble parameter, and the source term is written in terms of the gravitational potential $\Phi$ the conformal Newtonian gauge as 
\begin{align}
{\cal S}_{\vec k} =& \int \frac{\text{d}^3 q}{(2 \pi)^{3}} 
e_{ij}({\vec k }) q_i q_j 
\left[ 2\Phi_{\vec q}  \Phi_{\vec k -\vec q} 
+ \frac{4}{3(1+w)} 
\left( \mathcal{H}^{-1} \Phi'_{\vec q} + \Phi_{\vec q}\right) \left( \mathcal{H}^{-1} \Phi'_{\vec k -\vec q}  + \Phi_{\vec k -\vec q}  \right)  \right].
\end{align}
The evolution of the scalar perturbations is modified by the softening of the equation of state around the QCD era. For this reason, we solve numerically the evolution of $\Phi_{\vec k}$ given by (see e.g. ref.\,\cite{Mukhanov:2005sc})
\begin{align}
\Phi''_{\vec k} + 3 \mathcal{H} (1 + c_{\text{s}}^2) \Phi'_{\vec k} 
+ [
2 \mathcal{H}'+(1+3 c_{\text{s}}^2)\mathcal{H}^2 +c_{\text{s}}^2 k^2
] \Phi_{\vec k}  = 0, 
\label{EOM_Phi_complete}
\end{align}
only for spectral modes re-entering the horizon close to the QCD phase transition around $k \simeq 10^6$ Mpc$^{-1}$, 
while using the analytical solution assuming perfect radiation (i.e. eq.\,\eqref{eq:Variances}) otherwise.
Adopting the Green's function method, we solve for the tensor modes $h_{\vec k}$, accounting for the time-varying equation of state. 
The power spectrum of tensor modes becomes
\begin{align}
P_h (\eta, k) =  2 
 \int_0^\infty \text{d}t \int_{-1}^{1}\text{d} s 
 \left [ \frac{t(2+t)(s^2-1)}{(1-s+t)(1+s+t)} \right ]^2
 {\cal I}^2(t,s,\eta,k) 
 \nonumber \\
 \times P_\zeta \left ( k (t+s+1)/2 \right )
 P_\zeta \left ( k (t-s+1)/2 \right ),
 \label{P_h_ts}
\end{align}
and scales like the two powers of $P_\zeta$, due to the second order nature of the emission. 
We denote $P$ as the dimensionless power spectrum. The kernel function ${\cal I}(t,s,\eta,k)$ is computed by integrating over time the Green's function multiplied by the time-dependent source (see more details in ref.\,\cite{Abe:2020sqb}).

The current abundance of SGWB can be computed accounting for the propagation as free GWs after emission, whose energy density in the late time Universe is sensitive to the 
deviations from exact radiation dominated background due to the time dependence of $g_*$ and $g_{*s}$. One finds\,\cite{Espinosa:2018eve,Kohri:2018awv}
\begin{equation}
    \begin{aligned}\label{eq:OmegaGW}
        \Omega_{\rm GW}(k)h^2 
        =\Omega_{r,0}h^2
        \left (\frac{a_{\rm c}{\cal H}_{\rm c}}{a_{\rm f}{\cal H}_{\rm f}} \right )^2
        \frac{1}{24}\left (\frac{k}{{\cal H}_{\rm c}}\right )^2
        \overline{ P_h(k,\eta_{\rm c})}.
    \end{aligned}
\end{equation}
where
$\Omega_{r,0}$ stands for the current radiation density if the neutrino were massless and
we denoted as $\eta_c \gg 1/k $ the time after which GW emission of a given mode $k$ becomes negligible.
Here the oscillation average of the dimensionless power spectrum $\overline{\mathcal{P}_{h,\lambda}^2(k,\eta)}$ is present due to oscillatory behavior of the kernel $\overline{I(|\textbf{k}-\textbf{q}|,q,\eta)}$, where all the time dependencies of the induced GW are enclosed.
For concreteness, following the choice of ref.\,\cite{Abe:2020sqb}, we fix $\eta_c =400/k $.

Two different effects modulates the SGWB around the nano-Hz frequencies, beyond what is expected from our ${ P}_\zeta$ in a pure radiation background.
The pre-factor $\left ({a_{\rm c}{\cal H}_{\rm c}}/{a_{\rm f}{\cal H}_{\rm f}} \right )^2 = \left ({g_*}/{g_*^0} \right) \left({g_{*S}^0}/{g_{*S}}\right)^{4/3} $ 
is typically denoted  $c_g$ in the literature.
This accounts for the departure of the cosmological expansion from the solution in perfect radiation when there is a variation of effective relativistic degrees of freedom. In practice, it tracks the different  dilution of the energy density in the GW sector compared to the background, which is particularly relevant across the QCD phase. 
On top of this, the smaller $c_s$ encountered around the QCD era delays the oscillation of density perturbations right after its horizon re-entry, compared to pure radiation dominated background, as a consequence of the smaller sound horizon $c_s/H$. 
This leads to an enhancement of the SGWB around $f\approx {\rm few}$ nHz, a feature right in the frequency range observed by PTA experiments.

The importance of including the effects of the QCD phase transition in the gravitational wave signal produced will be addressed in Chapter \ref{cap:Models}. At the moment, we neglect this effect and see how, once a power spectrum is chosen and inserted into eq. \ref{eq:OmegaGW}, it is possible to determine the produced gravitational wave signal. As described at the end of Chapter \ref{cap:NGS}, it is always possible to tune the amplitude of the power spectrum in such a way as to mitigate the impact of non-gaussianity in the calculation of abundance. However, this has repercussions on the produced $\Omega_{\rm GW}$.

\begin{figure}[t]
	\begin{center}
\includegraphics[width=.99\textwidth]{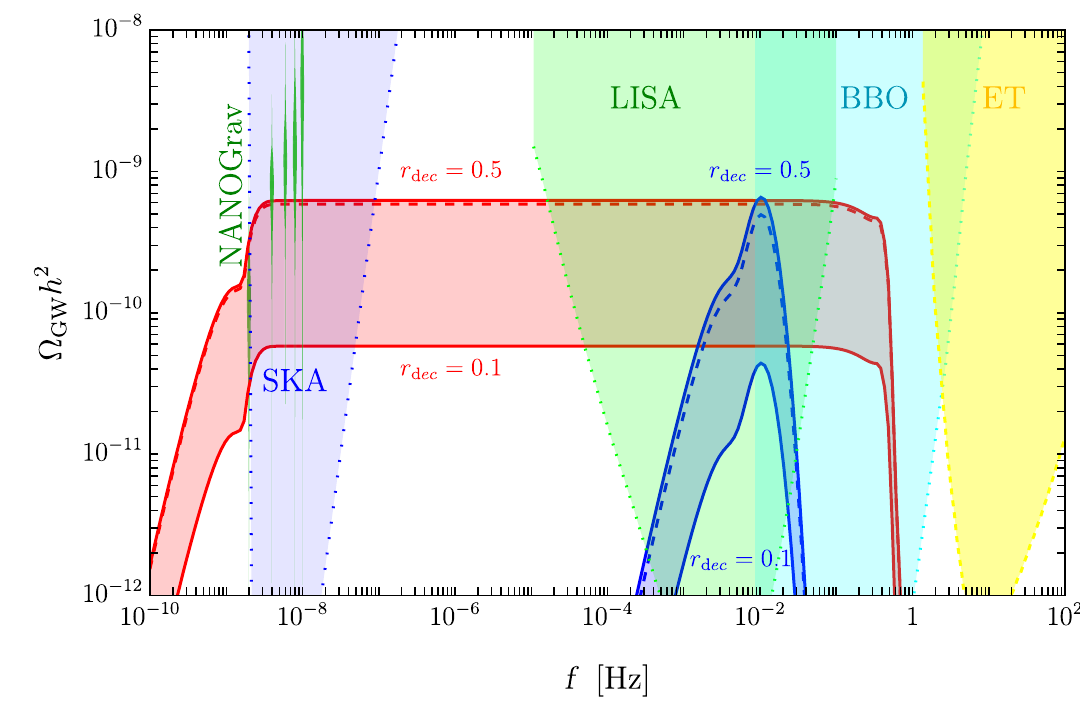}
		\caption{
Spectral density of gravitational waves computed with $r_{\rm dec}=0.1$ and $r_{\rm dec}=0.5$ both with the broad spectrum (red lines) and with the log-normal spectrum (blue lines). 
The coloured dashed lines indicate instead the results when only contributions coming from non-linearities are present. 
The plot also shows the constraints coming from the first 5 bins of the NANOGrav 15 yrs dataset \,\cite{NANOGrav:2023gor} and future sensitivity for planned experiments like SKA\,\cite{Zhao:2013bba}, LISA\,\cite{LISA:2022kgy}, BBO/DECIGO\,\cite{Yagi:2011wg} and ET
(power law integrated sensitivity curves as derived in ref.~\cite{Bavera:2021wmw}). 
}\label{fig:GW} 
	\end{center}
\end{figure}

As reported in fig.\,\ref{fig:fPBH1}, we show the PBHs abundance computed as described in the case of the curvaton models. 
We have taken $r_{{\rm dec}}=0.5$ and $r_{{\rm dec}}=0.1$ as benchmark values for primordial NGs, since they are typical values in expected in realistic curvaton models.
For each value of $r_{{\rm dec}}$, we re-scale the amplitude of the power spectrum in order to keep $f_{\rm PBH} \simeq 1$ fixed.  
Notice that the integral is always dominated by the asteroidal mass range, where there are no constraints on the abundance\,\cite{Carr:2020gox}. 
As shown in fig.\,\ref{fig:GW}, such a re-scaling reflects quadratically on the amplitude of $\Omega_{\rm GW}\approx A^2$ according to eq.\,(\ref{eq:OmegaGW}). 
As we can see from fig.\,\ref{fig:GW}, this tuning allow for different amplitudes of the signal both for a narrow and broad spectrum.
We choose the amplitude of the spectrum in order to obtain $f_{\rm PBH} \simeq 1$. 
Consequently, the corresponding signal of second-order GWs falls within the range of sensitivity of the future experiments LISA and BBO.
Overall, the effect of NGs can modify the predicted SGWB associated to PBH dark matter by one order if magnitude in curvaton models.
We will discuss how the inclusion of NGs are fundamental in order to match the PTA dataset.
\section{PBHs as a possible explanation of the PTA data}\label{sec:GW2}
The observation of a common spectrum process in the NANOGrav 12.5-year data~\cite{NANOGrav:2020bcs} sparked significant scientific interest and led to numerous interpretations of the signal as potential a stochastic gravitational wave background (SGWB) from cosmological sources, such as first order phase transitions~\cite{NANOGrav:2021flc,Xue:2021gyq,Nakai:2020oit,DiBari:2021dri,Sakharov:2021dim,Li:2021qer,Ashoorioon:2022raz,Benetti:2021uea,Barir:2022kzo,Hindmarsh:2022awe,Gouttenoire:2023naa}, cosmic strings and domain walls~\cite{Ellis:2020ena,Datta:2020bht,Samanta:2020cdk,Buchmuller:2020lbh,Blasi:2020mfx,Gorghetto:2021fsn,Buchmuller:2021mbb,Blanco-Pillado:2021ygr,Ferreira:2022zzo,An:2023idh,Qiu:2023wbs,Zeng:2023jut,King:2023cgv}, or scalar-induced gravitational waves (SIGWs) generated from primordial fluctuations~\cite{Vaskonen:2020lbd,Chen:2019xse,DeLuca:2020agl,Bhaumik:2020dor,Inomata:2020xad,Kohri:2020qqd,Domenech:2020ers,Vagnozzi:2020gtf,Namba:2020kij,Sugiyama:2020roc,Zhou:2020kkf,Lin:2021vwc,Rezazadeh:2021clf,Kawasaki:2021ycf,Ahmed:2021ucx,Yi:2022ymw,Yi:2022anu,Dandoy:2023jot,Zhao:2023xnh,Ferrante:2023bgz,Cai:2023uhc}. Consequently, observation of the common spectrum process was reported by other pulsar timing array (PTA) collaborations~\cite{Goncharov:2021oub, Chen:2021rqp, Antoniadis:2022pcn}. The recent PTA data release by the NANOGrav~\cite{NANOGrav:2023gor, NANOGrav:2023hde}, EPTA (in combination with InPTA)\,\cite{EPTA:2023fyk, EPTA:2023sfo, EPTA:2023xxk}, PPTA\,\cite{Reardon:2023gzh, Zic:2023gta, Reardon:2023zen} and CPTA\,\cite{Xu:2023wog} collaborations, shows evidence of a Hellings-Downs pattern in the angular correlations which is characteristic of gravitational waves (GW), with the most stringent constraints and largest statistical evidence arising from the NANOGrav 15-year data (NANOGrav15). The analysis of the NANOGrav 12.5 year data release suggested a nearly flat GW spectrum, $\Omega_{\rm GW}\propto f^{(-1.5, 0.5)}$ at $1\sigma$, in a narrow range of frequencies around $f=5.5$ nHz. In contrast, the recent 15-year data release finds a steeper slope, $\Omega_{\rm GW}\propto f^{(1.3, 2.4)}$ at $1\sigma$ (see Fig.~\ref{fig:fits}). Motivated by this finding, in this section we study whether the signal seen by PTAs may originate from GWs induced by large primordial perturbations. Such perturbations may be accompanied by a sizeable PBH abundance.
\subsection{Log-Likelihood Analysis of PTA experiments}
In the following, we aim to be as model-independent as possible and assume ans\"atze for spectral peaks applicable for classes of models.

A typical class of spectral peaks encountered, for instance, in single-field inflation and curvaton models can be described by a \emph{broken power-law} (BPL)
\be
    \mathcal{P}^{\rm BPL}_{\zeta}(k)
    =A \frac{\left(\alpha+\beta\right)^{\gamma}}{\left(\beta\left(k / k_*\right)^{-\alpha/\gamma}+\alpha\left(k / k_*\right)^{\beta/\gamma}\right)^{\gamma}},
\ee
where $\alpha, \beta>0$ describe respectively the growth and decay of the spectrum around the peak. One typically has $\alpha \lesssim 4$ \cite{Byrnes:2018txb}. The parameter $\gamma$ characterizes the flatness of the peak. Additionally, in quasi-inflection-point models producing stellar-mass PBHs, we expect $\beta \gtrsim 0.5$, while for curvaton models $\beta \gtrsim 2$.
Another broad class of spectra can be characterized by a \emph{log-normal} (LN) shape
\be\label{eq:PLN}
    \mathcal{P}^{\rm LN}_{\zeta}(k)
    = \frac{A}{\sqrt{2\pi}\Delta} \, \exp\left( -\frac{1}{2\Delta^2} \ln^2(k/k_*) \right)\,
\ee
Such spectra appear, e.g., in a subset of hybrid inflation and curvaton models. We find, however, that our conclusions are only weakly dependent on the details of peak shape.
To speed up the best likelihood analysis, we assume perfect radiation domination and do not account for the variation of sound speed during the QCD era (see, for example,~\cite{Hajkarim:2019nbx, Abe:2020sqb}) which also leads specific imprints in the low-frequency tail of any cosmological SGWB \cite{Franciolini:2023wjm}. 
On top of that, cosmic expansion may additionally be affected by unknown physics in the dark sector, which can, e.g., lead to a brief period of matter domination of kination~\cite{Ferreira:1997hj,Pallis:2005bb,Redmond:2018xty,Co:2021lkc,Gouttenoire:2021jhk,Chang:2021afa}. 
Both SIGW and PBH production can be strongly affected in such non-standard cosmologies~\cite{Dalianis:2019asr,Bhattacharya:2019bvk,Bhattacharya:2020lhc,Ireland:2023avg,Bhattacharya:2023ztw,Ghoshal:2023sfa}. 
Moreover, we neglect possible corrections due to primordial NGs. 
This is typically justified because, contrary to the PBH abundance which is extremely sensitive to the tail of the distribution, the GW emission is mostly controlled by the characteristic amplitude of perturbations, and thus well captured by the leading order.
In general, the computation of the SGWB remains the in the perturbative regime if $A (3f_{\rm NL}/5)^2 \ll 1$, where $f_{\rm NL}$ is the coefficient in front of the quadratic piece of the expansion (see Eq.~\eqref{eq:FirstExpansionQ} below).
For the type of NGs considered in this analysis, we always remain within this limit.  
Interestingly, however, both negative and positive $f_{\rm NL}$ increase the SIGW abundance, with the next to leading order correction $\Omega_{\rm GW}^{\rm NLO}/\Omega_{\rm GW} \propto A (3f_{\rm NL}/5)^2$ \cite{Cai:2018dig,Unal:2018yaa,Yuan:2020iwf,Atal:2021jyo,Adshead:2021hnm,Abe:2022xur,Chang:2022nzu,Garcia-Saenz:2022tzu,Li:2023qua} (see also \cite{Bartolo:2007vp}).

\begin{figure}
  \includegraphics[width=0.7\textwidth]{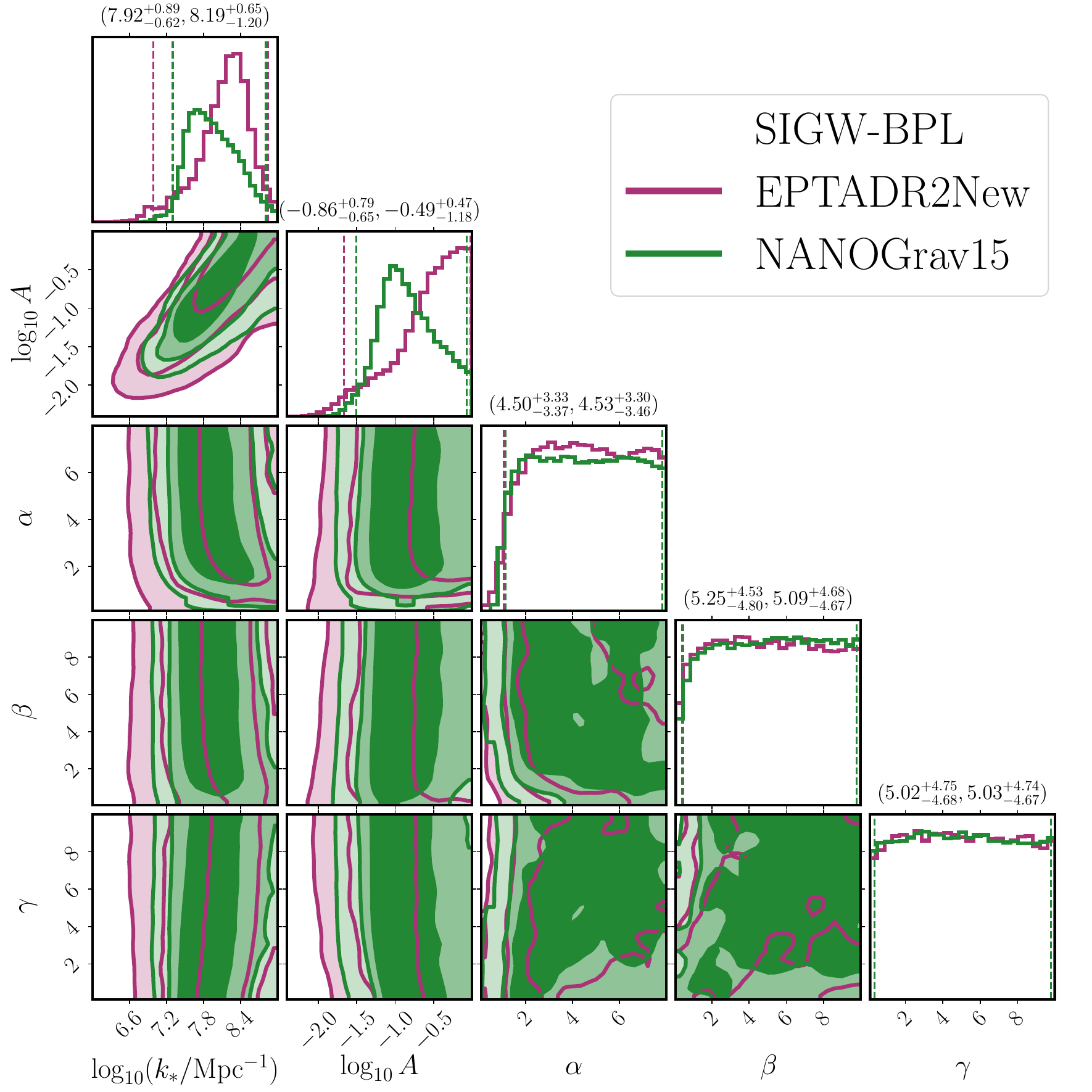}
  \includegraphics[width=0.45\textwidth]{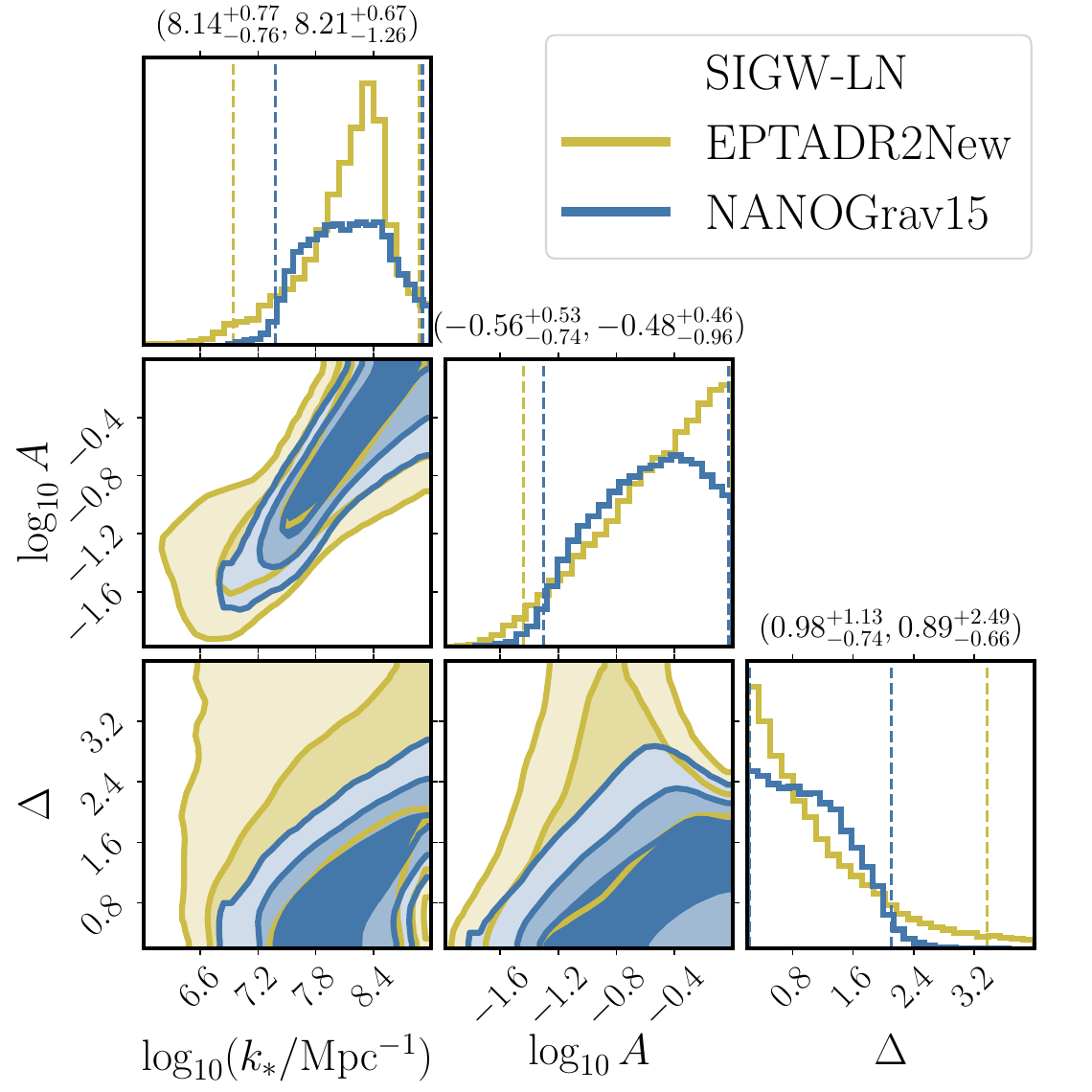}
  \caption{ \textit{\textbf{Top panel:}}
  Posterior for the parameters of a BPL model \eqref{eq:PPL} for SIGWs, assuming no other source of GWs is present in both EPTA and NANOGrav15 data. 
  The shaded regions in the off-diagonal panels show 2-D posteriors at the $1\sigma$, $2\sigma$, and $3\sigma$ confidence levels and the dashed lines in the 1-D posteriors indicate the $2\sigma$ confidence level.
  \textit{\textbf{Bottom Panel:}} Same as bottom panel but for the LN model \eqref{eq:PLN}.
  }
  \label{fig:cornerBPL}
\end{figure}

We perform a log-likelihood analysis of the NANOGrav15 and EPTA data, fitting, respectively, the posterior distributions for $\Omega_{\rm GW}$ for the 14 frequency bins reported in Ref.~\cite{NANOGrav:2023gor, NANOGrav:2023hde} and for the 9 frequency bin\,\cite{EPTA:2023fyk}, including only the last $10.3$ years of data. The results are shown in Fig.~\ref{fig:cornerBPL} for the BPL and LN scenarios, respectively. This analysis is simplified when compared to the one reported by PTA collaborations, which fits the PTA time delay data, modelling pulsar intrinsic noise as well as pulsar angular correlations. However, it provides fits consistent with the results of the NANOGrav~\cite{NANOGrav:2023hvm} and EPTA~\cite{EPTA:2023xxk} collaborations and thus suffices for the purposes of this letter. We neglect potential astrophysical foregrounds, by assuming that the signal arises purely from SIGWs. Around $A = \mathcal{O}(1)$ or flat low $k$ tails, the scenarios considered here are also constrained by CMB observations~\cite{Pagano:2015hma, Chluba:2012we}. However, these constraints tend to be less strict than PBH overproduction and we will neglect them here.

It is striking to see that the posterior distributions shown in Fig.~\ref{fig:cornerBPL} for both BPL and LN analyses indicate a rather weak dependence on the shape parameters, which are ($\alpha$,$\beta$,$\gamma$) and $\Delta$, respectively, as long as the spectra are sufficiently narrow in the IR, i.e. $\alpha \gtrsim 1.1$ and $\Delta \lesssim 2.1$ at $2\sigma$.  This is because the recent PTA data prefers blue-tilted spectra generated below frequencies of SIGW peak around $k_*$.

At small scales ($k \ll k_*$), the SIGW asymptotes to (for details, see the appendix)
\be
    \Omega_{\rm GW} (k \ll k_*) \, {\propto}\,  k^3(1 + \tilde A \ln^2(k/\tilde k))\, ,
\ee
where $\tilde  A$ and $\tilde k = \mathcal{O}(k_{*})$ are parameters that depend mildly on the shape of the curvature power spectrum, see App.\,\ref{supp:1}.
The asymptotic ``causality'' tail $\Omega_{\rm GW} \propto k^3$ is too steep to fit the NANOGrav15 well, being disfavoured by over $3\sigma$. However, this tension may be relieved by QCD effects~\cite{Franciolini:2023wjm}. As a result, the region providing the best fit typically lies between the peak and the causality tail, at scales slightly lower than $k_*$ at which the spectral slope is milder. Such a milder dependence can be observed in the $k_*-A$ panel of Figs.~\ref{fig:cornerBPL}, where $A$ in the $1\sigma$ region scales roughly linearly with $k_*$ indicating that $\Omega_{\rm GW}$  has an approximately quadratic dependence on $k$ in the frequency range relevant PTA experiments. Additionally, since $k_* \geq 2 \times 10^7$ at $2\sigma$, the peaks in the SIGW spectrum lie outside of the PTA frequency range. This can also be observed from Fig.~\ref{fig:fits}.

\subsection{The importance of non-gaussianities}
As we already discussed in Chap.\,\ref{cap:NGS} the NGs play a fundamental role in the precise determination of the abundance of PBHs. In this section we try to understand in which way the NGs can play a fundamental role to match the signal seen by PTA collaborations and PBH explanation.
Here we report again that two type of NGs are present when we compute the abundance. First, the \textit{non linearities} due to the non-linear relation between the curvature perturbation field $\zeta$ and the density contrast $\delta$. Second, the primordial NGs in the non gaussian curvature perturbation field. In this analysis we always include the non linearities and we consider three benchmark cases for primordial NGs:
\begin{itemize}
    \item \textit{Quadratic template}: a generic model-independent approach is to consider the quadratic template
\be\label{eq:FirstExpansionQ}
    \zeta = \zeta_{\rm G} + \frac{3}{5}f_{\rm NL}\zeta_{\rm G}^2\,,
\ee
with $f_{\rm NL}$ as a free parameter. 
\item \textit{Quasi inflection point (or USR)}: the peak in $\mathcal{P}_{\zeta}$ arises from a brief phase of ultra-slow-roll followed by constant-roll inflation dual to it~\cite{Atal:2018neu, Biagetti:2018pjj, Karam:2022nym}. In this case, the NGs can be related to the large $k$ spectral slope \cite{Atal:2019cdz,Tomberg:2023kli},
\be\label{eq:zeta_IP}
    \zeta = -\frac{2}{\beta}\log\left(1-\frac{\beta}{2}\zeta_{\rm G}\right).
\ee
\item \textit{Curvaton models}: a particular class in which the boost of the power spectrum is given by a spectator field, and the NG relation is given by
\be\label{eq:zeta_cur}
    \zeta = \log\big[X(r_{\rm dec},\zeta_{\rm G})\big], 
\ee
where $X(r_{\rm dec})$ is a function of $r_{\rm dec}$ (see Eq.~\eqref{eq:MasterX}) which we take to be the free parameter in our analysis.
\end{itemize}
We follow the prescription reported in sec.\ref{sec:C1A} to compute the abundance.

\begin{figure}[t]
  \centering
  \includegraphics[width=0.9\textwidth]{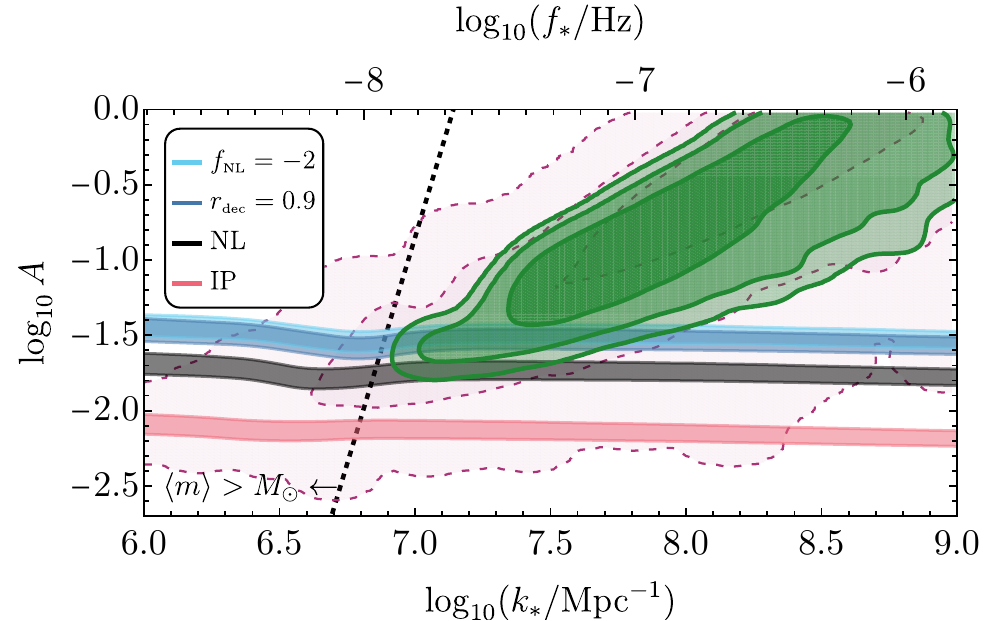}
  \caption{
   PBH abundance for different NG models: 
   non-linearities only (black), 
   quasi-inflection-point models with $\beta = 3$ (red), 
   curvaton models with $r_{\rm ec} = 0.9$ (blue) and negative $f_{\rm NL}$ (cyan).
   We assume a BPL power spectrum~\eqref{eq:PPL} with $\alpha =4$, $\beta=3$ and $\gamma = 1$. 
   The colored bands cover values of PBH abundance in the range $f_{\rm PBH} \in (1,10^{-3})$ from top to bottom. 
   The green and purple posterior comes from Fig.~\ref{fig:cornerBPL}, corresponding to NANOGrav15 and EPTA, respectively. 
   The dashed line indicates an average PBH mass $\langle m \rangle = M_\odot$.}
  \label{fig:abundance}
\end{figure}

The effect of NGs is illustrated in Fig.~\ref{fig:abundance} for a BPL model with $\beta=3$, $\alpha=4$, and $\gamma=1$. We find this scenario to be one of the more conservative ones, that is, changing the shape parameters or switching to an LN shape would yield similar or less optimistic conclusions for SIGW explanations of the recent PTA data.

Fig.~\ref{fig:abundance} shows that even in the absence of primordial NGs, the region avoiding overproduction of PBHs (black band and below) is excluded at over $2\sigma$ by NANOGrav15 while EPTA is currently less constraining.
This conclusion confirms the results obtained in Ref.~\cite{Dandoy:2023jot} based on IPTA-DR2 data \cite{Antoniadis:2022pcn}.
Existing constraints on the PBH abundance force $A$ to fall at the lower edge of the colored band, and slightly strengthen this conclusion. For quasi-inflection-point models, the situation is more dire as NGs tend to assist PBH production which pushes the overproduction limit below the $3\sigma$ region for NANOGrav15. Although both the slope and the NGs in the $\beta=3$ case, shown in red, are quite large, reducing the $\beta$ cannot bring these models above the black band. The flip-side of this conclusion is that NANOGrav15 does not impose additional constraints on the PBH abundance. Thus, a component of the signal may be related to the formation of subsolar mass PBHs that may be independently probed by future GW experiments~\cite{DeLuca:2021hde, Pujolas:2021yaw, Miller:2020kmv, Urrutia:2023mtk, Franciolini:2023opt}.

On the other hand, the tension between SIGWs and NANOGrav15 can be alleviated in models in which NGs suppress the PBH abundance. This is demonstrated by the blue bands in Fig.~\ref{fig:abundance}, which correspond to $f_{\rm PBH} \in (10^{-3},1)$ curvaton models~\eqref{eq:zeta_cur} with a large $r_{\rm dec}$ and for a generic quadratic ansatz~\eqref{eq:FirstExpansionQ} with a large negative $f_{\rm NL}$. It is important to stress, that both cases displayed in Fig.~\ref{fig:abundance} represent the most optimistic scenarios: increasing $r_{\rm dec}$ above $0.9$ would have an unnoticeable effect on $f_{\rm PBH}$ and decreasing $f_{\rm NL}$ below $-2$ has a positive effect on PBH formation and would shift the lines away from the best-fit region. This is because sizeable \emph{negative} curvature fluctuations can still generate large fluctuations in the compaction and seed sizeable abundance. 

The best-fit region for NANOGrav15 lies at scales $k_*>10^7 {\rm Mpc}^{-1}$ which corresponds to the production of sub-solar mass PBHs (see Fig.~\ref{fig:abundance}). Around $k_* \approx 10^{7} {\rm Mpc}^{-1}$, small dents in the colored bands in Fig.~\ref{fig:abundance} can be observed. These arise due to the effect of the QCD phase transition which promotes PBH formation. Thus, we find that the QCD-induced enhancement of $f_{\rm PBH}$ in the parameter space relevant for NANOGrav15 tends to be negligible.

Although our $f_{\rm PBH}$ estimates assume quite narrow curvature power spectra, we checked that our conclusions about PBH overproduction in single-field inflation persist also in the case of broad spectra.

As a last remark, limiting our analysis to the absence of NGs in the curvature perturbation field $\zeta$, we have found that our results differ from those published by the NANOGrav collaboration~\cite{NANOGrav:2023hvm}. These discrepancies arise because their analysis is subject to a few simplifications: the omission of critical collapse and the nonlinear relationship between curvature perturbations and density contrast, the adoption of a different value for the threshold (independently from the curvature power spectrum), and the use of a Gaussian window function (which is incompatible with their choice of threshold \cite{Young:2019osy}). Another minor limitation is that they disregard any corrections from the QCD equation of state, although we find that the result is minimally dependent on this aspect.
\subsection{Going beyond the average profile: Consequences on SIGWs}\label{sec:subGWS}
As discussed in sec.\ref{sec:C1B} the prescription based on going beyond the average profile provides much higher abundance of PBH compared to the prescription used in the previous section. This has important implications for the SGWB produced.
As a consequence even when correctly accounting for the impact of the curvature of the compaction function and calculating  all the relevant quantities on  superhorizon scales, thereby avoiding all concerns regarding non-linearities in the radiation transfer function and the determination of the true physical horizon, the tension  between the PTA dataset and the PBH hypothesis is even worse than what discussed in the previous section. We leave a complete discussion of the SGWB obtained using this description for a future work, probably when the IPTA dataset will be released in 2025 and it is not reported in this thesis.
\subsection*{Summary}
In this section we have analyzed the possibility that the signal reported by several PTA collaborations may originate from GWs induced by high-amplitude primordial curvature perturbations. This scenario is accompanied by the production of a sizeable abundance of PBHs. 
Our findings demonstrate that PBH formation models that feature Gaussian primordial perturbations, or positive NGs would overproduce PBHs unless the amplitude of the spectrum is much smaller than required to explain the GW signal.
For instance, most models relying on single-field inflation featuring an inflection point appear to be excluded at $3\sigma$ as the sole explanation of the NANOGrav 15-year data. 
However, this tension can be alleviated for models where large negative NGs suppress the PBH abundance. For instance, curvaton scenarios with a large $r_{\rm dec}$ and models exhibiting only large negative $f_{\rm NL}$. As a byproduct, however, we conclude that the PTA data does not impose constraints on the PBH abundance.
\section{PTA explanation: Are the PBHs the end of the story?}\label{sec:GW3}
When the PTA collaborations have released their dataset two classes of interpretation of this apparent discrepancy have been proposed. One is that the SMBH binaries may also be losing energy through some other mechanism, presumably through interactions with their environments, that would reduce the period over which they emit GWs in the frequency range measured \cite{NANOGrav:2023hfp,Ellis:2023dgf,Ghoshal:2023fhh,Bi:2023tib,Zhang:2023lzt}. A more radical interpretation is that the GWs are being emitted by some cosmological source whose origin is in fundamental physics. Candidate sources that have been considered include a network of cosmic (super)strings\,\cite{Ellis:2023tsl,Kitajima:2023vre,Wang:2023len,Lazarides:2023ksx,Eichhorn:2023gat,Chowdhury:2023opo,Servant:2023mwt,Antusch:2023zjk,Yamada:2023thl,Ge:2023rce,Basilakos:2023xof}, a first-order phase transition in the early Universe\,\cite{Fujikura:2023lkn,Addazi:2023jvg,Bai:2023cqj,Megias:2023kiy,Han:2023olf,Zu:2023olm,Megias:2023kiy,Ghosh:2023aum,Xiao:2023dbb,Li:2023bxy,DiBari:2023upq,Cruz:2023lnq,Gouttenoire:2023bqy,Ahmadvand:2023lpp,An:2023jxf}, domain walls\,\cite{Kitajima:2023cek,Guo:2023hyp,Blasi:2023sej,Gouttenoire:2023ftk,Barman:2023fad,Lu:2023mcz,Babichev:2023pbf,Gelmini:2023kvo,Zhang:2023nrs}, primordial fluctuations\,\cite{Franciolini:2023pbf,Vagnozzi:2023lwo,Franciolini:2023wjm,Inomata:2023zup,Cai:2023dls,Wang:2023ost,Ebadi:2023xhq,Gouttenoire:2023nzr,Liu:2023ymk,Abe:2023yrw,Unal:2023srk,Yi:2023mbm,Firouzjahi:2023lzg,Salvio:2023ynn,You:2023rmn,Bari:2023rcw,Ye:2023xyr,HosseiniMansoori:2023mqh,Cheung:2023ihl,Balaji:2023ehk,Jin:2023wri,Das:2023nmm,Ben-Dayan:2023lwd,Jiang:2023gfe,Liu:2023pau,Yi:2023tdk,Frosina:2023nxu,Bhaumik:2023wmw,Yuan:2023ofl,Gorji:2023sil}, ``audible" axions\,\cite{Figueroa:2023zhu,Geller:2023shn} and other more exotic scenarios\,\cite{Li:2023yaj,Lambiase:2023pxd,Borah:2023sbc,Datta:2023vbs,Murai:2023gkv,Niu:2023bsr,Choudhury:2023kam,Cannizzaro:2023mgc,Zhu:2023lbf,Aghaie:2023lan,He:2023ado}. If any such mechanism is in operation, one may expect there also to be an admixture of GWs from astrophysical SMBH binaries.~\footnote{See, for example,~\cite{Bian:2023dnv,Figueroa:2023zhu,Wu:2023hsa}}

In this section, we perform a Multi-Model Analysis (MMA), comparing the qualities of fits to the NANOGrav 15-year (NG15) data invoking SMBH binaries, driven either by GWs alone or together with environmental effects, with each of the proposed fundamental physics sources. We also explore fits postulating combinations of SMBH binaries with each of the fundamental physics mechanisms. Our MMA adopts a common statistical approach to all the proposed sources based directly upon the NANOGrav probability density function (PDF) in each frequency bin. 
Furthermore, in order to help disentangle cosmological and astrophysical sources, we highlight some experimental features that could in the future be used to distinguish between them. These include fluctuations in the GW signal between different frequency bins (including the possible observation of individual GW sources) and the behavior of the SGWB spectrum at higher frequencies. For example, whereas SMBH models predict observable fluctuations that increase at higher frequencies, generic fundamental physics sources would not predict large bin-by-bin fluctuations, and some predict an extended spectrum that would be detectable at higher frequencies, whereas others predict a sharp fall-off above the range where PTAs are sensitive.
\subsection{Supermassive astrophysical Black Holes}
\label{sec:SMBH}

The default interpretation of the SGWB postulates a population of tight astrophysical  SMBH binaries. We estimate the SMBH binary population as in~\cite{Ellis:2023dgf}: We use the extended Press-Schechter formalism~\cite{Press:1973iz,Bond:1990iw,Lacey:1993iv}, which depends on the rate $R_h$ of coalescences of galactic halos of masses $M_{1,2}$, the stellar mass-BH mass relation arising from observations of inactive galaxies~\cite{Kormendy:2013dxa} and the observed stellar mass-halo mass relation~\cite{Girelli:2020goz}. The merger rate of SMBHs of masses $m_{1,2}$ is
\bea
    \frac{\td R_{\rm BH}}{\td m_1 \td m_2} \approx \,p_{\rm BH} \int \td M_1 \td M_2 \, \frac{\td R_h}{\td M_1 \td M_2} \times p_{\rm occ}(m_1|M_1,z) p_{\rm occ}(m_2|M_2,z)  \,, 
\eea
where we incorporate a probability $p_{\rm BH}$ for the pair of SMBHs to approach each other sufficiently closely to merge~\cite{Ellis:2023owy} and $p_{\rm occ}$ relates the BH mass to the halo mass including the observed scatter~\cite{Ellis:2023dgf}. The interactions with their environments that allow the binaries to overcome the ``final parsec problem’’ are not fully understood~\cite{Begelman:1980vb}, and we regard $p_{\rm BH}$ as a phenomenological parameter to be determined using the PTA data. For simplicity, we assume $p_{\rm BH}$ to be constant.

\subsubsection{GW-driven binaries}

The spectrum of the GW signal from the SMBH population in general includes large fluctuations due to the small number of dominant and deviates from the naive power-law approximation that predicts~\cite{Phinney:2001di} the spectral index $\gamma = 13/3$. Following the analysis of Ref.~\cite{Ellis:2023dgf}, we compute the PDFs of the GW signal from SMBH binaries in the frequency bins used in the NG15 analysis and calculate their overlaps with the posterior PDFs (``violins") from the Hellings–Downs-correlated free spectrum analysis of the NG15 data~\cite{NANOGrav:2023gor}. Assuming that the evolution of the SMBH binaries is driven solely by energy loss due to GW emission, we find $p_{\rm BH} = 0.07_{-0.07}^{+0.05}$, which is consistent with an earlier analysis of IPTA data~\cite{Ellis:2023owy}, but the quality of the best fit is poor.

\begin{figure}
    \centering
    \includegraphics[width=0.6\columnwidth]{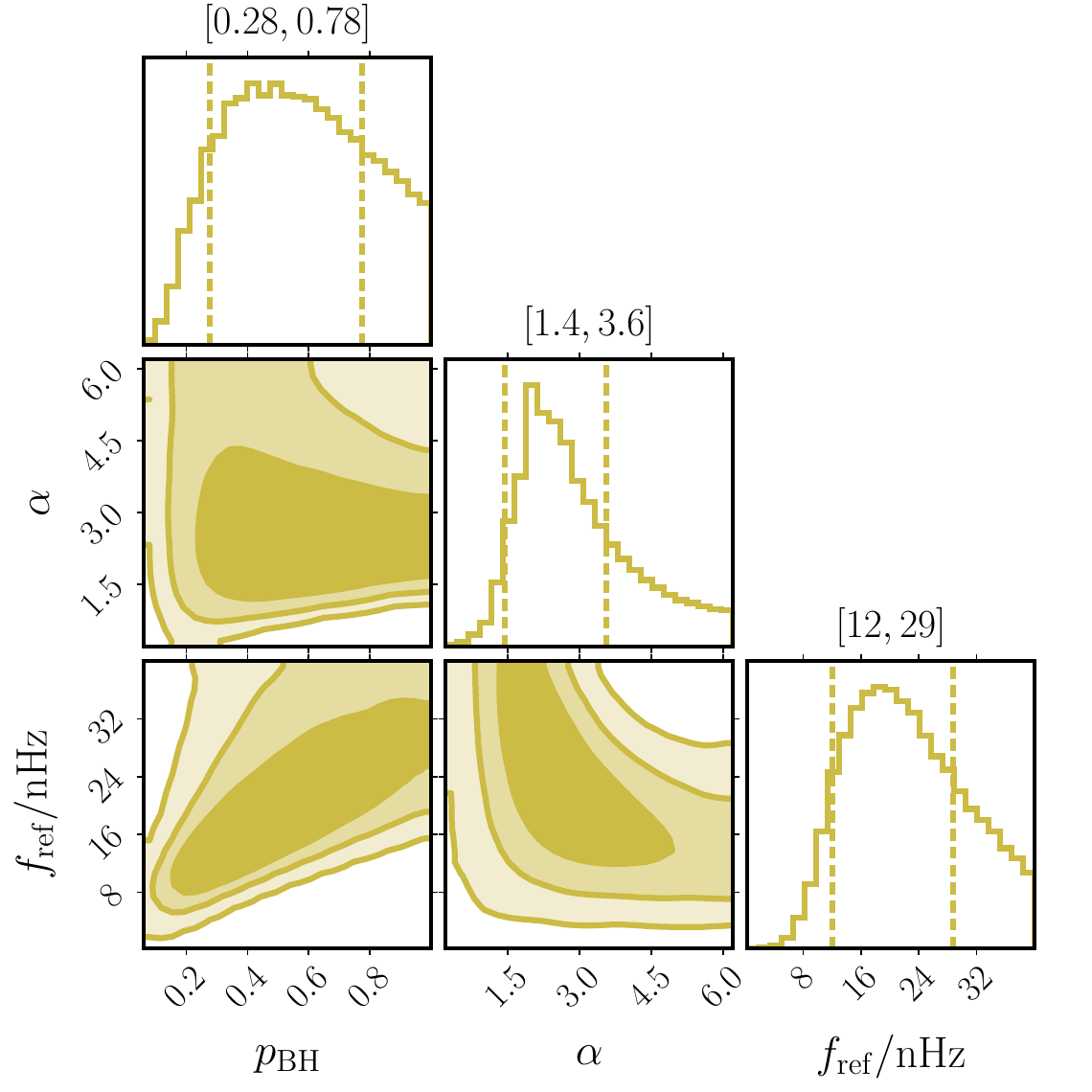}
    \caption{The posterior probability distribution of the fit of the SMBH binary model with environmental effects to the NG15 data. The contours enclose the $1\sigma$, $2\sigma$, and $3\sigma$ CL regions. On top of each column, we report $1\sigma$ CL ranges.}
    \label{fig:SMBH_posteriors}
\end{figure}

\subsubsection{Environmental effects}

It is natural to consider the possibility that environmental effects could affect the evolution of SMBH binaries while they are emitting GWs in the nHz frequency range. We parametrize the binary energy loss via environmental effects as~\cite{Ellis:2023dgf}
\be \label{eq:gas}
    \frac{t_{\rm env}}{t_{\rm GW}} = \left(\frac{f_r}{f_{\rm GW}}\right)^{\alpha} , \quad 
    f_{\rm GW} = f_{\rm ref} \left(\frac{\cM}{10^9 \Msun}\right)^{-\beta} , 
\ee
where $t_{\rm env}$ and $t_{\rm GW}$ are the timescales for energy loss via environmental effects and GWs, respectively, $f_{\rm ref}$ is a reference frequency, ${\cal M}$ is the binary chirp mass, and $\alpha$ and $\beta$ are phenomenological parameters. We take $\beta = 0.4$~\footnote{Our results are not sensitive to this choice: see~\cite{Ellis:2023dgf}.} and treat $p_{\rm BH}, \alpha$ and $f_{\rm ref}$ as parameters to be constrained using PTA data. This three-parameter model provides a significantly better fit to the NG15 data than the single-parameter GW-only model, with $-2 \Delta \ell = -11.3$ relative to the GW-driven SMBH model.\footnote{{We estimate the log-likelihood for observing the data $\vec{d}$ given a model characterized by $\vec{\theta}$ by
\be
    \ell(\vec{d}|\vec{\theta}) = \sum_j \ln \int \td \Omega P_{{\rm ex},j}(\Omega|\vec{d}) P_{{\rm th},j}(\Omega|\vec{\theta}) \,,
\ee
where the sum runs over the first 14 NG15 bins and $P_{{\rm ex},j}$ and $P_{{\rm th},j}$ denote the probability distributions of $\Omega$ in bin $j$ in the NG15 data and in the given model. We note that the {\tt Cefyll} package~\cite{lamb2023rapid} also computes the likelihood efficiently and gives a good approximation to fitting the full spectrum.}} The posterior probabilities for the three parameters, computed in Ref.~\cite{Ellis:2023dgf}, are shown in Fig.~\ref{fig:SMBH_posteriors}. The best fit is at 
\be
    p_{\rm BH} = 0.84,\quad \alpha = 2.0, \quad f_{\rm ref} = 34\,{\rm nHz} \,.
\ee
Throughout the rest of this section, this best-fit environmentally driven SMBH scenario sets the baseline to which other models are compared.

We note that our analysis assumes circular orbits that emit GWs at twice the orbital frequency. The eccentricity of the orbits would affect the GW spectrum by introducing higher harmonics, increasing the total power emitted in GWs and modifying the frequency spectral index~\cite{Enoki:2006kj}.\footnote{There is no indication of eccentric binaries in current (O3) LIGO/Virgo/KAGRA (LVK) data~\cite{LIGOScientific:2023lpe}.}  However, big eccentricities $e>0.9$ would lead to an attenuation of the background due to the acceleration of the binary inspiral~\cite{Kelley:2017lek}. The GW signal from SMBHs could also be affected by modifications to small-scale structures compared to standard $\Lambda$CDM, which could lead to the earlier formation of galaxies and SMBHs. The observation of high-redshift galaxies with surprisingly high stellar masses by the JWST~\cite{2022arXiv220802794N, 2022ApJ...940L..55F, 2023Natur.616..266L, 2023ApJS..265....5H} provides a potential hint for this possibility~\cite{Liu:2022bvr, Menci:2022wia, Biagetti:2022ode, Hutsi:2022fzw, Parashari:2023cui, Hassan:2023asd, Gouttenoire:2023nzr, Guo:2023hyp,Ellis:2024wdh}. 

\begin{figure}
    \centering
    \includegraphics[width=0.6\columnwidth]{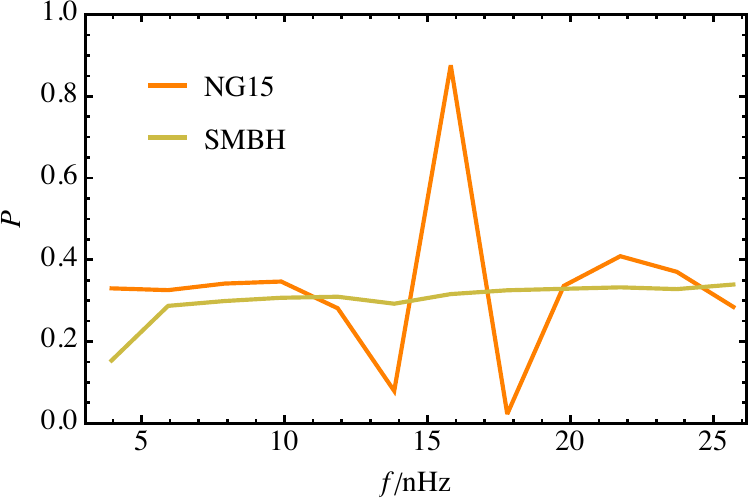}
    \caption{The probability~\eqref{eq:Pfluct} for upward fluctuations in the SMBH model with environmental effects (green line) and the fluctuations measured in the NG15 data (orange line).}
    \label{fig:Pfluct}
\end{figure}

\subsubsection{Fluctuations in the GW spectrum}

As already mentioned, the SMBH signal may exhibit significant fluctuations between frequency bins, which appear because around half of the signal strain is produced by ${\cal O}(10)$ sources. {Since the spectral fluctuations in the SGWB for every cosmological BSM model are expected to be negligible, the appearance of fluctuations would be a smoking gun for an astrophysical component in the signal.}

Spectral fluctuations can be quantified by the second finite difference
\bea\label{eq:Delta2_Omega}
    \Delta^2\Omega (f) 
    &\equiv \Omega(f+\delta f) + \Omega(f-\delta f) - 2\Omega(f) \, ,
\eea
where the step size $\delta f$ can, for instance, be taken to correspond to the width of a bin. The probability of an upward fluctuation in the $i$th bin can be estimated as~\footnote{Neglecting correlations between bins, this probability is given by
\bea
    P(-\Delta^2\Omega(f_i) > \Omega_{\rm th} )
    =  \int \td \Omega \td \Omega' P_{i+1}(\Omega-\Omega') P_{i-1}(\Omega') \times \int_{2\Omega_i - \Omega > \Omega_{\rm th}(\Omega,f_i)} \td \Omega_i P_i(\Omega_i)
\eea
where $P_{i}$ denote the theoretical $\Omega_{\rm GW}$ distributions in individual bins (see~\cite{Ellis:2023owy,Ellis:2023dgf}).}
\be \label{eq:Pfluct}
    P_{\rm fluct, i} \equiv P(-\Delta^2\Omega(f_i) > \Omega_{\rm th}(f_i) ) \, ,
\ee
where $f_i$ is the bin frequency and we choose
\be
    \Omega_{\rm th}(f_i) = \iota \left[\Omega (f_{i-1}) + \Omega (f_{i+1})\right] \,,
\ee
with $\iota = 0.2$. This choice of threshold can help distinguish SMBH binaries from cosmological GW sources, which are expected to produce smooth spectra. We find that for all of the cosmological sources discussed below, the probability~\eqref{eq:Pfluct} is zero in all NG15 bins. When the GW signal arises from an admixture of cosmological sources and SMBH binaries the probability of fluctuations will be smaller than in the pure SMBH model. The optimal threshold $\Omega_{\rm th}$ that can filter out cosmological models depends on the frequency binning -- while for smooth cosmological signals, the finite difference \eqref{eq:Delta2_Omega} decreases with increased spectral resolution (since $\Delta^2\Omega(f) \propto \delta f^2$ as $\delta f \to 0$), it can be sizeable for SMBH binaries as long as their frequency changes less than $\delta f$ during the observational period.
 
In Fig.~\ref{fig:Pfluct} we show the probability~\eqref{eq:Pfluct} for the best-fit parameters of the SMBH binary model with environmental effects. We find that $P\approx 0.3$ across all the bins and the probability of not having fluctuations that exceed this threshold in any of the 11 bins considered in Fig.~\ref{fig:Pfluct} is roughly $\prod P_{\rm fluct, i} \approx 1.3\%$, while the expected number of upward fluctuations in the NG15 bins is $\sum_i  P_{\rm fluct, i} \approx 4$ for both the NG15 data and the SMBH binary model. The NG15 data indicates that the 8th bin at $f\approx 16\,$nHz shows an upward fluctuation at the 91\% CL.

\subsubsection{Detecting individual binaries}

\begin{figure}
\centering
\includegraphics[width=0.6\columnwidth]{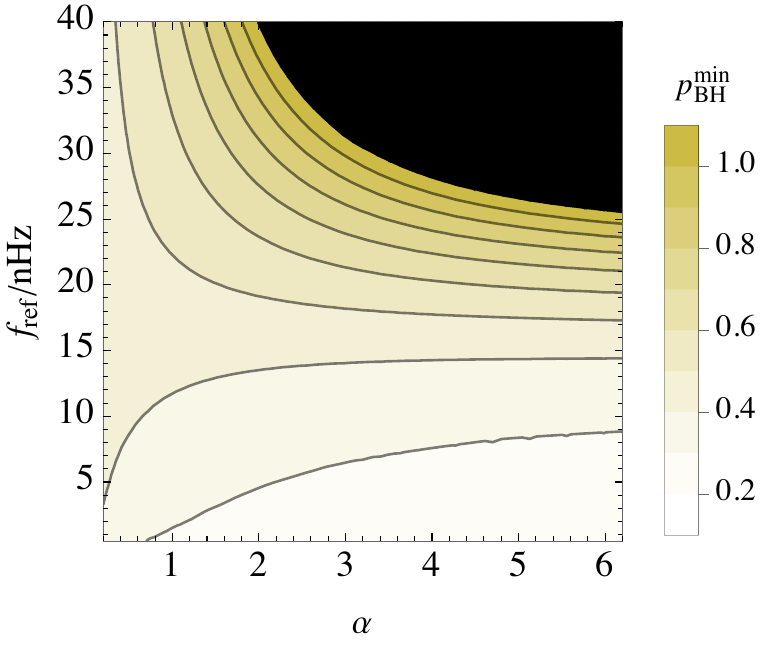}
\caption{Contours of the minimal merger efficiency $p_{\rm BH}^{\rm min}$ required for the probability of finding the candidate event at $f \sim 4 \rm{nHz}$ to exceed 5\% for different values of the parameters of the environmental effects. In the black region, the probability of the candidate event is less than 5\% for $p_{\rm BH}\leq 1$.}
\label{fig:Pmin}
\end{figure}

A nearby SMBH binary will stand out from the background as a resolvable binary. Intriguingly, the model predicts that the most probable frequency for an individual binary to be detectable is $\sim 4$~nHz~\cite{Ellis:2023dgf}, where there is evidence for an upward fluctuation in both NANOGrav and EPTA data~\cite{NANOGrav:2023pdq,EPTA:2023gyr}. If the SMBH binaries are driven by GW emission only the probability of observing such an event in the present NANOGrav data is less than$\sim 5\%$, but such an event is easier to accommodate when environmental energy loss is included in the SMBH binary model, where the probability is $\sim 23\%$. In Fig.~\ref{fig:Pmin} we show contours of the minimal merger efficiency, $p_{\rm BH}^{\rm min}$, for which the probability of finding such an event is larger than 5\% for different values of $\alpha$ and $f_{\rm ref}$.~\footnote{In addition to bin-to-bin fluctuations and resolvable binaries, other characteristic features of the SMBH models include the possibilities of observable circular polarization~\cite{Sato-Polito:2021efu,Ellis:2023owy} and anisotropies~\cite{Sato-Polito:2023spo}.}
\subsection{Cosmological sources}
\label{sec:cosmo}

In this Section, we discuss SGWB predictions in various cosmological scenarios invoking particle physics beyond the Standard Model (BSM). In these scenarios, the GWs are produced in the early Universe and their present abundance is
\bea\label{eq:OmegaGWtoday}
    &\Omega_{\rm GW} h^2 = \left[\frac{a(T)}{a_0}\right]^4 \left[\frac{H(T)}{H_0/h}\right]^2 \Omega_{\rm GW}(T) \\ 
    &\approx 1.6\times 10^{-5} \left[ \frac{g_*(T)}{100} \right] \!\left[ \frac{g_{*s}(T)}{100} \right]^{\!-\frac43} \Omega_{\rm GW}(T) \,,
\eea
where $T$ denotes the temperature at which the GWs were produced and $\Omega_{\rm GW}(T)$ their abundance at that moment, $a$ is the scale factor and $H$ the Hubble rate, the subscript $0$ refers to their present values, and $g_*$ and $g_{*s}$ are the effective numbers of relativistic energy and entropy degrees of freedom. The present frequency corresponding to the scale entering the Hubble horizon at temperature $T$ is
\bea\label{eq.fre}
    f_H(T)\!&=\frac{a(T)}{a_0} \frac{H(T)}{2\pi} \\
    &\approx\!2.6\!\times\!10^{-8}\,{\rm Hz} \left[ \frac{g_*(T)}{100} \right]^{\frac12}\!\left[ \frac{g_{*s}(T)}{100} \right]^{\!-\frac13}\!\frac{T}{\rm GeV} \,.
\eea
For short-lasting, so-called causality-limited GW sources, the GW spectrum at $f\ll f_H$ scales as $\Omega_{\rm GW}(f\ll f_H)= \Omega_{\rm CT}(f) \propto f^3$, assuming radiation domination~\cite{Caprini:2009fx,Domenech:2020kqm}. This behavior is affected by deviations from pure radiation domination, e.g., around the QCD phase transition. We use the results tabulated in Ref.~\cite{Franciolini:2023wjm} for the causality tail, $\Omega_{\rm CT}(f)$.

\subsubsection{Cosmic (super)strings}

Cosmic strings are one-dimensional topological defects that are predicted in many BSM scenarios~\cite{Hindmarsh:1994re,Jeannerot:2003qv,King:2020hyd}. Once produced, strings reach a scaling solution where their energy density follows the total cosmological density, due to the production of closed loops that then decay into GWs.
This process continues throughout the evolution of the Universe and results in a very broad and relatively flat spectrum. We focus here on (super)strings that may arise from string theory~\cite{Dvali:2003zj,Copeland:2003bj} or during  confinement in pure Yang-Mills theories~\cite{Yamada:2022aax,Yamada:2022imq}. Their main phenomenological difference from the standard case is a lower intercommutation probability $p$, which diminishes the loop production and increases the resulting string density and GW signal amplitude by a factor $p^{-1}$~\cite{Sakellariadou:2004wq,Blanco-Pillado:2017rnf}.  

Our calculation of the spectrum closely follows~\cite{Ellis:2023tsl}. We start with the velocity-dependent one-scale model~\cite{Martins:1995tg, Martins:1996jp, Martins:2000cs, Avelino:2012qy, Sousa:2013aaa}, which we use to compute the correlation length of the network $L$ and mean string velocity $\bar{v}$ by solving the equations
\bea \label{eq:vos}
    \frac{dL}{dt} &= (1+\bar{v}^2)\,HL + \frac{\tilde{c}\bar{v}}{2} \,, \\
    \frac{d \bar{v}}{dt} &= (1-\bar{v}^2)
    \left[\frac{k(\bar{v})}{L} - 2H\,\bar{v}\right] \,,
\eea
where
\be
    k(\bar{v}) = \frac{2\sqrt{2}}{\pi}(1-\bar{v}^2)(1+2\sqrt{2}\bar{v}^3)
\left(\frac{1-8\bar{v}^6}{1+8\bar{v}^6}\right) \ ,
\ee
and $\tilde{c} \simeq 0.23$ is the loop chopping efficiency~\cite{Martins:2000cs}.

Following recent numerical simulations~\cite{Blanco-Pillado:2011egf, Blanco-Pillado:2013qja, Blanco-Pillado:2015ana, Blanco-Pillado:2017oxo, Blanco-Pillado:2019vcs, Blanco-Pillado:2019tbi} we focus on large loops as the main GW sources. We include a prefactor ${\cal F} \sim 0.1$ to describe the fraction of energy in these loops and a factor $f_r=\sqrt{2}$ accounting for the energy going to peculiar velocities of loops~\cite{Auclair:2019wcv} and not contributing to radiation.

Loops cut from the network oscillate losing energy to GWs which reduces their length over time:
\be \label{eq:lloop}
    l (t) = \alpha_L L \left(t_i \right) - \Gamma G \mu \left(t -t_i \right) \ .
\ee
Where we use $\alpha_L \approx 0.37$, the total emission power $\Gamma \approx 50$~\cite{Burden:1985md, Blanco-Pillado:2017oxo, Blanco-Pillado:2017rnf} and $t_i$ is the time at which the given loop was created.

\begin{figure}
\includegraphics[width=0.33\textwidth]{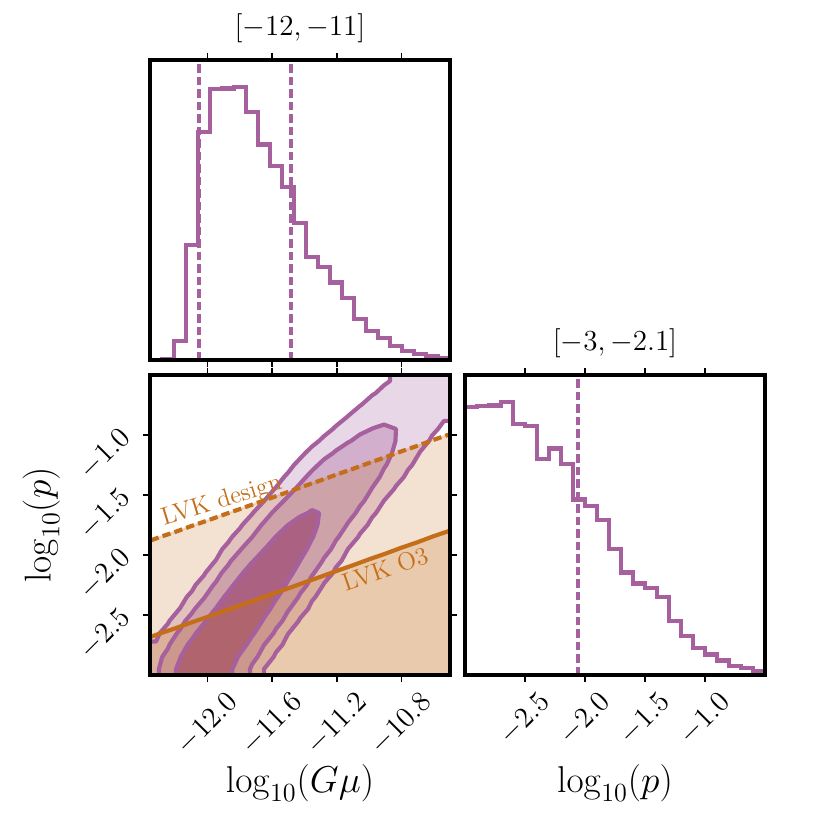}
\includegraphics[width=0.71\textwidth]{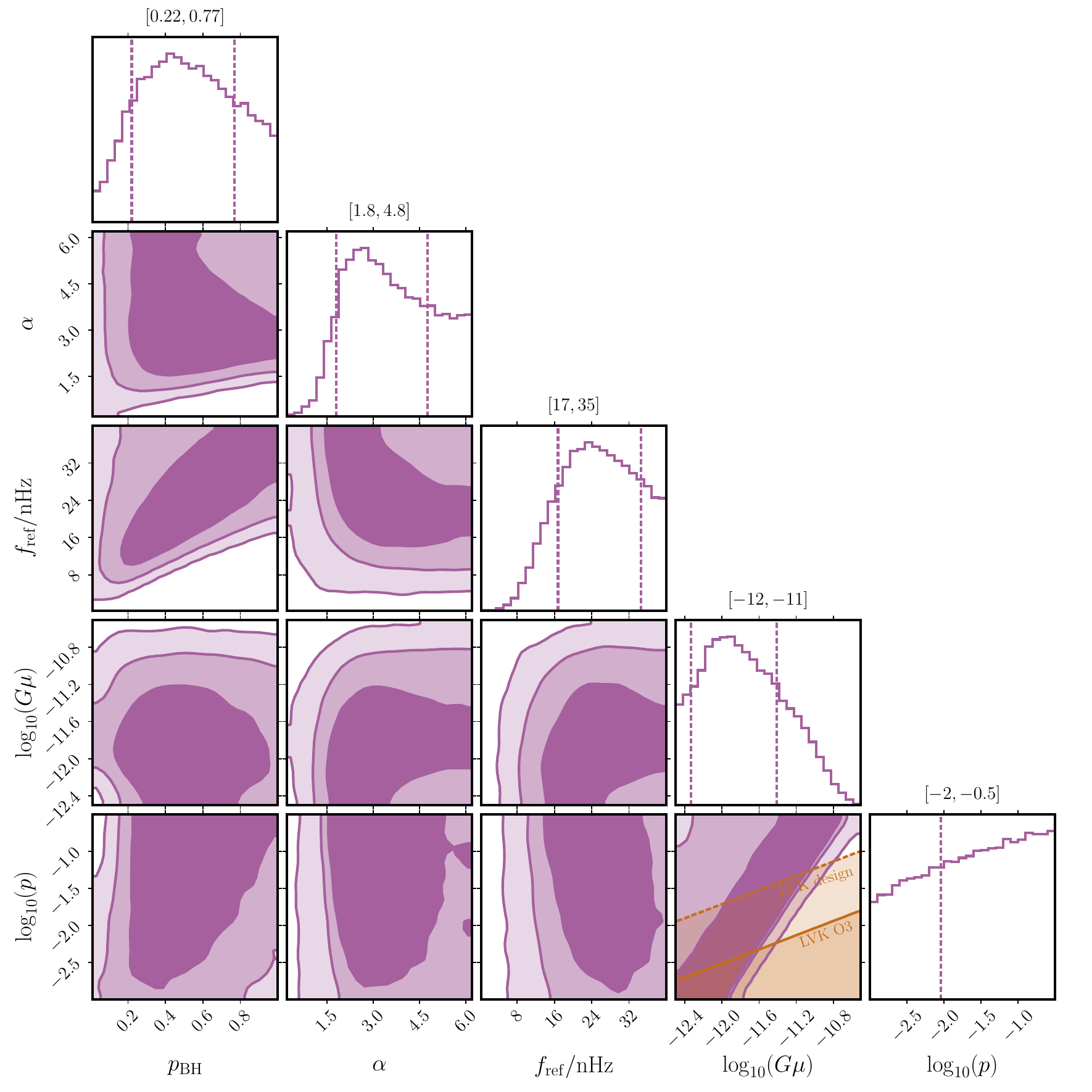}
\caption{\textit{\textbf{Left panel:}} The posterior probability distribution of the cosmic (super)string model fit to the NG15 data. The current (O3) LVK constraint and prospective future LVK sensitivity are shown by the solid and dashed lines. \textit{\textbf{Right panel:}} Same as the left panel but including also SMBH binaries with environmental effects.
}
\label{fig:CSfit}
\end{figure}

Calculation of the spectrum involves the integration of emissions from all contributing loops:
\be
\label{eq:GWdensity}
    \Omega_{\rm GW}^{(1)}(f) = \frac{16\pi}{3 f} \frac{\mathcal{F}}{f_r} \frac{G\mu^2 }{H_0^2} \frac{ \Gamma}{\zeta \alpha_L} \int_{t_F}^{t_0}\!d\tilde{t}\; n(l,\tilde{t}) \, , 
\ee
where
\be
    n(l,\tilde{t}) = \frac{\Theta(t_i - t_F)}{\alpha_L \dot{L}(t_i)+\Gamma G\mu} \frac{\tilde{c} v(t_i)}{L(t_i)^4}  \bigg[\frac{a(\tilde{t})}{a(t_0)}\bigg]^5 \bigg[\frac{a(t_i)}{a(\tilde{t})}\bigg]^3 \, .
\ee
The formation time of the loops has to be found solving Eq.~\eqref{eq:lloop} and restricting to loops emitting at frequency $f$ whose length is $l(\tilde{t},f)=\frac{2}{f}\frac{a(\tilde{t})}{a(t_0)}$.
The Heaviside function assures that we integrate from only the time $t_F$ when the network first reaches scaling after its production. We ensure that the total emitted power equals $\Gamma$ with the function $\zeta=\sum k^{-4/3}$, where we assume that the GW emission is dominated by cusps.
Finally, we sum over the emission modes using
\be
    \Omega_{\rm GW}^{(k)}(f) =k^{-q}\,\Omega_{\rm GW}^{(1)}(f/k) \, ,
\ee
and the prescription from ref. ~\cite{Cui:2019kkd} that allows us to sum over a very large number of modes, which is necessary to compute the spectrum accurately~\cite{Cui:2019kkd, Blasi:2020wpy, Gouttenoire:2019kij}.

In fig.~\ref{fig:CSfit} we show the posterior distributions for the string tension, $G\mu$, and the intercommutation probability, $p$, together with the constraints from current (O3) LIGO/Virgo/KAGRA (LVK) data~\cite{KAGRA:2021kbb} and the design sensitivity of future LVK data~\cite{LIGOScientific:2014pky,LIGOScientific:2016fpe,LIGOScientific:2019vic}, assuming that the SMBH background is negligible. The best pure (super)string fit not in conflict with current LVK data has
\be
    G\mu = 2\times 10^{-12} \,, \quad p = 6.3\times 10^{-3} \,, 
\ee
with $-2 \Delta \ell=1.5$ compared to the SMBH baseline model including environmental effects. 

Using a combined fit including the possibility of a SMBH background, the best fit is at
\bea
    &G\mu = 1.2\times 10^{-12} \,, \quad p = 3.3\times 10^{-3} \,, \\
    &p_{\rm BH}=0.42 \,, \quad f_{\rm ref}=24 \, {\rm nHz}\,, \quad \alpha=4.4 \,,  
\eea
with $-2 \Delta \ell=-0.7$. Each model individually can give a comparable fit, and indeed we find that a combined fit with significant contributions from both gives a noticeably better fit, although the improvement is not very significant statistically. A significant contribution from SMBH binaries also results in a non-negligible probability of finding a candidate event at $f \sim 4$~nHz, which in this case is 5\%.

\subsubsection{Phase transitions}

Cosmological first-order phase transitions generate GWs in bubble collisions and by motions of inhomogeneities in the fluid~\cite{Witten:1984rs,Hogan:1986qda}. Given the large amplitude of the NANOGrav signal, we focus on very strong transitions that accommodate this feature naturally. We assume that the energy of the transition dominates over the background, so that its strength  $\alpha \gg 1$, in which case its value does not play an important role in the calculation of the signal. It was recently shown in~\cite{Lewicki:2022pdb} that in such a case the energy budget of the transition does not play an important role, as fluid-related sources~\cite{Kamionkowski:1993fg,Hindmarsh:2015qta,Hindmarsh:2016lnk,Hindmarsh:2017gnf,Ellis:2018mja,Cutting:2019zws,Hindmarsh:2019phv,Ellis:2020awk,Nakai:2020oit} and the contributions from bubble collisions~\cite{Kosowsky:1992vn,Cutting:2018tjt,Ellis:2019oqb,Lewicki:2019gmv,Cutting:2020nla,Lewicki:2020jiv,Giese:2020znk,Ellis:2020nnr,Lewicki:2020azd} behave very similarly.

\begin{figure}
\includegraphics[width=0.33\columnwidth]{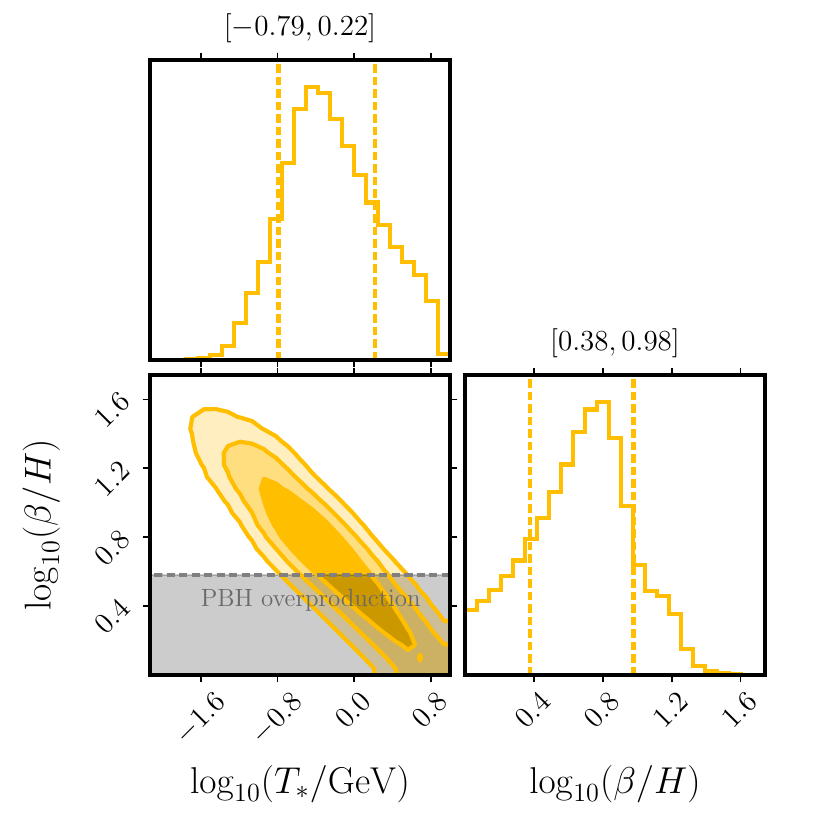}
\includegraphics[width=0.71\columnwidth]{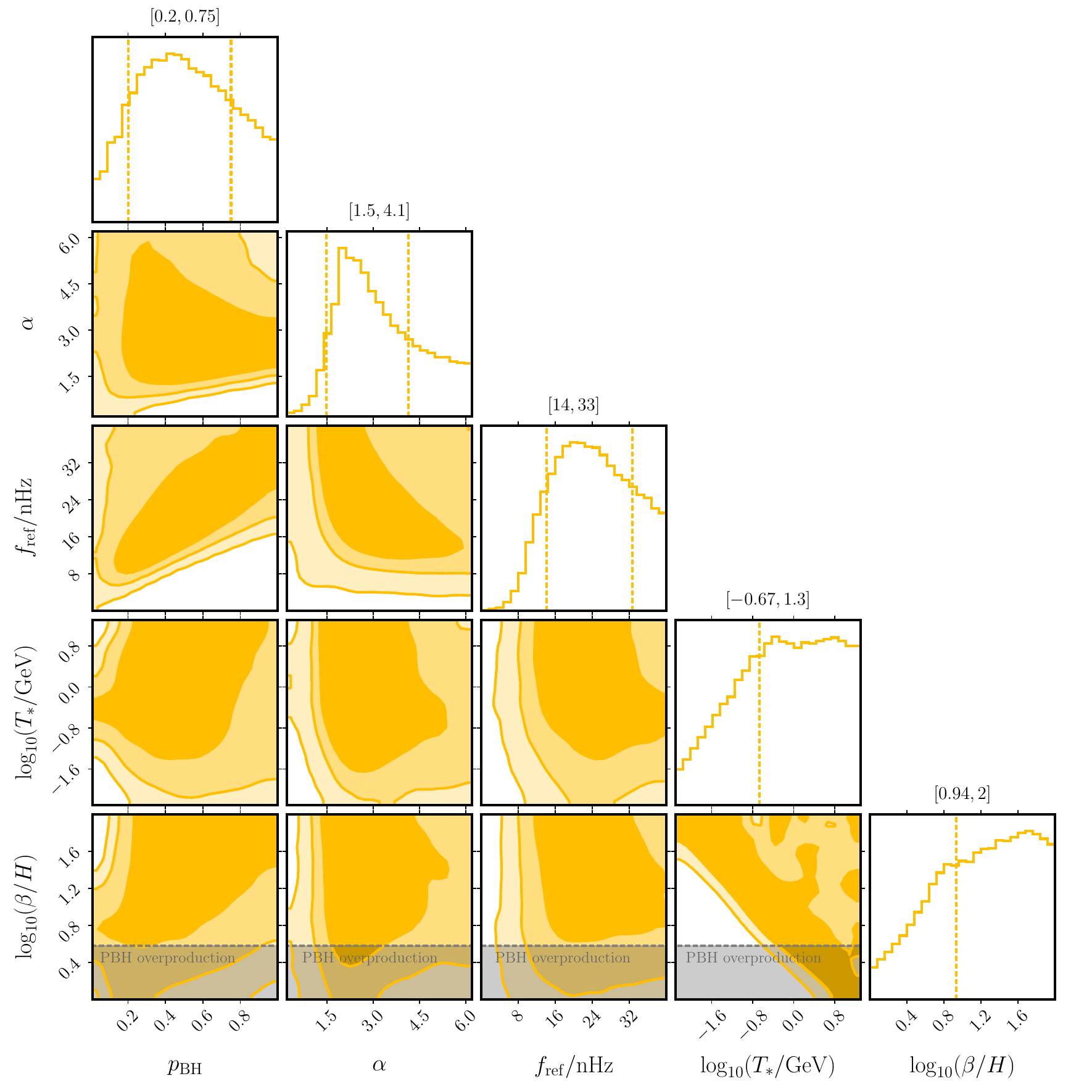}
\caption{\textit{\textbf{Left panel:}} The posterior probability distribution of the phase transition fit to the NG15 data. The constraint from the production of PBHs is shown in grey. \textit{\textbf{Right panel:}} Same as the left panel but including also SMBH binaries with environmental effects.
}
\label{fig:PTfit}
\end{figure}

The resulting GW spectrum is a broken power-law that may be parametrized as 
\be \label{eq:Omega_PT}
    \Omega_{\textrm{GW}}(f,T_*) = \left[\frac{\beta}{H}\right]^{-2} \frac{A (a+b)^c S_H(f,f_H(T_*))}{\left(b \left[\frac{f}{f_p}\right]^{\!-\frac{a}{c}} + a \left[\frac{f}{f_p}\right]^{\frac{b}{c}}\right)^{\!c}} \,,
\ee
where $\beta$ is the timescale of the transition, $T_*$ is the temperature reached after the transition,
$a=b=2.4$, $c=4$, $A=5.1\times 10^{-2}$ and the peak frequency $f_p = 0.7 f_H(T_*) \beta/H$~\cite{Lewicki:2022pdb}. The function
\be \label{eq:SH}
    S_H(f,f_H) = \left(1+\left[ \frac{\Omega_{\rm CT}(f)}{\Omega_{\rm CT}(f_H)}\right]^{-\frac{1}{\delta}}\left[\frac{f}{f_H}\right]^{\frac{a}{\delta}}\right)^{\!-\delta}
\ee
models the transition of the spectrum to the causality tail at scales larger than the horizon at the transition time~\cite{Caprini:2009fx}. The parameter $\delta$ determines how quickly the spectrum evolves towards the causality tail and could be determined by a simulation of the strong phase transition taking expansion into account. We fix $\delta = 1$, having verified its value does not have a large impact on the results. 

In Fig.~\ref{fig:PTfit} we show the posteriors of phase transition fit to the NG15 data with and without including also the background from SMBH binaries. The best fit to a pure phase transition signal is for
\be \label{eq:PTbestfit}
    T_* = 0.34 \, {\rm GeV}\,, \quad \beta/H = 6.0\,, 
\ee
resulting in $-2 \Delta \ell =-2.3$ compared to the SMBH baseline model. Including SMBHs with environmental effects, we find that the best fit is dominated by the phase transition signal with $p_{\rm BH} \approx 0$. 

Searching for points capable of providing the candidate event at $4$\,nHz by requiring its probability to be $\geq 5\%$ we find the best-fit point
\bea
    &T_*=2.6 \,{\rm GeV}\,, \quad \beta/H =4.0 \\
    &p_{\rm BH}=0.28\, , \; f_{\rm ref}=20 \, {\rm nHz}\, , \; \alpha=2.6\, ,  
\eea
which gives $-2 \Delta \ell =0.2$. 

Strong first-order phase transitions can also lead to the formation of primordial black holes (PBHs) if they are sufficiently slow~\cite{Hawking:1982ga, Kodama:1982sf, Liu:2021svg, Lewicki:2023ioy, Gouttenoire:2023naa}. In Figs.~\ref{fig:PTfit} we show the constraint $\beta/H \gtrsim 4$ from overproduction of PBHs~\cite{Lewicki:2023ioy}.\footnote{A slightly stronger bound, $\beta/H \gtrsim 6$, was found in~\cite{Gouttenoire:2023naa}.} We see that most of the 1$\sigma$ region of the fit including only the phase transition signal is allowed by this constraint. However, as pointed out in~\cite{Gouttenoire:2023bqy}, it is interesting that the NG15 fit is compatible with a large PBH abundance. These PBHs would be around the stellar mass range, where their abundance is constrained by optical lensing~\cite{EROS-2:2006ryy, Macho:2000nvd, Zumalacarregui:2017qqd,Gorton:2022fyb,Petac:2022rio,DeLuca:2022uvz}, GW observations~\cite{Raidal:2017mfl, Ali-Haimoud:2017rtz, Raidal:2018bbj, Vaskonen:2019jpv,  Hutsi:2020sol, Franciolini:2022tfm} and accretion~\cite{Ricotti:2007au, Horowitz:2016lib, Ali-Haimoud:2016mbv, Poulin:2017bwe, Hektor:2018qqw, Hutsi:2019hlw, Serpico:2020ehh} (for a review see e.g.~\cite{Carr:2020gox}). Nevertheless, a subdominant PBH abundance, $\mathcal{O}(0.1)$\% of DM, might be related to the BH binary mergers detected by LVK~\cite{Vaskonen:2019jpv,DeLuca:2020jug,Raidal:2017mfl,Raidal:2018bbj,Ali-Haimoud:2017rtz,DeLuca:2020qqa,Hutsi:2020sol,Franciolini:2021tla,Clesse:2020ghq,Franciolini:2022tfm} and can be further probed with future GW observatories~\cite{DeLuca:2021hde, Pujolas:2021yaw, Urrutia:2023mtk, Franciolini:2023opt}.
Assuming all the energy of the transition decays into the visible sector the only constraint is that the temperature reached after the transition $T_* > 5$\, MeV~\cite{Allahverdi:2020bys}. 
Models realising phase transitions at the low temperatures favoured by the PTA data would be subject to constraints from collider experiments. 
The simplest way to avoid these would be to assume they occur in decoupled dark sectors. However, in such cases, they would instead be subject to dark radiation constraints~\cite{Nakai:2020oit}.
The task of finding viable particle physics models realising such transitions lies beyond the scope of this thesis.

\subsubsection{Domain walls}

\begin{figure}
    \includegraphics[width=0.33\columnwidth]{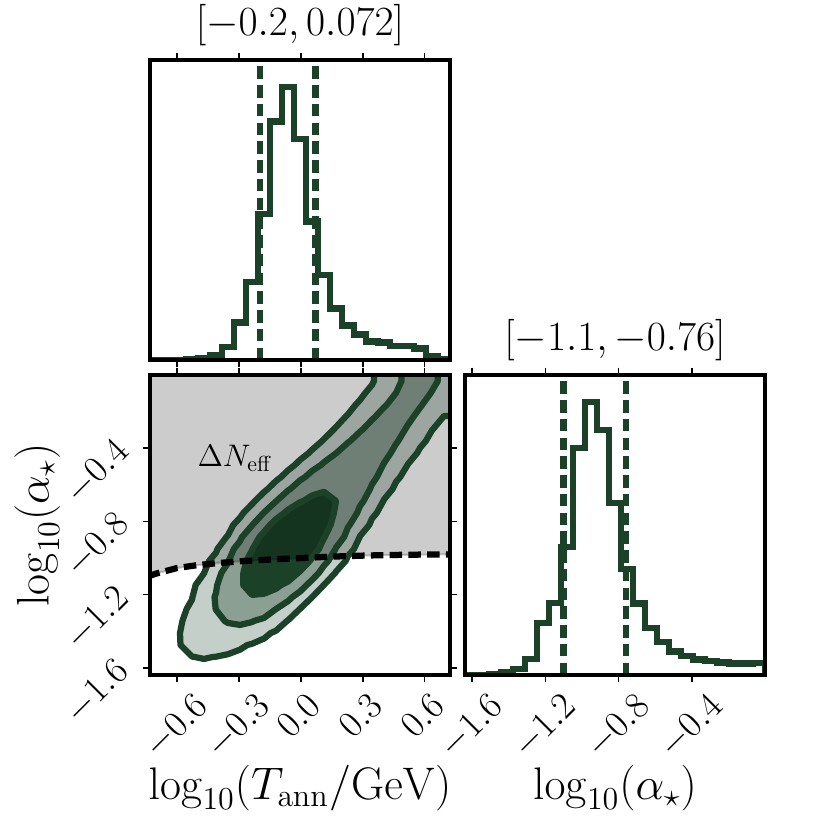}
    \includegraphics[width=0.71\columnwidth]{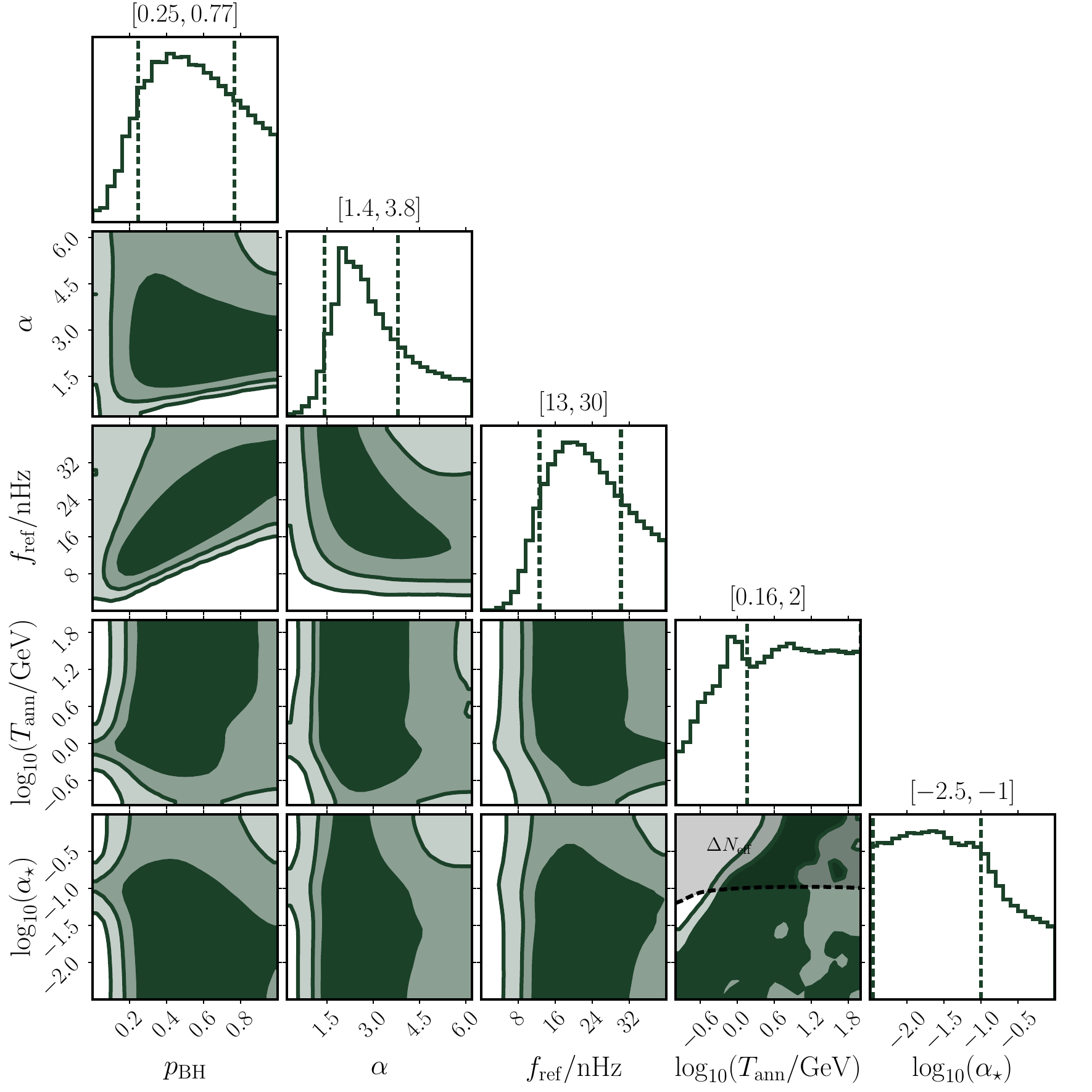}
    \caption{\textit{\textbf{Left panel:}} The posterior probability distribution of the DW fit to the NG15 data. The dashed line indicates the $\Delta N_{\rm eff}$ bound, which constrains the possibility of DWs annihilating completely into dark radiation. \textit{\textbf{Right panel:}} Same as the left panel but including also SMBH binaries with environmental effects.}
    \label{fig:new_DW}
\end{figure}

Domain walls (DWs)~\cite{Vilenkin:1984ib} are topological defects produced when a discrete symmetry in some BSM scenario is broken after inflation. During the scaling regime when the DW network expands together with its surroundings, the energy density is $\rho_{\rm DW}=c\, \sigma H$~\cite{Hiramatsu:2013qaa, Kibble:1976sj}, where $\sigma$ is the tension of the wall and $c = \mathcal{O}(1)$ is a scaling parameter. DWs emit GWs until they annihilate at a temperature $T = T_{\rm ann}$~\cite{Gelmini:1988sf,Coulson:1995nv,Larsson:1996sp,Preskill:1991kd}. The peak frequency of the resulting GW spectrum is given by the horizon size at the time of DW annihilation, $f_p = f_H(T_{\rm ann})$, and at frequencies $f\gg f_p$ the spectrum scales as $f^{-1}$. We approximate the GW spectrum at the formation time $T=T_{\rm ann}$ as~\cite{Hiramatsu:2013qaa,Ferreira:2022zzo}
\be
    \Omega_{\rm DW}(f,T_{\rm ann})
    = \frac{3 \epsilon \alpha_*^2}{8\pi} \!\left( \frac14\! \left[ \frac{\Omega_{\rm CT}(f_p)}{\Omega_{\rm CT}(f)}\right]^{\frac{1}{\delta}} \!\!+ \frac34\! \left[\frac{f}{f_p}\right]^{\!\frac{1}{\delta}}\right)^{\!\!-\delta} ,
\ee
where $\epsilon = \mathcal{O}(1)$ is an efficiency parameter and $\alpha_* \equiv \rho_{\rm DW}(T_{\rm ann})/\rho_r(T_{\rm ann})$ is the energy density in the domain walls relative to the radiation energy density $\rho_r$ at the annihilation moment. Following the results of numerical simulations~\cite{Hiramatsu:2013qaa}, we fix $\epsilon=0.7$ and $\delta = 1$. Changing the value of $\epsilon$ would simply rescale $\alpha_*$ and, in the same way as in the phase transition case, the value of $\delta$ does not affect significantly the results displayed.

The parameters of the DW model are the relative energy density in DWs, $\alpha_*$, and the temperature at which they annihilate, $T_{\rm ann}$. Since the DWs can constitute a big fraction of the total energy density, it is necessary to check that their annihilation respects the constraints imposed by BBN and the CMB.  If the DWs annihilate completely into dark radiation, the energy density can be expressed as the equivalent number of neutrino species~\cite{Ferreira:2022zzo}: $\Delta N_{\rm eff} = \rho_{\rm DW}(T_{\rm ann})/\rho_{\rm \nu}(T_{\rm ann}) = 13.6 \, g_{*}(T_{\rm ann})^{-\frac{1}{3}}\, \alpha_*$, which is constrained by BBN ($\Delta N_{\rm eff} < 0.33$)~\cite{Fields:2019pfx} and CMB ($\Delta N_{\rm eff} < 0.3$)~\cite{Planck:2018vyg,Ramberg:2022irf}. On the other hand, if the DWs annihilating into SM particles, BBN imposes $T_{\rm ann} > (4-5) \, {\rm MeV}$~\cite{Hasegawa:2019jsa} and CMB $T_{\rm ann} > 4.7 \, {\rm MeV}$~\cite{deSalas:2015glj} (see~\cite{Allahverdi:2020bys} for a review).

We have performed scans over the parameters of the DW model to fit the NG15 data. The posterior probability distributions for the DW model without a SMBH contribution are shown in Fig.~\ref{fig:new_DW}, and the best fit is at
\be
    T_{\rm ann} = 0.85\,{\rm GeV}\,, \quad \alpha_* = 0.11 \,, 
\ee
and has $-2 \Delta \ell=-3.1$ compared to the SMBH baseline model. These parameter values are allowed both if the DWs annihilate into SM particles. If they instead annihilate into dark radiation, the best fit is in tension with $\Delta N_{\rm eff}$ bound. However, the $1\sigma$ region of the posterior distribution extends to smaller values of $\alpha_*$ that are allowed by the $\Delta N_{\rm eff}$ bound.

The best fit of the combined model is dominated by the DW contribution with $p_{\rm BH} \approx 0$. Imposing a lower bound on $p_{\rm BH}$ from Fig.~\ref{fig:Pmin}, the best fit is 
\bea \label{FWSMBHfit}
    &T_{\rm ann} = 1.12 \, {\rm GeV}\,, \quad \alpha_* = 0.074\,, \\
    &p_{\rm BH}=0.51\, , \; f_{\rm ref} = 22 \, {\rm nHz}\, , \; \alpha = 2.6\, ,
\eea
with $-2 \Delta \ell=-0.5$. 

Similarly to first-order phase transitions, the annihilation of the DW network can lead to the copious formation of PBHs~\cite{Ferrer:2018uiu,Gelmini:2023ngs}. The PBH formation may be so efficient that it excludes the parameter region preferred by the NG15 data~\cite{Gouttenoire:2023ftk}. However, further studies are needed to confidently estimate the PBH formation from DW networks.

\subsubsection{Scalar-induced GWs}
\label{sec:SIGWth}

Primordial scalar curvature fluctuations can source a SGWB at the second order of perturbation theory~\cite{Tomita:1975kj,Matarrese:1993zf,Acquaviva:2002ud,Mollerach:2003nq,Ananda:2006af,Baumann:2007zm,Domenech:2021ztg,Yuan:2021qgz}. As we described in the previous section the most stringent constraint on a SIGW scenario for PTAs arises from the overproduction of PBHs associated with large peaks in the primordial curvature power spectrum~\cite{Dandoy:2023jot, Franciolini:2023pbf, Franciolini:2023wjm}. 

\begin{figure}
    \includegraphics[width=0.37\textwidth]{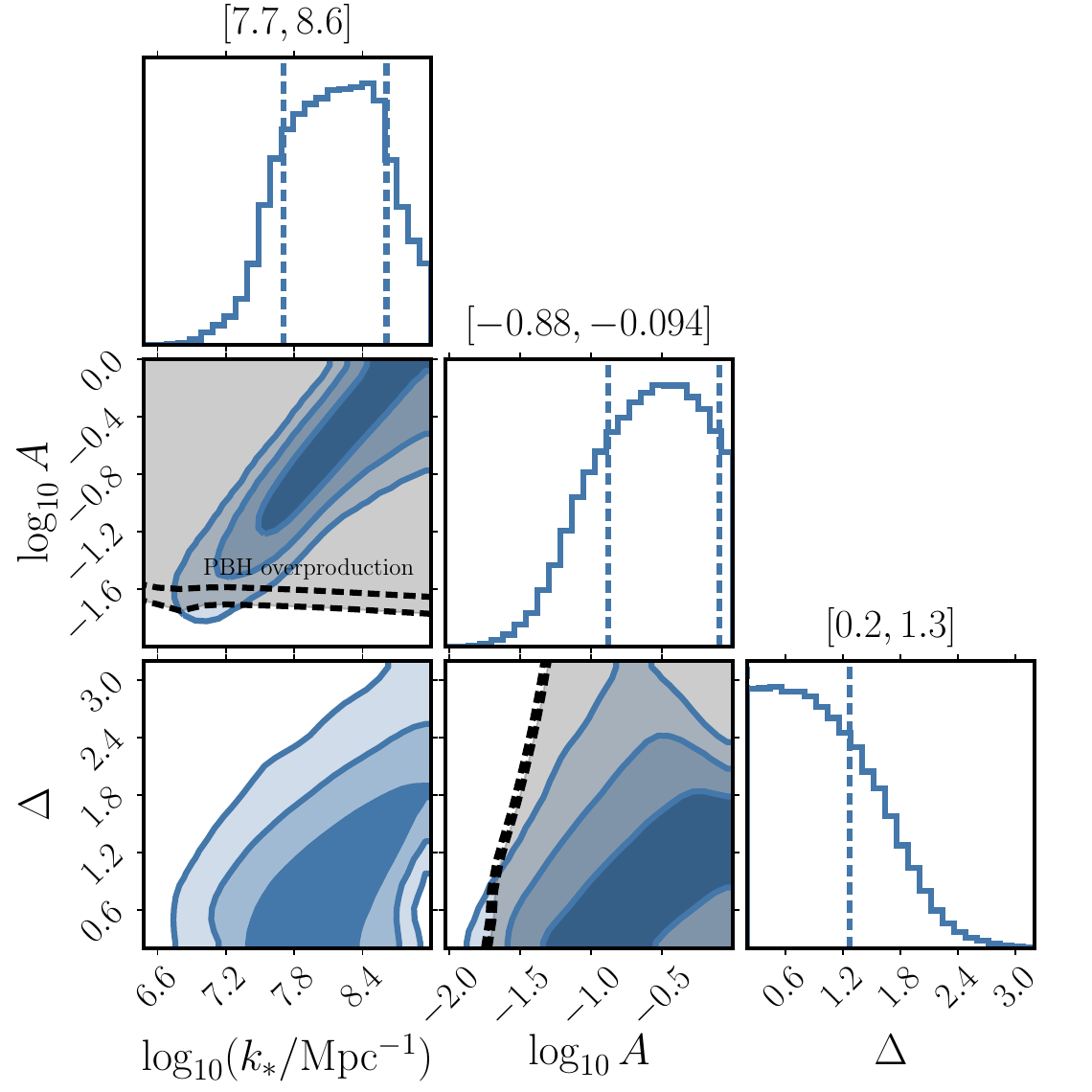}
     \includegraphics[width=0.69\textwidth]{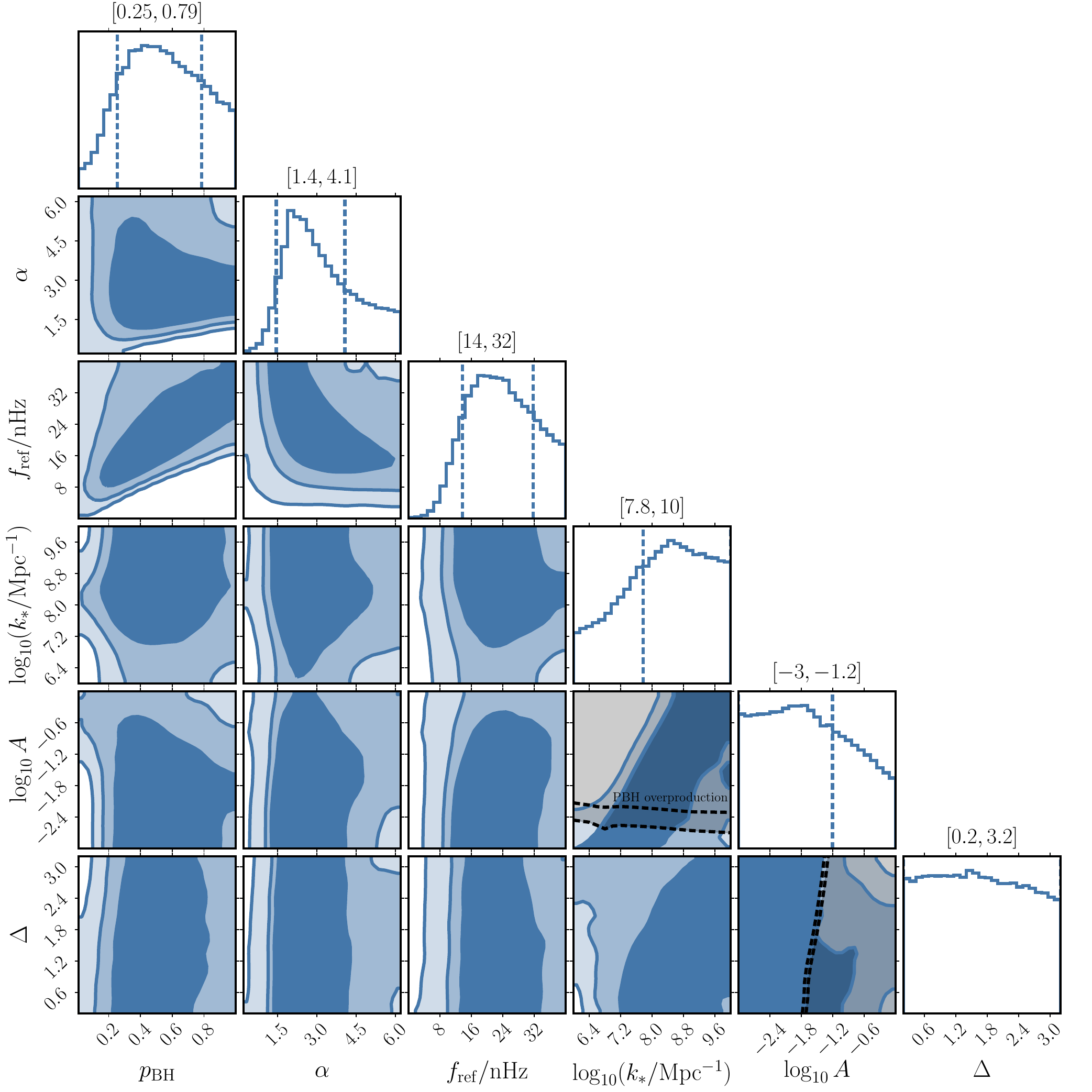}
    \caption{\textit{\textbf{Left panel:}} The posterior probability distribution of the SIGW fit to the NG15 data. The black shaded region indicates where $f_{\rm PBH}>1$, while the black dashed lines bracket the 1$\sigma$ ranges in the marginalised direction.\textit{\textbf{Right panel:}} Same as the left panel but including also SMBH binaries with environmental effects}
    \label{fig:SIGWonly_posteriors}
\end{figure}

The present-day SIGW background would have a spectrum given by Eq.~\eqref{eq:OmegaGWtoday}, where~\cite{Inomata:2019yww,DeLuca:2019ufz,Yuan:2019fwv,Domenech:2020xin}
\begin{align}
\Omega_{\rm GW}(T) 
= \frac{1}{3} 
\int_1^\infty \td t \int_{-1}^1 \td s 
\frac{\overline{J^2(u,v)}}{(u v)^2} 
{\cal P}_\zeta (v k)
{\cal P}_\zeta (u k) ,
\label{eq:P_h_ts}
\end{align}
where $u=(t + s)/2$, $v=(t - s)/2$ and the transfer function $\overline{J^2 (u,v)}$ depends on the cosmological background~\cite{Kohri:2018awv,Espinosa:2018eve}.
Including higher-order corrections to the SIGW signal that arise in the event of non-Gaussian curvature perturbations are reported in App.~\ref{app:NGSIGW}. 

For making a comparison with other possible explanation, here we assume that primordial curvature spectra can be characterized by a log-normal (LN) shape of the form
\be
    \mathcal{P}_{\zeta}(k)
    = \frac{A}{\sqrt{2\pi}\Delta} \, \exp\left( -\frac{1}{2\Delta^2} \ln^2(k/k_*) \right)\,.
\ee
The SIGW model therefore has three relevant parameters: the characteristic scale $k_*/{\rm Mpc}^{-1}$, the peak amplitude $A$, and the width $\Delta$. The distributions of their posterior values and two-parameter correlations for fits to the NG15 data assuming only scalar-induced second-order GWs are shown in Fig.~\ref{fig:SIGWonly_posteriors}. The best-fit values of the SIGW model parameters are:
\be\label{eq:bestfitSIGW}
    \log_{10} (k_*/{\rm Mpc}^{-1})=7.7\,, \quad \log_{10} A=-1.2 \,, \quad \Delta=0.21 
\ee
for which $-2 \Delta \ell = -2.1$.

We also show in Fig.~\ref{fig:SIGWonly_posteriors} the regions of parameter space where PBHs are overproduced. To compute the abundance more precisely we include the effects of the QCD phase transition and the shape of the power spectrum in the parameters, following Refs.\,\cite{Musco:2020jjb,Franciolini:2022tfm,Musco:2023dak}. As already pointed out in \cite{Franciolini:2023pbf}, the tension between the NANOGrav data and a scenario with the SIGW signal alone can be relaxed in models where large non-Gaussianities suppress the PBH abundance, or if the PBH formation takes place during non-standard cosmological phases deviating from radiation domination \cite{Balaji:2023ehk,Liu:2023pau}. Our results differ from those in Refs.\,\cite{Inomata:2023zup, Figueroa:2023zhu, Yi:2023mbm, Firouzjahi:2023lzg, You:2023rmn}. This discrepancy comes from several limitations of the analyses in these papers, such as the omission of critical collapse and the nonlinear relationship between curvature perturbations and density contrast, the adoption of a different value for the threshold (independently from the curvature power spectrum), and the use of a different window function, without properly recomputing the threshold values \cite{Young:2019osy}.
FOr the full fit, including the combination with SMBH binaries, the preferred values of the SIGW model parameters are similar to those in the SIGW-only fit, but their distributions are significantly broader, as was to be expected. The best-fit parameters of the combined model are
\bea
    &\log_{10} (k_*/{\rm Mpc}^{-1}) = 7.6 \,, \quad \log_{10} A = -1.2 \,, \quad \Delta = \, 0.30 \,, \\
    &p_{\rm BH}=0.06\,, \quad  f_{\rm ref}=26 \, {\rm nHz}\,, \quad \alpha=4.1\,.
\eea

The best-fit parameters describing the SIGW sector in the mixed scenario are very similar to those reported in Eq.~\eqref{eq:bestfitSIGW}, where SIGWs alone are assumed to explain PTA observations. This is because the SIGW model provides a better fit, and dominates the signal where the likelihood peaks. 
However, neither of the two models dominates over the other beyond the $\sim 2\sigma$ level, maintaining the large degeneracy of most parameters in the posterior distributions.

Imposing a lower bound on $p_{\rm BH}$ forcing the SMBHs model to produce the candidate event at $4$nHz with probability larger than $5\%$, we find instead that the astrophysical channel is required to dominate, and the best fit parameters are
\bea
    &\log_{10} (k_*/{\rm Mpc}^{-1}) = 6.3 \,, 
    \quad \log_{10} A = -2.2 \,, 
    \quad \Delta = \, 0.21 \,, \\
    &p_{\rm BH}=0.70\,, 
    \quad  f_{\rm ref}=24 \, {\rm nHz}\,, 
    \quad \alpha=3.8\,.
\eea
with a maximum likelihood similar to the SMBHs-only scenario (with environmental effects included), resulting in $-2 \Delta \ell=-0.4$. 

\subsubsection{First-order GWs}
\label{sec:FOGWs}

Another mechanism for generating a cosmological SGWB is during inflation via first-order GWs (FOGWs)~\cite{Grishchuk:1974ny,Starobinsky:1979ty}. We consider canonically normalized single-field slow-roll models, which predict a primordial tensor power spectrum that, in the relevant range of frequencies between the CMB and PTA scales, can be approximated by a pure power-law~\cite{NANOGrav:2023hvm}:
\be \label{eq:FO}
    \Omega_{\rm GW}(f) = \frac{r A_s }{24}\left(\frac{f}{f_{\rm CMB}}\right)^{n_t}\mathcal{T}(f) \, ,
\ee
where $f_{\rm CMB} = 7.7\times 10^{-17}$\,Hz, $A_s = 2.1\times 10^{-9}$~\cite{Planck:2018vyg}, $n_t$ is the tensor spectral index and $r$ is the tensor-to-scalar ratio. The transfer function that connects the radiation-dominated era to reheating can be approximated as~\cite{Kuroyanagi:2014nba,Kuroyanagi:2020sfw}
\be
    \mathcal{T}(f) \approx \frac{\Theta\left(f_{\rm {end }}-f\right)}{1-0.22\left(f / f_{\mathrm{rh}}\right)^{1.5}+0.65\left(f / f_{\mathrm{rh}}\right)^2} \,,
\ee
where $f_{\rm end}$ and $f_{\rm rh} \leq f_{\rm end}$ correspond, respectively, to the end of inflation and to the end of reheating. For simplicity, we assume that the cut-off scale is $f_{\rm end} \gg f_{\rm rh} = f_{H}(T_{\rm rh})$. In this way, the analysis is independent of the choice of the cutoff. 

\begin{figure}
    \includegraphics[width=0.38\columnwidth]{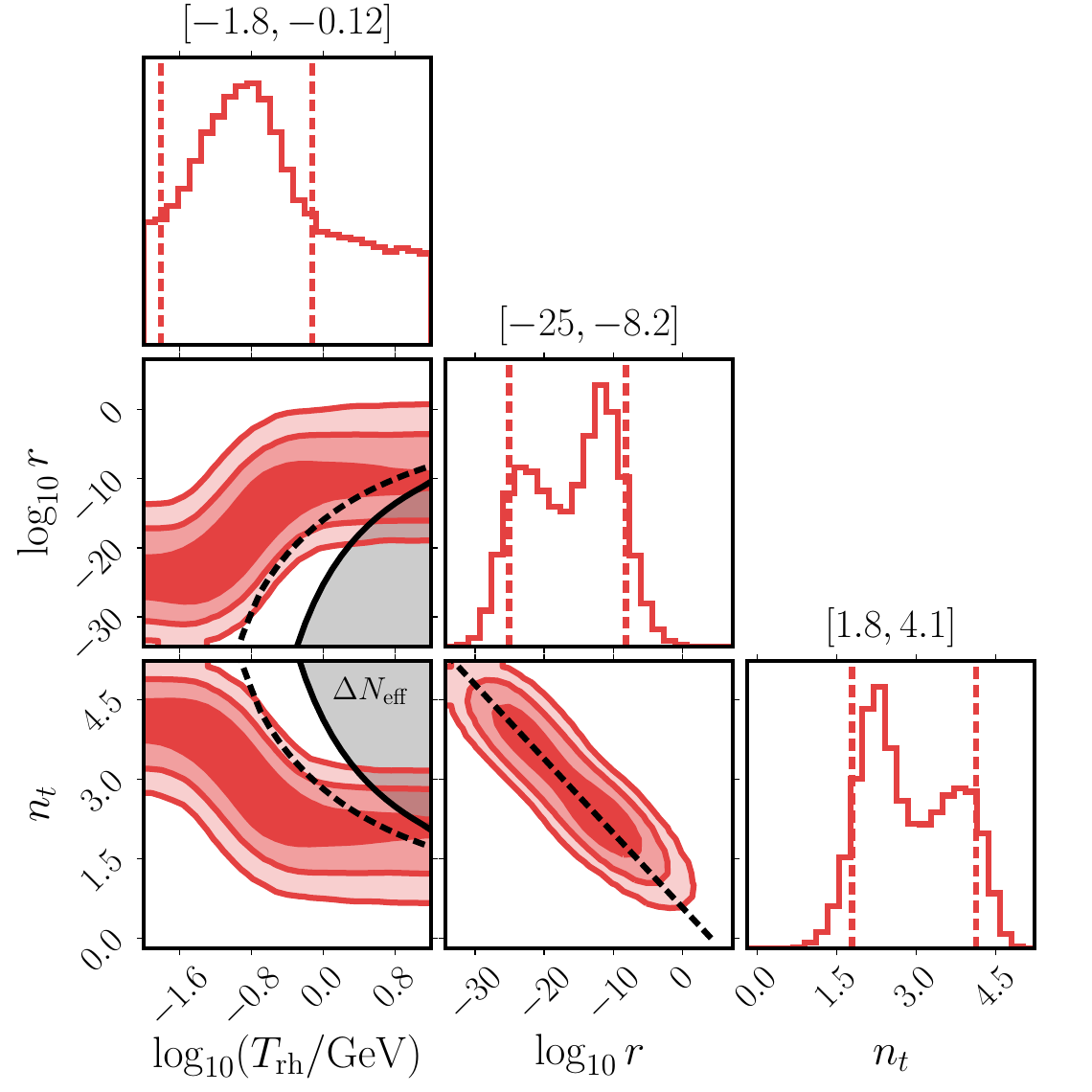}
    \includegraphics[width=0.7\columnwidth]{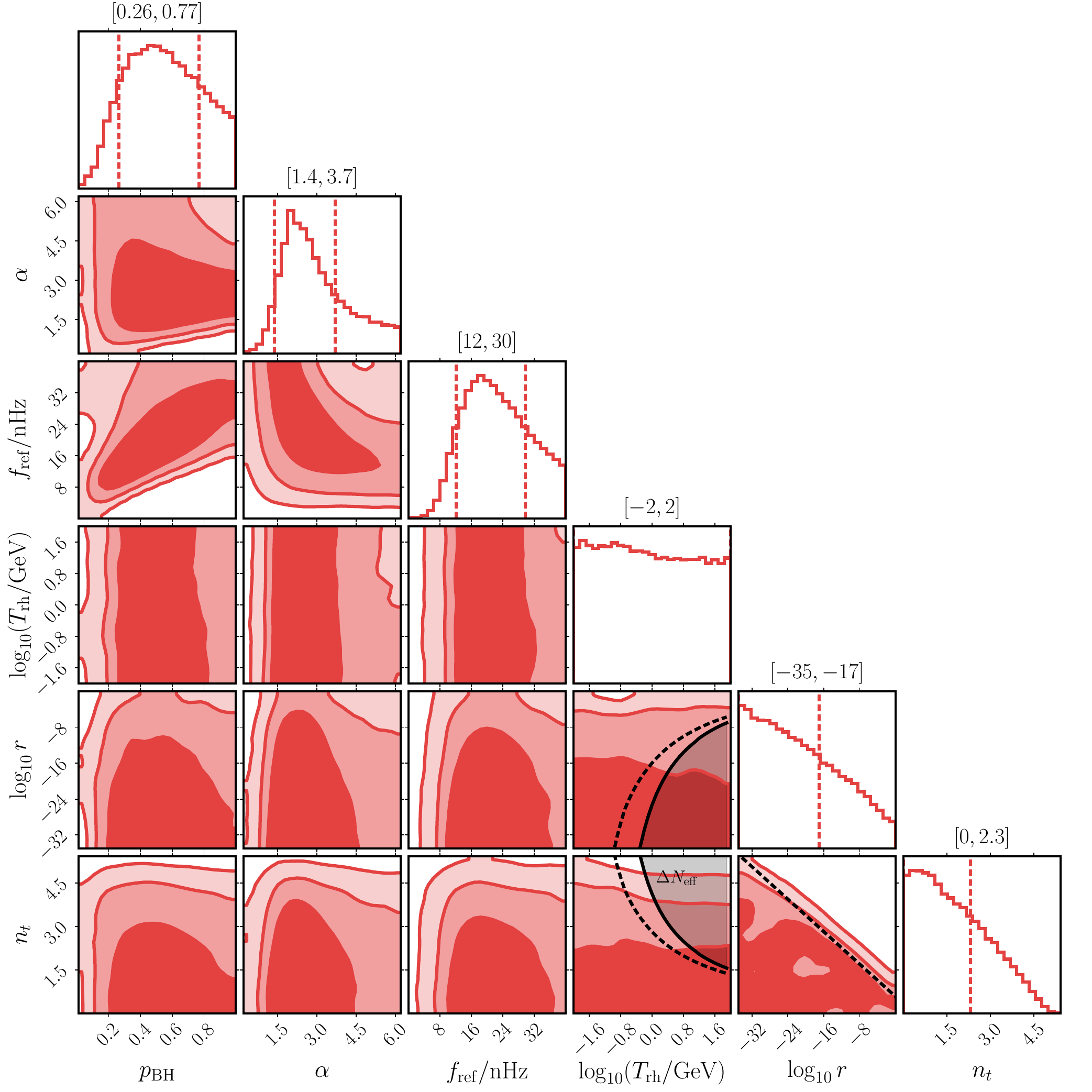}
    \caption{\textit{\textbf{Left panel:}} The posterior probability distribution of the FOGW fit to the NG15 data. The black-shaded region shows the constraints by $\Delta N_{\rm eff}$  bound in Eq.~\eqref{eq:deltaN}, assuming $f_{\rm end} = f_{\rm rh}$ (solid line) and $f_{\rm end} =10 f_{\rm rh}$ (dashed line). We also show the $r-n_t$ correlation \eqref{eq:nt-rcorr} in the corresponding panel. \textit{\textbf{Right panel:}} Same as the left panel but including also SMBH binaries with environmental effects. }
    \label{fig:FOGWonly_posteriors}
\end{figure}

The FOGW scenario is parametrized by the inflationary parameters $n_t$ and $r$, and the reheating temperature $T_{\rm rh}$. As seen in the left panel of Fig.~\ref{fig:FOGWonly_posteriors}, their values are very tightly correlated, and their best-fit values are  
\be\label{eq:figwalone_bestfit}
    \log_{10}{(T_{\rm rh}/\textrm{GeV})} = -0.67 \,, 
    \quad  \log_{10} r = -14 \,, 
    \quad  n_{t} \, = \, 2.6
\ee
for which  $-2 \Delta \ell=-2.0$. 

A constraint on the possible values of these parameters is given by the SGWB contribution to the radiation energy budget characterized by $\Delta N_{\rm eff}$, i.e. its contribution to the effective number of relativistic species. $\Delta N_{\rm eff}$ is constrained by BBN and CMB probes, and an approximate upper limit is given by $\Delta N_{\rm eff} \lesssim0.3$~\cite{Planck:2018vyg,Fields:2019pfx}. The GW spectrum, integrated from BBN scales to the cutoff frequency $f_{\rm end}$, must not exceed an upper limit that is set by the allowed amount of extra relativistic degrees of freedom at the time of BBN and recombination. This contribution is given~\cite{Smith:2006nka,Boyle:2007zx}
\be \label{eq:deltaN}
    1.8\times 10^{5}\int_{f_{\mathrm{BBN}}}^{f_{\mathrm{end}}} \frac{\mathrm{d} f}{f}  \Omega_{\mathrm{GW}}(f)h^2 \lesssim  \Delta N_{\mathrm{eff}}^{\max }
\ee
where we set $f_{\rm BBN}\simeq10^{-12}$~Hz, which is the frequency at horizon re-entry of tensor mode around the onset of BBN
at $T \simeq 10^{-4}$ GeV\,\cite{Caprini:2018mtu}. 
For a slope compatible with the NANOGrav data,  the signal cannot extend to excessively high frequencies, as doing so would violate constraints imposed by BBN\,\cite{Maggiore:1999vm,Pagano:2015hma}. 
For these reasons, we limit our analysis to the range $T_{\rm rh}\in [0.01,10]$ ${\rm GeV}$. 
As a consequence, this scenario does not produce observable signatures in experiments searching for GWs at higher frequencies than those observed with PTAs (see Section~\ref{sec:comp} for more details).
As can be seen from Fig.\ref{fig:FOGWonly_posteriors}, if the reheating temperature is $\lesssim 0.3$ GeV the preferred signal changes the tilt moving from $n_t$ to $n_t-2$. This is because, for such values of the temperature, the cut-off of the signal is at a lower frequency than the last NANOGrav bin, and the GW spectrum in the PTA band is composed of tensor modes that re-entered the horizon during reheating after inflation.
When deriving the constraint from the bound on $\Delta N_{\rm eff}$, we find that $n_t$ and $r$ are strongly correlated, and the equation that relates these quantities can be approximated by
\be \label{eq:nt-rcorr}
    n_t = -0.14\log_{10} r +0.58.
\ee
Using this last relation we constrain the parameter space as shown in Figs.~\ref{fig:FOGWonly_posteriors}.

Our results are in perfect agreement with Refs.~\cite{NANOGrav:2023hvm, Vagnozzi:2023lwo}.
The best fit of the combined models results in $-2 \Delta \ell=-1.9$ with $p_{\rm BH}\simeq 0$.
Thus, the FIGW contribution is dominant, and the best-fit parameter of the FIGW sector coincides with those reported in Eq.~\eqref{eq:figwalone_bestfit}, assuming no SMBH contribution. 
\\Imposing a lower bound on $p_{\rm BH}$ to produce the candidate SMBH merger at $4$nHz, 
we find instead
\bea
    &\log_{10}{(T_{\rm rh}/\textrm{GeV})} = 1.5 \,, \quad \log_{10} r = -9.1 \,, \quad n_{t} \, = \, 1.46 \, , \\
    &p_{\rm BH}=0.35\, , \quad  f_{\rm ref}=16 \, {\rm nHz} \,, \quad \alpha=4.7\, .
\eea
with resulting in $-2 \Delta \ell=-0.1$. 

\subsubsection{``Audible" axions}
\label{sec:axions}

In models in which an axion or an axion-like particle has a weak coupling to the SM, thanks to a coupling to a light dark photon, a SGWB can be produced~\cite{Machado:2018nqk, Machado:2019xuc,Co:2021rhi,Fonseca:2019ypl,Chatrchyan:2020pzh,Ratzinger:2020oct,Eroncel:2022vjg}.
After the axion begins to roll, it induces a tachyonic instability for one of the dark photon helicities, causing vacuum fluctuations to grow exponentially. This effect generates time-dependent anisotropic stress in the energy-momentum tensor, which ultimately sources GWs. In this scenario, the condition of having a coupling with a dark photon and not with the SM photon is due to the fact that the latter undergoes rapid thermalization that would destroy the conditions required for exponential particle production.\footnote{GWs can also be produced in kinetic misalignment scenarios (in which the axion field has a nonzero initial velocity) also in the absence of dark photons. As shown in \cite{Machado:2019xuc} the resulting signal in this scenario is, however, in general too small to be observable in the foreseeable future.} The GW formation ends when the tachyonic band closes at temperature~\cite{Machado:2018nqk}
\be
    T_* \approx \frac{1.2 \sqrt{m_a M_{\rm P}}}{g_*^{1/4} (\alpha \theta)^{2/3}} \,,
\ee
where $\alpha$ is the coupling with the dark photon, $\theta$ is the initial misalignment angle, and $m_a$ is the mass of the axion.
As shown in~\cite{Machado:2019xuc}, assuming that between the beginning of the axion oscillations and the end of the GW formation, the effective number of degrees of freedom does not change, the GW spectrum at temperature $T_*$ can be fitted by
\be \label{eq:OmegaAa}
    \Omega_{\mathrm{GW}}(f,T_*) = \frac{6.3 \left(\frac{f_a}{M_{\rm P}}\right)^4\left(\frac{\theta^2}{\alpha}\right)^{4/3} S_H(f,f_H(T_*))}{\left[\frac{f}{2.0 f_p}\right]^{\!-1.5}+\exp\!\left[12.9\left(\frac{f}{2.0 f_p}-1\right)\right]} \,,
\ee
where $f_a$ is the decay constant of the axion and the peak frequency of the spectrum is
\be
    f_p \approx 2.5 (\alpha \theta)^{4/3} f_H(T_*) \,.
\ee 
Compared to the fit in~\cite{Machado:2019xuc}, we have added the function $S_H(f,f_H)$ that accounts for the causality tail of the spectrum at frequencies $f<f_H(T_*)$ and is given by Eq.~\eqref{eq:SH} with $a=1.5$. We set the parameter $\delta$ in $S_H(f,f_H)$ to $\delta = 1$.
Typical values for the initial misalignment angle are of $\theta\sim\mathcal{O}(1)$ \,\cite{DiLuzio:2020wdo,Marsh:2015xka} and Eq.~\eqref{eq:OmegaAa} holds only for $\theta \sim 1$ and $\alpha \geq 10$, the range of values where the particle production process is efficient \cite{Agrawal:2017eqm}. For simplicity, we fix $\theta=1$  and $\alpha=20$ in the analysis. 

\begin{figure}
    \includegraphics[width=0.33\columnwidth]{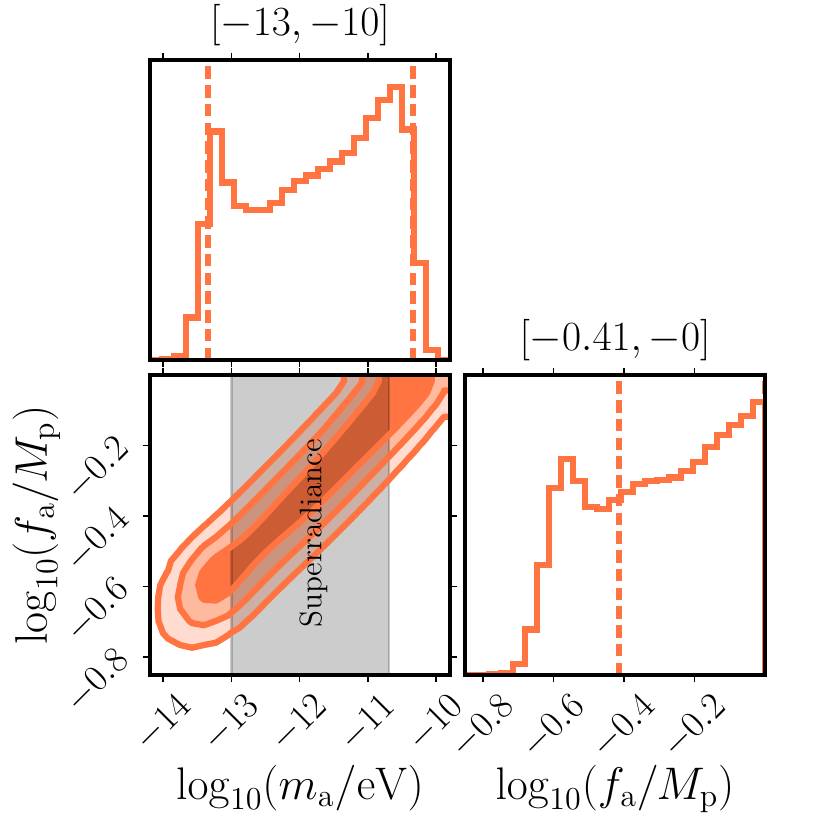}
    \includegraphics[width=0.71\columnwidth]{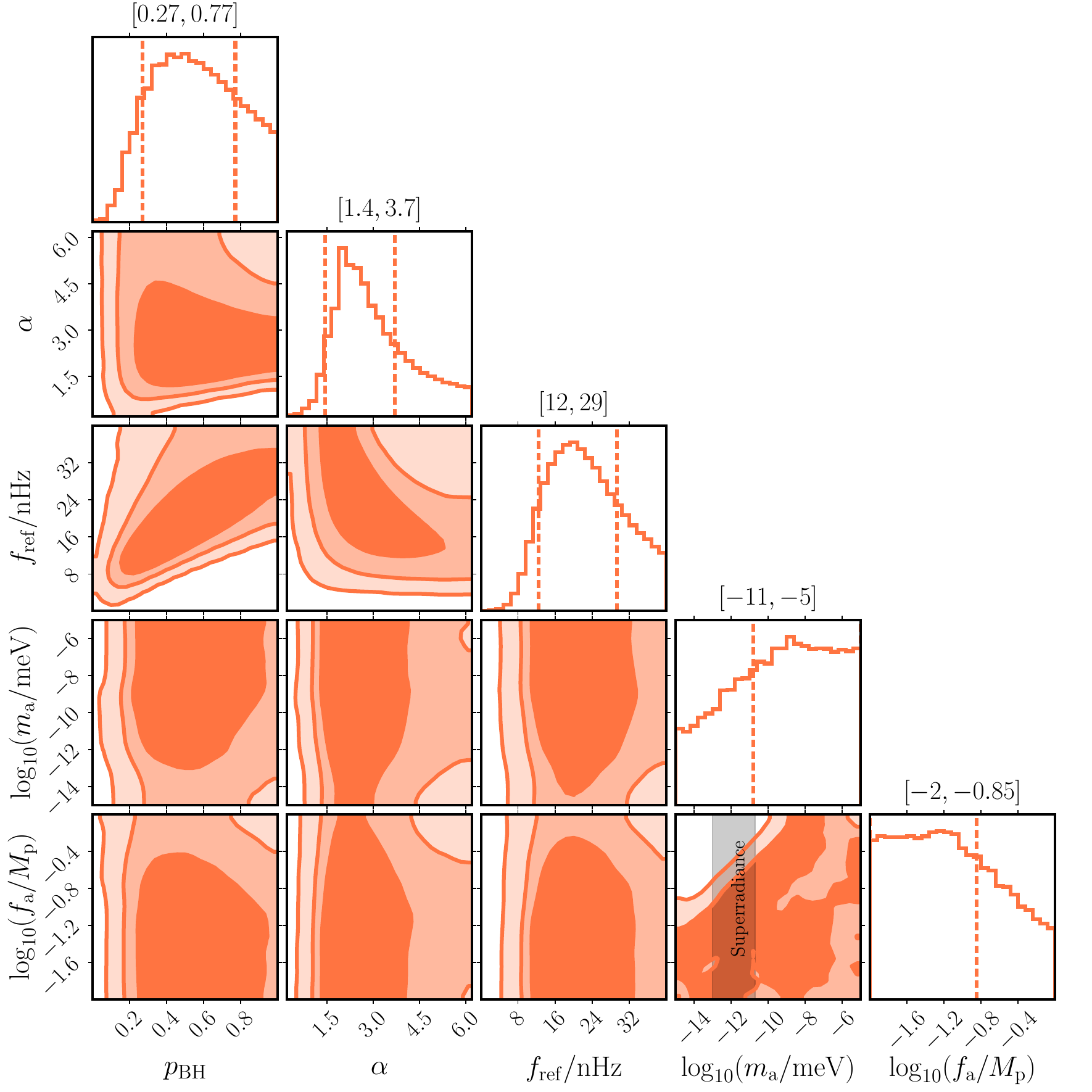}
    \caption{\textit{\textbf{Left panel:}} The posterior probability distribution of the ``audible" axion fit to the NG15 data. The shaded region is excluded by super-radiance constraints. \textit{\textbf{Right panel:}} Same as the left panel but including also SMBH binaries with environmental effects.}
    \label{fig:Axion_posteriors}
\end{figure}

We show in Fig.~\ref{fig:Axion_posteriors} the posteriors of the ``audible" axions two-parameter fit to the NG15 data. The best-fit values are achieved at
\be
    m_a= 3.1\times10^{-11} {\rm eV}\,,\quad f_a  =0.87\, M_{\rm P} \,,
\ee
corresponding to $-2 \Delta \ell = -1.6$ relative to the baseline. Differently from the analysis performed in ref.\,\cite{Figueroa:2023zhu} we find a larger possible range for the masses of the axions below super-Planckian $f_a$ values. Part of this region is currently ruled out by the super-radiance constraints\,\cite{Zhang:2021mks,Baryakhtar:2020gao}. Since the energy density of dark photons is diluted as dark radiation, it changes the number of effective relativistic degrees of freedom $\Delta N_{\rm eff}$. Following ref.\cite{Machado:2018nqk}, we can see that the parameter space over we are scanning gives us a contribution to $\Delta N_{\rm eff}$ which is much lower than the constraints from BBN ($\Delta N_{\rm eff} < 0.33$)~\cite{Fields:2019pfx} and CMB ($\Delta N_{\rm eff} < 0.3$)~\cite{Planck:2018vyg,Ramberg:2022irf}. Extra mechanisms should be invoked in order to avoid overshooting the relic abundance if there is back-scattering off the gauge fields\,\cite{Kitajima:2017peg}. However, the GW signal would remain unchanged. The study of these potential mechanisms goes beyond the scope of this study.  The discrepancy between our analysis and that in \cite{Figueroa:2023zhu} is twofold. First, the analysis in Ref.\,\cite{Figueroa:2023zhu} is based on both NANOGrav and EPTA data, with the latter dataset favoring larger amplitudes at higher frequencies. Heavier axions move the main peak to higher frequencies without changing the amplitude of the main peak. Consequently, including the EPTA dataset leads to stronger bounds on more massive axions compared to our analysis. Secondly, when the peak of the signal is at frequencies much higher than the PTA range, which occurs for the larger masses in Fig.\ref{fig:Axion_posteriors}, the causality tail dominates the signal compared to the Ansatz used to characterize the signal in \cite{Figueroa:2023zhu}.

The best fit of the combined fit is at the same point as without the SMBH, $p_{\rm BH} \approx 0$, so the inclusion of the SMBH binaries does not improve the fit. If we impose the posteriors from the candidate event at $4{\rm nHz}$, then the best-fit shifts to 
\bea 
    &p_{\rm BH}=0.58\, , \; f_{\rm ref}=20 \, {\rm nHz}\, , \; \alpha=5\, ,
\eea
and same axion parameters, corresponding to $-2 \Delta \ell = -0.3$. 

\subsection{Model comparisons} 
\label{sec:comp}

\begin{figure}
\centering
\includegraphics[width=0.95\textwidth]{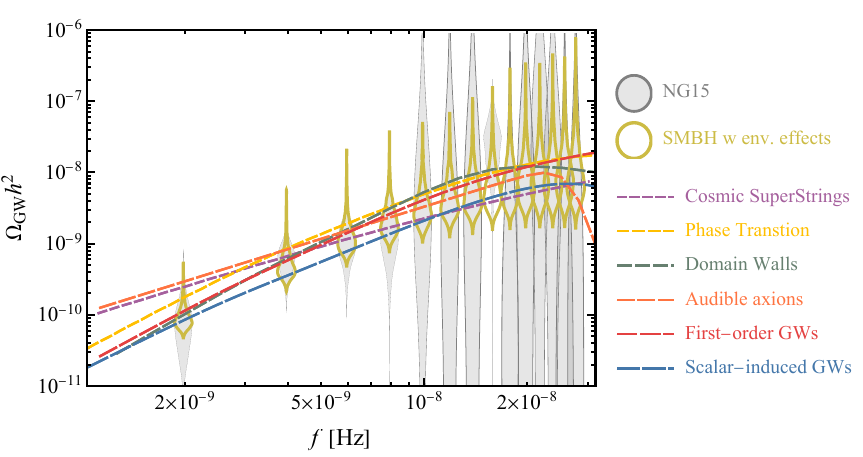}
\caption{Comparison of the best fits to the NG15 data for SMBH binaries with environmental effects and in the indicated BSM cosmological models.}
\label{fig:Bestfits}
\end{figure}

Fig.~\ref{fig:Bestfits} compares the best fits in the models we have studied in the frequency range measured by NANOGrav. We see that the cosmological models all capture to some extent the frequency dependence measured in the NANOGrav range, as does the astrophysical SMBH binary scenario if environmental effects are included. 
The cosmological models all fit the data slightly better than the astrophysical model, as seen quantitatively in Table~\ref{tab:results}, which collects the numerical results for our best fits in the astrophysical SMBH model with and without environmental effects and in the BSM cosmological scenarios we have studied. The best-fit values of the model parameters are shown in the second column, and the changes in the value of the Bayesian inference criterion relative to the baseline SMBH model with environmental effects, $\Delta {\rm BIC}$, shown in the third column are all negative. Among the cosmological sources, the cosmic (super)string scenario, first-order GWs and scalar-induced GWs all fit the data less well than domain walls, phase transitions and ``audible" axions, The numbers in parentheses in the third column indicate the Bayesian inference criterion for the best fit combining the cosmological model with the SMBH model, including the environmental effects. 
The combined fits are all worse than the SMBH model with environmental effects included ($ \Delta {\rm BIC} > 0$) because of the increase in the number of degrees of freedom.

\begin{figure}
\centering
\includegraphics[width=0.95\textwidth]{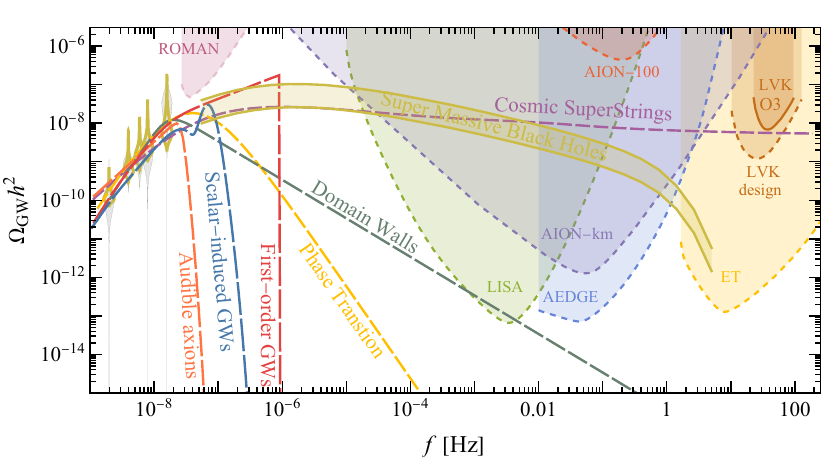}
\caption{Extension of Fig.~\ref{fig:Bestfits} to higher frequencies, indicating the prospective sensitivities of LVK and planned and proposed future detectors. For clarity, we include only the first four and the eighth bins which are the most narrow and have the largest impact on fit quality. 
The green band extends the ``violins" in the PTA  range to higher frequencies, and shows the mean GW energy density spectrum from SMBH binaries heavier than $10^3 M_\odot$ for $p_{\rm BH} = 0.25 - 1$. Individual SMBH binaries are expected to be measurable in this frequency range. The FOGW spectrum is expected to be cut off at some frequency between the ranges where PTAs and interferometers are sensitive.}
\label{fig:BestFits2}
\end{figure}

Finally, the fourth column of Table~\ref{tab:results} tabulates some prospective signatures of the best fits in the models considered, which could help discriminate between them. As discussed above, PTAs should be sensitive to bin-to-bin fluctuations if the observed SGWB is due to SMBH binaries. Moreover, there should be observable anisotropies in the SGWB~\cite{Sato-Polito:2023spo} and circular polarization could be observable~\cite{Sato-Polito:2021efu, Ellis:2023owy}. Also, future PTA data could reveal individual binary sources. We see from the fits combining the cosmological models with the SMBH model with environmental effects that the 1$\sigma$ ranges for the SMBH parameters are relatively broad. A consequence of this is that it is possible to find a mixed scenario where the fit is better than in the pure SMBH model but that can still accommodate observable individual binary sources.

Fig.~\ref{fig:BestFits2} extends the model comparisons to the higher frequencies accessible to other experiments including LISA~\cite{Bartolo:2016ami, Caprini:2019pxz, LISACosmologyWorkingGroup:2022jok}, The Einstein Telescope (ET)~\cite{Punturo:2010zz, Hild:2010id}, AION~\cite{Badurina:2019hst,Badurina:2021rgt}, AEDGE~\cite{AEDGE:2019nxb,Badurina:2021rgt}, the Nancy Roman telescope (ROMAN)~\cite{Wang:2022sxn} and the design sensitivity of LVK~\cite{LIGOScientific:2014pky, LIGOScientific:2016fpe, LIGOScientific:2019vic}. 
We see that LISA should be able to detect individual SMBH binaries, albeit with lower masses (${\cal O}(10^6 - 10^9)$ solar masses) than those responsible for the NANOGrav signal ($\gtrsim {\cal O}(10^9)$ solar masses~\cite{Ellis:2023owy}), at larger rates if environmental effects are important~\cite{Ellis:2023dgf}. Similarly, proposed experiments in the mid-frequency band such as AION and AEDGE would be sensitive to mergers of BHs with masses ${\cal O}(10^3 - 10^6)$ solar masses. On the other hand experiments such as LVK and ET are not sensitive to mergers of BHs with masses $\gtrsim 10^3$ solar masses.  In contrast, all these experiments should be able to detect GWs from cosmic (super)strings, even in the case of modified cosmological evolution~\cite{Ellis:2023tsl}.  On the other hand, the SGWB from a cosmological phase transition, SIGWs, FOGWs or ``audible" axions would be unobservable in any higher-frequency detector, whereas the SGWB from domain walls might be observable by LISA between $10^{-3} - 10^{-2}$~Hz, but probably not by AEDGE and other detectors operating at frequencies $\gtrsim 10^{-2}$~Hz.
\subsection*{Summary}
The evidence for the Hellings-Downs angular correlation reported by the NANOGrav, EPTA, PPTA, and CPTA collaborations sets an important milestone in gravitational-wave astronomy. One of the most pressing challenges to follow is to determine the nature of the signal: is it astrophysical or cosmological? 
\\Binary SMBH systems are the default astrophysical interpretation but remain to be established as the sources of the PTA signals. Many cosmological models invoking generic aspects of BSM physics have also been proposed as prospective sources. In this chapter, we have presented a comprehensive Multi-Model Analysis (MMA) that applies a common approach to assess the relative qualities of fits in these models, both with and without the inclusion of a SMBH binary background. We find that these models are capable of fitting the NANOGrav data at least as well as SMBH binaries alone (significantly better if environmental effects on the evolution of the binaries can be neglected). Future PTA datasets may be able to distinguish between the cosmological and astrophysical scenarios, e.g., by observing bin-to-bin fluctuations, anisotropies, circular polarization or individual sources. Observations at higher frequencies will also be invaluable: some cosmological scenarios predict a turndown of the frequency spectrum that precludes measurements by LISA, whereas astrophysical models suggest that BH mergers may be observable by LISA and mid-frequency experiments, and cosmic (super)strings predict that a SGWB background should be detectable by LISA and higher-frequency experiments.\\
The discovery of nHz GWs may be linked to the biggest bangs since the Big Bang, namely mergers of SMBHs, or be providing evidence for new fundamental physics that is inaccessible to laboratory experiments. Time and further data will tell.
\renewcommand{\arraystretch}{1.2}
\setlength{\tabcolsep}{3pt}
\begin{table*}[b]
\begin{center}
\textbf{Results from Multi-Model Analysis (MMA)}\\
\vspace{1mm}
\begin{tabular}{|p{0.28\textwidth}|p{0.2\textwidth}|c|l|}
\hline	
Scenario & Best-fit parameters & $\Delta {\rm BIC}$ & Signatures \\
\hline \hline
GW-driven SMBH binaries & $p_{\rm BH} = 0.07$ & 6.0 & FAPS, LISA, mid-$f$, \sout{LVK, ET} \\
\hline
GW + environment-driven & $p_{\rm BH} = 0.84$ & Baseline & FAPS, LISA, mid-$f$, \sout{LVK, ET} \\
SMBH binaries  & $\alpha = 2.0$ & (${\rm BIC} = 53.9$) & \\
 & $f_{\rm ref} = 34$~nHz & & \\
 \hline \hline
Cosmic (super)strings & $G \mu =2\times 10^{-12}$ & -1.2 &\sout{FAPS}, LISA, mid-$f$, {LVK, ET} \\
(CS) & $p = 6.3\times10^{-3}$ & (4.6) & \\
 \hline
Phase transition & $T_* = 0.34$~GeV & -4.9 &\sout{FAPS, LISA, mid-$f$, LVK, ET}\\
(PT) & $\beta/H = 6.0$ & (2.9) & \\
\hline
Domain walls & $T_{\rm ann} = 0.85$~GeV & -5.7 & \sout{FAPS}, LISA?, \sout{mid-$f$, LVK, ET}\\
(DWs) & $\alpha_* = 0.11$ & (2.2) & \\
 \hline
Scalar-induced GWs & $k_* = 10^{7.7}/{\rm Mpc}$ & -2.1 &
\sout{FAPS, LISA, mid-$f$, LVK, ET}\\
(SIGWs) & $A = 0.06$ & (5.8) & \\
& $\Delta = 0.21$ & & \\
\hline
First-order GWs & $\log_{10}{r}=-14$ & -2.0 & \sout{FAPS, LISA, mid-$f$, LVK, ET} \\
(FOGWs) & $n_{\rm t}= 2.6$ & (6.0) & \\
 & $\log_{10}{(T_{\rm rh}/\textrm{GeV})} = -0.67$&  & \\
\hline
``Audible" axions & $m_{a}= 3.1 \times 10^{-11}\, {\rm eV}$ & -4.2 & \sout{FAPS, LISA, mid-$f$, LVK, ET} \\
 & ${f_a}= 0.87\, M_{\rm P}$ & (3.7) & \\
\hline
\end{tabular} \\
\vspace{1mm}
{\footnotesize FAPS~$\equiv$~fluctuations, anisotropies, polarization, sources, mid-$f$~$\equiv$~mid-frequency experiment, e.g., AION~\cite{Badurina:2019hst}, AEDGE~\cite{AEDGE:2019nxb},LVK~$\equiv$~LIGO/Virgo/KAGRA~\cite{LIGOScientific:2014pky,LIGOScientific:2016fpe,LIGOScientific:2019vic}, ET~$\equiv$~Einstein Telescope~\cite{Sathyaprakash:2012jk,Reitze:2019iox}), \sout{signature}~$\equiv$~not detectable}
\end{center}
\vspace{-4mm}
\caption{
The parameters of the different models are defined in the text. For each model, we tabulate their best-fit values, and the Bayesian information criterion $BIC\equiv -2 \ell + k \ln 14$, where $k$ denotes the number of parameters, relative to that for the purely SMBH model with environmental effects that we take as the baseline. The quantity in the parentheses in the third column shows the $\Delta$BIC for the best-fit combined SMBH+cosmological scenario. The last column summarizes the prospective signatures for the best fits in the different models.}
\label{tab:results}
\end{table*}
\part{Primordial Black Holes: 
\\Common models for production}
\newtcolorbox{mynamedbox1}[1]{colback=venetianred!5!white,colframe=venetianred!80!black,title=#1}
\newtcolorbox{mynamedbox2}[1]{colback=azure!5!white,colframe=azure!80!black,title=#1}
\def\bz{\boldsymbol{\zeta}}
\def\RE{{\rm Re}}
\def\IM{{\rm Im}}
\def\lp{\left (}
\def\llp{\left [}
\def\rp{\right )}
\def\rrp{\right ]}
\definecolor{oucrimsonred}{rgb}{0.6, 0.0, 0.0}
\definecolor{persianblue}{rgb}{0.11, 0.22, 0.73}
\definecolor{forestgreen}{rgb}{0.13,0.35,0.13}
\definecolor{lightgray}{rgb}{0.83, 0.83, 0.83}
\def\hhref#1{\href{http://arxiv.org/abs/#1}{#1}} 
\definecolor{cornellred}{rgb}{0.0, 0.0, 0.5}
\definecolor{amethyst}{rgb}{0.6, 0.4, 0.8}
\definecolor{yellow}{rgb}{1.0, 1.0, 0.0}
\definecolor{firebrick}{rgb}{0.7, 0.13, 0.13}
\definecolor{tangerineyellow}{rgb}{1.0, 0.8, 0.0}
\definecolor{verdechiaro}{rgb}{0.6,1,0.6}
\definecolor{venetianred}{rgb}{0.78, 0.03, 0.08}
\definecolor{americanrose}{rgb}{1.0, 0.01, 0.24}
\definecolor{cobalt}{rgb}{0.0, 0.28, 0.67}
\definecolor{brandeisblue}{rgb}{0.0, 0.44, 1.0}
\definecolor{mycolor}{rgb}{0.0, 0.0, 0.5}
\definecolor{oxfordblue}{rgb}{0.0, 0.13, 0.28}
\definecolor{azure}{rgb}{0.0, 0.5, 1.0}
\definecolor{turquoiseblue}{rgb}{0.0, 1.0, 0.94}
\definecolor{deepfuchsia}{rgb}{0.76, 0.33, 0.76}
\definecolor{amber}{rgb}{1.0, 0.75, 0.0}
\definecolor{VioletRed4}{rgb}{0.55, 0.13, .32}
\definecolor{indiagreen}{rgb}{0.07, 0.53, 0.03}
\definecolor{VioletRed4}{rgb}{0.55, 0.13, .32}
\chapter{Primordial black hole production 
\\from single field models}\label{cap:Models}
\thispagestyle{plain}
Different mechanisms and models have been proposed as viable for forming PBHs. This represents a competing alternative explanation for the population of astrophysical black holes, and depending on the formation scenario considered and on the models for PBH formations, a variety of mass functions are predicted. In this chapter, we focus on the standard formation scenario that hypothesizes that PBHs form out of the gravitational collapse of large over-densities in the primordial density contrast field\,\cite{Ivanov:1994pa,GarciaBellido:1996qt,Ivanov:1997ia,Blinnikov:2016bxu}.
As we explained in the previous chapter, in the standard formation scenario, to achieve a significant amount of DM in the form of PBHs, it is necessary for the amplitude of the curvature power spectrum to be around $10^{-2}$ at the relevant range of scales. However, at the scales associated with the CMB, typically around $k\simeq 0.05$ $\textrm{Mpc}^{-1}$, the inflationary power spectrum has an amplitude around $10^{-9}$ \cite{Planck:2018jri}. Therefore, a mechanism is required to enhance the power spectrum at the relevant scales. In this chapter we will focus in cases in which the above-mentioned enhancement can be dynamically realized in the context of single-field
models of inflation. To be more precise, we will focus \textit{Ultra Slow-Roll (USR)} scenarios in which the enhancement is given by introducing a phase of ultra slow-roll (USR) during which the inflaton field, after the first conventional phase of slow-roll (SR) that is needed to fit large-scale cosmological observations, almost stops the descent along its potential (typically because of the presence of a quasi-stationary inflection point) before starting rolling down again in a final stage of SR dynamics that eventually ends inflation. 
In this chapter we briefly introduce the USR dynamics, then we analyze the 1-loop corrections in single field models, following closely ref.\,\cite{Franciolini:2023agm}. Then we see an explicit realization in an original model called \textit{double inflection point} and how there are some important observations related to the reheating for production of sub-solar mass PBHs in the context of such models (based on ref.\cite{Allegrini:2024ooy}).
\section{Ultra slow-roll dynamics} \label{Section_ref}
When the inflaton draws near the approximate stationary inflection point, its velocity suddenly decreases almost to zero. The inflaton almost stops close to the approximate stationary inflection point of the potential but it has just enough inertia to overcome the barrier. This part of dynamics is usually called ultra-slow roll (USR) phase\,\cite{Ballesteros:2017fsr,Germani:2017bcs}. The USR phase is formally defined by the condition $|\eta|>3$ on the Hubble parameter $\eta$ defined in eq.\,\ref{eq:eta}.

First, we set our conventions. Where phenomenologically irrelevant we set the reduced Planck mass to one; 
The Hubble-flow parameters $\epsilon_{i}$ (for $i\geqslant 1$) are defined by the recursive relation
\begin{align}
\epsilon_{i} \equiv \frac{\dot{\epsilon}_{i-1}}{H\epsilon_{i-1}}\,,~~~~~\textrm{with:}~~~
\epsilon_0 \equiv \frac{1}{H}\,.\label{eq:HubblePar1}
\end{align}
The relations with the usual inflationary  parameters are $\epsilon \equiv \epsilon_1 = -\dot{H}/H^2$ and instead of $\eta$ we have the second Hubble parameters\footnote{We remark that in ref.\,\cite{Kristiano:2022maq} the symbol $\eta$ refers to the second Hubble parameter $\epsilon_2$.}
\begin{align}
\eta \equiv - \frac{\ddot{H}}{2H\dot{H}} 
= \epsilon - 
\frac{1}{2}\frac{d\log\epsilon}{dN}\,,~~~~~~
\textrm{with:}~~~~~\epsilon_2 = 2\epsilon - 2\eta\,.\label{eq:HubblePar2}
\end{align}

We consider the theory described by the action 
\begin{align}
\mathcal{S} =
\int d^4x \sqrt{-g}\bigg[
\frac{1}{2}R(g)
 + \frac{1}{2}g^{\mu\nu}
 (\partial_{\mu}\phi)
 (\partial_{\nu}\phi) - 
 V(\phi)
\bigg]
\,.\label{eq:BackAction}
\end{align}
$R(g)$ is the scalar curvature associated with the space-time whose geometry is described by the metric $g$ with line element 
$ds^2 = g_{\mu\nu}dx^{\mu}dx^{\nu}$.
The classical background evolves in the flat FLRW Universe and the background value of the scalar field is a function of time, $\phi(t)$.  
We tacitly assume that the scalar potential features an approximate stationary inflection point so as to trigger the transition SR/USR/SR.

We focus on scalar perturbations. 
We consider the perturbed metric in the following generic form
\begin{align}
ds^2 = N^2dt^2 -
h_{ij}(N^i dt + dx^i)
(N^j dt + dx^j)\,,
\end{align}
and choose the gauge in which
\begin{align}
N = 1 + \delta N(\vec{x},t)\,,~~~~~~
N^i = \delta^{ij}\partial_jB(\vec{x},t)\,,~~~~~~
h_{ij} 
= a^2(t)e^{2\zeta(\vec{x},t)}\delta_{ij}\,,~~~~~~
\delta\phi(\vec{x},t) = 0\,.\label{eq:Gauge}
\end{align}
The field $\zeta(\vec{x},t)$ is the only independent scalar degree of freedom since 
$N$ and $N^i$ are Lagrange multipliers subject to the momentum and Hamiltonian constraints. 
It is important to stress that the variable $\zeta$ as defined in eq.\,(\ref{eq:Gauge}) is constant outside the horizon and represents the correct non-linear generalization of the Bardeen variable\,\cite{Maldacena:2002vr}.

At the quadratic order in the fluctuations, the action is 
\begin{align}
\mathcal{S}_2 = 
\int d^4 x\,\epsilon\,a^3
\left[
\dot{\zeta}^2 - \frac{(\partial_k\zeta)(\partial^k\zeta)}{a^2}
\right]\,.\label{eq:MainQuad}
\end{align}
Comoving curvature perturbations are quantized by introducing the free operator 
\begin{align}
\hat{\zeta}(\vec{x},\tau) = 
\int\frac{d^3\vec{k}}{(2\pi)^3}
\hat{\zeta}(\vec{k},\tau)e^{i\vec{x}\cdot\vec{k}}\,,
~~~~~\textrm{with:}~~~
\hat{\zeta}(\vec{k},\tau)=
\zeta_{k}(\tau)a_{\vec{k}}  + 
\zeta_{k}^*(\tau)a^{\dag}_{-\vec{k}}\,,\label{eq:MainDef}
\end{align}
and 
\begin{equation}\label{eq:Anni}
[a_{\vec{k}},a_{\vec{k}^{\prime}}] =[a_{\vec{k}}^{\dag},a_{\vec{k}^{\prime}}^{\dag}] =0\,,~~~~~~~[a_{\vec{k}},a_{\vec{k}^{\prime}}^{\dag}] = (2\pi)^3\delta^{(3)}(\vec{k}-\vec{k}^{\prime})\,,~~~~~~~a_{\vec{k}}|0\rangle = 0\,,
\end{equation}
where the last condition defines the vacuum of the free theory $|0\rangle$. We define the comoving wavenumber $k\equiv |\vec{k}|$. 
The scale factor in the FLRW Universe corresponds to a rescaling of the spatial coordinate; consequently, physically sensible results should be invariant under 
the rescaling\,\cite{Senatore:2009cf}
\begin{align}
a \to \lambda a\,,~~~~~~~
\vec{x} \to \vec{x}/\lambda\,,~~~~~~~
\vec{k} \to \lambda\vec{k}\,,~~~~~~
k \to |\lambda|k\,,~~~~~~
{\textrm{with}}~~\lambda \in \mathbb{R}\,.\label{eq:Rescaling}
\end{align}
Furthermore, if we consider the conformal time $\tau$ instead of the cosmic time $t$, we also have
\begin{align}
  \tau \to \tau/\lambda\,.  
\end{align}
Notice that, under the above rescaling, we have $a_{\vec{k}} \to a_{\vec{k}}/|\lambda|^{3/2}$ (from the scaling property of the three-dimensional $\delta$ function) and, consequently, $\zeta_k \to \zeta_k/|\lambda|^{3/2}$ so that $\hat{\zeta}(\vec{k},\tau) \to \hat{\zeta}(\vec{k},\tau)/|\lambda|^3$ and   $\hat{\zeta}(\vec{x},\tau)$ invariant.
In the case of free fields, we have 
\begin{align}
\langle 0|\hat{\zeta}(\vec{k}_1,\tau_1)
\hat{\zeta}(\vec{k}_2,\tau_2)|0\rangle =  
(2\pi)^3\delta(\vec{k}_1+\vec{k}_2)
\zeta_{k_1}(\tau_1)\zeta_{k_2}^*(\tau_2)\,.
\end{align}
In the presence of a time-derivative, we simply have 
\begin{align}
\langle 0|\hat{\zeta}^{\prime}(\vec{k}_1,\tau_1)
\hat{\zeta}(\vec{k}_2,\tau_2)|0\rangle =  
(2\pi)^3\delta(\vec{k}_1+\vec{k}_2)
\zeta^{\prime}_{k_1}(\tau_1)\zeta_{k_2}^*(\tau_2)\,.
\end{align}
Note that the time dependence occurs in the
mode function, not in the raising/lowering 
operator. The mode function $\zeta_{k}(\tau)$ is related to the linear-order Mukhanov-Sasaki (M-S) equation. More in detail, if we define $\zeta_k(\tau) = u_k(\tau)/z(\tau)$ with $z(\tau)\equiv a(\tau)\sqrt{2\epsilon(\tau)}$, the mode $u_k(\tau)$ verifies the equation 
\begin{align}
    \frac{d^2u_k}{d\tau^2} + \left(k^2 - \frac{1}{z}\frac{d^2z}{d\tau^2}\right)u_k = 0\,.\label{eq:EoM}
\end{align}
We are interested in the computation of the quantity 
\begin{align}
\lim_{\tau \to 0^-}\langle\hat{\zeta}(\vec{x}_1,\tau)
\hat{\zeta}(\vec{x}_2,\tau)\rangle = 
\int \frac{d^3\vec{k}}{(2\pi)^3}
P(k)e^{i\vec{k}\cdot (\vec{x}_1 - \vec{x}_2)}\,,\label{eq:TwoPointCorr}\end{align}
at some late time $\tau \to 0^-$ at which curvature perturbations become constant at super-horizon scales. Equivalently, we write
\begin{align}
\lim_{\tau \to 0^-}\langle\hat{\zeta}(\vec{x},\tau)
\hat{\zeta}(\vec{x},\tau)\rangle = 
\int \frac{dk}{k}\underbrace{\left[\frac{k^3}{2\pi^2}
P(k)\right]}_{\equiv \mathcal{P}(k)} 
= \int \frac{dk}{k}\mathcal{P}(k)\,,
\label{eq:CompaPS}
\end{align}
where $\mathcal{P}(k)$ is the a-dimensional power spectrum. 
At the level of the quadratic action, we find 
\begin{align}
\langle0|\hat{\zeta}(\vec{x}_1,\tau)
\hat{\zeta}(\vec{x}_2,\tau)|0\rangle & = 
\int\frac{d^3\vec{k}_1}{(2\pi)^3}\,
d^3\vec{k}_2 \delta(\vec{k}_1 + \vec{k}_2)\,
\zeta_{k_1}(\tau)\zeta_{k_2}^*(\tau)
e^{i(\vec{x}_1\cdot k_1 + \vec{x}_2\cdot k_2)} = \int\frac{d^3\vec{k}}{(2\pi)^3}
|\zeta_{k}(\tau)|^2 e^{i\vec{k}\cdot(\vec{x}_1 - \vec{x}_2)}\,,\label{eq:PSQuad}
\end{align}
which of course gives the familiar result
\begin{align}
\mathcal{P}(k) = 
\lim_{\tau \to 0^-}\frac{k^3}{2\pi^2}|\zeta_{k}(\tau)|^2\,.\label{eq:TreeLevel}
\end{align}

\begin{figure}[h!]
\begin{center}
\includegraphics[width=.9\textwidth]{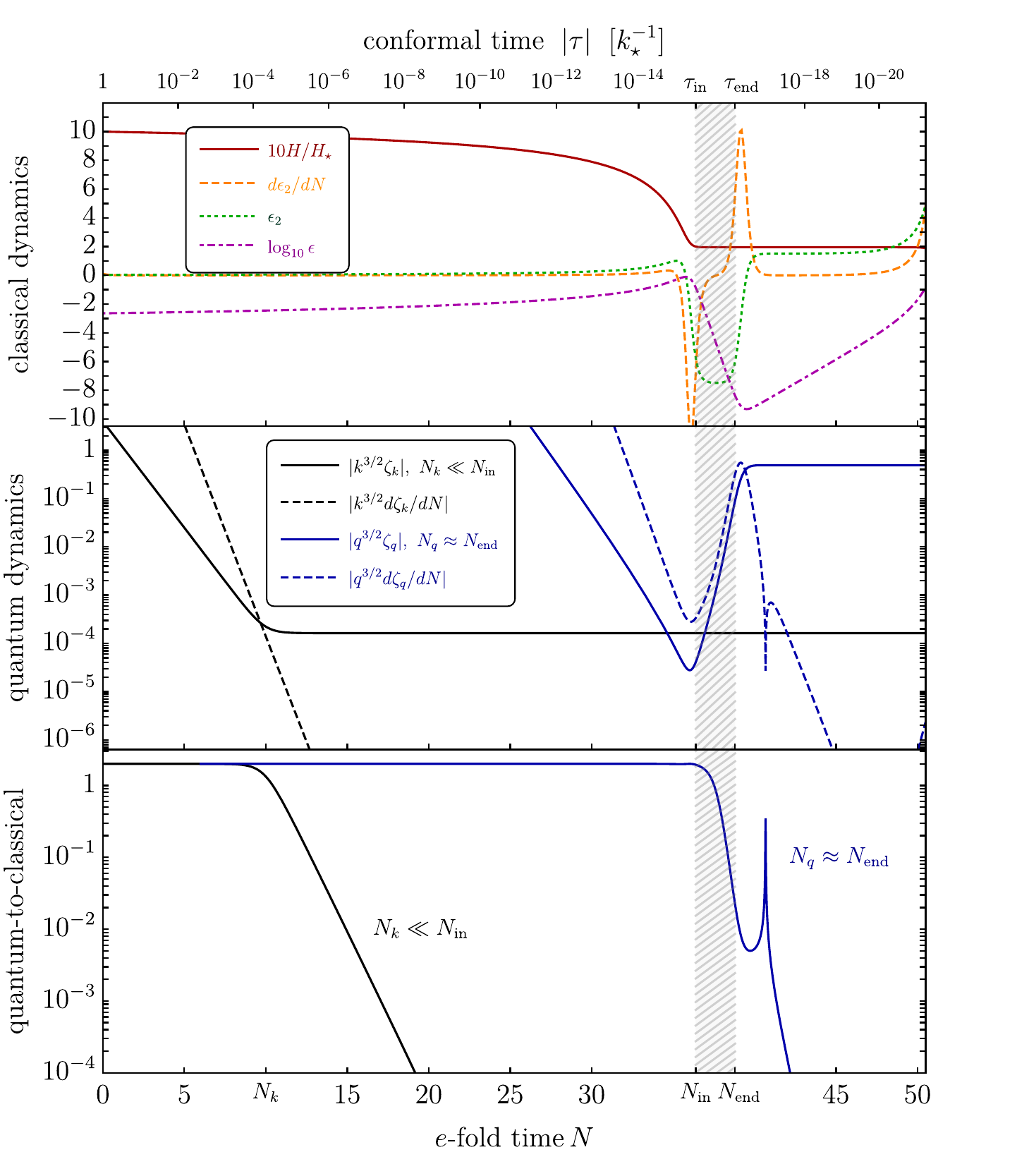}
\caption{
Classical (top panel) and quantum (central panel) dynamics in the context of an explicit single-field model of inflation  that exhibits the presence of a phase of USR in between the time interval $N_{\textrm{\rm in}} < N < N_{\textrm{\rm end}}$ (cf. ref.\,\cite{Ballesteros:2020qam}).  In this specific realization, 
we have $N_{\textrm{\rm in}} = 36.3$ and
$N_{\textrm{\rm end}} = 38.8$. 
\textbf{\textit{Top panel:}} we plot the evolution of the background quantities $\epsilon$, $\epsilon_2$ and $\epsilon_2^{\prime}$ (cf. eqs.\,(\ref{eq:HubblePar1},\,\ref{eq:HubblePar2})) together with the evolution of the Hubble rate (normalized with respect to the value $H_*\equiv H(N_*)$ and scaled by a factor 10 to ease of comparison).
\textbf{\textit{Central panel:}} we plot the solutions of the M-S equation in eq.\,(\ref{eq:M-S}) for two different curvature modes. 
The mode in black (blue) exits the Hubble horizon well before (during) the USR phase. 
\textbf{\textit{Bottom panel:}} we plot the  classicality parameter $C_k$ for the same two modes (cf. the main text for details). 
In the case of the black mode ($N_k\ll N_{\rm in}$) the classicality parameter quickly vanishes after horizon crossing, and remains negligible also during the USR phase. In the case of the blue mode ($N_q\approx N_{\rm end}$) the classicality parameter remains sizable during the USR phase, indicating that this mode retains its quantum nature during USR.   
 }\label{fig:Schematic}  
\end{center}
\end{figure}
In order to make our discussion more concrete, we can refer to fig.\,\ref{fig:Schematic}, see the caption for details. In this figure, we plot both the classical (top panel) and quantum (bottom panel) dynamics that characterize a realistic model of single-field inflation that features a phase of USR because of the presence of an approximate stationary inflection point in the inflationary potential. For more details about the model, see ref.,\cite{Ballesteros:2020qam}. Notably, without including loop corrections to the curvature power spectrum computation, this model is compatible with CMB constraints and predicts $\approx 100\%$ of dark matter in the form of asteroid-mass PBHs\footnote{The model predicts the  tensor-to-scalar ratio $r\simeq 0.037$ which is still (barely) compatible, at  95\% confidence level, with the latest results released by the BICEP and Keck collaboration\,\cite{BICEP:2021xfz}.}. 
In fig.\,\ref{fig:Schematic}, the relation between $e$-fold time $N$ (bottom $x$-axis) and conformal time $\tau$ (top $x$-axis) is given by the integral
\begin{align}
\tau = - \frac{1}{k_{\star}}
\int_{N}^{N_{0}}
\frac{H(N_{\star})}{H(N^{\prime})}e^{N_{\star}-N^{\prime}}dN^{\prime}\,,
\end{align}
where $N_0$ indicates the end of inflation, with $\tau$ conventionally set to $0$ at this time, and $N_{\star}$ the instant of time at which the comoving scale $k_{\star} = 0.05$ Mpc$^{-1}$ crosses the Hubble horizon. In fig.\,\ref{fig:Schematic}, we set $N_{\star} = 0$, and the model gives $N_0 - N_{\star} \simeq 52$.
We can highlight a few crucial properties of the dynamics presented above:
\begin{itemize}
    \item[{\it a)}] We start from the classical analysis. During USR, 
$\epsilon_2(\tau)$ changes according to the schematic 
\begin{align}
\epsilon_2\approx 0 
\ \ 
\overset{\textrm{SR/USR at time }\tau_{\textrm{in}}}{\Longrightarrow}
\ \ 
|\epsilon_2| > 3 
\ \ 
\overset{\textrm{USR/SR at time }\tau_{\textrm{end}}}
{\Longrightarrow} 
\ \ 
\epsilon_2\approx O(1)\,,
\end{align}
thus making $\epsilon_2^{\prime}(\tau)$ non-zero at around the two transitions at conformal times $\tau_{\textrm{in}}$ and $\tau_{\textrm{end}}$ (equivalently, at $e$-fold times $N_{\textrm{in}}$ and $N_{\textrm{end}}$). 
The evolution of $\epsilon_2$ and $\epsilon^{\prime}_2$ are shown in the top panel of fig.\,\ref{fig:Schematic}, with dotted green and dashed orange lines, respectively. 

\item[{\it b)}] During USR, 
the Hubble parameter $\epsilon$  decreases exponentially fast (the inflaton almost stops its classical motion). 
The evolution of $\epsilon$ is shown in the top panel of fig.\,\ref{fig:Schematic} (dot-dashed magenta line); in addition, we also plot the time evolution of the Hubble rate $H$. 

 \item[{\it c)}] We now consider the USR dynamics at the quantum level. 
 It is crucial to understand the typical behavior of curvature modes 
 (solid lines in the central panel of fig.\,\ref{fig:Schematic}) and their time derivatives (dashed lines). 
 In the central panel of fig.\,\ref{fig:Schematic}, we plot two representative cases: the black lines correspond to the case of a mode $\zeta_k$ that exits the Hubble horizon at some time $N_k$ well before the USR phase (like a CMB mode) while the blue lines correspond to a curvature mode $\zeta_q$ that exits the Hubble horizon at some time $N_q$ during the USR phase.
We notice that the derivative $|d\zeta_k/dN|$ decays exponentially fast, 
 and, soon after horizon crossing, becomes negligible, while $|\zeta_k|$ settles to a constant value. 
 Consequently, we expect that interaction terms that involve the time derivative of CMB modes will be strongly  suppressed. 

 \item[{\it d)}] Finally, we consider the issue of the quantum-to-classical transition.  We define the so-called classicality parameter\,\cite{Assassi:2012et} 
$C_k = |\zeta_k\dot{\zeta}_k^* -  
\zeta_k^*\dot{\zeta}_k|/
|\zeta_k\dot{\zeta}_k|$ which goes to zero in the classical limit. 
In the case of conventional SR dynamics, 
the  classicality parameter scales according to $C_k \sim 2k\tau$, and it vanishes exponentially fast right after the horizon crossing time. 
In the bottom panel of fig.\,\ref{fig:Schematic}, we plot the classicality parameter for two representative modes of our dynamics. 
The black modes experiences its horizon crossing well before the USR phase ($N_k\ll N_{\rm in}$). Its classicality parameter quickly vanishes and remains $\ll 1$ during the subsequent USR phase. The blue line, on the contrary, represents the classicality parameter for a mode that experiences its horizon crossing during the USR phase. Its classicality parameter remains sizable during the USR phase, indicating that this short mode can not be treated classically during USR.

\end{itemize}

As a warm-up exercise, we now introduce a simple semi-analytical model called \textit{reverse engineering approach}\,\cite{Byrnes:2018txb,Taoso:2021uvl,Franciolini:2022pav}. 
We define the hyperbolic tangent parametrization
\begin{align}
\eta(N)  & = \frac{1}{2}\left[
-\eta_{\rm II}+ \eta_{\rm II}\tanh\left(\frac{N-N_{\rm in}}{\delta N}\right)
\right] + \frac{1}{2}\left[
\eta_{\rm II} + \eta_{\rm III} + (\eta_{\rm III}-\eta_{\rm II})\tanh\left(\frac{N-N_{\rm end}}{\delta N}\right)
\right]\,,\label{eq:DynEta}
\end{align}
where the parameter $\delta N$ controls the width of the two transitions at $N_{\rm in}$ and $N_{\rm end}$. 
The limit $\delta N\to 0$ reproduces the step-function approximation. 
Using the definition
\begin{align}
\delta(x) = \lim_{\epsilon\to 0} \frac{1}{2\epsilon\cosh^2(x/\epsilon)}\,,
\end{align}
we find
\begin{align}
\lim_{\delta N \to 0}
\frac{d\eta}{dN} = 
(-\eta_{\rm II} 
+  \eta_{\rm III})\delta(N-N_{\rm end}) 
 +\eta_{\rm II}\delta(N-N_{\rm in})\,.\label{eq:DeltaDer}
\end{align}
Using $\eta \simeq  - (1/2)d\log\epsilon/dN$, we find the following expression
\begin{align}
\frac{\epsilon(N)}{\epsilon_{\textrm{ref}}} = & 
e^{-\eta_{\rm III}(N-N_{\rm ref})}\left[
\cosh\left(\frac{N-N_{\rm end}}{\delta N}\right)\cosh\left(\frac{N-N_{\rm in}}{\delta N}\right)
\right]^{-\frac{\delta N\eta_{\rm III}}{2}} \times \nn \\ & 
\left[
\cosh\left(\frac{N_{\rm ref}-N_{\rm end}}{\delta N}\right)\cosh\left(\frac{N_{\rm ref}-N_{\rm in}}{\delta N}\right)
\right]^{\frac{\delta N\eta_{\rm III}}{2}} \times \nn\\
&
\left[
\cosh\left(\frac{N-N_{\rm end}}{\delta N}\right){\rm sech}\left(\frac{N-N_{\rm in}}{\delta N}\right)
\right]^{\delta N\left(\eta_{\rm II} - \frac{\eta_{\rm III}}{2}\right)} \times \nn\\ &
\left[
\cosh\left(\frac{N_{\rm ref}-N_{\rm end}}{\delta N}\right){\rm sech}\left(\frac{N_{\rm ref}-N_{\rm in}}{\delta N}\right)
\right]^{\frac{\delta N}{2}(-2\eta_{\rm II} + \eta_{\rm III})},\,\label{eq:DynEps}
\end{align}
where $\epsilon_{\textrm{ref}}\ll 1$ is the value of $\epsilon$ at some initial reference time $N_{\textrm{ref}}$. 
For future reference, we define $\bar{\epsilon}(N)\equiv \epsilon(N)/\epsilon_{\textrm{ref}}$.
\begin{figure}[h!]
	\centering
\includegraphics[width=0.6\textwidth]{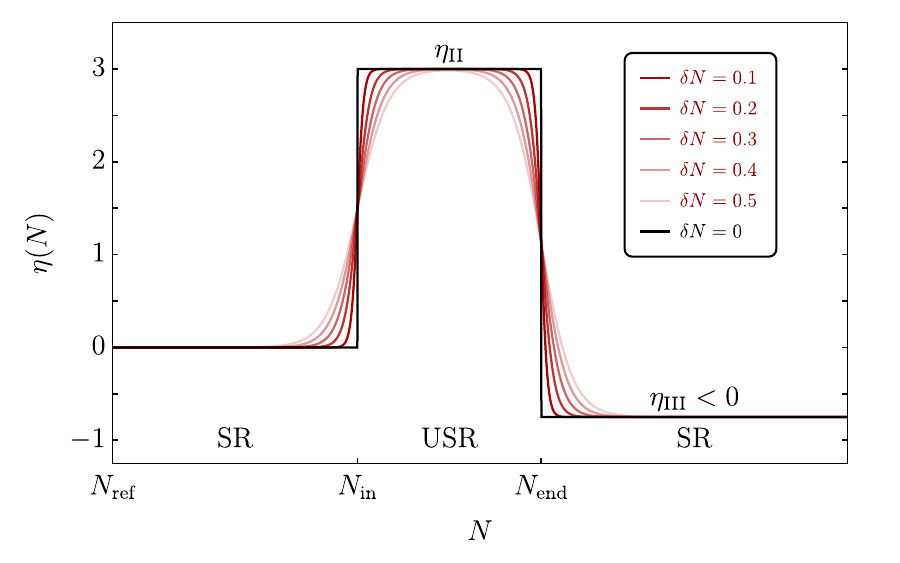}
\includegraphics[width=0.6\textwidth]{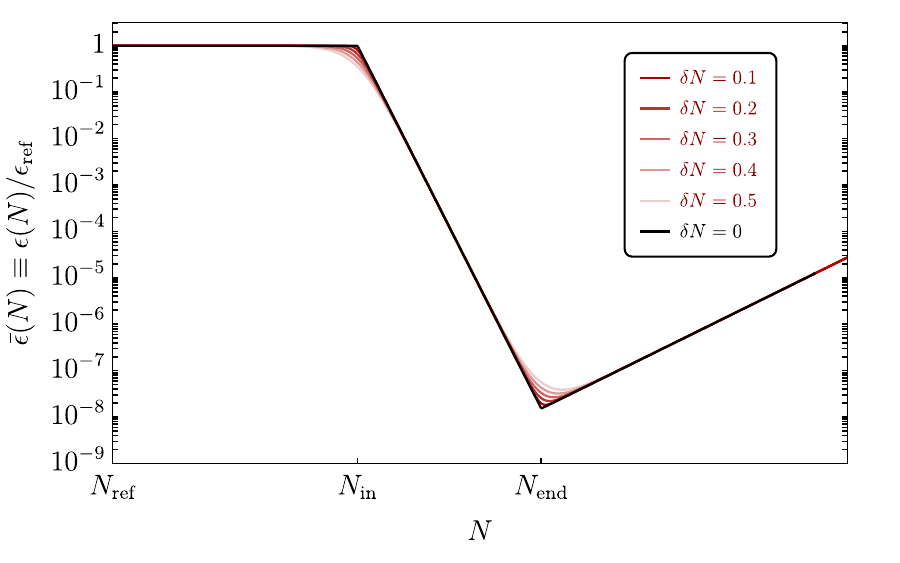}
\includegraphics[width=0.6\textwidth]{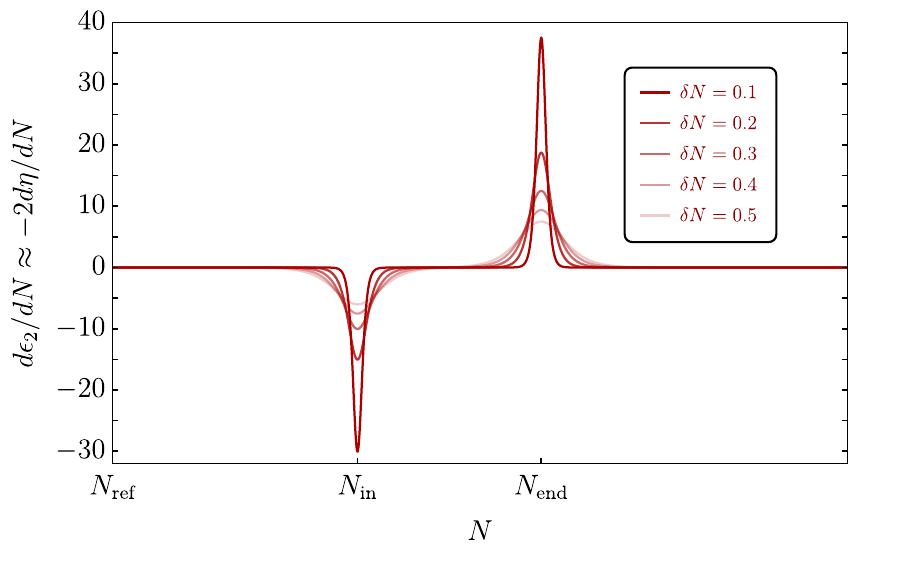}
\caption{
Schematic evolution of $\eta(N)$ in eq.\,(\ref{eq:DynEta}) (\textbf{\textit{top panel}}), $\epsilon(N)$ in eq.\,(\ref{eq:DynEps}) (\textbf{\textit{central panel}}) 
and $d\epsilon_2/dN$ (\textbf{\textit{bottom panel}})
as function of the number 
of $e$-folds $N$. We explore different values of $\delta N$ with the limit $\delta N\to 0$ that corresponds to instantaneous transitions SR/USR at $N=N_{\textrm{\rm in}}$ and USR/SR at $N=N_{\textrm{\rm end}}$. 
In the right panel, the limit $\delta N \to 0$ corresponds to  $\delta$ function transitions at $N_{\textrm{\rm in}}$ and $N_{\textrm{\rm end}}$. Furthermore, notice that the lines corresponding to $d\epsilon_2/dN$ and $-2d\eta/dN$ superimpose, showing that $-2d\eta/dN$ is a perfect approximation of $d\epsilon_2/dN$. 
 }\label{fig:Dyn}  
\end{figure}
In this way we have an analytical description of the background dynamics; most importantly, eqs.\,(\ref{eq:DynEta},\,\ref{eq:DynEps}) are almost all that we need to know to solve
the M-S equation \cite{Sasaki:1986hm,Mukhanov:1988jd}
\begin{align}\label{eq:M-S}
\frac{d^2 u_k}{dN^2} &+ (1-\epsilon)\frac{du_k}{dN} + 
\left[
\frac{k^2}{(aH)^2} + (1+\epsilon-\eta)(\eta - 2) - \frac{d}{dN}(\epsilon - \eta)
\right]u_k = 0\,.
\end{align}
For consistency with eq.\,(\ref{eq:DynEps}), we consider the Hubble rate as a function of time according to 
\begin{align}
H(N) = H(N_{\textrm{ref}})\exp\left[
-\int_{N_{\textrm{ref}}}^{N}
\epsilon(N^{\prime})dN^{\prime}
\right]\,.\label{eq:HubbleRatewithTime}
\end{align}
We shall use the short-hand notation 
$a(N_{\textrm{i}}) \equiv a_{\textrm{i}}$ and 
$H(N_{\textrm{i}}) \equiv H_{\textrm{i}}$. 
Consequently, 
we rewrite eq.\,(\ref{eq:M-S}) in the form 
\begin{align}\label{eq:MSimpl}
\frac{d^2 u_k}{dN^2} &+ (1-\epsilon)\frac{du_k}{dN} + 
\left[
\bar{k}^2 \bigg(\frac{H_{\textrm{in}}}{H}\bigg)^2 e^{2(N_{\textrm{in}} - N)} + (1+\epsilon-\eta)(\eta - 2) - \frac{d}{dN}(\epsilon - \eta)
\right]u_k = 0\,,
\end{align}
with $\bar{k} \equiv 
{k}/({a_{\textrm{in}}H_{\textrm{in}}})$.
We solve the above equation for different $\bar{k}$ with Bunch-Davies initial conditions
\begin{align}
 \sqrt{k}\,u_k(N) = \frac{1}{\sqrt{2}}\,,~~~~~~~~~~~~  
 \sqrt{k}\,\frac{d u_k}{dN}(N) = 
 -\frac{i}{\sqrt{2}}\frac{k}{a(N)H(N)}
 \,,
\end{align}
at some arbitrary time $N \ll N_k$ with $k = a(N_k)H(N_k)$. 
Modes with $\bar{k} \approx O(1)$ are modes that exit the Hubble horizon at about the beginning of the USR phase,
modes with $\bar{k} \ll 1$ are modes that exit the Hubble horizon well before the beginning of the USR phase, 
modes with $\bar{k} \gg 1$ are modes that exit the Hubble horizon well after the beginning of the USR phase.
In the left panel of fig.\,\ref{fig:TestPS}, 
we show the tree-level power spectrum that we obtain by numerically solving eq.\,(\ref{eq:MSimpl}) and using eq.\,(\ref{eq:TreeLevel}). 
Thanks to our parametrization in eq.\,(\ref{eq:DynEta}) we control the sharpness of the transition varying $\delta N$.
\begin{figure}[h]
\begin{center}
$$\includegraphics[width=.495\textwidth]{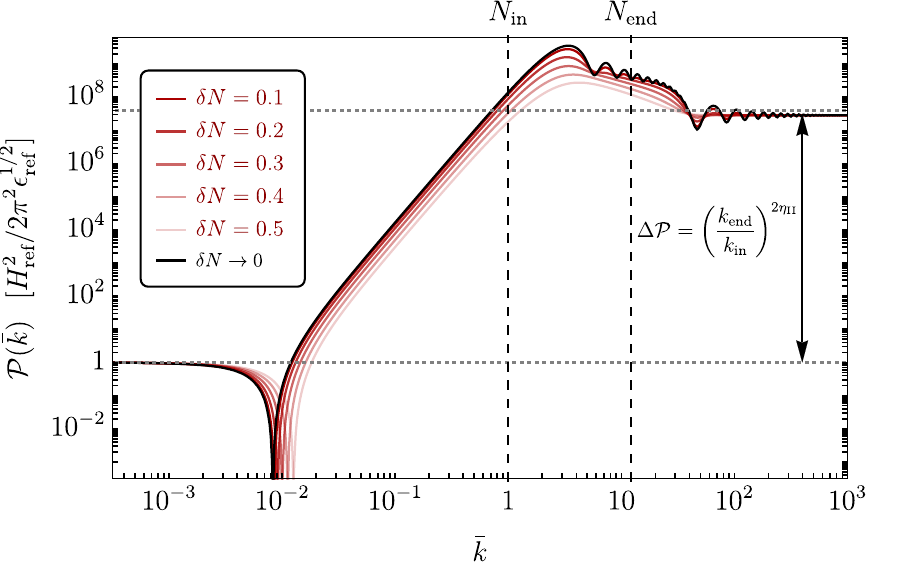}~~
\includegraphics[width=.495\textwidth]{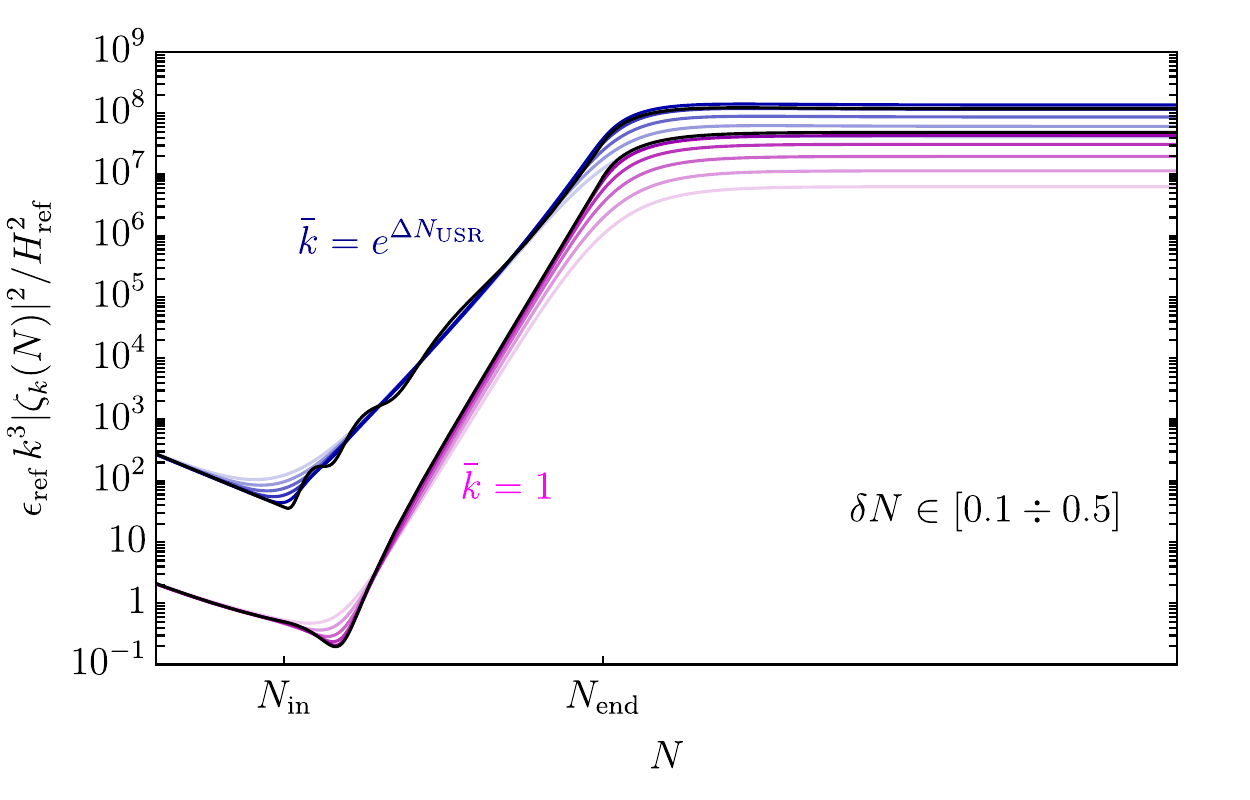}$$\vspace{-0.5cm}
\caption{
\textit{\textbf{Left panel:}}
Tree-level power spectrum using the reverse engineering approach.  
The numerical values of the other parameters are 
$\eta_{\textrm{\rm II}} = 3.5$, 
$\eta_{\textrm{\rm III}} = 0$ and 
$N_{\textrm{\rm end}} - N_{\textrm{\rm in}} = 2.5$. 
In our parametrization, we go beyond the instantaneous transition approximation and 
we explore different values of $\delta N$.
The vertical double-arrow indicates the growth of the power spectrum 
given by the na\"{\i}ve scaling 
$\Delta\mathcal{P} = (k_{\rm end}/k_{\rm in})^{2\eta_{\rm II}} =  e^{2\eta_{\rm II}\Delta N_{\rm{USR}}}$.  
This scaling captures well the amplitude of the transition from the initial to the final SR phase but it does not give a reliable estimate of the peak amplitude of the power spectrum, which can easily be one order of magnitude larger.
\textit{\textbf{Right panel:}}
Time evolution of two representative modes with $\bar{k} = 1$ and $\bar{k} = e^{\Delta N_{\rm{USR}}}$ for $\delta N \in [0.1- 0.5]$ (from darker to lighter colors, respectively). 
The black lines represent the limit $\delta N\to 0$.   
 }\label{fig:TestPS}  
\end{center}
\end{figure}
 
\section{One-loop issue in USR models}
A very legitimate question is whether the 
USR dynamics is consistent with perturbativity.  
Technically speaking, the dimensionless power spectrum of curvature perturbation $\mathcal{P}(k)$ is typically computed within the free theory. However, curvature perturbations, being gravitational in nature, feature an intricate architecture of non-linear interactions. The effect of non-linear interactions is twofold. 
On the one hand, they generate, in addition to the variance, non-zero higher-order cumulants that may leave a peculiar non-Gaussian pattern to the statistics of the curvature field. 
On the other hand, the variance itself gets corrected with respect to the value computed within the free theory. 
We focus on the second effect and, on very general grounds, we define the perturbative criterion
\begin{align}
\mathcal{P}(k) \equiv \mathcal{P}_{\rm tree}(k)
\left[1 + 
\Delta {\cal P}_{\rm 1-loop}(k) \right] ~~~\Longrightarrow~~~
\Delta {\cal P}_{\rm 1-loop}(k)\overset{!}{<} 1\,,\label{eq:Pertu}
\end{align}
meaning that the power spectrum computed within the free theory (the ``tree-level'' power spectrum in the above equation) must be larger than the corrections $\Delta\mathcal{P}$ introduced by the presence of interactions. Such corrections can be organized in a formal series expansion, and  
we will focus in particular on the first-order term, dubbed ``1-loop'' in the  above equation.

The gut feeling is that, unless one is led to consider $\mathcal{P}_{\rm tree}(k) = O(1)$, perturbativity should be under control.  
However, in
ref.\,\cite{Kristiano:2022maq} when it firstly appeared on the ArXiv in November 2022, the authors made the bold claim that, in the presence of USR, the perturbativity condition in eq.\,(\ref{eq:Pertu}) could be violated at scales $k$ relevant for 
CMB observations; even more strikingly, initially ref.\,\cite{Kristiano:2022maq} argues that USR dynamics  
tailored to generate a sizable abundance of asteroid- or solar-mass PBHs are ruled-out\footnote{In the published version of ref.\cite{Kristiano:2022maq}, appeared one year later than ref.\,\cite{Franciolini:2023agm} the authors change the claim of the paper stressing that there is a constraints on the USR models but in realistic scenarios they are not ruled out, finding, where possible, comparable results respect the analysis presented in this chapter.}. 
What makes the initial claim of ref.\,\cite{Kristiano:2022maq} so hard to accept is that it basically says that 
loop of short modes alters the correlation of long CMB modes. 
This is counter-intuitive since it clashes with the intuition that physics at different scales should be decoupled. 

Given the above, it is not surprising that ref.\,\cite{Kristiano:2022maq} sparked an intense debate mostly exclusively polarized on defending or disproving the claim that PBH formation from single-field inflation is ruled out\,\cite{Riotto:2023hoz,Kristiano:2023scm,Riotto:2023gpm,Firouzjahi:2023aum,Firouzjahi:2023ahg,Choudhury:2023vuj,Choudhury:2023jlt,Choudhury:2023rks,Choudhury:2023hvf}. 
Here we critically examine the consequences of  
eq.\,(\ref{eq:Pertu}) in the presence of single-field inflation with USR dynamics, taking into account a proper computations of the PBH abundance and going beyond the instantaneous transition approximation used in ref.\,\cite{Kristiano:2022maq}.

In order to make contact with the analysis of ref.\,\cite{Kristiano:2022maq}, we use the \textit{reverse engineering approach} (eq.\ref{eq:DynEta}) and we set $\eta_{\rm III} = 0$.
However,  it should be noted that 
in more realistic models we need $\eta_{\textrm{III}}\neq 0$ and negative so that the power spectrum decreases for modes with $\bar{k} \gg 1$. This feature is necessary if we want to  connect the USR phase to a subsequent SR phase that ends inflation.  
Since we are considering single-field models of inflation, in our analysis this is a necessary requirement.
Consequently, the power spectrum at small scales -- both before and after the  peak --  does not respect the property of scale invariance. 
Before the peak, the power spectrum 
of the short modes grows with a maximum slope given by $\mathcal{P}(\bar{k}) \sim \bar{k}^4$; after the peak, the power spectrum 
of the short modes decays approximately as $\mathcal{P}(\bar{k}) \sim \bar{k}^{2\eta_{\textrm{III}}}$. 
After the peak, therefore, the power spectrum becomes approximately scale invariant only 
if we take $\eta_{\textrm{III}} \approx 0$; 
however, in such case $\epsilon$ remains anchored to the tiny value reached during the USR phase and inflation never ends.

In ref.\,\cite{Kristiano:2022maq}, 
the loop integration is restricted to the interval of modes $\bar{k} \in [\bar{k}_{\textrm{in}},
\bar{k}_{\textrm{end}}]$  where 
$\bar{k}_{\textrm{in}} = 1$  
and 
$\bar{k}_{\textrm{end}} = e^{\Delta N_{\textrm{USR}}}(H_{\textrm{end}}/H_{\textrm{in}}) 
\simeq e^{\Delta N_{\textrm{USR}}}$ 
with $\Delta N_{\textrm{USR}} \equiv N_{\textrm{end}} - 
N_{\textrm{in}}$.  
This interval of modes is limited by the two vertical dashed lines in the left panel of fig.\,\ref{fig:TestPS}. 
In ref.\,\cite{Kristiano:2022maq}, limiting the integration to the range $\bar{k} \in [\bar{k}_{\textrm{in}},
\bar{k}_{\textrm{end}}]$ is justified by the fact that the power spectrum of short modes peaks in this window of modes.

For future reference, let us stress one more important point. In the left panel of fig.\,\ref{fig:TestPS} we indicate the growth of the power spectrum given by the scaling $\Delta\mathcal{P} = (k_{\rm end}/k_{\rm in})^{2\eta_{\rm II}}$.  
This result immediately follows from the application of the SR formula 
$\mathcal{P}(k) = H^2/8\pi^2\epsilon$ if one accounts for the exponential decay  $\epsilon \sim e^{-2\eta_{\rm II}N}$ during USR and converts $N$ into $k$ by means of  the horizon-crossing condition $k=aH$. 
Therefore, not surprisingly, the scaling 
$\Delta\mathcal{P} = (k_{\rm end}/k_{\rm in})^{2\eta_{\rm II}}$ captures well the growth of the power spectrum if one directly jumps from the initial to the final SR phase.  
However, as shown in the left panel of fig.\,\ref{fig:TestPS},  the above estimate does not accurately describe the amplitude of the power spectrum at the  position of its peak; the latter can easily be one order of magnitude larger than what suggested by $\Delta\mathcal{P} = (k_{\rm end}/k_{\rm in})^{2\eta_{\rm II}}$.
This features has important consequences when estimating the PBH abundance, which rather sensitive to the spectral amplitude.

Finally, it is possible to check numerically that neglecting the time dependence of the Hubble rate as in eq.\,(\ref{eq:HubbleRatewithTime}) has a negligible impact. 
In the following, therefore, we shall keep $H$ constant (that is, $H=H_{\textrm{ref}}$ does not evolve in time). 
Furthermore, if we take $H$ constant and in the limit $\delta N = 0$, it is possible to get, for some special values of $\eta_{\textrm{II}}$ and $\eta_{\textrm{III}}$, a complete analytical description of the SR/USR/SR dynamics\,\cite{Byrnes:2018txb,Ballesteros:2020qam}.
\subsection{The interaction terms}
The goal is to compute corrections that arise from the presence of interactions. 
This means that, in 
eq.\,(\ref{eq:TwoPointCorr}), the vacuum expectation value should refer to the vacuum $|\Omega\rangle$ of the interacting theory and the dynamics of the operator $\hat{\zeta}(\vec{x},\tau)$ are described by the full action that also includes interactions.

We compute the left-hand side of eq.\,(\ref{eq:TwoPointCorr}) by means of the ``{\it in}-{\it in}'' formalism (see e.g.~\cite{Calzetta:1986ey,Jordan:1986ug,Adshead:2009cb}). Correlators are given by
\begin{align}
\langle\Omega|\hat{\mathcal{O}}(\tau)|\Omega\rangle
\equiv \langle\hat{\mathcal{O}}(\tau)\rangle = 
\langle 0|\left\{
\bar{T}\exp\left[{i\int_{-\infty(1+i\epsilon)}^{\tau}
d\tau^{\prime}\hat{H}_{\textrm{int}}(\tau^{\prime})}\right]
\right\}\hat{\mathcal{O}}_I(\tau)\left\{
T\exp\left[{-i\int_{-\infty(1-i\epsilon)}^{\tau}
d\tau^{\prime}\hat{H}_{\textrm{int}}(\tau^{\prime})}\right]
\right\}|0\rangle\,,\label{eq:MainInIn}
\end{align}
where on the right-hand side all fields appearing in $\hat{\mathcal{O}}_I(\tau)$ and $\hat{H}_{\textrm{int}}(\tau^{\prime})$ are free fields in the interaction picture. 
We shall indicate free fields in the interaction picture with the additional subscript $_{I}$. 
It should be noted that the latter are nothing but the operators of the free theory that we quantized in eq.\,(\ref{eq:MainDef}).
 
$T$ and $\bar{T}$ are the time and anti-time ordering operator, respectively.
As customary, the small imaginary deformation in the integration contour 
guarantees that $|\Omega\rangle\to |0\rangle$ as $\tau\to -\infty$ where $|\Omega\rangle$ is the vacuum of the interacting theory.
On the left-hand side, the operator $\hat{\mathcal{O}}(\tau)$ is the equal-time product of operators at different space
points, precisely like in eq.\,(\ref{eq:TwoPointCorr}). 
We expand in the interaction Hamiltonian, so we use the Dyson series
\begin{align}
T\exp\left[-i\int_{-\infty_-}^{\tau}
d\tau^{\prime}\hat{H}_{\textrm{int}}(\tau^{\prime})\right] = 1 -i\int_{-\infty_-}^{\tau}
d\tau^{\prime}\hat{H}_{\textrm{int}}(\tau^{\prime}) +
i^2
\int_{-\infty_-}^{\tau}d\tau^{\prime}
\int_{-\infty_-}^{\tau^{\prime}}d\tau^{\prime\prime}\hat{H}_{\textrm{int}}(\tau^{\prime})\hat{H}_{\textrm{int}}(\tau^{\prime\prime}) + \dots\,,\label{eq:DysonSeries}
\end{align}
where, for simplicity, we introduce the short-hand notation $\infty_{\pm}\equiv \infty(1\pm i\epsilon)$.
Each order in $\hat{H}_{\textrm{int}}$ is an interaction vertex, and carries both a time integral and the space integral (enclosed in the definition of $\hat{H}_{\textrm{int}}$) which in Fourier space enforces momentum conservation.

It is crucial to correctly identify the interaction Hamiltonian. Before proceeding in this direction, let us clarify our notation. We expand the action in the form
\begin{align}
\mathcal{S} = 
\int d^3\vec{x}dt\,
\underbrace{\mathcal{L}[\zeta(\vec{x},t),\dot{\zeta}(\vec{x},t),
\partial_k\zeta(\vec{x},t)]}_{
\equiv\,
\mathcal{L}[\zeta(\vec{x},t)]
} = 
\underbrace{\int d^3\vec{x}dt\,
\mathcal{L}_2(\vec{x},t)}_{\equiv\,\mathcal{S}_2} + 
\underbrace{\int d^3\vec{x}dt\,
\mathcal{L}_3[\zeta(\vec{x},t)]}_{\equiv\,\mathcal{S}_3} + 
\underbrace{\int d^3\vec{x}dt\,
\mathcal{L}_4[\zeta(\vec{x},t)]}_{\equiv\,\mathcal{S}_4}+\dots\,,\label{eq:DDefL}
\end{align}
with $\mathcal{S}_2$ defined in eq.\,(\ref{eq:MainQuad}). 
We also define (as a function of conformal time) 
\begin{align}
H_{\textrm{int}}^{(k)}(\tau) 
\equiv 
 \int d^3\vec{x}
 \,\mathcal{H}_k[\zeta(\vec{x},\tau)]
 ~~~~~\Longrightarrow~~~~~
 \hat{H}_{\textrm{int}}^{(k)}(\tau) 
\equiv 
 \int d^3\vec{x}
 \,\mathcal{H}_k[\hat{\zeta}_I(\vec{x},\tau)]
\,.\label{eq:Hk}
\end{align}
At the cubic order, we simply have
\begin{align}
\mathcal{H}_3[\zeta(\vec{x},\tau)] 
= - \mathcal{L}_3[\zeta(\vec{x},\tau)]\,.
\end{align}
At the quartic order, simply writing $\mathcal{H}_4 = - \mathcal{L}_4$ does not capture the correct result if the cubic Lagrangian features interactions that depend on the time derivative of $\zeta$ since the latter modify the definition of the conjugate momentum. 

Using, at the operator level, the notation introduced in eq.\,(\ref{eq:Hk}), we schematically write at the first order in the Dyson series expansion
 \begin{align}
\langle\hat{\zeta}&(\vec{x}_1,\tau)
\hat{\zeta}(\vec{x}_2,\tau)\rangle_{1^{\textrm{st}}} = \nn\\
&
\langle 0|\hat{\zeta}_I(\vec{x}_1,\tau)
\hat{\zeta}_I(\vec{x}_2,\tau)\bigg[
-i\int_{-\infty_-}^{\tau}
d\tau^{\prime}\hat{H}^{(4)}_{\textrm{int}}(\tau^{\prime})
\bigg]|0\rangle + 
\langle 0|
\bigg[
i\int_{-\infty_+}^{\tau}
d\tau^{\prime}\hat{H}^{(4)}_{\textrm{int}}(\tau^{\prime})
\bigg]
\hat{\zeta}_I(\vec{x}_1,\tau)
\hat{\zeta}_I(\vec{x}_2,\tau)|0\rangle\,.\label{eq:Shem1}
\end{align}
At the first order, therefore, the first non-zero quantum  correction involves the quartic Hamiltonian.  
At the second order in the Dyson series expansion and considering again terms with up to eight fields in the vacuum expectation values, we write schematically
\begin{align}
\langle\hat{\zeta}(\vec{x}_1,\tau)
\hat{\zeta}(\vec{x}_2,\tau)\rangle_{2^{\textrm{nd}}} & =  \langle 0|\hat{\zeta}_I(\vec{x}_1,\tau)
\hat{\zeta}_I(\vec{x}_2,\tau)\bigg[-
\int_{-\infty_-}^{\tau}d\tau^{\prime}
\int_{-\infty_-}^{\tau^{\prime}}d\tau^{\prime\prime}\hat{H}^{(3)}_{\textrm{int}}(\tau^{\prime})
\hat{H}^{(3)}_{\textrm{int}}(\tau^{\prime\prime})\bigg]|0\rangle 
\nn\\
& +
\langle 0|
\bigg[-
\int_{-\infty_+}^{\tau}d\tau^{\prime}
\int_{-\infty_+}^{\tau^{\prime}}d\tau^{\prime\prime}\hat{H}^{(3)}_{\textrm{int}}(\tau^{\prime\prime})
\hat{H}^{(3)}_{\textrm{int}}(\tau^{\prime})\bigg]
\hat{\zeta}_I(\vec{x}_1,\tau)
\hat{\zeta}_I(\vec{x}_2,\tau)|0\rangle
\nn\\
& +
\langle 0|
\bigg[
i\int_{-\infty_+}^{\tau}
d\tau^{\prime}\hat{H}^{(3)}_{\textrm{int}}(\tau^{\prime})
\bigg]
\hat{\zeta}_I(\vec{x}_1,\tau)
\hat{\zeta}_I(\vec{x}_2,\tau)
\bigg[
-i\int_{-\infty_-}^{\tau}
d\tau^{\prime\prime}\hat{H}^{(3)}_{\textrm{int}}(\tau^{\prime\prime})
\bigg]
|0\rangle\,.\label{eq:Shem2}
\end{align}
The vacuum expectation values of interacting-picture fields can be computed using Wick's theorem.
Schematically, eqs.\,(\ref{eq:Shem1},\,\ref{eq:Shem2}) 
give rise to the following 
connected diagrams.
\begin{align}
\begin{adjustbox}{max width=0.99\textwidth}
\raisebox{-8mm}{
 \scalebox{1}{
 \begin{tikzpicture}
	\draw[thick,dashed][thick] (-3.5,0)--(3.5,0);
    \node at (4,0.2) {\scalebox{1}{$\tau$}};
    \node at (-3,0.5) {\scalebox{1}{$(\vec{x}_1,\tau)$}};
    \node at (+3,0.5) {\scalebox{1}{$(\vec{x}_2,\tau)$}};
    \draw[thick,dashed][thick] (-3.5,-2.19)--(3.5,-2.19);
    \node at (4,-2) {\scalebox{1}{$\tau_1$}};
    \draw[thick][thick] (-3,0)--(0.5,-2.19);
    \draw[thick][thick] (3,0)--(0.5,-2.19);
    \draw[black,thick] (0.5,-3.74)circle(45pt); 
    \draw[black,fill=venetianred,thick] (0.5,-2.19)circle(3pt);
        \node at (0.4,-2.8) {\scalebox{1}{{\color{venetianred}{$\hat{H}^{(4)}_{\textrm{int}}(\tau_1)$}}}};    
    \draw[black,fill=verdechiaro,thick] (-3,0)circle(3pt);
    \draw[black,fill=verdechiaro,thick] (+3,0)circle(3pt); 
	\end{tikzpicture}
    \hspace{1.cm}
	\begin{tikzpicture}
	\draw[thick,dashed][thick] (-3.5,0)--(3.5,0);
    \node at (4,0.2) {\scalebox{1}{$\tau$}};
    \node at (-3,0.5) {\scalebox{1}{$(\vec{x}_1,\tau)$}};
    \node at (+3,0.5) {\scalebox{1}{$(\vec{x}_2,\tau)$}};
    \draw[thick,dashed][thick] (-3.5,-2)--(3.5,-2);
    \node at (4,-2+0.28) {\scalebox{1}{$\tau_1$}};
    \draw[thick,dashed][thick] (-3.5,-4.5)--(3.5,-4.5);
    \node at (4,-4.5+0.28) {\scalebox{1}{$\tau_2$}};
    \draw[thick][thick] (-3,0)--(-2.1,-2);
    \draw[thick][thick] (3,0)--(+2.3,-4.5);
   \draw[thick](-2.1,-2)..controls(1,-2.5)and(1,-2.5)..(+2.3,-4.5);
   \draw[thick](-2.1,-2)..controls(-1,-4)and(-1,-4)..(+2.3,-4.5);
    \draw[black,fill=verdechiaro,thick] (-3,0)circle(3pt);
    \draw[black,fill=verdechiaro,thick] (+3,0)circle(3pt);
    \draw[black,fill=azure,thick] (-2.1,-2)circle(3pt);
    \draw[black,fill=azure,thick] (+2.3,-4.5)circle(3pt);   
    \node at (-2.65,-2.5) {\scalebox{1}{{\color{azure}{$\hat{H}^{(3)}_{\textrm{int}}(\tau_1)$}}}};
        \node at (2.5,-5) {\scalebox{1}{{\color{azure}{$\hat{H}^{(3)}_{\textrm{int}}(\tau_2)$}}}};
	\end{tikzpicture}
    \hspace{1.cm}
 	\begin{tikzpicture}
	\draw[thick,dashed][thick] (-3.5,0)--(3.5,0);
    \node at (4,0.2) {\scalebox{1}{$\tau$}};
    \node at (-3,0.5) {\scalebox{1}{$(\vec{x}_1,\tau)$}};
    \node at (+3,0.5) {\scalebox{1}{$(\vec{x}_2,\tau)$}};
    \draw[thick,dashed][thick] (-3.5,-1.75)--(3.5,-1.75);
    \node at (4,-2+0.28) {\scalebox{1}{$\tau_1$}};
    \draw[thick,dashed][thick] (-3.5,-3)--(3.5,-3);
    \node at (4,-3) {\scalebox{1}{$\tau_2$}}; 
    \draw[thick][thick] (-3,0)--(0,-1.75);
    \draw[thick][thick] (+3,0)--(0,-1.75);
    \draw[thick][thick] (0,-1.75)--(0,-3); 
    \draw[black,thick] (0,-4.2)circle(35pt);    
    \draw[black,fill=azure,thick] 
    (0,-1.75)circle(3pt);     \draw[black,fill=azure,thick] 
    (0,-3)circle(3pt); 
    \node at (-0.85,-2.1) {\scalebox{1}{{\color{azure}{$\hat{H}^{(3)}_{\textrm{int}}(\tau_1)$}}}};
    \node at (0.,-3.5) {\scalebox{1}{{\color{azure}{$\hat{H}^{(3)}_{\textrm{int}}(\tau_2)$}}}};
    \draw[black,fill=verdechiaro,thick] (-3,0)circle(3pt);
    \draw[black,fill=verdechiaro,thick] (+3,0)circle(3pt);
	\end{tikzpicture}
 }
 } \nn
\end{adjustbox} \label{eq:LoopsSchematic}
\end{align}

From the above classification, we see that, at the same loop order, we have three classes of connected diagrams that, in principle, should be discussed together. Notice that, contrary to the first two, the last diagram is not of 1-Particle-Irreducible (1PI) type since it consists of a tadpole attached to a two-point propagator. To proceed further, we need to specify the background dynamics, that shape the time evolution of the Hubble parameters, and the interaction Hamiltonian, that specifies which terms in the Dyson expansion contribute at a given perturbative order.   
\subsubsection{Cubic order}
At the cubic order in the fluctuations, 
the action is
\begin{align}
\mathcal{S}_3 &= 
\int d^4 x\bigg\{
\epsilon^2a^3\dot{\zeta}^2\zeta 
+ \epsilon^2a \zeta(\partial_k\zeta)(\partial^k\zeta) 
-2\epsilon^2a^3\dot{\zeta}
(\partial_k\zeta)\partial^k(\partial^{-2}\dot{\zeta}) 
+ \frac{\epsilon\dot{\epsilon_2}}{2}a^3\dot{\zeta}\zeta^2
\nn\\
&-\frac{a^3\epsilon^3}{2}\big[
\dot{\zeta}^2\zeta - \zeta\partial_{k}\partial_l(\partial^{-2}\dot{\zeta})
\partial^{k}\partial^l
(\partial^{-2}\dot{\zeta})
\big] +\bigg[
\frac{d}{dt}\left(\epsilon a^3 \dot{\zeta}\right)  - 
\epsilon a \partial_k\partial^k\zeta
\bigg] \times
 \nn\\
&+\bigg[
\frac{\epsilon_2\zeta^2}{2}
+ \frac{2\dot{\zeta}\zeta }{H}
 -\frac{(\partial_k\zeta)(\partial^k\zeta) }{2a^2 H^2}
+\frac{1}{2a^2 H^2}\partial^{-2}
\partial_k\partial_l(\partial^k\zeta
\partial^l\zeta)
\frac{\epsilon}{H}
(\partial_k\zeta)\partial^k(\partial^{-2}\dot{\zeta}) 
-
\frac{\epsilon}{H}\partial^{-2}
\partial_{k}\partial_{l}
\partial^{k}\zeta 
\partial^{l}(\partial^{-2}\dot{\zeta})
\bigg]\bigg\}\,.
\end{align}
As shown in ref.\,\cite{Maldacena:2002vr}, it is possible to simplify the cubic action by means of a field redefinition that introduces a non-linear shift in the original field. 
Concretely, if we define
\begin{align}
\zeta \equiv \zeta_n + f(\zeta_n)\,,\label{eq:FieldRed}  
\end{align}
with
\begin{align}
f(&\zeta)
\equiv
\frac{1}{2}\bigg[
\frac{\epsilon_2\zeta^2}{2}
+ \frac{2\dot{\zeta}\zeta }{H}
 -\frac{(\partial_k\zeta)(\partial^k\zeta) }{2a^2 H^2}
+\frac{1}{2a^2 H^2}\partial^{-2}
\partial_k\partial_l(\partial^k\zeta
\partial^l\zeta)+
\frac{\epsilon}{H}
(\partial_k\zeta)\partial^k(\partial^{-2}\dot{\zeta}) 
-
\frac{\epsilon}{H}\partial^{-2}
\partial_{k}\partial_{l}
\partial^{k}\zeta 
\partial^{l}(\partial^{-2}\dot{\zeta})
\bigg]\,,\label{eq:FR1}
\end{align}
we find, by direct computation, that at the quadratic order the field 
$\zeta_n$ is described by the action 
\begin{align}
\mathcal{S}_2(\zeta_n) = 
\int d^4 x\,\epsilon\,a^3
\left[
\dot{\zeta}_n^2 - \frac{(\partial_k\zeta_n)(\partial^k\zeta_n)}{a^2}
\right]\,,\label{eq:QuadraticAction}
\end{align}
which has the same structure as the quadratic action for the original variable $\zeta$. 
However, at the cubic order, we find 
\begin{align}
\mathcal{S}_3(\zeta_n) = 
\int d^4 x\bigg\{&
\epsilon^2a^3\dot{\zeta}_n^2\zeta_n
+ \epsilon^2a \zeta_n(\partial_k\zeta_n)(\partial^k\zeta_n) 
-2\epsilon^2a^3\dot{\zeta}_n
(\partial_k\zeta_n)\partial^k(\partial^{-2}\dot{\zeta}_n) 
+ \frac{\epsilon\dot{\epsilon_2}}{2}a^3\dot{\zeta}_n\zeta_n^2\nn\\
&
-\frac{a^3\epsilon^3}{2}
\big[
\dot{\zeta}_n^2\zeta_n 
- \zeta_n\partial_{k}\partial_l(\partial^{-2}\dot{\zeta}_n)
\partial^{k}\partial^l
(\partial^{-2}\dot{\zeta}_n)
\big]\bigg\}\,,  \label{eq:Cubic2}
\end{align}
in which, thanks to the above field redefinition, the second and third lines in the initial cubic action cancel out.

If we neglect terms with spatial derivatives and interactions suppressed by two or more powers of the Hubble parameter $\epsilon$, we find 
\begin{align}
\mathcal{S}_3(\zeta_n)  \ni 
\int d^4x\,\frac{\epsilon\dot{\epsilon_2}}{2}a^3\dot{\zeta}_n\zeta_n^2\,.\label{eq:Yoko}
\end{align}
Notice that we do not count the coupling  $\epsilon_2$ as a slow-roll suppression since we are interested in the USR phase during which $|\epsilon_2| > 3$ and $\dot{\epsilon}_2 \neq 0$.
Eq.\,(\ref{eq:Yoko}) is the only interaction included in 
ref.\,\cite{Kristiano:2022maq}. 
This means that, implicitly, ref.\,\cite{Kristiano:2022maq} computes the two-point function for the field $\zeta_n$.  
This is because, in terms of the dynamical variable $\zeta$, there is another interaction of order $\epsilon\epsilon_2$ that should be included.

However, as stressed in ref.\cite{Maldacena:2002vr}, $\zeta_n$ is not the right dynamical variable to consider since it is not conserved outside the horizon. 
This is a trivial consequence of eq.\,(\ref{eq:FieldRed}). 
Since $\zeta$ is conserved outside the horizon, $\zeta_n$ can not be conserved simply because various coefficients in the non-linear relation in eq.\,(\ref{eq:FieldRed}) are time-dependent. 
Alternatively, as discussed in ref.\cite{Maldacena:2002vr}, 
the above fact is also evident from the very same structure of the interactions that appear in eq.\,(\ref{eq:Cubic2}). 
The interaction $\epsilon\dot{\epsilon}_2\dot{\zeta}_n\zeta_n^2$ only has one time-derivative acting on the field $\zeta_n$; consequently, 
it alters 
the value of $\zeta_n$ on super-horizon scales (if one computes the equation of motion for $\zeta_n$, it is easy to see that the constant solution is not stable).
Let us make the above considerations more concrete.
Eventually, we are interested in the computation of the two-point function for the original curvature field. 
Given the field redefinition in eq.\,(\ref{eq:FieldRed}), we write
\begin{align}
\langle\hat{\zeta}(\vec{x}_1,\tau)
\hat{\zeta}(\vec{x}_2,\tau)\rangle = &  
\langle
\big\{
\hat{\zeta}_n(\vec{x}_1,\tau) 
+ f[\hat{\zeta}_n(\vec{x}_1,\tau)]
\big\}
\big\{
\hat{\zeta}_n(\vec{x}_2,\tau) 
+ f[\hat{\zeta}_n(\vec{x}_2,\tau)]
\big\}
\rangle \nn\\
= & 
\langle\hat{\zeta}_n(\vec{x}_1,\tau)
\hat{\zeta}_n(\vec{x}_2,\tau)\rangle + \label{eq:Expa1}\\
& \langle\hat{\zeta}_n(\vec{x}_1,\tau)
f[\hat{\zeta}_n(\vec{x}_2,\tau)]\rangle +
\langle
f[\hat{\zeta}_n(\vec{x}_1,\tau)]
\hat{\zeta}_n(\vec{x}_1,\tau)
\rangle + \label{eq:Expa2}\\
& \langle
f[\hat{\zeta}_n(\vec{x}_1,\tau)]
f[\hat{\zeta}_n(\vec{x}_2,\tau)]
\rangle\,,\label{eq:Expa3}
\end{align}
The first term, eq.\,(\ref{eq:Expa1}), corresponds to the two-point function for the shifted curvature field whose  cubic action is given by eq.\,(\ref{eq:Cubic2}); 
$\langle\hat{\zeta}_n(\vec{x}_1,\tau)
\hat{\zeta}_n(\vec{x}_2,\tau)\rangle$ can be computed perturbatively by means of 
the ``{\it in}-{\it in}'' formalism.
Eqs.\,(\ref{eq:Expa2},\,\ref{eq:Expa3})  account for the difference between $\zeta$ and $\zeta_n$ at the non-linear level.
Notice that the first term in the 
functional form in eq.\,(\ref{eq:FR1}) does not die off in the late-time limit $\tau\to 0^-$ (in which the power spectrum must be eventually evaluated) 
if we consider the case in which 
$\epsilon_2 \neq 0$ after the USR phase (as expected in realistic single-field models. 
However, if we limit to the case in which 
$\eta_{\rm III} = 0$ the contribution from the field redefinition vanishes. 
This limit was considered in 
ref.\,\cite{Kristiano:2022maq}. 
In order to make contact with the analysis presented in ref.\,\cite{Kristiano:2022maq}, we shall also adopt in the bulk of this analysis the assumption $\eta_{\rm III}  = 0$.

Let us  now come back to the schematic in eq.\,(\ref{eq:LoopsSchematic}). 
The cubic Hamiltonian interaction that follows from eq.\,(\ref{eq:Yoko}) gives rise to the last two topologies of connected diagrams illustrated in eq.\,(\ref{eq:LoopsSchematic}). 
As in ref.\,\cite{Kristiano:2022maq},  we will only focus on the 1PI diagram, that is, the central diagram in eq.\,(\ref{eq:LoopsSchematic}).
The last diagram in eq.\,(\ref{eq:LoopsSchematic}) consists of a tadpole 
that is attached to a $\zeta$-propagator and affects at one-loop its two-point correlation function. 
The correct way to deal with tadpoles is by changing the background solution, cf. ref.\,\cite{Sloth:2006nu} for a discussion in the case of ordinary SR inflation and ref.\,\cite{Senatore:2009cf} for the case in which there are additional spectator fields.  Recently, ref.\,\cite{Inomata:2022yte} estimated the tadpole correction to the background evolution in the context of a model in which there is a resonant amplification of field fluctuations.  
Imposing the condition that such modification is negligible could give rise to an additional perturbativity bound. In this analysis, for simplicity, we neglect the contribution of such tadpole.

\subsubsection{Beyond the cubic action}\label{sec:Qua}
Before proceeding, we comment about quartic interactions since, as qualitatively discussed in eq.\,(\ref{eq:LoopsSchematic}), they give rise to one-loop corrections which are of the same order if compared to those generated by cubic interaction terms.  
The derivation of the  fourth-order action has been discussed in ref.\,\cite{Jarnhus:2007ia}. 
Based on this result, ref.\,\cite{Kristiano:2022maq,Kristiano:2023scm} claims that the relevant quartic interaction in the case of USR dynamics (that is, the quartic interaction proportional to $\dot{\epsilon_2}$) gives a vanishing contribution when inserted in eq.\,(\ref{eq:Shem1}). 
Ref.\,\cite{Firouzjahi:2023aum} adopts an approach based on the effective field theory of inflation and includes cubic and quartic interactions. It finds that the latter gives a non-trivial contribution, and finds a loop-corrected power spectrum different from the one in ref.\,\cite{Kristiano:2022maq,Kristiano:2023scm}. It would be important to perform a consistent comparison between these two approaches, including the full cubic and quartic interactions in both cases.
Generally speaking, we expect cubic and quartic interactions to be inextricably linked\,\cite{Kristiano:2024ngc,Kristiano:2024vst}. 
For instance, the quartic Hamiltonian receives a
contribution that arises from the modification of the conjugate momentum if there are cubic interactions which depend on $\dot{\zeta}$. 
Similarly, cubic interactions with spatial derivatives are paired with quartic interactions induced by a residual spatial conformal symmetry of the perturbed metric\,\cite{Senatore:2009cf}. 
En route, we notice  that  interactions with spatial derivatives are usually neglected for modes that are  super-horizon. However, in the spirit of the loop computation in ref.\,\cite{Kristiano:2022maq}, the momenta over which the loop is integrated cross the horizon during the USR phase, and, na\"{\i}vely, their spatial derivatives do not pay any  super-horizon suppression.  

In this section, as a preliminary step towards a more complete analysis and in order to compare our results with the claim made in refs.\,\cite{Riotto:2023hoz,Kristiano:2023scm,Riotto:2023gpm,Firouzjahi:2023aum,Firouzjahi:2023ahg,Choudhury:2023vuj,Choudhury:2023jlt,Choudhury:2023rks,Choudhury:2023hvf},  
we only  focus on the cubic interaction in eq.\,(\ref{eq:Yoko}).
However, we stress  that  
all the arguments listed above motivate the need for a more comprehensive analysis. 
\subsection{One-loop computation}\label{sec:1LoopSec}

We consider in this section the cubic interaction Hamiltonian given by (we omit the subscript $_{I}$ in the interaction-picture fields)
\begin{align}
\hat{H}_{\rm int}^{(3)}(\tau) = 
\frac{1}{2}\int d^3\vec{x}\,
\epsilon(\tau)\epsilon_2^{\prime}(\tau)a^2(\tau)
\zeta^{\prime}(\vec{x},\tau)\zeta(\vec{x},\tau)^2\,.\label{eq:MainHami}
\end{align}
We consider 
eq.\,(\ref{eq:Shem2});
this can be written in the compact form 
\begin{align}
\langle\hat{\zeta}(\vec{x}_1,\tau)&
\hat{\zeta}(\vec{x}_2,\tau)\rangle_{2^{\rm nd}} = 
\langle\hat{\zeta}(\vec{x}_1,\tau)
\hat{\zeta}(\vec{x}_2,\tau)\rangle_{2^{{\rm nd}}}^{(1,1)} 
-2 \RE \llp \langle\hat{\zeta}(\vec{x}_1,\tau)
\hat{\zeta}(\vec{x}_2,\tau)\rangle_{2^{{\rm{nd}}}}^{(0,2)} \rrp
\end{align}
where
\begin{align}
\langle\hat{\zeta}(\vec{x}_1,\tau)
\hat{\zeta}(\vec{x}_2,\tau)\rangle_{2^{{\rm{nd}}}}^{(1,1)}  \equiv 
& \int_{-\infty(1+i\epsilon)}^{\tau}
d\tau_{1}
\int_{-\infty(1-i\epsilon)}^{\tau}
d\tau_2
\langle 0|
\hat{H}_{{\rm int}}(\tau_1)
\hat{\zeta}_I(\vec{x}_1,\tau)
\hat{\zeta}_I(\vec{x}_2,\tau)
\hat{H}_{{\rm{int}}}(\tau_2)|0\rangle\,,\label{eq:OneLoop11}
\\
\langle\hat{\zeta}(\vec{x}_1,\tau)
\hat{\zeta}(\vec{x}_2,\tau)\rangle_{2^{\textrm{nd}}}^{(0,2)}  \equiv &   
\int_{-\infty(1-i\epsilon)}^{\tau}d\tau_1
\int_{-\infty(1-i\epsilon)}^{\tau_1}d\tau_2\langle 0|
\hat{\zeta}_I(\vec{x}_1,\tau)
\hat{\zeta}_I(\vec{x}_2,\tau)\hat{H}_{\textrm{int}}(\tau_1)\hat{H}_{\textrm{int}}(\tau_2)|0\rangle\,.\label{eq:OneLoop02} 
\end{align}
This expansion is consistent with Eq.~(16) of Ref.~\cite{Senatore:2009cf}. 
Consider the first contribution in eq.\,(\ref{eq:OneLoop11}), one finds
\begin{align}
\langle\hat{\zeta}(\vec{x}_1,\tau)
\hat{\zeta}(\vec{x}_2,\tau)\rangle_{2^{\textrm{nd}}}^{(1,1)} =  \frac{1}{4}&\int_{-\infty_+}^{\tau}
d\tau_{1}\epsilon(\tau_1)\epsilon_2^{\prime}(\tau_1)
  a^2(\tau_1)
\int_{-\infty_-}^{\tau}
d\tau_2\epsilon(\tau_2)\epsilon_2^{\prime}(\tau_2)
  a^2(\tau_2)\int d^3\vec{y}d^3\vec{z}\nn\\
  &
  \int\left[\prod_{i=1}^8\frac{d^3\vec{k}_i}{(2\pi)^3}\right]
  e^{i\vec{y}\cdot (\vec{k}_1 + \vec{k}_2 + \vec{k}_3)}
  e^{i(\vec{x}_1\cdot \vec{k}_4+\vec{x}_2\cdot \vec{k}_5)}
  e^{i\vec{z}\cdot (\vec{k}_6 + \vec{k}_7 + \vec{k}_8)}\nn\\
  &
  \langle 0|
\hat{\zeta}^{\prime}_I(\vec{k}_1,\tau_1)
\hat{\zeta}_I(\vec{k}_2,\tau_1)
\hat{\zeta}_I(\vec{k}_3,\tau_1)
\hat{\zeta}_I(\vec{k}_4,\tau)
\hat{\zeta}_I(\vec{k}_5,\tau)
\hat{\zeta}^{\prime}_I(\vec{k}_6,\tau_2)
\hat{\zeta}_I(\vec{k}_7,\tau_2)
\hat{\zeta}_I(\vec{k}_8,\tau_2)
  |0\rangle\,.\label{eq:Main11}
\end{align}
The $36$ connected Wick contractions can be expressed as 
\begin{align}
\langle\hat{\zeta}(\vec{x}_1,\tau)
\hat{\zeta}(\vec{x}_2,\tau)\rangle_{2^{{\rm nd}}}^{(1,1)}   = 
&
\int_{-\infty_+}^{\tau}
d\tau_{1}\epsilon(\tau_1)\epsilon_2^{\prime}(\tau_1)
  a^2(\tau_1)
\int_{-\infty_-}^{\tau}
d\tau_2\epsilon(\tau_2)\epsilon_2^{\prime}(\tau_2)
  a^2(\tau_2) 
\int\frac{d^3\vec{k}}{(2\pi)^3} 
\frac{d^3\vec{q}}{(2\pi)^3}\,e^{i(\vec{x}_1 - \vec{x}_2)\cdot(\vec{k}+\vec{q})}
\nn\\
&\big [ \left|\zeta_{k+q}(\tau)\right|^2
\big\{
\zeta_{k}(\tau_1)
\zeta^{\prime}_{k+q}(\tau_1)
\zeta_{q}(\tau_1)
\zeta_{k}^*(\tau_2)
\zeta_{k+q}^{\prime\,*}(\tau_2)
\zeta_q^*(\tau_2) + \nn\\
&\hspace{2cm}
\zeta_k(\tau_1)\zeta^{\prime}_{k+q}(\tau_1)\zeta_q(\tau_1)\zeta_{k+q}^*(\tau_2)\big[
\zeta_k^{\prime\,*}(\tau_2)\zeta_q^*(\tau_2) +
\zeta_k^*(\tau_2)\zeta_q^{\prime\,*}(\tau_2)
\big]+ \nn\\
&\hspace{2cm}
\zeta_k^*(\tau_2)\zeta^{\prime\,*}_{k+q}(\tau_2)\zeta_q^*(\tau_2)\zeta_{k+q}(\tau_1)\big[
\zeta_k^{\prime}(\tau_1)\zeta_q(\tau_1) +
\zeta_k(\tau_1)\zeta_q^{\prime}(\tau_1)
\big]+ \nn\\
&\hspace{2cm}
\zeta_k^{\prime}(\tau_1)\zeta_{k+q}(\tau_1)\zeta_q(\tau_1)\zeta_{k+q}^*(\tau_2)
\big[
\zeta_q^*(\tau_2)\zeta_k^{\prime\,*}(\tau_2) +
\zeta_k^*(\tau_2)\zeta_q^{\prime\,*}(\tau_2)
\big]+ \nn\\
&\hspace{2cm}
\zeta_k^{\prime\,*}(\tau_2)\zeta_{k+q}^*(\tau_2)\zeta_q^*(\tau_2)\zeta_{k+q}(\tau_1)
\big[
\zeta_q(\tau_1)\zeta_k^{\prime}(\tau_1) +
\zeta_k(\tau_1)\zeta_q^{\prime}(\tau_1)
\big]\big\}
\big ]\,.\label{eq:1stContra}
\end{align}

Consider now eq.\,(\ref{eq:OneLoop02}). 
One can write it in the form
\begin{align}
\langle\hat{\zeta}(\vec{x}_1,\tau)
\hat{\zeta}(\vec{x}_2,\tau)\rangle_{2^{\textrm{nd}}}^{(0,2)} =  
\frac{1}{4}
&\int_{-\infty_-}^{\tau}
d\tau_{1}\epsilon(\tau_1)\epsilon_2^{\prime}(\tau_1)
  a^2(\tau_1)
\int_{-\infty_-}^{\tau_1}
d\tau_2\epsilon(\tau_2)\epsilon_2^{\prime}(\tau_2)
  a^2(\tau_2)\int d^3\vec{y}d^3\vec{z}\nn\\
  &
  \int\left[\prod_{i=1}^8\frac{d^3\vec{k}_i}{(2\pi)^3}\right]
  e^{i\vec{y}\cdot (\vec{k}_1 + \vec{k}_2 + \vec{k}_3)}
  e^{i(\vec{x}_1\cdot \vec{k}_4+\vec{x}_2\cdot \vec{k}_5)}
  e^{i\vec{z}\cdot (\vec{k}_6 + \vec{k}_7 + \vec{k}_8)}\nn\\
  &
  \langle 0|
  \hat{\zeta}_I(\vec{k}_4,\tau)
\hat{\zeta}_I(\vec{k}_5,\tau)
\hat{\zeta}^{\prime}_I(\vec{k}_1,\tau_1)
\hat{\zeta}_I(\vec{k}_2,\tau_1)
\hat{\zeta}_I(\vec{k}_3,\tau_1)
\hat{\zeta}^{\prime}_I(\vec{k}_6,\tau_2)
\hat{\zeta}_I(\vec{k}_7,\tau_2)
\hat{\zeta}_I(\vec{k}_8,\tau_2)
  |0\rangle\,.\label{eq:Main02}
\end{align}
After Wick contractions, we find
\begin{align}
 \langle\hat{\zeta}(\vec{x}_1,\tau)
\hat{\zeta}(\vec{x}_2,\tau)\rangle_{2^{\textrm{nd}}}^{(0,2)} 
=
&\int_{-\infty_-}^{\tau}
d\tau_{1}\epsilon(\tau_1)\epsilon_2^{\prime}(\tau_1)
  a^2(\tau_1)
\int_{-\infty_-}^{\tau_1}
d\tau_2\epsilon(\tau_2)\epsilon_2^{\prime}(\tau_2)
  a^2(\tau_2) 
\int\frac{d^3\vec{k}}{(2\pi)^3} 
\frac{d^3\vec{q}}{(2\pi)^3}
e^{i(\vec{x}_1 - \vec{x}_2)\cdot(\vec{k}+\vec{q})}
\nn\\
&\big[
\zeta_{k+q}^2(\tau)
\big\{
\zeta_{k}(\tau_1)
\zeta^{\prime\,*}_{k+q}(\tau_1)
\zeta_{q}(\tau_1)
\zeta_{k}^*(\tau_2)
\zeta_{k+q}^{\prime\,*}(\tau_2)
\zeta_q^*(\tau_2) + \nn\\
& \hspace{1.5cm}
\zeta_k(\tau_1)\zeta^{\prime\,*}_{k+q}(\tau_1)\zeta_q(\tau_1)\zeta_{k+q}^*(\tau_2)\big[
\zeta_k^{\prime\,*}(\tau_2)\zeta_q^*(\tau_2) +
\zeta_k^*(\tau_2)\zeta_q^{\prime\,*}(\tau_2)
\big] + \nn\\
&\hspace{1.5cm}
\zeta_k^*(\tau_2)\zeta^{\prime\,*}_{k+q}(\tau_2)\zeta_q^*(\tau_2)\zeta_{k+q}^*(\tau_1)\big[
\zeta_k^{\prime}(\tau_1)\zeta_q(\tau_1) +
\zeta_k(\tau_1)\zeta_q^{\prime}(\tau_1)
\big]+ \nn\\
&\hspace{1.5cm}
\zeta_k^{\prime}(\tau_1)\zeta_{k+q}^*(\tau_1)\zeta_q(\tau_1)\zeta_{k+q}^*(\tau_2)
\big[
\zeta_q^*(\tau_2)\zeta_k^{\prime\,*}(\tau_2) +
\zeta_k^*(\tau_2)\zeta_q^{\prime\,*}(\tau_2)
\big]+ \nn\\
&\hspace{1.5cm}
\zeta_k^{\prime\,*}(\tau_2)\zeta_{k+q}^*(\tau_2)\zeta_q^*(\tau_2)\zeta_{k+q}^*(\tau_1)
\big[
\zeta_q(\tau_1)\zeta_k^{\prime}(\tau_1) +
\zeta_k(\tau_1)\zeta_q^{\prime}(\tau_1)
\big]\big\}
\big]\,.\label{eq:2nd3rdContra}
\end{align}
At this point we shift the momentum following the prescription $	k \rightarrow k-q$ in such a way that $k$ is identified with the external momentum.
The power spectrum at one loop can be therefore written as 
\begin{equation}
\mathcal{P}(k) = \lim_{\tau \to 0^-}\left(\frac{k^3}{2\pi^2}\right)
\left\{
\left|\zeta_k(\tau)\right|^2 + 
\frac{1}{(4\pi)^2}\left[\Delta P_1(k,\tau) + \Delta P_2(k,\tau)\right]
\right\}\,,\label{eq:MasterOne}
\end{equation}
with
\begingroup
\allowdisplaybreaks
\begin{align}
\Delta P_1(k,\tau) & \equiv 
4\int_{-\infty_+}^{\tau}
d\tau_{1}\epsilon(\tau_1)\epsilon_2^{\prime}(\tau_1)
  a^2(\tau_1)
\int_{-\infty_-}^{\tau}
d\tau_2\epsilon(\tau_2)\epsilon_2^{\prime}(\tau_2)
  a^2(\tau_2) 
\int_0^{\infty}dq\,q^2\,d(\cos\theta)
\left|\zeta_{k}(\tau)\right|^2 
\nn
\\
&\times \big\{
\zeta_{k-q}(\tau_1)
\zeta^{\prime}_{k}(\tau_1)
\zeta_{q}(\tau_1)
\zeta_{k-q}^*(\tau_2)
\zeta_{k}^{\prime\,*}(\tau_2)
\zeta_q^*(\tau_2) + 
\nn\\ 
& \hspace{.62cm}
\zeta_{k-q}(\tau_1)\zeta^{\prime}_{k}(\tau_1)\zeta_q(\tau_1)\zeta_{k}^*(\tau_2)\big[
\zeta_{k-q}^{\prime\,*}(\tau_2)\zeta_q^*(\tau_2) +
\zeta_{k-q}^*(\tau_2)\zeta_q^{\prime\,*}(\tau_2)
\big]+ \nn\\
&\hspace{.62cm}
\zeta_{k-q}^*(\tau_2)\zeta^{\prime\,*}_{k}(\tau_2)\zeta_q^*(\tau_2)\zeta_{k}(\tau_1)\big[
\zeta_{k-q}^{\prime}(\tau_1)\zeta_q(\tau_1) +
\zeta_{k-q}(\tau_1)\zeta_q^{\prime}(\tau_1)
\big]+ \nn\\
&\hspace{.62cm}
\zeta_{k-q}^{\prime}(\tau_1)\zeta_{k}(\tau_1)\zeta_q(\tau_1)\zeta_{k}^*(\tau_2)
\big[
\zeta_q^*(\tau_2)\zeta_{k-q}^{\prime\,*}(\tau_2) +
\zeta_{k-q}^*(\tau_2)\zeta_q^{\prime\,*}(\tau_2)
\big]+ \nn\\
&\hspace{.62cm}
\zeta_{k-q}^{\prime\,*}(\tau_2)\zeta_{k}^*(\tau_2)\zeta_q^*(\tau_2)\zeta_{k}(\tau_1)
\big[
\zeta_q(\tau_1)\zeta_{k-q}^{\prime}(\tau_1) +
\zeta_{k-q}(\tau_1)\zeta_q^{\prime}(\tau_1)
\big]\big\}\,,\label{eq:DeltaP1Full}
\\
\Delta P_2(k,\tau) & \equiv 
-8 \RE \Big [ \int_{-\infty_-}^{\tau}
d\tau_{1}\epsilon(\tau_1)\epsilon_2^{\prime}(\tau_1)
  a^2(\tau_1)
\int_{-\infty_-}^{\tau_1}
d\tau_2\epsilon(\tau_2)\epsilon_2^{\prime}(\tau_2)
  a^2(\tau_2) 
\int_0^{\infty}dq\,q^2\,d(\cos\theta) \zeta_{k}^2(\tau)
\nn \\
& \times 
\big\{
\zeta_{k-q}(\tau_1)
\zeta^{\prime\,*}_{k}(\tau_1)
\zeta_{q}(\tau_1)
\zeta_{k-q}^*(\tau_2)
\zeta_{k}^{\prime\,*}(\tau_2)
\zeta_q^*(\tau_2) + \nn\\
& \hspace{0.62cm}
\zeta_{k-q}(\tau_1)\zeta^{\prime\,*}_{k}(\tau_1)\zeta_q(\tau_1)\zeta_{k}^*(\tau_2)\big[
\zeta_{k-q}^{\prime\,*}(\tau_2)\zeta_q^*(\tau_2) +
\zeta_{k-q}^*(\tau_2)\zeta_q^{\prime\,*}(\tau_2)
\big] + \nn\\
&\hspace{0.62cm}
\zeta_{k-q}^*(\tau_2)\zeta^{\prime\,*}_{k}(\tau_2)\zeta_q^*(\tau_2)\zeta_{k}^*(\tau_1)\big[
\zeta_{k-q}^{\prime}(\tau_1)\zeta_q(\tau_1) +
\zeta_{k-q}(\tau_1)\zeta_q^{\prime}(\tau_1)
\big]+ \nn\\
&\hspace{0.62cm}
\zeta_{k-q}^{\prime}(\tau_1)\zeta_{k}^*(\tau_1)\zeta_q(\tau_1)\zeta_{k}^*(\tau_2)
\big[
\zeta_q^*(\tau_2)\zeta_{k-q}^{\prime\,*}(\tau_2) +
\zeta_{k-q}^*(\tau_2)\zeta_q^{\prime\,*}(\tau_2)
\big]+ \nn\\
&\hspace{0.62cm}
\zeta_{k-q}^{\prime\,*}(\tau_2)\zeta_{k}^*(\tau_2)\zeta_q^*(\tau_2)\zeta_{k}^*(\tau_1)
\big[
\zeta_q(\tau_1)\zeta_{k-q}^{\prime}(\tau_1) +
\zeta_{k-q}(\tau_1)\zeta_q^{\prime}(\tau_1)
\big]\big\}
\Big ]
\,.\label{eq:DeltaP2Full}
\end{align}
\endgroup

\subsubsection{Loop correction with a large hierarchy of scales}\label{sec:LoopHier}

First, we will be concerned with external momenta that describe the large CMB scales, while the USR takes place when modes $k_{\rm USR} \gg k$ cross the horizon. 
The situation is summarized in the following schematic
\begin{align}
\begin{adjustbox}{max width=0.9\textwidth}
\raisebox{-8mm}{
	\begin{tikzpicture}
      \draw[thick,dotted,color=magenta][thick] (-5.12,1)--(-5.12,-6.2);  
\draw[draw=white,fill=magenta!25] (1,1) rectangle ++(0.7,-7);
\draw[draw=white,fill=blue!25] (-6,-4) rectangle ++(+12,-0.4);
	\draw[->,>=Latex,thick][thick] (-6,-6)--(6,-6);
 \draw[->,>=Latex,thick][thick] (-6,-6)--(-6,1);
  \draw[red,thick][thick] (-6,0)--(4.5,-6);
  	\draw[thick,dashed][thick] (-6,-0.5)--(6,-0.5);
   \draw[thick,dotted,color=magenta][thick] (2.6,1)--(2.6,-6.2);  
  	\draw[thick,dotted,color=blue][thick] (-6,-4)--(6,-4);  
     	\draw[thick,dotted,color=blue][thick] (-6,-4.4)--(6,-4.4);
    \node at (-6.5,1.5) {\scalebox{1}{$\textrm{comoving length}\,\lambda$}}; 
    \node at (7.2,-6.5) {\scalebox{1}{$\textrm{comoving time}$}};
 \node at (5.4,-0.1) {\scalebox{1}{$\textrm{long mode}\,\lambda =1/k$}};
  \node at (4.5,-4.2) {\scalebox{1}{{\color{blue}{
  $\textrm{short modes}$}}}};
    \node at (-6.4,-3.9) {\scalebox{1}{{\color{blue}{$q_{\textrm{in}}$}}}};
    \node at (-6.4,-4.4) {\scalebox{1}{{\color{blue}{$q_{\textrm{end}}$}}}};
    \node at (2.6,-6.5) {\scalebox{1}{{\color{magenta}{$\tau$}}}}; 
    \node at (4.62,-5.5) {\scalebox{1}{{\color{red}{$1/aH$}}}};   
    \node at (-6.75,-6.5) {\scalebox{1}{{\color{magenta}{$\tau\to -\infty$}}}};  
     \node at (-5.12,-6.5) {\scalebox{1}{{\color{magenta}{$\tau_k$}}}};  
     \node at (5,-6.5) {\scalebox{1}{{\color{magenta}{$\tau\to 0^-$}}}};
      \node at (1.35,-2.3) {\scalebox{0.9}{{\color{magenta}{$\textrm{USR}$}}}};
      \node at (0.97,-6.25) {\scalebox{1}{{\color{magenta}{$\tau_{\textrm{in}}$}}}}; 
    \node at (1.75,-6.25) {\scalebox{1}{{\color{magenta}{$\tau_{\textrm{end}}$}}}};
	\end{tikzpicture}
 }
\end{adjustbox} 
\label{eq:TimeEvo}
\end{align}
in which the blue horizontal band represents the interval of modes that cross the horizon during the USR phase, the vertical band shaded in magenta. 
In other words, as we will restrict the integration over momenta $q  \in [q_{\rm in},q_{\rm end}]$ that are enhanced by the USR phase, we can assume $q\gg k$.
 Consequently, as in ref.\,\cite{Kristiano:2022maq}, 
we approximate
\begin{align}
 k - q = \sqrt{k^2 + q^2 - 2kq\cos(\theta)} \approx q\,,~~~~~~~
 \int_{-1}^{+1} d(\cos\theta) = 2\,.\label{eq:ApproxQ}
\end{align}
With these assumptions, we can further simplify the expressions. 
We collect each contribution depending on the number of time derivatives acting on the long mode $\zeta_k$.
In each expression, the first line indicates terms with no derivative on the long modes, the second one those with one derivative, while the last with two.
One finds 
\begin{align}
\Delta P_1(k,\tau) & \equiv 
8\int_{\tau_{\rm in}}^{\tau}
d\tau_{1}\epsilon(\tau_1)\epsilon_2^{\prime}(\tau_1)
  a^2(\tau_1)
\int_{\tau_{\rm in}}^{\tau}
d\tau_2\epsilon(\tau_2)\epsilon_2^{\prime}(\tau_2)
  a^2(\tau_2) 
\int_{q_{\rm in}}^{q_{\rm end}}dq\,q^2\,
\left|\zeta_{k}(\tau)\right|^2 
 \nn\\
&
\times \big\{
4 \zeta_{k}(\tau_1)
\zeta_{k}^*(\tau_2) 
\zeta_q(\tau_1)
\zeta_{q}^{\prime}(\tau_1)
\zeta_q^*(\tau_2)
\zeta_{q}^{\prime\,*}(\tau_2) +
\nn\\ 
& \hspace{.62cm}
2 
\zeta^{\prime}_{k}(\tau_1)
\zeta_{k}^*(\tau_2)
\zeta_{q}(\tau_1)
\zeta_q(\tau_1)
\zeta_{q}^{\prime\,*}(\tau_2)
\zeta_q^*(\tau_2)
+
2
\zeta_{k}(\tau_1)
\zeta^{\prime\,*}_{k}(\tau_2)
\zeta_q(\tau_1) 
\zeta_{q}^{\prime}(\tau_1)
\zeta_{q}^*(\tau_2)
\zeta_q^*(\tau_2)
+
\nn
\\
& \hspace{.62cm}
\zeta^{\prime}_{k}(\tau_1)
\zeta_{k}^{\prime\,*}(\tau_2)
\zeta_{q}(\tau_1)
\zeta_{q}(\tau_1)
\zeta_{q}^*(\tau_2)
\zeta_q^*(\tau_2) 
\big\}\,,
\label{eq:int_1}
\\
\Delta P_2(k,\tau) & \equiv 
-16 \RE \Big [ \int_{\tau_{\rm in}}^{\tau}
d\tau_{1}\epsilon(\tau_1)\epsilon_2^{\prime}(\tau_1)
  a^2(\tau_1)
\int_{\tau_{\rm in}}^{\tau_1}
d\tau_2\epsilon(\tau_2)\epsilon_2^{\prime}(\tau_2)
  a^2(\tau_2) 
\int_{q_{\rm in}}^{q_{\rm end}}dq\,q^2\, \zeta_{k}(\tau)^2
\nn \\
& \times 
\big\{
4 
\zeta_{k}^*(\tau_1)
\zeta_{k}^*(\tau_2)
\zeta_{q}^{\prime}(\tau_1)
\zeta_q(\tau_1)
\zeta_q^*(\tau_2)
\zeta_{q}^{\prime\,*}(\tau_2)
+
\nn\\
& \hspace{0.62cm}
2 
\zeta^{\prime\,*}_{k}(\tau_1)
\zeta^*_{k}(\tau_2)
\zeta_{q}(\tau_1)^2
\zeta_{q}^{\prime\,*}(\tau_2)
\zeta_q^*(\tau_2) 
+ 
2 
\zeta^*_{k}(\tau_1)
\zeta^{\prime\,*}_{k}(\tau_2)
\zeta_{q}^{\prime}(\tau_1)
\zeta_q(\tau_1)
\zeta_{q}^*(\tau_2)^2
+ \nn\\
&\hspace{0.62cm}
\zeta^{\prime\,*}_{k}(\tau_1)
\zeta_{k}^{\prime\,*}(\tau_2)
\zeta_{q}(\tau_1)^2
\zeta_{q}^*(\tau_2)^2
\big\}
\Big ]
\,.
\label{eq:int_2}
\end{align}
We can combine the two contributions using the properties of symmetric integrals for holomorphic symmetric functions
$f(\tau_1,\tau_2) = f(\tau_2,\tau_1)$ \cite{Adshead:2008gk}
\begin{equation}
\int_{\tau_{\rm in}}^{\tau} d \tau_1 \int_{\tau_{\rm in}}^{\tau_1} d \tau_2 f\left(\tau_1, \tau_2\right)=
\frac{1}{2} \int_{\tau_{\rm in}}^{\tau} d \tau_1 \int_{\tau_{\rm in}}^{\tau} d \tau_2 f\left(\tau_1, \tau_2\right).\label{eq:InteIde}
\end{equation}
To shorten the notation, we introduce $\Delta P(k,\tau) = \Delta P_1(k,\tau)+\Delta P_2(k,\tau)$ and collect the individual contribution order by order in derivatives:
\begin{itemize}
\item {\bf 0th order in time derivatives of the long mode.}
For ease of reading, we introduce 
the short-hand notation 
$\epsilon(\tau)\epsilon_2^{\prime}(\tau)
  a^2(\tau) \equiv g(\tau)$.
Consider the sum of the two integrals 
\begin{align}
\Delta P_{0{\rm th}}&(k,\tau) =
32\int_{\tau_{\rm in}}^{\tau}
d\tau_{1}
g(\tau_1)
\int_{\tau_{\rm in}}^{\tau}
d\tau_2 g(\tau_2) 
\int_{q_{\rm in}}^{q_{\rm end}}dq\,q^2\,
\left|\zeta_{k}(\tau)\right|^2 \zeta_{k}(\tau_1)
\zeta_{k}^*(\tau_2) 
\zeta_q(\tau_1)
\zeta_{q}^{\prime}(\tau_1)
\zeta_q^*(\tau_2)
\zeta_{q}^{\prime\,*}(\tau_2) \label{eq:FirstLine}\\&
-64 \RE \Big [ \int_{\tau_{\rm in}}^{\tau}
d\tau_{1}  g(\tau_1)
\int_{\tau_{\rm in}}^{\tau_1}
d\tau_2 g(\tau_2) 
\int_{q_{\rm in}}^{q_{\rm end}}dq\,q^2\, \zeta_{k}(\tau)^2
\zeta_{k}^*(\tau_1)
\zeta_{k}^*(\tau_2)
\zeta_{q}^{\prime}(\tau_1)
\zeta_q(\tau_1)
\zeta_q^*(\tau_2)
\zeta_{q}^{\prime\,*}(\tau_2)\Big ]\,.
\end{align}
We notice that, in the first integral in eq.\,(\ref{eq:FirstLine}), the exchange $\tau_1 \leftrightarrow \tau_2$ transforms 
\begin{align}  
\zeta_{k}(\tau_1)
\zeta_{k}^*(\tau_2) 
\zeta_q(\tau_1)
\zeta_{q}^{\prime}(\tau_1)
\zeta_q^*(\tau_2)
\zeta_{q}^{\prime\,*}(\tau_2) 
\overset{
\tau_1 \leftrightarrow \tau_2
}{\Longrightarrow} &
\zeta_{k}(\tau_2)
\zeta_{k}^*(\tau_1) 
\zeta_q(\tau_2)
\zeta_{q}^{\prime}(\tau_2)
\zeta_q^*(\tau_1)
\zeta_{q}^{\prime\,*}(\tau_1) = \nn \\&  
[\zeta_{k}(\tau_1)
\zeta_{k}^*(\tau_2) 
\zeta_q(\tau_1)
\zeta_{q}^{\prime}(\tau_1)
\zeta_q^*(\tau_2)
\zeta_{q}^{\prime\,*}(\tau_2) ]^*\,.
\end{align}
Therefore, the first integral in eq.\,(\ref{eq:FirstLine}) is fully symmetric under the exchange $\tau_1 \leftrightarrow \tau_2$, and
 we rewrite $\Delta P_{0{\rm th}}(k,\tau)$ as 
\begin{align}
\Delta P_{0{\rm th}}&(k,\tau)  = 
\nn \\&
32\int_{\tau_{\rm in}}^{\tau}
d\tau_{1}
g(\tau_1)
\int_{\tau_{\rm in}}^{\tau}
d\tau_2 g(\tau_2) 
\int_{q_{\rm in}}^{q_{\rm end}}dq\,q^2\,
\left|\zeta_{k}(\tau)\right|^2
\RE\big[\zeta_{k}(\tau_1)
\zeta_{k}^*(\tau_2) 
\zeta_q(\tau_1)
\zeta_{q}^{\prime}(\tau_1)
\zeta_q^*(\tau_2)
\zeta_{q}^{\prime\,*}(\tau_2)\big]\label{eq:FirstLine2}\\&
-64 \int_{\tau_{\rm in}}^{\tau}
d\tau_{1}  g(\tau_1)
\int_{\tau_{\rm in}}^{\tau_1}
d\tau_2 g(\tau_2) 
\int_{q_{\rm in}}^{q_{\rm end}}dq\,q^2\, 
\RE\big[\zeta_{k}(\tau)^2 
\zeta_{k}^*(\tau_1)
\zeta_{k}^*(\tau_2)
\zeta_{q}^{\prime}(\tau_1)
\zeta_q(\tau_1)
\zeta_q^*(\tau_2)
\zeta_{q}^{\prime\,*}(\tau_2)\big]\,.
\end{align}
and apply to the first integral in eq.\,(\ref{eq:FirstLine2})
the identity in eq.\,(\ref{eq:InteIde}). 
We arrive  at 
\begin{align}
\Delta P_{0{\rm th}}(k,\tau) & =   64\int_{\tau_{\rm in}}^{\tau}
d\tau_{1}
g(\tau_1)
\int_{\tau_{\rm in}}^{\tau_1}
d\tau_2 g(\tau_2) 
\int_{q_{\rm in}}^{q_{\rm end}}dq\,q^2 \times
\nn\\
&
\{
\RE\big[
\zeta_{k}(\tau)
\zeta_{k}^*(\tau)
\zeta_{k}(\tau_1)
\zeta_{k}^*(\tau_2) 
\zeta_q(\tau_1)
\zeta_{q}^{\prime}(\tau_1)
\zeta_q^*(\tau_2)
\zeta_{q}^{\prime\,*}(\tau_2)\big] \nn \\ 
&- 
\RE\big[\zeta_{k}(\tau)^2 
\zeta_{k}^*(\tau_1)
\zeta_{k}^*(\tau_2)
\zeta_{q}^{\prime}(\tau_1)
\zeta_q(\tau_1)
\zeta_q^*(\tau_2)
\zeta_{q}^{\prime\,*}(\tau_2)\big]
\}\,.\label{eq:GUH}
\end{align}
We are now in the position of combining  the two integrand functions. 
Schematically, we define the two combinations 
\begin{align}
X \equiv  \zeta_{k}^*(\tau)\zeta_{k}(\tau_1)\,,~~~~~~~~~
Y \equiv \zeta_{k}(\tau)
\zeta_q(\tau_1)
\zeta_{q}^{\prime}(\tau_1)
\zeta_{k}^*(\tau_2)
\zeta_q^*(\tau_2)
\zeta_{q}^{\prime\,*}(\tau_2)\,,
\end{align}
such that the integrand in eq.\,(\ref{eq:GUH}) becomes 
\begin{align}\label{eq:identityRE}
\RE(XY) - \RE(X^*Y) 
= -2\IM(X)\IM(Y)\,.
\end{align}
We thus arrive at the result
\begin{mynamedbox2}{Contribution with no time derivatives on the long mode  $k$}
\vspace{-.35cm}
\begin{align}
\Delta P_{0{\rm th}}(k,\tau)  = &
-128
\int_{\tau_{\rm in}}^{\tau}
d\tau_{1}\epsilon(\tau_1)\epsilon_2^{\prime}(\tau_1) a^2(\tau_1)
\int_{\tau_{\rm in}}^{\tau_1}
d\tau_2\epsilon(\tau_2)\epsilon_2^{\prime}(\tau_2) a^2(\tau_2) 
\int_{q_{\rm in}}^{q_{\rm end}}dq\,q^2
\nn \\
&\times 
\IM \llp\zeta_{k}^*(\tau)  
\zeta_{k}(\tau_1)  \rrp
\IM \llp \zeta_{k}(\tau)
\zeta_q(\tau_1)
\zeta_{q}^{\prime}(\tau_1)
\zeta_{k}^*(\tau_2)
\zeta_q^*(\tau_2)
\zeta_{q}^{\prime\,*}(\tau_2) 
\rrp.
\label{0th}
\end{align}
\end{mynamedbox2}
Given that we are interested in modes $k$ that are much smaller than the USR-enhanced ones, they are super-horizon at the time of USR phase. Thus, for any time $\tau\gtrsim \tau_{\rm in}$ of relevance for both time integrations, one has that
\begin{equation}
    \IM\llp\zeta_k (\tau)\zeta_k^*(\tau_1)\rrp 
	\simeq \IM\llp |\zeta_k(\tau)|^2\rrp = 0,
	\label{eq:immod}
\end{equation}
which makes the above contribution negligible.

\item {\bf 1st order in time derivatives of the long mode.}
Starting from the second lines of eqs.~(\ref{eq:int_1},\ref{eq:int_2}), we now consider the sum 
\begin{align}
\Delta P_{\rm 1st}(k,\tau) = & 16\int_{\tau_{\rm in}}^{\tau}
d\tau_{1} g(\tau_1)
\int_{\tau_{\rm in}}^{\tau}
d\tau_2 g(\tau_2) 
\int_{q_{\rm in}}^{q_{\rm end}}dq\,q^2\,
\left|\zeta_{k}(\tau)\right|^2 
\nn\\
&~~~~ 
\big[
\zeta^{\prime}_{k}(\tau_1)
\zeta_{k}^*(\tau_2)
\zeta_{q}(\tau_1)
\zeta_q(\tau_1)
\zeta_{q}^{\prime\,*}(\tau_2)
\zeta_q^*(\tau_2)
+
\zeta_{k}(\tau_1)
\zeta^{\prime\,*}_{k}(\tau_2)
\zeta_q(\tau_1) 
\zeta_{q}^{\prime}(\tau_1)
\zeta_{q}^*(\tau_2)
\zeta_q^*(\tau_2)\big] \nn\\
-&  32 \RE \Big \{ \int_{\tau_{\rm in}}^{\tau}
d\tau_{1}g(\tau_1)
\int_{\tau_{\rm in}}^{\tau_1}
d\tau_2g(\tau_2) 
\int_{q_{\rm in}}^{q_{\rm end}}dq\,q^2
\zeta_{k}(\tau)^2
\nn \\
&~~~~\big[ 
\zeta^{\prime\,*}_{k}(\tau_1)
\zeta^*_{k}(\tau_2)
\zeta_{q}(\tau_1)^2
\zeta_{q}^{\prime\,*}(\tau_2)
\zeta_q^*(\tau_2) 
+ 
\zeta^*_{k}(\tau_1)
\zeta^{\prime\,*}_{k}(\tau_2)
\zeta_{q}^{\prime}(\tau_1)
\zeta_q(\tau_1)
\zeta_{q}^*(\tau_2)^2\big]\Big \}\,.
\end{align}
Manipulations analogue to those discussed in the previous point allow one to combine the two integrals together. We find 
\begin{align}
\Delta P_{\rm 1st}(k,\tau) = & -64\int_{\tau_{\rm in}}^{\tau}
d\tau_{1}g(\tau_1)
\int_{\tau_{\rm in}}^{\tau_1}
d\tau_2g(\tau_2) 
\int_{q_{\rm in}}^{q_{\rm end}}dq\,q^2
\nn \\
& \{
\IM\llp
\zeta_{k}^*(\tau)
 \zeta_{k}(\tau_1)
 \rrp
 \IM\llp
 \zeta_{k}(\tau)
\zeta_{k}^*(\tau_2)^2
\zeta_q(\tau_1)
\zeta_{k}^{\prime\,*}(\tau_2)
\zeta_q^{\prime}(\tau_1)
\rrp +
\IM\llp\zeta_{k}^*(\tau)
 \zeta_{k}^{\prime}(\tau_1)\rrp \nn \\
 &\IM\llp
 \zeta_{k}(\tau)
\zeta_q(\tau_1)^2
\zeta_{k}^*(\tau_2)
\zeta_q^*(\tau_2)
\zeta_{q}^{\prime\,*}(\tau_2)\rrp
\}.
\end{align}
Again, since we are interested in modes $k$ that are much smaller than the USR-enhanced ones, and are super-horizon at the time of USR phase, the contribution within the curly brackets in the first line vanishes thanks to eq.~\eqref{eq:immod}. This leaves us with
\begin{mynamedbox2}{Contribution with one time derivatives on the long mode   $k$}
\vspace{-.35cm}
\begin{align}
\Delta P_{\rm 1st}(k,\tau) 
& 
\equiv 
-64
\int_{\tau_{\rm in}}^{\tau}
d\tau_{1}\epsilon(\tau_1)\epsilon_2^{\prime}(\tau_1)a^2(\tau_1)
\int_{\tau_{\rm in}}^{\tau_1}
d\tau_2\epsilon(\tau_2)\epsilon_2^{\prime}(\tau_2)a^2(\tau_2) 
\int_{q_{\rm in}}^{q_{\rm end}}dq\,q^2\,
\nn\\
&
\times
 \IM\llp\zeta_{k}^*(\tau)
 \zeta_{k}^{\prime}(\tau_1)\rrp
 \IM\llp
 \zeta_{k}(\tau)
\zeta_q(\tau_1)
\zeta_q(\tau_1)
\zeta_{k}^*(\tau_2)
\zeta_q^*(\tau_2)
\zeta_{q}^{\prime\,*}(\tau_2)\rrp.
 \label{1th}
\end{align}
\end{mynamedbox2}

\item {\bf 2nd order in time derivatives of the long mode.}
Analogue manipulations give
\begin{mynamedbox2}{Contribution with 
two time derivatives on the long mode $k$}
\vspace{-.35cm}
\begin{align}
\Delta P_{\rm 2nd}(k,\tau) 
& 
\equiv 
-32
\int_{\tau_{\rm in}}^{\tau}
d\tau_{1}\epsilon(\tau_1)\epsilon_2^{\prime}(\tau_1)a^2(\tau_1)
\int_{\tau_{\rm in}}^{\tau_1}
d\tau_2\epsilon(\tau_2)\epsilon_2^{\prime}(\tau_2)a^2(\tau_2) 
\int_{q_{\rm in}}^{q_{\rm end}}dq\,q^2\,
\nn\\
&
\times
\IM\llp
\zeta_{k}^*(\tau)  \zeta_{k}^{\prime}(\tau_1)
\rrp
\IM\llp
\zeta_{k}(\tau)
\zeta_q(\tau_1)
\zeta_{q}(\tau_1)
\zeta_{k}^{\prime\,*}(\tau_2)
\zeta_q^*(\tau_2)
\zeta_{q}^{*}(\tau_2)
\rrp.
\label{2th}
\end{align}
\end{mynamedbox2}
\end{itemize}
We stress that the only approximation employed so far is to take the external momentum to be much smaller than the one in the loop, i.e. $k\ll q$, which is justified in presence of a large hierarchy between the CMB and the USR scales.

\subsubsection{Loop correction 
at any scales}\label{sec:any}
It will be useful in the following to remove the assumption that the external momentum is much smaller than the modes in the loop, i.e. the large separation of scales $k\ll q$.
Starting again from eqs.~(\ref{eq:MasterOne},\ref{eq:DeltaP1Full},\ref{eq:DeltaP2Full}), we can proceed with analogous steps as in the previous section and define
\begin{align}
X_1 &\equiv  \zeta_{k}^*(\tau)\zeta_{k}'(\tau_1)\,,
~~~~~~~~~
Y_1 \equiv \zeta_{k}(\tau)
\zeta_{k-q}(\tau_1)
\zeta_{q}(\tau_1)
\zeta_{k-q}^*(\tau_2)
\zeta_{k}^{\prime\,*}(\tau_2)
\zeta_q^*(\tau_2) \,,
\nn
\\
X_2 &\equiv  \zeta_{k}^*(\tau)\zeta_{k}'(\tau_1)\,,
~~~~~~~~~
Y_2 \equiv \zeta_{k}(\tau)
\zeta_{k-q}(\tau_1)
\zeta_q(\tau_1)\zeta_{k}^*(\tau_2)\big[
\zeta_{k-q}^{\prime\,*}(\tau_2)\zeta_q^*(\tau_2) +
\zeta_{k-q}^*(\tau_2)\zeta_q^{\prime\,*}(\tau_2)
\big]
\,,
\nn
\\
X_3 &\equiv \zeta_{k}^*(\tau)\zeta_{k}(\tau_1)\,,
~~~~~~~~~
Y_3 \equiv 
\zeta_{k}(\tau)
\zeta_{k-q}^*(\tau_2)\zeta^{\prime\,*}_{k}(\tau_2)\zeta_q^*(\tau_2)\big[
\zeta_{k-q}^{\prime}(\tau_1)\zeta_q(\tau_1) +
\zeta_{k-q}(\tau_1)\zeta_q^{\prime}(\tau_1)
\nn
\\
X_4 &\equiv \zeta_{k}^*(\tau)\zeta_{k}(\tau_1)\,,
~~~~~~~~~
Y_4 \equiv 
\zeta_{k}(\tau)
\zeta_{k-q}^{\prime}(\tau_1)\zeta_q(\tau_1)\zeta_{k}^*(\tau_2)
\big[
\zeta_q^*(\tau_2)\zeta_{k-q}^{\prime\,*}(\tau_2) +
\zeta_{k-q}^*(\tau_2)\zeta_q^{\prime\,*}(\tau_2)
\big]
\nn
\\
X_5 &\equiv \zeta_{k}^*(\tau)\zeta_{k}(\tau_1)\,,
~~~~~~~~~
Y_5 \equiv 
\zeta_{k}(\tau)
\zeta_{k-q}^{\prime\,*}(\tau_2)\zeta_{k}^*(\tau_2)\zeta_q^*(\tau_2)
\big[
\zeta_q(\tau_1)\zeta_{k-q}^{\prime}(\tau_1) +
\zeta_{k-q}(\tau_1)\zeta_q^{\prime}(\tau_1)
\big]
\end{align}
in such a way that $\Delta P \equiv \Delta P_1 + \Delta P_2 $ can be written in the schematic form
\begin{mynamedbox2}{Generic loop correction at any scale $k$}
\vspace{-.35cm}
\begin{align}\label{eq:GenSc}
\Delta P(k,\tau) & \equiv 
-16 \int_{\tau_{\rm in}}^{\tau}
d\tau_{1}
g(\tau_1)
\int_{\tau_{\rm in}}^{\tau_1}
d\tau_2
g(\tau_2)
\int_{q_{\rm in }}^{q_{\rm end}}dq\,q^2\,
\int_{-1}^{1} d(\cos\theta) 
\times \sum_{i = 1}^5 
 \Im (X_i)  \Im (Y_i),
\end{align}
\end{mynamedbox2}
\noindent
thanks to the identity in eq.~\eqref{eq:identityRE} and where we again introduced $\epsilon(\tau)\epsilon_2^{\prime}(\tau)
  a^2(\tau) \equiv g(\tau)$.
This expression is much more intricate than the one obtained in the limit of a large hierarchy of scales between the mode $k$ and the USR loop momenta. It will allow us to seize the loop correction to the power spectrum also at the USR scales where the peak of the power spectrum is generated. 

\subsection{Time integration beyond the instantaneous transition and at any scales}\label{sec:Pheno}

\subsubsection{Loop evaluation at the CMB scales}\label{sec:LoopCMB}

Let us try to simplify the structure of eq.\,(\ref{eq:MasterOne}) in light of the approximations introduced so far. First of all, let us write eq.\,(\ref{eq:MasterOne}) 
in the form
\begin{align}
    \mathcal{P}(k) = \frac{H^2}{8\pi^2\epsilon_{\textrm{ref}}} 
\left\{
1 + 
\lim_{\tau \to 0^-}\frac{4\epsilon_{\textrm{ref}}k^3}{H^2(4\pi)^2}\Delta P_{\rm 1st}(k,\tau) + 
\lim_{\tau \to 0^-}\frac{4\epsilon_{\textrm{ref}}k^3}{H^2(4\pi)^2}\Delta P_{\rm 2nd}(k,\tau)
\right\}
\,,\label{eq:MasterOne2}
\end{align}
where we used the slow-roll approximation for the first term in 
eq.\,(\ref{eq:MasterOne}) given that 
$k$ is of the order of the CMB pivot scale. 
We focus on the leading correction given by $\Delta P_{\rm 1st}(k,\tau)$. 
Using the number of $e$-folds as the time variable,
we find that it can be written in the compact form (cf. our definition in eq.\,(\ref{eq:Pertu}))
\begin{align}
&
\Delta\mathcal{P}_{\rm 1-loop}(k_*)
\equiv 
\lim_{\tau \to 0^-}\frac{4\epsilon_{\textrm{ref}}k^3}{H^2(4\pi)^2}\Delta P_{\rm 1st}(k,\tau)  
= \nn\\
&
 32\left(
\frac{H^2}{8\pi^2\epsilon_{\textrm{ref}}}
\right)
\int_{N_{\rm in} -  
\Delta N}^{N_{\rm end} +  \Delta N} dN_1  
\frac{d\eta}{dN}(N_1) 
\int_{N_{\rm in} -  
\Delta N}^{N_1} dN_2 
\bar{\epsilon}(N_2)
\frac{d\eta}{dN}(N_2)
e^{3(N_2 - N_{\rm in})}
\int\frac{d\bar{q}}{\bar{q}^4}
\IM\left[
\bar{\zeta}_q(N_1)^2
\bar{\zeta}_q^*(N_2)
\frac{d\bar{\zeta}_q^*}{dN}(N_2)
\right]
\label{eq:PS1Simpl}
\end{align}
where we introduced the following manupulations:
\begin{itemize}
\item[{\it i)}] We use the approximation $\epsilon_2(N) \approx -2\eta(N)$. 
This is because in the relevant range of $N$ over which we integrate 
$\epsilon \ll 1$ while $\eta = O(1)$, cf. the right panel of fig.\,\ref{fig:Dyn}.
\item[{\it ii)}]  
We define $\bar{q} \equiv q/a_{\textrm{in}}H$. 
Furthermore, we use the two relations
\begin{align}
\frac{a(N_1)H}{k} = e^{N_1 - N_k}\,,~~~~\textrm{with:}~~~~
a(N_k)H = k\,,
~~~~~~~\textrm{and}~~~~
\frac{a(N_2)H}{q} = \frac{e^{N_2 - N_{\rm in}}}{\bar{q}}\,.\label{eq:CrossingDef}
\end{align}
\item[{\it iii)}] We introduce the short-hand notation
\begin{align}
  \bar{\zeta}_q(N) \equiv 
\frac{
\epsilon_{\textrm{ref}}^{1/2}q^{3/2}\zeta_q(N)}{H}\,.\label{eq:BarField}
\end{align}
The virtue of this definition is that  $\bar{\zeta}_q(N)$ is precisely the quantity we compute numerically by solving the M-S equation, cf. the right panel of fig.\,\ref{fig:TestPS}. 
Furthermore, it should be noted that 
the definition in eq.\,(\ref{eq:BarField}) is automatically invariant under the rescaling in 
eq.\,(\ref{eq:Rescaling}). The same comment applies to the definition of $\bar{q}$ and the ratios in eq.\,(\ref{eq:CrossingDef}). Consequently, an expression entirely written in terms of barred quantities is automatically invariant under the rescaling in 
eq.\,(\ref{eq:Rescaling}).
\item[{\it iv)}] 
Importantly, in the  derivation of 
eq.\,(\ref{eq:PS1Simpl}) we use (cf. appendix\,\ref{app:TimeDer})
\begin{align}\label{eq:wroskcond}
\textrm{Im}\bigg[
 \bar{\zeta}_k^*(N)
 \frac{d\bar{\zeta}_k}{dN}(N_1)
 \bigg] \simeq 
 \textrm{Im}\bigg[
 \bar{\zeta}_k^*(N_1)
 \frac{d\bar{\zeta}_k}{dN}(N_1)
 \bigg]  =  -\frac{\bar{k}^3 }{4\bar{\epsilon}(N_1)}e^{3(N_{\textrm{in}}- N_1)}\,,
\end{align}
with $\epsilon(N)$ given by eq.\,(\ref{eq:DynEps}) for generic $\delta N$. 
This is because $N_{1,2}$
vary at around $N_{\rm end}$, 
and in this time interval modes with comoving wavenumbers $k \approx k_*$
are way outside the horizon and stay constant.
For the  very same reason, we also use the slow-roll approximation
\begin{align}
   \bar{\zeta}_k(N_1)
   \bar{\zeta}^*_k(N_2) = 
   \frac{1}{4}\,.
\end{align}
\item[{\it v)}] The range of integration in 
eq.\,(\ref{eq:PS1Simpl}) is as follows.
In the case of a smooth transition, we take 
$N_{1}\in 
[N_{\textrm{in}} - \Delta N, N_{\textrm{end}} + \Delta N]$ and 
$N_{2}\in 
[N_{\textrm{in}} - \Delta N,  N_1]$
where $\Delta N$ should be large enough to complete the SR/USR/SR transition (that is, $\Delta N \gtrsim \delta N$). 
In the limit of instantaneous transition, we set 
$N_1  = N_2 = N_{\rm end}$, which corresponds to consider the dominant contribution given by the first $\delta$ function in eq.\,(\ref{eq:DeltaDer}). Moreover, we include a factor $1/2$ since, with respect to the integration over $N_2$, the argument of the $\delta$ function in eq.\,(\ref{eq:DeltaDer}) picks up the upper limit  
of the integration interval at $N_{\rm end}$. 
The integration over $q$, on the contrary, is limited by 
$\bar{q} \in [1,e^{\Delta N_{\textrm{USR}}}]$.
\end{itemize}

\subsubsection{The instantaneous transition}\label{sec:Insta}

We consider the instantaneous limit (dubbed $\delta N\to 0$ in the following) of eq.\,(\ref{eq:PS1Simpl}). We find
\begin{align}
\lim_{\delta N \to 0}\Delta\mathcal{P}_{\rm 1-loop}(k_*) =
\left(\frac{H^2}{8\pi^2\epsilon_{\textrm{ref}}} \right)\eta_{\textrm{II}}^2
\left(\frac{k_{\textrm{end}}}{k_{\textrm{in}}}\right)^{-2\eta_{\textrm{II}} + 3}
\lim_{\delta N\to 0}
(16)\int_1^{e^{\Delta N_{\textrm{USR}}}}
\frac{d\bar{q}}{\bar{q}^4}
|\bar{\zeta}_q(N_{\rm end})|^2\IM\left[
\bar{\zeta}_q(N_{\rm end})
\frac{d\bar{\zeta}_q^*}{dN}(N_{\rm end})
\right]\,.
\end{align}
This expression can be further simplified using (cf. eq.\,(\ref{eq:Wrowro}))
\begin{align}
\IM\left[
\bar{\zeta}_q(N_{\rm end})
\frac{d\bar{\zeta}_q^*}{dN}(N_{\rm end})
\right]  = 
\frac{\bar{q}^3}{4}
e^{(2\eta_{\textrm{II}}-3)(N_{\textrm{end}}-
N_{\textrm{in}})}
=  
\frac{\bar{q}^3}{4}\left(
\frac{k_{\rm end}}{k_{\rm in}}
\right)^{2\eta_{\rm II}-3}
\,,
\end{align}
so that we write
\begin{align}
\lim_{\delta N \to 0}\Delta\mathcal{P}_{\rm 1-loop}(k_*) =
\left(\frac{H^2}{8\pi^2\epsilon_{\textrm{ref}}} \right)\,4\eta_{\textrm{II}}^2\,
\lim_{\delta N\to 0}
\int_1^{e^{\Delta N_{\textrm{USR}}}}
\frac{d\bar{q}}{\bar{q}}
|\bar{\zeta}_q(N_{\rm end})|^2\,.\label{eq:FinInte}
\end{align}
We remark that this expression is valid for generic values of $\eta_{\rm II}$ during USR.
 
We consider now the computation of the last integral. 
The factor
$\bar{\zeta}_q$ grows exponentially during the 
USR phase. 
In the case of sub-horizon modes, we have $\bar{\zeta}_q(N) \sim e^{-(1-\eta_{\rm II})N}$ while in the case of super-horizon modes we find $\bar{\zeta}_q(N) \sim  e^{-(3-2\eta_{\rm II})N}$ (cf. appendix\,\ref{app:TimeDer}).
However, the precise estimate of  the integral in eq.\,(\ref{eq:FinInte}) is complicated by the fact that 
curvature modes $\bar{\zeta}_q$ with $\bar{q}\in [1,\exp(\Delta N_{\textrm{USR}})]$ are neither sub- nor super-horizon but they exit the horizon during the USR phase, thus making the analytical estimate of the argument of their exponential growth more challenging.

The situation simplifies if we consider some special values of $\eta_{\textrm{II}}$. 
We consider the case $\eta_{\textrm{II}} = 3$ 
(that is $\epsilon_2 = -6$, it should be noted that this is also the case studied in ref.\,\cite{Kristiano:2022maq}). 
In this case, everything can be computed analytically. 
We find the scaling 
\begin{align}
\lim_{\delta N\to 0}
\int_1^{e^{\Delta N_{\textrm{USR}}}}
\frac{d\bar{q}}{\bar{q}}
|\bar{\zeta}_q(N_{\rm end})|^2 
\approx  \frac{e^{6\Delta N_{\rm USR}}}{4}(1 + \Delta N_{\rm USR})
= 
\frac{1}{4}\left(
\frac{k_{\rm end}}{k_{\rm in}}
\right)^6\left[
1 + \log\left(
\frac{k_{\rm end}}{k_{\rm in}}
\right)
\right]\,,\label{eq:ModeInte}
\end{align}
which becomes more and more accurate for larger $k_{\rm end}/k_{\rm in}$.
The final result is 
\begin{mynamedbox1}{
Leading one-loop correction at CMB scales in the instantaneous SR/USR/SR transition}
\vspace{-0.35cm}
\begin{align}
\lim_{\delta N \to 0}
\Delta\mathcal{P}_{\rm 1-loop}(k_*) \approx 
\left(\frac{H^2}{8\pi^2\epsilon_{\textrm{ref}}} \right) \eta_{\rm II}^2
\left(\frac{k_{\textrm{end}}}{k_{\textrm{in}}}\right)^{6}
\left[
1 + \log\left(
\frac{k_{\rm end}}{k_{\rm in}}
\right)
\right]
\,,~~~~~~~\textrm{with}~~\eta_{\textrm{II}} = 3,\,\,\eta_{\rm III} = 0\label{eq:AhiAhi}
\end{align}
\end{mynamedbox1}
\noindent 
which perfectly agrees with the findings of ref.\,\cite{Kristiano:2022maq} in the same limit.

The above result has a number of limitations, which we address separately:
\begin{itemize}
\item[$\circ$] \textbf{Dynamics during USR.} 
We modify the assumption $\eta_{\rm II} =  3$ and we take $\delta N\to 0$ and $\eta_{\rm III} = 0$.  
\item[$\circ$] \textbf{Dynamics at the SR/USR/SR transition.}
We consider $\delta N \neq 0$, with generic $\eta_{\rm II}$ but $\eta_{\rm III} = 0$. 
Considering a non-zero value of $\delta N$ is very  important because it corresponds to a more realistic smooth SR/USR/SR transition, as opposed to the instantaneous limit with $\delta N = 0$.

\end{itemize}

\subsubsection{Dynamics during USR}\label{sec:GenericUSR}

We compute eq.\,(\ref{eq:FinInte}) for generic values of $\eta_{\rm II}$, still keeping $\delta N\to 0$ and $\eta_{\rm III} = 0$.  
From the computation of the tree-level power spectrum (see fig.\,\ref{fig:TestPS}) we define
\begin{align}
\frac{\mathcal{P}_{\rm USR}}{
\mathcal{P}_{\rm CMB}
} \equiv 
\frac{\mathcal{P}(\bar{k}_{\rm max})}{\mathcal{P}(\bar{k}\ll 1)}\,,\label{eq:PSMax}
\end{align}
where $\bar{k}_{\rm max}$ represent the position of the max of $\mathcal{P}(\bar{k})$ after the growth 
due to the USR dynamics.
\begin{figure}[h]

$$\includegraphics[width=.495\textwidth]{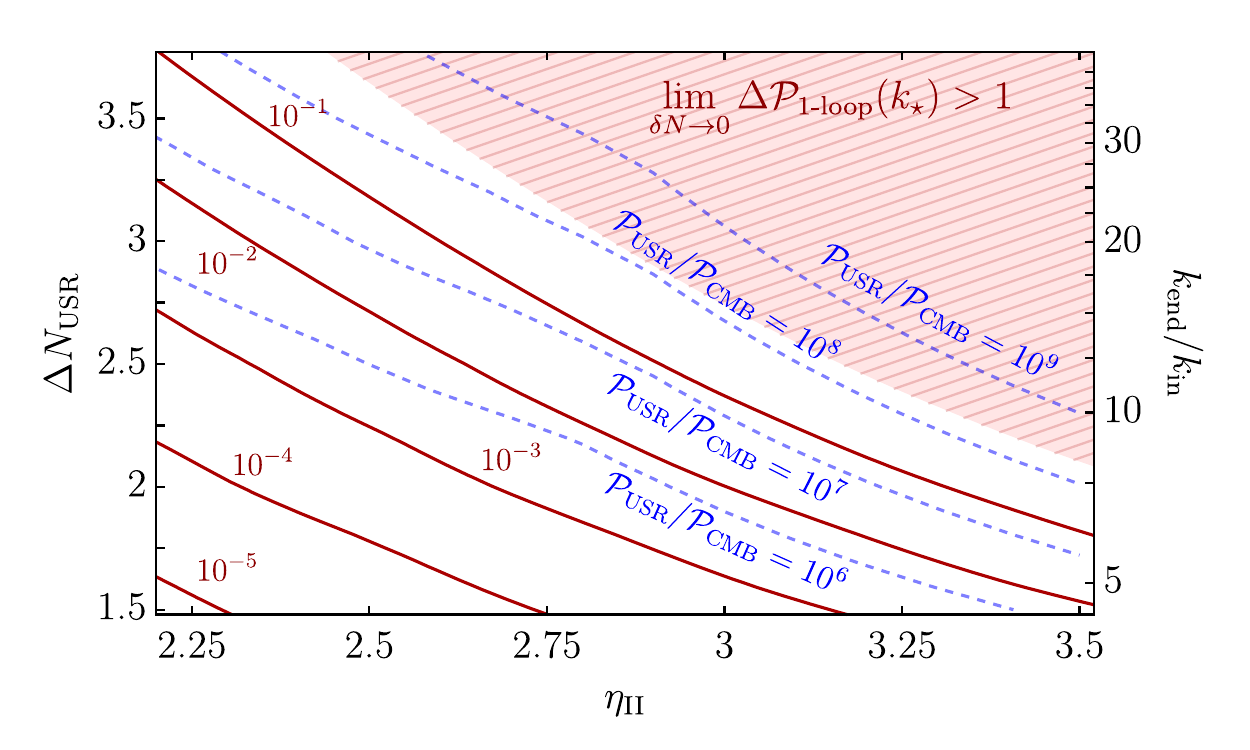}~
\includegraphics[width=.495\textwidth]{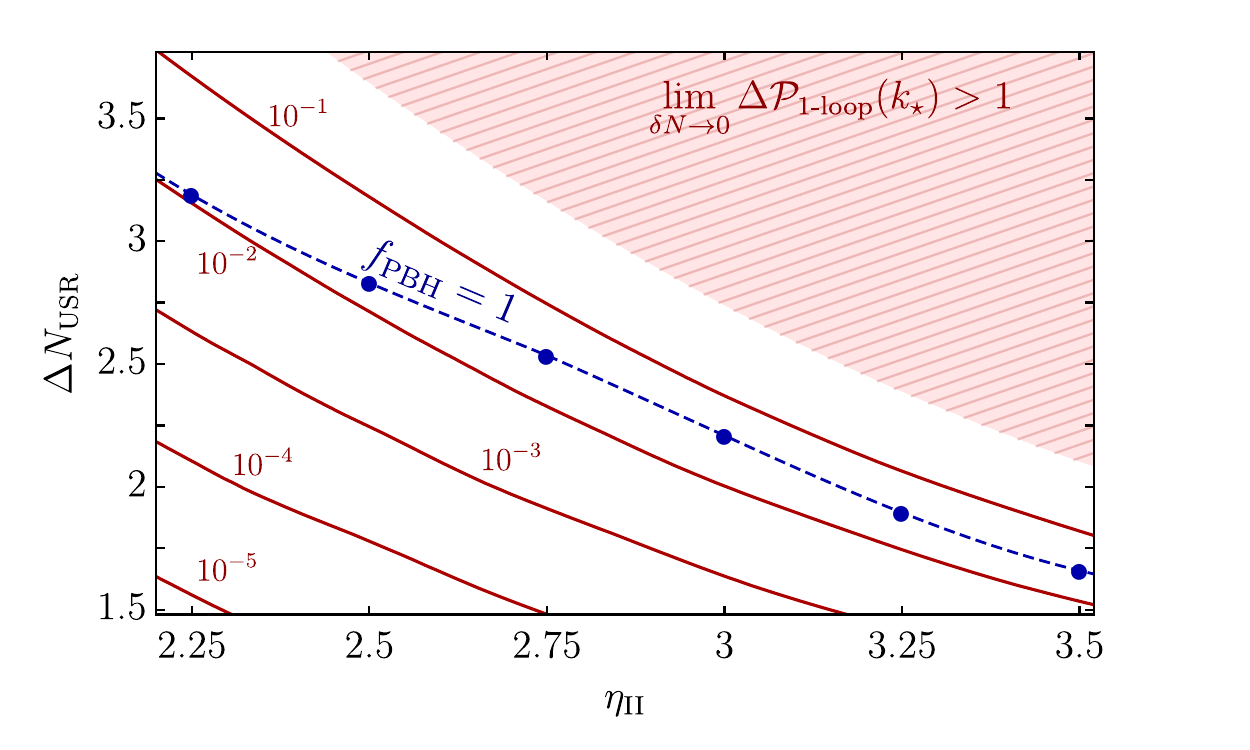}$$
\caption{  
In both panels,  
we consider a generic USR dynamics with varying $\eta_{\rm II}$ ($x$-axis) and $\Delta N_{\rm USR}$ ($y$-axis). We take $\eta_{\rm III} = 0$ and the instantaneous limit $\delta N = 0$.
We plot in solid red
contours of constant $\lim_{\delta N \to 0}
\Delta\mathcal{P}_{\rm 1-loop}(k_*),$
defined in eq.\,(\ref{eq:PS1Simpl}) and
computed according to eq.\,(\ref{eq:FinInte}) with $H^2/8\pi^2\epsilon_{\rm ref} = 2.1\times 10^{-9}$. 
 The region hatched in red 
 is defined by the condition  $\lim_{\delta N \to 0}
\Delta\mathcal{P}_{\rm 1-loop}(k_*) > 1$. 
\textit{\textbf{Left panel:}} We superimpose contours 
of constant $\mathcal{P}_{\rm USR}/\mathcal{P}_{\rm CMB}$ as defined in eq.\,(\ref{eq:PSMax}) (dashed blue). 
\textit{\textbf{Right panel:}} We superimpose the line defined by the condition $f_{\rm PBH} = 1$. 
Along this line, we get 100\% of DM in the form of asteroid-mass PBHs.
 }\label{fig:RegionBound}  
\end{figure}
We compare in the left panel of  fig.\,\ref{fig:RegionBound} contours 
of constant $\mathcal{P}_{\rm USR}/\mathcal{P}_{\rm CMB}$ (dashed blue) and constant $\lim_{\delta N \to 0}
\Delta\mathcal{P}_{\rm 1-loop}(k_*) $ (solid red). 
We take $H^2/8\pi^2\epsilon_{\rm ref} = 2.1\times 10^{-9}$.
Our analysis shows that enhancements  
$\mathcal{P}_{\rm USR}/\mathcal{P}_{\rm CMB} \gtrsim 10^8$  are barely compatible with the perturbativity condition
$\lim_{\delta N \to 0}
\Delta\mathcal{P}_{\rm 1-loop}(k_*) < 1$, which roughly means ``loops $<$ tree level''. The region $\lim_{\delta N \to 0}
\Delta\mathcal{P}_{\rm 1-loop}(k_*) > 1$ is hatched in red in  fig.\,\ref{fig:RegionBound}.

We can actually do better and compare with a careful computation of the PBH abundance.  
The parameters of the dynamics, using the reverse engineering approach, (with $\eta_{\rm III} = 0$ and $\delta N\to 0$) are chosen in such a way that the integral 
\begin{align}
f_{\rm PBH} \equiv \frac{\Omega_{\rm PBH}}{\Omega_{\rm CDM}} = \int f_{\rm PBH}(M_{\rm PBH}) d\log M_{\rm PBH} \approx 1\,,\label{eq:fPBHinte}
\end{align}
which means that we get $\approx 100\%$ of DM in the form of PBHs. 
More in detail, we tune, for each $\eta_{\rm II}$, the value of $\Delta N_{\rm USR}$ so to get  $f_{\rm PBH} \approx 1$; 
we choose 
the numerical value of $k_{\rm in}$ in such a way that the peak of the PBH mass distribution $f_{\rm PBH}(M_{\rm PBH})$ falls within the interval $M_{\rm PBH}/M_{\odot} \in [10^{-14},\, 10^{-13}]$ in which the condition 
$f_{\rm PBH} \approx 1$ is experimentally allowed, the so-called asteroid-mass PBHs \cite{Carr:2020xqk}. 
We compute eq.\,(\ref{eq:fPBHinte}) using threshold statistics and including the full non-linear relation between the curvature and the density contrast fields (cf.\,\cite{Young:2019yug,DeLuca:2019qsy}) (see chapter.2) but assuming a gaussian curvature field $\zeta$\footnote{It is possible to further improve our analysis by including the presence of primordial non-gaussianity (e.g. \cite{Namjoo:2012aa,Chen:2013eea,Cai:2018dkf,Pi:2022ysn,Passaglia:2018ixg,Figueroa:2020jkf,Biagetti:2021eep,Figueroa:2021zah,Tomberg:2023kli}). 
In the case of local non-guassianity parametrized by a positive non-Guassian parameter $f_{\rm NL}$, as expected in the case of USR, we generically expect a larger abundance of PBHs compared to the Gaussian case\,\cite{Atal:2019cdz,Ferrante:2022mui,Gow:2022jfb,Ferrante:2023bgz,Ianniccari:2024bkh}. 
This means that, in order to achieve the same abundance of PBHs, one needs a power spectrum with a smaller peak amplitude. This argument implies that the presence of primordial non-Gaussianity will tend to decrease the relevance of the one-loop corrections.
}.
In the right panel of fig.\,\ref{fig:RegionBound} we 
plot the line defined by the condition $f_{\rm PBH} \approx 1$. The comparison 
between the left- and right-hand side of fig.\,\ref{fig:RegionBound}  shows that, in order to fulfil the condition $f_{\rm PBH} \approx 1$, one needs $\mathcal{P}_{\rm USR}/\mathcal{P}_{\rm CMB}  = O(10^7)$.\footnote{There is some difference between peak theory and threshold statistics in
the computation of the abundance, already present at the Gaussian level (see, e.g., refs.\,\cite{Green:2004wb,Young:2014ana,DeLuca:2019qsy}). The approach based on peak theory usually requires slightly smaller values of $\mathcal{P}_{\rm USR}/\mathcal{P}_{\rm CMB}$ in order to get the same abundance of PBHs, thus making our findings, based on threshold statistics, even stronger.} 

We conclude that the condition $f_{\rm PBH} \approx 1$ lies within the region in which perturbativity is still applicable. 
This is in contrast with the conclusion reached in refs.\,\cite{Kristiano:2022maq,Kristiano:2023scm,Firouzjahi:2023aum} in the limit of instantaneous SR/USR/SR transition.
The origin of the difference is the more accurate calculation of the PBHs abundance performed in our analysis. In particular, in previous analyses, estimates of $f_{\rm PBH}\approx 1$ are based on requiring $\mathcal{P}_{\rm USR}\simeq10^{-2}$, and on the scaling $\Delta\mathcal{P} = (k_{\rm end}/k_{\rm in})^{2\eta_{\rm II}}$ in order to capture the growth of the power spectrum at small scales. However this scaling does not accurately describe the amplitude of the power spectrum at its peak, see the left panel of fig.\,\ref{fig:TestPS}.
Ref.\,\cite{Motohashi:2023syh} computed one-loop corrections in the limit of instantaneous SR/USR/SR transitions in scenarios with $\eta_{\rm II}\leq 3$, finding that the perturbativity bound is relaxed for $\eta_{\rm II}$ smaller than 3. At a qualitative level, similar results are obtained in the left panel of fig.\,\ref{fig:RegionBound}.
In ref.\,\cite{Motohashi:2023syh}, the perturbativity bound is traduced in an upper limit on the power-spectrum at the scale $k_{\rm end}.$ However, as explained above, this procedure underestimates the maximum amplitude of the power spectrum, see again the left panel of fig.\,\ref{fig:TestPS}.

\subsubsection{Dynamics at the SR/USR/SR transition}\label{sec:trans}

We go beyond the instantaneous transition to check if there are cancellations that affect the order-of-magnitude of the result in eq.\,(\ref{eq:AhiAhi}). 
There are indeed compelling reasons to believe that this is the case, as advocated in refs.\,\cite{Riotto:2023gpm,Firouzjahi:2023aum,Firouzjahi:2023ahg}.
The story goes as follows (the original argument was proposed in ref.\,\cite{Cai:2018dkf} in which the role of non-Gaussianity from non-attractor inflation models was considered). 
From the Hubble parameters in eq.\,(\ref{eq:HubblePar1}) and the background dynamics that follow from the action in eq.\,(\ref{eq:BackAction}), it is possible to calculate the potential and its derivatives exactly. Up to the third order in field derivatives, we find (see also ref.\,\cite{Leach:2002ar})
\begin{align}
V(\phi) & = H^2(3-\epsilon)\,,\\
V^{\prime}(\phi) & = 
\frac{H^2}{\sqrt{2}}
\epsilon^{1/2}\left(
6-2\epsilon +  \epsilon_2
\right)\,,\\
\end{align}
\begin{align}
V^{\prime\prime}(\phi) & = H^2\left(
6 \epsilon - 2 \epsilon^2 - \frac{3 \epsilon_2}{2} + 
 \frac{5\epsilon \epsilon_2}{2} - \frac{\epsilon_2^2}{4} - 
 + \frac{\epsilon_2 \epsilon_3}{2}
\right)\,,\\
V^{\prime\prime\prime}(\phi) & = \frac{H^2}{
2\sqrt{2\epsilon}
}
\left[
-8 \epsilon^3 + 
 6 \epsilon^2 (4 + 3 \epsilon_2) - \epsilon \epsilon_2(18 + 6 \epsilon_2 + 7 \epsilon_3) + \epsilon_2 \epsilon_3 (3 + \epsilon_2 +\epsilon_3 + \epsilon_4)
\right] \\
& = 
\frac{1}{
2\sqrt{2\epsilon}
}\left\{
H^2\left[
\frac{\ddot{\epsilon_2}}{H^2}+
(3+\epsilon_2)
\frac{\dot{\epsilon_2}}{H}
\right]
+O(\epsilon)
\right\}\,,\label{eq:ThirdDerV}
\end{align}
where in eq.\,(\ref{eq:ThirdDerV}) we expanded in the parameter $\epsilon$ and wrote $\epsilon_{3,4}$ in terms of $\epsilon_2$.  
Consider the flat gauge in which curvature perturbations are entirely encoded into field fluctuations $\delta\phi$ by means of the relation $\zeta = H\delta\phi/\dot{\phi} =  -\delta\phi/\sqrt{2\epsilon}$. 
In this gauge, the interactions come from Taylor-expanding the quadratic action in field fluctuations and, at the cubic order, one expects 
\begin{align}
\mathcal{L}_3 \supset 
\frac{a^3}{6}
V^{\prime\prime\prime}\delta\phi^3 = 
- \frac{a^3\epsilon}{3}
(\sqrt{2\epsilon}V^{\prime\prime\prime})\zeta^3 = 
- \frac{a^3\epsilon}{6}
\left\{
H^2\left[
\frac{\ddot{\epsilon_2}}{H^2}+
(3+\epsilon_2)
\frac{\dot{\epsilon_2}}{H}
\right]
+O(\epsilon)
\right\}\zeta^3\,.\label{eq:FlatGauCu}
\end{align}
As shown in ref.\,\cite{Chen:2013eea}, 
the above interaction agrees (modulo a surface term) with eq.\,(\ref{eq:Yoko}) if we integrate by parts
\begin{align}
\int d^4x\,\frac{\epsilon\dot{\epsilon_2}}{2}a^3\dot{\zeta}\zeta^2 ~\to~  
-\int d^4x\,
\frac{1}{6}\frac{d}{dt}\left(
\epsilon\dot{\epsilon_2}a^3
\right)\zeta^3 = 
-\int d^4x\,
\frac{a^3\epsilon}{6}H^2\left[
\frac{\ddot{\epsilon_2}}{H^2}+
(3+\epsilon_2)
\frac{\dot{\epsilon_2}}{H}
\right]\zeta^3\,,\label{eq:CuboIBP}
\end{align}
where in the last step we used the exact identity
\begin{align}
\frac{d}{dt}\left(
\epsilon\dot{\epsilon_2}a^3
\right) = 
 a^3\epsilon  H^2\left[
\frac{\ddot{\epsilon_2}}{H^2}+
(3+\epsilon_2)
\frac{\dot{\epsilon_2}}{H}
\right]\,.\label{eq:EffectiveCoupling}
\end{align}
The cubic interaction in eq.\,(\ref{eq:CuboIBP}) agrees with eq.\,(\ref{eq:FlatGauCu}) up to $\epsilon$-suppressed terms.  
Rewriting the interaction as in eq.\,(\ref{eq:FlatGauCu}) is quite instructive. 
From eq.\,
(\ref{eq:EffectiveCoupling}),  it seems plausible that drastic variations in time of $\epsilon_2$ could enhance the cubic interaction. However, eq.\,(\ref{eq:FlatGauCu}) shows that these interactions are ultimately controlled by $V^{\prime\prime\prime}$ so that  in the case with a smooth SR/USR/SR transition  
in which $V^{\prime\prime\prime}$ is expected to be ``small'',
there must be cancellations at work within the combination in eq.\,(\ref{eq:EffectiveCoupling}) so that the relevant coupling in  eq.\,(\ref{eq:FlatGauCu}) reduces to the term that is SR suppressed. 
This is the main argument that was put forth in refs.\,\cite{Riotto:2023gpm,Firouzjahi:2023aum,Firouzjahi:2023ahg}. 

\begin{figure}[h]
\begin{center}
$$\includegraphics[width=.495\textwidth]{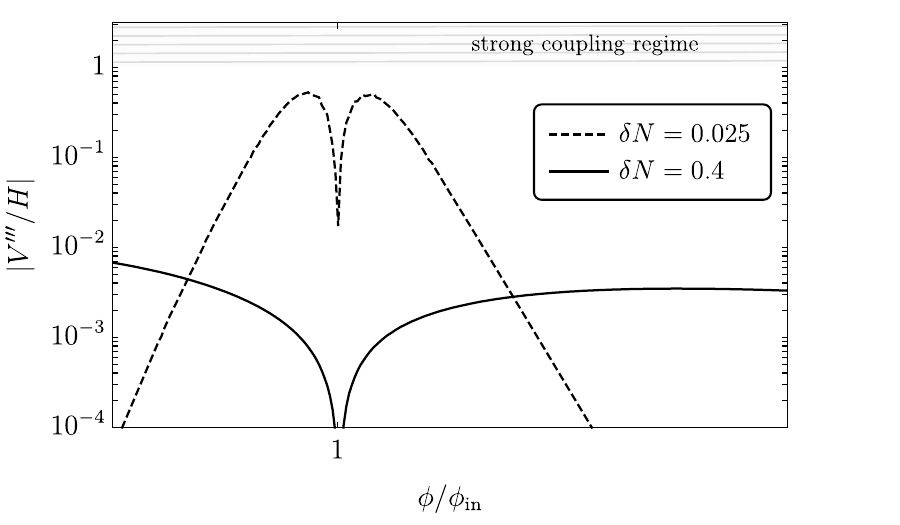}~
\includegraphics[width=.495\textwidth]{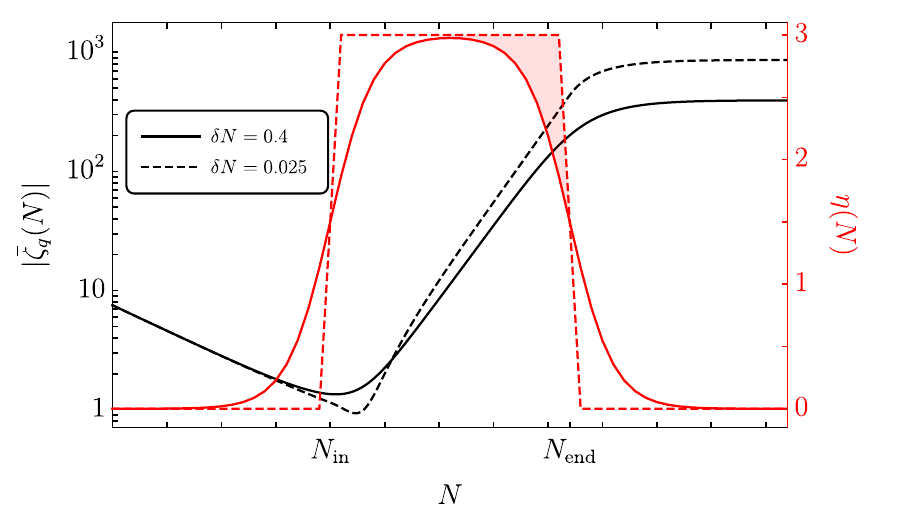}$$\vspace{-0.5cm}
\caption{ 
\textit{\textbf{Left panel:}} 
Graph of $V^{\prime\prime\prime}/H$ as function of the background field value $\phi$ for two representative dynamics with, respectively, $\delta N = 0.025$ (dashed black) and $\delta N = 0.4$ (solid black). 
We compute the potential by means of the reverse engineering approach described in ref.\,\cite{Franciolini:2022pav}. 
The values $V_{\star}$ and $H_{\star}$ of, respectively, the potential and the Hubble rate at CMB scales are chosen in such a way that both dynamics are consistent with CMB observations (namely, $V_{\star} \simeq  3\times 10^{-9}$ and $H_{\star}  \simeq  3\times 10^{-5}$ with the reduced Planck mass set to 1). 
On the right (left) side of the field value $\phi = \phi_{\rm in}$,
$V^{\prime\prime\prime}/H$ is negative (positive).
\textit{\textbf{Right panel:}}  
Left-side $y$-axis: time evolution of the curvature modes $|\bar{\zeta}_q(N)|$ for $\bar{q}  = 2$ in the case $\delta N = 0.025$ (dashed black line) and $\delta N = 0.4$ (solid black line). 
Right-side $y$-axis: profile of $\eta$ in the case $\delta N = 0.025$ (dashed red line) and $\delta N = 0.4$ (solid red line). The region shaded in red highlights the  difference between the sharp and the smooth transition in terms of $\eta$: in the case of a sharp transition, the curvature mode has more time to grow under the  effect of the  negative friction phase implied by the condition $\eta > 3/2$.
 }\label{fig:CancDetail}  
\end{center}
\end{figure}

We shall elaborate further on this point. 
First  of all, let us clarify what ``$V^{\prime\prime\prime}$  small'' means.  
We rewrite eq.\,(\ref{eq:ThirdDerV}) as follows (we omit the $O(\epsilon)$ terms and, for clarity's sake, we write explicitly the reduced Planck mass)
\begin{align}
\frac{V^{\prime\prime\prime}}{H}  =  
\left(\frac{H}{\bar{M}_{\rm Pl}}\right)
\frac{1}{
2\sqrt{2\epsilon}
}
\left[
\frac{\ddot{\epsilon_2}}{H^2}+
(3+\epsilon_2)
\frac{\dot{\epsilon_2}}{H}
\right]\,.\label{eq:Intu}
\end{align}
On the left-hand side, the quantity $V^{\prime\prime\prime}/H$ has the dimension of a coupling. Consequently, imposing 
the condition $V^{\prime\prime\prime}/H < 1$ corresponds to a weak coupling regime while $V^{\prime\prime\prime}/H > 1$ corresponds to a strongly coupled one.  
Said differently, from the perspective of the right-hand side of eq.\,(\ref{eq:Intu}), the condition $V^{\prime\prime\prime}/H > 1$ corresponds to a situation in which the a-dimensional factor in front of $H/\bar{M}_{\rm Pl}$ becomes so large that it overcomes the natural suppression given by $H/\bar{M}_{\rm Pl} \ll 1$. 
In the left panel of fig.\,\ref{fig:CancDetail}, we compute the ratio $V^{\prime\prime\prime}/H$ for two benchmark SR/USR/SR dynamics with different values of $\delta N$. 
In the case in which $\delta N \to 0$ (sharp transition), we  observe that $V^{\prime\prime\prime}/H$ dangerously grows towards the strongly coupled regime while in the case of a smooth transition it safely takes 
$O(\ll 1)$ values.  
As anticipated at the beginning of this section, 
this argument confirms that in the case of a smooth transition we expect a reduction in the size of the trilinear interaction controlled by the factor in eq.\,(\ref{eq:EffectiveCoupling}).

With this motivation in mind, we go back to the analysis in section\,\ref{sec:Insta} and we perform the following calculation.

We compute numerically the integral in eq.\,(\ref{eq:PS1Simpl}) in order to check the validity of the scaling in  
eq.\,(\ref{eq:AhiAhi}) beyond the limit of instantaneous  transition.  
We define the quantity 
\begin{align}
\mathcal{J}_{\delta N}&(\eta_{\rm II},\Delta N_{\rm USR}) \equiv 
\Delta\mathcal{P}_{\rm 1-loop}(k_*) /\mathcal{P}_{\rm tree}(k_*) 
=
\nn\\
&
 32
\int_{N_{\rm in}-\Delta N}^{
N_{\rm end} + \Delta N}
dN_1  
\frac{d\eta}{dN}(N_1) 
\int_{N_{\rm in}-\Delta N}^{N_1}
dN_2
\bar{\epsilon}(N_2)
\frac{d\eta}{dN}(N_2)
e^{3(N_2 - N_{\rm in})}
\int_1^{e^{\Delta N_{\rm USR}}}
\frac{d\bar{q}}{\bar{q}^4}
\IM\left[
\bar{\zeta}_q(N_1)^2
\bar{\zeta}_q^*(N_2)
\frac{d\bar{\zeta}_q^*}{dN}(N_2)
\right]\,,\label{eq:FullIntegral}
\end{align}
that we can directly compare, in the case $\eta_{\rm II} = 3$, with 
$\eta_{\rm II}^2(k_{\rm end}/k_{\rm in})^6[1 + \log(k_{\rm end}/k_{\rm in})]$ in  
eq.\,(\ref{eq:AhiAhi}) using the fact that 
$k_{\rm end}/k_{\rm in} =  e^{\Delta N_{\rm USR}}$.
First, we set $\delta N$ to a very small number, in order to mimic the limit $\delta N \to 0$, and evaluate $\mathcal{J}_{\delta N}(3,\Delta N_{\rm USR})$ as function 
of $\Delta N_{\rm USR}$.
\begin{figure}[h]
\begin{center}
$$\includegraphics[width=.495\textwidth]{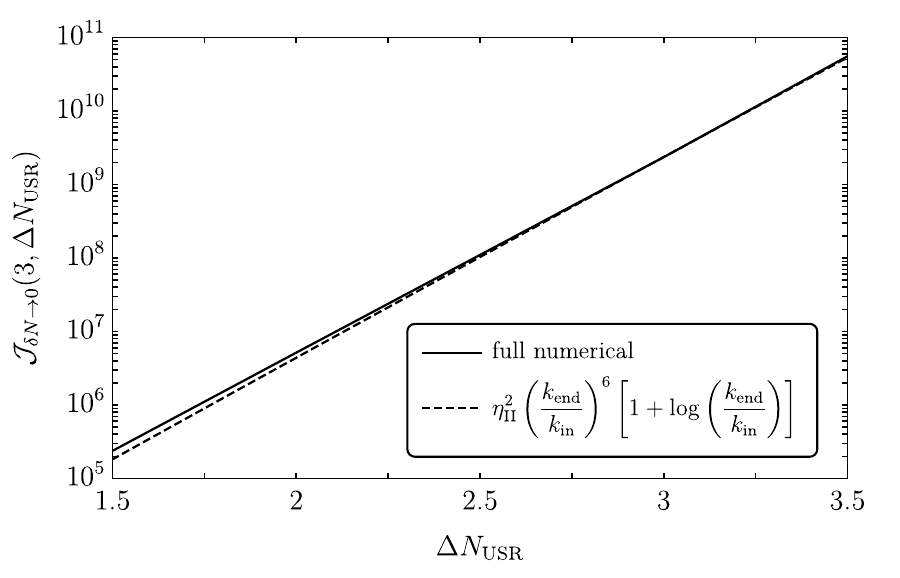}~
\includegraphics[width=.495\textwidth]{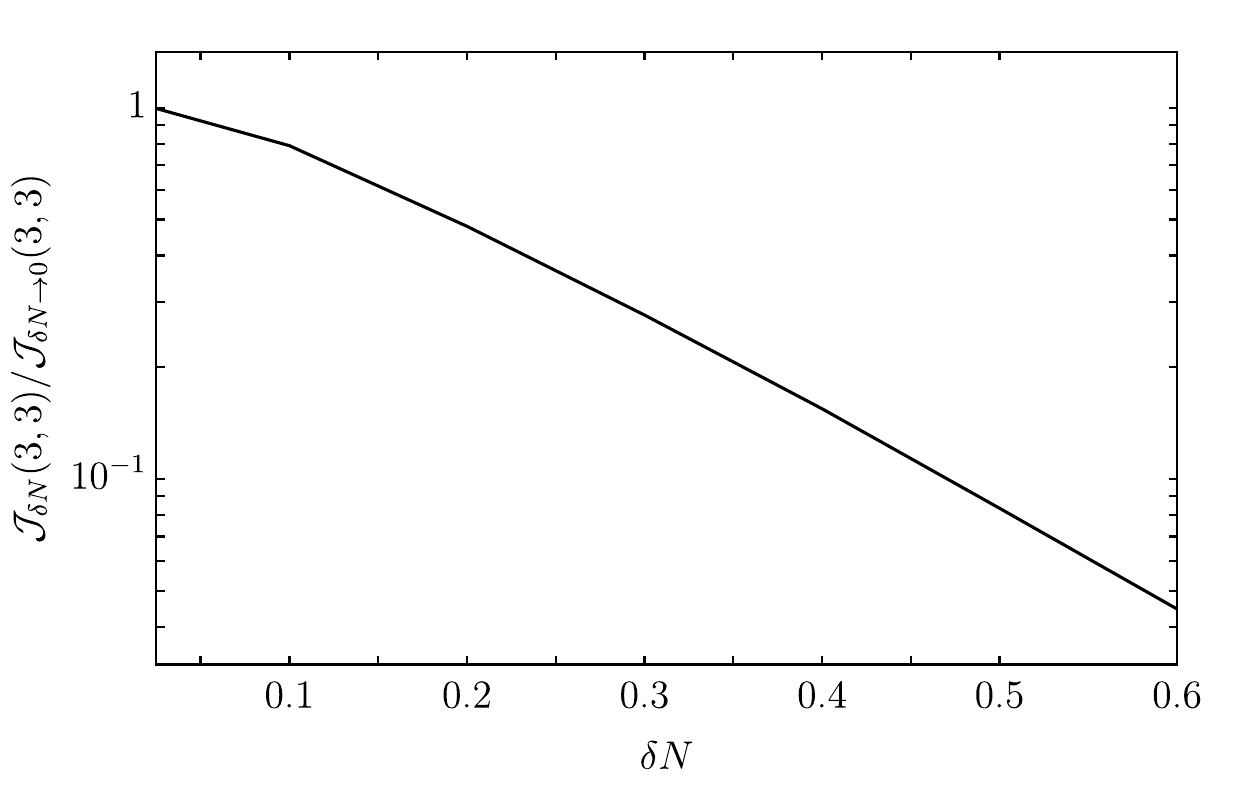}$$\vspace{-0.5cm}
\caption{
\textit{\textbf{Left panel:}} Comparison between 
the value of the full integral in eq.\,(\ref{eq:FullIntegral}) and the analytical estimate in eq.\,(\ref{eq:AhiAhi}). 
To mimic the instantaneous transition we take $\delta N = 0.025$. 
\textit{\textbf{Right panel:}}
We plot the ratio $\mathcal{J}_{\delta N}(3,\Delta N_{\rm USR})$ as function of $\delta N$. 
In both figures we take $\eta_{\rm II} = 3$.
 }\label{fig:SmoothTransition2}  
\end{center}
\end{figure}
The comparison is shown in the left panel of fig.\,\ref{fig:SmoothTransition2}. We find an excellent agreement in particular for large $\Delta N_{\rm USR}.$ This is expected, since the approximation in eq.\,(\ref{eq:ModeInte}) is more accurate for larger $k_{\rm end}/k_{\rm in}$. 
Then, we set $\Delta N_{\rm USR} = 3$ and 
compare the value of  
$\mathcal{J}_{\delta N \to 0}(3,3)$ with 
$\mathcal{J}_{\delta N}(3,3)$ as function of 
$\delta N$.  
We plot the ratio  
$\mathcal{J}_{\delta N}(3,3)/\mathcal{J}_{\delta N \to 0}(3,3)$ 
in the right panel of fig.\,\ref{fig:SmoothTransition2}.

Realistic single-field models that  
feature the presence of a phase of USR dynamics typically have $\delta N = 0.4-0.5$ (cf., e.g., ref.\,\cite{Ballesteros:2020qam,Taoso:2021uvl}). 
This means that, according to our result in the right panel of fig.\,\ref{fig:SmoothTransition2}, 
we expect that in realistic models the size of the loop correction gets reduced by  one order of magnitude with respect to what is obtained in the limit of instantaneous SR/USR/SR  transition.
This confirms the intuition presented in refs.\,\cite{Riotto:2023gpm}.

\begin{figure}[h]
\begin{center}
\includegraphics[width=.8\textwidth]{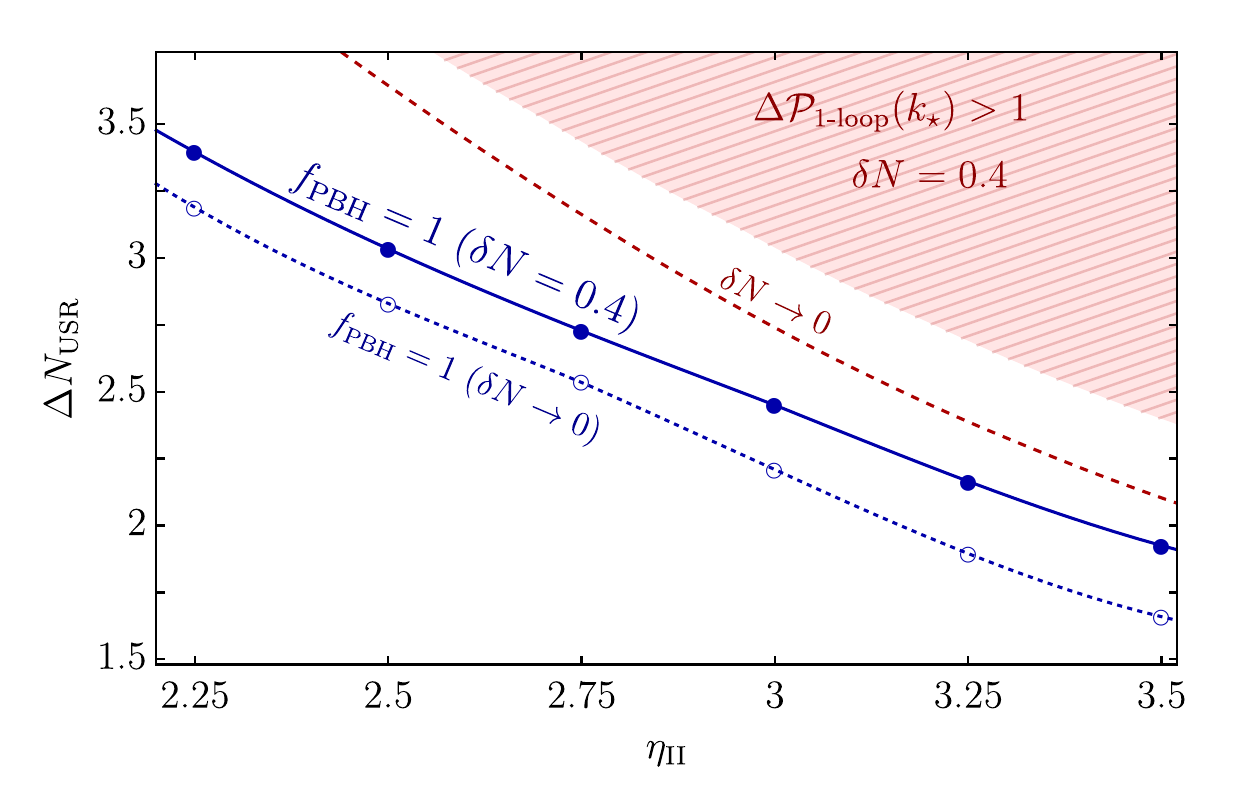}\vspace{-0.5cm}
\caption{
We consider a generic USR dynamics with varying $\eta_{\rm II}$ ($x$-axis) and $\Delta N_{\rm USR}$ ($y$-axis). We take $\eta_{\rm III} = 0$ and the smooth limit $\delta N = 0.4$.
The region hatched in red corresponds to $
\Delta\mathcal{P}_{\rm 1-loop}(k_*) > 0$. 
Along the line defined by the condition $f_{\rm PBH} = 1$, we get 100\% of DM in the form of asteroid-mass PBHs.
The dotted blue line and  the red dashed line correspond, respectively, to the conditions $f_{\rm PBH} = 1$ and 
$\lim_{\delta N\to 0}\Delta\mathcal{P}_{\rm 1-loop}(k_*) > 0$ as derived in the limit of   instantaneous transition.
 }\label{fig:FinalPlot}  
\end{center}
\end{figure}

It should be noted that in the case of smooth SR/USR/SR transition the amplitude of the power spectrum gets reduced with respect to 
the $\delta N \to 0$ limit (cf. the left panel of fig.\,\ref{fig:TestPS}).
The origin of this effect becomes evident if we consider the right panel of fig.\,\ref{fig:CancDetail}. In this figure, we plot the time evolution of the curvature mode $|\bar{\zeta}_{q}|$ with $\bar{q} = 2$ in the two cases of a sharp and smooth transition (dashed and solid lines, respectively -- see caption for details). In the case of a sharp transition, the curvature mode experiences a longer USR phase, and its final amplitude is larger with respect to the case of a smooth transition.
As a consequence, therefore, we expect that the smaller size of the loop correction will be, at least partially, compensated by the fact that 
finite $\delta N$ also reduces the  amplitude of the tree-level power spectrum.   
In order to quantify this information, we repeat the analysis done in section\,\ref{sec:GenericUSR} but now for finite $\delta N$.
We plot our result in fig.\,\ref{fig:FinalPlot}. 
For definiteness, we consider the benchmark value $\delta N = 0.4$ while we keep 
$\eta_{\rm II}$ and $\Delta N_{\rm USR}$ generic as in fig.\,\ref{fig:RegionBound}.

Our numerical analysis mirrors the previous intuition. 
The perturbativity bound (the region hatched in red corresponds to the condition $\Delta\mathcal{P}_{\rm 1-loop}(k_*) > 0$) gets weaker because of the partial cancellation illustrated in the right panel of fig.\,\ref{fig:SmoothTransition2}.  
However, as previously discussed, the drawback is that taking  $\delta N \neq 0$ also reduces the peak amplitude of  the power spectrum. Consequently, the condition $f_{\rm PBH} = 1$ requires, for fixed $\eta_{\rm II}$, larger $\Delta N_{\rm USR}$.

\begin{figure}[h]
\begin{center}
$$\includegraphics[width=.495\textwidth]{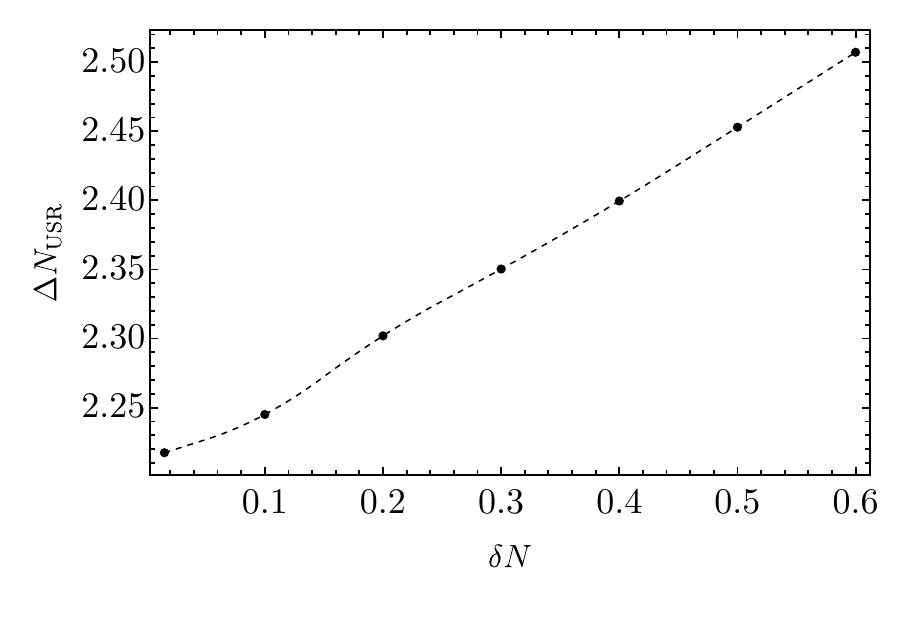}~
\includegraphics[width=.495\textwidth]{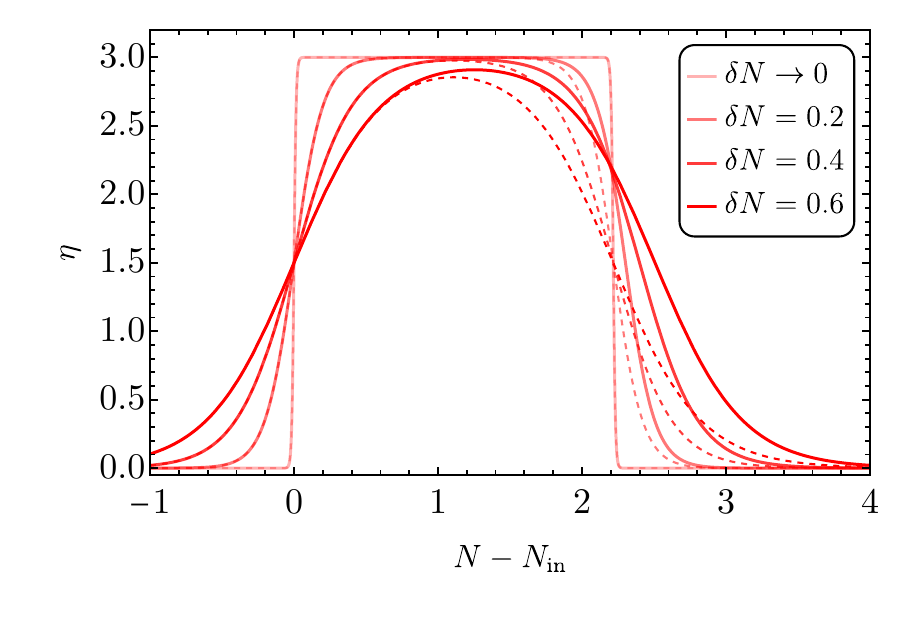}$$\vspace{-0.5cm}
\caption{
\textit{\textbf{Left:}}
Value of $\Delta N_{\rm USR}$ required in order to have $f_{\rm PBH}=1$ for $\eta_{\rm II} = 3$.
\textit{\textbf{Right:}}
Different examples of evolution of $\eta(N)$ responsible for the USR, assuming various $\delta N$ and fixing $\eta_{\rm II} = 3$.
Dashed lines reports the scenario where $\delta N$ is increased while $\Delta N_{\rm USR}$ is kept fixed to the value imposed to have unit PBH abundance in the limit $\delta N \to 0$.
Solid lines report the result when $\Delta N_{\rm USR}$ is instead adjusted to keep $f_{\rm PBH} = 1$ fixed. We see that smoother transitions results in longer USR phases.
 }\label{fig:deltaNsmoothT}  
\end{center}
\end{figure}

As for the limit of instantaneous transition, 
the condition $f_{\rm PBH} = 1$ does not 
violate the perturbativity  bound since the  two above-mentioned effects nearly compensate each other. 
However, our analysis reveals an interesting aspect: modelling the SR/USR/SR transition (and, in particular, the final USR/SR one) beyond the instantaneous limit reduces the impact of the loop correction but, at the same time, lowers the peak amplitude of the  tree-level power spectrum, which must be compensated by a larger $\Delta N_{\rm USR}$ see fig.~\,\ref{fig:deltaNsmoothT}. 
As illustrated in fig.\,\ref{fig:FinalPlot}, both these effects must be considered together in order to  properly quantify the impact of loop corrections and the consequent perturbativity bound.

This is an interesting point. Refs.\,\cite{Riotto:2023gpm,Firouzjahi:2023aum,Firouzjahi:2023ahg} argue that if one goes beyond the limit  of instantaneous transition then the loop correction to the CMB power spectrum becomes effectively  harmless. Technically speaking, in our analysis the role of the parameter $-6 < h < 0$ that in \cite{Riotto:2023gpm,Firouzjahi:2023aum,Firouzjahi:2023ahg} (see also ref.\,\cite{Cai:2018dkf}) controls the sharpness of the transition is played by our parameter $\delta N$ (with $h\to -6$ that corresponds to our $\delta N \to 0$ and $h\to 0$ that corresponds to increasing values of $\delta N$).
 
In light of our analysis, a very important remark naturally arises:
There is a non-trivial and crucial interplay between  the detail of the USR/SR transition and the  amplitude of the tree-level power spectrum that must be properly included before drawing any conclusion about the relative size of the loop corrections. 
On the one hand, it is true that 
a smooth USR/SR transition reduces the size of the loop correction; 
on the other one, the same smoothing also reduces the amplitude of the power spectrum so that, in order to keep $f_{\rm PBH}$ fixed, one is forced to either increase the duration of the USR phase or the magnitude of $\eta$ during the latter. In the end, the two effects tend to compensate each other if one imposes the condition $f_{\rm PBH} = 1$ (cf. fig.\,\ref{fig:FinalPlot}).
\subsubsection{Loop evaluation at any scales}\label{sec:LoopUSR}
We evaluate the loop correction at a generic external momentum $k$, thus alleviating the assumption $k\ll q$.
The dominant modes contributing to the loop integration remain 
the ones that cross the horizon during the USR phase $q\in [k_{\rm in}, k_{\rm end}]$. 
As done in the previous section, we are interested in comparing
the one-loop correction with the tree level power spectrum at the end of inflation, and therefore we perform the late time limit $\tau \to 0^-$.
Following the notation introduced in eq.\,\eqref{eq:MasterOne}, we define
\begin{equation}
\mathcal{P}(k) = \lim_{\tau \to 0^-}\left(\frac{k^3}{2\pi^2}\right)
\left [
\left|\zeta_k(\tau)\right|^2 + 
\frac{1}{(4\pi)^2}
\Delta P(k,\tau) 
\right ] 
\equiv 
\mathcal{P}_{\rm tree}(k)
\left ( 1 + \Delta {\cal P}_{\rm 1-loop} \right)
\,,\label{eq:MasterOne_2}
\end{equation}
In order to simplify the computation
we consider the instantaneous limit  $\delta N\to 0$ of 
eq.\,(\ref{eq:GenSc}). 
We perform both time integrations
keeping the dominant contribution given by the first Dirac delta in eq.\,\eqref{eq:DeltaDer}. 
This implies that we evaluate
the integrand function at 
$\tau_1 = \tau_2 = \tau_{\rm end}$. 
Notice also that, since the second integration only gets contributions from half of the Dirac delta domain, we additionally include a factor of $1/2$.
Finally, the jump in $\epsilon_2$ leaves a factor $(2 \eta_{\rm II})$ for each time integration. 
Therefore, we find
\begingroup
\begin{align}\label{eq:genmomcorr}
\Delta P(k,\tau)  \equiv 
-32 \eta_{\rm II}^2
&
[\epsilon(\tau_{\rm end}) a^2(\tau_{\rm end})]^2
\int_{q_{\rm in}}^{q_{\rm end}}
dq\,q^2\,
\int_{-1}^{1} d(\cos\theta) \times\Big \{
\nn \\ 
{\rm Im}\big[\zeta_{k}^*(\tau)\zeta_{k}'(\tau_{\rm end}) \big]
\times  \Big [\rm  &  {\rm Im} \big[\zeta_{k}(\tau)\zeta_{k}^{\prime*}(\tau_{\rm end}) |\zeta_{q}(\tau_{\rm end})|^2| \zeta_{k-q}(\tau_{\rm end})|^2\big]+
\nn \\ 
 &{\rm Im}\big[\zeta_{k}(\tau) \zeta_{k}^{*}(\tau_{\rm end})\big( 
|\zeta_q(\tau_{\rm end})|^2\zeta_{k-q}(\tau_{\rm end})\zeta_{k-q}^{\prime\,*}(\tau_{\rm end}) +
 |\zeta_{k-q}(\tau_{\rm end})|^2\zeta_q(\tau_{\rm end})\zeta_q^{\prime\,*}(\tau_{\rm end}) \big)\big]\Big ] 
 \rm+
 \nn \\
 {\rm Im} [\zeta_{k}^*(\tau)\zeta_{k}(\tau_{\rm end})]
\times \Big [ 
\rm& {\rm Im}\big[\zeta_{k}(\tau) \zeta_{k}^{\prime*}(\tau_{\rm end})\big( 
|\zeta_q(\tau_{\rm end})|^2\zeta_{k-q}^{*}(\tau_{\rm end})\zeta_{k-q}^{\prime}(\tau_{\rm end}) +
 |\zeta_{k-q}(\tau_{\rm end})|^2\zeta^{*}_q(\tau_{\rm end})\zeta_q^{\prime}(\tau_{\rm end}) \big)\big]+
  \nn \\
& {\rm Im}\big[\zeta_{k}(\tau) \zeta_{k}^{*}(\tau_{\rm end})\big(  |\zeta_{k-q}^{\prime}(\tau_{\rm end}) |^2
|\zeta_q(\tau_{\rm end})|^2
+
\zeta_{k-q}^\prime (\tau_{\rm end})\zeta_{k-q}^*(\tau_{\rm end})
\zeta_q(\tau_{\rm end}) \zeta_q^{\prime\,*}(\tau_{\rm end}) \big)\big]+
  \nn \\
&{\rm Im}\big[\zeta_{k}(\tau) \zeta_{k}^{*}(\tau_{\rm end})\big(  |\zeta_{k-q}^{\prime}(\tau_{\rm end}) |^2
|\zeta_q(\tau_{\rm end})|^2
+
\zeta_{k-q}^{\prime\,*} (\tau_{\rm end})\zeta_{k-q}(\tau_{\rm end})
\zeta^*_q(\tau_{\rm end}) \zeta_q^{\prime}(\tau_{\rm end}) \big)\big]
\Big ]
\Big \}\rm.
\end{align}
\endgroup
We have collected the pieces such that each line corresponds to the $i$-th term in the sum of eq.\,\eqref{eq:GenSc} and 
$ k - q \equiv  \sqrt{k^2 + q^2 - 2kq\cos(\theta)}$ as in the previous section.

\begin{figure}[h]
\begin{center}
$$\includegraphics[width=.495\textwidth]{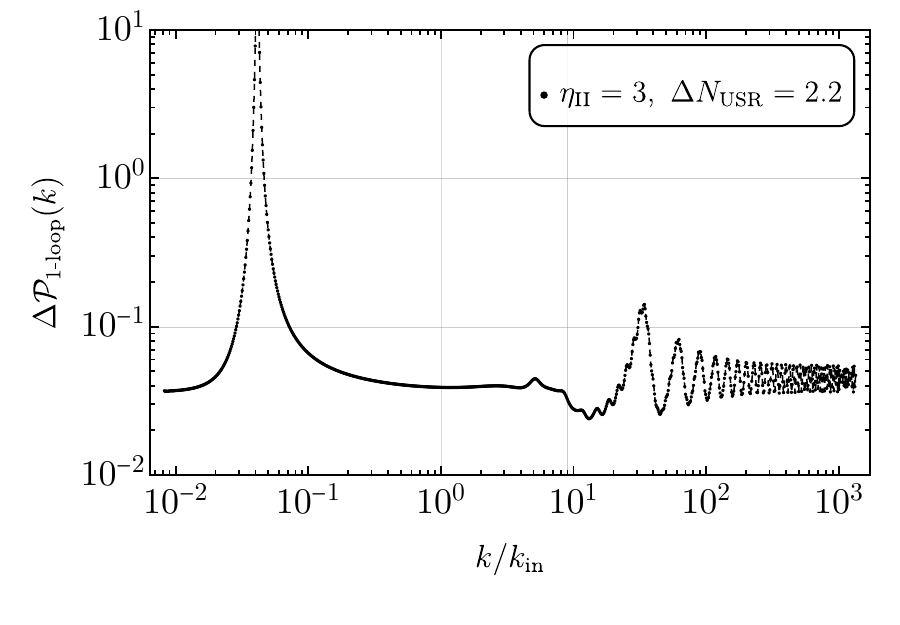}~
\includegraphics[width=.495\textwidth]{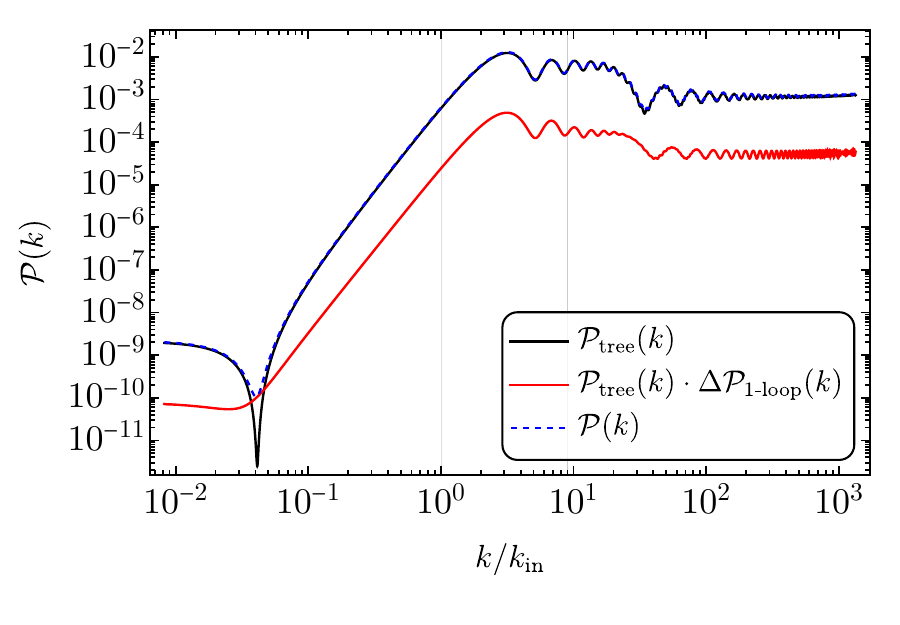}$$
\vspace{-0.5cm}
\caption{  
In both panels,  
we consider a USR dynamics with 
$\eta_{\rm II} = 3$,
$\Delta N_{\rm USR} = 2.2$, 
$\eta_{\rm III} = 0$ 
and the instantaneous limit $\delta N = 0$.
These values corresponds to a scenario producing $f_{\rm PBH} \simeq 1$.
The vertical gridlines corresponds to $k = k_{\rm in}$ and $k_{\rm end}$ in both panels. 
\textit{\textbf{Left panel:}}
correction to the tree level power spectrum as a function of $k$ in the limit of $\tau\to 0^-$. 
\textit{\textbf{Right panel:}} 
tree level power spectrum (black) compared to the 1-loop correction (red line) and 
their sum (blue dashed line).
}\label{fig:RegionBound2}  
\end{center}
\end{figure}

In the left panel of fig.~\ref{fig:RegionBound2}, we show the resulting 1-loop correction as a function of the wavenumber $k$ for a representative set of parameters leading to $f_{\rm BH}\approx1$:
$\eta_{\rm II} = 3$ and  $\Delta N_{\rm USR} = 2.2$.
We find values of $\Delta{\cal P}_{\rm 1-loop}$ of the order of few percent, barring small oscillatory features. A notable exception is the scale where the tree level power spectrum presents a dip, see fig.\,\ref{fig:TestPS}, $k_{\rm dip}/k_{\rm in} \approx  \sqrt{5/4} e^{-3 \Delta N_{\rm USR}/2}$~\cite{Byrnes:2018txb}. 
At that scale the 1-loop correction dominates, resulting in a spike in $\Delta{\cal P}_{\rm 1-loop}$. 
As a consequence, the dip is only realized if the 1-loop correction is neglected, see the right panel of fig.~\ref{fig:RegionBound2}.
We also observe that in the limit of small $k\ll k_{\rm in}$ the result quickly  converges towards the one discussed in the previous section, as expected.
Finally, it is also interesting to notice that the correction $\Delta{\cal P}_{\rm 1-loop}$ stays almost the same at any scale, except around $k_{\rm dip}$.
For this reason, we expect that a generalization of this calculation to the case for $\delta N \neq 0$ will lead to results similar to ones presented in the previous section for $k\ll q$.

At first sight, our result  that  loop corrections impact the tree-level power spectrum at the percent level seems at odds with the findings of ref.\,\cite{Inomata:2022yte} in which it was found that the one-loop power spectrum could dominate over the tree-level one, thus indicating the breakdown of the perturbation theory.  
Upon a closer look, however, there is no contradiction. 
Ref.\,\cite{Inomata:2022yte} considers a particular instance of background dynamics in which curvature perturbations are resonantly amplified due to a specific pattern of oscillatory features in the inflaton potential. 
In such a model, we checked that the condition $V^{\prime\prime\prime}/H \ll 1$ (cf. eq.\,(\ref{eq:Intu}) and the related discussion) is not verified and, therefore, it is not unexpected to find an amplification of loop effects.

It is instructive to consider also a different limit. Since we are assuming that the USR is followed by a second period of slow roll, characterized by a negligible $\eta_{\rm III}$ and a small $\epsilon$, modes in the range $q \in [k_{\rm in},k_{\rm end}]$ freeze around $\tau_{\rm end}$.
Therefore, the loop correction at $\tau_{\rm end}$ is very close to its limit at $\tau \to 0^-$, as we verified through a numerical calculation.
For this reason, we set $\tau \to \tau_{\rm end}$ in eq.\,\eqref{eq:genmomcorr} and drop the factors proportional to ${\rm Im} [\zeta_{k}^*(\tau_{\rm end})\zeta_{k}(\tau_{\rm end})]$ which vanish identically. 
Next, we switch to the barred fields and momenta notation introduced in sec.~\ref{sec:LoopCMB} and simplify the expression using the Wroskian identity \eqref{eq:wroskcond}.
Finally, we arrive at the expression
\begin{mynamedbox1}{
One-loop correction at generic scales in the instantaneous SR/USR/SR transition}
\vspace{-0.35cm}
\begin{align}
&\lim_{\delta N \to 0}
\Delta\mathcal{P}_{\rm 1-loop}(k,\tau_{\rm end}) 
\approx 
4 \eta_{\rm II}^2
\left(
\frac{H^2}{ 8 \pi^2 \epsilon_{\rm ref}}
\right)
\frac{1}{|\bar \zeta_k(N_{\rm end})|^2}
\int_1^{e^{\Delta N_{\rm USR}}}
\frac{d \bar q}{\bar q}
\int_{-1}^{+1} d (\cos \theta) \times 
\nn \\
&\left \{
\frac{\bar k ^3}{(\bar k - \bar q)^3}
 |\bar\zeta_{q}(N_{\rm end})|^2
| \bar\zeta_{k-q}(N_{\rm end})|^2
+
|\bar\zeta_{k}(N_{\rm end})|^2
\left [
|\bar\zeta_q(N_{\rm end})|^2
+
\frac{\bar q ^3}{(\bar k - \bar q)^3}
|\bar \zeta_{k-q}(N_{\rm end})|^2
\right ]
\right \}\label{eq:OurMainEq}
\end{align}
\end{mynamedbox1}
\noindent 
\noindent
We stress that this result is exact in the limit of sharp transition $\delta N \to 0$ and only neglects the contribution from the integration of the step-function in $N_{\rm in}$, which is numerically sub-leading.

A number of important comments are in order.
\begin{itemize}
\item[{\it i)}] 
In the case in which $k$ is a long CMB mode ($k\ll q$), the first term in the curly brackets  is suppressed by the factor $(k/q)^3$. 
This is nothing but the number of independent  Hubble patches of size $q^{-1}$ in a box of radius $k^{-1}$.
Intuitively, therefore, this contribution represents the situation in which random small-scale fluctuations lead by chance to a large-scale fluctuation, and the suppression factor $(k/q)^3$ simply indicates that it is very unlucky for short-mode to be coherent over long scales.  The meaning of this term is very clear, and the above argument is so compelling that it forces the intuition to believe that there is no way in which CMB modes can be affected by small-scale ones.

It is worth emphasizing that the computation of the one-loop correction to the correlation of long-wavelength modes $k$ due to short modes $q$ running in the loop can be thought of as solving the non-linear evolution equation for the long mode, cf. ref.\,\cite{Senatore:2009cf,Riotto:2023hoz}. 
In the language of this {\it source method}, the first term in the curly brackets of eq.\,(\ref{eq:OurMainEq})  corresponds to the so-called cut-in-the-middle  diagrams, cf. ref.\,\cite{Pimentel:2012tw}.
It also corresponds to the Poisson-suppressed term identified in ref.\,\cite{Riotto:2023hoz} while, in the language of ref.\,\cite{Kristiano:2023scm}, it corresponds to the correlation of two inhomogeneous solutions. 
More in detail, the evolution of the long mode in the presence of interactions reads 
\begin{align}
\hat{\zeta}(\vec{k},N_{\textrm{f}})  = & 
\underbrace{
\hat{\zeta}(\vec{k},N_{\textrm{in}}) 
+ 
a^3(N_{\textrm{in}})\epsilon(N_{\textrm{in}})
\frac{d\hat{\zeta}}{d N}(\vec{k},N_{\textrm{in}})
\int_{N_{\textrm{in}}}^{N_{\textrm{f}}}
dN\frac{1}{a^3(N)\epsilon(N)}}_{\textrm{homogeneous solution (free evolution)}}\label{eq:FirstL}\\ &
\underbrace{-\frac{\eta_{\rm II}}{2}\int \frac{d^3\vec{q}}{(2\pi)^3}
 \hat{\zeta}(\vec{q},N_{\rm end})\hat{\zeta}(-\vec{q},N_{\rm end})
 +
 \frac{\eta_{\rm II}}{3}
 \int \frac{d^3\vec{q}}{(2\pi)^3}
\frac{d\hat{\zeta}}{d N}(\vec{q},N_{\rm end})
 \hat{\zeta}(-\vec{q},N_{\rm end})}_{
\textrm{inhomogeneous solution (interactions)
 }}\,,\label{eq:SecondL}
\end{align}
where $N_{\rm f}$ represents some final $e$-fold time after the end of the USR phase. Consider the terms in the first line, eq.\,(\ref{eq:FirstL}). 
This is the standard result in the absence of interactions (the homogeneous solution in ref.\,\cite{Kristiano:2023scm}). 
Eq.\,(\ref{eq:FirstL}) tells us that the long mode stays constant unless the duration of the USR phase is so long to overcome the smallness of the time derivative of the long mode, which decayed exponentially fast during the phase preceding the USR. 
Eq.\,(\ref{eq:FirstL}), therefore, gives the standard tree-level power spectrum if one computes the correlator $\langle \hat{\zeta}(\vec{k},N_{\textrm{f}})
\hat{\zeta}(-\vec{k},N_{\textrm{f}})\rangle$. 
The two terms in eq.\,(\ref{eq:SecondL}) corresponds to the inhomogeneous solution in ref.\,\cite{Kristiano:2023scm}, and encode the  effect of the interactions in the evolution of the long mode. 
The first term in the curly brackets of our eq.\,(\ref{eq:OurMainEq})
corresponds to the correlator of two inhomogeneous solutions.
The equivalence between the source method and the ``{\it in}-{\it in}'' formalism has been discussed explicitly in ref.\,\cite{Kristiano:2023scm}. 

\item[{\it ii)}] Consider now the second term in the curly brackets of eq.\,(\ref{eq:OurMainEq}). 
Notice that this term always factorises $|\zeta_k|^2$ that cancels the denominator in front of the integral, which is present because of our definition of $\Delta\mathcal{P}_{\rm 1-loop}$, cf. eq.\,(\ref{eq:MasterOne_2}).  
In the case in which $k$ is a long CMB mode ($k\ll q$), 
this term does not pay any $(k/q)^3$ suppression.  
In the language of the source method, it corresponds to the so-called 
cut-in-the-side diagrams, cf. ref.\,\cite{Senatore:2009cf}. These diagrams represent the evolution of the long mode due to the effect that the long mode itself has on the expectation value of quadratic operators made of short modes\,\cite{Senatore:2009cf}. 
In ref.\,\cite{Kristiano:2023scm}, the second term in the curly brackets of eq.\,(\ref{eq:OurMainEq}) follows from the correlation between inhomogeneous and homogeneous solutions.  
\end{itemize} 
\section{Possible physical nature of the 1-loop corrections}
At the conceptual level, it remains true that, in the presence of USR dynamics, loop corrections of short modes may sizably affect the power spectrum at CMB scales. 
This result echoes an issue of naturalness -- an infrared quantity (the amplitude of the curvature power spectrum at CMB scales) appears to be sensitive, via loop effects, to physics that takes place at much shorter scales (those related to PBH formation) --  and clashes with  the intuition that  physics at  such vastly different scales should decouple.

\begin{figure}[h]
\begin{center}
\includegraphics[width=1\textwidth]{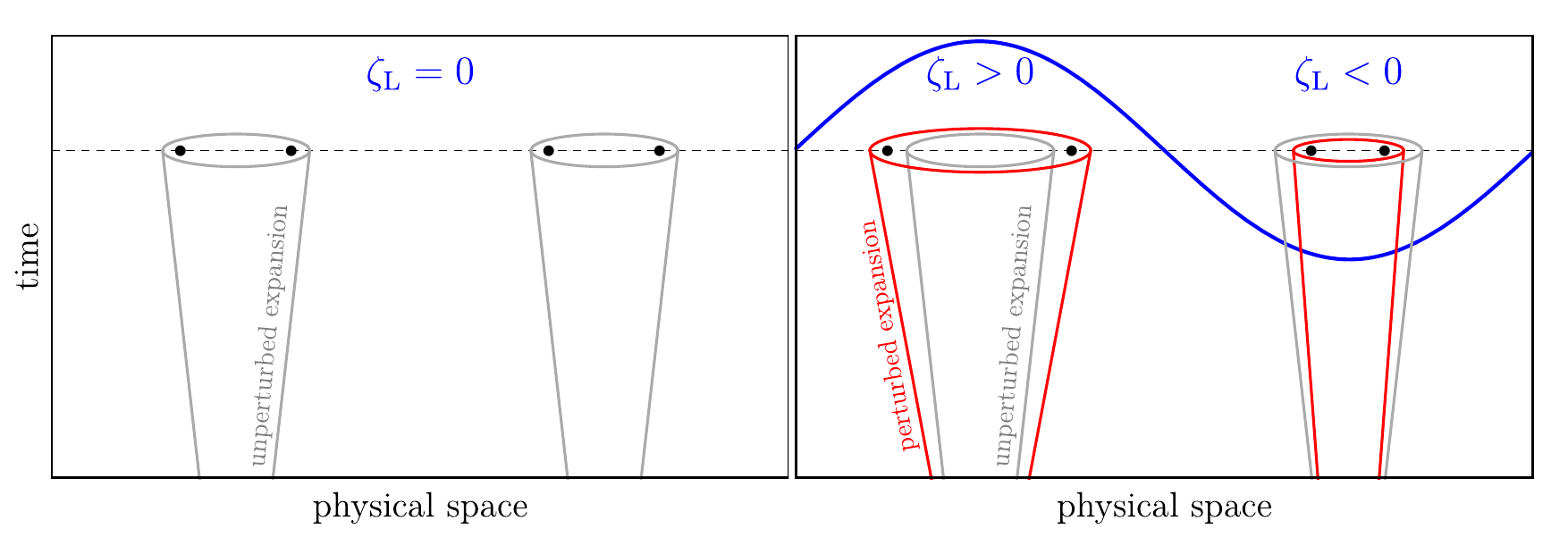}\vspace{-0.5cm}
\caption{
\textit{\textbf{Left:}}
Expansion in time
of the unperturbed Universe (time passes by along the $y$-axis); the Universe expands by the same amount at every point.
\textit{\textbf{Right:}} 
Expansion in time
of the perturbed Universe. 
The long mode ($\zeta_{\rm L}$, blue) acts as a local rescaling of the scale factor, and short scales are modulated accordingly. 
More specifically, if we consider the black dots we see that they experience a different amount of expansion depending on the value of $\zeta_{\rm L}$.
 }\label{fig:CurvatureGauge}  
\end{center}
\end{figure}

The coupling between short and long modes gives a physical effect for the following reason.
The relevant loop correction to the power spectrum at CMB scales comes from the correlation between homogeneous and inhomogeneous solutions.  
This is most easily seen within the source method in which one considers the correlation between a freely evolving long mode and 
a second long mode which evolves in the presence of interactions, cf. eq.\,(\ref{eq:SecondL}). Borrowing from ref.\,\cite{Pimentel:2012tw} (see also ref.\,\cite{Riotto:2023hoz}), 
we write the formal solution of the non-linear evolution equations for
a long wavelength mode $\zeta_{\rm L}$ as 
$\zeta_{\rm L} = 
\hat{O}^{-1}[S[\zeta_{\rm S},\zeta_{\rm S},\zeta_{\rm L}]]$, where $S$ represents a generic sum of operators that are quadratic in the short wavelength mode $\zeta_{\rm S}$ and that can also depend on $\zeta_{\rm L}$ if one considers the short modes in the background perturbed by the long mode. 
More concretely, in our case such a solution is the one given by eq.\,(\ref{eq:SecondL}). 
The one-loop power spectrum is  given by
\begin{align}
\langle\zeta_{\rm L}\zeta_{\rm L}\rangle \sim  
\langle
\hat{O}^{-1}[S[\zeta_{\rm S},\zeta_{\rm S},\zeta_{\rm L} =  0]]\,
\hat{O}^{-1}[S[\zeta_{\rm S},\zeta_{\rm S},\zeta_{\rm L} =  0]]
\rangle + 
\langle
\hat{O}^{-1}[S[\zeta_{\rm S},\zeta_{\rm S},\zeta_{\rm L}]]\,\zeta_{\rm L}
\rangle\,.\label{eq:Zelda}
\end{align}
The  first term represents the effect of the short-scale modes in their unperturbed state (that is, with $\zeta_{\rm L} = 0$)
directly on the power spectrum of the long wavelength mode. 
This is our first term in eq.\,(\ref{eq:OurMainEq}). 
As discussed in section\,\ref{sec:LoopUSR}, this term does not alter the long-wavelength correlation 
since it is very improbable that random short-scale fluctuations coherently add up to induce a long-wavelength correlation.
The second  term in eq.\,(\ref{eq:Zelda}), on the contrary, correlates a freely evolving long mode $\zeta_{\rm L}$ with the effect that the long mode itself has on the expectation value of quadratic operators made of short modes. 
Let us explain this point, which is crucial.
Consider the schematic in fig.\,\ref{fig:CurvatureGauge}.
The key point is the following.
In the comoving gauge, the short modes evolve in the background that is perturbed by the long mode.
In the limit in which the long mode $\zeta_{\rm L}$ has
a wavelength much longer than the horizon, it simply acts as a rescaling of the coordinates 
since it enters as a local change of the scale factor. 
This is schematically illustrated in fig.\,\ref{fig:CurvatureGauge}. 
This figure shows intuitively that 
the short scales are modulated by the presence of the long mode.  
The  presence of the long mode acts as a rescaling of the
coordinates and we can absorb it by rescaling the short-scale momenta 
$q\to \tilde{q} = e^{\zeta_{\rm L}}q$\,\cite{Riotto:2023hoz}.  
If the power spectrum of the short modes is scale-invariant, this rescaling does nothing. However, if the  power spectrum of the short modes breaks scale invariance, we schematically have in the loop 
integral over the short modes, expanding at the first  order
\begin{align}
\int\frac{dq}{q}\mathcal{P}(q) 
\overset{q\to \tilde{q} = e^{\zeta_{\rm L}}q}{~~~~\Longrightarrow~~~~}
\int\frac{d\tilde{q}}{\tilde{q}}\mathcal{P}(\tilde{q}) =
\int\frac{dq}{q}\mathcal{P}(e^{\zeta_{\rm L}}q)
= 
\int\frac{dq}{q}\left[
\mathcal{P}(q) + \zeta_{\rm L} 
\frac{d\mathcal{P}}{dq}q
\right]  =
\int\frac{dq}{q}\left[
\mathcal{P}(q) + \zeta_{\rm L}\,\mathcal{P}(q)\,
\frac{d\log\mathcal{P}}{d\log q}
\right]\,,
\end{align}
so that the presence of the long mode affects the correlation of short modes when their power spectrum is not scale invariant.
The second term in the above equation describes precisely the effect put forth before: the presence of the long mode alters the expectation value of quadratic operators made of short modes, in this case the short-mode two-point function.
This result seems to violate the separate Universe conjecture as also shown in ref.~\cite{Jackson:2023obv,Domenech:2023dxx}. This conjecture states that, in single field inflation models, the curvature perturbation in the superhorizon limit only acts as a rescaling of the spatial coordinates (see e.g. ref.~\cite{Pajer:2013ana,Creminelli:2004yq}) and therefore a local observer in a Hubble horizon patch cannot measure the superhorizon-limit curvature perturbations because it can be absorbed into a rescaling of the spatial coordinates.
Indeed the separate Universe conjecture is limited to the case of single-clock inflation. 
The term single-clock inflation usually refers to the most general
form for the inflationary action (typically constructed through the effective field theory approach) that is consistent with unbroken spatial diffeomorphisms and the presence of a preferred temporal coordinate that represents the ``clock'' during inflation (time-diffeomorphisms are spontaneously broken). 
Single-field slow-roll inflation represents the prototypical example of  single-clock inflation.
The single-clock background is an attractor, and long-wavelength perturbations appear in short-wavelength modes as a constant renormalization of the scale factor that  does not affect the local physics.

\begin{figure}[h]
\begin{center}
$$\includegraphics[width=.495\textwidth]{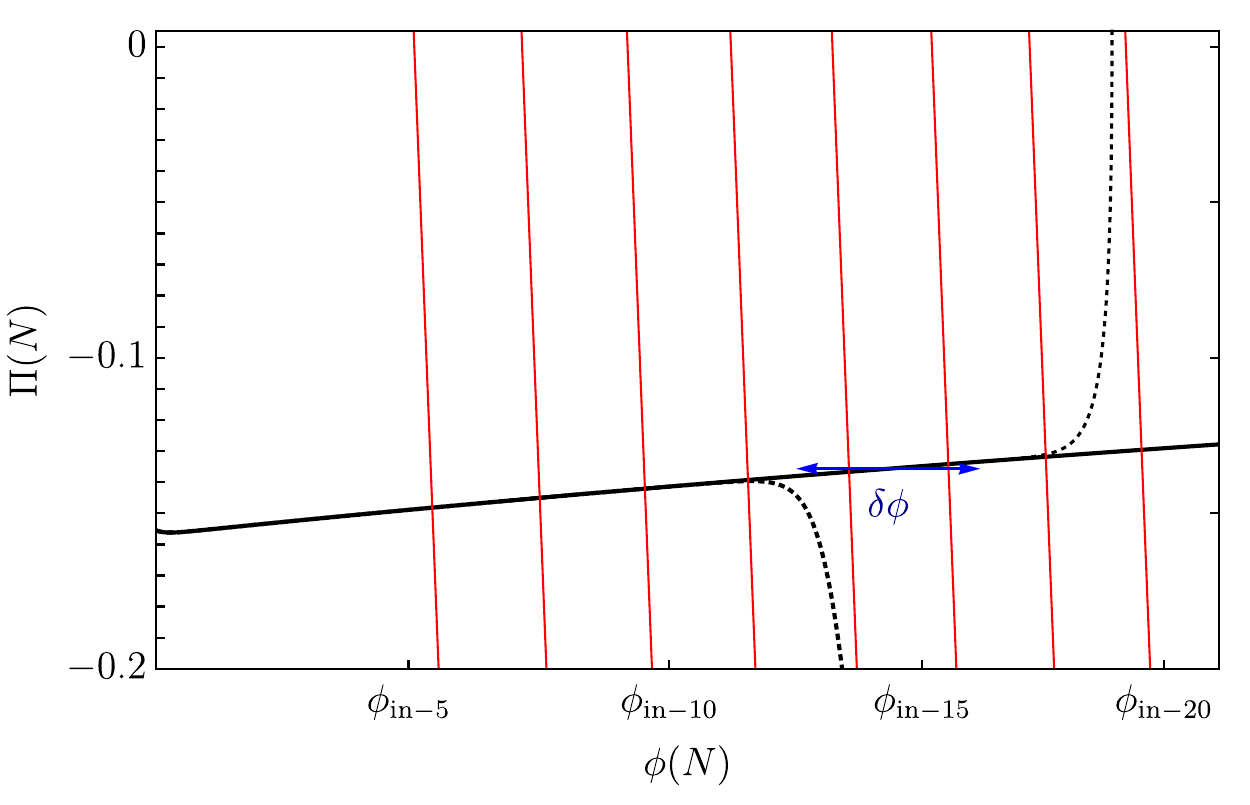}~
\includegraphics[width=.495\textwidth]{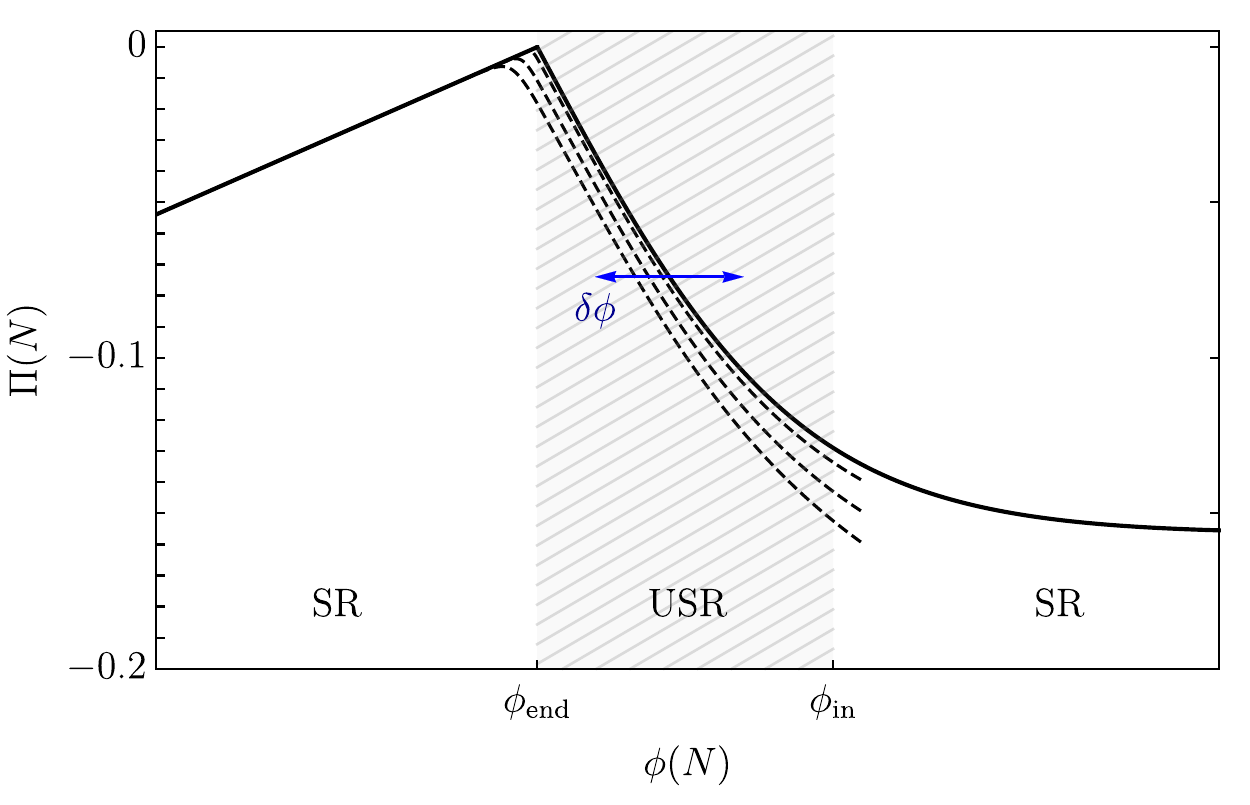}$$\vspace{-0.5cm}
\caption{\textit{\textbf{Left:}} Dynamics evolution (from right to the left) of the initial SR phase.
The black dotted lines represent two benchmark solutions with large initial velocities, rapidly attracted by the SR trajectory (solid black line). 
\textit{\textbf{Right:}} Dynamics evolution in presence of a USR phase. The background trajectory ceases to be an attractor. Here, the perturbation $\delta\phi$ in the field direction has the effect of altering the background trajectory in phase-space, as indicated by the dashed black lines.
 }\label{fig:PhaseSpace}  
\end{center}
\end{figure}

However USR violates the assumption of an inflationary attractor solution which underlies single-clock inflation. 
In USR, the field velocity is no longer uniquely determined by the field position and the background is no longer an attractor. 
To be concrete, we consider in fig.\,\ref{fig:PhaseSpace} the phase-space analysis of the SR/USR/SR dynamics, see also ref.~\cite{Passaglia:2018ixg} for a similar discussion.
First, from the time-evolution of $\epsilon$ and $\eta$ we reconstruct the inflationary potential $V(\phi)$ by means of the reverse engineering approach described in ref.\,\cite{Franciolini:2022pav}. We then solve the inflaton equation of motion $\ddot{\phi} + 3H\dot{\phi} + V^{\prime}(\phi) = 0$ and plot the corresponding phase space trajectory (for different initial conditions) in the plane $(\phi,\Pi)$ with $\Pi \equiv d\phi/dN$. 
The dynamics evolves from right to left in fig.\,\ref{fig:PhaseSpace}. 
In the left panel of fig.\,\ref{fig:PhaseSpace} we plot the initial SR phase. The attractor nature of SR is evident. 
The black dotted lines correspond to two benchmark solutions with large initial velocities. As shown in the plot, they are attracted exponentially fast by the SR trajectory (black solid line). 
Consequently, if we consider some perturbation $\delta\phi$ in the field direction
(which can be thought as a long-wavelength curvature perturbation in the flat gauge) we remain anchored to the background trajectory since small deviations in momentum are quickly reabsorbed. As a result, the perturbation $\delta\phi$ can be simply traded for a shift in the number of $e$-folds (red lines in the right panel of fig.\,\ref{fig:PhaseSpace}) that allows one to move on the background trajectory. A shift in the number of $e$-folds is nothing but a constant renormalization of the scale factor. The situation is different when we enter in the USR phase, right panel in fig.\,\ref{fig:PhaseSpace}. 
In this case, the background trajectory is no-longer an attractor and the perturbation $\delta\phi$ in the field direction has the effect of changing the background trajectory in phase-space (dashed black lines).

\begin{figure}[h!]
\begin{center}
\includegraphics[width=.975\textwidth]{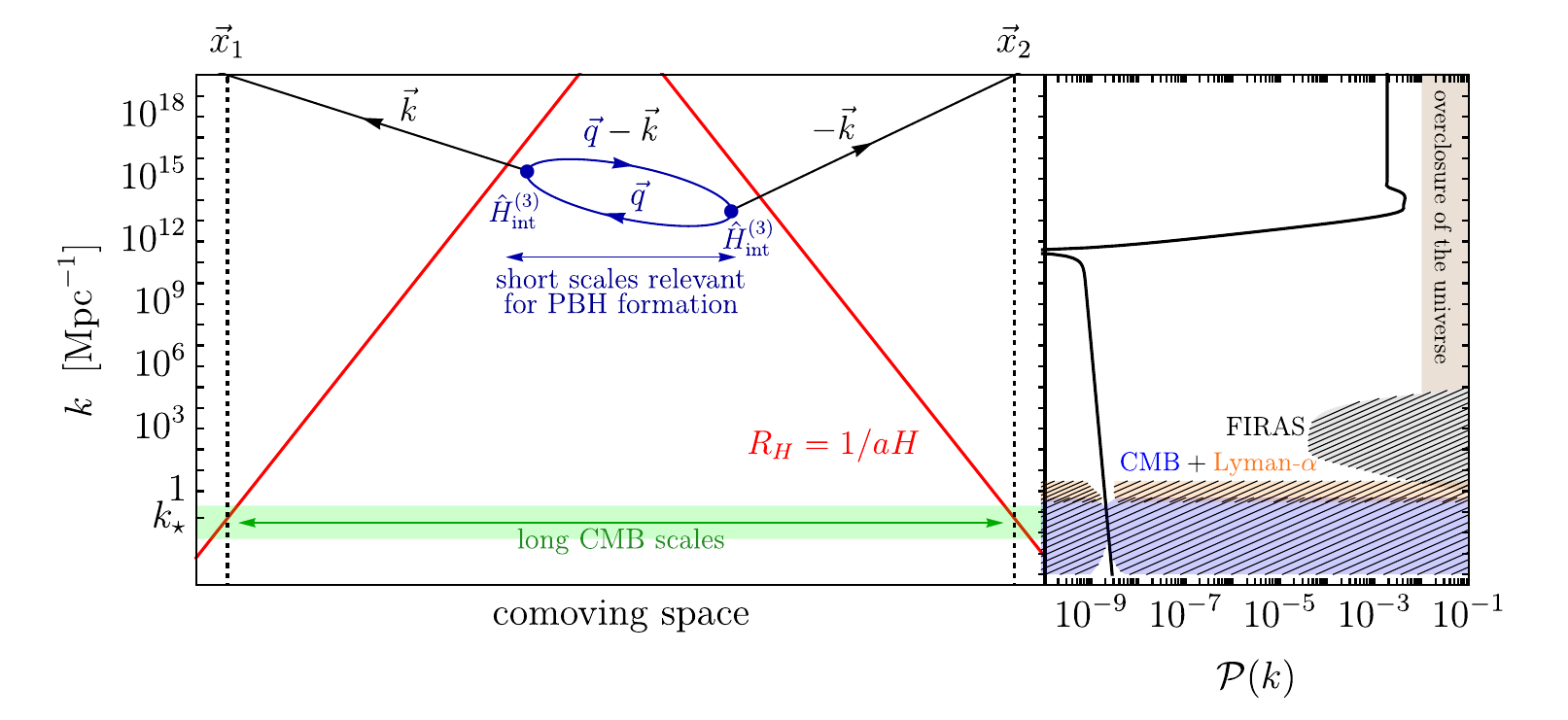}\vspace{-0.5cm}
\caption{
Illustrative schematic of the
correction induced  on the two-point correlator of long modes by a loop of short modes. 
On the right side, we plot the prototypical tree-level
power spectrum of curvature  perturbations as a function of the comoving wavenumber $k$ in the presence of SR/USR/SR dynamics (with $\eta_{\rm III}  = 0$ in the language of the reverse engineering approach. 
The power spectrum features a strong violation of scale invariance at small scales which is needed in order to produce a sizable abundance of PBHs. For illustration, 
we plot the region excluded by CMB anisotropy measurements, ref.\,\cite{Planck:2018jri}, the FIRAS bound on CMB spectral distortions, ref.\,\cite{Fixsen:1996nj,Bianchini:2022dqh}  and the bound obtained from Lyman-$\alpha$ forest data \cite{Bird:2010mp}. If $\mathcal{P}(k) \gtrsim 10^{-2}$,
the abundance of PBHs overcloses the Universe. The plot is rotated in such a way as
to share the same $y$-axis with the left part of the figure. 
On the left side, we schematically plot the evolution of the comoving Hubble
horizon $R_H = 1/aH$ during inflation. 
Observable CMB modes (horizontal green band) cross the Hubble horizon earlier (bottom-end of the figure) and, at the tree  level, their correlation remains frozen from this time on.
At a  much later time, 
the dynamics experience a phase of USR. Modes that 
cross the horizon during the  USR phase have their tree-level power spectrum greatly enhanced and the latter strongly violates scale invariance.
Loop of such short modes may induce a sizable correction to the tree-level correlation of long modes, cf. eq.\,(\ref{eq:Squeezed}).
 }\label{fig:SummarySchematic}  
\end{center}
\end{figure}

Back to eq.\,(\ref{eq:Zelda}), one expects the one-loop correction \cite{Riotto:2023hoz}
\begin{align}
\Delta\mathcal{P}_{\rm 1-loop}(k) \sim 
\mathcal{P}(k)\int\frac{dq}{q}
\,\mathcal{P}(q)\,
\frac{d\log\mathcal{P}}{d\log q}\,.\label{eq:Squeezed}
\end{align} 
The above discussion shows that the one-loop corrections on long modes do not decouple when the power spectrum of the short modes is not scale invariant.
This explains why our correction vanishes in the limit $\eta_{\rm II} =  0$ in which indeed the power spectrum does become scale invariant.
The breaking of scale invariance is the hallmark of the 
USR dynamics and, more importantly, a necessary feature in all models of single-field inflation that generate an order-one abundance of PBHs (cf. the intuitive schematic in fig.\,\ref{fig:SummarySchematic}). 

\section{Double inflection point potential}
We consider a scalar field theory described, in the absence of gravity, by the following Lagrangian density
\begin{align}
\mathcal{L} = \frac{1}{2}(\partial_{\mu}\phi)(\partial^{\mu}\phi) - V(\phi)\,,~~~
V(\phi) = \sum_{k=2}^{n}\frac{a_k g^{k-2}}{k! M^{k-4}}\phi^k\,,\label{eq:FlatSpaceTheory}
\end{align}
where $a_k$ are dimensionless numbers, $g$ some  fundamental coupling and $M$ a mass scale.  
We rewrite the potential as
\begin{align}
V(\phi) = \frac{M^4}{g^2} \sum_{k=2}^{n}
\frac{a_k}{k!}\left(\frac{g\phi}{M}\right)^k\,.
\end{align}
It is convenient to rescale the field in units of the mass-to-coupling ratio $M/g$, and we define 
$x \equiv g\phi/M$. 
We introduce two approximate stationary inflection points in the above potential. To this end, we find that we need at least $n=6$. We write the potential in the following form
\begin{align}
V(x) = \frac{c_4M^4}{g^2}
\left(
\bar{c}_2 x^2 + 
\bar{c}_3 x^3 +
x^4 +
\bar{c}_5 x^5 + 
\bar{c}_6 x^6
\right)\,,\label{eq:Pot1}
\end{align}
with $c_k \equiv a_k/k!$ and 
$\bar{c}_k \equiv c_k/c_4 = a_k4!/a_4k!$. 
Clearly, the potential in eq.\,(\ref{eq:Pot1})  renders our theory non-renormalizable and, as such, it should be interpreted as an effective field theory. We discuss the validity of the EFT approach and the fine tuning beyond this model in appendix\,\ref{app:EFT}.
In order to enforce the presence of a double inflection point, we introduce the following parametrization. 
Before proceeding, we remark that this step is not really needed, and one could as well work directly with the potential in eq.\,(\ref{eq:Pot1}). 
However, the parametrization we are about to introduce will be helpful in the course of the numerical analysis.
We impose the four conditions
\begin{align}
V^{\prime}(x_0) =
V^{\prime\prime}(x_0) = 0\,,
~~~
V^{\prime}(x_1) =
V^{\prime\prime}(x_1) = 0\,,\label{eq:ExactStationaryInflectionPoint}
\end{align}
that identify $x_{0,1}$ as exact stationary inflection points. 
Solving for $\bar{c}_k$, we find
\begin{equation}
V(x) = \frac{c_4 M^4}{g^2} 
\left\{
x^4 + \frac{2}{x_0^2+4x_0x_1 + x_1^2}
\left[
x_0^2x_1^2 x^2 -
\frac{4}{3}x_0x_1(x_0+x_1)x^3 -
\frac{4}{5}(x_0 + x_1)x^5 + \frac{x^6}{3}
\right]\right\}\,.\label{eq:ExactPotential}
\end{equation}
The presence of two inflection points results from a balance between coefficients of opposite signs, specifically those of quadratic/cubic and fifth/sixth order, respectively.

It is known that a shallow local minimum instead of an exact stationary inflection point helps the production of PBHs, cf. refs.\,\cite{Ballesteros:2017fsr,Ballesteros:2020qam}. 
For this reason, we opt for a slight generalization of the parametrization given above. 
The simplest possibility is to 
perturb the conditions in eq.\,(\ref{eq:ExactStationaryInflectionPoint}) and write 
$V^{\prime}(x_0) = \kappa\alpha_1$,
$V^{\prime\prime}(x_0) = \kappa\alpha_2$ and 
$V^{\prime}(x_1) = \kappa\alpha_3$, 
$V^{\prime\prime}(x_1) = \kappa\alpha_4$, with $\kappa \equiv c_4M^4/g^2$ (so that $\alpha_i$ are dimensionless numbers).  
Solving again for $\bar{c}_k$, one can then generalize eq.\,(\ref{eq:ExactPotential}) to the case of approximate stationary inflection points. 
However, the resulting expression is quite cumbersome. For simplicity's sake, therefore, we adopt in our analysis a different parametrization.
We simply write, instead of eq.\,(\ref{eq:ExactPotential}), the expression 
\begin{align}
V(x) = &\frac{c_4 M^4}{g^2} 
\left\{
x^4 + \frac{2}{x_0^2+4x_0x_1 + x_1^2}\times\right. \nn\\
&\left[
x_0^2x_1^2(1+\beta_2)x^2 -
\frac{4}{3}x_0x_1(x_0+x_1)(1+\beta_3)x^3\right. \nn\\ 
&\left.\left. -
\frac{4}{5}(x_0 + x_1)(1+\beta_5)x^5 + \frac{(1+\beta_6)x^6}{3}
\right]
\right\}\,,
\end{align}
where the dimensionless coefficients $\beta_i$ parametrize deviations from the situation in which the stationary inflection  points are exact.
For completeness, the mapping between the two representations (that is, between $\alpha_i$ and $\beta_i$) is given in appendix\,\ref{app:Para}.

We minimally couple the theory in eq.\,(\ref{eq:FlatSpaceTheory}) with Einstein gravity.
However the addition of a non-minimal coupling does not change the conclusions regarding the reheating. In fact, as shown in ref.\,\cite{Frosina:2023nxu}, where a similar potential is used but with a non-minimal coupling and the parameters are specifically tuned to achieve a peaked mass function in the subsolar mass range, the minimum number of e-folds remains around $60$ with the same value of the Hubble rate $H$. We aim to structure our discussion in a broad context, but we will directly present the outcomes of different analyses along with corresponding figures focusing on the best realization of our model. In the latter case, the parameter values are reported in Tab.\,\ref{Tab:XXX}.
\begin{center}\label{Tab:XXX}
\begin{tabular}{||c||c|c|c|c|c|c|c||}
\hline Parameter & $x_0$ & $x_1$ & $\tilde{\beta}_2$ & $\tilde{\beta}_3$ & $\tilde{\beta}_5$ & $\tilde{\beta}_6$ & $c_4 M^4 / g^2$ \\
\hline Value & $1.44$ & $6.00$ & $0.00$ & $0.44$ & $-0.70$ & $-0.95$ & $5\times 10^{-12}$  \\
\hline \hline
\end{tabular}
\end{center}
TABLE: Numerical values (with two digits precision) for the benchmark realization of our model that is explicitly studied in the course of this analysis. Here for a matter of clarity we are introducing the notation $\tilde{\beta}_i=10^{3} \beta_i$.

\begin{figure*}[!t]
	\centering
\includegraphics[width=0.995\textwidth]{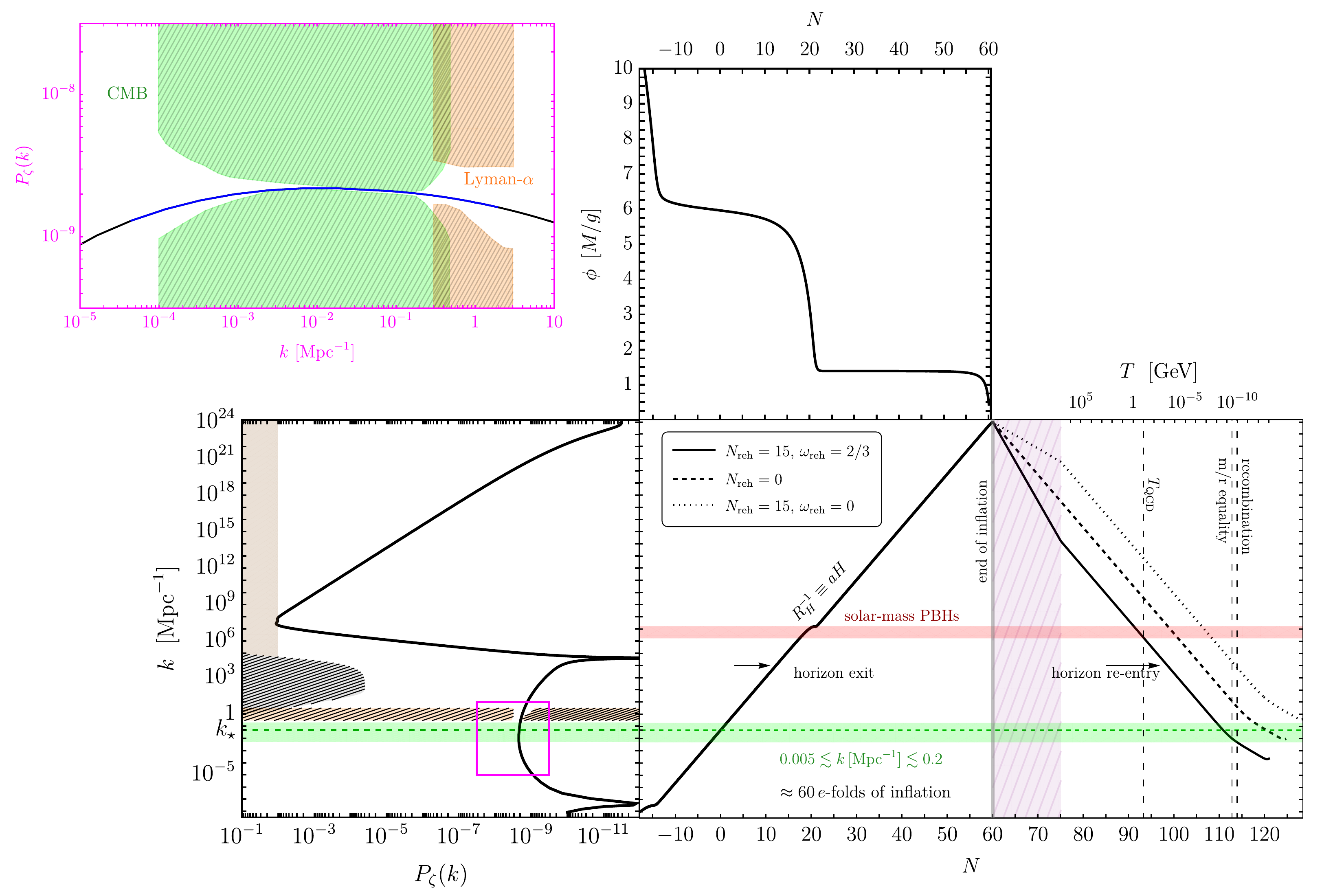}
	\caption{ \textbf{\textit{Left panel:}} Power spectrum computed using the double inflection point potential with the parameters in tab.\,\ref{Tab:XXX}. In the pink panel it is the same power spectrum but zoomed on the relevant CMB scales. \textbf{\textit{Central panel:}} Evolution of the horizon and the related mode $k$ at a given e-folds time $N$. The green band is the one related to the CMB modes. As we can see assuming instant reheating the crossing between the black line and the green band happens after the recombination. \textbf{\textit{Top panel:}} Dynamics of the inflaton field $\phi$ as a function of the e-folds time $N$.
 }
\label{fig:FullUniEvo}
\end{figure*}
\begin{figure}[!t]
\includegraphics[width=0.495\textwidth]{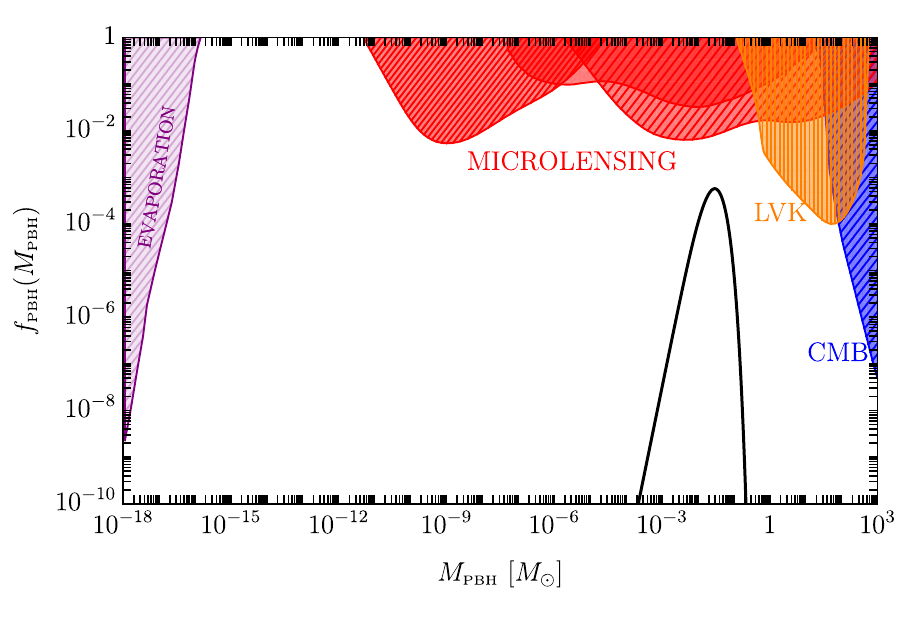}
\includegraphics[width=0.495\textwidth]{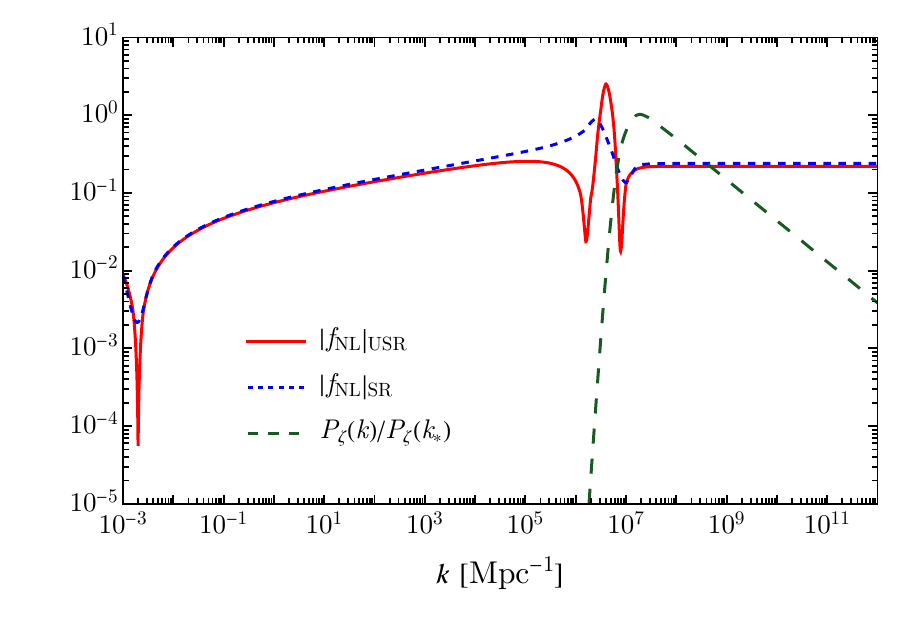}
	\caption{\textbf{\textit{Left panel:}} $f_{\rm PBH}(M_{\rm PBH})$ computed starting from the corresponding power spectrum in the Fig.\ref{fig:FullUniEvo}. Constraints on $f_{\rm PBH}$ shown in the plot are addressed in sec.\,\ref{sec:Clas}.\textbf{\textit{Right panel:}} Evolution of the non-gaussian parameter $f_{\rm NL}$ computed using the Maldacena's approximation (eq.\,\ref{eq:FNLMaldacena}), valid only in Slow-roll approximation (blue dashed line), and the full relation (eq.\,\ref{eq:FNLatal}) (red solid line). The green dashed line is the power spectrum, normalized respect to its peak. } 
\label{fig:PBHAbuIov}
\end{figure}

The obtained power spectrum is reported in the left panel of Fig.\,\ref{fig:FullUniEvo}. In the pink panel a zoom on the CMB scales is reported, where the inflationary power spectrum satisfied the constraints from Planck\,\cite{Planck:2018jri}.
We followed the prescription presented in sec.\ref{sec:C1A} based on the threshold statistics on the compaction function in order to compute the abundance of PBHs.
We followed the prescription given in ref.\,\cite{Musco:2020jjb} to compute the values of $\mathcal{C}_{\rm th}$ and $r_m$, which depend on the shape of the power spectrum.
The results are shown in the left panel Fig.\,\ref{fig:PBHAbuIov}. As we can see from the latter, for peaked distribution in the sub-solar mass range the PBH explanation for the totality of the dark matter is not possible, but they can form maximum $1\%$ of it. However, as previously mentioned in the introduction, the sub-solar mass range is particularly intriguing since standard stellar evolution processes do not account for the formation of sub-solar mass black holes. The detection of even a single sub-solar mass black hole would provide crucial evidence supporting the existence of PBHs. Before discussing the horizon crossing problem in the presence of a USR phase, we comment about the approximation made in the computation of the abundance.
We computed the quantity $f_{\mathrm{PBH}}\left(M_{\mathrm{PBH}}\right)$ assuming Gaussian statistics for the comoving curvature perturbation field $\zeta$. However, the presence of local non-Gaussianity of primordial origin - typically parameterized through the parameter $f_{\mathrm{NL}}$ is unavoidable since the curvature perturbation field is governed by non-linear dynamics\footnote{In general when $\zeta \geq 5/(6f_{\rm NL})$ there is also a contribution from PBHs formed by bubbles of trapped vacuum. In realistic USR scenarios, $f_{\rm NL}$ is of order $\mathcal{O}(0.1)$ and the contribution from these bubbles is negligible\,\cite{Escriva:2023uko,Uehara:2024yyp}.}. In presence of a dynamics entirely governed by an attractor phase, as in the case of USR, the $f_{\rm NL}$ is purely determined by the Maldacena relation\,\cite{Maldacena:2002vr}
\begin{equation}\label{eq:FNLMaldacena}
   \left| f_{\rm NL}\right|_{\rm SR}=\left|\frac{5}{12}\left(1-n_s\right)\right|
\end{equation}
While in presence of a non-attractor phase the previous relation is modified as follow\,\cite{Atal:2018neu}
\begin{equation}\label{eq:FNLatal}
    \left| f_{\rm NL}\right|_{\rm USR}=\left|\frac{5}{12}\left(-3+\sqrt{9-12 \eta_V}\right)\right|
\end{equation}
We plot these two values along with the properly normalized power spectrum from Fig.\,\ref{fig:FullUniEvo} in the right panel Fig.\,\ref{fig:PBHAbuIov}.
Eq.\,\ref{eq:FNLatal} is valid for all the modes that exit the horizon during USR or during the subsequent constant roll phase. For the modes before USR, the two relations give the same result since also if the dynamics is not purely slow-roll, it is still show an attractor behavior.
As we can see from the right panel of Fig.\,\ref{fig:PBHAbuIov} next to the peak of the power spectrum, where the biggest contribution to the abundance comes, the primordial non-gaussian parameter $f_{\rm NL}$ is very small. As a consequence the impact of primordial non-gaussianity can be easily reabsorbed modifying slightly the amplitude of the curvature power spectrum\,\cite{Ferrante:2022mui,Gow:2022jfb,Ianniccari:2024bkh}. Since these modifications are achievable with an irrelevant change of the parameters\,\cite{Frosina:2023nxu}, the conclusions remains the same as in the case of gaussian curvature perturbation field.

\subsection{The Reheating issue for sub-solar mass PBH}
The initial step of our discussion pertains to the evolution of the post-inflationary Universe. In this case, we adopt the standard $\Lambda$CDM cosmology. 
We use the numerical values of astrophysical constants and parameters as reviewed in 
ref.\,\cite{ParticleDataGroup:2022pth} (cf. also 
\href{https://pdg.lbl.gov/2023/reviews/contents_sports.html}{pdg tables}).

We focus our attention on the time-evolution of the comoving Hubble radius $R_H\equiv (aH)^{-1}$. In this expression, $a=a(t)$ is the scale factor of the flat Friedmann-Robertson-Walker (FRW) metric whose line element  is commonly expressed as $ds^2 = dt^2 - a(t)^2[dr^2+ r^2(d\theta^2+sin^2\theta d\phi^2)]$ with $t$ the cosmic time and $(r,\theta,\phi)$ comoving spherical coordinates.
The Hubble expansion rate is defined by $H \equiv \dot{a}/a$ where the overdot denotes a derivative with respect to the cosmic time, i.e. $\dot{a}=da/dt$. 
We use subscripts $_{0}$ to denote quantities evaluated today, at $t = t_0$. 
The evolution of the Hubble expansion rate in $\Lambda_{\rm CDM}$ cosmology follows from the Friedmann equation, and takes the form  
\begin{align}
H(a) = H_0\sqrt{
\Omega_{r,0}\left(\frac{a_0}{a}\right)^4 +
\Omega_{m,0}\left(\frac{a_0}{a}\right)^3 +
\Omega_{\Lambda,0}
}\,,
\end{align}
where 
$H_0 = 100\,h\,\textrm{km}\,\textrm{s}^{-1}\, \textrm{Mpc}^{-1}$ is the 
present-day Hubble expansion rate while $\Omega_{i,0}\equiv \rho_{i,0}/\rho_{\textrm{crit},0}$ are the dimensionless density parameters with $i=r,m,\Lambda$ denoting, respectively, radiation, matter and vacuum energy. The critical density of the Universe today is given by $\rho_{\textrm{crit},0} = 3H_0^2/8\pi G_N$ with $G_N$ the Newtonian constant of gravitation. 
In this section, we employ the standard normalization for the scale factor, assuming $a_0 = 1$.
We introduce the conformal time $\tau$ defined by $d\tau = dt/a(t)$. After integration, we get
\begin{align}
\tau(a)-\tau_{\textrm{in}} = 
\int_{a_{\textrm{in}}}^{a}
\frac{d a^{\prime}}{a^{\prime 2}
H(a^{\prime})}\,.\label{eq:ConformalTimeInte}
\end{align}
In the standard cosmological evolution without inflation, we integrate from the initial big bang singularity at $\tau_{\textrm{in}} = 0$ with $a_{\textrm{in}} = 0$. 
Through eq.\,(\ref{eq:ConformalTimeInte}), it is possible, for a given value of the scale factor $a$ and thus for a fixed value of the comoving Hubble radius, to compute the corresponding conformal time. 
Shortly after recombination, photon decoupling occurred---an epoch when photons became free to travel through space. Before photon decoupling, photons were tightly coupled to charged particles in the ionized plasma, constantly scattering and preventing light from freely propagating through space. After recombination, photons were able to decouple and travel freely, giving rise to the CMB radiation we observe today. Recombination and photon decoupling occurred around the same redshift, approximately at $z_{\textrm{rec}} = 1090$. The relation between redshift and scale factor is $1+z = a(t_0)/a(t)$. 

The aim now is to utilize this information to extract insights and potentially impose constraints on inflationary models. 
In the standard big bang cosmology, the Universe underwent rapid expansion during its early phases (driven first by radiation then by matter). This expansion led to non-overlapping past light cones for many regions in the observable Universe, implying that these regions never had the opportunity for direct causal contact. According to conventional cosmological principles, such regions, widely separated and never in causal contact, should not exhibit similar temperatures. 

For a Universe dominated by a fluid with equation of state $P = \omega \rho$, the evolution of the comoving Hubble radius is dictated by 
\begin{align}
\frac{1}{aH} \propto a^{(1+3\omega)/2}\,.\label{eq:PerfectFluid}
\end{align}
Inflation addresses the horizon problem by proposing a phase of decreasing comoving Hubble radius in the early Universe, where $\frac{d}{dt}(1/aH) < 0$. Eq.\,(\ref{eq:PerfectFluid}) thus implies that one needs $\omega < -1/3$. 
This is because pressure and energy density are given by 
\begin{align}
P_{\phi} = \frac{1}{2}\dot{\phi}^2 - V(\phi)\,,~~~
\rho_{\phi} = \frac{1}{2}\dot{\phi}^2 + 
V(\phi)\,,\label{eq:InflaField}
\end{align}
where consistency with the symmetries of the FRW spacetime requires that the value of the inflaton only depends on time, $\phi=\phi(t)$. Under the assumption that the potential term dominates over the kinetic energy one gets $P_{\phi} \approx -\rho_{\phi}$.
From eq.\,(\ref{eq:PerfectFluid}) with $\omega = -1$, we find that the Hubble rate remains approximately constant during inflation, while the scale factor undergoes exponential expansion, $a(t) \propto e^{Ht}$. 

What happens is that the big bang singularity at $\tau = 0$ is formally extended to $\tau = -\infty$. The radiation era is preceded by an inflationary phase dominated by the dynamics of the inflaton field during which there is a shrinking of the comoving Hubble sphere. 
What happens is that a certain value of $k$, which exits the comoving Hubble sphere during the radiation era (tracing back in time), will re-enter it at some point during the inflationary phase if it persists for a sufficiently long duration.

The transition between the inflationary phase and the radiation era is referred to as the reheating phase. 
The mechanism through which reheating occurs  depends on the specific details of the particle physics involved in the early Universe. During reheating, the inflaton field oscillates around its minimum, and its energy is transferred to other particles, leading to the creation of a hot and dense bath of particles. This process is crucial for connecting the inflationary phase to the subsequent radiation era, setting the stage for the formation of light elements and the evolution of the Universe as we observe it today.
Ideally, reheating is treated as an instantaneous process, while more realistically it requires further modeling. In a more detailed description, reheating is a complex process involving the conversion of the energy stored in the inflaton field into the particles that constitute the standard model of particle physics. The traditional treatment of reheating considers the dynamics of inflaton decay, particle production, and thermalization.

In this section, we essentially remain agnostic about the details of reheating (although in the discussion of our results, we will strive to be as concrete as possible). For the time being, we will simply operate under the assumption that between the end of inflation and the onset of the radiation era, the Universe undergoes a phase during which it is dominated by a fluid with a constant equation of state, $\omega_{\textrm{reh}}$ (for recent review on reheating see also refs.\,\cite{Bassett:2005xm,Allahverdi:2010xz,Amin:2014eta}).

We frequently employ the following shorthand notation for the time evolution of the Hubble rate and scale factor
\begin{align}
H(N_{i}) \equiv H_i\,,~~~a(N_i) \equiv a_i\,.   
\end{align}
 
We consider some value of comoving wavenumber $k$ and compare it with the inverse comoving Hubble radius at the time of matter-radiation equality, $k/a_{\textrm{eq}}H_{\textrm{eq}}$. 
We introduce the $e$-fold time $N_k$ defined through the equation
\begin{align}
k=a(N_k)H(N_k) = a_kN_k\,.    
\end{align}
Following from our previous discussion, $N_k$ indicates the moment in time during inflation when $k$ transitions from being sub-horizon to super-horizon. We thus write
\begin{equation}
\log\left(
\frac{k}{a_{\textrm{eq}}H_{\textrm{eq}}}
\right) = 
-\Delta N_k - \Delta N_{\textrm{reh}} - \Delta N_{\textrm{RD}} + \log\left(
\frac{H_k}{H_{\textrm{eq}}}
\right)\,.\label{eq:Master1}
\end{equation}
In this expression 
\begin{itemize}
\item[{\it i)}] $\Delta N_k \equiv N_{\textrm{end}}-N_k$ indicates the number of $e$-fold between the moment in time during inflation when $k$ transitions from being sub-horizon to super-horizon and the end of inflation.
\item[{\it ii)}] $\Delta N_{\textrm{reh}}\equiv N_{\textrm{reh}}-N_{\textrm{end}}$ denotes the duration of the reheating phase in terms of the number of $e$-folds.
\item[{\it iii)}] $\Delta N_{\textrm{RD}}\equiv N_{\textrm{eq}}-N_{\textrm{reh}}$  
denotes the duration of the radiation era that elapses between the end of reheating and matter-radiation equality in terms of the number of $e$-folds. 
\end{itemize}
According to these definitions, 
$\Delta N_{k}$, $\Delta N_{\textrm{reh}}$ and $\Delta N_{\textrm{RD}}$ are positive quantities. 

\begin{figure}[!t]
	\centering
\includegraphics[width=0.6\textwidth]{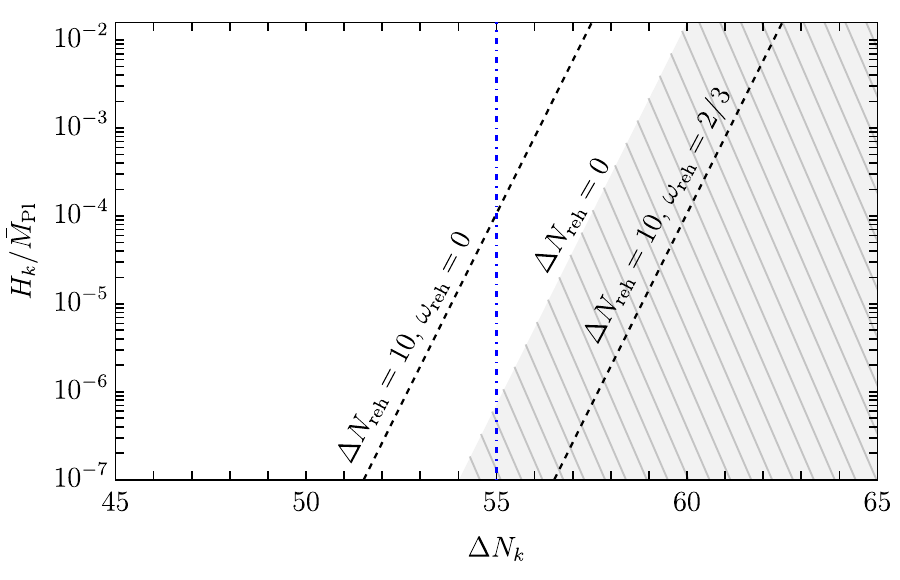}
\includegraphics[width=0.6\textwidth]{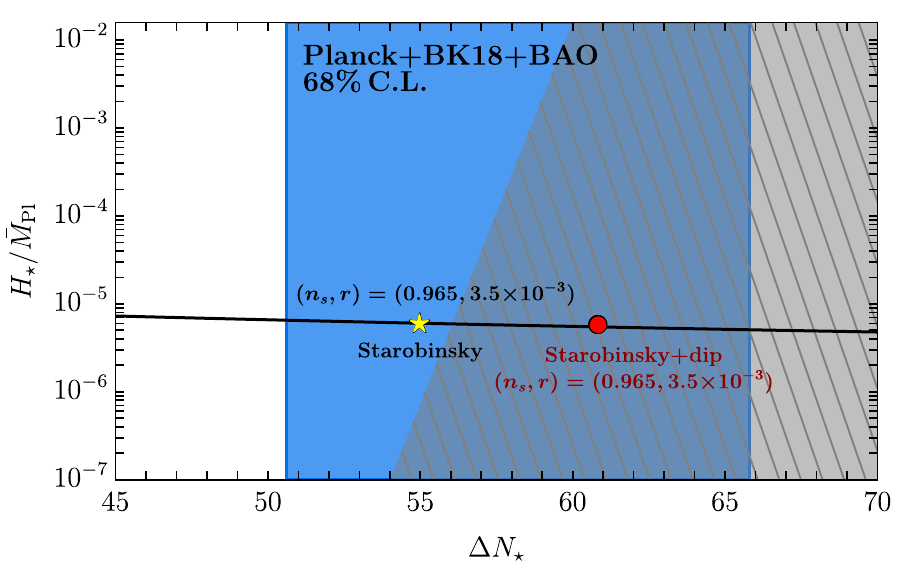}
\includegraphics[width=0.6\textwidth]{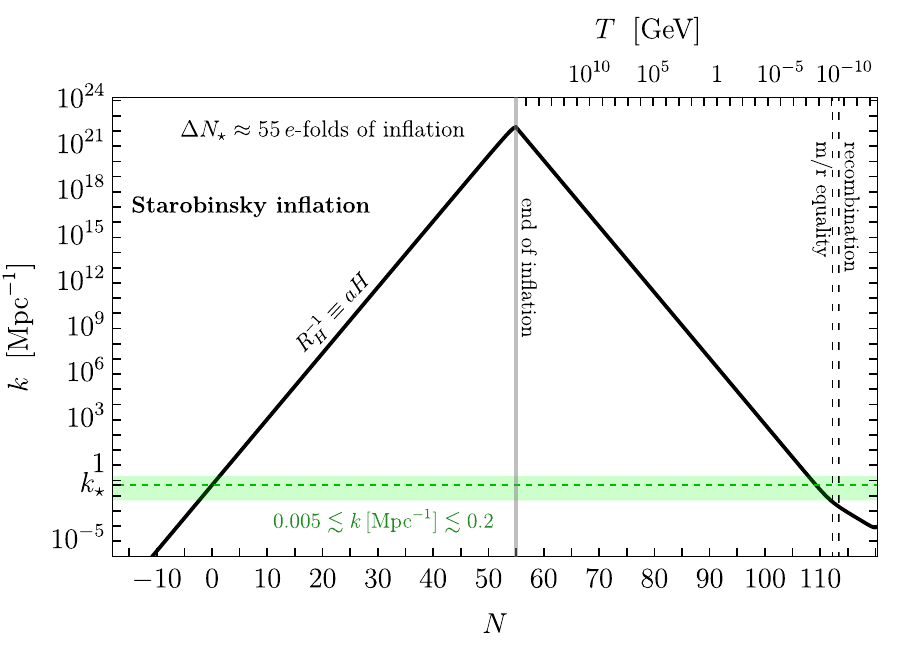}

\caption{\textbf{\textit{Top panel:}} Values of the Hubble rate $H_{k}$ when the related mode $k$ crosses the horizon with different duration of the reheating $\Delta N_{\rm reh}$ and $w_{\rm reh}$.The region shaded in gray corresponds to the condition $\log(k_{\star}/
a_{\textrm{eq}}H_{\textrm{eq}}) < 1$ computed according to eq.\,(\ref{eq:MainLog}) assuming instantaneous reheating.\textbf{\textit{Middle panel:}}. 
Same as top panel but focus on the CMB scale $k_*=0.05$ ${\rm Mpc}^{-1}$. The solid black line corresponds to eq.\,(\ref{eq:HStarStaro}) with $A_s = 2.1\times 10^{-9}$. 
The region shaded in blue corresponds to the 68\% C.L. contour on $(n_s,r)$ that we get using the Planck
2018 ref.\,\cite{BICEP:2021xfz}.  
\textbf{\textit{Bottom panel:}} Evolution of the cosmological horizon as a function of the e-folds time $N$ for the Starobinsky model assuming instant reheating..
 }
\label{fig:InflaBound}
\end{figure}

As previously mentioned, we characterize the reheating period as a time when the Universe is governed by a fluid exhibiting a constant equation of state, denoted as $\omega_{\textrm{reh}}$. Consequently, the continuity equation $\dot{\rho} + 3H\rho(1+\omega_{\textrm{reh}}) = 0$ reads 
$d\log\rho = -3(1+\omega_{\textrm{reh}})d\log a$ from which we get 
\begin{align}
\rho(N_{\textrm{reh}}) = \rho(N_{\textrm{end}}) 
e^{-3(1+\omega_{\textrm{reh}})(N_{\textrm{reh}}-N_{\textrm{end}})}\,,\label{eq:reheatingdensity}
\end{align}
where we integrated between the end of inflation and the end of the reheating phase (cf. point {\it ii)} above). 
In eq.\,(\ref{eq:reheatingdensity}), 
$\rho(N_{\textrm{end}}) \equiv \rho_{\textrm{end}}$ is the energy density at the end of inflation, and can be computed for a given explicit model of inflation. 
On the other hand, 
$\rho(N_{\textrm{reh}})\equiv \rho_{\textrm{reh}}$ is the energy density at the end of the reheating phase. 
The latter can be related to the reheating temperature of the Universe $T_{\textrm{reh}}$ through 
\begin{align}
\rho_{\textrm{reh}} = \left(
\frac{\pi^2}{30}
\right)g_{\textrm{reh}}T_{\textrm{reh}}^4\,,    
\end{align}
where $g_{\textrm{reh}}$ is the effective number of relativistic species upon thermalization. 
The conservation of comoving entropy gives a relation between the reheating temperature and the present-day temperature $T_0$. 
Consequently, eq.\,(\ref{eq:reheatingdensity}) reads
\begin{equation}
\frac{3(1+\omega_{\textrm{reh}})}{4}\Delta N_{\textrm{reh}}= \frac{1}{4}\log\left(
\frac{30}{g_{\textrm{reh}}\pi^2}
\right) + 
\frac{1}{4}\log\left(
\frac{\rho_{\textrm{end}}}{T_0^4}
\right) +\frac{1}{3}\log\left(
\frac{11g_{s,\textrm{reh}}}{43}
\right) + \log\left(
\frac{a_{\textrm{eq}}}{a_0}
\right)- \Delta N_{\textrm{RD}}\,,\label{eq:Masterrd}
\end{equation}
where $g_{s,\textrm{reh}}$ is the effective number of relativistic degrees of freedom contributing to the Universe's entropy at reheating. Both $g_{s,\textrm{reh}}$ and $g_{\textrm{reh}}$ depends on temperature. 
Since they enter only logarithmically, for the sake of simplicity, we choose to fix their values to the fiducial value $g_{s,\textrm{reh}} = g_{\textrm{reh}} = 100$.
We use eq.\,(\ref{eq:Masterrd}) to eliminate the 
$\Delta N_{\textrm{RD}}$-dependence in  eq.\,(\ref{eq:Master1}). Then 
\begin{align}
\log\left(
\frac{k}{a_{\textrm{eq}}H_{\textrm{eq}}}
\right) = &  +\frac{(3\omega_{\textrm{reh}}-1)}{4}\Delta N_{\textrm{reh}} 
-
\frac{1}{4}\log\left(
\frac{30}{g_{\textrm{reh}}\pi^2}
\right)-
\frac{1}{3}\log\left(
\frac{11g_{s,\textrm{reh}}}{43}
\right) \nn\\ &
-\Delta N_{k} -\log\left(
\frac{a_{\textrm{eq}}}{a_0}
\right)\underbrace{-\log\left(
\frac{\rho_{\textrm{end}}^{1/4}}{T_0}
\right) + \log\left(
\frac{H_k}{H_{\textrm{eq}}}
\right)}_{=\,\frac{1}{2}\log\left(
\frac{H_k}{\sqrt{3}\MPl}
\right) + \log\left(
\frac{T_0}{H_{\textrm{eq}}}
\right)}\,.
\end{align}
As far as the last two terms are concerned, we applied the following simplification. 
On the one hand, we have $\rho_{\textrm{end}}=3\MPl^2H_{\textrm{end}}^2$. 
However, since during inflation $H$ is almost constant (see discussion below eq.\,(\ref{eq:InflaField})), we simply 
write $\rho_{\textrm{end}}=3\MPl^2H_{k}^2$.
All in all, we arrive at 
\begin{align}
\log\left(
\frac{k}{a_{\textrm{eq}}H_{\textrm{eq}}}
\right) = & + \frac{1}{2}
\log\left(\frac{H_k}{\sqrt{3}\MPl}\right)
+ \log\left(\frac{T_0}{H_{\textrm{eq}}}\right)\ +\frac{(3\omega_{\textrm{reh}}-1)}{4}\Delta N_{\textrm{reh}} -
\frac{1}{4}\log\left(
\frac{30}{g_{\textrm{reh}}\pi^2}
\right)
\nn\\
&-\Delta N_{k}  -
\frac{1}{3}\log\left(
\frac{11g_{s,\textrm{reh}}}{43}
\right) - \log\left(\frac{a_{\textrm{eq}}}{a_0}\right)\,.\label{eq:MainLog}
\end{align}
Consequently, if we want the comoving wavenumber $k$ to be sub-horizon at the time of matter-radiation equality, we shall have $\log(k/a_{\textrm{eq}}H_{\textrm{eq}}) > 0$.
Through the preceding equation, this condition translates into a constraint on the parameters $\Delta N_k$, $H_k$, $\omega_{\textrm{reh}}$, and $\Delta N_{\textrm{reh}}$. 
Consider, for instance, the case of instantaneous reheating, $\Delta N_{\textrm{reh}} = 0$. 
In the top panel of Fig.\,\ref{fig:InflaBound}, the gray shaded region corresponds to the condition $\log(k/a_{\textrm{eq}}H_{\textrm{eq}}) < 0$. For specificity, let's consider a mode that exits the horizon approximately $\Delta N_k = 55$ $e$-folds before the end of inflation (vertical dot-dashed blue line in Fig.\,\ref{fig:InflaBound}). If, at the time of its horizon crossing, we have $H_k/\MPl \lesssim 10^{-6}$, then the mode will still be super-horizon at the time of matter-radiation equality. Conversely, if $H_k/\MPl \gtrsim 10^{-6}$, the mode will be sub-horizon at the time of matter-radiation equality.

The inclusion of reheating changes the picture depending on its duration $\Delta N_{\textrm{reh}}$ and the sign of the factor $(3\omega_{\textrm{reh}}-1)$. 
\begin{itemize}
    \item[$\circ$] If  $\omega_{\textrm{reh}} < 1/3$, the reheating stage gives a negative contribution to the right-hand side of eq.\,(\ref{eq:MainLog}). 
    Consequently, the keep the mode $k$ sub-horizon at the time of matter-radiation equality one needs to compensate with a larger value of $H_k$ or a smaller $\Delta N_k$.
    \item[$\circ$] If  
    $\omega_{\textrm{reh}} > 1/3$, the reheating stage gives a positive contribution to the right-hand side of eq.\,(\ref{eq:MainLog}). This consequently allows for the extension of the duration of the interval $\Delta N_k$ or the reduction of the value of $H_k$ without violating the condition $\log(k/a_{\textrm{eq}}H_{\textrm{eq}}) > 0$.
\end{itemize}

We streamline the main argument of this section with the help of a simple model of inflation. Therefore, we put aside for the moment the formation of PBHs and consider the Starobinsky model of inflation\,\cite{Starobinsky:1980te}.
The latter is based on the scalar potential 
\begin{align}
V_{\textrm{Staro}}(\phi) = 
\frac{3M^2\MPl^2}{4}
\left[1 - 
\exp\left(
-\sqrt{\frac{2}{3}}
\frac{\phi}{\MPl}
\right)
\right]^2\,.\label{eq:PoteStaro}
\end{align}
where $M$ is a fundamental mass scale that,
in the chosen normalization for the potential in eq.\,(\ref{eq:PoteStaro}), coincides with the mass of the inflaton.

In the slow-roll approximation, we find (cf. appendix\,\ref{app:Staro}) 
\begin{align}
H_{\star}^2 = \frac{M^2}{4}
\left[
1+\frac{1}{W_{-1}(f_{\Delta N_{\star}})}
\right]^2\,,\label{eq:HStarStaro}
\end{align}
where $W_{-1}(z)$ is the branch with 
$k=-1$ of the Lambert W function $W_k(z)$ and $f_{\Delta N_{\star}}$ is defined in eq.\,(\ref{eq:ShortHandfx}). 
The mass scale $M$ is fixed by the amplitude of the scalar power spectrum measured at the CMB pivot scale. We find
\begin{align}
A_s = \frac{3M^2[1+W_{-1}(f_{\Delta N_{\star}})]^4}{128\pi^2 \MPl^2 
W_{-1}(f_{\Delta N_{\star}})^2}\,.
\end{align}
On the other hand, the amount of inflation $\Delta N_{\star}$ must be compatible with 
the constraints on the scalar spectral index $n_s$ and the tensor-to-scalar ratio $r$ since these can be expressed as
\begin{align}
n_s & = 1-\frac{16}{3[1+W_{-1}(f_{\Delta N_{\star}})]^2} 
+ \frac{8}{
3[1+W_{-1}(f_{\Delta N_{\star}})]
}\,,\nn\\
r & = 
\frac{64}{
3[1+W_{-1}(f_{\Delta N_{\star}})]^2
}\,.
\end{align}
In  the middle panel of Fig.\,\ref{fig:InflaBound} we reproduce the same figure of the top panel but now focus on the CMB pivot scale $k_{\star}$. 
The solid black line corresponds to eq.\,(\ref{eq:HStarStaro}) with $A_s = 2.1\times 10^{-9}$. 
The region shaded in blue corresponds to the 68\% C.L. contour on $(n_s,r)$ that we get using the Planck
2018 baseline analysis including BICEP/Keck and BAO data, cf. ref.\,\cite{BICEP:2021xfz} (see also \href{http://bicepkeck.org/bk18_2021_release.html}{BK18 Data Products}). 
The region shaded in gray corresponds to the condition $\log(k_{\star}/
a_{\textrm{eq}}H_{\textrm{eq}}) < 1$ computed according to eq.\,(\ref{eq:MainLog}) assuming instantaneous reheating. 

The upshot of the analysis is that it is possible to appropriately choose the value of $\Delta N_{\star}$ in order to satisfy the cosmological constraints from Planck+BICEP/Keck without violating the condition that small scales re-enter the horizon before matter-radiation equality. 
For illustration, the yellow star corresponds to $\Delta N_{\star} = 55$. As evident from the plot, this inflationary solution does not create any tension. 

To further substantiate this conclusion beyond the slow-roll approximation, we numerically solved the inflationary dynamics by fixing $\Delta N_{\star} = 55$. 
In the bottom panel of the same figure we show the time evolution of the inverse comoving Hubble radius from inflation to the present day. 
The range of comoving wavenumbers that are relevant for CMB observations are indicated with an horizontal green band. 
In particular, $k_{\star} = 0.05$ Mpc$^{-1}$ is indicated with an horizontal green dashed line.
The CMB pivot scale exits the horizon  $\Delta N_{\star} = 55$ $e$-folds before the end of inflation and re-enters the horizon before matter-radiation equality. We assume instantaneous reheating.

We now come back to the issue of PBH formation from single-field inflation. 
The point is as follows. We would now like our inflationary model, in addition to resolving the horizon problem and being compatible with the observed Universe, to also produce a sizable population of PBHs. This will impose an additional constraint on the model, which, as we will see, essentially removes the freedom of choice on $\Delta N_{\star}$ that was crucial in the previous example for adapting the inflationary model to observational constraints.

To proceed with our analysis, it is necessary to consider a single-field inflation model that has the capability to generate a sizable abundance of PBHs. In our investigation, we focus specifically on PBHs with solar or sub-solar masses. 
These PBHs are potentially highly interesting from a phenomenological perspective. On one hand, considering a binary system of PBHs, a merger event could be detectable by the gravitational interferometers of the LIGO/Virgo/KAGRA collaboration. On the other hand, the same perturbations that would generate the PBHs could serve as a source for a stochastic background of gravitational waves at the nHz frequency, potentially consistent with that recently observed by the PTA experiments such as NANOGrav~\cite{NANOGrav:2023gor, NANOGrav:2023hde}, EPTA (in combination with InPTA)\,\cite{EPTA:2023fyk, EPTA:2023sfo, EPTA:2023xxk}, PPTA\,\cite{Reardon:2023gzh, Zic:2023gta, Reardon:2023zen} and CPTA\,\cite{Xu:2023wog}. 

As a former exercise, we take a reference model where we generate a sizable abundance of (sub)solar-mass PBHs through the presence of an ultra slow-roll phase. For instance, following ref.\,\cite{Mishra:2019pzq}, we perturb the potential of the Starobinsky model
in eq.\,(\ref{eq:PoteStaro}) by introducing a dip-like feature of the following form 
\begin{align}
V_{\textrm{Staro+dip}}(\phi) \equiv 
V_{\textrm{Staro}}(\phi)\left[
1-A\cosh^{-2}\left(\frac{\phi-\phi_0}{\sigma}\right)
\right]\,.\label{eq:StaroDip}   
\end{align}
The hyperbolic function---characterized by  height $A$, position $\phi_0$ and width $\sigma$ that was added to the Starobinsky model---plays the role of a 
speed-breaker term 
for the inflaton. 
As a consequence, the amplitude of the power spectrum of scalar perturbations is significantly enhanced for modes exiting the horizon during the phase when the inflaton is compelled to slow down its classical motion. 
This marks the presence of an ultra slow-roll phase. We tune  the values of the parameters 
$(A,\phi_0,\sigma)$ in such a way that the power spectrum of scalar curvature perturbations peaks at around $k_{\textrm{peak}} = 10^6$ Mpc$^{-1}$ with a peak amplitude of the order $10^{-2}$ (cf. the caption of fig.\,\ref{fig:InflaBound}). These are the typical values that are needed to form a sizable abundance of (sub)solar-mass PBHs. 
We keep fixed the same reference value $n_{s} = 0.965$ that we used in our discussion of the Starobinsky model, cf. fig.\,\ref{fig:InflaBound}.
The key observation is that, to accommodate the presence of a sufficiently long ultra-slow-roll phase, the value of $\Delta N_{\star}$ is forced to increase. 
This is shown in the same fig.\,\ref{fig:InflaBound}. The red dot corresponds to the case 
in which the  Starobinsky potential is perturbed by the presence of a dip-like feature as in eq.\,(\ref{eq:StaroDip}).  
The resulting value of $\Delta N_{\star}$ now exceeds the bound given by eq.\,(\ref{eq:MainLog}).
Note that, on the other hand, in this case the value of $H_{k}$ remains essentially unaltered.
We do not consider this phenomenological toy model to realistically describe the effects of an ultra-slow-roll phase in single-field inflation models. For instance, it is possible to arbitrarily adjust the position and height of the dip without altering the underlying base potential. 
For this reason, we consider now a realistic model of single-field inflation that implements a phase of ultra slow-roll. We take as the model under consideration the double inflection point potential presented in the discussion of the previous section. As showed in the right panel of fig.\,\ref{fig:FullUniEvo} for the realistic scenarios in which we are producing sub-solar mass PBHs with the USR models, in some cases we need to introduce a reheating stage with a stiff equation of state in order to have that the CMB modes cross the horizon before the recombination.

\subsection*{Summary and comparison with the recent literature}
In this chapter, we focused on some peculiar aspects and issues related to the PBH formation in the single field scenario, also generally called Ultra-slow roll models.
First we discussed the implications of perturbativity in the context of single-field inflationary models that feature the presence of a transient phase of USR. 
More in detail, we defined the perturbativity condition
\begin{align}
\mathcal{P}(k) \equiv \mathcal{P}_{\rm tree}(k)
\left[1 + 
\Delta {\cal P}_{\rm 1-loop}(k) \right] ~~~\Longrightarrow~~~
\Delta {\cal P}_{\rm 1-loop}(k)\overset{!}{<} 1\,,\label{eq:PertuConclu}
\end{align}
in which  the one-loop correction is integrated over the  short modes that are enhanced by the USR dynamics. 
We explored the consequences of 
eq.\,(\ref{eq:PertuConclu}) at any scale $k$ even though the main motivation for our analysis was the recent claim of ref.\,\cite{Kristiano:2022maq} according to which the relative size of the loop correction at scales relevant for CMB observations (that is, $k  = O(k_{\star})$ with $k_{\star} = 0.05$ Mpc$^{-1}$) threatens the validity of perturbativity in USR dynamics in single-field inflation.

In this section, we summarize the main results and limitations of our analysis and we will discuss future prospects. 

\begin{itemize}
\item[$\circ$]  
In the limit of instantaneous SR/USR/SR  transition, we confirm the computation of the 1-loop corrections of ref.\,\cite{Kristiano:2022maq}. 
However, 
we provide a more detailed and precise discussion of the theoretical bound that can be obtained by imposing the perturbativity condition 
in eq.\,(\ref{eq:PertuConclu}) on the power spectrum of curvature perturbations at CMB scales.  

As far as this part of the analysis is concerned, the key difference with  respect to 
ref.\,\cite{Kristiano:2022maq} is that we compare the size of the loop correction with an accurate computation of the PBH abundance rather that with the order-of-magnitude estimate of the enhancement of the power spectrum, based on the SR formula, used in ref.\,\cite{Kristiano:2022maq}. 
Our approach, therefore, includes the following effects. 
{\it i)} First of all, we generalize the USR dynamics  for generic values  $\eta_{\rm II} \neq 3$;
{\it  ii)} the  enhancement of the power spectrum at scales relevant for PBH formation is accurately computed by numerically solving the M-S equation rather than using the SR scaling; {\it  iii)} by computing the PBH abundance $f_{\rm PBH}$, we automatically account for the fact that the correct variable that describes PBH formation in the standard scenario of gravitational collapse is the smoothed density contrast rather than the curvature perturbation field, and we include in our computation the full non-linear relation between the two.

As for this part of the analysis, our findings are summarized in fig.\,\ref{fig:RegionBound}.  
We find that loop corrections remain of the order of few percent and therefore it is not possible to make the bold claim that PBH formation from single-field inflation is ruled out -- not even in the limit of instantaneous SR/USR/SR  transition. 

\item[$\circ$] We extend the analysis of 
ref.\,\cite{Kristiano:2022maq} by considering a more realistic USR dynamics. 
In particular, we implement 
a smooth description of the SR/USR/SR  
transition. 
Recently, refs.\,\cite{Riotto:2023gpm,Firouzjahi:2023ahg,Firouzjahi:2023aum} claimed  that the presence of a smooth transition in the final USR/SR transition makes 
the loop correction 
effectively harmless.  
Our analysis shows that this conclusion could be invalidated by the fact that there is an interplay between the size of  the loop correction and the amplitude of  the tree-level power spectrum that is needed to generate a sizable abundance of PBHs. On the one hand, it is true that 
a smooth USR/SR transition reduces the size of the loop correction; 
on the other one, the same smoothing also reduces the amplitude of the tree-level power spectrum so that, in order to keep $f_{\rm PBH}$ fixed, one is forced to either increase the duration of the USR phase or the magnitude of $\eta$ during the latter. In the end, the two effects tend to compensate each other.  
As for this part of the analysis, our findings are summarized in fig.\,\ref{fig:FinalPlot}.

\item[$\circ$] We consider the one-loop correction of short modes to the tree-level power spectrum at any scale. 
We find that perturbativity is always satisfied in models that account for the condition $f_{\rm PBH} = 1$. 

More quantitatively, we find that the relative size of the loop correction with respect to the tree-level value of the  power spectrum does not exceed the level of a few percent. As for this part of the analysis, our findings are summarized in fig.\,\ref{fig:RegionBound2}.
We  point out one notable exception of phenomenological relevance. 
A generic feature of the USR dynamics is that it produces a characteristic dip in the  tree-level power spectrum, as the one observed in the left panel of fig. \ref{fig:TestPS}. 
The phenomenological consequences of  such a putative dip range from CMB
$\mu$-space distortions\,\cite{Ozsoy:2021pws} to 21-cm signals\,\cite{Balaji:2022zur}.  
Our analysis shows that  the existence of the dip is nothing but an artifact of the tree-level computation, and it is significantly reduced after including loop corrections. This is  because, due to the smallness of the tree-level power spectrum around the characteristic wavenumbers of the dip, the non-vanishing loop correction gives the dominant contribution. This is illustrated in the right panel of fig.\,\ref{fig:RegionBound2}.

\end{itemize}
Several important directions can be followed in the next future. 
{\it i)} More realistic modelling of the USR dynamics. 
In realistic single-field inflationary models we expect $\eta_{\rm III} < 0$ and sizable; this is because at the end of the USR we are left with $\epsilon \ll 1$ but we need $\epsilon = O(1)$ to end inflation. 
Since $\epsilon \sim e^{-2\eta N}$, we need $\eta$ large and negative after USR. 
Consequently, after USR we do not expect a scale-invariant power spectrum and eq.\,(\ref{eq:Squeezed}) applies.
{\it ii)} Understanding the role of quartic interactions, tadpoles and interactions with spatial derivatives. So far, most of the attention has been focused on the role of the cubic interaction Hamiltonian in eq.\,(\ref{eq:MainHami}). However, as schematically shown in eq.\,(\ref{eq:LoopsSchematic}), quartic interactions and non-1PI diagrams involving tadpoles are also present. 
In particular, the schematic  in eq.\,(\ref{eq:LoopsSchematic}) shows that tadpole diagrams may be relevant because, by attaching them to propagators, they modify the two-point correlator. 
The correct way to deal with tadpoles is by changing the background solution (cf. ref.\,\cite{Senatore:2009cf}; see also ref.\,\cite{Inomata:2022yte}). 
Since it is well-known that background solutions in USR models for PBH formation suffer a high-level of parametric tuning (cf. ref.\,\cite{Cole:2023wyx}), the role of tadpole corrections may have some relevance. 
Furthermore, all interactions with spatial derivatives have been so far discarded. However, the short modes running in the loop cross the horizon precisely during the USR phase and, therefore, their spatial derivatives do not pay any super-horizon suppression. 
{\it iii)} Renormalization. An essential future step is to implement a thorough renormalization procedure in the context of USR dynamics, a topic that has not yet been addressed in the existing literature.
Eventually we discussed the issue of producing primordial black holes in the sub-solar mass range within reheating scenarios with $w\leq 1/3$. As we can see in the next chapter this can be easily achieved with the help of a spectator field.
Following the publication of ref.,\cite{Franciolini:2023agm}, numerous papers on 1-loop corrections have appeared in the literature ,\cite{Tasinato:2023ukp,Fumagalli:2023hpa,Firouzjahi:2024psd,Iacconi:2023ggt,Davies:2023hhn,Tasinato:2023ioq,Firouzjahi:2023bkt,Jackson:2023obv,Tada:2023rgp,Inomata:2024lud,Kristiano:2024vst}. However, the situation remains somewhat unclear. It seems that these 1-loop corrections might violate the separate Universe conjecture, which asserts that in single-field inflation models, the curvature perturbation in the superhorizon limit merely rescales the spatial coordinates, making it non-physical. 'Non-physical' here means that a local observer within a Hubble horizon patch cannot detect superhorizon curvature perturbations, as they can be absorbed into the spatial coordinate definition (i.e., a rescaling).

USR dynamics are intrinsically non-attractive. Couplings that are suppressed during SR are no longer suppressed in USR. Consequently, while the separate Universe approach holds in SR, it is violated in USR. Reference \cite{Jackson:2023obv} demonstrates that a transition between SR and USR induces non-adiabatic behavior, driven by gradient terms, which leads to a violation of the separate Universe approach. However, this violation seems to arise from perturbations at horizon scales (see, for instance, the discussion around eq. 3.15). Therefore, we expect the separate Universe approach to describe the dynamics before and after the USR phase adequately.

Furthermore, ref.,\cite{Domenech:2023dxx} shows that in the presence of a USR phase, the standard $\delta N$ formalism only becomes valid after a sufficient period following a sudden transition. Thus, while the separate Universe approach remains valid before and after the USR phase, it breaks down for modes affected by the USR phase itself.

At the background level, the separate Universe approach is applicable to modes related to the CMB. However, this reasoning holds only at the background level. As demonstrated in this chapter, at the loop level, all scales are involved, resulting in corrections that also impact CMB scales. The origin of these 1-loop corrections remains incompletely understood. Moreover, we emphasize that the primary goal of the analysis presented before is to show that in the original scenario proposed by Kristiano and Yokoyama in ref.,\cite{Kristiano:2022maq}, the 1-loop corrections do not lead to a violation of perturbativity.

Significant progress in understanding the role of renormalization and its impact on the 1-loop issue has been made in ref.,\cite{Ballesteros:2024zdp}. It has been shown that even when including all relevant cubic and quartic interactions, along with the necessary counterterms to renormalize the ultraviolet divergences (regularized by a cutoff), perturbation theory does not necessarily break down in models where the duration of USR does not lead to overproduction of PBHs.


\chapter{Primordial black hole production \\from a spectator field model: 
\\Axion-like curvaton model}\label{cap:Models2}
An additional class of models in which a boost of the power spectrum can be achieved at the relevant scales, is given by introducing a customary spectator field. In this chapter we focus, following closely ref.\,\cite{Ferrante:2023bgz}, on the \textit{Curvaton models.} These models are a sub-class of models with spectator field. In such models a complex scalar field $\Phi=\varphi e^{i \vartheta}$ is introduced and its angular component $\vartheta$ is dubbed \textit{curvaton}. The inflationary dynamics of the radial field $\varphi$ enhances by many orders of magnitude angular perturbations $\delta\vartheta$ at small scales, while scales associated with CMB observations are dominated by the inflaton contribution. After inflation, assuming a coupling between the curvaton and photons, the perturbations $\delta\vartheta_k$ are transferred into radiation, and the boost of the perturbation in $\delta\vartheta_k$ becomes a boost in the curvature perturbation field $\zeta$.

We study PBH production in the framework of a specific axion-like curvaton model\,\cite{Kawasaki:2012wr,Ando:2017veq,Inomata:2020xad}, following closely ref.\cite{Ferrante:2023bgz}, where a complex scalar field $\Phi=\varphi e^{i \vartheta}$ is introduced and its angular component $\vartheta$ is dubbed \textit{curvaton}.
The rolling of the radial component $\varphi$ down the potential during inflation enhances by many orders of magnitude the angular perturbations $\delta \theta$ at small scales, while,  at scales associated with CMB observations, $\delta \theta$ is suppressed and the main contribution to density fluctuations comes from the inflaton field, which lives in an uncoupled sector.
Exploiting these features, we are able to obtain a broad power spectrum of curvature perturbations, i.e. a power spectrum which spans over many orders of magnitude of scales $k$, in a concrete model of the early Universe. 
This is crucial in order to make connection between observables related to PBH dark matter and those associated with scalar-induced GWs. 
\section{Model dynamics}
\subsection{The inflationary dynamics}
For simplicity, we work in the limit in which the Hubble rate during inflation is constant (pure de Sitter background), 
and we indicate its value with $H_{\rm inf}$. 
We consider the axion-like curvaton model explored in refs.\,\cite{Kawasaki:2012wr,Ando:2017veq,Inomata:2020xad}.  
In these models the radial field is subject to the quadratic potential
\begin{align}\label{eq:EffectiveV}
V(\varphi) = \frac{c}{2}H_{\rm inf}^2(\varphi - f)^2\,,
\end{align}
which has a minimum at $\varphi = f$. Refs.\,\cite{Dine:1995kz,Kawasaki:2012wr,Ando:2017veq,Inomata:2020xad} justify the potential in eq.\,(\ref{eq:EffectiveV}) in the context of supergravity models.

If the radial field $\varphi$ during inflation rolls down the quadratic potential in eq.\,(\ref{eq:EffectiveV}), starting from some large field value of Planckian order, the angular perturbations get a large enhancement. This is easy to see explicitly. 
The equation of motion for $\varphi$ (written in terms of the number of $e$-folds $N$, defined by $dN = H dt$, as time variable)
\begin{align}
\frac{d^2\varphi}{dN^2} + 3\frac{d\varphi}{dN} + \frac{1}{H_{\rm inf}^2}\frac{dV(\varphi)}{d\varphi} =0\,,
\end{align}
admits the analytical solution 
\begin{align}\label{eq:NaiveSolution}
\varphi_H(N) = f_H + c_1 e^{N(-3 - \sqrt{9-4c})/2} + c_2 e^{N(-3 + \sqrt{9-4c})/2}\,,
\end{align}
where we introduce the a-dimensional field $\varphi_H \equiv \varphi/H_{\rm inf}$ and we define $f_H \equiv f/H_{\rm inf}$; the constants $c_{1,2}$ are fixed by some initial condition $\varphi_H(N_*) = \varphi_*$ and, for simplicity, vanishing initial velocity. We consider the case in which $0 < c < 9/4$.
In this case, the field value in eq.\,(\ref{eq:NaiveSolution}) decreases exponentially fast starting from the value $\varphi_*$.

From the above expression we can easily compute the value $\varphi_H(N_k)$, that is 
the value of the field $\varphi_H$ at the time $N_k$ at which the mode with comoving wavenumber $k$ crosses the Hubble horizon
 $a(N_k)H = k$. We define
\begin{align}\label{eq:HorizonExit}
N_k = N_* + \log\left (\frac{k}{k_*} \right)\,,
\end{align}
where $k_*$ is the comoving wavenumber that crosses the horizon at time $N_*$, that is $a(N_*)H = k_*$. 
To fix ideas, we consider $k_* = 0.05$ Mpc$^{-1}$ as a reference scale with $N_* = \mathcal{O}(60)$.
We find
\begin{align}
\varphi_H(N_k) &= \frac{1}{2\bar{c}}\bigg(\frac{k}{k_*}\bigg)^{-(3+\bar{c})/2}\Bigg\{
\varphi_*\left[
-3+\bar{c}+(3+\bar{c})\bigg(\frac{k}{k_*}\bigg)^{\bar{c}}
\right] 
\nonumber 
\\
&- f_H\left[
-3+\bar{c} - 2\bar{c}\bigg(\frac{k}{k_*}\bigg)^{(3+\bar{c})/2} + (3+\bar{c})\bigg(\frac{k}{k_*}\bigg)^{\bar{c}}
\right]
\Bigg\}\,,
\end{align}
with $\bar{c}\equiv \sqrt{9-4c}$. 
This expression is important because it shows how the factor $1/|\varphi_H(N_k)|^2$ enters in the determination of the amplitude of the angular perturbations once they exit the 
horizon. Consequently, one finds the analytical result  
\begin{align}\label{eq:AngularPert}
k^{3/2}|\delta\vartheta_k| = \frac{1}{\sqrt{2}\varphi_H(N_k)}~~~~~~~~~
\Longrightarrow~~~~~~~~~
P_{\delta\theta}(k) = \frac{k^3|\delta\vartheta_k|^2}{2\pi^2} 
= \frac{1}{4\pi^2|\varphi_H(N_k)|^2}\,.
\end{align} 
The angular power spectrum grows as a power-law 
$P_{\delta\theta}(k) \propto (k/k_*)^{n_{\theta}}$ with spectral index given by 
$n_{\theta} = 3 - \sqrt{9-4c}$. 
If the field $\varphi$ rolls from Planckian values down to  $\varphi_H = \mathcal{O}(f_H)$, one gets many orders-of-magnitude of 
power-law enhancement which is eventually crucial for the formation of PBHs or the 
generation of a sizable GW signal. 
More concretely, the power spectrum  of angular fluctuations ranges in between the two limiting values
\begin{align}\label{eq:PowerSpectrumRange}
 \frac{1}{4\pi^2 \varphi_*^2}~[{\rm for\,} k/k_*=1]~
 \leqslant~P_{\delta\theta}(k)~\leqslant~\frac{1}{4\pi^2 f_H^2}~[{\rm for\,} k/k_* \gg 1]
 ~~~\Longrightarrow~~~
 \Delta P_{\delta\theta} \equiv \frac{\varphi_*^2}{f_H^2}\,,
\end{align}
so that the  enhancement is controlled 
by the ratio $\varphi_*/f_H$.

Another relevant information is the time it takes for the power spectrum to fully grow from its initial value 
at $k/k_* = 1$
up to $1/4\pi^2 f_H^2$. 
Let us define this quantity as $\Delta N$. If we approximate the power spectrum with a power-law (since we are only interested 
in the growing part), we find
\begin{align}
P_{\delta\theta}(k) \approx \frac{1}{4\pi^2\varphi_*^2}\left(
\frac{k}{k_*}
\right)^{3 - \sqrt{9-4c}}
\end{align}
which leads to 
\begin{align}\label{eq:TimeScaling}
\Delta N = \frac{1}{3-\sqrt{9-4c}}\log\left(\frac{\varphi_*^2}{f_H^2}\right) = 
\frac{\log(\Delta P_{\delta\theta})}{3-\sqrt{9-4c}}\,.
\end{align}
The result depends on $c$ since it controls the slope of the growing power spectrum.

\begin{figure}[!htb!]
\begin{center}
$$\includegraphics[width=.49\textwidth]{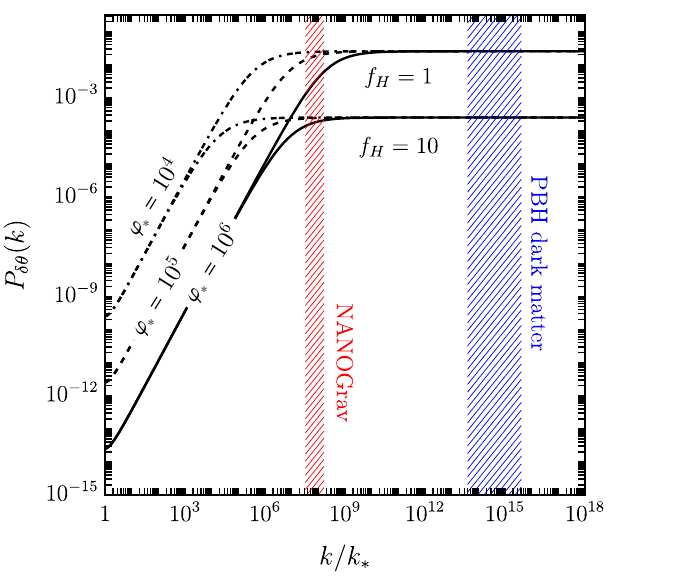}
\quad\includegraphics[width=.49\textwidth]{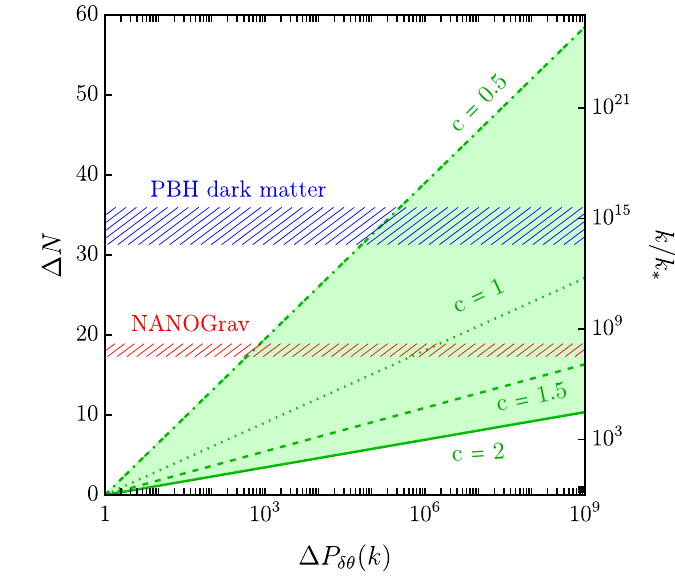}$$
\caption{ \label{fig:PSboosted} 
Left panel: 
angular power spectrum for different choices of $\varphi_*$ and $f_H$ and fixed 
$c=27/16$ 
(leading to a spectral growth $n_\theta =1.5 $).
Right panel:
different spectral growth as a function of $\Delta N$ depending on the assumed initial conditions for the radial field dynamics, i.e. different $c$.
For illustration, in both panels, we show the frequency range (converted in units of $k/k_*$) that is relevant for NANOGrav and other PTA experiments and the mass range (converted in units of $k/k_*$) in which 
the totality of the present-day dark matter abundance 
can be compatible with the PBH hypothesis.
 }
\end{center}
\end{figure}

We show the power spectrum in eq.\,(\ref{eq:AngularPert}) in the left panel of fig.\,\ref{fig:PSboosted} for 
fixed $c$ but different choices of $\varphi_*$ and $f_H$ as function of $k/k_*$ (see caption for details). In the right panel of the same figure, we plot the function $\Delta N$ in eq.\,(\ref{eq:TimeScaling}) 
as function of $\Delta P_{\delta\theta}$ for different values of $c$.
For illustration we superimpose {\it i)} in red, the range of frequencies $2.5\times 10^{-9} \lesssim f\,\,[{\rm Hz}] \lesssim  1.2\times 10^{-8}$
that is relevant for the observation of a potential SGWB signatures by PTA experiment\,\cite{NANOGrav:2020bcs} and {\it ii)} in blue the range of mass in which PBHs may constitute the totality of dark matter observed in the present-day Universe (given existing constraints  \cite{Carr:2020gox}), $10^{18} \lesssim M_{\rm PBH}\,[{\rm g}] \lesssim 10^{21}$.

\begin{figure}[!htb!]
\begin{center}
\includegraphics[width=.85\textwidth]{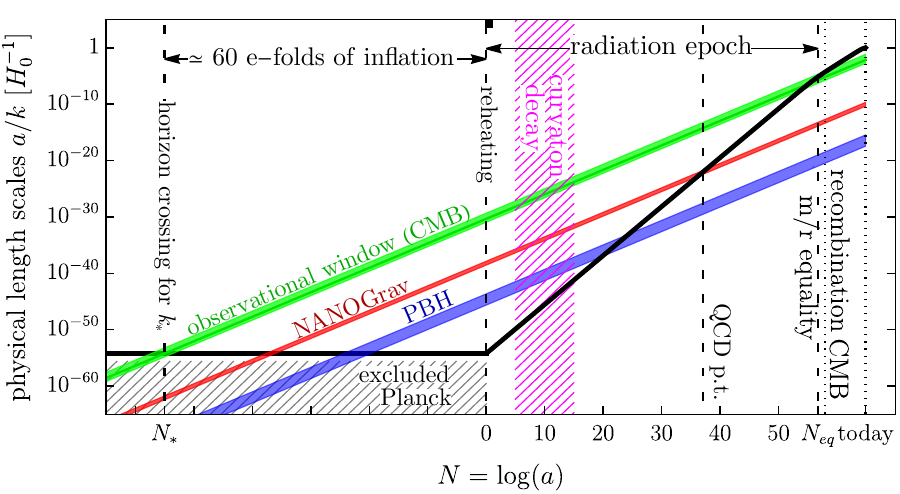}
\caption{ \label{fig:TaleOfScales}The black line shows the evolution of the physical Hubble horizon $1/H$ as a function of the number of $e$-folds $N$, computed assuming slow-roll inflation and standard $\Lambda$CDM model. The green region shows the evolution of the scales associated with CMB observations. The scales of phenomenological interest for PBHs are those associated with a signal of second-order GWs compatible with NANOGrav and PTA data (red region) and those at which PBHs account for the totality of dark matter (blue region). In both cases it is crucial that such scales re-enter the Hubble horizon after all the curvaton field has decayed into radiation otherwise PBHs production cannot occur by means of gravitational collapse of the radiation fluid, as expected in curvaton models. Interestingly enough, the horizon crossing of scales relevant for a detection by the NANOGrav collaboration occurs around the time of QCD phase transition. This means that the phase transition due to the confinement of the strong sector is of phenomenological relevance and has to be taken into account when computing observables of the model.}
\end{center}
\end{figure}

Starting from ad-hoc initial conditions to get the above result, i.e. $\varphi_*$ of planckian order, is not entirely satisfactory and we address the degree of fine tuning required for setting our initial conditions in appendix\,\ref{app:Ini}.
At this stage, we can already make a number of relevant comments.  
\begin{itemize}
\item[{\it i)}]
The power spectrum of angular perturbations $\delta\vartheta_k$ can be enhanced during inflation -- triggered by the 
classical rolling of its radial counterpart -- by several orders of magnitude. The key equation is eq.\,(\ref{eq:TimeScaling}).  
The a-dimensional ratio $\Delta P_{\delta\theta} \equiv \varphi_*^2/f_H^2$ controls the 
enhancement of the power spectrum and the parameter $c$ controls the $e$-fold time 
$\Delta N$ that is needed to complete such a growth, see fig.\,\ref{fig:PSboosted}. 
This is because the larger $c$ the faster the classical rolling of the radial field (see eq.\,(\ref{eq:NaiveSolution})). 
In particular, we remark that, as follows from the right-hand side of the inequality in eq.\,(\ref{eq:PowerSpectrumRange}), the smaller $f_H$ the larger the upper value reachable by   $P_{\delta\theta}(k)$.
\item[{\it ii)}] By requiring the condition $9-4c>0$, we  consequently fix the maximum spectral growth to $3$ and we can see from the left  panel of fig.\,\ref{fig:PSboosted} that $P_{\delta \theta}$ does not exhibit oscillatory behavior when reaching the plateau.
In principle, violating the condition $9-4c>0$ would cause $\varphi$ to reach the minimum of the quadratic potential and oscillate around it, thereby altering the shape of the angular power spectrum as described by eq.\,(\ref{eq:AngularPert}). 
In our case, this does not occur due to the dominant role of the Hubble friction term, which brings $\varphi$ to $f$ asymptotically and damps any oscillations.
As a result, we will not generate any significant oscillatory features on the PBH abundance nor on the SGWB spectrum
\item[{\it iii)}]
In fig.\,\ref{fig:TaleOfScales} we show the evolution of physical length-scales $\lambda_{\rm phys} = a/k$ 
(in units of the present-day Hubble length $1/H_0$) throughout 
the history of the Universe from the inflationary epoch until today. 
We assume instantaneous reheating and, after inflation, standard $\Lambda$CDM cosmology.
Experimental constraint on the tensor-to-scalar ratio can be translated into an upper bound on the energy scale of inflation; in turn, this implies an upper bound on the Hubble parameter during inflation\,\cite{Planck:2018jri}
\begin{align}
\frac{H_{\rm inf}}{\bar{M}_{\rm Pl}} < 2.5\times 10^{-5}\,,~~~~~~{\rm at}\,\,\,95\%\,\,{\rm C.L.}\,.
\end{align}
$\bar{M}_{\rm Pl}$ is the reduced Planck mass ($\bar{M}_{\rm Pl}^2 = 1/8\pi G_N \simeq 2.4\times 10^{18}$ GeV).
For us, inflation takes place in a decoupled sector, and we only need to specify $H_{\rm inf}$ and $N_*$. 
We consider a high-scale inflationary model with $H_{\rm inf}/\bar{M}_{\rm Pl} = 10^{-6}$ and we take $N_* = 55$ as the $e$-fold time at which the pivot scale $k_* = 0.05$ Mpc$^{-1}$ exits the Hubble horizon during inflation.

The comoving scale $k$ re-enters the horizon after inflation at the temperature 
\begin{align}
T_k = \left(\frac{90 \bar{M}_{\rm Pl}^2 H_{\rm inf}^2}{\pi^2 g_*}\right)^{1/4}e^{-N_*}\left(
\frac{k}{k_*}
\right)\,,
\end{align} 
with $g_*$ the relativistic degrees of freedom.
It is also interesting to mention here that, if we convert the comoving wavenumber into frequency by means of $k=2\pi f$, one finds
\begin{align}
T_{f}\,
\simeq 
61 \, {\rm MeV}\, 
\left (\frac{g_*}{10.75} \right )^{1/6}\left(
\frac{f}{10^{-9} {\rm Hz}}
\right).
\end{align} 
The above estimate implies that comoving wavenumbers corresponding to frequencies of order $\mathcal{O}(10^{-9})$ Hz re-enter the Hubble horizon when the temperature is of the order of the QCD phase transition (see fig.\,\ref{fig:TaleOfScales}).
Implications of this idea are discussed in sec \ref{Sec:SGWB}.
 \end{itemize}
 
\subsection{The dynamics after inflation}
\label{sec:AfterInfla}
The angular field $\vartheta$ constitutes a sub-dominant component of the energy-density budget during inflation. 
In order to make the perturbations $\delta\vartheta_k$ phenomenologically relevant, we need to transfer them into radiation after the end of inflation. 
For this to happen, we introduce -- following the spirit of conventional curvaton models -- a coupling between the field $\vartheta$ and photons.
After the end of inflation, the inflaton energy density is converted into radiation. 
We assume instantaneous reheating.  The reheating temperature is then given by
\begin{align}\label{eq:Reheat}
T_{\rm RH} = \left(\frac{90}{\pi^2 g_*}\right)^{1/4}\bar{M}_{\rm Pl}^{1/2}H_{\rm inf}^{1/2}
\simeq 10^{15}
{\rm GeV}
\left(\frac{106.75}{g_*}\right)^{1/4}\left(\frac{H_{\rm inf}}{10^{-6}\,\bar{M}_{\rm Pl}}\right)^{1/2}
\,.
\end{align}
In eq.\,(\ref{eq:Reheat}) we take $g_*=  106.75$ (but we remind that $g_*$ is a function of $T$).
We assume that the global $U(1)$ symmetry is broken explicitly by some non-perturbative effect and  the curvaton has the following potential
\begin{align}\label{eq:CurvatonMass}
\mathcal{V}(\phi) = \Lambda^4\left(1 - \cos\frac{\phi}{f}\right) \simeq \frac{1}{2}m_{\phi}^2\phi^2\,,
\end{align}
where the curvaton mass is $m_{\phi} = \Lambda^2/f$. 
We now have $\vartheta = \phi/f$. We simplify the analysis and consider in eq.\,(\ref{eq:CurvatonMass})
the quadratic approximation\footnote{If we go beyond the quadratic potential, including for instance anharmonic corrections caused by curvaton self interactions, the equation of motion of the curvaton becomes non-linear and NG acquires a scale-dependence \,\cite{Enqvist_2010}.}  in which we only have a mass term for $\phi$. 

The curvaton field remains approximately constant until  the Hubble parameter falls below the curvaton mass when the  curvaton starts oscillating around the minimum of its potential. 
Notice that the potential in eq.\,(\ref{eq:CurvatonMass}) is generated only after the temperature $T$ of the radiation bath drops below some value of the order of the confinement scale $\Lambda$. For instance, in the case of the QCD axion we have
\begin{align}
m_{\phi}^2(T) =
\left\{
\begin{array}{ccc}
 c_0(\Lambda_{\rm QCD}^4/f^2) & {\rm for} & T \lesssim T_0    \\
 c_T(\Lambda_{\rm QCD}^4/f^2)\left(\Lambda_{\rm QCD}/T\right)^{n}  & {\rm for} & T \gtrsim T_0    
\end{array}
\right. ,
\end{align}
with the  parameters $c_0$, $c_T$ and $n$ that can  be determined  using  the dilute 
instanton gas approximation valid at high temperatures. 
Typical values are $c_0 \simeq 10^{-3}$, $c_T \simeq 10^{-7}$ and $n\simeq 7$ with $T_0 \simeq 100$ MeV and $\Lambda_{\rm QCD}\simeq 400$ MeV. 
In principle, we can postulate a situation that mimics the one of QCD, and consider  the temperature-dependent  curvaton mass
\begin{align}\label{eq:ThermalALPmass}
m_{\phi}^2(T) =
\left\{
\begin{array}{ccc}
 m_{\phi}^2 & {\rm for} & T \lesssim \Lambda    \\
 m_{\phi}^2\left(\Lambda/T\right)^{n}  & {\rm for} & T \gtrsim \Lambda    
\end{array}
\right. ,
\end{align}
with zero-temperature value $m_{\phi} = \Lambda^2/f$. 
We remark that  the  temperature  that controls the curvaton mass does not need to equal the temperature of the SM bath. 
For instance, this is the case if the curvaton mass is generated by couplings to a hidden sector that is not in kinetic equilibrium with the SM.  
For simplicity, we take $T$ in eq.\,(\ref{eq:ThermalALPmass}) to be the temperature of the SM bath (see ref.\,\cite{Blinov:2019rhb} for a critical discussion). 
For definiteness, we set $n=8$ in eq.\,(\ref{eq:ThermalALPmass}).  

During curvaton oscillations, $\rho_{\phi} \propto a^{-3}$ and $\rho_{\gamma} \propto a^{-4}$.
Therefore,  the  curvaton  component  of the energy density $\rho_{\phi}$ increases with respect to the radiation component $\rho_{\gamma}$.
This stage of the dynamics lasts until the Hubble rate $H$ becomes of the order of the decay width $\Gamma_{\phi}$ of the curvaton. After this time, the decay of the curvaton into radiation becomes quickly efficient. 
Schematically, the dynamics after the end of inflation is summarized in the following sketch (with $N = \int Hdt$)
{\small
\begin{align}\label{eq:Sketch}
	\begin{tikzpicture}
	 {\scalebox{1}{
    \draw[->][thick] (-7,0)--(7.,0);
    \draw[thick][thick] (-7,0.1)--(-7,-0.1);
    \node at (-7,-0.4) {\scalebox{1}{$N = 0$}};
    \node at (-7,+0.4) {\scalebox{0.8}{reheating}};
    \node at (-5,+0.2) {\scalebox{0.8}{curvaton constant}};
    \node at (-5,-0.3) {\scalebox{0.8}{{\color{indiagreen}{``first'' radiation stage}}}};
    \node at (-5,+0.75) {\scalebox{0.9}{{\color{cornellred}{Phase I}}}};
    \draw[thick][thick] (-3,0.1)--(-3,-0.1);
    \node at (-3,+0.4) {\scalebox{1}{$H \sim m_{\phi}$}};
    \node at (-3,-0.4) {\scalebox{1}{$N_{\rm osc}$}};
    \node at (-0.5,+0.2) {\scalebox{0.8}{curvaton oscillates}};
    \node at (-0.5,+0.75) {\scalebox{0.9}{{\color{cornellred}{Phase II}}}};
    \node at (-0.5,-0.25) {\scalebox{0.8}{$P_{\phi}\simeq 0$, $\rho_{\phi}\sim a^{-3}$ and $\rho_{\gamma}\sim a^{-4}$}};
    \draw[thick][thick] (2.,0.1)--(2.,-0.1);
    \node at (2.,+0.4) {\scalebox{1}{$H \sim \Gamma_{\phi}$}};
    \node at (2.,-0.4) {\scalebox{1}{$N_{\rm dec}$}};
    \node at (4.5,+0.75) {\scalebox{0.9}{{\color{cornellred}{Phase III}}}};    
    \node at (4.5,+0.2) {\scalebox{0.8}{curvaton decays to radiation}};
    \node at (4.5,-0.3) {\scalebox{0.8}{{\color{indiagreen}{``second'' radiation stage}}}};
    \node at (-5,1.2) {\scalebox{0.8}{{\color{VioletRed4}{\textbf{Schematic of background evolution}}}}};
    }}
	\end{tikzpicture}
\end{align}
}
We assume that $\phi$ decays as consequence of the following dimension-5 effective operator 
\begin{align}
\mathcal{L}_{\phi F\tilde{F}} = \frac{g_{a\gamma\gamma}}{4f}\,\phi F_{\mu\nu}\tilde{F}^{\mu\nu}\,,
\end{align}
which gives the decay rate $\Gamma_{\phi}$ 
\begin{align}\label{eq:Gamma}
\Gamma_{\phi} &= \frac{g_{a\gamma\gamma}^2 m_{\phi}^3}{64\pi f^2} \simeq 
g_{a\gamma\gamma}\left(\frac{\Lambda}{10^4\,{\rm GeV}}\right)^6
\left(\frac{10^{11}\,{\rm GeV}}{f}\right)^5\left(\frac{10^{8}}{\tau_{\rm U}}\right) 
\,,
\end{align}
that we have written in erms of the present age of the Universe $\tau_{\rm U} \simeq 13.8\times 10^{9}$ yr, while the mass $m_{\phi}$ is
\begin{align}
m_{\phi} & \simeq 10^6 \,{\rm eV}
\left(\frac{\Lambda}{10^4\,{\rm GeV}}\right)^2
\left(\frac{10^{11}\,{\rm GeV}}{f}\right)\,.
\end{align}
The lifetime $\tau_{\phi} = 1/\Gamma_{\phi}$ can be either much longer or much shorter than $\tau_{\rm U}$.
Typically, the cosmologically long-lived option $\tau_{\phi} \gg \tau_{\rm U}$ is preferred  
since in this case $\phi$ can naturally serve as dark matter whose abundance 
is generated by means of the misalignment mechanism (see ref.\,\cite{Marsh:2015xka,DiLuzio:2020wdo} for a review), but in the opposite regime, as we are considering in this analysis, $\phi$ is not cosmologically stable and decays into radiation.
 
In order to describe the dynamics of the model after the end of inflation, we study Einstein and fluid equations.
Consider first the unperturbed background. 
The Friedmann equation and the continuity equation are 
\begin{align}\label{eq:Fluids}
{\rm Friedmann\,equation:}&~~~~3\bar{M}_{\rm Pl}^2 H^2 = \rho_{\gamma} + \rho_{\phi}\,,
\nn
\\
{\rm continuity\,equation\,for\,each\,fluid:}&~~~~\dot{\rho}_{\alpha}  = -3H(\rho_{\alpha} + P_{\alpha}) + Q_{\alpha}\,,
\end{align}
where in this case we have two background fluids, that are the radiation fluid 
with energy density $\rho_{\gamma} = \rho_{\gamma}(t)$ and pressure $P_{\gamma} = P_{\gamma}(t) = \rho_{\gamma}/3$ and the homogeneous field $\phi = \phi(t)$ whose energy density and pressure are 
given by\footnote{The energy-momentum tensor in the FLRW geometry is given by 
$T_{\alpha}^{\mu\nu} = {\rm diag}(\rho_{\alpha},P_{\alpha}/a^2,P_{\alpha}/a^2,P_{\alpha}/a^2)$ 
and for the scalar field $\phi$ energy density and pressure can be identified if one computes the components 
of $T_{\phi}^{\mu\nu} = (\partial^{\mu}\phi)(\partial^{\nu}\phi) - g^{\mu\nu}[(\partial_{\rho}\phi)(\partial^{\rho}\phi)/2 + \mathcal{V}(\phi)]$. 
The continuity equation $\dot{\rho}_{\alpha} = -3H(\rho_{\alpha} + P_{\alpha})$ (in the absence of energy transfer, $\Gamma_{\phi} = 0$) 
simply follows from the temporal component of $T_{\alpha~~;\mu}^{\mu\nu} = 0$. 
In the presence of interactions between the fluids, notice that in general we have $T_{\alpha~~;\mu}^{\mu\nu} \neq 0$ for each individual fluid but, as a consequence of the Bianchi identity, we must have the conservation equation 
$\sum_{\alpha}T_{\alpha~~;\mu}^{\mu\nu} = 0$ for their sum. 
This condition forces the relation $Q_{\gamma} = - Q_{\phi}$. 
We refer to appendix\,\ref{app:Pert} for a more detailed discussion.
}
\begin{align}\label{eq:EnergyAndPressure}
\rho_{\phi} = \frac{1}{2}\dot{\phi}^2 + \mathcal{V}(\phi)\,,~~~~~P_{\phi} = \frac{1}{2}\dot{\phi}^2 - \mathcal{V}(\phi)\,.
\end{align}
In eq.\,(\ref{eq:Fluids}) $Q_{\alpha}$ represents the energy transfer per unit time to the $\alpha$-fluid. 
This transfer of energy is due to the decay $\phi\to \gamma\gamma$ previously introduced, and 
we have $Q_{\gamma} = - Q_{\phi} = \Gamma_{\phi}\rho_{\phi}$. 
All in all, the three relevant equations are 
\begin{align}
3\bar{M}_{\rm Pl}^2 H^2 & = \rho_{\gamma} + \rho_{\phi}\equiv \rho\,,\label{eq:DynBG1}\\
\dot{\rho}_{\gamma} + 4H\rho_{\gamma} & = +\Gamma_{\phi}\rho_{\phi}\,,\label{eq:DynBG2}\\
\dot{\rho}_{\phi} + 3H(\rho_{\phi} + P_{\phi}) & = - \Gamma_{\phi}\rho_{\phi}\,.\label{eq:DynBG3}
\end{align}
During phase II (see eq.\,(\ref{eq:Sketch})) the curvaton oscillates and one can take the 
approximation $P_{\phi} \simeq 0$. 
This is because during this phase the curvaton field oscillates with a typical timescale set by its mass, $t_{\phi} = 1/m_{\phi}$, which is smaller than the inverse Hubble rate, $t_{\phi} < 1/H$.
One can, therefore, average over an oscillation and estimate $\dot{\phi} \simeq m_{\phi}\phi$; 
from eq.\,(\ref{eq:EnergyAndPressure}) it follows that $P_{\phi}\simeq 0$. 
This is the approximation used in ref.\,\cite{Firouzjahi:2012iz}. 
This is a good description of the classical field dynamics when the Hubble rate drops below the mass of the curvaton.\footnote{In the limit $P_{\phi} \to 0$ the continuity equation 
$\dot{\rho}_{\phi} + 3H(\rho_{\phi} + P_{\phi}) = - \Gamma_{\phi}\rho_{\phi}$ can be written in the form $\ddot{\phi} + (3H + \Gamma_{\phi})\dot{\phi} + \mathcal{V}^{\prime}(\phi) = 0$ and the factor $\Gamma_{\phi}$ enters in the Klein-Gordon equation in the form of a damping term.}

\begin{figure}[!h!]
\begin{center}
$$
\includegraphics[width=.33\textwidth]{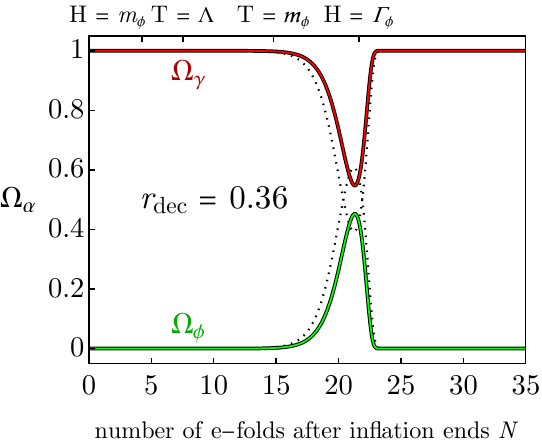}
\includegraphics[width=.33\textwidth]{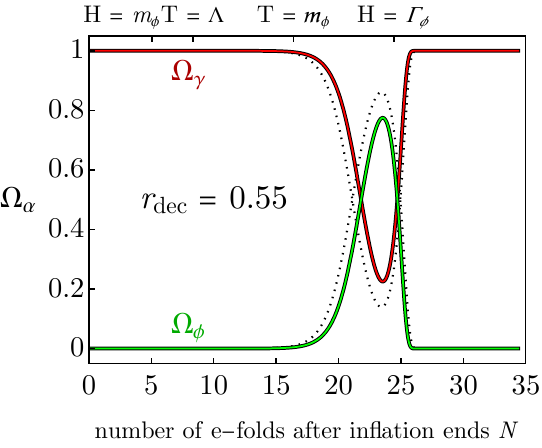}
\includegraphics[width=.33\textwidth]{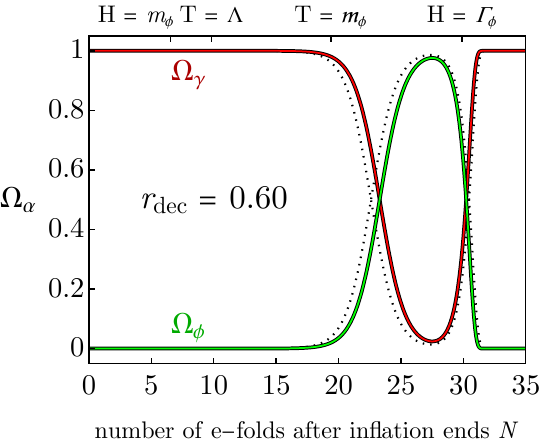}$$
\caption{ \label{fig:EvoAfterInfla} 
Here we set $f = 10^{15}$ GeV, $H_{\rm inf}/\bar{M}_{\rm Pl} = 10^{-6}$, $\vartheta_0 = 0.01$. 
In the left panel we have $m_{\phi} = 5\times10^8$ GeV, in the central panel $m_{\phi} = 10^8$ GeV and 
in the right panel $m_{\phi} = 5\times 10^7$ GeV.
The $e$-fold time difference $\Delta N$ between $T=m_{\phi}$ and $H=\Gamma_{\phi}$ increases from left to right.
This implies that the curvaton energy density has progressively more time to increase  with respect to radiation (since during this time interval it redshifts as non-relativistic matter) before decaying into the latter. 
In each panel we quote the corresponding value of the parameter $r_{\rm dec}$ defined in eq.\,(\ref{eq:rDef}). 
The numerical value of $r_{\rm dec}$ is computed at the time $N_{\rm dec}$ at which $H = \Gamma_{\phi}$, as indicated on the labels of the top $x$-axis.
Dashed lines represent the model in the simplified setup (further details are provided in the main text). } 
\end{center}
\end{figure}

We solve the system in eqs.\,(\ref{eq:DynBG1}-\ref{eq:DynBG3}) in two steps. 
First, we consider the evolution during phase I. 
We set the initial conditions $\rho_{\gamma}(N=0) = 3\bar{M}_{\rm Pl}^{2}H_{\rm inf}^{2}$ 
and $\vartheta(N=0) = \vartheta_0$ (with vanishing initial velocity). We rewrite eq.\,(\ref{eq:DynBG3}) in terms of an evolution equation for $\vartheta(N)$ with energy density and pressure given by eq.\,(\ref{eq:EnergyAndPressure}), and we use the number of $e$-folds as time variable. 
To be more precise, the system we solve is
\begin{align}
 \frac{dH}{dN} + \frac{3H}{2} + \frac{H\rho_{\gamma}}{2\rho} + 
\frac{3HP_{\phi}}{2\rho} & = 0\,, \label{eq:DynBGSim1}
\\
\frac{d\rho_{\gamma}}{dN} + 4\rho_{\gamma} - \frac{f^2\Gamma_{\phi}H}{2}\bigg[
\left(\frac{d\vartheta}{dN}\right)^2 + 
\frac{m_{\phi}(T)^2}{H^2}\vartheta^2
\bigg] &= 0\,,
\\
\frac{d\vartheta}{dN}\frac{d^2\vartheta}{dN^2} 
+ \bigg(\frac{d\vartheta}{dN}\bigg)^2\bigg[3+\frac{d}{dN}\log H\bigg] 
+
\frac{m_{\phi}(T)^2}{H^2}\vartheta\bigg[\frac{d\vartheta}{dN} 
-\frac{n\vartheta}{2}\frac{d}{dN}\log T\bigg]&
\nn \\
+
\frac{\Gamma_{\phi}}{2H}\left[
\left(\frac{d\vartheta}{dN}\right)^2 + 
\frac{m_{\phi}(T)^2}{H^2}\vartheta^2
\right] & = 0\,. \label{eq:DynBGSim3}
\end{align}
Notice that, as in the case of the conventional misalignment mechanism when the global $U(1)$ symmetry gets spontaneously broken during inflation, $\vartheta_0$ is a free parameter.
We follow the evolution of the system until the field $\vartheta$ and its pressure $P_{\phi}$ start oscillating on timescales shorter than the inverse Hubble rate. At this point, as discussed before, the pressure can be safely neglected. Numerically, this first part of the dynamics lasts until $T = \Lambda$ at $e$-fold time $N_{\Lambda}$.\footnote{Numerically, we checked that the exact value of $N$ at which we switch to phase II does not impact our results as long as $N \gtrsim N_{\Lambda}$.}
The subsequent evolution can be described by the same system in eqs.\,(\ref{eq:DynBG1}-\ref{eq:DynBG3}) in which we now set $P_{\phi} = 0$.  
In this limit the system simplifies to 
\begin{align}
\frac{dH}{dN} + \frac{3H}{2} + \frac{H\Omega_{\gamma}}{2} & = 0\,,\label{eq:DynSim1}\\
\frac{d\Omega_{\gamma}}{dN} + \Omega_{\phi}\Omega_{\gamma} - \frac{\Gamma_{\phi}\Omega_{\phi}}{H} & = 0\,,\\
\frac{d\Omega_{\phi}}{dN} - \Omega_{\phi}\Omega_{\gamma} + \frac{\Gamma_{\phi}\Omega_{\phi}}{H} & = 0\,,\label{eq:DynSim3}
\end{align}
where we introduce the time-dependent quantities $\Omega_{\gamma} \equiv \rho_{\gamma}/\rho$ and 
$\Omega_{\phi} \equiv \rho_{\phi}/\rho$ (with $\Omega_{\gamma}+\Omega_{\phi} = 1$ by definition). 
We solve the system in eqs.\,(\ref{eq:DynSim1}-\ref{eq:DynSim3}) with initial conditions given by the solutions of eqs.\,(\ref{eq:DynBGSim1}-\ref{eq:DynBGSim3}) at time $N = N_{\Lambda}$.
We show in fig.\,\ref{fig:EvoAfterInfla} the background dynamics in terms of $\Omega_{\gamma}$ and $\Omega_{\phi}$ after the end of inflation for three benchmark sets of parameters. 

It is also possible to describe the dynamics in a simplified way. 
During phase I, we can set $\Gamma_{\phi} = 0$ and $\rho_{\phi} = 0$ (massless field frozen at some initial constant value). Furthermore, we neglect the time-dependence of $g_*$ and set $g_* = 106.75$. This is correct as long as this part of the dynamics takes palace before the electroweak phase transition.  
Under these assumptions, we find the following analytical solutions for the temperature and the Hubble rate
\begin{align}
T(N) = \left(\frac{90 \bar{M}_{\rm Pl}^2 H_{\rm inf}^2}{\pi^2 g_*}\right)^{1/4}e^{-N}\,,~~~~~~~~H(N) = H_{\rm inf}e^{-2N}\,.
\end{align}
In this simplified setup, we can compute the $e$-fold time $N_{\Lambda}$ defined by $T(N_{\Lambda}) \equiv \Lambda$ and 
the $e$-fold time $N_{\phi}$ defined by $H(N_{\phi}) \equiv m_{\phi}$. We find
\begin{align}
N_{\Lambda} = 
\frac{1}{4}\log\left(\frac{90\bar{M}_{\rm Pl}^2 H_{\rm inf}^2}
{\pi^2g_*\Lambda^4}\right)\,,~~~~~~~
N_{\phi} = \frac{1}{2}\log\left(\frac{H_{\rm inf}}{m_{\phi}}\right)\,.
\end{align}
Notice that if we consider the case $f \ll \bar{M}_{\rm Pl}$ then the previous relations (we remind that $m_{\phi} = \Lambda^2/f$) imply that $N_{\phi} < N_{\Lambda}$ (see fig.\,\ref{fig:EvoAfterInfla}). 
This means that at time $N_{\phi}$ when $H(N_{\phi}) = m_{\phi}$ the temperature is of order $T(N_{\phi}) \simeq (\bar{M}_{\rm Pl}/f)^{1/2}\Lambda \gg \Lambda$ and the field $\vartheta$ does not fill yet the effect of the non-zero mass. This means that identifying the time $N_{\rm osc}$ with $N_{\phi}$ would not be entirely correct. 
On the other hand, at time $N_{\Lambda}$ the temperature is $T(N_{\Lambda}) = \Lambda$ and the mass $m_{\phi}(T)$ fully formed (see eq.\,(\ref{eq:ThermalALPmass})). 
It is, therefore, more realistic to identify $N_{\rm osc}$ with $N_{\Lambda}$.
At this time, we estimate the initial condition for the energy density $\Omega_{\phi}$ to be  $\Omega_{\phi}(N_{\rm osc}) = \Lambda^4\vartheta_0^2/3\bar{M}_{\rm Pl}^2H(N_{\rm osc})^2$ while  for the radiation bath we have $\Omega_{\gamma}(N_{\rm osc}) = 1 - \Omega_{\phi}(N_{\rm osc})$. Given these initial conditions, we solve the system in eqs.\,(\ref{eq:DynSim1}-\ref{eq:DynSim3}) for $N > N_{\rm osc}$. 
The result is also shown in fig.\,\ref{fig:EvoAfterInfla} (dotted black lines).

Notice that it is important to have good control over the initial value of $\Omega_{\phi}$ that is used to solve the system in eqs.\,(\ref{eq:DynSim1}-\ref{eq:DynSim3}) since this value gets exponentially modified during the dynamical evolution in phase II. 
In the so-called sudden decay limit (that is $\Gamma_{\phi} = 0$) we indeed have the analytical solutions\,\cite{Firouzjahi:2012iz}
\begin{align}\label{eq:SuddenDecay}
\Omega_{\gamma}(N) = \frac{1}{1+pe^{N}}\,,~~~~~~~\Omega_{\phi}(N) = \frac{pe^{N}}{1+pe^{N}}\,,
\end{align}
with $p\equiv \Omega_{\phi}/\Omega_{\gamma}$ at $N = N_{\rm osc}$. 

\begin{figure}[h]
\begin{center}
$$\includegraphics[width=.445\textwidth]{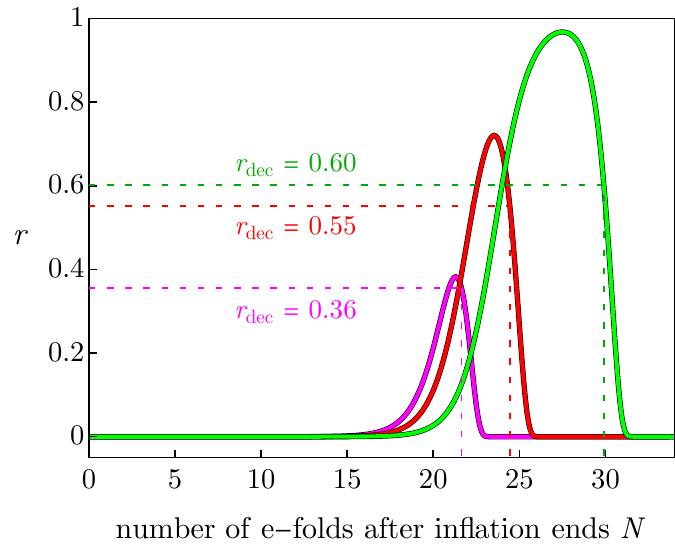}
\qquad\includegraphics[width=.45\textwidth]{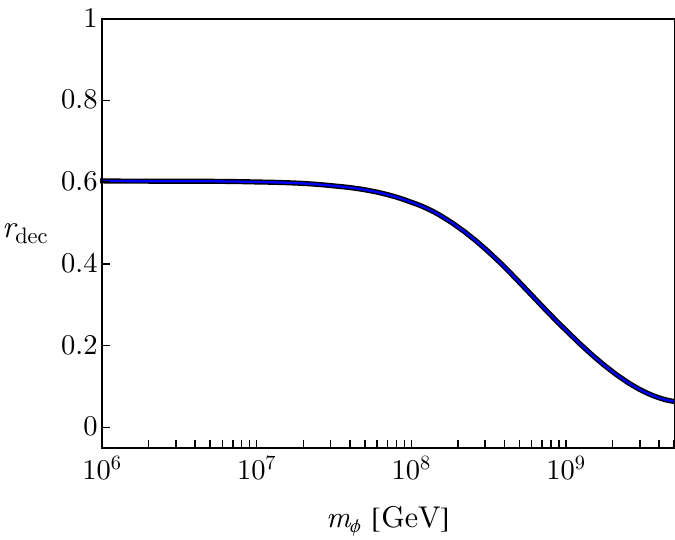}$$
\caption{
We fix $f = 10^{15}$ GeV, $H_{\rm inf}/\bar{M}_{\rm Pl} = 10^{-6}$, $\vartheta_0 = 0.01$.  
In the left panel, we show the evolution of $r$ for the three values of $m_{\phi}$ shown in fig.\,\ref{fig:EvoAfterInfla}; we also identify the value of $r_{\rm dec}$, defined as in eq.\,(\ref{eq:rDef}). 
In the right panel, we show the value of $r_{\rm dec}$ as function of $m_{\phi}$.
 }\label{fig:rDec}  
\end{center}
\end{figure}

If we define the time-dependent quantity $r(N) \equiv 3\rho_{\phi}/(3\rho_{\phi} + 4\rho_{\gamma})$, a key parameter in the analysis of the following sections will be the weighted fraction of the curvaton energy density to the total energy density at the time of curvaton decay, defined by\footnote{Notice that in refs.\,\cite{Ando:2017veq,Inomata:2020xad} this parameter is differently defined as the ratio between the energy densities of the curvaton and radiation. More concretely, refs.\,\cite{Ando:2017veq,Inomata:2020xad} define $r$ to be $\kappa \equiv \rho_{\phi}/\rho_{\gamma}$, and the relation with $r$ is given by $r = 3\kappa/(4+3\kappa)$. 
We, therefore, adopt the more conventional definition of $r \equiv 3\rho_{\phi}/(3\rho_{\phi} + 4\rho_{\gamma})$ used in the curvaton literature.}
\begin{align}\label{eq:rDef}
r_{\rm dec} \equiv r(N_{\rm dec}) = 
\left.
\frac{3\rho_{\phi}}{
3\rho_{\phi} + 4\rho_{\gamma}
}\right|_{N = N_{\rm dec}}
= \left.\frac{3\Omega_{\phi}}{4 - \Omega_{\phi}}\right|_{N = N_{\rm dec}}\,,~~~~~
{\rm with}\,\,\,H = \Gamma_{\phi}\,\,\,{\rm at\,\,}N=N_{\rm dec}\,.
\end{align}
We show the evolution of the parameter $r$ in fig.\,\ref{fig:rDec}. 
More in detail, the left panel of fig.\,\ref{fig:rDec} shows the evolution of $r$ corresponding to the three benchmark models discussed in fig.\,\ref{fig:EvoAfterInfla}; the right-hand panel, on the contrary, shows the value of $r_{\rm dec}$ as function of $m_{\phi}$ (with the parameters $H_{\rm inf}$, $f$ and $\vartheta_0$ fixed to the values quoted in the caption and in fig.\,\ref{fig:EvoAfterInfla}).
The time evolution of $r$ clearly retraces the time evolution of $\Omega_{\phi}$, and features a maximum during the oscillating phase of the curvaton.
Notice that the numerical value of $r_{\rm dec}$ depends on its specific definition. 
If we define $r_{\rm dec}$ strictly as in eq.\,(\ref{eq:rDef}), then we find $r_{\rm dec} \lesssim 0.6$ even for cases in which the energy density of the curvaton happens to dominate the total energy density of the Universe before decaying (see the plateau in the right panel of fig.\,\ref{fig:rDec}). This is due to the fact that the decay of the curvaton is not an instantaneous process that happens at $H = \Gamma_{\phi}$ but, as evident from fig.\,\ref{fig:EvoAfterInfla} and fig.\,\ref{fig:rDec}, when $H = \Gamma_{\phi}$ part of the curvaton energy density was already converted back into radiation.

The importance of the parameter $r_{\rm dec}$ is twofold. On the one hand, at the linear level, 
it controls the fraction of perturbations that are transferred to radiation;
we shall discuss in more detail this effect in section\,\ref{eq:Per}.  
On the other hand, it also controls the impact of non-gaussian corrections, as we will discuss later.

At this stage of the analysis, we can pause to try a first exploration of the parameter space.
Cosmologically, the relevant parameters are $N_{\rm dec}$ and $r_{\rm dec}$. 
As far as the fundamental parameters are concerned, we focus our attention primarily on $m_{\phi}$ and $f$.
\begin{figure}[!h!]
\begin{center}
$$\includegraphics[width=.46\textwidth]{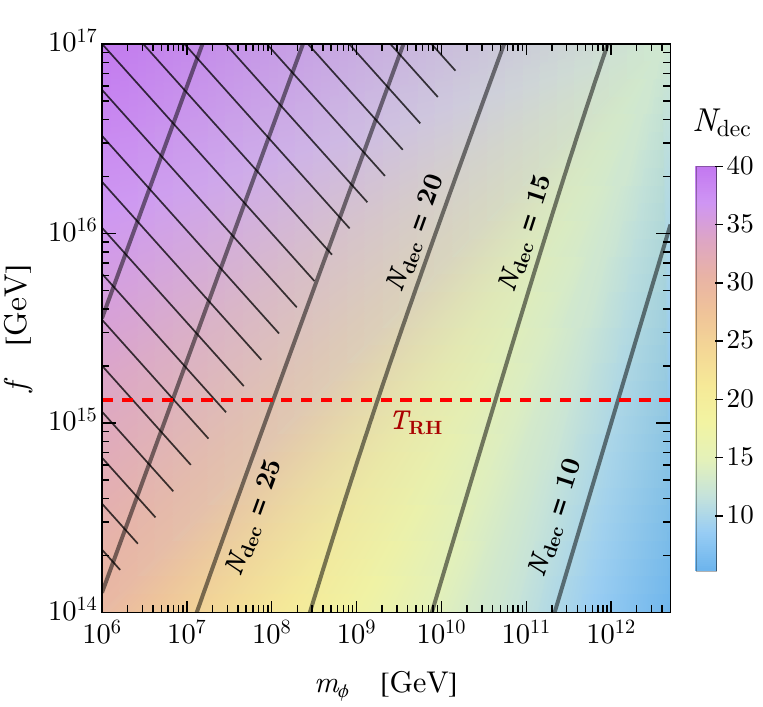}
\qquad\includegraphics[width=.46\textwidth]{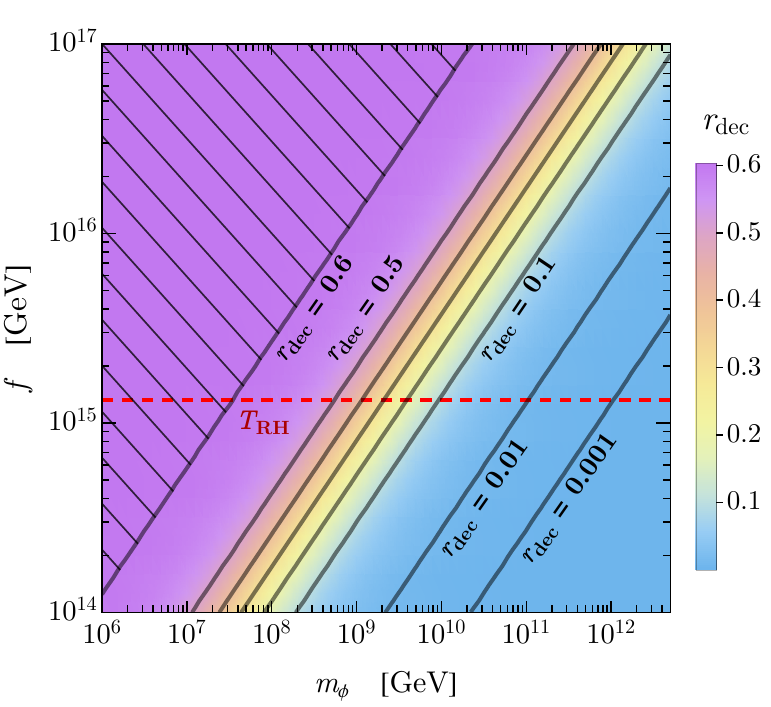}$$
\caption{ 
$\vartheta_0 = 0.01$.  
Scan over the parameter space of the model in light of the background analysis discussed in section\,\ref{sec:AfterInfla}.  
The horizontal dashed red line indicates the assumed re-heating temperature scale \eqref{eq:Reheat}. The black hatched region identifies the parameter space for which $\tau_{\phi} \geq \tau_{\rm U}$.
 }\label{fig:MiniScan}  
\end{center}
\end{figure}
In fig.\,\ref{fig:MiniScan} we fix $H_{\rm inf}/\bar{M}_{\rm Pl} = 10^{-6}$ and, for the moment, we keep $\vartheta_0 = 0.01$.  

\begin{figure}[!h!]
\begin{center}
$$\includegraphics[width=.45\textwidth]{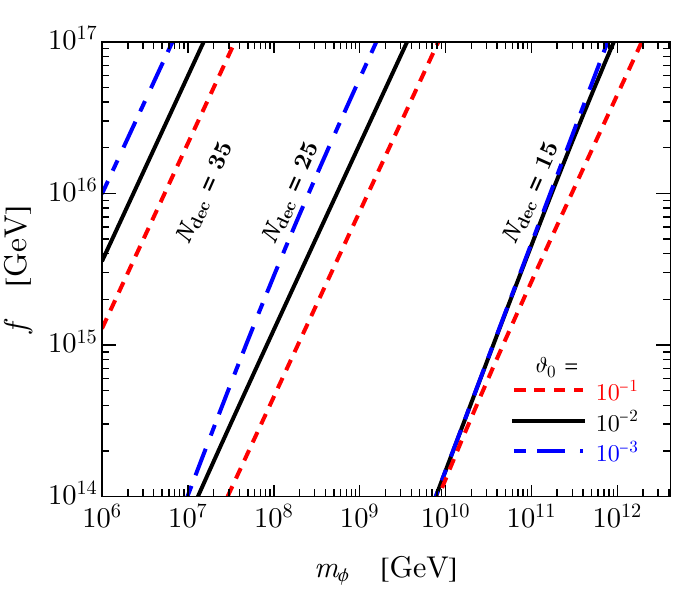}
\qquad\includegraphics[width=.45\textwidth]{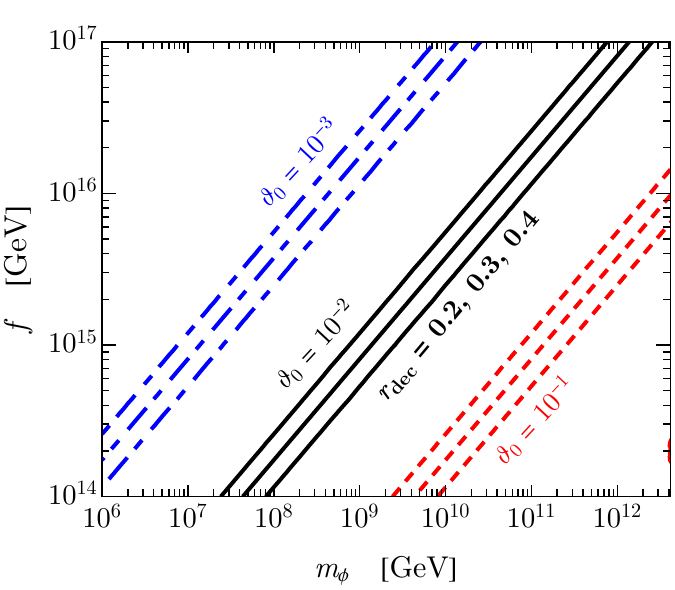}$$
\caption{
Same as in fig.\,\ref{fig:MiniScan} but for different values of $\vartheta_0$ 
($\vartheta_0 = 10^{-1}$, dashed red lines; $\vartheta_0 = 10^{-3}$, dot-dashed blue lines; solid black lines 
refer to $\vartheta_0 = 10^{-2}$). As far as $N_{\rm dec}$ is concerned, on the left panel we only show contours 
corresponding to $N_{\rm dec} = 15,25,35$; similarly, on the right panel, we only show contours 
corresponding to $r_{\rm dec} = 0.2,0.3,0.4$ (from the right- to the left-most lines in each of the three groups).
 }\label{fig:MiniScanTheta}  
\end{center}
\end{figure}

We comment about the $\vartheta_0$-dependence in fig.\,\ref{fig:MiniScanTheta}. 
Intuitively, we do not expect a strong $\vartheta_0$-dependence for what concerns $N_{\rm dec}$ since 
the latter is determined by the equation $H = \Gamma_{\phi}$. In this equation, the decay width $\Gamma_{\phi}$ 
does not depend on $\vartheta_0$; the value of the Hubble parameter, on the contrary, does 
depend on $\vartheta_0$ since the size of $\vartheta_0$ controls the (initial value of the) curvaton energy density 
which enters in the Friedmann equation, eq.\,(\ref{eq:DynBG1}).  
However, as long as the curvaton does not dominate the energy budget of the Universe for long time, this change 
does not alter much the evolution of $H$ and, consequently, the time at which $H = \Gamma_{\phi}$.  
This is confirmed by the numerical analysis shown in the left panel of fig.\,\ref{fig:MiniScanTheta} (see caption for details). 
Consider now the value of $r_{\rm dec}$. Contrary to the previous argument, we do expect a sizable dependence on $\vartheta_0$.
The reason follows from what we already noticed above eq.\,(\ref{eq:SuddenDecay}):
the value of the curvaton energy density at the beginning of the oscillating phase (controlled by $\vartheta_0$) 
gets exponentially modified by the dynamics during Phase II.

\subsection{Perturbations at linear order: curvature power spectrum }\label{eq:Per}

At the linear order, the gauge invariant curvature perturbation on spatial slices of uniform energy
density reads\,\cite{Kodama:1984ziu}
\begin{equation}
\zeta \equiv -\psi - H\frac{\delta\rho}{\dot{\rho}}\,,
\end{equation}
with $\rho = \rho_{\phi} + \rho_{\gamma}$, 
$\dot{\rho} = \dot{\rho}_{\phi} + \dot{\rho}_{\gamma}$
 and $\delta\rho = \delta\rho_{\phi} + \delta\rho_{\gamma}$; 
 $\psi$ is the gauge-dependent curvature perturbation that enters in the 
 linear perturbations about a spatially-flat Friedmann-Robertson-Walker background, as defined in eq.\,(\ref{eq:MetricPerturbation}).
 We refer to appendix\,\ref{app:Pert} for a more detailed discussion.

It is customary to introduce individual curvature perturbations $\zeta_{\alpha}$ each of which is associated with a single energy density component, and 
similarly defined by 
$\zeta_{\alpha} \equiv -\psi - H(\delta\rho_{\alpha}/\dot{\rho}_{\alpha})$. 
Consequently, the curvature perturbation $\zeta$ can be equivalently written as  
\begin{equation}\label{eq:TotalZeta1}
\zeta = -\psi + \frac{[\dot{\rho}_{\phi}(\zeta_{\phi} + \psi) + 
\dot{\rho}_{\gamma}(\zeta_{\gamma} + \psi)
]}{\dot{\rho}} = \frac{\dot{\rho}_{\phi}}{\dot{\rho}}\,\zeta_{\phi} + 
\frac{\dot{\rho}_{\gamma}}{\dot{\rho}}\,\zeta_{\gamma}\,.
\end{equation}
This equation is exact.  
We are interested in the evolution of $\zeta$ in Fourier space, and more concretely in its power spectrum.  
In other words, the goal is computing the quantity 
\begin{align}\label{eq:FinaalPS}
P_{\zeta}(k) = \frac{k^3}{2\pi^2}|\zeta_k(N_{\rm f})|^2\,,
\end{align}
where $\zeta_k(N)$ is the time-dependent Fourier mode of $\zeta$ for a given comoving wavenumber $k \equiv |\vec{k}|$ 
and the notation in eq.\,(\ref{eq:FinaalPS}) means that $\zeta_k(N)$ should be evaluated at
some appropriate time $N_{\rm f}$ after the mode $\zeta_k$ settles to its final value, that is conserved until horizon re-entry.  
Importantly, the dynamics of $\zeta$ in controlled by the equation
\begin{align}\label{eq:KeyDynamics}
\frac{d\zeta_k}{dN} = -\frac{\delta P_{{\rm nad},k}}{\rho + P} + \frac{k^2}{3(aH)^2}\big(\Psi_k - \mathcal{R}_k\big)\,,
\end{align}
where $\mathcal{R}$ is the total comoving curvature perturbation, $\Psi$ the curvature perturbation on uniform shear hypersurfaces and $\delta P_{\rm nad}$ the total non-adiabatic pressure perturbation 
(see appendix\,\ref{app:Pert}). This equation implies that $\zeta_k$ is conserved on super-horizon scales (that is when $k\ll aH$ and the last term in eq.\,(\ref{eq:KeyDynamics}) can be safely neglected) and in the absence of non-adiabatic pressure perturbations (that is the first term on the right-hand side of eq.\,(\ref{eq:KeyDynamics})).
Since during phase II+III we have a non-zero $\delta P_{\rm nad}$ (because of the relative entropy perturbation between the curvaton and the radiation fluid, see eq.\,(\ref{eq:Pnad})), the time $N_{\rm f}$ is given by {\it i)} anytime in between $N_{\rm dec} < N_{\rm f} < N_{k}$ if the mode $k$ re-enters the horizon (at time $N_k$ such that $k/aH = 1$) after the curvaton decay takes place or {\it ii)} simply by $N_{\rm f} \simeq N_{k} < N_{\rm dec}$ if the mode $k$ re-enters the horizon before the curvaton decay.
\begin{figure}[!h!]
\begin{center}
$$\includegraphics[width=.45\textwidth]{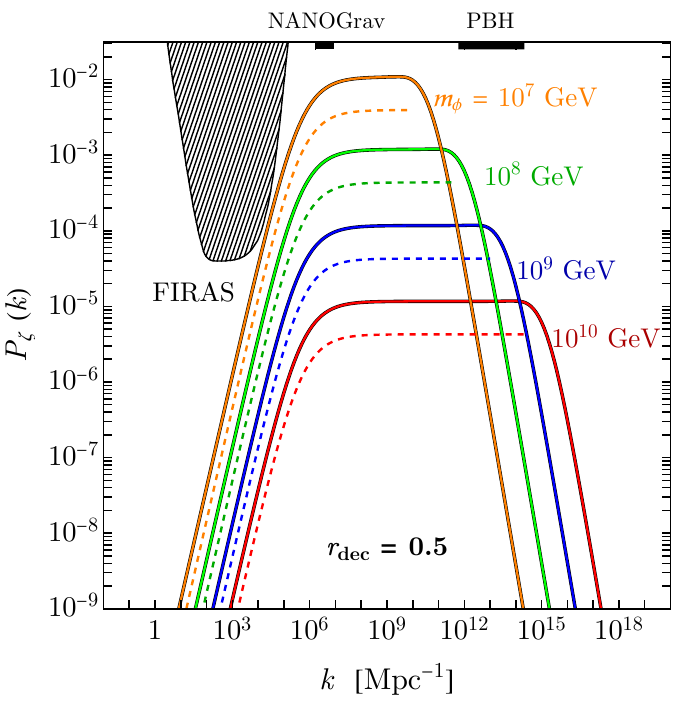}
\qquad\includegraphics[width=.45\textwidth]{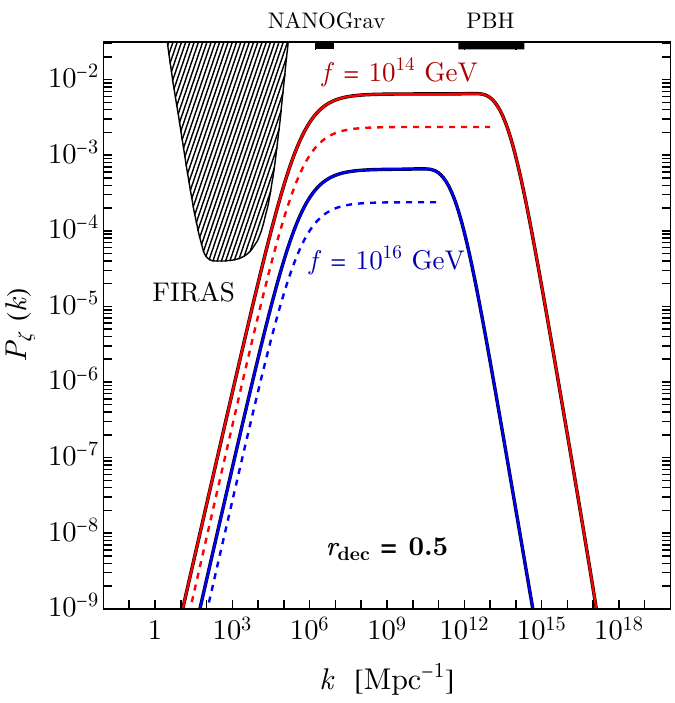}$$
\caption{ 
Left panel. Power spectra $P_{\zeta}(k)$ in eq.\,(\ref{eq:FinaalPS}) 
for four different representative values of $m_{\phi}$. 
We fix $f = 3\times 10^{15}$ GeV and tune $\vartheta_0$ to get $r_{\rm dec} = 0.5$ 
in each of the four spectra. The dashed lines correspond to the sudden decay approximation while the solid lines 
include the dynamics of phase II+III (see the schematic in eq.\,(\ref{eq:Sketch}) and discussion in section\,\ref{eq:Per}).
Right panel. Same as in the left panel but with $m_{\phi} = 10^8$ GeV fixed and for two representative values of 
the decay constant $f$. 
To guide the eye, on the top x-axis we indicate the $k$-range that is relevant for the generation of second-order GWs detectable by the PTA experiments and 
the $k$-range that is viable for the identification of the totality of dark matter with PBHs. The dashed regions shows the experimental constraints coming from the
analysis of CMB spectral distortion by the FIRAS collaboration\,\cite{Fixsen:1996nj,Chluba:2012gq,Chluba:2012we,Bianchini:2022dqh}. }\label{fig:PS}  
\end{center}
\end{figure}

Schematically, we describe the dynamics of perturbations for a given $k$-mode as summarized in the following sketch:
{\small
\begin{align}\label{eq:Sketch2}
	\begin{tikzpicture}
	 {\scalebox{1}{
    \draw[->][thick] (-8.5,0)--(6.2,0);
    \draw[thick][thick] (-5,0.1)--(-5,-0.1);
    \node at (-5,-0.4) {\scalebox{1}{$N = 0$}};
    \node at (-6.75,+0.75) {\scalebox{0.9}{{\color{cornellred}{Inflation}}}};
    \node at (-7.1,-0.3) {\scalebox{0.9}{dynamics described}};
    \node at (-6.85-0.5,-0.7) {\scalebox{0.9}{in terms of $\delta\vartheta_k$}};
    \node at (-6.8,+0.3) {\scalebox{0.9}{{\color{cornellred}{$\varphi$ dynamics boosts $\delta\vartheta_k$}}}};
    \node at (-2-0.6,+0.75) {\scalebox{0.9}{{\color{cornellred}{Phase I}}}};
    \node at (-2.6,+0.3) {\scalebox{0.9}{\color{cornellred}{No curvaton decay, $\Gamma_{\phi} = 0$}}}; 
    \node at (-2.9,-0.3) {\scalebox{0.9}{dynamics of $\delta\vartheta_k$}}; 
    \node at (-2.35,-0.7) {\scalebox{0.9}{bridges $N=0$ and $N_{\rm osc}$}}; 
    \node at (-2.85,-1.1) {\scalebox{0.9}{eqs.\,(\ref{eq:Zetaaxion},\,\ref{eq:Raxion})}};  
    \draw[thick][thick] (-0.3,0.1)--(-0.3,-0.1); 
    \node at (-0.3,-0.4) {\scalebox{1}{$N_{\rm osc}$}};
    \node at (2.4,-0.3) {\scalebox{0.9}{dynamics described in terms of}};
    \node at (3.15-0.5,-0.7) {\scalebox{0.9}{$\zeta_k$, $\zeta_{\phi,k}$, $\mathcal{R}_k$ and $\mathcal{R}_{\phi,k}$ with $\Gamma_{\phi} \neq 0$}};
    \node at (2.2,-1.1) {\scalebox{0.9}{eqs.\,(\ref{eq:Final1},\,\ref{eq:Final2},\,\ref{eq:Final3},\,\ref{eq:Final4})}};
    \node at (2.65,+0.75) {\scalebox{0.9}{{\color{cornellred}{Phase II + Phase III}}}};    
    \node at (2.65,+0.3) {\scalebox{0.9}{{\color{cornellred}{curvaton oscillates and decays to radiation}}}};
    \draw[thick][thick] (5.6,0.1)--(5.6,-0.1);        
    \node at (5.6,-0.4) {\scalebox{1}{$N_{\rm f}$}};    
    \node at (-5.4,1.2) {\scalebox{0.8}{{\color{VioletRed4}{\textbf{Dynamics of $k$-perturbation, 
    from $\delta\vartheta_k$ to $\zeta_k$}}}}};
    }}
	\end{tikzpicture}
\end{align}
}

Fig.\,\ref{fig:PS} shows $P_\zeta(k)$, computed by choosing the proper $N_{\rm f}$ for each $k$, for different values of $f$ and $m_\phi$. 
We can deduce the qualitative behaviour of the power spectrum shape under a change of parameters. We observe that decreasing $f$ has the effect of increasing the amplitude and enlarging the range of $k$ where $P_{\zeta}(k)$ is enhanced. This can be explained by considering that the fast-rolling dynamics of $\varphi$ lasts longer for smaller values of $f$ (see also eq.\,(\ref{eq:Gamma})). Hence, angular perturbations have more time to grow (see also eq.\,(\ref{eq:PowerSpectrumRange}). Moreover, as the curvaton decay width scales as $\Gamma_{\phi}\propto f^{-2}$, decreasing $f$ means anticipating decay into radiation. An early-time decay interests larger scales, resulting in a broader power spectrum. Analogously, decreasing $m_{\phi}$ shrinks the power spectrum and pushes it up to higher amplitude values. The reason to this has to be found, again, in the scaling of the decay width $\Gamma_{\phi}\propto m_{\phi}^3$. Also, decreasing the curvaton mass has the effect of increasing the $e$-fold time difference $\Delta N$ between $T = m_\phi$ and $H = \Gamma_\phi$ implying that now the angular perturbations, and hence the amplitude of the power spectrum, have more time to grow (see also fig.\,\ref{fig:EvoAfterInfla}).

\begin{figure}[!h!]
\begin{center}
$$\includegraphics[width=.47\textwidth]{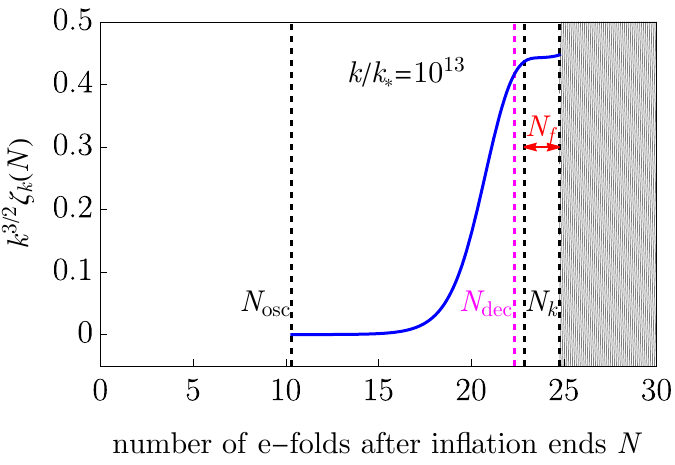}
\qquad\includegraphics[width=.47\textwidth]{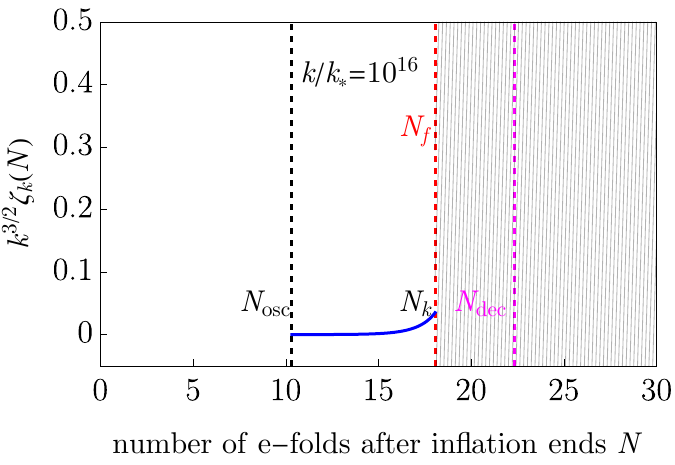}$$
\caption{  
Growth of the curvature perturbation for different modes. Left panel: the mode $k$ in exam re-enters the horizon at $N_k > N_{\rm dec}$ meaning that it still has some time to grow after the time of curvaton decay. This reflects in an enhancement in the amplitude of the power spectrum with respect to the sudden decay approximation. Right panel: the mode $k$ in exam re-enters the horizon before the curvaton decay and its value should be chosen at $N_f \simeq N_{k}$.
 }\label{fig:EvoZeta}  
\end{center}
\end{figure}

For each set of parameters, we show the power spectrum computed, as customary in the literature, in the sudden decay approximation (see ref.\,\cite{Ando:2017veq} for a detailed analysis). Additionally, in fig.\,\ref{fig:PS} we show that, by computing $P_\zeta(k)$ studying the full evolution of $\zeta_k$, one gets an enhancement of roughly a factor 2 in the amplitude with respect to the sudden decay approximation. 
This is a consequence of the fact that in the sudden decay approximation one computes the plateau of the power spectrum at $N_{\rm dec}$ but, as shown in fig.\,\ref{fig:EvoZeta}, $\zeta_k$ continues growing after that and then $\zeta_k(N_{\rm dec}) < \zeta_k(N_{\rm f})$.

It is worth noticing that, by properly tuning the initial parameters of the model, we are able to obtain a broad power spectrum which spans from the scales relevant to produce a signal compatible with NANOGrav (and other PTA experiments) to the ones at which PBHs can comprise the totality of dark matter, realising the scenario proposed in Ref.~\cite{DeLuca:2020agl} by means of an explicit particle physics model of inflation.

Since, in the curvaton scenario discussed here, the enhancement of curvature perturbations at small scales is due to the presence of a spectator field, while the power spectrum at the scales relevant for the CMB is governed by a decoupled sector, this scenario does not suffer of the loop issue corrections to the power spectrum of curvature perturbations due to an USR phase.

\subsection{Primordial non-gaussianities}\label{sub:X}
Requiring that the total energy density is uniform, conservation of energy allow us to relate the curvature perturbation to the energy density fraction of the curvaton
\begin{equation}
\label{eq:zeta_zetaG}
    e^{4 \zeta} - \Omega_\phi e^{3\zeta_\phi}e^\zeta-\left(1-\Omega_\phi\right)=0\,.
\end{equation}
Here we have also assumed, as customary in curvaton models, that the contribution to curvature perturbation coming from the radiation fluid is subdominant with respect to the curvaton one at small scales. By solving the above equation perturbatively, we are able to find an order-by-order relation between $\zeta$ and $\zeta_\phi$ where, in the flat gauge, 
\begin{equation}
    \zeta_\phi=\frac{1}{3}\log\left(1+\frac{\delta\rho_\phi}{\rho_\phi}\right)\,.
\end{equation}
After defining $\zeta_{\rm G} = r_{\rm dec} \zeta_{\phi}^{(1)}$, where $\zeta_{\phi}^{(1)}$ is the first-order term in the expansion, the power series for $\zeta_\phi$ can be resummed to the exact relation
\begin{equation}
    \zeta_\phi = \frac{2}{3}\log\left( 1+\frac{3}{2r_{\rm dec}}\zeta_{\rm G}\right)\,.
\end{equation}
By plugging the above relation into eq.\,(\ref{eq:zeta_zetaG}), and after defining 
\begin{align}\label{eq:MasterXM}
    \zeta = \log\big[X(r_{\rm dec},\zeta_{\rm G})\big]\,,
\end{align}
we get the forth-order polynomial equation
\begin{equation}
    X^4 - \Omega_\phi\left(1+\frac{3}{2 r_{\rm dec}}\zeta_{\rm G}\right)^2 X - \left( 1-\Omega_\phi\right)=0\,,
\end{equation}
which can be solved to find
\begin{align}
&X(r_{\rm dec},\zeta_{\rm G}) \equiv \frac{1}{\sqrt{2 (3+r_{\rm dec})^{1/3}}}
\Bigg\{
\sqrt{
\frac{
-3 + r_{\rm dec}(2+r_{\rm dec}) + [(3+r_{\rm dec})P(r_{\rm dec},\zeta_{\rm G})]^{2/3}
}{
(3+r_{\rm dec})P^{1/3}(r_{\rm dec},\zeta_{\rm G})
}
}+   
\nn \\
&
\sqrt{
\frac{(1-r_{\rm dec})}
{P^{1/3}(r_{\rm dec},\zeta_{\rm G})} -
\frac{P^{1/3}(r_{\rm dec},\zeta_{\rm G})}
{(3+r_{\rm dec})^{1/3}} 
+ 
\frac{
(2r_{\rm dec} + 3\zeta_G)^{2}
P^{1/6}(r_{\rm dec},\zeta_{\rm G})
}{r_{\rm dec}
\sqrt{-3 + r_{\rm dec}(2+r_{\rm dec}) +
[(3+r_{\rm dec})P(r_{\rm dec},\zeta_{\rm G})]^{2/3}
}}} 
\Bigg\}\,,
\end{align}
and 
\begin{align}
P(r_{\rm dec},\zeta_{\rm G}) \equiv 
\frac{(2r_{\rm dec} + 3\zeta_{\rm G})^{4}}{16 r_{\rm dec}^2} + \sqrt{
(1-r_{\rm dec})^3(3+r_{\rm dec}) + \frac{(2r_{\rm dec} + 3\zeta_{\rm G})^8}{256 r_{\rm dec}^4}
}\,.
\end{align}
It will be useful in the following section to extract the leading order expansion of eq.\,(\ref{eq:MasterXM}), which can be approximated at the second order in the Gaussian component $\zeta_{\rm G}$ as
\begin{align}\label{eq:ZetaQ2}
\zeta_{2} \equiv \zeta_{\rm G} + \frac{3}{5}f_{\rm NL}(r_{\rm dec})\zeta_{\rm G}^2\,,
~~~~~~~{\rm with}~~~~~~~
f_{\rm NL}(r_{\rm dec}) \equiv  \frac{5}{3}\bigg(
\frac{3}{4r_{\rm dec}} - 1 - \frac{r_{\rm dec}}{2}
\bigg)\,.
\end{align}
It is typically assumed that computing the predictions of the model assuming only the leading order correction to the Gaussian case provides a sufficiently good approximation. 
However, we show in the next section that this is not the case when one calculates the abundance of PBHs, which are extremely sensitive to non-gaussian corrections. 
\section{Phenomenology: PBHs and SIGWs from the curvaton model}
\subsection{Primordial black hole
formation}
We followed the prescription presented in sec.\ref{sec:C1A} based on the threshold statistics on the compaction function in order to compute the abundance of PBHs.
We have followed the prescription given in ref.\,\cite{Musco:2020jjb} to compute the values of $\mathcal{C}_{\rm th}$ and $r_m$, which depend on the shape of the power spectrum.
As the power spectrum we obtained in the axion-curvaton scenario we consider in this analysis is nearly scale-invariant, one gets $\mathcal{C}_{\rm th}=0.55$ and $kr_m= 4.49\equiv \kappa$.
The presence of the QCD phase transitions is taken into account by considering, as shown in refs.\,\cite{Franciolini:2022tfm}, that $\gamma(M_H)$, $\mathcal{K}(M_H)$, $\mathcal{C}_{th}(M_H)$ and $\Phi(M_H)$ are functions of the horizon mass for $M_{\rm PBH}=\mathcal{O}(M_{\odot})$.

\begin{figure}[!h!]
\begin{center}
$$\includegraphics[width=.40\textwidth]{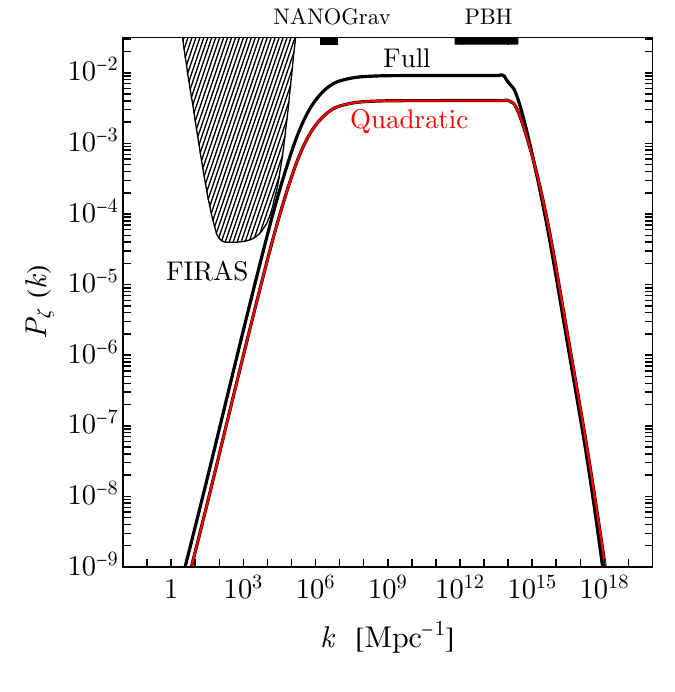}
~~\includegraphics[width=.59\textwidth]{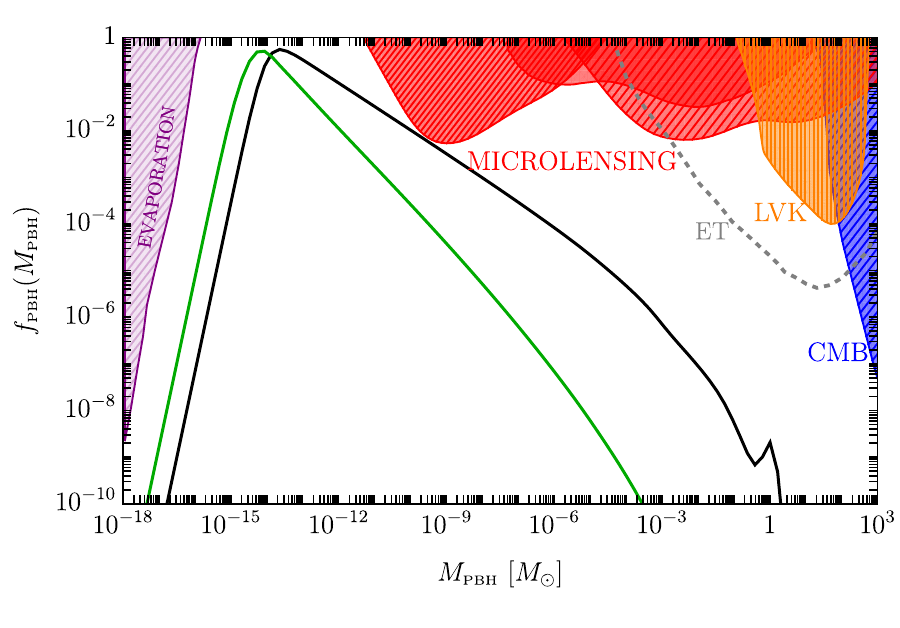}$$
\caption{
\textbf{\textit{Left panel:}} broad power spectrum of the curvature perturbation obtained with the axion-curvaton model assuming $N_{*}=58$. The black solid line refers to the power spectrum needed to obtain $f_{\rm PBH}\simeq 1$ when computing the abundance of PBHs with a non-perturbative treatment of NGs. 
We chose $m_{\phi}=5.15\times10^7$ ${\rm GeV}$,\;$f=4.20\times10^{13}$ ${\rm GeV}$,\;$\vartheta_0=5.00\times 10^{-2}$. 
For comparison, we also show an analogous scenario (red line) providing $f_{\rm PBH}=1$  in the quadratic approximation and corresponding to the choice of $m_{\phi}=1.00\times10^8$ ${\rm GeV}$$,\;f=6.30\times10^{13}$ GeV, $\vartheta_0=5.00\times 10^{-2}$. For both cases we get $r_{\rm dec}=0.5$. The shaded region shows the experimental
constraints coming from the analysis of CMB spectral distortion by the FIRAS collaboration\,\cite{Fixsen:1996nj,Bianchini:2022dqh}. \textbf{\textit{Right panel:}} $f_{\rm PBH}(M_{\rm PBH})$ computed starting from the corresponding power spectrum in the Full (black solid line) and in the quadratic (green solid line) computation, as in the left panel. In both cases PBHs comprise all the dark matter in the Universe. 
Constraints on $f_{\rm PBH}$ shown in the plot are addressed in sec.\,\ref{sec:Clas}.
 }\label{fig:TypeIDM}  
\end{center}
\end{figure}

The integrated abundance of PBHs is given by the integral 
\begin{equation}
f_{\rm PBH}=\int f_{\rm PBH}(M_{\rm PBH})d\log M_{\rm PBH}.
\end{equation}
We tune the parameters of the axion-curvaton model 
requiring PBHs to account for the totality of dark matter, i.e. fixing $f_{\rm PBH}\simeq1$.
It is instructive to compute the mass fraction of PBHs including NGs both with the quadratic approximation (eq.\,\eqref{eq:ZetaQ2}) and the exact functional form (eq.\,(\ref{eq:MasterXM})).
This allows us to investigate the relevance of the non-perturbative treatment of NGs which we adopt here based on ref.~\cite{Ferrante:2022mui}.
Fig.\,\ref{fig:TypeIDM} shows $f_{\rm PBH}$ so computed, together with the corresponding power spectrum obtained through the mechanism presented in before.

In fig.~\ref{fig:TypeIDM} one can see that the power spectrum of curvature perturbations grows from its small value at large scales to the enhanced plateau at $k\gtrsim 10^6$ Mpc$^{-1}$. We choose $c = 1.6$ in eq.~\eqref{eq:EffectiveV}, which implies a
power law growth ${\cal P}_\zeta\approx k^{n_\theta}$ 
of the curvature spectrum
with an index  $n_\theta = 1.4$. 
It is interesting to notice that this growth is shallower than what it is typically achieved in single field models of inflation characterised by an USR phase\,\cite{Byrnes:2018txb}.

In the right panel of fig.\,\ref{fig:TypeIDM} we show the most stringent experimental constraints on $f_{\rm PBH}$ (see section\,\ref{sec:Clas}. 

In accordance with what was observed in sec.\,\ref{sec:C1A}, 
we find that for $r_{\rm dec}=0.5$, the effect of primordial NGs truncated to second order is that of enhancing PBHs abundance with respect to the exact computation. Therefore, requiring no PBH overproduction (that would surpass the dark matter abundance) imposes to decrease the amplitude of the power spectrum in this approximate case. 
Such a difference between the approximated and exact spectra
has an important impact on the signal of second-order GWs and could in principle be explored at GWs detectors, as shown next.

It is important to note that, as we can see from the right plot of fig.\,\ref{fig:TypeIDM}, there is a second peak in $f_{\rm PBH}(M_{\rm PBH})$ around solar masses, 
which is caused by the softening of the equation of state during the QCD cross-over phase transition\,\cite{Jedamzik:1998hc,Byrnes:2018clq,Franciolini:2022tfm,Escriva:2022bwe}.
While this enhancement may be sizeable, it is not sufficient to generate a large enough abundance leading to a sizeable merger rate of stellar mass PBH mergers in this scenario. 
In practice, the properties of this model, in combination with the FIRAS bound that force the enhancement of perturbations to be placed at scales larger than around $\approx 10^5 / {\rm Mpc}$, cause the PBH abundance in the stellar mass range to be small.\footnote{Changing the assumed value of $c$ to larger numbers, i.e. larger $n_\theta$, would slightly alleviate such a conclusion due to a steeper spectral growth around $10^4$/Mpc.} 
Furthermore, the reason why the QCD peak is only visible in the exact computation has to be found in the left-hand plot in fig.\,\ref{fig:TypeIDM}. The quadratic approximation leads to an overproduction of PBHs, which has to be compensated by a decrease in the amplitude of the power spectrum if one wants not to overshoot the limiting value $f_{\rm PBH}=1$. 
Readjusting the parameters to decrease the amplitude of the power spectrum leads to a slight shift of the rising slope towards larger $k$, resulting in a smaller abundances at high masses where the QCD transition would have an impact. 
 
\subsection{Scalar-induced gravitational waves}\label{Sec:SGWB}
We compute the emission of GWs by accounting for the softening of equation of state at the QCD era, which has an important role in shaping the spectral tilt in the PTA frequency range, following sec.\ref{sec:GW1}.
On the other hand, we will neglect higher order contributions to the SGWB. This is because, differently from what happens in the case of PBH formation which are extremely sensitive to the non-gaussian tail of the curvature distribution, 
the emission of GWs is dominated by the leading order in our case. Indeed, in the scenario we consider, we have $r_{\rm dec} = 0.5$ that corresponds to $f_{\rm NL}= 0.42$ from eq.\,\eqref{eq:ZetaQ2}. 
Therefore, as shown in refs.\,\cite{Cai:2018dig,Unal:2018yaa,Ragavendra:2021qdu,Adshead:2021hnm,Abe:2022xur,Garcia-Saenz:2022tzu}, 
corrections from higher orders terms only amount to a negligible contribution to the SGWB.

\begin{figure}[!t!]
	\begin{center}
$$\includegraphics[width=.49\textwidth]{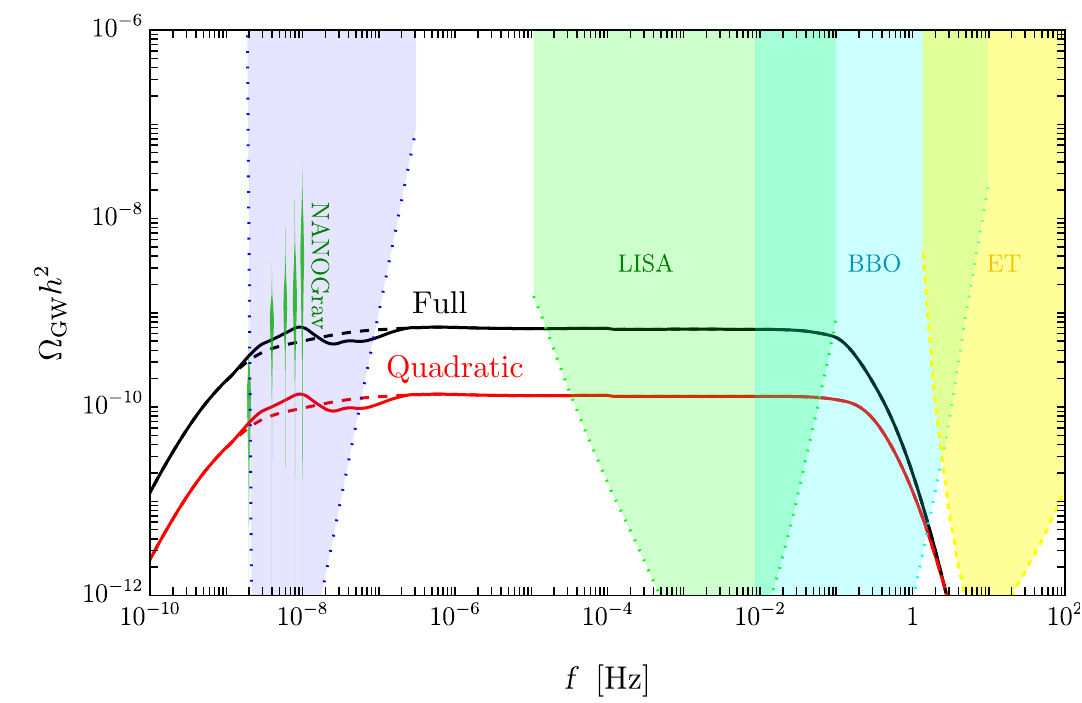}
~~\includegraphics[width=.49\textwidth]
{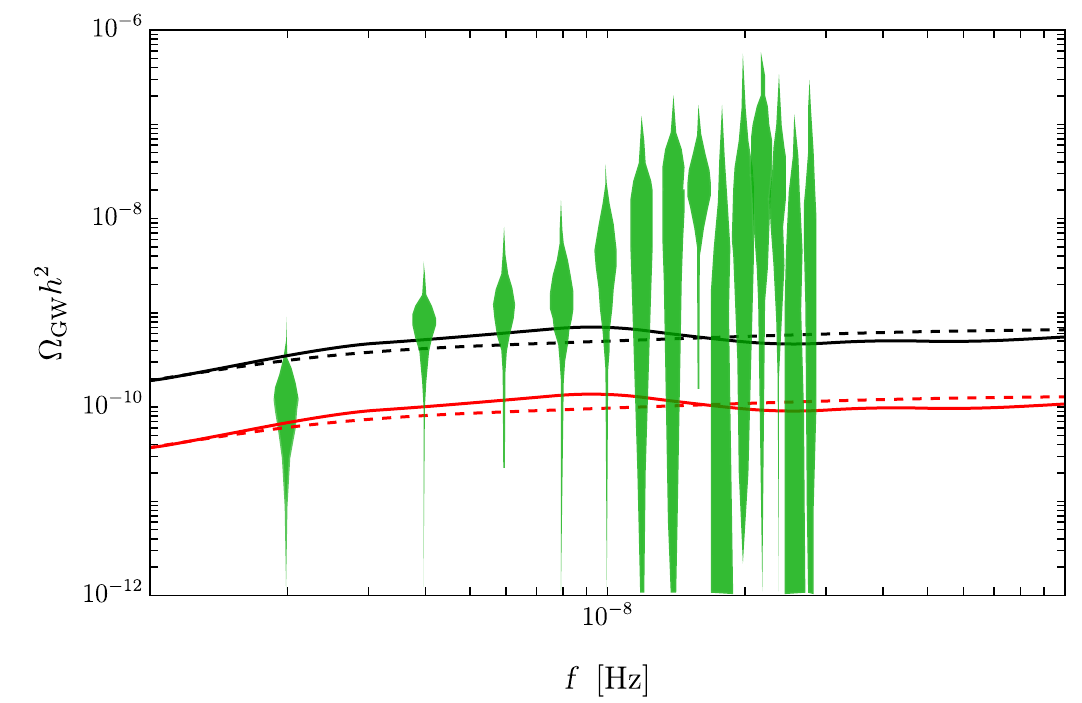}$$
\caption{
		Signal of second order GWs associated with the broad power spectrum obtained within the curvaton model. 
		When computing abundance of PBHs in the quadratic approximation (red line) and requiring $f_{\rm PBH}\simeq 1$, 
		one would find an amplitude which is smaller than the one required to explain the tentative signal detected by the NANOGrav 15 yrs.
However, correctly adopting the non-perturbative computation of the PBH abundance (black line), we find that it is possible to explain at the same time dark matter in the form of asteroidal mass PBHs and the SGWB signal from NANOGrav within the curvaton scenario. 
We plot constraints on the signal of second-order GWs coming from NANOGrav 15 yrs experiment\,\cite{NANOGrav:2023gor} and future sensitivity for planned experiments like SKA\,\cite{Zhao:2013bba}, LISA\,\cite{LISA:2022kgy}, BBO/DECIGO\,\cite{Yagi:2011wg} and ET (power law integrated sensitivity curves as derived in ref.~\cite{Bavera:2021wmw}). 
The dashed line report the SGWB obtained neglecting the variation of sound speed during the QCD era (around $f\approx 10^{-8}$Hz) but accounting for a temperature dependent overall factor $c_g$.
}\label{fig:GWCurv}  
	\end{center}
\end{figure}

Finally, we compute the signal of GWs associated with the abundance of PBHs in fig.\,\ref{fig:TypeIDM}, by plugging the corresponding power spectrum into eq.\,(\ref{eq:OmegaGW}). Results are shown in fig.\,\ref{fig:GWCurv}, where $\Omega_{\rm GW}h^2$ is given as a function of the frequency $f= k/2\pi$.\footnote{Notice that before the curvaton decays, its isocurvature perturbations may also induce second order GWs\,\cite{Bartolo:2007vp,Kawasaki:2013xsa}. We neglect the isocurvature contribution and only focus on the adiabatic source, as it would only affect the tail at large frequencies of the SGWB, while also being suppressed by powers of the small ratio $\Gamma_\phi/m_\phi$ considered in this analysis (see eq.~\eqref{eq:Gamma}).}
This plot shows that the curvaton scenario discussed in this chapter is able to produce an enhanced and flat power spectrum, that interestingly connects the asteroidal mass PBH dark matter and SGWB at PTA frequencies, providing a concrete realization
of the scenario proposed in ref.~\cite{DeLuca:2020agl}.

One of the take-home messages of this study is also contained in  fig.\,\ref{fig:GWCurv}, that shows how primordial NGs have important phenomenological relevance and must be considered with care. 
Indeed, relying on the quadratic approximation when computing the PBH abundance would already exclude the possibility of explaining both the totality of dark matter and the tentative signal by NANOGrav 15 years observations within a unified PBH formation scenario based on the curvaton model. As stressed already in chapter\,\ref{cap:GWs}.
On the other hand, as we have shown, the non-perturbative treatment of the NGs inherently induced in curvaton scenarios invalidates this conclusion. 
As a by-product of this result, we also show that the GW amplitude of PBH dark matter in the LISA frequency band is also modified when considering the specific NG formation scenario discussed in this chapter.

As already pointed out in sec.\ref{sec:GW2}, including primordial NGs (here derived within the curvaton model) in the calculation of the abundance of PBHs can significantly affect the amplitude of the power spectrum in order to explain the totality of dark matter, alleviating the tension with the NANOGRav dataset.
In the example presented in this section, with $r_{\rm dec} = 0.5$, the spectral amplitude required to obtain $f_{\rm PBH} = 1$ is close to the one obtained with Gaussian perturbations (and the only including the effect on non-linearities), while it would be reduced if smaller values of $r_{\rm dec}$ were considered (see also \cite{Ferrante:2022mui}).

\subsubsection{Summary}\label{sec:Finale}
 
Firstly, we showed that it is feasible to generate a broad enhanced power spectrum of the curvature perturbations in a specific curvaton model. This was achieved by carefully selecting the initial parameters of the model, which resulted in a broad power spectrum that encompasses both the scales required for the totality of the dark matter in the form of PBH and those relevant for GW signal that is consistent with the one potentially hinted by NANOGrav and other PTA experiments.
Secondly, we showed in a concrete example the phenomenological relevance of going beyond perturbative approaches when computing the PBH abundance in presence of NGs.

We found that the quadratic approximation of NGs leads to  an overestimation of the PBH abundance compared to the results obtained using the full non-Gaussian relation, 
due to the relevant impact of higher-order terms and a violation of perturbativity in the computation of PBH abundance observed for broad spectra \cite{Ferrante:2022mui}. 
This leads to an interesting phenomenological outcome: the amplitude of $\Omega_{\rm GW}$ obtained by considering the exact relation $\zeta\left(\zeta_{\rm G}\right) = \log\left[X\left(r_{\rm dec},\zeta_{\rm G}\right)\right]$ can better fit the tentative signal observed by NANOGrav15, as demonstrated in fig.\,\ref{fig:GWCurv}, only if the NGs are correctly computed beyond the quadratic approximation, highlighting the importance of a precise calculation of the impact of non-Gaussian corrections when considering both PBH abundance and scalar-induced GWs.

\chapter{Epilog}
After three remarkable years, our academic voyage reaches its conclusion.

We find ourselves in what may be the most thrilling era for studying PBHs. The field has garnered immense interest, leading to significant advancements in both theoretical understanding of PBH formation and empirical investigations, especially within the burgeoning realm of GW astronomy.

In this thesis, we explore various aspects of the physics pertaining to PBHs. After a brief introduction in Chapter,\ref{cap:intro}, which covers the phenomenological aspects related to PBHs, in Chapter \ref{cap:NGS} we delve into the critical role of non-Gaussianity in accurately determining the abundance of PBHs. Chapter \ref{sec:Clas}
provides a brief overview of the categories of constraints on the PBH abundance and then a detailed description on how we update the constraints utilizing both LVK O3 data and the PTA datasets.

Chapter \ref{cap:GWs} examines the gravitational wave signatures produced from the cosmological perturbations that could lead to PBH formation, and how current and future observations might help us distinguish between different PBH production models. Finally, Chapters \ref{cap:Models} and \ref{cap:Models2} describe two distinct classes of models capable of generating PBHs.

Our work highlighted the following key take-home messages and outlook:
\begin{itemize}
    \item The essential roles played by both types of non-Gaussianities in calculating PBH abundance. We have found that power series expansions for primordial non-Gaussianities  yield unreliable results when the curvature power spectrum is not very narrow. We stress that one may be tempted to think that one can always reabsorb NG corrections to the computation of the PBH abundance by means of small retuning of model parameters. While generically true, this is not harmless, as it would spoil the model predictions which depend on the power spectrum (like the amplitude of the induced GW signal) and are directly sensitive to the inclusion of NG corrections. Furthermore, with the aim of constraining PBHs with future GW detections, missing a precise description of NG corrections may prevent us from setting reliable bounds on their abundance and the fraction of DM they can amount for.
    \item Current observations of merger rates by the LVK collaboration can be utilized to constrain PBH abundance. The current observation excludes the possibility that the DM today is entirely composed by PBH mass distribution with an average mass around the (sub)solar mass range.
    The future detection of a sub-solar mass candidate could serve as a definitive 'smoking gun' for the discovery of PBHs.
    \item The evidence for the Hellings-Downs angular correlation reported by the PTA collaborations sets an important milestone in gravitational-wave astronomy. One of the most pressing challenges to follow is to determine the nature signal. We have scrutinized the prospect that the signal detected by various PTA collaborations might stem from GWs induced by high-amplitude primordial curvature perturbations. Our analysis indicates that models of PBH formation with Gaussian primordial perturbations, or positive non-Gaussianities, would excessively produce PBHs unless the spectrum's amplitude is significantly lower than what is necessary to account for the GW signal. As a consequence, in order to relax the tension between the PTA dataset and PBH explanation we need to introduce models with large negative non-gaussianities. However, alternative explanations for the GW signals registered by PTA collaborations, both cosmological and astronomical, are possible. The identification of nHz GWs could be associated with the most colossal events since the Big Bang—the mergers of Supermassive Black Holes—or it might signal new fundamental physics beyond the reach of terrestrial experiments. Only time and additional data will unravel these mysteries.
    \item In the standard formation scenario, to achieve a significant amount of DM in the form of PBHs, it is necessary to boost the amplitude of the curvature power spectrum to be around $10^{-2}$ at the relevant range of scales. This enhancement can be achieved in numerous inflationary models. We focused on two types of models: Ultra-slow Roll (USR) and the Curvaton model. In the USR models, we showed that if we compute properly the abundance of PBHs the perturbativity criterium is not violated in presence of 1-loop corrections. Then we introduce an explicit realization of an USR phase in an original model called \textit{double inflection point}.
    For the curvaton models we study PBH production in the frame of a specific axion-like curvaton model showing that we are able to obtain a broad power spectrum of curvature perturbations, i.e. a power spectrum which spans over many orders of magnitude of scales $k$. A feature that is crucial in order to make connection between observables related to PBH dark matter and those associated with scalar-induced GWs.
    
\end{itemize}
As we stand on the cusp of the golden age of gravitational waves, the potential observations in the coming years could unveil the existence of these enigmatic entities known as PBHs, opening new horizons in our quest to understand the Universe.
\vspace{1cm}

\begin{minipage}[t]{0.5\textwidth}
\raggedright
\textit{Noi ci allegrammo, e tosto tornò in pianto, 
\\ché de la nova terra un turbo nacque, 
\\e percosse del legno il primo canto.
\vspace{0.5cm}
\\Tre volte il fé girar con tutte l’acque; 
\\a la quarta levar la poppa in suso 
\\e la prora ire in giù, com’altrui piacque.     
\vspace{0.5cm}
\\infin che ’l mar fu sovra noi richiuso.}
\vspace{0.5cm}

Fine.
\end{minipage}
\hfill
\begin{minipage}[t]{0.5\textwidth}
\raggedleft
\textit{We rejoiced, and soon it turned to lamentation,
\\for from the new land a whirlwind rose
\\and struck the fore part of the vessel.
\vspace{0.5cm}
\\Three times it made her whirl with all the waters,
\\the fourth it made her stern lift up
\\ and the prow go down, as pleased Another,
\vspace{0.5cm}
\\till the sea had closed over us.} 
\vspace{0.5cm}
\\The end.
\end{minipage}
\thispagestyle{plain}

\cleardoublepage
\newpage  
\addcontentsline{toc}{chapter}{Bibliography}
\bibliography{Include/Backmatter/Bibliography}

\begin{thebibliography}{100}

\bibitem{Ferrante:2022mui}
Giacomo Ferrante, Gabriele Franciolini, Antonio Iovino, Junior., and Alfredo Urbano.
\newblock {Primordial non-Gaussianity up to all orders: Theoretical aspects and implications for primordial black hole models}.
\newblock {\em Phys. Rev. D}, 107(4):043520, 2023.
\newblock \href {https://arxiv.org/abs/2211.01728} {\path{arXiv:2211.01728}}, \href {https://doi.org/10.1103/PhysRevD.107.043520} {\path{doi:10.1103/PhysRevD.107.043520}}.

\bibitem{Franciolini:2023agm}
Gabriele Franciolini, Antonio Iovino, Junior., Marco Taoso, and Alfredo Urbano.
\newblock {Perturbativity in the presence of ultraslow-roll dynamics}.
\newblock {\em Phys. Rev. D}, 109(12):123550, 2024.
\newblock \href {https://arxiv.org/abs/2305.03491} {\path{arXiv:2305.03491}}, \href {https://doi.org/10.1103/PhysRevD.109.123550} {\path{doi:10.1103/PhysRevD.109.123550}}.

\bibitem{Ferrante:2023bgz}
G.~Ferrante, G.~Franciolini, A.~J. Iovino, and A.~Urbano.
\newblock {Primordial black holes in the curvaton model: possible connections to pulsar timing arrays and dark matter}.
\newblock {\em JCAP}, 06:057, 2023.
\newblock \href {https://doi.org/10.1088/1475-7516/2023/06/057} {\path{doi:10.1088/1475-7516/2023/06/057}}.

\bibitem{Franciolini:2023pbf}
Gabriele Franciolini, Antonio Iovino, Junior., Ville Vaskonen, and Hardi Veermae.
\newblock {Recent Gravitational Wave Observation by Pulsar Timing Arrays and Primordial Black Holes: The Importance of Non-Gaussianities}.
\newblock {\em Phys. Rev. Lett.}, 131(20):201401, 2023.
\newblock \href {https://arxiv.org/abs/2306.17149} {\path{arXiv:2306.17149}}, \href {https://doi.org/10.1103/PhysRevLett.131.201401} {\path{doi:10.1103/PhysRevLett.131.201401}}.

\bibitem{Ellis:2023oxs}
John Ellis, Malcolm Fairbairn, Gabriele Franciolini, Gert H\"utsi, Antonio Iovino, Marek Lewicki, Martti Raidal, Juan Urrutia, Ville Vaskonen, and Hardi Veerm\"ae.
\newblock {What is the source of the PTA GW signal?}
\newblock {\em Phys. Rev. D}, 109(2):023522, 2024.
\newblock \href {https://arxiv.org/abs/2308.08546} {\path{arXiv:2308.08546}}, \href {https://doi.org/10.1103/PhysRevD.109.023522} {\path{doi:10.1103/PhysRevD.109.023522}}.

\bibitem{Ianniccari:2024bkh}
Andrea Ianniccari, Antonio~J. Iovino, Alex Kehagias, Davide Perrone, and Antonio Riotto.
\newblock {Primordial black hole abundance: The importance of broadness}.
\newblock {\em Phys. Rev. D}, 109(12):123549, 2024.
\newblock \href {https://arxiv.org/abs/2402.11033} {\path{arXiv:2402.11033}}, \href {https://doi.org/10.1103/PhysRevD.109.123549} {\path{doi:10.1103/PhysRevD.109.123549}}.

\bibitem{Andres-Carcasona:2024wqk}
M.~Andr\'es-Carcasona, A.~J. Iovino, V.~Vaskonen, H.~Veerm\"ae, M.~Mart\'\i{}nez, O.~Pujol\`as, and Ll.~M. Mir.
\newblock {Constraints on primordial black holes from LIGO-Virgo-KAGRA O3 events}.
\newblock {\em Phys. Rev. D}, 110(2):023040, 2024.
\newblock \href {https://arxiv.org/abs/2405.05732} {\path{arXiv:2405.05732}}, \href {https://doi.org/10.1103/PhysRevD.110.023040} {\path{doi:10.1103/PhysRevD.110.023040}}.

\bibitem{Iovino:2024tyg}
A.~J. Iovino, G.~Perna, A.~Riotto, and H.~Veerm\"ae.
\newblock {Curbing PBHs with PTAs}.
\newblock {\em JCAP}, 10:050, 2024.
\newblock \href {https://arxiv.org/abs/2406.20089} {\path{arXiv:2406.20089}}, \href {https://doi.org/10.1088/1475-7516/2024/10/050} {\path{doi:10.1088/1475-7516/2024/10/050}}.

\bibitem{Allegrini:2024ooy}
Sasha Allegrini, Loris Del~Grosso, Antonio~J. Iovino, and Alfredo Urbano.
\newblock {Is the formation of primordial black holes from single-field inflation compatible with standard cosmology?}
\newblock 12 2024.
\newblock \href {https://arxiv.org/abs/2412.14049} {\path{arXiv:2412.14049}}.

\bibitem{Ianniccari:2024eza}
Andrea Ianniccari, Antonio~J. Iovino, Alex Kehagias, Davide Perrone, and Antonio Riotto.
\newblock {Black Hole Formation\textemdash{}Null Geodesic Correspondence}.
\newblock {\em Phys. Rev. Lett.}, 133(8):081401, 2024.
\newblock \href {https://doi.org/10.1103/PhysRevLett.133.081401} {\path{doi:10.1103/PhysRevLett.133.081401}}.

\bibitem{Ianniccari:2024ysv}
A.~Ianniccari, A.~J. Iovino, A.~Kehagias, P.~Pani, G.~Perna, D.~Perrone, and A.~Riotto.
\newblock {Deciphering the Instability of the Black Hole Ringdown Quasinormal Spectrum}.
\newblock {\em Phys. Rev. Lett.}, 133(21):211401, 2024.
\newblock \href {https://arxiv.org/abs/2407.20144} {\path{arXiv:2407.20144}}, \href {https://doi.org/10.1103/PhysRevLett.133.211401} {\path{doi:10.1103/PhysRevLett.133.211401}}.

\bibitem{Iovino:2024sgs}
A.~J. Iovino, S.~Matarrese, G.~Perna, A.~Ricciardone, and A.~Riotto.
\newblock {How Well Do We Know the Scalar-Induced Gravitational Waves?}
\newblock 12 2024.
\newblock \href {https://arxiv.org/abs/2412.06764} {\path{arXiv:2412.06764}}.

\bibitem{Planck:2018vyg}
N.~Aghanim et~al.
\newblock {Planck 2018 results. VI. Cosmological parameters}.
\newblock {\em Astron. Astrophys.}, 641:A6, 2020.
\newblock [Erratum: Astron.Astrophys. 652, C4 (2021)].
\newblock \href {https://arxiv.org/abs/1807.06209} {\path{arXiv:1807.06209}}, \href {https://doi.org/10.1051/0004-6361/201833910} {\path{doi:10.1051/0004-6361/201833910}}.

\bibitem{Guth:1980zm}
Alan~H. Guth.
\newblock {The Inflationary Universe: A Possible Solution to the Horizon and Flatness Problems}.
\newblock {\em Phys. Rev. D}, 23:347--356, 1981.
\newblock \href {https://doi.org/10.1103/PhysRevD.23.347} {\path{doi:10.1103/PhysRevD.23.347}}.

\bibitem{Riotto:2002yw}
Antonio Riotto.
\newblock {Inflation and the theory of cosmological perturbations}.
\newblock {\em ICTP Lect. Notes Ser.}, 14:317--413, 2003.
\newblock \href {https://arxiv.org/abs/hep-ph/0210162} {\path{arXiv:hep-ph/0210162}}.

\bibitem{Sasaki:1986hm}
Misao Sasaki.
\newblock {Large Scale Quantum Fluctuations in the Inflationary Universe}.
\newblock {\em Prog. Theor. Phys.}, 76:1036, 1986.
\newblock \href {https://doi.org/10.1143/PTP.76.1036} {\path{doi:10.1143/PTP.76.1036}}.

\bibitem{Mukhanov:1988jd}
Viatcheslav~F. Mukhanov.
\newblock {Quantum Theory of Gauge Invariant Cosmological Perturbations}.
\newblock {\em Sov. Phys. JETP}, 67:1297--1302, 1988.

\bibitem{Zeldovich:1967lct}
Ya.~B. Zel'dovich and I.~D. Novikov.
\newblock {The Hypothesis of Cores Retarded during Expansion and the Hot Cosmological Model}.
\newblock {\em Soviet Astron. AJ (Engl. Transl. ),}, 10:602, 1967.

\bibitem{Hawking:1971ei}
Stephen Hawking.
\newblock {Gravitationally collapsed objects of very low mass}.
\newblock {\em Mon. Not. Roy. Astron. Soc.}, 152:75, 1971.

\bibitem{Hawking:1975vcx}
S.~W. Hawking.
\newblock {Particle Creation by Black Holes}.
\newblock {\em Commun. Math. Phys.}, 43:199--220, 1975.
\newblock [Erratum: Commun.Math.Phys. 46, 206 (1976)].
\newblock \href {https://doi.org/10.1007/BF02345020} {\path{doi:10.1007/BF02345020}}.

\bibitem{Hawking:1974rv}
S.~W. Hawking.
\newblock {Black hole explosions}.
\newblock {\em Nature}, 248:30--31, 1974.
\newblock \href {https://doi.org/10.1038/248030a0} {\path{doi:10.1038/248030a0}}.

\bibitem{Carr:1974nx}
Bernard~J. Carr and S.~W. Hawking.
\newblock {Black holes in the early Universe}.
\newblock {\em Mon. Not. Roy. Astron. Soc.}, 168:399--415, 1974.

\bibitem{Carr:1975qj}
Bernard~J. Carr.
\newblock {The Primordial black hole mass spectrum}.
\newblock {\em Astrophys. J.}, 201:1--19, 1975.
\newblock \href {https://doi.org/10.1086/153853} {\path{doi:10.1086/153853}}.

\bibitem{Chapline:1975ojl}
George~F. Chapline.
\newblock {Cosmological effects of primordial black holes}.
\newblock {\em Nature}, 253(5489):251--252, 1975.
\newblock \href {https://doi.org/10.1038/253251a0} {\path{doi:10.1038/253251a0}}.

\bibitem{Meszaros:1975ef}
P.~Meszaros.
\newblock {Primeval black holes and galaxy formation}.
\newblock {\em Astron. Astrophys.}, 38:5--13, 1975.

\bibitem{MacGibbon:2007yq}
Jane~H. MacGibbon, Bernard~J. Carr, and Don~N. Page.
\newblock {Do Evaporating Black Holes Form Photospheres?}
\newblock {\em Phys. Rev. D}, 78:064043, 2008.
\newblock \href {https://arxiv.org/abs/0709.2380} {\path{arXiv:0709.2380}}, \href {https://doi.org/10.1103/PhysRevD.78.064043} {\path{doi:10.1103/PhysRevD.78.064043}}.

\bibitem{Duechting:2004dk}
Norbert Duechting.
\newblock {Supermassive black holes from primordial black hole seeds}.
\newblock {\em Phys. Rev. D}, 70:064015, 2004.
\newblock \href {https://arxiv.org/abs/astro-ph/0406260} {\path{arXiv:astro-ph/0406260}}, \href {https://doi.org/10.1103/PhysRevD.70.064015} {\path{doi:10.1103/PhysRevD.70.064015}}.

\bibitem{Kawasaki:2012kn}
Masahiro Kawasaki, Alexander Kusenko, and Tsutomu~T. Yanagida.
\newblock {Primordial seeds of supermassive black holes}.
\newblock {\em Phys. Lett. B}, 711:1--5, 2012.
\newblock \href {https://arxiv.org/abs/1202.3848} {\path{arXiv:1202.3848}}, \href {https://doi.org/10.1016/j.physletb.2012.03.056} {\path{doi:10.1016/j.physletb.2012.03.056}}.

\bibitem{Bernal:2017nec}
Jos\'e~Luis Bernal, Alvise Raccanelli, Licia Verde, and Joseph Silk.
\newblock {Signatures of primordial black holes as seeds of supermassive black holes}.
\newblock {\em JCAP}, 05:017, 2018.
\newblock [Erratum: JCAP 01, E01 (2020)].
\newblock \href {https://arxiv.org/abs/1712.01311} {\path{arXiv:1712.01311}}, \href {https://doi.org/10.1088/1475-7516/2018/05/017} {\path{doi:10.1088/1475-7516/2018/05/017}}.

\bibitem{Ivanov:1994pa}
P.~Ivanov, P.~Naselsky, and I.~Novikov.
\newblock {Inflation and primordial black holes as dark matter}.
\newblock {\em Phys. Rev. D}, 50:7173--7178, 1994.
\newblock \href {https://doi.org/10.1103/PhysRevD.50.7173} {\path{doi:10.1103/PhysRevD.50.7173}}.

\bibitem{GarciaBellido:1996qt}
Juan Garcia-Bellido, Andrei~D. Linde, and David Wands.
\newblock {Density perturbations and black hole formation in hybrid inflation}.
\newblock {\em Phys. Rev. D}, 54:6040--6058, 1996.
\newblock \href {https://arxiv.org/abs/astro-ph/9605094} {\path{arXiv:astro-ph/9605094}}, \href {https://doi.org/10.1103/PhysRevD.54.6040} {\path{doi:10.1103/PhysRevD.54.6040}}.

\bibitem{Ivanov:1997ia}
P.~Ivanov.
\newblock {Nonlinear metric perturbations and production of primordial black holes}.
\newblock {\em Phys. Rev. D}, 57:7145--7154, 1998.
\newblock \href {https://arxiv.org/abs/astro-ph/9708224} {\path{arXiv:astro-ph/9708224}}, \href {https://doi.org/10.1103/PhysRevD.57.7145} {\path{doi:10.1103/PhysRevD.57.7145}}.

\bibitem{Blinnikov:2016bxu}
S.~Blinnikov, A.~Dolgov, N.~K. Porayko, and K.~Postnov.
\newblock {Solving puzzles of GW150914 by primordial black holes}.
\newblock {\em JCAP}, 11:036, 2016.
\newblock \href {https://arxiv.org/abs/1611.00541} {\path{arXiv:1611.00541}}, \href {https://doi.org/10.1088/1475-7516/2016/11/036} {\path{doi:10.1088/1475-7516/2016/11/036}}.

\bibitem{Green:1997sz}
Anne~M. Green and Andrew~R. Liddle.
\newblock {Constraints on the density perturbation spectrum from primordial black holes}.
\newblock {\em Phys. Rev. D}, 56:6166--6174, 1997.
\newblock \href {https://arxiv.org/abs/astro-ph/9704251} {\path{arXiv:astro-ph/9704251}}, \href {https://doi.org/10.1103/PhysRevD.56.6166} {\path{doi:10.1103/PhysRevD.56.6166}}.

\bibitem{Green:1999xm}
Anne~M. Green and Andrew~R. Liddle.
\newblock {Critical collapse and the primordial black hole initial mass function}.
\newblock {\em Phys. Rev. D}, 60:063509, 1999.
\newblock \href {https://arxiv.org/abs/astro-ph/9901268} {\path{arXiv:astro-ph/9901268}}, \href {https://doi.org/10.1103/PhysRevD.60.063509} {\path{doi:10.1103/PhysRevD.60.063509}}.

\bibitem{Yokoyama:1998xd}
Jun'ichi Yokoyama.
\newblock {Cosmological constraints on primordial black holes produced in the near critical gravitational collapse}.
\newblock {\em Phys. Rev. D}, 58:107502, 1998.
\newblock \href {https://arxiv.org/abs/gr-qc/9804041} {\path{arXiv:gr-qc/9804041}}, \href {https://doi.org/10.1103/PhysRevD.58.107502} {\path{doi:10.1103/PhysRevD.58.107502}}.

\bibitem{Khlopov:1980mg}
M.~Yu. Khlopov and A.~G. Polnarev.
\newblock {Primordial black holes as a cosmological test of grand unification}.
\newblock {\em Phys. Lett. B}, 97:383--387, 1980.
\newblock \href {https://doi.org/10.1016/0370-2693(80)90624-3} {\path{doi:10.1016/0370-2693(80)90624-3}}.

\bibitem{Polnarev:1985btg}
A.~G. Polnarev and M.~Yu. Khlopov.
\newblock {COSMOLOGY, PRIMORDIAL BLACK HOLES, AND SUPERMASSIVE PARTICLES}.
\newblock {\em Sov. Phys. Usp.}, 28:213--232, 1985.
\newblock \href {https://doi.org/10.1070/PU1985v028n03ABEH003858} {\path{doi:10.1070/PU1985v028n03ABEH003858}}.

\bibitem{Green:1997pr}
Anne~M. Green, Andrew~R. Liddle, and Antonio Riotto.
\newblock {Primordial black hole constraints in cosmologies with early matter domination}.
\newblock {\em Phys. Rev. D}, 56:7559--7565, 1997.
\newblock \href {https://arxiv.org/abs/astro-ph/9705166} {\path{arXiv:astro-ph/9705166}}, \href {https://doi.org/10.1103/PhysRevD.56.7559} {\path{doi:10.1103/PhysRevD.56.7559}}.

\bibitem{Harada:2016mhb}
Tomohiro Harada, Chul-Moon Yoo, Kazunori Kohri, Ken-ichi Nakao, and Sanjay Jhingan.
\newblock {Primordial black hole formation in the matter-dominated phase of the Universe}.
\newblock {\em Astrophys. J.}, 833(1):61, 2016.
\newblock \href {https://arxiv.org/abs/1609.01588} {\path{arXiv:1609.01588}}, \href {https://doi.org/10.3847/1538-4357/833/1/61} {\path{doi:10.3847/1538-4357/833/1/61}}.

\bibitem{Alabidi:2013lya}
Laila Alabidi, Kazunori Kohri, Misao Sasaki, and Yuuiti Sendouda.
\newblock {Observable induced gravitational waves from an early matter phase}.
\newblock {\em JCAP}, 05:033, 2013.
\newblock \href {https://arxiv.org/abs/1303.4519} {\path{arXiv:1303.4519}}, \href {https://doi.org/10.1088/1475-7516/2013/05/033} {\path{doi:10.1088/1475-7516/2013/05/033}}.

\bibitem{Barrow:1996jk}
John~D. Barrow and Bernard~J. Carr.
\newblock {Formation and evaporation of primordial black holes in scalar - tensor gravity theories}.
\newblock {\em Phys. Rev. D}, 54:3920--3931, 1996.
\newblock \href {https://doi.org/10.1103/PhysRevD.54.3920} {\path{doi:10.1103/PhysRevD.54.3920}}.

\bibitem{Papanikolaou:2021uhe}
Theodoros Papanikolaou, Charalampos Tzerefos, Spyros Basilakos, and Emmanuel~N. Saridakis.
\newblock {Scalar induced gravitational waves from primordial black hole Poisson fluctuations in f(R) gravity}.
\newblock {\em JCAP}, 10:013, 2022.
\newblock \href {https://arxiv.org/abs/2112.15059} {\path{arXiv:2112.15059}}, \href {https://doi.org/10.1088/1475-7516/2022/10/013} {\path{doi:10.1088/1475-7516/2022/10/013}}.

\bibitem{Mann:2021mnc}
Robert~B. Mann, Sebastian Murk, and Daniel~R. Terno.
\newblock {Black holes and their horizons in semiclassical and modified theories of gravity}.
\newblock {\em Int. J. Mod. Phys. D}, 31(09):2230015, 2022.
\newblock \href {https://arxiv.org/abs/2112.06515} {\path{arXiv:2112.06515}}, \href {https://doi.org/10.1142/S0218271822300154} {\path{doi:10.1142/S0218271822300154}}.

\bibitem{Khlopov:1985fch}
M.~Yu. Khlopov, B.~A. Malomed, Ia.~B. Zeldovich, and Ya.~B. Zeldovich.
\newblock {Gravitational instability of scalar fields and formation of primordial black holes}.
\newblock {\em Mon. Not. Roy. Astron. Soc.}, 215(4):575--589, 1985.
\newblock \href {https://doi.org/10.1093/mnras/215.4.575} {\path{doi:10.1093/mnras/215.4.575}}.

\bibitem{Cotner:2016cvr}
Eric Cotner and Alexander Kusenko.
\newblock {Primordial black holes from supersymmetry in the early universe}.
\newblock {\em Phys. Rev. Lett.}, 119(3):031103, 2017.
\newblock \href {https://arxiv.org/abs/1612.02529} {\path{arXiv:1612.02529}}, \href {https://doi.org/10.1103/PhysRevLett.119.031103} {\path{doi:10.1103/PhysRevLett.119.031103}}.

\bibitem{Cotner:2019ykd}
Eric Cotner, Alexander Kusenko, Misao Sasaki, and Volodymyr Takhistov.
\newblock {Analytic Description of Primordial Black Hole Formation from Scalar Field Fragmentation}.
\newblock {\em JCAP}, 10:077, 2019.
\newblock \href {https://arxiv.org/abs/1907.10613} {\path{arXiv:1907.10613}}, \href {https://doi.org/10.1088/1475-7516/2019/10/077} {\path{doi:10.1088/1475-7516/2019/10/077}}.

\bibitem{Polnarev:1988dh}
Alexander Polnarev and Robert Zembowicz.
\newblock {Formation of Primordial Black Holes by Cosmic Strings}.
\newblock {\em Phys. Rev. D}, 43:1106--1109, 1991.
\newblock \href {https://doi.org/10.1103/PhysRevD.43.1106} {\path{doi:10.1103/PhysRevD.43.1106}}.

\bibitem{Hawking:1987bn}
S.~W. Hawking.
\newblock {Black Holes From Cosmic Strings}.
\newblock {\em Phys. Lett. B}, 231:237--239, 1989.
\newblock \href {https://doi.org/10.1016/0370-2693(89)90206-2} {\path{doi:10.1016/0370-2693(89)90206-2}}.

\bibitem{Garriga:1993gj}
Jaume Garriga and Maria Sakellariadou.
\newblock {Effects of friction on cosmic strings}.
\newblock {\em Phys. Rev. D}, 48:2502--2515, 1993.
\newblock \href {https://arxiv.org/abs/hep-th/9303024} {\path{arXiv:hep-th/9303024}}, \href {https://doi.org/10.1103/PhysRevD.48.2502} {\path{doi:10.1103/PhysRevD.48.2502}}.

\bibitem{Caldwell:1995fu}
R.~R. Caldwell and Paul Casper.
\newblock {Formation of black holes from collapsed cosmic string loops}.
\newblock {\em Phys. Rev. D}, 53:3002--3010, 1996.
\newblock \href {https://arxiv.org/abs/gr-qc/9509012} {\path{arXiv:gr-qc/9509012}}, \href {https://doi.org/10.1103/PhysRevD.53.3002} {\path{doi:10.1103/PhysRevD.53.3002}}.

\bibitem{MacGibbon:1997pu}
Jane~H. MacGibbon, Robert~H. Brandenberger, and U.~F. Wichoski.
\newblock {Limits on black hole formation from cosmic string loops}.
\newblock {\em Phys. Rev. D}, 57:2158--2165, 1998.
\newblock \href {https://arxiv.org/abs/astro-ph/9707146} {\path{arXiv:astro-ph/9707146}}, \href {https://doi.org/10.1103/PhysRevD.57.2158} {\path{doi:10.1103/PhysRevD.57.2158}}.

\bibitem{Helfer:2018qgv}
Thomas Helfer, Josu~C. Aurrekoetxea, and Eugene~A. Lim.
\newblock {Cosmic String Loop Collapse in Full General Relativity}.
\newblock {\em Phys. Rev. D}, 99(10):104028, 2019.
\newblock \href {https://arxiv.org/abs/1808.06678} {\path{arXiv:1808.06678}}, \href {https://doi.org/10.1103/PhysRevD.99.104028} {\path{doi:10.1103/PhysRevD.99.104028}}.

\bibitem{Ghoshal:2023fhh}
Anish Ghoshal and Alessandro Strumia.
\newblock {Probing the Dark Matter density with gravitational waves from super-massive binary black holes}.
\newblock {\em JCAP}, 02:054, 2024.
\newblock \href {https://arxiv.org/abs/2306.17158} {\path{arXiv:2306.17158}}, \href {https://doi.org/10.1088/1475-7516/2024/02/054} {\path{doi:10.1088/1475-7516/2024/02/054}}.

\bibitem{Rubin:2000dq}
S.~G. Rubin, M.~Yu. Khlopov, and A.~S. Sakharov.
\newblock {Primordial black holes from nonequilibrium second order phase transition}.
\newblock {\em Grav. Cosmol.}, 6:51--58, 2000.
\newblock \href {https://arxiv.org/abs/hep-ph/0005271} {\path{arXiv:hep-ph/0005271}}.

\bibitem{Rubin:2001yw}
Sergei~G. Rubin, Alexander~S. Sakharov, and Maxim~Yu. Khlopov.
\newblock {The Formation of primary galactic nuclei during phase transitions in the early universe}.
\newblock {\em J. Exp. Theor. Phys.}, 91:921--929, 2001.
\newblock \href {https://arxiv.org/abs/hep-ph/0106187} {\path{arXiv:hep-ph/0106187}}, \href {https://doi.org/10.1134/1.1385631} {\path{doi:10.1134/1.1385631}}.

\bibitem{Garriga:2015fdk}
Jaume Garriga, Alexander Vilenkin, and Jun Zhang.
\newblock {Black holes and the multiverse}.
\newblock {\em JCAP}, 02:064, 2016.
\newblock \href {https://arxiv.org/abs/1512.01819} {\path{arXiv:1512.01819}}, \href {https://doi.org/10.1088/1475-7516/2016/02/064} {\path{doi:10.1088/1475-7516/2016/02/064}}.

\bibitem{Deng:2016vzb}
Heling Deng, Jaume Garriga, and Alexander Vilenkin.
\newblock {Primordial black hole and wormhole formation by domain walls}.
\newblock {\em JCAP}, 04:050, 2017.
\newblock \href {https://arxiv.org/abs/1612.03753} {\path{arXiv:1612.03753}}, \href {https://doi.org/10.1088/1475-7516/2017/04/050} {\path{doi:10.1088/1475-7516/2017/04/050}}.

\bibitem{Liu:2019lul}
Jing Liu, Zong-Kuan Guo, and Rong-Gen Cai.
\newblock {Primordial Black Holes from Cosmic Domain Walls}.
\newblock {\em Phys. Rev. D}, 101(2):023513, 2020.
\newblock \href {https://arxiv.org/abs/1908.02662} {\path{arXiv:1908.02662}}, \href {https://doi.org/10.1103/PhysRevD.101.023513} {\path{doi:10.1103/PhysRevD.101.023513}}.

\bibitem{Kusenko:2020pcg}
Alexander Kusenko, Misao Sasaki, Sunao Sugiyama, Masahiro Takada, Volodymyr Takhistov, and Edoardo Vitagliano.
\newblock {Exploring Primordial Black Holes from the Multiverse with Optical Telescopes}.
\newblock {\em Phys. Rev. Lett.}, 125:181304, 2020.
\newblock \href {https://arxiv.org/abs/2001.09160} {\path{arXiv:2001.09160}}, \href {https://doi.org/10.1103/PhysRevLett.125.181304} {\path{doi:10.1103/PhysRevLett.125.181304}}.

\bibitem{Gouttenoire:2023gbn}
Yann Gouttenoire and Edoardo Vitagliano.
\newblock {Primordial black holes and wormholes from domain wall networks}.
\newblock {\em Phys. Rev. D}, 109(12):123507, 2024.
\newblock \href {https://arxiv.org/abs/2311.07670} {\path{arXiv:2311.07670}}, \href {https://doi.org/10.1103/PhysRevD.109.123507} {\path{doi:10.1103/PhysRevD.109.123507}}.

\bibitem{Crawford:1982yz}
Matt Crawford and David~N. Schramm.
\newblock {Spontaneous Generation of Density Perturbations in the Early Universe}.
\newblock {\em Nature}, 298:538--540, 1982.
\newblock \href {https://doi.org/10.1038/298538a0} {\path{doi:10.1038/298538a0}}.

\bibitem{Kodama:1982sf}
Hideo Kodama, Misao Sasaki, and Katsuhiko Sato.
\newblock {Abundance of Primordial Holes Produced by Cosmological First Order Phase Transition}.
\newblock {\em Prog. Theor. Phys.}, 68:1979, 1982.
\newblock \href {https://doi.org/10.1143/PTP.68.1979} {\path{doi:10.1143/PTP.68.1979}}.

\bibitem{Jedamzik:1996mr}
Karsten Jedamzik.
\newblock {Primordial black hole formation during the QCD epoch}.
\newblock {\em Phys. Rev. D}, 55:5871--5875, 1997.
\newblock \href {https://arxiv.org/abs/astro-ph/9605152} {\path{arXiv:astro-ph/9605152}}, \href {https://doi.org/10.1103/PhysRevD.55.R5871} {\path{doi:10.1103/PhysRevD.55.R5871}}.

\bibitem{Kanemura:2024pae}
Shinya Kanemura, Masanori Tanaka, and Ke-Pan Xie.
\newblock {Primordial black holes from slow phase transitions: a model-building perspective}.
\newblock {\em JHEP}, 06:036, 2024.
\newblock \href {https://arxiv.org/abs/2404.00646} {\path{arXiv:2404.00646}}, \href {https://doi.org/10.1007/JHEP06(2024)036} {\path{doi:10.1007/JHEP06(2024)036}}.

\bibitem{Lewicki:2024ghw}
Marek Lewicki, Piotr Toczek, and Ville Vaskonen.
\newblock {Black holes and gravitational waves from slow phase transitions}.
\newblock 2 2024.
\newblock \href {https://arxiv.org/abs/2402.04158} {\path{arXiv:2402.04158}}.

\bibitem{Hawking:1982ga}
S.~W. Hawking, I.~G. Moss, and J.~M. Stewart.
\newblock {Bubble Collisions in the Very Early Universe}.
\newblock {\em Phys. Rev. D}, 26:2681, 1982.
\newblock \href {https://doi.org/10.1103/PhysRevD.26.2681} {\path{doi:10.1103/PhysRevD.26.2681}}.

\bibitem{Moss:1994iq}
I.~G. Moss.
\newblock {Singularity formation from colliding bubbles}.
\newblock {\em Phys. Rev. D}, 50:676--681, 1994.
\newblock \href {https://doi.org/10.1103/PhysRevD.50.676} {\path{doi:10.1103/PhysRevD.50.676}}.

\bibitem{Kitajima:2020kig}
Naoya Kitajima and Fuminobu Takahashi.
\newblock {Primordial Black Holes from QCD Axion Bubbles}.
\newblock {\em JCAP}, 11:060, 2020.
\newblock \href {https://arxiv.org/abs/2006.13137} {\path{arXiv:2006.13137}}, \href {https://doi.org/10.1088/1475-7516/2020/11/060} {\path{doi:10.1088/1475-7516/2020/11/060}}.

\bibitem{Kasai:2023qic}
Kentaro Kasai, Masahiro Kawasaki, Naoya Kitajima, Kai Murai, Shunsuke Neda, and Fuminobu Takahashi.
\newblock {Primordial origin of supermassive black holes from axion bubbles}.
\newblock {\em JCAP}, 05:092, 2024.
\newblock \href {https://arxiv.org/abs/2310.13333} {\path{arXiv:2310.13333}}, \href {https://doi.org/10.1088/1475-7516/2024/05/092} {\path{doi:10.1088/1475-7516/2024/05/092}}.

\bibitem{Escriva:2023uko}
Albert Escriv\`a, Vicente Atal, and Jaume Garriga.
\newblock {Formation of trapped vacuum bubbles during inflation, and consequences for PBH scenarios}.
\newblock {\em JCAP}, 10:035, 2023.
\newblock \href {https://arxiv.org/abs/2306.09990} {\path{arXiv:2306.09990}}, \href {https://doi.org/10.1088/1475-7516/2023/10/035} {\path{doi:10.1088/1475-7516/2023/10/035}}.

\bibitem{Espinosa:2017sgp}
J.~R. Espinosa, D.~Racco, and A.~Riotto.
\newblock {Cosmological Signature of the Standard Model Higgs Vacuum Instability: Primordial Black Holes as Dark Matter}.
\newblock {\em Phys. Rev. Lett.}, 120(12):121301, 2018.
\newblock \href {https://arxiv.org/abs/1710.11196} {\path{arXiv:1710.11196}}, \href {https://doi.org/10.1103/PhysRevLett.120.121301} {\path{doi:10.1103/PhysRevLett.120.121301}}.

\bibitem{Espinosa:2018euj}
Jos\'e~Ram\'on Espinosa, Davide Racco, and Antonio Riotto.
\newblock {Primordial Black Holes from Higgs Vacuum Instability: Avoiding Fine-tuning through an Ultraviolet Safe Mechanism}.
\newblock {\em Eur. Phys. J. C}, 78(10):806, 2018.
\newblock \href {https://arxiv.org/abs/1804.07731} {\path{arXiv:1804.07731}}, \href {https://doi.org/10.1140/epjc/s10052-018-6274-2} {\path{doi:10.1140/epjc/s10052-018-6274-2}}.

\bibitem{Alonso-Monsalve:2023brx}
Elba Alonso-Monsalve and David~I. Kaiser.
\newblock {Primordial Black Holes with QCD Color Charge}.
\newblock {\em Phys. Rev. Lett.}, 132(23):231402, 2024.
\newblock \href {https://arxiv.org/abs/2310.16877} {\path{arXiv:2310.16877}}, \href {https://doi.org/10.1103/PhysRevLett.132.231402} {\path{doi:10.1103/PhysRevLett.132.231402}}.

\bibitem{Choptuik:1992jv}
Matthew~W. Choptuik.
\newblock {Universality and scaling in gravitational collapse of a massless scalar field}.
\newblock {\em Phys. Rev. Lett.}, 70:9--12, 1993.
\newblock \href {https://doi.org/10.1103/PhysRevLett.70.9} {\path{doi:10.1103/PhysRevLett.70.9}}.

\bibitem{Niemeyer:1997mt}
Jens~C. Niemeyer and K.~Jedamzik.
\newblock {Near-critical gravitational collapse and the initial mass function of primordial black holes}.
\newblock {\em Phys. Rev. Lett.}, 80:5481--5484, 1998.
\newblock \href {https://arxiv.org/abs/astro-ph/9709072} {\path{arXiv:astro-ph/9709072}}, \href {https://doi.org/10.1103/PhysRevLett.80.5481} {\path{doi:10.1103/PhysRevLett.80.5481}}.

\bibitem{Yokoyama:1998qw}
Jun'ichi Yokoyama.
\newblock {Formation of primordial black holes in the inflationary universe}.
\newblock {\em Phys. Rept.}, 307:133--139, 1998.
\newblock \href {https://doi.org/10.1016/S0370-1573(98)00044-1} {\path{doi:10.1016/S0370-1573(98)00044-1}}.

\bibitem{Niemeyer:1999ak}
Jens~C. Niemeyer and K.~Jedamzik.
\newblock {Dynamics of primordial black hole formation}.
\newblock {\em Phys. Rev. D}, 59:124013, 1999.
\newblock \href {https://arxiv.org/abs/astro-ph/9901292} {\path{arXiv:astro-ph/9901292}}, \href {https://doi.org/10.1103/PhysRevD.59.124013} {\path{doi:10.1103/PhysRevD.59.124013}}.

\bibitem{Shibata:1999zs}
Masaru Shibata and Misao Sasaki.
\newblock {Black hole formation in the Friedmann universe: Formulation and computation in numerical relativity}.
\newblock {\em Phys. Rev. D}, 60:084002, 1999.
\newblock \href {https://arxiv.org/abs/gr-qc/9905064} {\path{arXiv:gr-qc/9905064}}, \href {https://doi.org/10.1103/PhysRevD.60.084002} {\path{doi:10.1103/PhysRevD.60.084002}}.

\bibitem{Gundlach:1999cu}
Carsten Gundlach.
\newblock {Critical phenomena in gravitational collapse}.
\newblock {\em Living Rev. Rel.}, 2:4, 1999.
\newblock \href {https://arxiv.org/abs/gr-qc/0001046} {\path{arXiv:gr-qc/0001046}}.

\bibitem{Musco:2004ak}
Ilia Musco, John~C. Miller, and Luciano Rezzolla.
\newblock {Computations of primordial black hole formation}.
\newblock {\em Class. Quant. Grav.}, 22:1405--1424, 2005.
\newblock \href {https://arxiv.org/abs/gr-qc/0412063} {\path{arXiv:gr-qc/0412063}}, \href {https://doi.org/10.1088/0264-9381/22/7/013} {\path{doi:10.1088/0264-9381/22/7/013}}.

\bibitem{Polnarev:2006aa}
Alexander~G. Polnarev and Ilia Musco.
\newblock {Curvature profiles as initial conditions for primordial black hole formation}.
\newblock {\em Class. Quant. Grav.}, 24:1405--1432, 2007.
\newblock \href {https://arxiv.org/abs/gr-qc/0605122} {\path{arXiv:gr-qc/0605122}}, \href {https://doi.org/10.1088/0264-9381/24/6/003} {\path{doi:10.1088/0264-9381/24/6/003}}.

\bibitem{Musco:2008hv}
Ilia Musco, John~C. Miller, and Alexander~G. Polnarev.
\newblock {Primordial black hole formation in the radiative era: Investigation of the critical nature of the collapse}.
\newblock {\em Class. Quant. Grav.}, 26:235001, 2009.
\newblock \href {https://arxiv.org/abs/0811.1452} {\path{arXiv:0811.1452}}, \href {https://doi.org/10.1088/0264-9381/26/23/235001} {\path{doi:10.1088/0264-9381/26/23/235001}}.

\bibitem{Uehara:2024yyp}
Koichiro Uehara, Albert Escriv\`a, Tomohiro Harada, Daiki Saito, and Chul-Moon Yoo.
\newblock {Numerical simulation of type II primordial black hole formation}.
\newblock 1 2024.
\newblock \href {https://arxiv.org/abs/2401.06329} {\path{arXiv:2401.06329}}.

\bibitem{Ianniccari:2024ltb}
Andrea Ianniccari, Antonio~J. Iovino, Alex Kehagias, Davide Perrone, and Antonio Riotto.
\newblock {The Black Hole Formation -- Null Geodesic Correspondence}.
\newblock {\em Phys. Rev. Lett.}, 133(8):081401, 2024.
\newblock \href {https://arxiv.org/abs/2404.02801} {\path{arXiv:2404.02801}}, \href {https://doi.org/10.1103/PhysRevLett.133.081401} {\path{doi:10.1103/PhysRevLett.133.081401}}.

\bibitem{PhysRevLett.26.331}
B.~Carter.
\newblock Axisymmetric black hole has only two degrees of freedom.
\newblock {\em Phys. Rev. Lett.}, 26:331--333, Feb 1971.
\newblock URL: \url{https://link.aps.org/doi/10.1103/PhysRevLett.26.331}, \href {https://doi.org/10.1103/PhysRevLett.26.331} {\path{doi:10.1103/PhysRevLett.26.331}}.

\bibitem{PhysRev.164.1776}
Werner Israel.
\newblock Event horizons in static vacuum space-times.
\newblock {\em Phys. Rev.}, 164:1776--1779, Dec 1967.
\newblock URL: \url{https://link.aps.org/doi/10.1103/PhysRev.164.1776}, \href {https://doi.org/10.1103/PhysRev.164.1776} {\path{doi:10.1103/PhysRev.164.1776}}.

\bibitem{Banados:2017unc}
Eduardo Banados et~al.
\newblock {An 800-million-solar-mass black hole in a significantly neutral Universe at redshift 7.5}.
\newblock {\em Nature}, 553(7689):473--476, 2018.
\newblock \href {https://arxiv.org/abs/1712.01860} {\path{arXiv:1712.01860}}, \href {https://doi.org/10.1038/nature25180} {\path{doi:10.1038/nature25180}}.

\bibitem{Patruno:2006bw}
Alessandro Patruno, S.~F.~Portegies Zwart, J.~Dewi, and C.~Hopman.
\newblock {The ultraluminous x-ray source in m82: an intermediate mass black hole with a giant companion}.
\newblock {\em Mon. Not. Roy. Astron. Soc.}, 370:L6--L9, 2006.
\newblock \href {https://arxiv.org/abs/astro-ph/0602230} {\path{arXiv:astro-ph/0602230}}, \href {https://doi.org/10.1111/j.1745-3933.2006.00176.x} {\path{doi:10.1111/j.1745-3933.2006.00176.x}}.

\bibitem{Maccarone:2007dd}
Thomas~J. Maccarone, Arunav Kundu, Stephen~E. Zepf, and Katherine~L. Rhode.
\newblock {A black hole in a globular cluster}.
\newblock {\em Nature}, 445:183--185, 2007.
\newblock \href {https://arxiv.org/abs/astro-ph/0701310} {\path{arXiv:astro-ph/0701310}}, \href {https://doi.org/10.1038/nature05434} {\path{doi:10.1038/nature05434}}.

\bibitem{PhysRev.55.374}
J.~R. Oppenheimer and G.~M. Volkoff.
\newblock On massive neutron cores.
\newblock {\em Phys. Rev.}, 55:374--381, Feb 1939.
\newblock URL: \url{https://link.aps.org/doi/10.1103/PhysRev.55.374}, \href {https://doi.org/10.1103/PhysRev.55.374} {\path{doi:10.1103/PhysRev.55.374}}.

\bibitem{Gao:2015xle}
He~Gao, Bing Zhang, and Hou-Jun L\"u.
\newblock {Constraints on binary neutron star merger product from short GRB observations}.
\newblock {\em Phys. Rev. D}, 93(4):044065, 2016.
\newblock \href {https://arxiv.org/abs/1511.00753} {\path{arXiv:1511.00753}}, \href {https://doi.org/10.1103/PhysRevD.93.044065} {\path{doi:10.1103/PhysRevD.93.044065}}.

\bibitem{Shibata:2019ctb}
Masaru Shibata, Enping Zhou, Kenta Kiuchi, and Sho Fujibayashi.
\newblock {Constraint on the maximum mass of neutron stars using GW170817 event}.
\newblock {\em Phys. Rev. D}, 100(2):023015, 2019.
\newblock \href {https://arxiv.org/abs/1905.03656} {\path{arXiv:1905.03656}}, \href {https://doi.org/10.1103/PhysRevD.100.023015} {\path{doi:10.1103/PhysRevD.100.023015}}.

\bibitem{Bailyn:1997xt}
Charles~D. Bailyn, Raj~K. Jain, Paolo Coppi, and Jerome~A. Orosz.
\newblock {The Mass distribution of stellar black holes}.
\newblock {\em Astrophys. J.}, 499:367, 1998.
\newblock \href {https://arxiv.org/abs/astro-ph/9708032} {\path{arXiv:astro-ph/9708032}}, \href {https://doi.org/10.1086/305614} {\path{doi:10.1086/305614}}.

\bibitem{Thompson:2018ycv}
Todd~A. Thompson et~al.
\newblock {Discovery of a Candidate Black Hole - Giant Star Binary System in the Galactic Field}.
\newblock 6 2018.
\newblock \href {https://arxiv.org/abs/1806.02751} {\path{arXiv:1806.02751}}, \href {https://doi.org/10.1126/science.aau4005} {\path{doi:10.1126/science.aau4005}}.

\bibitem{LIGOScientific:2018jsj}
B.~P. Abbott et~al.
\newblock {Binary Black Hole Population Properties Inferred from the First and Second Observing Runs of Advanced LIGO and Advanced Virgo}.
\newblock {\em Astrophys. J. Lett.}, 882(2):L24, 2019.
\newblock \href {https://arxiv.org/abs/1811.12940} {\path{arXiv:1811.12940}}, \href {https://doi.org/10.3847/2041-8213/ab3800} {\path{doi:10.3847/2041-8213/ab3800}}.

\bibitem{LIGOScientific:2020iuh}
R.~Abbott et~al.
\newblock {GW190521: A Binary Black Hole Merger with a Total Mass of $150 M_{\odot}$}.
\newblock {\em Phys. Rev. Lett.}, 125(10):101102, 2020.
\newblock \href {https://arxiv.org/abs/2009.01075} {\path{arXiv:2009.01075}}, \href {https://doi.org/10.1103/PhysRevLett.125.101102} {\path{doi:10.1103/PhysRevLett.125.101102}}.

\bibitem{LIGOScientific:2020zkf}
R.~Abbott et~al.
\newblock {GW190814: Gravitational Waves from the Coalescence of a 23 Solar Mass Black Hole with a 2.6 Solar Mass Compact Object}.
\newblock {\em Astrophys. J. Lett.}, 896(2):L44, 2020.
\newblock \href {https://arxiv.org/abs/2006.12611} {\path{arXiv:2006.12611}}, \href {https://doi.org/10.3847/2041-8213/ab960f} {\path{doi:10.3847/2041-8213/ab960f}}.

\bibitem{LIGOScientific:2024elc}
{Observation of Gravitational Waves from the Coalescence of a $2.5-4.5~M_\odot$ Compact Object and a Neutron Star}.
\newblock 4 2024.
\newblock \href {https://arxiv.org/abs/2404.04248} {\path{arXiv:2404.04248}}.

\bibitem{Huang:2024wse}
Qing-Guo Huang, Chen Yuan, Zu-Cheng Chen, and Lang Liu.
\newblock {GW230529\_181500: a potential primordial binary black hole merger in the mass gap}.
\newblock {\em JCAP}, 08:030, 2024.
\newblock \href {https://arxiv.org/abs/2404.05691} {\path{arXiv:2404.05691}}, \href {https://doi.org/10.1088/1475-7516/2024/08/030} {\path{doi:10.1088/1475-7516/2024/08/030}}.

\bibitem{Dhani:2024jja}
Arnab Dhani, Sebastian V\"olkel, Alessandra Buonanno, Hector Estelles, Jonathan Gair, Harald~P. Pfeiffer, Lorenzo Pompili, and Alexandre Toubiana.
\newblock {Systematic Biases in Estimating the Properties of Black Holes Due to Inaccurate Gravitational-Wave Models}.
\newblock 4 2024.
\newblock \href {https://arxiv.org/abs/2404.05811} {\path{arXiv:2404.05811}}.

\bibitem{Fragione:2020han}
Giacomo Fragione, Abraham Loeb, and Frederic~A. Rasio.
\newblock {On the Origin of GW190521-like events from repeated black hole mergers in star clusters}.
\newblock {\em Astrophys. J. Lett.}, 902(1):L26, 2020.
\newblock \href {https://arxiv.org/abs/2009.05065} {\path{arXiv:2009.05065}}, \href {https://doi.org/10.3847/2041-8213/abbc0a} {\path{doi:10.3847/2041-8213/abbc0a}}.

\bibitem{Zevin:2020gbd}
Michael Zevin, Simone~S. Bavera, Christopher P.~L. Berry, Vicky Kalogera, Tassos Fragos, Pablo Marchant, Carl~L. Rodriguez, Fabio Antonini, Daniel~E. Holz, and Chris Pankow.
\newblock {One Channel to Rule Them All? Constraining the Origins of Binary Black Holes Using Multiple Formation Pathways}.
\newblock {\em Astrophys. J.}, 910(2):152, 2021.
\newblock \href {https://arxiv.org/abs/2011.10057} {\path{arXiv:2011.10057}}, \href {https://doi.org/10.3847/1538-4357/abe40e} {\path{doi:10.3847/1538-4357/abe40e}}.

\bibitem{Franciolini:2021tla}
Gabriele Franciolini, Vishal Baibhav, Valerio De~Luca, Ken K.~Y. Ng, Kaze W.~K. Wong, Emanuele Berti, Paolo Pani, Antonio Riotto, and Salvatore Vitale.
\newblock {Searching for a subpopulation of primordial black holes in LIGO-Virgo gravitational-wave data}.
\newblock {\em Phys. Rev. D}, 105(8):083526, 2022.
\newblock \href {https://arxiv.org/abs/2105.03349} {\path{arXiv:2105.03349}}, \href {https://doi.org/10.1103/PhysRevD.105.083526} {\path{doi:10.1103/PhysRevD.105.083526}}.

\bibitem{Deluca:2020sae}
V.~De~Luca, V.~Desjacques, G.~Franciolini, P.~Pani, and A.~Riotto.
\newblock {GW190521 Mass Gap Event and the Primordial Black Hole Scenario}.
\newblock {\em Phys. Rev. Lett.}, 126(5):051101, 2021.
\newblock \href {https://arxiv.org/abs/2009.01728} {\path{arXiv:2009.01728}}, \href {https://doi.org/10.1103/PhysRevLett.126.051101} {\path{doi:10.1103/PhysRevLett.126.051101}}.

\bibitem{McClintock:2013vwa}
Jeffrey~E. McClintock, Ramesh Narayan, and James~F. Steiner.
\newblock {Black Hole Spin via Continuum Fitting and the Role of Spin in Powering Transient Jets}.
\newblock {\em Space Sci. Rev.}, 183:295--322, 2014.
\newblock \href {https://arxiv.org/abs/1303.1583} {\path{arXiv:1303.1583}}, \href {https://doi.org/10.1007/s11214-013-0003-9} {\path{doi:10.1007/s11214-013-0003-9}}.

\bibitem{Pretorius:2007jn}
Frans Pretorius and Deepak Khurana.
\newblock {Black hole mergers and unstable circular orbits}.
\newblock {\em Class. Quant. Grav.}, 24:S83--S108, 2007.
\newblock \href {https://arxiv.org/abs/gr-qc/0702084} {\path{arXiv:gr-qc/0702084}}, \href {https://doi.org/10.1088/0264-9381/24/12/S07} {\path{doi:10.1088/0264-9381/24/12/S07}}.

\bibitem{Campanelli:2005dd}
Manuela Campanelli, C.~O. Lousto, P.~Marronetti, and Y.~Zlochower.
\newblock {Accurate evolutions of orbiting black-hole binaries without excision}.
\newblock {\em Phys. Rev. Lett.}, 96:111101, 2006.
\newblock \href {https://arxiv.org/abs/gr-qc/0511048} {\path{arXiv:gr-qc/0511048}}, \href {https://doi.org/10.1103/PhysRevLett.96.111101} {\path{doi:10.1103/PhysRevLett.96.111101}}.

\bibitem{Chiba:2017rvs}
Takeshi Chiba and Shuichiro Yokoyama.
\newblock {Spin Distribution of Primordial Black Holes}.
\newblock {\em PTEP}, 2017(8):083E01, 2017.
\newblock \href {https://arxiv.org/abs/1704.06573} {\path{arXiv:1704.06573}}, \href {https://doi.org/10.1093/ptep/ptx087} {\path{doi:10.1093/ptep/ptx087}}.

\bibitem{DeLuca:2019buf}
V.~De~Luca, V.~Desjacques, G.~Franciolini, A.~Malhotra, and A.~Riotto.
\newblock {The initial spin probability distribution of primordial black holes}.
\newblock {\em JCAP}, 05:018, 2019.
\newblock \href {https://arxiv.org/abs/1903.01179} {\path{arXiv:1903.01179}}, \href {https://doi.org/10.1088/1475-7516/2019/05/018} {\path{doi:10.1088/1475-7516/2019/05/018}}.

\bibitem{Mirbabayi:2019uph}
Mehrdad Mirbabayi, Andrei Gruzinov, and Jorge Nore\~na.
\newblock {Spin of Primordial Black Holes}.
\newblock {\em JCAP}, 03:017, 2020.
\newblock \href {https://arxiv.org/abs/1901.05963} {\path{arXiv:1901.05963}}, \href {https://doi.org/10.1088/1475-7516/2020/03/017} {\path{doi:10.1088/1475-7516/2020/03/017}}.

\bibitem{DeLuca:2020bjf}
V.~De~Luca, G.~Franciolini, P.~Pani, and A.~Riotto.
\newblock {The evolution of primordial black holes and their final observable spins}.
\newblock {\em JCAP}, 04:052, 2020.
\newblock \href {https://arxiv.org/abs/2003.02778} {\path{arXiv:2003.02778}}, \href {https://doi.org/10.1088/1475-7516/2020/04/052} {\path{doi:10.1088/1475-7516/2020/04/052}}.

\bibitem{LIGOScientific:2018mvr}
B.~P. Abbott et~al.
\newblock {GWTC-1: A Gravitational-Wave Transient Catalog of Compact Binary Mergers Observed by LIGO and Virgo during the First and Second Observing Runs}.
\newblock {\em Phys. Rev. X}, 9(3):031040, 2019.
\newblock \href {https://arxiv.org/abs/1811.12907} {\path{arXiv:1811.12907}}, \href {https://doi.org/10.1103/PhysRevX.9.031040} {\path{doi:10.1103/PhysRevX.9.031040}}.

\bibitem{Rodriguez:2016kxx}
Carl~L. Rodriguez, Sourav Chatterjee, and Frederic~A. Rasio.
\newblock {Binary Black Hole Mergers from Globular Clusters: Masses, Merger Rates, and the Impact of Stellar Evolution}.
\newblock {\em Phys. Rev. D}, 93(8):084029, 2016.
\newblock \href {https://arxiv.org/abs/1602.02444} {\path{arXiv:1602.02444}}, \href {https://doi.org/10.1103/PhysRevD.93.084029} {\path{doi:10.1103/PhysRevD.93.084029}}.

\bibitem{OLeary:2016ayz}
Ryan~M. O'Leary, Yohai Meiron, and Bence Kocsis.
\newblock {Dynamical formation signatures of black hole binaries in the first detected mergers by LIGO}.
\newblock {\em Astrophys. J. Lett.}, 824(1):L12, 2016.
\newblock \href {https://arxiv.org/abs/1602.02809} {\path{arXiv:1602.02809}}, \href {https://doi.org/10.3847/2041-8205/824/1/L12} {\path{doi:10.3847/2041-8205/824/1/L12}}.

\bibitem{Belczynski:2001uc}
Krzysztof Belczynski, Vassiliki Kalogera, and Tomasz Bulik.
\newblock {A Comprehensive study of binary compact objects as gravitational wave sources: Evolutionary channels, rates, and physical properties}.
\newblock {\em Astrophys. J.}, 572:407--431, 2001.
\newblock \href {https://arxiv.org/abs/astro-ph/0111452} {\path{arXiv:astro-ph/0111452}}, \href {https://doi.org/10.1086/340304} {\path{doi:10.1086/340304}}.

\bibitem{Marchant:2016wow}
Pablo Marchant, Norbert Langer, Philipp Podsiadlowski, Thomas~M. Tauris, and Takashi~J. Moriya.
\newblock {A new route towards merging massive black holes}.
\newblock {\em Astron. Astrophys.}, 588:A50, 2016.
\newblock \href {https://arxiv.org/abs/1601.03718} {\path{arXiv:1601.03718}}, \href {https://doi.org/10.1051/0004-6361/201628133} {\path{doi:10.1051/0004-6361/201628133}}.

\bibitem{Sasaki:2018dmp}
Misao Sasaki, Teruaki Suyama, Takahiro Tanaka, and Shuichiro Yokoyama.
\newblock {Primordial black holes\textemdash{}perspectives in gravitational wave astronomy}.
\newblock {\em Class. Quant. Grav.}, 35(6):063001, 2018.
\newblock \href {https://arxiv.org/abs/1801.05235} {\path{arXiv:1801.05235}}, \href {https://doi.org/10.1088/1361-6382/aaa7b4} {\path{doi:10.1088/1361-6382/aaa7b4}}.

\bibitem{Franciolini:2021nvv}
Gabriele Franciolini.
\newblock {\em {Primordial Black Holes: from Theory to Gravitational Wave Observations}}.
\newblock PhD thesis, Geneva U., Dept. Theor. Phys., 2021.
\newblock \href {https://arxiv.org/abs/2110.06815} {\path{arXiv:2110.06815}}, \href {https://doi.org/10.13097/archive-ouverte/unige:156136} {\path{doi:10.13097/archive-ouverte/unige:156136}}.

\bibitem{Nakamura:1997sm}
Takashi Nakamura, Misao Sasaki, Takahiro Tanaka, and Kip~S. Thorne.
\newblock {Gravitational waves from coalescing black hole MACHO binaries}.
\newblock {\em Astrophys. J. Lett.}, 487:L139--L142, 1997.
\newblock \href {https://arxiv.org/abs/astro-ph/9708060} {\path{arXiv:astro-ph/9708060}}, \href {https://doi.org/10.1086/310886} {\path{doi:10.1086/310886}}.

\bibitem{Ali-Haimoud:2017rtz}
Y.~Ali-Ha\"\i{}moud, E.~D. Kovetz, and M.~Kamionkowski.
\newblock {Merger rate of primordial black-hole binaries}.
\newblock {\em Phys. Rev. D}, 96(12):123523, 2017.
\newblock \href {https://doi.org/10.1103/PhysRevD.96.123523} {\path{doi:10.1103/PhysRevD.96.123523}}.

\bibitem{Bird:2016dcv}
Simeon Bird, Ilias Cholis, Julian~B. Muñoz, Yacine Ali-Haïmoud, Marc Kamionkowski, Ely~D. Kovetz, Alvise Raccanelli, and Adam~G. Riess.
\newblock {Did LIGO detect dark matter?}
\newblock {\em Phys. Rev. Lett.}, 116(20):201301, 2016.
\newblock \href {https://arxiv.org/abs/1603.00464} {\path{arXiv:1603.00464}}, \href {https://doi.org/10.1103/PhysRevLett.116.201301} {\path{doi:10.1103/PhysRevLett.116.201301}}.

\bibitem{Korol:2019jud}
Valeriya Korol, Ilya Mandel, M.~Coleman Miller, Ross~P. Church, and Melvyn~B. Davies.
\newblock {Merger rates in primordial black hole clusters without initial binaries}.
\newblock {\em Mon. Not. Roy. Astron. Soc.}, 496(1):994--1000, 2020.
\newblock \href {https://arxiv.org/abs/1911.03483} {\path{arXiv:1911.03483}}, \href {https://doi.org/10.1093/mnras/staa1644} {\path{doi:10.1093/mnras/staa1644}}.

\bibitem{Clesse:2016vqa}
Sebastien Clesse and Juan Garc\'\i{}a-Bellido.
\newblock {The clustering of massive Primordial Black Holes as Dark Matter: measuring their mass distribution with Advanced LIGO}.
\newblock {\em Phys. Dark Univ.}, 15:142--147, 2017.
\newblock \href {https://arxiv.org/abs/1603.05234} {\path{arXiv:1603.05234}}, \href {https://doi.org/10.1016/j.dark.2016.10.002} {\path{doi:10.1016/j.dark.2016.10.002}}.

\bibitem{Bagui:2021dqi}
Eleni Bagui and Sebastien Clesse.
\newblock {A boosted gravitational wave background for primordial black holes with broad mass distributions and thermal features}.
\newblock {\em Phys. Dark Univ.}, 38:101115, 2022.
\newblock \href {https://arxiv.org/abs/2110.07487} {\path{arXiv:2110.07487}}, \href {https://doi.org/10.1016/j.dark.2022.101115} {\path{doi:10.1016/j.dark.2022.101115}}.

\bibitem{Raidal:2017mfl}
Martti Raidal, Ville Vaskonen, and Hardi Veerm\"ae.
\newblock {Gravitational Waves from Primordial Black Hole Mergers}.
\newblock {\em JCAP}, 09:037, 2017.
\newblock \href {https://arxiv.org/abs/1707.01480} {\path{arXiv:1707.01480}}, \href {https://doi.org/10.1088/1475-7516/2017/09/037} {\path{doi:10.1088/1475-7516/2017/09/037}}.

\bibitem{DeLuca:2020qqa}
V.~De~Luca, G.~Franciolini, P.~Pani, and A.~Riotto.
\newblock {Primordial Black Holes Confront LIGO/Virgo data: Current situation}.
\newblock {\em JCAP}, 06:044, 2020.
\newblock \href {https://arxiv.org/abs/2005.05641} {\path{arXiv:2005.05641}}, \href {https://doi.org/10.1088/1475-7516/2020/06/044} {\path{doi:10.1088/1475-7516/2020/06/044}}.

\bibitem{Ioka:1998nz}
Kunihito Ioka, Takeshi Chiba, Takahiro Tanaka, and Takashi Nakamura.
\newblock {Black hole binary formation in the expanding universe: Three body problem approximation}.
\newblock {\em Phys. Rev. D}, 58:063003, 1998.
\newblock \href {https://arxiv.org/abs/astro-ph/9807018} {\path{arXiv:astro-ph/9807018}}, \href {https://doi.org/10.1103/PhysRevD.58.063003} {\path{doi:10.1103/PhysRevD.58.063003}}.

\bibitem{Mouri:2002mc}
Hideaki Mouri and Yoshiaki Taniguchi.
\newblock {Runaway merging of black holes: analytical constraint on the timescale}.
\newblock {\em Astrophys. J. Lett.}, 566:L17--L20, 2002.
\newblock \href {https://arxiv.org/abs/astro-ph/0201102} {\path{arXiv:astro-ph/0201102}}, \href {https://doi.org/10.1086/339472} {\path{doi:10.1086/339472}}.

\bibitem{Jedamzik:2020ypm}
Karsten Jedamzik.
\newblock {Primordial Black Hole Dark Matter and the LIGO/Virgo observations}.
\newblock {\em JCAP}, 09:022, 2020.
\newblock \href {https://arxiv.org/abs/2006.11172} {\path{arXiv:2006.11172}}, \href {https://doi.org/10.1088/1475-7516/2020/09/022} {\path{doi:10.1088/1475-7516/2020/09/022}}.

\bibitem{Binnington:2009bb}
Taylor Binnington and Eric Poisson.
\newblock {Relativistic theory of tidal Love numbers}.
\newblock {\em Phys. Rev. D}, 80:084018, 2009.
\newblock \href {https://arxiv.org/abs/0906.1366} {\path{arXiv:0906.1366}}, \href {https://doi.org/10.1103/PhysRevD.80.084018} {\path{doi:10.1103/PhysRevD.80.084018}}.

\bibitem{Damour:2009vw}
Thibault Damour and Alessandro Nagar.
\newblock {Relativistic tidal properties of neutron stars}.
\newblock {\em Phys. Rev. D}, 80:084035, 2009.
\newblock \href {https://arxiv.org/abs/0906.0096} {\path{arXiv:0906.0096}}, \href {https://doi.org/10.1103/PhysRevD.80.084035} {\path{doi:10.1103/PhysRevD.80.084035}}.

\bibitem{Chia:2020yla}
Horng~Sheng Chia.
\newblock {Tidal deformation and dissipation of rotating black holes}.
\newblock {\em Phys. Rev. D}, 104(2):024013, 2021.
\newblock \href {https://arxiv.org/abs/2010.07300} {\path{arXiv:2010.07300}}, \href {https://doi.org/10.1103/PhysRevD.104.024013} {\path{doi:10.1103/PhysRevD.104.024013}}.

\bibitem{Crescimbeni:2024cwh}
F.~Crescimbeni, G.~Franciolini, P.~Pani, and A.~Riotto.
\newblock {Primordial black holes or else? Tidal tests on subsolar mass gravitational-wave observations}.
\newblock 2 2024.
\newblock \href {https://arxiv.org/abs/2402.18656} {\path{arXiv:2402.18656}}.

\bibitem{Cardoso:2019rvt}
Vitor Cardoso and Paolo Pani.
\newblock {Testing the nature of dark compact objects: a status report}.
\newblock {\em Living Rev. Rel.}, 22(1):4, 2019.
\newblock \href {https://arxiv.org/abs/1904.05363} {\path{arXiv:1904.05363}}, \href {https://doi.org/10.1007/s41114-019-0020-4} {\path{doi:10.1007/s41114-019-0020-4}}.

\bibitem{Coleman:1985ki}
Sidney~R. Coleman.
\newblock {Q-balls}.
\newblock {\em Nucl. Phys. B}, 262(2):263, 1985.
\newblock [Addendum: Nucl.Phys.B 269, 744 (1986)].
\newblock \href {https://doi.org/10.1016/0550-3213(86)90520-1} {\path{doi:10.1016/0550-3213(86)90520-1}}.

\bibitem{Lee:1986tr}
T.~D. Lee and Y.~Pang.
\newblock {Fermion Soliton Stars and Black Holes}.
\newblock {\em Phys. Rev. D}, 35:3678, 1987.
\newblock \href {https://doi.org/10.1103/PhysRevD.35.3678} {\path{doi:10.1103/PhysRevD.35.3678}}.

\bibitem{DelGrosso:2023trq}
Loris Del~Grosso, Gabriele Franciolini, Paolo Pani, and Alfredo Urbano.
\newblock {Fermion soliton stars}.
\newblock {\em Phys. Rev. D}, 108(4):044024, 2023.
\newblock \href {https://arxiv.org/abs/2301.08709} {\path{arXiv:2301.08709}}, \href {https://doi.org/10.1103/PhysRevD.108.044024} {\path{doi:10.1103/PhysRevD.108.044024}}.

\bibitem{Liebling:2012fv}
Steven~L. Liebling and Carlos Palenzuela.
\newblock {Dynamical boson stars}.
\newblock {\em Living Rev. Rel.}, 26(1):1, 2023.
\newblock \href {https://arxiv.org/abs/1202.5809} {\path{arXiv:1202.5809}}, \href {https://doi.org/10.1007/s41114-023-00043-4} {\path{doi:10.1007/s41114-023-00043-4}}.

\bibitem{Barack:2018yly}
Leor Barack et~al.
\newblock {Black holes, gravitational waves and fundamental physics: a roadmap}.
\newblock {\em Class. Quant. Grav.}, 36(14):143001, 2019.
\newblock \href {https://arxiv.org/abs/1806.05195} {\path{arXiv:1806.05195}}, \href {https://doi.org/10.1088/1361-6382/ab0587} {\path{doi:10.1088/1361-6382/ab0587}}.

\bibitem{Volonteri:2010wz}
Marta Volonteri.
\newblock {Formation of Supermassive Black Holes}.
\newblock {\em Astron. Astrophys. Rev.}, 18:279--315, 2010.
\newblock \href {https://arxiv.org/abs/1003.4404} {\path{arXiv:1003.4404}}, \href {https://doi.org/10.1007/s00159-010-0029-x} {\path{doi:10.1007/s00159-010-0029-x}}.

\bibitem{Madau:2001sc}
Piero Madau and Martin~J. Rees.
\newblock {Massive black holes as Population III remnants}.
\newblock {\em Astrophys. J. Lett.}, 551:L27--L30, 2001.
\newblock \href {https://arxiv.org/abs/astro-ph/0101223} {\path{arXiv:astro-ph/0101223}}, \href {https://doi.org/10.1086/319848} {\path{doi:10.1086/319848}}.

\bibitem{Namikawa:2016edr}
Toshiya Namikawa, Atsushi Nishizawa, and Atsushi Taruya.
\newblock {Detecting Black-Hole Binary Clustering via the Second-Generation Gravitational-Wave Detectors}.
\newblock {\em Phys. Rev. D}, 94(2):024013, 2016.
\newblock \href {https://arxiv.org/abs/1603.08072} {\path{arXiv:1603.08072}}, \href {https://doi.org/10.1103/PhysRevD.94.024013} {\path{doi:10.1103/PhysRevD.94.024013}}.

\bibitem{Desjacques:2018wuu}
Vincent Desjacques and Antonio Riotto.
\newblock {Spatial clustering of primordial black holes}.
\newblock {\em Phys. Rev. D}, 98(12):123533, 2018.
\newblock \href {https://arxiv.org/abs/1806.10414} {\path{arXiv:1806.10414}}, \href {https://doi.org/10.1103/PhysRevD.98.123533} {\path{doi:10.1103/PhysRevD.98.123533}}.

\bibitem{Ballesteros:2018swv}
Guillermo Ballesteros, Pasquale~D. Serpico, and Marco Taoso.
\newblock {On the merger rate of primordial black holes: effects of nearest neighbours distribution and clustering}.
\newblock {\em JCAP}, 10:043, 2018.
\newblock \href {https://arxiv.org/abs/1807.02084} {\path{arXiv:1807.02084}}, \href {https://doi.org/10.1088/1475-7516/2018/10/043} {\path{doi:10.1088/1475-7516/2018/10/043}}.

\bibitem{MoradinezhadDizgah:2019wjf}
Azadeh Moradinezhad~Dizgah, Gabriele Franciolini, and Antonio Riotto.
\newblock {Primordial Black Holes from Broad Spectra: Abundance and Clustering}.
\newblock {\em JCAP}, 11:001, 2019.
\newblock \href {https://arxiv.org/abs/1906.08978} {\path{arXiv:1906.08978}}, \href {https://doi.org/10.1088/1475-7516/2019/11/001} {\path{doi:10.1088/1475-7516/2019/11/001}}.

\bibitem{DeLuca:2020jug}
V.~De~Luca, V.~Desjacques, G.~Franciolini, and A.~Riotto.
\newblock {The clustering evolution of primordial black holes}.
\newblock {\em JCAP}, 11:028, 2020.
\newblock \href {https://arxiv.org/abs/2009.04731} {\path{arXiv:2009.04731}}, \href {https://doi.org/10.1088/1475-7516/2020/11/028} {\path{doi:10.1088/1475-7516/2020/11/028}}.

\bibitem{DeLuca:2022uvz}
Valerio De~Luca, Gabriele Franciolini, Antonio Riotto, and Hardi Veerm\"ae.
\newblock {Ruling Out Initially Clustered Primordial Black Holes as Dark Matter}.
\newblock {\em Phys. Rev. Lett.}, 129(19):191302, 2022.
\newblock \href {https://arxiv.org/abs/2208.01683} {\path{arXiv:2208.01683}}, \href {https://doi.org/10.1103/PhysRevLett.129.191302} {\path{doi:10.1103/PhysRevLett.129.191302}}.

\bibitem{Tada:2015noa}
Yuichiro Tada and Shuichiro Yokoyama.
\newblock {Primordial black holes as biased tracers}.
\newblock {\em Phys. Rev. D}, 91(12):123534, 2015.
\newblock \href {https://arxiv.org/abs/1502.01124} {\path{arXiv:1502.01124}}, \href {https://doi.org/10.1103/PhysRevD.91.123534} {\path{doi:10.1103/PhysRevD.91.123534}}.

\bibitem{Young:2015kda}
Sam Young and Christian~T. Byrnes.
\newblock {Signatures of non-gaussianity in the isocurvature modes of primordial black hole dark matter}.
\newblock {\em JCAP}, 04:034, 2015.
\newblock \href {https://arxiv.org/abs/1503.01505} {\path{arXiv:1503.01505}}, \href {https://doi.org/10.1088/1475-7516/2015/04/034} {\path{doi:10.1088/1475-7516/2015/04/034}}.

\bibitem{DeLuca:2021hcf}
V.~De~Luca, G.~Franciolini, and A.~Riotto.
\newblock {Constraining the initial primordial black hole clustering with CMB distortion}.
\newblock {\em Phys. Rev. D}, 104(6):063526, 2021.
\newblock \href {https://arxiv.org/abs/2103.16369} {\path{arXiv:2103.16369}}, \href {https://doi.org/10.1103/PhysRevD.104.063526} {\path{doi:10.1103/PhysRevD.104.063526}}.

\bibitem{Raidal:2018bbj}
Martti Raidal, Christian Spethmann, Ville Vaskonen, and Hardi Veerm\"ae.
\newblock {Formation and Evolution of Primordial Black Hole Binaries in the Early Universe}.
\newblock {\em JCAP}, 02:018, 2019.
\newblock \href {https://arxiv.org/abs/1812.01930} {\path{arXiv:1812.01930}}, \href {https://doi.org/10.1088/1475-7516/2019/02/018} {\path{doi:10.1088/1475-7516/2019/02/018}}.

\bibitem{Ding:2019tjk}
Qianhang Ding, Tomohiro Nakama, Joseph Silk, and Yi~Wang.
\newblock {Detectability of Gravitational Waves from the Coalescence of Massive Primordial Black Holes with Initial Clustering}.
\newblock {\em Phys. Rev. D}, 100(10):103003, 2019.
\newblock \href {https://arxiv.org/abs/1903.07337} {\path{arXiv:1903.07337}}, \href {https://doi.org/10.1103/PhysRevD.100.103003} {\path{doi:10.1103/PhysRevD.100.103003}}.

\bibitem{1916SPAW.......688E}
Albert {Einstein}.
\newblock {N{\"a}herungsweise Integration der Feldgleichungen der Gravitation}.
\newblock {\em Sitzungsberichte der K\&ouml;niglich Preussischen Akademie der Wissenschaften}, pages 688--696, January 1916.

\bibitem{Hewish:1968bj}
A.~Hewish, S.~J. Bell, J.~D.~H Pilkington, P.~F. Scott, and R.~A. Collins.
\newblock {Observation of a rapidly pulsating radio source}.
\newblock {\em Nature}, 217:709--713, 1968.
\newblock \href {https://doi.org/10.1038/217709a0} {\path{doi:10.1038/217709a0}}.

\bibitem{Gold:1968zf}
T.~Gold.
\newblock {Rotating neutron stars as the origin of the pulsating radio sources}.
\newblock {\em Nature}, 218:731--732, 1968.
\newblock \href {https://doi.org/10.1038/218731a0} {\path{doi:10.1038/218731a0}}.

\bibitem{Hulse:1974eb}
R.~A. Hulse and J.~H. Taylor.
\newblock {Discovery of a pulsar in a binary system}.
\newblock {\em Astrophys. J. Lett.}, 195:L51--L53, 1975.
\newblock \href {https://doi.org/10.1086/181708} {\path{doi:10.1086/181708}}.

\bibitem{LIGOScientific:2016aoc}
B.~P. Abbott et~al.
\newblock {Observation of Gravitational Waves from a Binary Black Hole Merger}.
\newblock {\em Phys. Rev. Lett.}, 116(6):061102, 2016.
\newblock \href {https://arxiv.org/abs/1602.03837} {\path{arXiv:1602.03837}}, \href {https://doi.org/10.1103/PhysRevLett.116.061102} {\path{doi:10.1103/PhysRevLett.116.061102}}.

\bibitem{NANOGrav:2023gor}
Gabriella Agazie et~al.
\newblock {The NANOGrav 15 yr Data Set: Evidence for a Gravitational-wave Background}.
\newblock {\em Astrophys. J. Lett.}, 951(1):L8, 2023.
\newblock \href {https://arxiv.org/abs/2306.16213} {\path{arXiv:2306.16213}}, \href {https://doi.org/10.3847/2041-8213/acdac6} {\path{doi:10.3847/2041-8213/acdac6}}.

\bibitem{NANOGrav:2023hde}
Gabriella Agazie et~al.
\newblock {The NANOGrav 15 yr Data Set: Observations and Timing of 68 Millisecond Pulsars}.
\newblock {\em Astrophys. J. Lett.}, 951(1):L9, 2023.
\newblock \href {https://arxiv.org/abs/2306.16217} {\path{arXiv:2306.16217}}, \href {https://doi.org/10.3847/2041-8213/acda9a} {\path{doi:10.3847/2041-8213/acda9a}}.

\bibitem{EPTA:2023fyk}
J.~Antoniadis et~al.
\newblock {The second data release from the European Pulsar Timing Array - III. Search for gravitational wave signals}.
\newblock {\em Astron. Astrophys.}, 678:A50, 2023.
\newblock \href {https://arxiv.org/abs/2306.16214} {\path{arXiv:2306.16214}}, \href {https://doi.org/10.1051/0004-6361/202346844} {\path{doi:10.1051/0004-6361/202346844}}.

\bibitem{EPTA:2023sfo}
J.~Antoniadis et~al.
\newblock {The second data release from the European Pulsar Timing Array - I. The dataset and timing analysis}.
\newblock {\em Astron. Astrophys.}, 678:A48, 2023.
\newblock \href {https://arxiv.org/abs/2306.16224} {\path{arXiv:2306.16224}}, \href {https://doi.org/10.1051/0004-6361/202346841} {\path{doi:10.1051/0004-6361/202346841}}.

\bibitem{EPTA:2023xxk}
J.~Antoniadis et~al.
\newblock {The second data release from the European Pulsar Timing Array - IV. Implications for massive black holes, dark matter, and the early Universe}.
\newblock {\em Astron. Astrophys.}, 685:A94, 2024.
\newblock \href {https://arxiv.org/abs/2306.16227} {\path{arXiv:2306.16227}}, \href {https://doi.org/10.1051/0004-6361/202347433} {\path{doi:10.1051/0004-6361/202347433}}.

\bibitem{Reardon:2023gzh}
Daniel~J. Reardon et~al.
\newblock {Search for an Isotropic Gravitational-wave Background with the Parkes Pulsar Timing Array}.
\newblock {\em Astrophys. J. Lett.}, 951(1):L6, 2023.
\newblock \href {https://arxiv.org/abs/2306.16215} {\path{arXiv:2306.16215}}, \href {https://doi.org/10.3847/2041-8213/acdd02} {\path{doi:10.3847/2041-8213/acdd02}}.

\bibitem{Zic:2023gta}
Andrew Zic et~al.
\newblock {The Parkes Pulsar Timing Array third data release}.
\newblock {\em Publ. Astron. Soc. Austral.}, 40:e049, 2023.
\newblock \href {https://arxiv.org/abs/2306.16230} {\path{arXiv:2306.16230}}, \href {https://doi.org/10.1017/pasa.2023.36} {\path{doi:10.1017/pasa.2023.36}}.

\bibitem{Reardon:2023zen}
Daniel~J. Reardon et~al.
\newblock {The Gravitational-wave Background Null Hypothesis: Characterizing Noise in Millisecond Pulsar Arrival Times with the Parkes Pulsar Timing Array}.
\newblock {\em Astrophys. J. Lett.}, 951(1):L7, 2023.
\newblock \href {https://arxiv.org/abs/2306.16229} {\path{arXiv:2306.16229}}, \href {https://doi.org/10.3847/2041-8213/acdd03} {\path{doi:10.3847/2041-8213/acdd03}}.

\bibitem{Xu:2023wog}
Heng Xu et~al.
\newblock {Searching for the Nano-Hertz Stochastic Gravitational Wave Background with the Chinese Pulsar Timing Array Data Release I}.
\newblock {\em Res. Astron. Astrophys.}, 23(7):075024, 2023.
\newblock \href {https://arxiv.org/abs/2306.16216} {\path{arXiv:2306.16216}}, \href {https://doi.org/10.1088/1674-4527/acdfa5} {\path{doi:10.1088/1674-4527/acdfa5}}.

\bibitem{Zhao:2013bba}
Wen Zhao, Yang Zhang, Xiao-Peng You, and Zong-Hong Zhu.
\newblock {Constraints of relic gravitational waves by pulsar timing arrays: Forecasts for the FAST and SKA projects}.
\newblock {\em Phys. Rev. D}, 87(12):124012, 2013.
\newblock \href {https://arxiv.org/abs/1303.6718} {\path{arXiv:1303.6718}}, \href {https://doi.org/10.1103/PhysRevD.87.124012} {\path{doi:10.1103/PhysRevD.87.124012}}.

\bibitem{KAGRA:2021kbb}
R.~Abbott et~al.
\newblock {Upper limits on the isotropic gravitational-wave background from Advanced LIGO and Advanced Virgo\textquoteright{}s third observing run}.
\newblock {\em Phys. Rev. D}, 104(2):022004, 2021.
\newblock \href {https://arxiv.org/abs/2101.12130} {\path{arXiv:2101.12130}}, \href {https://doi.org/10.1103/PhysRevD.104.022004} {\path{doi:10.1103/PhysRevD.104.022004}}.

\bibitem{LIGOScientific:2014pky}
J.~Aasi et~al.
\newblock {Advanced LIGO}.
\newblock {\em Class. Quant. Grav.}, 32:074001, 2015.
\newblock \href {https://arxiv.org/abs/1411.4547} {\path{arXiv:1411.4547}}, \href {https://doi.org/10.1088/0264-9381/32/7/074001} {\path{doi:10.1088/0264-9381/32/7/074001}}.

\bibitem{LIGOScientific:2016fpe}
B.~P. Abbott et~al.
\newblock {GW150914: Implications for the stochastic gravitational wave background from binary black holes}.
\newblock {\em Phys. Rev. Lett.}, 116(13):131102, 2016.
\newblock \href {https://arxiv.org/abs/1602.03847} {\path{arXiv:1602.03847}}, \href {https://doi.org/10.1103/PhysRevLett.116.131102} {\path{doi:10.1103/PhysRevLett.116.131102}}.

\bibitem{LIGOScientific:2019vic}
B.~P. Abbott et~al.
\newblock {Search for the isotropic stochastic background using data from Advanced LIGO\textquoteright{}s second observing run}.
\newblock {\em Phys. Rev. D}, 100(6):061101, 2019.
\newblock \href {https://arxiv.org/abs/1903.02886} {\path{arXiv:1903.02886}}, \href {https://doi.org/10.1103/PhysRevD.100.061101} {\path{doi:10.1103/PhysRevD.100.061101}}.

\bibitem{LISA:2022kgy}
K.~G. Arun et~al.
\newblock {New horizons for fundamental physics with LISA}.
\newblock {\em Living Rev. Rel.}, 25(1):4, 2022.
\newblock \href {https://arxiv.org/abs/2205.01597} {\path{arXiv:2205.01597}}, \href {https://doi.org/10.1007/s41114-022-00036-9} {\path{doi:10.1007/s41114-022-00036-9}}.

\bibitem{Yagi:2011wg}
Kent Yagi and Naoki Seto.
\newblock {Detector configuration of DECIGO/BBO and identification of cosmological neutron-star binaries}.
\newblock {\em Phys. Rev. D}, 83:044011, 2011.
\newblock [Erratum: Phys.Rev.D 95, 109901 (2017)].
\newblock \href {https://arxiv.org/abs/1101.3940} {\path{arXiv:1101.3940}}, \href {https://doi.org/10.1103/PhysRevD.83.044011} {\path{doi:10.1103/PhysRevD.83.044011}}.

\bibitem{Yagi:2013awa}
Kent Yagi and Nicolas Yunes.
\newblock {I-Love-Q Relations in Neutron Stars and their Applications to Astrophysics, Gravitational Waves and Fundamental Physics}.
\newblock {\em Phys. Rev. D}, 88(2):023009, 2013.
\newblock \href {https://arxiv.org/abs/1303.1528} {\path{arXiv:1303.1528}}, \href {https://doi.org/10.1103/PhysRevD.88.023009} {\path{doi:10.1103/PhysRevD.88.023009}}.

\bibitem{Branchesi:2023mws}
Marica Branchesi et~al.
\newblock {Science with the Einstein Telescope: a comparison of different designs}.
\newblock {\em JCAP}, 07:068, 2023.
\newblock \href {https://arxiv.org/abs/2303.15923} {\path{arXiv:2303.15923}}, \href {https://doi.org/10.1088/1475-7516/2023/07/068} {\path{doi:10.1088/1475-7516/2023/07/068}}.

\bibitem{Aggarwal:2020olq}
Nancy Aggarwal et~al.
\newblock {Challenges and opportunities of gravitational-wave searches at MHz to GHz frequencies}.
\newblock {\em Living Rev. Rel.}, 24(1):4, 2021.
\newblock \href {https://arxiv.org/abs/2011.12414} {\path{arXiv:2011.12414}}, \href {https://doi.org/10.1007/s41114-021-00032-5} {\path{doi:10.1007/s41114-021-00032-5}}.

\bibitem{Detweiler:1979wn}
Steven~L. Detweiler.
\newblock {Pulsar timing measurements and the search for gravitational waves}.
\newblock {\em Astrophys. J.}, 234:1100--1104, 1979.
\newblock \href {https://doi.org/10.1086/157593} {\path{doi:10.1086/157593}}.

\bibitem{Hellings:1983fr}
R.~w. Hellings and G.~s. Downs.
\newblock {Upper limits on the isotropic gravitational radiation background from pulsar timing analysis}.
\newblock {\em Astrophys. J. Lett.}, 265:L39--L42, 1983.
\newblock \href {https://doi.org/10.1086/183954} {\path{doi:10.1086/183954}}.

\bibitem{Sesana:2004sp}
Alberto Sesana, Francesco Haardt, Piero Madau, and Marta Volonteri.
\newblock {Low - frequency gravitational radiation from coalescing massive black hole binaries in hierarchical cosmologies}.
\newblock {\em Astrophys. J.}, 611:623--632, 2004.
\newblock \href {https://arxiv.org/abs/astro-ph/0401543} {\path{arXiv:astro-ph/0401543}}, \href {https://doi.org/10.1086/422185} {\path{doi:10.1086/422185}}.

\bibitem{Abramovici:1992ah}
Alex Abramovici et~al.
\newblock {LIGO: The Laser interferometer gravitational wave observatory}.
\newblock {\em Science}, 256:325--333, 1992.
\newblock \href {https://doi.org/10.1126/science.256.5055.325} {\path{doi:10.1126/science.256.5055.325}}.

\bibitem{Giazotto:1988gw}
A.~Giazotto.
\newblock {The Virgo Project: A Wide Band Antenna for Gravitational Wave Detection}.
\newblock {\em Nucl. Instrum. Meth. A}, 289:518--525, 1990.
\newblock \href {https://doi.org/10.1016/0168-9002(90)91525-G} {\path{doi:10.1016/0168-9002(90)91525-G}}.

\bibitem{KAGRA:2018plz}
T.~Akutsu et~al.
\newblock {KAGRA: 2.5 Generation Interferometric Gravitational Wave Detector}.
\newblock {\em Nature Astron.}, 3(1):35--40, 2019.
\newblock \href {https://arxiv.org/abs/1811.08079} {\path{arXiv:1811.08079}}, \href {https://doi.org/10.1038/s41550-018-0658-y} {\path{doi:10.1038/s41550-018-0658-y}}.

\bibitem{LIGOScientific:2020ibl}
R.~Abbott et~al.
\newblock {GWTC-2: Compact Binary Coalescences Observed by LIGO and Virgo During the First Half of the Third Observing Run}.
\newblock {\em Phys. Rev. X}, 11:021053, 2021.
\newblock \href {https://arxiv.org/abs/2010.14527} {\path{arXiv:2010.14527}}, \href {https://doi.org/10.1103/PhysRevX.11.021053} {\path{doi:10.1103/PhysRevX.11.021053}}.

\bibitem{KAGRA:2021vkt}
R.~Abbott et~al.
\newblock {GWTC-3: Compact Binary Coalescences Observed by LIGO and Virgo during the Second Part of the Third Observing Run}.
\newblock {\em Phys. Rev. X}, 13(4):041039, 2023.
\newblock \href {https://arxiv.org/abs/2111.03606} {\path{arXiv:2111.03606}}, \href {https://doi.org/10.1103/PhysRevX.13.041039} {\path{doi:10.1103/PhysRevX.13.041039}}.

\bibitem{KAGRA:2021duu}
R.~Abbott et~al.
\newblock {Population of Merging Compact Binaries Inferred Using Gravitational Waves through GWTC-3}.
\newblock {\em Phys. Rev. X}, 13(1):011048, 2023.
\newblock \href {https://arxiv.org/abs/2111.03634} {\path{arXiv:2111.03634}}, \href {https://doi.org/10.1103/PhysRevX.13.011048} {\path{doi:10.1103/PhysRevX.13.011048}}.

\bibitem{Mandel:2018hfr}
Ilya Mandel and Alison Farmer.
\newblock {Merging stellar-mass binary black holes}.
\newblock {\em Phys. Rept.}, 955:1--24, 2022.
\newblock \href {https://arxiv.org/abs/1806.05820} {\path{arXiv:1806.05820}}, \href {https://doi.org/10.1016/j.physrep.2022.01.003} {\path{doi:10.1016/j.physrep.2022.01.003}}.

\bibitem{Woosley:2002zz}
S.~E. Woosley, A.~Heger, and T.~A. Weaver.
\newblock {The evolution and explosion of massive stars}.
\newblock {\em Rev. Mod. Phys.}, 74:1015--1071, 2002.
\newblock \href {https://doi.org/10.1103/RevModPhys.74.1015} {\path{doi:10.1103/RevModPhys.74.1015}}.

\bibitem{LIGOScientific:2018glc}
B.~P. Abbott et~al.
\newblock {Search for Subsolar-Mass Ultracompact Binaries in Advanced LIGO\textquoteright{}s First Observing Run}.
\newblock {\em Phys. Rev. Lett.}, 121(23):231103, 2018.
\newblock \href {https://arxiv.org/abs/1808.04771} {\path{arXiv:1808.04771}}, \href {https://doi.org/10.1103/PhysRevLett.121.231103} {\path{doi:10.1103/PhysRevLett.121.231103}}.

\bibitem{LIGOScientific:2019kan}
B.~P. Abbott et~al.
\newblock {Search for Subsolar Mass Ultracompact Binaries in Advanced LIGO\textquoteright{}s Second Observing Run}.
\newblock {\em Phys. Rev. Lett.}, 123(16):161102, 2019.
\newblock \href {https://arxiv.org/abs/1904.08976} {\path{arXiv:1904.08976}}, \href {https://doi.org/10.1103/PhysRevLett.123.161102} {\path{doi:10.1103/PhysRevLett.123.161102}}.

\bibitem{Nitz:2020bdb}
Alexander~Harvey Nitz and Yi-Fan Wang.
\newblock {Search for Gravitational Waves from High-Mass-Ratio Compact-Binary Mergers of Stellar Mass and Subsolar Mass Black Holes}.
\newblock {\em Phys. Rev. Lett.}, 126(2):021103, 2021.
\newblock \href {https://arxiv.org/abs/2007.03583} {\path{arXiv:2007.03583}}, \href {https://doi.org/10.1103/PhysRevLett.126.021103} {\path{doi:10.1103/PhysRevLett.126.021103}}.

\bibitem{Nitz:2021mzz}
Alexander~H. Nitz and Yi-Fan Wang.
\newblock {Search for gravitational waves from the coalescence of sub-solar mass and eccentric compact binaries}.
\newblock 2 2021.
\newblock \href {https://arxiv.org/abs/2102.00868} {\path{arXiv:2102.00868}}, \href {https://doi.org/10.3847/1538-4357/ac01d9} {\path{doi:10.3847/1538-4357/ac01d9}}.

\bibitem{Nitz:2021vqh}
Alexander~H. Nitz and Yi-Fan Wang.
\newblock {Search for Gravitational Waves from the Coalescence of Subsolar-Mass Binaries in the First Half of Advanced LIGO and Virgo\textquoteright{}s Third Observing Run}.
\newblock {\em Phys. Rev. Lett.}, 127(15):151101, 2021.
\newblock \href {https://arxiv.org/abs/2106.08979} {\path{arXiv:2106.08979}}, \href {https://doi.org/10.1103/PhysRevLett.127.151101} {\path{doi:10.1103/PhysRevLett.127.151101}}.

\bibitem{Nitz:2022ltl}
Alexander~H. Nitz and Yi-Fan Wang.
\newblock {Broad search for gravitational waves from subsolar-mass binaries through LIGO and Virgo\textquoteright{}s third observing run}.
\newblock {\em Phys. Rev. D}, 106(2):023024, 2022.
\newblock \href {https://arxiv.org/abs/2202.11024} {\path{arXiv:2202.11024}}, \href {https://doi.org/10.1103/PhysRevD.106.023024} {\path{doi:10.1103/PhysRevD.106.023024}}.

\bibitem{Miller:2020kmv}
Andrew~L. Miller, S\'ebastien Clesse, Federico De~Lillo, Giacomo Bruno, Antoine Depasse, and Andres Tanasijczuk.
\newblock {Probing planetary-mass primordial black holes with continuous gravitational waves}.
\newblock {\em Phys. Dark Univ.}, 32:100836, 2021.
\newblock \href {https://arxiv.org/abs/2012.12983} {\path{arXiv:2012.12983}}, \href {https://doi.org/10.1016/j.dark.2021.100836} {\path{doi:10.1016/j.dark.2021.100836}}.

\bibitem{Miller:2021knj}
Andrew~L. Miller, Nancy Aggarwal, S\'ebastien Clesse, and Federico De~Lillo.
\newblock {Constraints on planetary and asteroid-mass primordial black holes from continuous gravitational-wave searches}.
\newblock {\em Phys. Rev. D}, 105(6):062008, 2022.
\newblock \href {https://arxiv.org/abs/2110.06188} {\path{arXiv:2110.06188}}, \href {https://doi.org/10.1103/PhysRevD.105.062008} {\path{doi:10.1103/PhysRevD.105.062008}}.

\bibitem{Andres-Carcasona:2022prl}
M.~Andres-Carcasona, A.~Menendez-Vazquez, M.~Martinez, and Ll.~M. Mir.
\newblock {Searches for mass-asymmetric compact binary coalescence events using neural networks in the LIGO/Virgo third observation period}.
\newblock {\em Phys. Rev. D}, 107(8):082003, 2023.
\newblock \href {https://arxiv.org/abs/2212.02829} {\path{arXiv:2212.02829}}, \href {https://doi.org/10.1103/PhysRevD.107.082003} {\path{doi:10.1103/PhysRevD.107.082003}}.

\bibitem{Hutsi:2020sol}
Gert H\"utsi, Martti Raidal, Ville Vaskonen, and Hardi Veerm\"ae.
\newblock {Two populations of LIGO-Virgo black holes}.
\newblock {\em JCAP}, 03:068, 2021.
\newblock \href {https://arxiv.org/abs/2012.02786} {\path{arXiv:2012.02786}}, \href {https://doi.org/10.1088/1475-7516/2021/03/068} {\path{doi:10.1088/1475-7516/2021/03/068}}.

\bibitem{Hall:2020daa}
Alex Hall, Andrew~D. Gow, and Christian~T. Byrnes.
\newblock {Bayesian analysis of LIGO-Virgo mergers: Primordial vs. astrophysical black hole populations}.
\newblock {\em Phys. Rev. D}, 102:123524, 2020.
\newblock \href {https://arxiv.org/abs/2008.13704} {\path{arXiv:2008.13704}}, \href {https://doi.org/10.1103/PhysRevD.102.123524} {\path{doi:10.1103/PhysRevD.102.123524}}.

\bibitem{Wong:2020yig}
Kaze W.~K. Wong, Gabriele Franciolini, Valerio De~Luca, Vishal Baibhav, Emanuele Berti, Paolo Pani, and Antonio Riotto.
\newblock {Constraining the primordial black hole scenario with Bayesian inference and machine learning: the GWTC-2 gravitational wave catalog}.
\newblock {\em Phys. Rev. D}, 103(2):023026, 2021.
\newblock \href {https://arxiv.org/abs/2011.01865} {\path{arXiv:2011.01865}}, \href {https://doi.org/10.1103/PhysRevD.103.023026} {\path{doi:10.1103/PhysRevD.103.023026}}.

\bibitem{DeLuca:2021wjr}
V.~De~Luca, G.~Franciolini, P.~Pani, and A.~Riotto.
\newblock {Bayesian Evidence for Both Astrophysical and Primordial Black Holes: Mapping the GWTC-2 Catalog to Third-Generation Detectors}.
\newblock {\em JCAP}, 05:003, 2021.
\newblock \href {https://arxiv.org/abs/2102.03809} {\path{arXiv:2102.03809}}, \href {https://doi.org/10.1088/1475-7516/2021/05/003} {\path{doi:10.1088/1475-7516/2021/05/003}}.

\bibitem{Vaskonen:2019jpv}
Ville Vaskonen and Hardi Veerm\"ae.
\newblock {Lower bound on the primordial black hole merger rate}.
\newblock {\em Phys. Rev. D}, 101(4):043015, 2020.
\newblock \href {https://arxiv.org/abs/1908.09752} {\path{arXiv:1908.09752}}, \href {https://doi.org/10.1103/PhysRevD.101.043015} {\path{doi:10.1103/PhysRevD.101.043015}}.

\bibitem{Romero-Rodriguez:2021aws}
Alba Romero-Rodriguez, Mario Martinez, Oriol Pujol\`as, Mairi Sakellariadou, and Ville Vaskonen.
\newblock {Search for a Scalar Induced Stochastic Gravitational Wave Background in the Third LIGO-Virgo Observing Run}.
\newblock {\em Phys. Rev. Lett.}, 128(5):051301, 2022.
\newblock \href {https://arxiv.org/abs/2107.11660} {\path{arXiv:2107.11660}}, \href {https://doi.org/10.1103/PhysRevLett.128.051301} {\path{doi:10.1103/PhysRevLett.128.051301}}.

\bibitem{Franciolini:2022tfm}
Gabriele Franciolini, Ilia Musco, Paolo Pani, and Alfredo Urbano.
\newblock {From inflation to black hole mergers and back again: Gravitational-wave data-driven constraints on inflationary scenarios with a first-principle model of primordial black holes across the QCD epoch}.
\newblock {\em Phys. Rev. D}, 106(12):123526, 2022.
\newblock \href {https://arxiv.org/abs/2209.05959} {\path{arXiv:2209.05959}}, \href {https://doi.org/10.1103/PhysRevD.106.123526} {\path{doi:10.1103/PhysRevD.106.123526}}.

\bibitem{Punturo:2010zz}
M.~Punturo et~al.
\newblock {The Einstein Telescope: A third-generation gravitational wave observatory}.
\newblock {\em Class. Quant. Grav.}, 27:194002, 2010.
\newblock \href {https://doi.org/10.1088/0264-9381/27/19/194002} {\path{doi:10.1088/0264-9381/27/19/194002}}.

\bibitem{Hild:2010id}
S.~Hild et~al.
\newblock {Sensitivity Studies for Third-Generation Gravitational Wave Observatories}.
\newblock {\em Class. Quant. Grav.}, 28:094013, 2011.
\newblock \href {https://arxiv.org/abs/1012.0908} {\path{arXiv:1012.0908}}, \href {https://doi.org/10.1088/0264-9381/28/9/094013} {\path{doi:10.1088/0264-9381/28/9/094013}}.

\bibitem{Reitze:2019iox}
David Reitze et~al.
\newblock {Cosmic Explorer: The U.S. Contribution to Gravitational-Wave Astronomy beyond LIGO}.
\newblock {\em Bull. Am. Astron. Soc.}, 51(7):035, 2019.
\newblock \href {https://arxiv.org/abs/1907.04833} {\path{arXiv:1907.04833}}.

\bibitem{Weltman:2018zrl}
A.~Weltman et~al.
\newblock {Fundamental physics with the Square Kilometre Array}.
\newblock {\em Publ. Astron. Soc. Austral.}, 37:e002, 2020.
\newblock \href {https://arxiv.org/abs/1810.02680} {\path{arXiv:1810.02680}}, \href {https://doi.org/10.1017/pasa.2019.42} {\path{doi:10.1017/pasa.2019.42}}.

\bibitem{Bartolo:2016ami}
Nicola Bartolo et~al.
\newblock {Science with the space-based interferometer LISA. IV: Probing inflation with gravitational waves}.
\newblock {\em JCAP}, 12:026, 2016.
\newblock \href {https://arxiv.org/abs/1610.06481} {\path{arXiv:1610.06481}}, \href {https://doi.org/10.1088/1475-7516/2016/12/026} {\path{doi:10.1088/1475-7516/2016/12/026}}.

\bibitem{Caprini:2019pxz}
Chiara Caprini, Daniel~G. Figueroa, Raphael Flauger, Germano Nardini, Marco Peloso, Mauro Pieroni, Angelo Ricciardone, and Gianmassimo Tasinato.
\newblock {Reconstructing the spectral shape of a stochastic gravitational wave background with LISA}.
\newblock {\em JCAP}, 11:017, 2019.
\newblock \href {https://arxiv.org/abs/1906.09244} {\path{arXiv:1906.09244}}, \href {https://doi.org/10.1088/1475-7516/2019/11/017} {\path{doi:10.1088/1475-7516/2019/11/017}}.

\bibitem{LISACosmologyWorkingGroup:2022jok}
Pierre Auclair et~al.
\newblock {Cosmology with the Laser Interferometer Space Antenna}.
\newblock 4 2022.
\newblock \href {https://arxiv.org/abs/2204.05434} {\path{arXiv:2204.05434}}.

\bibitem{Kawamura:2019jqt}
Seiji Kawamura.
\newblock {Primordial gravitational wave and DECIGO}.
\newblock {\em PoS}, KMI2019:019, 2019.
\newblock \href {https://doi.org/10.22323/1.356.0019} {\path{doi:10.22323/1.356.0019}}.

\bibitem{TianQin:2015yph}
Jun Luo et~al.
\newblock {TianQin: a space-borne gravitational wave detector}.
\newblock {\em Class. Quant. Grav.}, 33(3):035010, 2016.
\newblock \href {https://arxiv.org/abs/1512.02076} {\path{arXiv:1512.02076}}, \href {https://doi.org/10.1088/0264-9381/33/3/035010} {\path{doi:10.1088/0264-9381/33/3/035010}}.

\bibitem{TianQin:2020hid}
Jianwei Mei et~al.
\newblock {The TianQin project: current progress on science and technology}.
\newblock {\em PTEP}, 2021(5):05A107, 2021.
\newblock \href {https://arxiv.org/abs/2008.10332} {\path{arXiv:2008.10332}}, \href {https://doi.org/10.1093/ptep/ptaa114} {\path{doi:10.1093/ptep/ptaa114}}.

\bibitem{AEDGE:2019nxb}
Yousef~Abou El-Neaj et~al.
\newblock {AEDGE: Atomic Experiment for Dark Matter and Gravity Exploration in Space}.
\newblock {\em EPJ Quant. Technol.}, 7:6, 2020.
\newblock \href {https://arxiv.org/abs/1908.00802} {\path{arXiv:1908.00802}}, \href {https://doi.org/10.1140/epjqt/s40507-020-0080-0} {\path{doi:10.1140/epjqt/s40507-020-0080-0}}.

\bibitem{Badurina:2021rgt}
Leonardo Badurina, Oliver Buchmueller, John Ellis, Marek Lewicki, Christopher McCabe, and Ville Vaskonen.
\newblock {Prospective sensitivities of atom interferometers to gravitational waves and ultralight dark matter}.
\newblock {\em Phil. Trans. A. Math. Phys. Eng. Sci.}, 380(2216):20210060, 2021.
\newblock \href {https://arxiv.org/abs/2108.02468} {\path{arXiv:2108.02468}}, \href {https://doi.org/10.1098/rsta.2021.0060} {\path{doi:10.1098/rsta.2021.0060}}.

\bibitem{Badurina:2019hst}
L.~Badurina et~al.
\newblock {AION: An Atom Interferometer Observatory and Network}.
\newblock {\em JCAP}, 05:011, 2020.
\newblock \href {https://arxiv.org/abs/1911.11755} {\path{arXiv:1911.11755}}, \href {https://doi.org/10.1088/1475-7516/2020/05/011} {\path{doi:10.1088/1475-7516/2020/05/011}}.

\bibitem{Harada:2015yda}
Tomohiro Harada, Chul-Moon Yoo, Tomohiro Nakama, and Yasutaka Koga.
\newblock {Cosmological long-wavelength solutions and primordial black hole formation}.
\newblock {\em Phys. Rev. D}, 91(8):084057, 2015.
\newblock \href {https://arxiv.org/abs/1503.03934} {\path{arXiv:1503.03934}}, \href {https://doi.org/10.1103/PhysRevD.91.084057} {\path{doi:10.1103/PhysRevD.91.084057}}.

\bibitem{DeLuca:2019qsy}
V.~De~Luca, G.~Franciolini, A.~Kehagias, M.~Peloso, A.~Riotto, and C.~\"Unal.
\newblock {The Ineludible non-Gaussianity of the Primordial Black Hole Abundance}.
\newblock {\em JCAP}, 07:048, 2019.
\newblock \href {https://arxiv.org/abs/1904.00970} {\path{arXiv:1904.00970}}, \href {https://doi.org/10.1088/1475-7516/2019/07/048} {\path{doi:10.1088/1475-7516/2019/07/048}}.

\bibitem{Young:2019yug}
Sam Young, Ilia Musco, and Christian~T. Byrnes.
\newblock {Primordial black hole formation and abundance: contribution from the non-linear relation between the density and curvature perturbation}.
\newblock {\em JCAP}, 11:012, 2019.
\newblock \href {https://arxiv.org/abs/1904.00984} {\path{arXiv:1904.00984}}, \href {https://doi.org/10.1088/1475-7516/2019/11/012} {\path{doi:10.1088/1475-7516/2019/11/012}}.

\bibitem{Germani:2019zez}
Cristiano Germani and Ravi~K. Sheth.
\newblock {Nonlinear statistics of primordial black holes from Gaussian curvature perturbations}.
\newblock {\em Phys. Rev. D}, 101(6):063520, 2020.
\newblock \href {https://arxiv.org/abs/1912.07072} {\path{arXiv:1912.07072}}, \href {https://doi.org/10.1103/PhysRevD.101.063520} {\path{doi:10.1103/PhysRevD.101.063520}}.

\bibitem{Motohashi:2017kbs}
Hayato Motohashi and Wayne Hu.
\newblock {Primordial Black Holes and Slow-Roll Violation}.
\newblock {\em Phys. Rev. D}, 96(6):063503, 2017.
\newblock \href {https://arxiv.org/abs/1706.06784} {\path{arXiv:1706.06784}}, \href {https://doi.org/10.1103/PhysRevD.96.063503} {\path{doi:10.1103/PhysRevD.96.063503}}.

\bibitem{Inomata:2016rbd}
Keisuke Inomata, Masahiro Kawasaki, Kyohei Mukaida, Yuichiro Tada, and Tsutomu~T. Yanagida.
\newblock {Inflationary primordial black holes for the LIGO gravitational wave events and pulsar timing array experiments}.
\newblock {\em Phys. Rev. D}, 95(12):123510, 2017.
\newblock \href {https://arxiv.org/abs/1611.06130} {\path{arXiv:1611.06130}}, \href {https://doi.org/10.1103/PhysRevD.95.123510} {\path{doi:10.1103/PhysRevD.95.123510}}.

\bibitem{Garcia-Bellido:2017mdw}
Juan Garcia-Bellido and Ester Ruiz~Morales.
\newblock {Primordial black holes from single field models of inflation}.
\newblock {\em Phys. Dark Univ.}, 18:47--54, 2017.
\newblock \href {https://arxiv.org/abs/1702.03901} {\path{arXiv:1702.03901}}, \href {https://doi.org/10.1016/j.dark.2017.09.007} {\path{doi:10.1016/j.dark.2017.09.007}}.

\bibitem{Ballesteros:2017fsr}
Guillermo Ballesteros and Marco Taoso.
\newblock {Primordial black hole dark matter from single field inflation}.
\newblock {\em Phys. Rev. D}, 97(2):023501, 2018.
\newblock \href {https://arxiv.org/abs/1709.05565} {\path{arXiv:1709.05565}}, \href {https://doi.org/10.1103/PhysRevD.97.023501} {\path{doi:10.1103/PhysRevD.97.023501}}.

\bibitem{Hertzberg:2017dkh}
Mark~P. Hertzberg and Masaki Yamada.
\newblock {Primordial Black Holes from Polynomial Potentials in Single Field Inflation}.
\newblock {\em Phys. Rev. D}, 97(8):083509, 2018.
\newblock \href {https://arxiv.org/abs/1712.09750} {\path{arXiv:1712.09750}}, \href {https://doi.org/10.1103/PhysRevD.97.083509} {\path{doi:10.1103/PhysRevD.97.083509}}.

\bibitem{Kannike:2017bxn}
Kristjan Kannike, Luca Marzola, Martti Raidal, and Hardi Veerm\"ae.
\newblock {Single Field Double Inflation and Primordial Black Holes}.
\newblock {\em JCAP}, 09:020, 2017.
\newblock \href {https://arxiv.org/abs/1705.06225} {\path{arXiv:1705.06225}}, \href {https://doi.org/10.1088/1475-7516/2017/09/020} {\path{doi:10.1088/1475-7516/2017/09/020}}.

\bibitem{Dalianis:2018frf}
Ioannis Dalianis, Alex Kehagias, and George Tringas.
\newblock {Primordial black holes from \ensuremath{\alpha}-attractors}.
\newblock {\em JCAP}, 01:037, 2019.
\newblock \href {https://arxiv.org/abs/1805.09483} {\path{arXiv:1805.09483}}, \href {https://doi.org/10.1088/1475-7516/2019/01/037} {\path{doi:10.1088/1475-7516/2019/01/037}}.

\bibitem{Inomata:2018cht}
Keisuke Inomata, Masahiro Kawasaki, Kyohei Mukaida, and Tsutomu~T. Yanagida.
\newblock {Double inflation as a single origin of primordial black holes for all dark matter and LIGO observations}.
\newblock {\em Phys. Rev. D}, 97(4):043514, 2018.
\newblock \href {https://arxiv.org/abs/1711.06129} {\path{arXiv:1711.06129}}, \href {https://doi.org/10.1103/PhysRevD.97.043514} {\path{doi:10.1103/PhysRevD.97.043514}}.

\bibitem{Cheong:2019vzl}
Dhong~Yeon Cheong, Sung~Mook Lee, and Seong~Chan Park.
\newblock {Primordial black holes in Higgs-$R^2$ inflation as the whole of dark matter}.
\newblock {\em JCAP}, 01:032, 2021.
\newblock \href {https://arxiv.org/abs/1912.12032} {\path{arXiv:1912.12032}}, \href {https://doi.org/10.1088/1475-7516/2021/01/032} {\path{doi:10.1088/1475-7516/2021/01/032}}.

\bibitem{Ballesteros:2020qam}
Guillermo Ballesteros, Juli\'an Rey, Marco Taoso, and Alfredo Urbano.
\newblock {Primordial black holes as dark matter and gravitational waves from single-field polynomial inflation}.
\newblock {\em JCAP}, 07:025, 2020.
\newblock \href {https://arxiv.org/abs/2001.08220} {\path{arXiv:2001.08220}}, \href {https://doi.org/10.1088/1475-7516/2020/07/025} {\path{doi:10.1088/1475-7516/2020/07/025}}.

\bibitem{Iacconi:2021ltm}
Laura Iacconi, Hooshyar Assadullahi, Matteo Fasiello, and David Wands.
\newblock {Revisiting small-scale fluctuations in \ensuremath{\alpha}-attractor models of inflation}.
\newblock {\em JCAP}, 06(06):007, 2022.
\newblock \href {https://arxiv.org/abs/2112.05092} {\path{arXiv:2112.05092}}, \href {https://doi.org/10.1088/1475-7516/2022/06/007} {\path{doi:10.1088/1475-7516/2022/06/007}}.

\bibitem{Kawai:2021edk}
Shinsuke Kawai and Jinsu Kim.
\newblock {Primordial black holes from Gauss-Bonnet-corrected single field inflation}.
\newblock {\em Phys. Rev. D}, 104(8):083545, 2021.
\newblock \href {https://arxiv.org/abs/2108.01340} {\path{arXiv:2108.01340}}, \href {https://doi.org/10.1103/PhysRevD.104.083545} {\path{doi:10.1103/PhysRevD.104.083545}}.

\bibitem{Lyth:2002my}
David~H. Lyth, Carlo Ungarelli, and David Wands.
\newblock {The Primordial density perturbation in the curvaton scenario}.
\newblock {\em Phys. Rev. D}, 67:023503, 2003.
\newblock \href {https://arxiv.org/abs/astro-ph/0208055} {\path{arXiv:astro-ph/0208055}}, \href {https://doi.org/10.1103/PhysRevD.67.023503} {\path{doi:10.1103/PhysRevD.67.023503}}.

\bibitem{Malik:2002jb}
Karim~A. Malik, David Wands, and Carlo Ungarelli.
\newblock {Large scale curvature and entropy perturbations for multiple interacting fluids}.
\newblock {\em Phys. Rev. D}, 67:063516, 2003.
\newblock \href {https://arxiv.org/abs/astro-ph/0211602} {\path{arXiv:astro-ph/0211602}}, \href {https://doi.org/10.1103/PhysRevD.67.063516} {\path{doi:10.1103/PhysRevD.67.063516}}.

\bibitem{Bugaev:2013vba}
E.~V. Bugaev and P.~A. Klimai.
\newblock {Primordial black hole constraints for curvaton models with predicted large non-Gaussianity}.
\newblock {\em Int. J. Mod. Phys. D}, 22:1350034, 2013.
\newblock \href {https://arxiv.org/abs/1303.3146} {\path{arXiv:1303.3146}}, \href {https://doi.org/10.1142/S021827181350034X} {\path{doi:10.1142/S021827181350034X}}.

\bibitem{Nakama:2016gzw}
Tomohiro Nakama, Joseph Silk, and Marc Kamionkowski.
\newblock {Stochastic gravitational waves associated with the formation of primordial black holes}.
\newblock {\em Phys. Rev. D}, 95(4):043511, 2017.
\newblock \href {https://arxiv.org/abs/1612.06264} {\path{arXiv:1612.06264}}, \href {https://doi.org/10.1103/PhysRevD.95.043511} {\path{doi:10.1103/PhysRevD.95.043511}}.

\bibitem{Byrnes:2012yx}
Christian~T. Byrnes, Edmund~J. Copeland, and Anne~M. Green.
\newblock {Primordial black holes as a tool for constraining non-Gaussianity}.
\newblock {\em Phys. Rev. D}, 86:043512, 2012.
\newblock \href {https://arxiv.org/abs/1206.4188} {\path{arXiv:1206.4188}}, \href {https://doi.org/10.1103/PhysRevD.86.043512} {\path{doi:10.1103/PhysRevD.86.043512}}.

\bibitem{Young:2013oia}
Sam Young and Christian~T. Byrnes.
\newblock {Primordial black holes in non-Gaussian regimes}.
\newblock {\em JCAP}, 08:052, 2013.
\newblock \href {https://arxiv.org/abs/1307.4995} {\path{arXiv:1307.4995}}, \href {https://doi.org/10.1088/1475-7516/2013/08/052} {\path{doi:10.1088/1475-7516/2013/08/052}}.

\bibitem{Yoo:2018kvb}
Chul-Moon Yoo, Tomohiro Harada, Jaume Garriga, and Kazunori Kohri.
\newblock {Primordial black hole abundance from random Gaussian curvature perturbations and a local density threshold}.
\newblock {\em PTEP}, 2018(12):123E01, 2018.
\newblock \href {https://arxiv.org/abs/1805.03946} {\path{arXiv:1805.03946}}, \href {https://doi.org/10.1093/ptep/pty120} {\path{doi:10.1093/ptep/pty120}}.

\bibitem{Kawasaki:2019mbl}
Masahiro Kawasaki and Hiromasa Nakatsuka.
\newblock {Effect of nonlinearity between density and curvature perturbations on the primordial black hole formation}.
\newblock {\em Phys. Rev. D}, 99(12):123501, 2019.
\newblock \href {https://arxiv.org/abs/1903.02994} {\path{arXiv:1903.02994}}, \href {https://doi.org/10.1103/PhysRevD.99.123501} {\path{doi:10.1103/PhysRevD.99.123501}}.

\bibitem{Yoo2}
Chul-Moon Yoo, Tomohiro Harada, Shin’ichi Hirano, and Kazunori Kohri.
\newblock Abundance of primordial black holes in peak theory for an arbitrary power spectrum.
\newblock {\em Progress of Theoretical and Experimental Physics}, 2021, 10 2020.
\newblock \href {https://doi.org/10.1093/ptep/ptaa155} {\path{doi:10.1093/ptep/ptaa155}}.

\bibitem{Riccardi:2021rlf}
Flavio Riccardi, Marco Taoso, and Alfredo Urbano.
\newblock {Solving peak theory in the presence of local non-gaussianities}.
\newblock {\em JCAP}, 08:060, 2021.
\newblock \href {https://arxiv.org/abs/2102.04084} {\path{arXiv:2102.04084}}, \href {https://doi.org/10.1088/1475-7516/2021/08/060} {\path{doi:10.1088/1475-7516/2021/08/060}}.

\bibitem{Taoso:2021uvl}
Marco Taoso and Alfredo Urbano.
\newblock {Non-gaussianities for primordial black hole formation}.
\newblock {\em JCAP}, 08:016, 2021.
\newblock \href {https://arxiv.org/abs/2102.03610} {\path{arXiv:2102.03610}}, \href {https://doi.org/10.1088/1475-7516/2021/08/016} {\path{doi:10.1088/1475-7516/2021/08/016}}.

\bibitem{Meng:2022ixx}
D.~S. Meng, C.~Yuan, and Q.~G. Huang.
\newblock {One-loop correction to the enhanced curvature perturbation with local-type non-Gaussianity for the formation of primordial black holes}.
\newblock {\em Phys. Rev. D}, 106(6):063508, 2022.
\newblock \href {https://doi.org/10.1103/PhysRevD.106.063508} {\path{doi:10.1103/PhysRevD.106.063508}}.

\bibitem{Escriva:2022pnz}
Albert Escriv\`a, Yuichiro Tada, Shuichiro Yokoyama, and Chul-Moon Yoo.
\newblock {Simulation of primordial black holes with large negative non-Gaussianity}.
\newblock {\em JCAP}, 05(05):012, 2022.
\newblock \href {https://arxiv.org/abs/2202.01028} {\path{arXiv:2202.01028}}, \href {https://doi.org/10.1088/1475-7516/2022/05/012} {\path{doi:10.1088/1475-7516/2022/05/012}}.

\bibitem{Atal:2019cdz}
Vicente Atal, Jaume Garriga, and Airam Marcos-Caballero.
\newblock {Primordial black hole formation with non-Gaussian curvature perturbations}.
\newblock {\em JCAP}, 09:073, 2019.
\newblock \href {https://arxiv.org/abs/1905.13202} {\path{arXiv:1905.13202}}, \href {https://doi.org/10.1088/1475-7516/2019/09/073} {\path{doi:10.1088/1475-7516/2019/09/073}}.

\bibitem{Tomberg:2023kli}
Eemeli Tomberg.
\newblock {Stochastic constant-roll inflation and primordial black holes}.
\newblock {\em Phys. Rev. D}, 108(4):043502, 2023.
\newblock \href {https://arxiv.org/abs/2304.10903} {\path{arXiv:2304.10903}}, \href {https://doi.org/10.1103/PhysRevD.108.043502} {\path{doi:10.1103/PhysRevD.108.043502}}.

\bibitem{Sasaki:2006kq}
Misao Sasaki, Jussi Valiviita, and David Wands.
\newblock {Non-Gaussianity of the primordial perturbation in the curvaton model}.
\newblock {\em Phys. Rev. D}, 74:103003, 2006.
\newblock \href {https://arxiv.org/abs/astro-ph/0607627} {\path{arXiv:astro-ph/0607627}}, \href {https://doi.org/10.1103/PhysRevD.74.103003} {\path{doi:10.1103/PhysRevD.74.103003}}.

\bibitem{Pi:2021dft}
Shi Pi and Misao Sasaki.
\newblock {Primordial black hole formation in nonminimal curvaton scenarios}.
\newblock {\em Phys. Rev. D}, 108(10):L101301, 2023.
\newblock \href {https://arxiv.org/abs/2112.12680} {\path{arXiv:2112.12680}}, \href {https://doi.org/10.1103/PhysRevD.108.L101301} {\path{doi:10.1103/PhysRevD.108.L101301}}.

\bibitem{Enqvist_2010}
Kari Enqvist, Sami Nurmi, Olli Taanila, and Tomo Takahashi.
\newblock Non-gaussian fingerprints of self-interacting curvaton.
\newblock {\em Journal of Cosmology and Astroparticle Physics}, 2010(04):009–009, April 2010.
\newblock URL: \url{http://dx.doi.org/10.1088/1475-7516/2010/04/009}, \href {https://doi.org/10.1088/1475-7516/2010/04/009} {\path{doi:10.1088/1475-7516/2010/04/009}}.

\bibitem{Fonseca:2011aa}
Jose Fonseca and David Wands.
\newblock {Non-Gaussianity and Gravitational Waves from Quadratic and Self-interacting Curvaton}.
\newblock {\em Phys. Rev. D}, 83:064025, 2011.
\newblock \href {https://arxiv.org/abs/1101.1254} {\path{arXiv:1101.1254}}, \href {https://doi.org/10.1103/PhysRevD.83.064025} {\path{doi:10.1103/PhysRevD.83.064025}}.

\bibitem{Bardeen:1985tr}
James~M. Bardeen, J.~R. Bond, Nick Kaiser, and A.~S. Szalay.
\newblock {The Statistics of Peaks of Gaussian Random Fields}.
\newblock {\em Astrophys. J.}, 304:15--61, 1986.
\newblock \href {https://doi.org/10.1086/164143} {\path{doi:10.1086/164143}}.

\bibitem{Musco:2020jjb}
Ilia Musco, Valerio De~Luca, Gabriele Franciolini, and Antonio Riotto.
\newblock {Threshold for primordial black holes. II. A simple analytic prescription}.
\newblock {\em Phys. Rev. D}, 103(6):063538, 2021.
\newblock \href {https://arxiv.org/abs/2011.03014} {\path{arXiv:2011.03014}}, \href {https://doi.org/10.1103/PhysRevD.103.063538} {\path{doi:10.1103/PhysRevD.103.063538}}.

\bibitem{Evans:1994pj}
Charles~R. Evans and Jason~S. Coleman.
\newblock {Observation of critical phenomena and selfsimilarity in the gravitational collapse of radiation fluid}.
\newblock {\em Phys. Rev. Lett.}, 72:1782--1785, 1994.
\newblock \href {https://arxiv.org/abs/gr-qc/9402041} {\path{arXiv:gr-qc/9402041}}, \href {https://doi.org/10.1103/PhysRevLett.72.1782} {\path{doi:10.1103/PhysRevLett.72.1782}}.

\bibitem{Musco:2012au}
Ilia Musco and John~C. Miller.
\newblock {Primordial black hole formation in the early universe: critical behaviour and self-similarity}.
\newblock {\em Class. Quant. Grav.}, 30:145009, 2013.
\newblock \href {https://arxiv.org/abs/1201.2379} {\path{arXiv:1201.2379}}, \href {https://doi.org/10.1088/0264-9381/30/14/145009} {\path{doi:10.1088/0264-9381/30/14/145009}}.

\bibitem{Musco:2023dak}
Ilia Musco, Karsten Jedamzik, and Sam Young.
\newblock {Primordial black hole formation during the QCD phase transition: Threshold, mass distribution, and abundance}.
\newblock {\em Phys. Rev. D}, 109(8):083506, 2024.
\newblock \href {https://arxiv.org/abs/2303.07980} {\path{arXiv:2303.07980}}, \href {https://doi.org/10.1103/PhysRevD.109.083506} {\path{doi:10.1103/PhysRevD.109.083506}}.

\bibitem{Yoo:2019pma}
Chul-Moon Yoo, Jinn-Ouk Gong, and Shuichiro Yokoyama.
\newblock {Abundance of primordial black holes with local non-Gaussianity in peak theory}.
\newblock {\em JCAP}, 09:033, 2019.
\newblock \href {https://arxiv.org/abs/1906.06790} {\path{arXiv:1906.06790}}, \href {https://doi.org/10.1088/1475-7516/2019/09/033} {\path{doi:10.1088/1475-7516/2019/09/033}}.

\bibitem{Young:2014ana}
Sam Young, Christian~T. Byrnes, and Misao Sasaki.
\newblock {Calculating the mass fraction of primordial black holes}.
\newblock {\em JCAP}, 07:045, 2014.
\newblock \href {https://arxiv.org/abs/1405.7023} {\path{arXiv:1405.7023}}, \href {https://doi.org/10.1088/1475-7516/2014/07/045} {\path{doi:10.1088/1475-7516/2014/07/045}}.

\bibitem{Franciolini:2018vbk}
G.~Franciolini, A.~Kehagias, S.~Matarrese, and A.~Riotto.
\newblock {Primordial Black Holes from Inflation and non-Gaussianity}.
\newblock {\em JCAP}, 03:016, 2018.
\newblock \href {https://arxiv.org/abs/1801.09415} {\path{arXiv:1801.09415}}, \href {https://doi.org/10.1088/1475-7516/2018/03/016} {\path{doi:10.1088/1475-7516/2018/03/016}}.

\bibitem{Biagetti:2021eep}
Matteo Biagetti, Valerio De~Luca, Gabriele Franciolini, Alex Kehagias, and Antonio Riotto.
\newblock {The formation probability of primordial black holes}.
\newblock {\em Phys. Lett. B}, 820:136602, 2021.
\newblock \href {https://arxiv.org/abs/2105.07810} {\path{arXiv:2105.07810}}, \href {https://doi.org/10.1016/j.physletb.2021.136602} {\path{doi:10.1016/j.physletb.2021.136602}}.

\bibitem{Kitajima:2021fpq}
Naoya Kitajima, Yuichiro Tada, Shuichiro Yokoyama, and Chul-Moon Yoo.
\newblock {Primordial black holes in peak theory with a non-Gaussian tail}.
\newblock {\em JCAP}, 10:053, 2021.
\newblock \href {https://arxiv.org/abs/2109.00791} {\path{arXiv:2109.00791}}, \href {https://doi.org/10.1088/1475-7516/2021/10/053} {\path{doi:10.1088/1475-7516/2021/10/053}}.

\bibitem{Musco:2018rwt}
Ilia Musco.
\newblock {Threshold for primordial black holes: Dependence on the shape of the cosmological perturbations}.
\newblock {\em Phys. Rev. D}, 100(12):123524, 2019.
\newblock \href {https://arxiv.org/abs/1809.02127} {\path{arXiv:1809.02127}}, \href {https://doi.org/10.1103/PhysRevD.100.123524} {\path{doi:10.1103/PhysRevD.100.123524}}.

\bibitem{Young:2022phe}
Sam Young.
\newblock {Peaks and primordial black holes: the~effect of non-Gaussianity}.
\newblock {\em JCAP}, 05(05):037, 2022.
\newblock \href {https://arxiv.org/abs/2201.13345} {\path{arXiv:2201.13345}}, \href {https://doi.org/10.1088/1475-7516/2022/05/037} {\path{doi:10.1088/1475-7516/2022/05/037}}.

\bibitem{Escriva:2019phb}
Albert Escriv\`a, Cristiano Germani, and Ravi~K. Sheth.
\newblock {Universal threshold for primordial black hole formation}.
\newblock {\em Phys. Rev. D}, 101(4):044022, 2020.
\newblock \href {https://arxiv.org/abs/1907.13311} {\path{arXiv:1907.13311}}, \href {https://doi.org/10.1103/PhysRevD.101.044022} {\path{doi:10.1103/PhysRevD.101.044022}}.

\bibitem{Germani:2018jgr}
Cristiano Germani and Ilia Musco.
\newblock {Abundance of Primordial Black Holes Depends on the Shape of the Inflationary Power Spectrum}.
\newblock {\em Phys. Rev. Lett.}, 122(14):141302, 2019.
\newblock \href {https://arxiv.org/abs/1805.04087} {\path{arXiv:1805.04087}}, \href {https://doi.org/10.1103/PhysRevLett.122.141302} {\path{doi:10.1103/PhysRevLett.122.141302}}.

\bibitem{Kehagias:2019eil}
Alex Kehagias, Ilia Musco, and Antonio Riotto.
\newblock {Non-Gaussian Formation of Primordial Black Holes: Effects on the Threshold}.
\newblock {\em JCAP}, 12:029, 2019.
\newblock \href {https://arxiv.org/abs/1906.07135} {\path{arXiv:1906.07135}}, \href {https://doi.org/10.1088/1475-7516/2019/12/029} {\path{doi:10.1088/1475-7516/2019/12/029}}.

\bibitem{DeLuca:2020agl}
V.~De~Luca, G.~Franciolini, and A.~Riotto.
\newblock {NANOGrav Data Hints at Primordial Black Holes as Dark Matter}.
\newblock {\em Phys. Rev. Lett.}, 126(4):041303, 2021.
\newblock \href {https://arxiv.org/abs/2009.08268} {\path{arXiv:2009.08268}}, \href {https://doi.org/10.1103/PhysRevLett.126.041303} {\path{doi:10.1103/PhysRevLett.126.041303}}.

\bibitem{Franciolini:2022pav}
Gabriele Franciolini and Alfredo Urbano.
\newblock {Primordial black hole dark matter from inflation: The reverse engineering approach}.
\newblock {\em Phys. Rev. D}, 106(12):123519, 2022.
\newblock \href {https://arxiv.org/abs/2207.10056} {\path{arXiv:2207.10056}}, \href {https://doi.org/10.1103/PhysRevD.106.123519} {\path{doi:10.1103/PhysRevD.106.123519}}.

\bibitem{Inomata:2020xad}
Keisuke Inomata, Masahiro Kawasaki, Kyohei Mukaida, and Tsutomu~T. Yanagida.
\newblock {NANOGrav Results and LIGO-Virgo Primordial Black Holes in Axionlike Curvaton Models}.
\newblock {\em Phys. Rev. Lett.}, 126(13):131301, 2021.
\newblock \href {https://arxiv.org/abs/2011.01270} {\path{arXiv:2011.01270}}, \href {https://doi.org/10.1103/PhysRevLett.126.131301} {\path{doi:10.1103/PhysRevLett.126.131301}}.

\bibitem{PhysRevD.81.104019}
B.~J. Carr, Kazunori Kohri, Yuuiti Sendouda, and Jun'ichi Yokoyama.
\newblock New cosmological constraints on primordial black holes.
\newblock {\em Phys. Rev. D}, 81:104019, May 2010.
\newblock URL: \url{https://link.aps.org/doi/10.1103/PhysRevD.81.104019}, \href {https://doi.org/10.1103/PhysRevD.81.104019} {\path{doi:10.1103/PhysRevD.81.104019}}.

\bibitem{DeLuca:2021hde}
Valerio De~Luca, Gabriele Franciolini, Paolo Pani, and Antonio Riotto.
\newblock {The minimum testable abundance of primordial black holes at future gravitational-wave detectors}.
\newblock {\em JCAP}, 11:039, 2021.
\newblock \href {https://arxiv.org/abs/2106.13769} {\path{arXiv:2106.13769}}, \href {https://doi.org/10.1088/1475-7516/2021/11/039} {\path{doi:10.1088/1475-7516/2021/11/039}}.

\bibitem{Ng:2022agi}
Ken K.~Y. Ng, Gabriele Franciolini, Emanuele Berti, Paolo Pani, Antonio Riotto, and Salvatore Vitale.
\newblock {Constraining High-redshift Stellar-mass Primordial Black Holes with Next-generation Ground-based Gravitational-wave Detectors}.
\newblock {\em Astrophys. J. Lett.}, 933(2):L41, 2022.
\newblock \href {https://arxiv.org/abs/2204.11864} {\path{arXiv:2204.11864}}, \href {https://doi.org/10.3847/2041-8213/ac7aae} {\path{doi:10.3847/2041-8213/ac7aae}}.

\bibitem{Martinelli:2022elq}
Matteo Martinelli, Francesca Scarcella, Natalie~B. Hogg, Bradley~J. Kavanagh, Daniele Gaggero, and Pierre Fleury.
\newblock {Dancing in the dark: detecting a population of distant primordial black holes}.
\newblock {\em JCAP}, 08(08):006, 2022.
\newblock \href {https://arxiv.org/abs/2205.02639} {\path{arXiv:2205.02639}}, \href {https://doi.org/10.1088/1475-7516/2022/08/006} {\path{doi:10.1088/1475-7516/2022/08/006}}.

\bibitem{Byrnes:2018clq}
Christian~T. Byrnes, Mark Hindmarsh, Sam Young, and Michael R.~S. Hawkins.
\newblock {Primordial black holes with an accurate QCD equation of state}.
\newblock {\em JCAP}, 08:041, 2018.
\newblock \href {https://arxiv.org/abs/1801.06138} {\path{arXiv:1801.06138}}, \href {https://doi.org/10.1088/1475-7516/2018/08/041} {\path{doi:10.1088/1475-7516/2018/08/041}}.

\bibitem{Pujolas:2021yaw}
Oriol Pujolas, Ville Vaskonen, and Hardi Veerm\"ae.
\newblock {Prospects for probing gravitational waves from primordial black hole binaries}.
\newblock {\em Phys. Rev. D}, 104(8):083521, 2021.
\newblock \href {https://arxiv.org/abs/2107.03379} {\path{arXiv:2107.03379}}, \href {https://doi.org/10.1103/PhysRevD.104.083521} {\path{doi:10.1103/PhysRevD.104.083521}}.

\bibitem{Gow:2022jfb}
Andrew~D. Gow, Hooshyar Assadullahi, Joseph H.~P. Jackson, Kazuya Koyama, Vincent Vennin, and David Wands.
\newblock {Non-perturbative non-Gaussianity and primordial black holes}.
\newblock {\em EPL}, 142(4):49001, 2023.
\newblock \href {https://arxiv.org/abs/2211.08348} {\path{arXiv:2211.08348}}, \href {https://doi.org/10.1209/0295-5075/acd417} {\path{doi:10.1209/0295-5075/acd417}}.

\bibitem{Hoffman:1985pu}
Y.~Hoffman and J.~Shaham.
\newblock {Local denisty maxima: Progenitors of structure}.
\newblock {\em Astrophys. J.}, 297:16--22, 1985.
\newblock \href {https://doi.org/10.1086/163498} {\path{doi:10.1086/163498}}.

\bibitem{Atal:2018neu}
Vicente Atal and Cristiano Germani.
\newblock {The role of non-gaussianities in Primordial Black Hole formation}.
\newblock {\em Phys. Dark Univ.}, 24:100275, 2019.
\newblock \href {https://arxiv.org/abs/1811.07857} {\path{arXiv:1811.07857}}, \href {https://doi.org/10.1016/j.dark.2019.100275} {\path{doi:10.1016/j.dark.2019.100275}}.

\bibitem{Biagetti:2018pjj}
Matteo Biagetti, Gabriele Franciolini, Alex Kehagias, and Antonio Riotto.
\newblock {Primordial Black Holes from Inflation and Quantum Diffusion}.
\newblock {\em JCAP}, 07:032, 2018.
\newblock \href {https://arxiv.org/abs/1804.07124} {\path{arXiv:1804.07124}}, \href {https://doi.org/10.1088/1475-7516/2018/07/032} {\path{doi:10.1088/1475-7516/2018/07/032}}.

\bibitem{Karam:2022nym}
Alexandros Karam, Niko Koivunen, Eemeli Tomberg, Ville Vaskonen, and Hardi Veerm\"ae.
\newblock {Anatomy of single-field inflationary models for primordial black holes}.
\newblock {\em JCAP}, 03:013, 2023.
\newblock \href {https://arxiv.org/abs/2205.13540} {\path{arXiv:2205.13540}}, \href {https://doi.org/10.1088/1475-7516/2023/03/013} {\path{doi:10.1088/1475-7516/2023/03/013}}.

\bibitem{Firouzjahi:2023xke}
Hassan Firouzjahi and Antonio Riotto.
\newblock {Sign of non-Gaussianity and the primordial black holes abundance}.
\newblock {\em Phys. Rev. D}, 108(12):123504, 2023.
\newblock \href {https://arxiv.org/abs/2309.10536} {\path{arXiv:2309.10536}}, \href {https://doi.org/10.1103/PhysRevD.108.123504} {\path{doi:10.1103/PhysRevD.108.123504}}.

\bibitem{Harada:2013epa}
Tomohiro Harada, Chul-Moon Yoo, and Kazunori Kohri.
\newblock {Threshold of primordial black hole formation}.
\newblock {\em Phys. Rev. D}, 88(8):084051, 2013.
\newblock [Erratum: Phys.Rev.D 89, 029903 (2014)].
\newblock \href {https://arxiv.org/abs/1309.4201} {\path{arXiv:1309.4201}}, \href {https://doi.org/10.1103/PhysRevD.88.084051} {\path{doi:10.1103/PhysRevD.88.084051}}.

\bibitem{Ali-Haimoud:2018dau}
Yacine Ali-Ha\"\i{}moud.
\newblock {Correlation Function of High-Threshold Regions and Application to the Initial Small-Scale Clustering of Primordial Black Holes}.
\newblock {\em Phys. Rev. Lett.}, 121(8):081304, 2018.
\newblock \href {https://arxiv.org/abs/1805.05912} {\path{arXiv:1805.05912}}, \href {https://doi.org/10.1103/PhysRevLett.121.081304} {\path{doi:10.1103/PhysRevLett.121.081304}}.

\bibitem{Bringmann:2018mxj}
Torsten Bringmann, Paul~Frederik Depta, Valerie Domcke, and Kai Schmidt-Hoberg.
\newblock {Towards closing the window of primordial black holes as dark matter: The case of large clustering}.
\newblock {\em Phys. Rev. D}, 99(6):063532, 2019.
\newblock \href {https://arxiv.org/abs/1808.05910} {\path{arXiv:1808.05910}}, \href {https://doi.org/10.1103/PhysRevD.99.063532} {\path{doi:10.1103/PhysRevD.99.063532}}.

\bibitem{Suyama:2019cst}
Teruaki Suyama and Shuichiro Yokoyama.
\newblock {Clustering of primordial black holes with non-Gaussian initial fluctuations}.
\newblock {\em PTEP}, 2019(10):103E02, 2019.
\newblock \href {https://arxiv.org/abs/1906.04958} {\path{arXiv:1906.04958}}, \href {https://doi.org/10.1093/ptep/ptz105} {\path{doi:10.1093/ptep/ptz105}}.

\bibitem{Balaji:2024hpu}
Shyam Balaji, Guillem Dom\`enech, Gabriele Franciolini, Alexander Ganz, and Jan Tr\"ankle.
\newblock {Probing modified Hawking evaporation with gravitational waves from the primordial black hole dominated universe}.
\newblock 3 2024.
\newblock \href {https://arxiv.org/abs/2403.14309} {\path{arXiv:2403.14309}}.

\bibitem{Arbey:2019vqx}
Alexandre Arbey, J\'er\'emy Auffinger, and Joseph Silk.
\newblock {Constraining primordial black hole masses with the isotropic gamma ray background}.
\newblock {\em Phys. Rev. D}, 101(2):023010, 2020.
\newblock \href {https://arxiv.org/abs/1906.04750} {\path{arXiv:1906.04750}}, \href {https://doi.org/10.1103/PhysRevD.101.023010} {\path{doi:10.1103/PhysRevD.101.023010}}.

\bibitem{Boudaud:2018hqb}
Mathieu Boudaud and Marco Cirelli.
\newblock {Voyager 1 $e^\pm$ Further Constrain Primordial Black Holes as Dark Matter}.
\newblock {\em Phys. Rev. Lett.}, 122(4):041104, 2019.
\newblock \href {https://arxiv.org/abs/1807.03075} {\path{arXiv:1807.03075}}, \href {https://doi.org/10.1103/PhysRevLett.122.041104} {\path{doi:10.1103/PhysRevLett.122.041104}}.

\bibitem{DeRocco:2019fjq}
William DeRocco and Peter~W. Graham.
\newblock {Constraining Primordial Black Hole Abundance with the Galactic 511 keV Line}.
\newblock {\em Phys. Rev. Lett.}, 123(25):251102, 2019.
\newblock \href {https://arxiv.org/abs/1906.07740} {\path{arXiv:1906.07740}}, \href {https://doi.org/10.1103/PhysRevLett.123.251102} {\path{doi:10.1103/PhysRevLett.123.251102}}.

\bibitem{Laha:2019ssq}
Ranjan Laha.
\newblock {Primordial Black Holes as a Dark Matter Candidate Are Severely Constrained by the Galactic Center 511 keV $\gamma$ -Ray Line}.
\newblock {\em Phys. Rev. Lett.}, 123(25):251101, 2019.
\newblock \href {https://arxiv.org/abs/1906.09994} {\path{arXiv:1906.09994}}, \href {https://doi.org/10.1103/PhysRevLett.123.251101} {\path{doi:10.1103/PhysRevLett.123.251101}}.

\bibitem{Laha:2020ivk}
Ranjan Laha, Julian~B. Mu\~noz, and Tracy~R. Slatyer.
\newblock {INTEGRAL constraints on primordial black holes and particle dark matter}.
\newblock {\em Phys. Rev. D}, 101(12):123514, 2020.
\newblock \href {https://arxiv.org/abs/2004.00627} {\path{arXiv:2004.00627}}, \href {https://doi.org/10.1103/PhysRevD.101.123514} {\path{doi:10.1103/PhysRevD.101.123514}}.

\bibitem{Ballesteros:2019exr}
Guillermo Ballesteros, Javier Coronado-Bl\'azquez, and Daniele Gaggero.
\newblock {X-ray and gamma-ray limits on the primordial black hole abundance from Hawking radiation}.
\newblock {\em Phys. Lett. B}, 808:135624, 2020.
\newblock \href {https://arxiv.org/abs/1906.10113} {\path{arXiv:1906.10113}}, \href {https://doi.org/10.1016/j.physletb.2020.135624} {\path{doi:10.1016/j.physletb.2020.135624}}.

\bibitem{Dasgupta:2019cae}
Basudeb Dasgupta, Ranjan Laha, and Anupam Ray.
\newblock {Neutrino and positron constraints on spinning primordial black hole dark matter}.
\newblock {\em Phys. Rev. Lett.}, 125(10):101101, 2020.
\newblock \href {https://arxiv.org/abs/1912.01014} {\path{arXiv:1912.01014}}, \href {https://doi.org/10.1103/PhysRevLett.125.101101} {\path{doi:10.1103/PhysRevLett.125.101101}}.

\bibitem{Inomata:2020lmk}
Keisuke Inomata, Masahiro Kawasaki, Kyohei Mukaida, Takahiro Terada, and Tsutomu~T. Yanagida.
\newblock {Gravitational Wave Production right after a Primordial Black Hole Evaporation}.
\newblock {\em Phys. Rev. D}, 101(12):123533, 2020.
\newblock \href {https://arxiv.org/abs/2003.10455} {\path{arXiv:2003.10455}}, \href {https://doi.org/10.1103/PhysRevD.101.123533} {\path{doi:10.1103/PhysRevD.101.123533}}.

\bibitem{Niikura:2017zjd}
Hiroko Niikura et~al.
\newblock {Microlensing constraints on primordial black holes with Subaru/HSC Andromeda observations}.
\newblock {\em Nature Astron.}, 3(6):524--534, 2019.
\newblock \href {https://arxiv.org/abs/1701.02151} {\path{arXiv:1701.02151}}, \href {https://doi.org/10.1038/s41550-019-0723-1} {\path{doi:10.1038/s41550-019-0723-1}}.

\bibitem{Smyth:2019whb}
Nolan Smyth, Stefano Profumo, Samuel English, Tesla Jeltema, Kevin McKinnon, and Puragra Guhathakurta.
\newblock {Updated Constraints on Asteroid-Mass Primordial Black Holes as Dark Matter}.
\newblock {\em Phys. Rev. D}, 101(6):063005, 2020.
\newblock \href {https://arxiv.org/abs/1910.01285} {\path{arXiv:1910.01285}}, \href {https://doi.org/10.1103/PhysRevD.101.063005} {\path{doi:10.1103/PhysRevD.101.063005}}.

\bibitem{Macho:2000nvd}
R.~A. Allsman et~al.
\newblock {MACHO project limits on black hole dark matter in the 1-30 solar mass range}.
\newblock {\em Astrophys. J. Lett.}, 550:L169, 2001.
\newblock \href {https://arxiv.org/abs/astro-ph/0011506} {\path{arXiv:astro-ph/0011506}}, \href {https://doi.org/10.1086/319636} {\path{doi:10.1086/319636}}.

\bibitem{Oguri:2017ock}
Masamune Oguri, Jose~M. Diego, Nick Kaiser, Patrick~L. Kelly, and Tom Broadhurst.
\newblock {Understanding caustic crossings in giant arcs: characteristic scales, event rates, and constraints on compact dark matter}.
\newblock {\em Phys. Rev. D}, 97(2):023518, 2018.
\newblock \href {https://arxiv.org/abs/1710.00148} {\path{arXiv:1710.00148}}, \href {https://doi.org/10.1103/PhysRevD.97.023518} {\path{doi:10.1103/PhysRevD.97.023518}}.

\bibitem{Kawai:2024bni}
Hiroki Kawai and Masamune Oguri.
\newblock {Constraints on primordial black holes from the observed number of Icarus-like ultrahigh magnification events}.
\newblock 11 2024.
\newblock \href {https://arxiv.org/abs/2411.13816} {\path{arXiv:2411.13816}}.

\bibitem{Mroz:2024wag}
Przemek Mr\'oz et~al.
\newblock {Microlensing Optical Depth and Event Rate toward the Large Magellanic Cloud Based on 20 yr of OGLE Observations}.
\newblock {\em Astrophys. J. Suppl.}, 273(1):4, 2024.
\newblock \href {https://arxiv.org/abs/2403.02398} {\path{arXiv:2403.02398}}, \href {https://doi.org/10.3847/1538-4365/ad452e} {\path{doi:10.3847/1538-4365/ad452e}}.

\bibitem{Franciolini:2022htd}
Gabriele Franciolini, Anshuman Maharana, and Francesco Muia.
\newblock {Hunt for light primordial black hole dark matter with ultrahigh-frequency gravitational waves}.
\newblock {\em Phys. Rev. D}, 106(10):103520, 2022.
\newblock \href {https://arxiv.org/abs/2205.02153} {\path{arXiv:2205.02153}}, \href {https://doi.org/10.1103/PhysRevD.106.103520} {\path{doi:10.1103/PhysRevD.106.103520}}.

\bibitem{Franciolini:2023opt}
Gabriele Franciolini, Francesco Iacovelli, Michele Mancarella, Michele Maggiore, Paolo Pani, and Antonio Riotto.
\newblock {Searching for primordial black holes with the Einstein Telescope: Impact of design and systematics}.
\newblock {\em Phys. Rev. D}, 108(4):043506, 2023.
\newblock \href {https://arxiv.org/abs/2304.03160} {\path{arXiv:2304.03160}}, \href {https://doi.org/10.1103/PhysRevD.108.043506} {\path{doi:10.1103/PhysRevD.108.043506}}.

\bibitem{Ali-Haimoud:2016mbv}
Yacine Ali-Ha\"\i{}moud and Marc Kamionkowski.
\newblock {Cosmic microwave background limits on accreting primordial black holes}.
\newblock {\em Phys. Rev. D}, 95(4):043534, 2017.
\newblock \href {https://arxiv.org/abs/1612.05644} {\path{arXiv:1612.05644}}, \href {https://doi.org/10.1103/PhysRevD.95.043534} {\path{doi:10.1103/PhysRevD.95.043534}}.

\bibitem{Poulin:2017bwe}
Vivian Poulin, Pasquale~D. Serpico, Francesca Calore, Sebastien Clesse, and Kazunori Kohri.
\newblock {CMB bounds on disk-accreting massive primordial black holes}.
\newblock {\em Phys. Rev. D}, 96(8):083524, 2017.
\newblock \href {https://arxiv.org/abs/1707.04206} {\path{arXiv:1707.04206}}, \href {https://doi.org/10.1103/PhysRevD.96.083524} {\path{doi:10.1103/PhysRevD.96.083524}}.

\bibitem{Serpico:2020ehh}
Pasquale~D. Serpico, Vivian Poulin, Derek Inman, and Kazunori Kohri.
\newblock {Cosmic microwave background bounds on primordial black holes including dark matter halo accretion}.
\newblock {\em Phys. Rev. Res.}, 2(2):023204, 2020.
\newblock \href {https://arxiv.org/abs/2002.10771} {\path{arXiv:2002.10771}}, \href {https://doi.org/10.1103/PhysRevResearch.2.023204} {\path{doi:10.1103/PhysRevResearch.2.023204}}.

\bibitem{Inman:2019wvr}
Derek Inman and Yacine Ali-Ha\"\i{}moud.
\newblock {Early structure formation in primordial black hole cosmologies}.
\newblock {\em Phys. Rev. D}, 100(8):083528, 2019.
\newblock \href {https://arxiv.org/abs/1907.08129} {\path{arXiv:1907.08129}}, \href {https://doi.org/10.1103/PhysRevD.100.083528} {\path{doi:10.1103/PhysRevD.100.083528}}.

\bibitem{Hutsi:2019hlw}
Gert H\"utsi, Martti Raidal, and Hardi Veerm\"ae.
\newblock {Small-scale structure of primordial black hole dark matter and its implications for accretion}.
\newblock {\em Phys. Rev. D}, 100(8):083016, 2019.
\newblock \href {https://arxiv.org/abs/1907.06533} {\path{arXiv:1907.06533}}, \href {https://doi.org/10.1103/PhysRevD.100.083016} {\path{doi:10.1103/PhysRevD.100.083016}}.

\bibitem{Manshanden:2018tze}
Julien Manshanden, Daniele Gaggero, Gianfranco Bertone, Riley M.~T. Connors, and Massimo Ricotti.
\newblock {Multi-wavelength astronomical searches for primordial black holes}.
\newblock {\em JCAP}, 06:026, 2019.
\newblock \href {https://arxiv.org/abs/1812.07967} {\path{arXiv:1812.07967}}, \href {https://doi.org/10.1088/1475-7516/2019/06/026} {\path{doi:10.1088/1475-7516/2019/06/026}}.

\bibitem{Hektor:2018qqw}
Andi Hektor, Gert H\"utsi, Luca Marzola, Martti Raidal, Ville Vaskonen, and Hardi Veerm\"ae.
\newblock {Constraining Primordial Black Holes with the EDGES 21-cm Absorption Signal}.
\newblock {\em Phys. Rev. D}, 98(2):023503, 2018.
\newblock \href {https://arxiv.org/abs/1803.09697} {\path{arXiv:1803.09697}}, \href {https://doi.org/10.1103/PhysRevD.98.023503} {\path{doi:10.1103/PhysRevD.98.023503}}.

\bibitem{Inoue:2017csr}
Yoshiyuki Inoue and Alexander Kusenko.
\newblock {New X-ray bound on density of primordial black holes}.
\newblock {\em JCAP}, 10:034, 2017.
\newblock \href {https://arxiv.org/abs/1705.00791} {\path{arXiv:1705.00791}}, \href {https://doi.org/10.1088/1475-7516/2017/10/034} {\path{doi:10.1088/1475-7516/2017/10/034}}.

\bibitem{Lu:2020bmd}
Philip Lu, Volodymyr Takhistov, Graciela~B. Gelmini, Kohei Hayashi, Yoshiyuki Inoue, and Alexander Kusenko.
\newblock {Constraining Primordial Black Holes with Dwarf Galaxy Heating}.
\newblock {\em Astrophys. J. Lett.}, 908(2):L23, 2021.
\newblock \href {https://arxiv.org/abs/2007.02213} {\path{arXiv:2007.02213}}, \href {https://doi.org/10.3847/2041-8213/abdcb6} {\path{doi:10.3847/2041-8213/abdcb6}}.

\bibitem{Takhistov:2021aqx}
Volodymyr Takhistov, Philip Lu, Graciela~B. Gelmini, Kohei Hayashi, Yoshiyuki Inoue, and Alexander Kusenko.
\newblock {Interstellar gas heating by primordial black holes}.
\newblock {\em JCAP}, 03(03):017, 2022.
\newblock \href {https://arxiv.org/abs/2105.06099} {\path{arXiv:2105.06099}}, \href {https://doi.org/10.1088/1475-7516/2022/03/017} {\path{doi:10.1088/1475-7516/2022/03/017}}.

\bibitem{LIGOScientific:2020kqk}
R.~Abbott et~al.
\newblock {Population Properties of Compact Objects from the Second LIGO-Virgo Gravitational-Wave Transient Catalog}.
\newblock {\em Astrophys. J. Lett.}, 913(1):L7, 2021.
\newblock \href {https://arxiv.org/abs/2010.14533} {\path{arXiv:2010.14533}}, \href {https://doi.org/10.3847/2041-8213/abe949} {\path{doi:10.3847/2041-8213/abe949}}.

\bibitem{Thrane:2018qnx}
Eric Thrane and Colm Talbot.
\newblock {An introduction to Bayesian inference in gravitational-wave astronomy: parameter estimation, model selection, and hierarchical models}.
\newblock {\em Publ. Astron. Soc. Austral.}, 36:e010, 2019.
\newblock [Erratum: Publ.Astron.Soc.Austral. 37, e036 (2020)].
\newblock \href {https://arxiv.org/abs/1809.02293} {\path{arXiv:1809.02293}}, \href {https://doi.org/10.1017/pasa.2019.2} {\path{doi:10.1017/pasa.2019.2}}.

\bibitem{Mandel:2018mve}
Ilya Mandel, Will~M. Farr, and Jonathan~R. Gair.
\newblock {Extracting distribution parameters from multiple uncertain observations with selection biases}.
\newblock {\em Mon. Not. Roy. Astron. Soc.}, 486(1):1086--1093, 2019.
\newblock \href {https://arxiv.org/abs/1809.02063} {\path{arXiv:1809.02063}}, \href {https://doi.org/10.1093/mnras/stz896} {\path{doi:10.1093/mnras/stz896}}.

\bibitem{Mastrogiovanni:2023zbw}
Simone Mastrogiovanni, Gr\'egoire Pierra, St\'ephane Perri\`es, Danny Laghi, Giada~Caneva Santoro, Archisman Ghosh, Rachel Gray, Christos Karathanasis, and Konstantin Leyde.
\newblock {ICAROGW: A python package for inference of astrophysical population properties of noisy, heterogeneous and incomplete observations}.
\newblock 5 2023.
\newblock \href {https://arxiv.org/abs/2305.17973} {\path{arXiv:2305.17973}}.

\bibitem{Talbot:2023pex}
Colm Talbot and Jacob Golomb.
\newblock {Growing pains: understanding the impact of likelihood uncertainty on hierarchical Bayesian inference for gravitational-wave astronomy}.
\newblock {\em Mon. Not. Roy. Astron. Soc.}, 526(3):3495--3503, 2023.
\newblock \href {https://arxiv.org/abs/2304.06138} {\path{arXiv:2304.06138}}, \href {https://doi.org/10.1093/mnras/stad2968} {\path{doi:10.1093/mnras/stad2968}}.

\bibitem{LIGOScientific:2021aug}
R.~Abbott et~al.
\newblock {Constraints on the Cosmic Expansion History from GWTC\textendash{}3}.
\newblock {\em Astrophys. J.}, 949(2):76, 2023.
\newblock \href {https://arxiv.org/abs/2111.03604} {\path{arXiv:2111.03604}}, \href {https://doi.org/10.3847/1538-4357/ac74bb} {\path{doi:10.3847/1538-4357/ac74bb}}.

\bibitem{Farr:2019rap}
Will~M. Farr.
\newblock {Accuracy Requirements for Empirically-Measured Selection Functions}.
\newblock {\em Research Notes of the AAS}, 3(5):66, 2019.
\newblock \href {https://arxiv.org/abs/1904.10879} {\path{arXiv:1904.10879}}, \href {https://doi.org/10.3847/2515-5172/ab1d5f} {\path{doi:10.3847/2515-5172/ab1d5f}}.

\bibitem{Vaskonen:2020lbd}
Ville Vaskonen and Hardi Veerm\"ae.
\newblock {Did NANOGrav see a signal from primordial black hole formation?}
\newblock {\em Phys. Rev. Lett.}, 126(5):051303, 2021.
\newblock \href {https://arxiv.org/abs/2009.07832} {\path{arXiv:2009.07832}}, \href {https://doi.org/10.1103/PhysRevLett.126.051303} {\path{doi:10.1103/PhysRevLett.126.051303}}.

\bibitem{Raidal:2024bmm}
Ville Vaskonen and Hardi Veerm\"ae.
\newblock {Formation of primordial black hole binaries and their merger rates}.
\newblock 4 2024.
\newblock \href {https://arxiv.org/abs/2404.08416} {\path{arXiv:2404.08416}}.

\bibitem{Franciolini:2022ewd}
G.~Franciolini, K.~Kritos, E.~Berti, and J.~Silk.
\newblock {Primordial black hole mergers from three-body interactions}.
\newblock {\em Phys. Rev. D}, 106(8):083529, 2022.
\newblock \href {https://doi.org/10.1103/PhysRevD.106.083529} {\path{doi:10.1103/PhysRevD.106.083529}}.

\bibitem{Stone:2019qvl}
Nicholas~C. Stone and Nathan W.~C. Leigh.
\newblock {A statistical solution to the chaotic, non-hierarchical three-body problem}.
\newblock {\em Nature}, 576(7787):406--410, 2019.
\newblock \href {https://arxiv.org/abs/1909.05272} {\path{arXiv:1909.05272}}, \href {https://doi.org/10.1038/s41586-019-1833-8} {\path{doi:10.1038/s41586-019-1833-8}}.

\bibitem{Carr:2017jsz}
B.~Carr, M.~Raidal, T.~Tenkanen, V.~Vaskonen, and H.~Veerm\"ae.
\newblock {Primordial black hole constraints for extended mass functions}.
\newblock {\em Phys. Rev. D}, 96(2):023514, 2017.
\newblock \href {https://doi.org/10.1103/PhysRevD.96.023514} {\path{doi:10.1103/PhysRevD.96.023514}}.

\bibitem{Byrnes:2018txb}
Christian~T. Byrnes, Philippa~S. Cole, and Subodh~P. Patil.
\newblock {Steepest growth of the power spectrum and primordial black holes}.
\newblock {\em JCAP}, 06:028, 2019.
\newblock \href {https://arxiv.org/abs/1811.11158} {\path{arXiv:1811.11158}}, \href {https://doi.org/10.1088/1475-7516/2019/06/028} {\path{doi:10.1088/1475-7516/2019/06/028}}.

\bibitem{Ezquiaga:2018gbw}
Jose~Mar\'\i{}a Ezquiaga and Juan Garc\'\i{}a-Bellido.
\newblock {Quantum diffusion beyond slow-roll: implications for primordial black-hole production}.
\newblock {\em JCAP}, 08:018, 2018.
\newblock \href {https://arxiv.org/abs/1805.06731} {\path{arXiv:1805.06731}}, \href {https://doi.org/10.1088/1475-7516/2018/08/018} {\path{doi:10.1088/1475-7516/2018/08/018}}.

\bibitem{Ezquiaga:2019ftu}
Jose~Mar\'\i{}a Ezquiaga, Juan Garc\'\i{}a-Bellido, and Vincent Vennin.
\newblock {The exponential tail of inflationary fluctuations: consequences for primordial black holes}.
\newblock {\em JCAP}, 03:029, 2020.
\newblock \href {https://arxiv.org/abs/1912.05399} {\path{arXiv:1912.05399}}, \href {https://doi.org/10.1088/1475-7516/2020/03/029} {\path{doi:10.1088/1475-7516/2020/03/029}}.

\bibitem{Bhaumik:2019tvl}
Nilanjandev Bhaumik and Rajeev~Kumar Jain.
\newblock {Primordial black holes dark matter from inflection point models of inflation and the effects of reheating}.
\newblock {\em JCAP}, 01:037, 2020.
\newblock \href {https://arxiv.org/abs/1907.04125} {\path{arXiv:1907.04125}}, \href {https://doi.org/10.1088/1475-7516/2020/01/037} {\path{doi:10.1088/1475-7516/2020/01/037}}.

\bibitem{Motohashi:2019rhu}
Hayato Motohashi, Shinji Mukohyama, and Michele Oliosi.
\newblock {Constant Roll and Primordial Black Holes}.
\newblock {\em JCAP}, 03:002, 2020.
\newblock \href {https://arxiv.org/abs/1910.13235} {\path{arXiv:1910.13235}}, \href {https://doi.org/10.1088/1475-7516/2020/03/002} {\path{doi:10.1088/1475-7516/2020/03/002}}.

\bibitem{Rasanen:2018fom}
Syksy Rasanen and Eemeli Tomberg.
\newblock {Planck scale black hole dark matter from Higgs inflation}.
\newblock {\em JCAP}, 01:038, 2019.
\newblock \href {https://arxiv.org/abs/1810.12608} {\path{arXiv:1810.12608}}, \href {https://doi.org/10.1088/1475-7516/2019/01/038} {\path{doi:10.1088/1475-7516/2019/01/038}}.

\bibitem{Balaji:2022rsy}
Shyam Balaji, Joseph Silk, and Yi-Peng Wu.
\newblock {Induced gravitational waves from the cosmic coincidence}.
\newblock {\em JCAP}, 06(06):008, 2022.
\newblock \href {https://arxiv.org/abs/2202.00700} {\path{arXiv:2202.00700}}, \href {https://doi.org/10.1088/1475-7516/2022/06/008} {\path{doi:10.1088/1475-7516/2022/06/008}}.

\bibitem{Frolovsky:2023hqd}
Daniel Frolovsky and Sergei~V. Ketov.
\newblock {Production of Primordial Black Holes in Improved E-Models of Inflation}.
\newblock {\em Universe}, 9(6):294, 2023.
\newblock \href {https://arxiv.org/abs/2304.12558} {\path{arXiv:2304.12558}}, \href {https://doi.org/10.3390/universe9060294} {\path{doi:10.3390/universe9060294}}.

\bibitem{Dimopoulos:2017ged}
Konstantinos Dimopoulos.
\newblock {Ultra slow-roll inflation demystified}.
\newblock {\em Phys. Lett. B}, 775:262--265, 2017.
\newblock \href {https://arxiv.org/abs/1707.05644} {\path{arXiv:1707.05644}}, \href {https://doi.org/10.1016/j.physletb.2017.10.066} {\path{doi:10.1016/j.physletb.2017.10.066}}.

\bibitem{Germani:2017bcs}
Cristiano Germani and Tomislav Prokopec.
\newblock {On primordial black holes from an inflection point}.
\newblock {\em Phys. Dark Univ.}, 18:6--10, 2017.
\newblock \href {https://arxiv.org/abs/1706.04226} {\path{arXiv:1706.04226}}, \href {https://doi.org/10.1016/j.dark.2017.09.001} {\path{doi:10.1016/j.dark.2017.09.001}}.

\bibitem{Choudhury:2013woa}
Sayantan Choudhury and Anupam Mazumdar.
\newblock {Primordial blackholes and gravitational waves for an inflection-point model of inflation}.
\newblock {\em Phys. Lett. B}, 733:270--275, 2014.
\newblock \href {https://arxiv.org/abs/1307.5119} {\path{arXiv:1307.5119}}, \href {https://doi.org/10.1016/j.physletb.2014.04.050} {\path{doi:10.1016/j.physletb.2014.04.050}}.

\bibitem{Ragavendra:2023ret}
H.~V. Ragavendra and L.~Sriramkumar.
\newblock {Observational Imprints of Enhanced Scalar Power on Small Scales in Ultra Slow Roll Inflation and Associated Non-Gaussianities}.
\newblock {\em Galaxies}, 11(1):34, 2023.
\newblock \href {https://arxiv.org/abs/2301.08887} {\path{arXiv:2301.08887}}, \href {https://doi.org/10.3390/galaxies11010034} {\path{doi:10.3390/galaxies11010034}}.

\bibitem{Cheng:2021lif}
Shu-Lin Cheng, Da-Shin Lee, and Kin-Wang Ng.
\newblock {Power spectrum of primordial perturbations during ultra-slow-roll inflation with back reaction effects}.
\newblock {\em Phys. Lett. B}, 827:136956, 2022.
\newblock \href {https://arxiv.org/abs/2106.09275} {\path{arXiv:2106.09275}}, \href {https://doi.org/10.1016/j.physletb.2022.136956} {\path{doi:10.1016/j.physletb.2022.136956}}.

\bibitem{Karam:2023haj}
A.~Karam, N.~Koivunen, E.~Tomberg, A.~Racioppi, and H.~Veerm\"ae.
\newblock {Primordial black holes and inflation from double-well potentials}.
\newblock 5 2023.
\newblock \href {https://arxiv.org/abs/2305.09630} {\path{arXiv:2305.09630}}.

\bibitem{Mishra:2023lhe}
S.~S. Mishra, E.~J. Copeland, and A.~M. Green.
\newblock {Primordial black holes and stochastic inflation beyond slow roll: I -- noise matrix elements}.
\newblock 3 2023.
\newblock \href {https://arxiv.org/abs/2303.17375} {\path{arXiv:2303.17375}}.

\bibitem{Cole:2023wyx}
P.~S. Cole, A.~D. Gow, C.~T. Byrnes, and S.~P. Patil.
\newblock {Primordial black holes from single-field inflation: a fine-tuning audit}.
\newblock {\em JCAP}, 08:031, 2023.
\newblock \href {https://doi.org/10.1088/1475-7516/2023/08/031} {\path{doi:10.1088/1475-7516/2023/08/031}}.

\bibitem{Frosina:2023nxu}
Lorenzo Frosina and Alfredo Urbano.
\newblock {Inflationary interpretation of the nHz gravitational-wave background}.
\newblock {\em Phys. Rev. D}, 108(10):103544, 2023.
\newblock \href {https://arxiv.org/abs/2308.06915} {\path{arXiv:2308.06915}}, \href {https://doi.org/10.1103/PhysRevD.108.103544} {\path{doi:10.1103/PhysRevD.108.103544}}.

\bibitem{Choudhury:2024one}
Sayantan Choudhury, Ahaskar Karde, Sudhakar Panda, and M.~Sami.
\newblock {Realisation of the ultra-slow roll phase in Galileon inflation and PBH overproduction}.
\newblock 1 2024.
\newblock \href {https://arxiv.org/abs/2401.10925} {\path{arXiv:2401.10925}}.

\bibitem{Wang:2024vfv}
Xinpeng Wang, Ying-li Zhang, and Misao Sasaki.
\newblock {Enhanced Curvature Perturbation and Primordial Black Hole Formation in Two-stage Inflation with a break}.
\newblock 4 2024.
\newblock \href {https://arxiv.org/abs/2404.02492} {\path{arXiv:2404.02492}}.

\bibitem{Enqvist:2001zp}
K.~Enqvist and M.~S. Sloth.
\newblock {Adiabatic CMB perturbations in pre - big bang string cosmology}.
\newblock {\em Nucl. Phys. B}, 626:395--409, 2002.
\newblock \href {https://doi.org/10.1016/S0550-3213(02)00043-3} {\path{doi:10.1016/S0550-3213(02)00043-3}}.

\bibitem{Lyth:2001nq}
D.~H. Lyth and D.~Wands.
\newblock {Generating the curvature perturbation without an inflaton}.
\newblock {\em Phys. Lett. B}, 524:5--14, 2002.
\newblock \href {https://doi.org/10.1016/S0370-2693(01)01366-1} {\path{doi:10.1016/S0370-2693(01)01366-1}}.

\bibitem{Sloth:2002xn}
M.~S. Sloth.
\newblock {Superhorizon curvaton amplitude in inflation and pre - big bang cosmology}.
\newblock {\em Nucl. Phys. B}, 656:239--251, 2003.
\newblock \href {https://doi.org/10.1016/S0550-3213(03)00114-7} {\path{doi:10.1016/S0550-3213(03)00114-7}}.

\bibitem{Dimopoulos:2003ii}
K.~Dimopoulos, G.~Lazarides, D.~Lyth, and R.~Ruiz~de Austri.
\newblock {The Peccei-Quinn field as curvaton}.
\newblock {\em JHEP}, 05:057, 2003.
\newblock \href {https://doi.org/10.1088/1126-6708/2003/05/057} {\path{doi:10.1088/1126-6708/2003/05/057}}.

\bibitem{Kohri:2012yw}
K.~Kohri, C.~M. Lin, and T.~Matsuda.
\newblock {Primordial black holes from the inflating curvaton}.
\newblock {\em Phys. Rev. D}, 87(10):103527, 2013.
\newblock \href {https://doi.org/10.1103/PhysRevD.87.103527} {\path{doi:10.1103/PhysRevD.87.103527}}.

\bibitem{Kawasaki:2012wr}
M.~Kawasaki, N.~Kitajima, and T.~T. Yanagida.
\newblock {Primordial black hole formation from an axionlike curvaton model}.
\newblock {\em Phys. Rev. D}, 87(6):063519, 2013.
\newblock \href {https://doi.org/10.1103/PhysRevD.87.063519} {\path{doi:10.1103/PhysRevD.87.063519}}.

\bibitem{Kawasaki:2013xsa}
M.~Kawasaki, N.~Kitajima, and S.~Yokoyama.
\newblock {Gravitational waves from a curvaton model with blue spectrum}.
\newblock {\em JCAP}, 08:042, 2013.
\newblock \href {https://doi.org/10.1088/1475-7516/2013/08/042} {\path{doi:10.1088/1475-7516/2013/08/042}}.

\bibitem{Carr:2017edp}
B.~Carr, T.~Tenkanen, and V.~Vaskonen.
\newblock {Primordial black holes from inflaton and spectator field perturbations in a matter-dominated era}.
\newblock {\em Phys. Rev. D}, 96(6):063507, 2017.
\newblock \href {https://doi.org/10.1103/PhysRevD.96.063507} {\path{doi:10.1103/PhysRevD.96.063507}}.

\bibitem{Ando:2017veq}
K.~Ando, K.~Inomata, M.~Kawasaki, K.~Mukaida, and T.~T. Yanagida.
\newblock {Primordial black holes for the LIGO events in the axionlike curvaton model}.
\newblock {\em Phys. Rev. D}, 97(12):123512, 2018.
\newblock \href {https://doi.org/10.1103/PhysRevD.97.123512} {\path{doi:10.1103/PhysRevD.97.123512}}.

\bibitem{Ando:2018nge}
K.~Ando, M.~Kawasaki, and H.~Nakatsuka.
\newblock {Formation of primordial black holes in an axionlike curvaton model}.
\newblock {\em Phys. Rev. D}, 98(8):083508, 2018.
\newblock \href {https://doi.org/10.1103/PhysRevD.98.083508} {\path{doi:10.1103/PhysRevD.98.083508}}.

\bibitem{Chen:2019zza}
C.~Chen and Y.~F. Cai.
\newblock {Primordial black holes from sound speed resonance in the inflaton-curvaton mixed scenario}.
\newblock {\em JCAP}, 10:068, 2019.
\newblock \href {https://doi.org/10.1088/1475-7516/2019/10/068} {\path{doi:10.1088/1475-7516/2019/10/068}}.

\bibitem{Liu:2020zzv}
Lei-Hua Liu and Tomislav Prokopec.
\newblock {Non-minimally coupled curvaton}.
\newblock {\em JCAP}, 06:033, 2021.
\newblock \href {https://arxiv.org/abs/2005.11069} {\path{arXiv:2005.11069}}, \href {https://doi.org/10.1088/1475-7516/2021/06/033} {\path{doi:10.1088/1475-7516/2021/06/033}}.

\bibitem{Cai:2021wzd}
R.~G. Cai, C.~Chen, and C.~Fu.
\newblock {Primordial black holes and stochastic gravitational wave background from inflation with a noncanonical spectator field}.
\newblock {\em Phys. Rev. D}, 104(8):083537, 2021.
\newblock \href {https://doi.org/10.1103/PhysRevD.104.083537} {\path{doi:10.1103/PhysRevD.104.083537}}.

\bibitem{Liu:2021rgq}
L.~H. Liu.
\newblock {The primordial black hole from running curvaton}.
\newblock {\em Chin. Phys. C}, 47:1, 2023.
\newblock \href {https://doi.org/10.1088/1674-1137/ac9d28} {\path{doi:10.1088/1674-1137/ac9d28}}.

\bibitem{Chen:2023lou}
C.~Chen, A.~Ghoshal, Z.~Lalak, Y.~Luo, and A.~Naskar.
\newblock {Growth of curvature perturbations for PBH formation \& detectable GWs in non-minimal curvaton scenario revisited}.
\newblock {\em JCAP}, 08:041, 2023.
\newblock \href {https://doi.org/10.1088/1475-7516/2023/08/041} {\path{doi:10.1088/1475-7516/2023/08/041}}.

\bibitem{Torrado:2017qtr}
J.~Torrado, C.~T. Byrnes, R.~J. Hardwick, V.~Vennin, and D.~Wands.
\newblock {Measuring the duration of inflation with the curvaton}.
\newblock {\em Phys. Rev. D}, 98(6):063525, 2018.
\newblock \href {https://doi.org/10.1103/PhysRevD.98.063525} {\path{doi:10.1103/PhysRevD.98.063525}}.

\bibitem{Cable:2023lca}
A.~Cable and A.~Wilkins.
\newblock {Spectators no more! How even unimportant fields can ruin your Primordial Black Hole model}.
\newblock 6 2023.
\newblock \href {https://arxiv.org/abs/2306.09232} {\path{arXiv:2306.09232}}.

\bibitem{Inomata:2023drn}
Keisuke Inomata, Masahiro Kawasaki, Kyohei Mukaida, and Tsutomu~T. Yanagida.
\newblock {Axion curvaton model for the gravitational waves observed by pulsar timing arrays}.
\newblock {\em Phys. Rev. D}, 109(4):043508, 2024.
\newblock \href {https://arxiv.org/abs/2309.11398} {\path{arXiv:2309.11398}}, \href {https://doi.org/10.1103/PhysRevD.109.043508} {\path{doi:10.1103/PhysRevD.109.043508}}.

\bibitem{EROS-2:2006ryy}
P.~Tisserand et~al.
\newblock {Limits on the Macho Content of the Galactic Halo from the EROS-2 Survey of the Magellanic Clouds}.
\newblock {\em Astron. Astrophys.}, 469:387--404, 2007.
\newblock \href {https://arxiv.org/abs/astro-ph/0607207} {\path{arXiv:astro-ph/0607207}}, \href {https://doi.org/10.1051/0004-6361:20066017} {\path{doi:10.1051/0004-6361:20066017}}.

\bibitem{Mroz:2024mse}
P.~Mroz et~al.
\newblock {No massive black holes in the Milky Way halo}.
\newblock 3 2024.
\newblock \href {https://arxiv.org/abs/2403.02386} {\path{arXiv:2403.02386}}.

\bibitem{Koushiappas:2017chw}
Savvas~M. Koushiappas and Abraham Loeb.
\newblock {Dynamics of Dwarf Galaxies Disfavor Stellar-Mass Black Holes as Dark Matter}.
\newblock {\em Phys. Rev. Lett.}, 119(4):041102, 2017.
\newblock \href {https://arxiv.org/abs/1704.01668} {\path{arXiv:1704.01668}}, \href {https://doi.org/10.1103/PhysRevLett.119.041102} {\path{doi:10.1103/PhysRevLett.119.041102}}.

\bibitem{Brandt:2016aco}
Timothy~D. Brandt.
\newblock {Constraints on MACHO Dark Matter from Compact Stellar Systems in Ultra-Faint Dwarf Galaxies}.
\newblock {\em Astrophys. J. Lett.}, 824(2):L31, 2016.
\newblock \href {https://arxiv.org/abs/1605.03665} {\path{arXiv:1605.03665}}, \href {https://doi.org/10.3847/2041-8205/824/2/L31} {\path{doi:10.3847/2041-8205/824/2/L31}}.

\bibitem{Monroy-Rodriguez:2014ula}
Miguel~A. Monroy-Rodr\'\i{}guez and Christine Allen.
\newblock {The end of the MACHO era- revisited: new limits on MACHO masses from halo wide binaries}.
\newblock {\em Astrophys. J.}, 790(2):159, 2014.
\newblock \href {https://arxiv.org/abs/1406.5169} {\path{arXiv:1406.5169}}, \href {https://doi.org/10.1088/0004-637X/790/2/159} {\path{doi:10.1088/0004-637X/790/2/159}}.

\bibitem{Murgia:2019duy}
Riccardo Murgia, Giulio Scelfo, Matteo Viel, and Alvise Raccanelli.
\newblock {Lyman-\ensuremath{\alpha} Forest Constraints on Primordial Black Holes as Dark Matter}.
\newblock {\em Phys. Rev. Lett.}, 123(7):071102, 2019.
\newblock \href {https://arxiv.org/abs/1903.10509} {\path{arXiv:1903.10509}}, \href {https://doi.org/10.1103/PhysRevLett.123.071102} {\path{doi:10.1103/PhysRevLett.123.071102}}.

\bibitem{Zumalacarregui:2017qqd}
Miguel Zumalacarregui and Uros Seljak.
\newblock {Limits on stellar-mass compact objects as dark matter from gravitational lensing of type Ia supernovae}.
\newblock {\em Phys. Rev. Lett.}, 121(14):141101, 2018.
\newblock \href {https://arxiv.org/abs/1712.02240} {\path{arXiv:1712.02240}}, \href {https://doi.org/10.1103/PhysRevLett.121.141101} {\path{doi:10.1103/PhysRevLett.121.141101}}.

\bibitem{Chen:2019xse}
Zu-Cheng Chen, Chen Yuan, and Qing-Guo Huang.
\newblock {Pulsar Timing Array Constraints on Primordial Black Holes with NANOGrav 11-Year Dataset}.
\newblock {\em Phys. Rev. Lett.}, 124(25):251101, 2020.
\newblock \href {https://arxiv.org/abs/1910.12239} {\path{arXiv:1910.12239}}, \href {https://doi.org/10.1103/PhysRevLett.124.251101} {\path{doi:10.1103/PhysRevLett.124.251101}}.

\bibitem{Kohri:2020qqd}
Kazunori Kohri and Takahiro Terada.
\newblock {Solar-Mass Primordial Black Holes Explain NANOGrav Hint of Gravitational Waves}.
\newblock {\em Phys. Lett. B}, 813:136040, 2021.
\newblock \href {https://arxiv.org/abs/2009.11853} {\path{arXiv:2009.11853}}, \href {https://doi.org/10.1016/j.physletb.2020.136040} {\path{doi:10.1016/j.physletb.2020.136040}}.

\bibitem{Domenech:2020ers}
Guillem Dom\`enech and Shi Pi.
\newblock {NANOGrav hints on planet-mass primordial black holes}.
\newblock {\em Sci. China Phys. Mech. Astron.}, 65(3):230411, 2022.
\newblock \href {https://arxiv.org/abs/2010.03976} {\path{arXiv:2010.03976}}, \href {https://doi.org/10.1007/s11433-021-1839-6} {\path{doi:10.1007/s11433-021-1839-6}}.

\bibitem{Dandoy:2023jot}
Virgile Dandoy, Valerie Domcke, and Fabrizio Rompineve.
\newblock {Search for scalar induced gravitational waves in the international pulsar timing array data release 2 and NANOgrav 12.5 years datasets}.
\newblock {\em SciPost Phys. Core}, 6:060, 2023.
\newblock \href {https://arxiv.org/abs/2302.07901} {\path{arXiv:2302.07901}}, \href {https://doi.org/10.21468/SciPostPhysCore.6.3.060} {\path{doi:10.21468/SciPostPhysCore.6.3.060}}.

\bibitem{Fixsen:1996nj}
D.~J. Fixsen, E.~S. Cheng, J.~M. Gales, John~C. Mather, R.~A. Shafer, and E.~L. Wright.
\newblock {The Cosmic Microwave Background spectrum from the full COBE FIRAS data set}.
\newblock {\em Astrophys. J.}, 473:576, 1996.
\newblock \href {https://arxiv.org/abs/astro-ph/9605054} {\path{arXiv:astro-ph/9605054}}, \href {https://doi.org/10.1086/178173} {\path{doi:10.1086/178173}}.

\bibitem{Chluba:2012gq}
Jens Chluba, Rishi Khatri, and Rashid~A. Sunyaev.
\newblock {CMB at 2x2 order: The dissipation of primordial acoustic waves and the observable part of the associated energy release}.
\newblock {\em Mon. Not. Roy. Astron. Soc.}, 425:1129--1169, 2012.
\newblock \href {https://arxiv.org/abs/1202.0057} {\path{arXiv:1202.0057}}, \href {https://doi.org/10.1111/j.1365-2966.2012.21474.x} {\path{doi:10.1111/j.1365-2966.2012.21474.x}}.

\bibitem{Chluba:2012we}
Jens Chluba, Adrienne~L. Erickcek, and Ido Ben-Dayan.
\newblock {Probing the inflaton: Small-scale power spectrum constraints from measurements of the CMB energy spectrum}.
\newblock {\em Astrophys. J.}, 758:76, 2012.
\newblock \href {https://arxiv.org/abs/1203.2681} {\path{arXiv:1203.2681}}, \href {https://doi.org/10.1088/0004-637X/758/2/76} {\path{doi:10.1088/0004-637X/758/2/76}}.

\bibitem{Chluba:2013dna}
Jens Chluba and Daniel Grin.
\newblock {CMB spectral distortions from small-scale isocurvature fluctuations}.
\newblock {\em Mon. Not. Roy. Astron. Soc.}, 434:1619--1635, 2013.
\newblock \href {https://arxiv.org/abs/1304.4596} {\path{arXiv:1304.4596}}, \href {https://doi.org/10.1093/mnras/stt1129} {\path{doi:10.1093/mnras/stt1129}}.

\bibitem{Bianchini:2022dqh}
Federico Bianchini and Giulio Fabbian.
\newblock {CMB spectral distortions revisited: A new take on \ensuremath{\mu} distortions and primordial non-Gaussianities from FIRAS data}.
\newblock {\em Phys. Rev. D}, 106(6):063527, 2022.
\newblock \href {https://arxiv.org/abs/2206.02762} {\path{arXiv:2206.02762}}, \href {https://doi.org/10.1103/PhysRevD.106.063527} {\path{doi:10.1103/PhysRevD.106.063527}}.

\bibitem{Perna:2024ehx}
Gabriele Perna, Chiara Testini, Angelo Ricciardone, and Sabino Matarrese.
\newblock {Fully non-Gaussian Scalar-Induced Gravitational Waves}.
\newblock {\em JCAP}, 05:086, 2024.
\newblock \href {https://arxiv.org/abs/2403.06962} {\path{arXiv:2403.06962}}, \href {https://doi.org/10.1088/1475-7516/2024/05/086} {\path{doi:10.1088/1475-7516/2024/05/086}}.

\bibitem{Kehagias:2024kgk}
Alex Kehagias, Davide Perrone, and Antonio Riotto.
\newblock {Why the Universal Threshold for Primordial Black Hole Formation is Universal}.
\newblock 5 2024.
\newblock \href {https://arxiv.org/abs/2405.05208} {\path{arXiv:2405.05208}}.

\bibitem{Agius:2024ecw}
Dominic Agius, Rouven Essig, Daniele Gaggero, Francesca Scarcella, Gregory Suczewski, and Mauro Valli.
\newblock {Feedback in the dark: a critical examination of CMB bounds on primordial black holes}.
\newblock {\em JCAP}, 07:003, 2024.
\newblock \href {https://arxiv.org/abs/2403.18895} {\path{arXiv:2403.18895}}, \href {https://doi.org/10.1088/1475-7516/2024/07/003} {\path{doi:10.1088/1475-7516/2024/07/003}}.

\bibitem{Facchinetti:2022kbg}
Ga\'etan Facchinetti, Matteo Lucca, and S\'ebastien Clesse.
\newblock {Relaxing CMB bounds on primordial black holes: The role of ionization fronts}.
\newblock {\em Phys. Rev. D}, 107(4):043537, 2023.
\newblock \href {https://arxiv.org/abs/2212.07969} {\path{arXiv:2212.07969}}, \href {https://doi.org/10.1103/PhysRevD.107.043537} {\path{doi:10.1103/PhysRevD.107.043537}}.

\bibitem{Bird:2010mp}
Simeon Bird, Hiranya~V. Peiris, Matteo Viel, and Licia Verde.
\newblock {Minimally Parametric Power Spectrum Reconstruction from the Lyman-alpha Forest}.
\newblock {\em Mon. Not. Roy. Astron. Soc.}, 413:1717--1728, 2011.
\newblock \href {https://arxiv.org/abs/1010.1519} {\path{arXiv:1010.1519}}, \href {https://doi.org/10.1111/j.1365-2966.2011.18245.x} {\path{doi:10.1111/j.1365-2966.2011.18245.x}}.

\bibitem{Cyr:2023pgw}
Bryce Cyr, Thomas Kite, Jens Chluba, J.~Colin Hill, Donghui Jeong, Sandeep~Kumar Acharya, Boris Bolliet, and Subodh~P. Patil.
\newblock {Disentangling the primordial nature of stochastic gravitational wave backgrounds with CMB spectral distortions}.
\newblock {\em Mon. Not. Roy. Astron. Soc.}, 528(1):883--897, 2024.
\newblock \href {https://arxiv.org/abs/2309.02366} {\path{arXiv:2309.02366}}, \href {https://doi.org/10.1093/mnras/stad3861} {\path{doi:10.1093/mnras/stad3861}}.

\bibitem{Tagliazucchi:2023dai}
Matteo Tagliazucchi, Matteo Braglia, Fabio Finelli, and Mauro Pieroni.
\newblock {The quest of CMB spectral distortions to probe the scalar-induced gravitational wave background interpretation in PTA data}.
\newblock 10 2023.
\newblock \href {https://arxiv.org/abs/2310.08527} {\path{arXiv:2310.08527}}.

\bibitem{Sharma:2024img}
Devanshu Sharma, Julien Lesgourgues, and Christian~T. Byrnes.
\newblock {Spectral distortions from acoustic dissipation with non-Gaussian (or not) perturbations}.
\newblock {\em JCAP}, 07:090, 2024.
\newblock \href {https://arxiv.org/abs/2404.18474} {\path{arXiv:2404.18474}}, \href {https://doi.org/10.1088/1475-7516/2024/07/090} {\path{doi:10.1088/1475-7516/2024/07/090}}.

\bibitem{Kormendy:1995er}
John Kormendy and Douglas Richstone.
\newblock {Inward bound: The Search for supermassive black holes in galactic nuclei}.
\newblock {\em Ann. Rev. Astron. Astrophys.}, 33:581, 1995.
\newblock \href {https://doi.org/10.1146/annurev.aa.33.090195.003053} {\path{doi:10.1146/annurev.aa.33.090195.003053}}.

\bibitem{Magorrian:1997hw}
John Magorrian et~al.
\newblock {The Demography of massive dark objects in galaxy centers}.
\newblock {\em Astron. J.}, 115:2285, 1998.
\newblock \href {https://arxiv.org/abs/astro-ph/9708072} {\path{arXiv:astro-ph/9708072}}, \href {https://doi.org/10.1086/300353} {\path{doi:10.1086/300353}}.

\bibitem{Richstone:1998ky}
D.~Richstone et~al.
\newblock {Supermassive black holes and the evolution of galaxies}.
\newblock {\em Nature}, 395:A14--A19, 1998.
\newblock \href {https://arxiv.org/abs/astro-ph/9810378} {\path{arXiv:astro-ph/9810378}}.

\bibitem{Kohri:2014lza}
Kazunori Kohri, Tomohiro Nakama, and Teruaki Suyama.
\newblock {Testing scenarios of primordial black holes being the seeds of supermassive black holes by ultracompact minihalos and CMB $\mu$-distortions}.
\newblock {\em Phys. Rev. D}, 90(8):083514, 2014.
\newblock \href {https://arxiv.org/abs/1405.5999} {\path{arXiv:1405.5999}}, \href {https://doi.org/10.1103/PhysRevD.90.083514} {\path{doi:10.1103/PhysRevD.90.083514}}.

\bibitem{Fukugita:1997bi}
M.~Fukugita, C.~J. Hogan, and P.~J.~E. Peebles.
\newblock {The Cosmic baryon budget}.
\newblock {\em Astrophys. J.}, 503:518, 1998.
\newblock \href {https://arxiv.org/abs/astro-ph/9712020} {\path{arXiv:astro-ph/9712020}}, \href {https://doi.org/10.1086/306025} {\path{doi:10.1086/306025}}.

\bibitem{Unal:2020mts}
Caner \"Unal, Ely~D. Kovetz, and Subodh~P. Patil.
\newblock {Multimessenger probes of inflationary fluctuations and primordial black holes}.
\newblock {\em Phys. Rev. D}, 103(6):063519, 2021.
\newblock \href {https://arxiv.org/abs/2008.11184} {\path{arXiv:2008.11184}}, \href {https://doi.org/10.1103/PhysRevD.103.063519} {\path{doi:10.1103/PhysRevD.103.063519}}.

\bibitem{Nakama:2017xvq}
Tomohiro Nakama, Bernard Carr, and Joseph Silk.
\newblock {Limits on primordial black holes from $\mu$ distortions in cosmic microwave background}.
\newblock {\em Phys. Rev. D}, 97(4):043525, 2018.
\newblock \href {https://arxiv.org/abs/1710.06945} {\path{arXiv:1710.06945}}, \href {https://doi.org/10.1103/PhysRevD.97.043525} {\path{doi:10.1103/PhysRevD.97.043525}}.

\bibitem{Hooper:2023nnl}
D.~Hooper, A.~Ireland, G.~Krnjaic, and A.~Stebbins.
\newblock {Supermassive Primordial Black Holes From Inflation}.
\newblock 8 2023.
\newblock \href {https://arxiv.org/abs/2308.00756} {\path{arXiv:2308.00756}}.

\bibitem{Byrnes:2024vjt}
Christian~T. Byrnes, Julien Lesgourgues, and Devanshu Sharma.
\newblock {Robust \ensuremath{\mu}-distortion constraints on primordial supermassive black holes from non-Gaussian perturbations}.
\newblock {\em JCAP}, 09:012, 2024.
\newblock \href {https://arxiv.org/abs/2404.18475} {\path{arXiv:2404.18475}}, \href {https://doi.org/10.1088/1475-7516/2024/09/012} {\path{doi:10.1088/1475-7516/2024/09/012}}.

\bibitem{Domenech:2021ztg}
Guillem Dom\`enech.
\newblock {Scalar Induced Gravitational Waves Review}.
\newblock {\em Universe}, 7(11):398, 2021.
\newblock \href {https://arxiv.org/abs/2109.01398} {\path{arXiv:2109.01398}}, \href {https://doi.org/10.3390/universe7110398} {\path{doi:10.3390/universe7110398}}.

\bibitem{Cai:2018dig}
Rong-gen Cai, Shi Pi, and Misao Sasaki.
\newblock {Gravitational Waves Induced by non-Gaussian Scalar Perturbations}.
\newblock {\em Phys. Rev. Lett.}, 122(20):201101, 2019.
\newblock \href {https://arxiv.org/abs/1810.11000} {\path{arXiv:1810.11000}}, \href {https://doi.org/10.1103/PhysRevLett.122.201101} {\path{doi:10.1103/PhysRevLett.122.201101}}.

\bibitem{Unal:2018yaa}
Caner Unal.
\newblock {Imprints of Primordial Non-Gaussianity on Gravitational Wave Spectrum}.
\newblock {\em Phys. Rev. D}, 99(4):041301, 2019.
\newblock \href {https://arxiv.org/abs/1811.09151} {\path{arXiv:1811.09151}}, \href {https://doi.org/10.1103/PhysRevD.99.041301} {\path{doi:10.1103/PhysRevD.99.041301}}.

\bibitem{Ragavendra:2021qdu}
H.~V. Ragavendra.
\newblock {Accounting for scalar non-Gaussianity in secondary gravitational waves}.
\newblock {\em Phys. Rev. D}, 105(6):063533, 2022.
\newblock \href {https://arxiv.org/abs/2108.04193} {\path{arXiv:2108.04193}}, \href {https://doi.org/10.1103/PhysRevD.105.063533} {\path{doi:10.1103/PhysRevD.105.063533}}.

\bibitem{Adshead:2021hnm}
Peter Adshead, Kaloian~D. Lozanov, and Zachary~J. Weiner.
\newblock {Non-Gaussianity and the induced gravitational wave background}.
\newblock {\em JCAP}, 10:080, 2021.
\newblock \href {https://arxiv.org/abs/2105.01659} {\path{arXiv:2105.01659}}, \href {https://doi.org/10.1088/1475-7516/2021/10/080} {\path{doi:10.1088/1475-7516/2021/10/080}}.

\bibitem{Abe:2022xur}
Katsuya~T. Abe, Ryoto Inui, Yuichiro Tada, and Shuichiro Yokoyama.
\newblock {Primordial black holes and gravitational waves induced by exponential-tailed perturbations}.
\newblock {\em JCAP}, 05:044, 2023.
\newblock \href {https://arxiv.org/abs/2209.13891} {\path{arXiv:2209.13891}}, \href {https://doi.org/10.1088/1475-7516/2023/05/044} {\path{doi:10.1088/1475-7516/2023/05/044}}.

\bibitem{Garcia-Saenz:2022tzu}
Sebastian Garcia-Saenz, Lucas Pinol, S\'ebastien Renaux-Petel, and Denis Werth.
\newblock {No-go theorem for scalar-trispectrum-induced gravitational waves}.
\newblock {\em JCAP}, 03:057, 2023.
\newblock \href {https://arxiv.org/abs/2207.14267} {\path{arXiv:2207.14267}}, \href {https://doi.org/10.1088/1475-7516/2023/03/057} {\path{doi:10.1088/1475-7516/2023/03/057}}.

\bibitem{Tomita:1975kj}
Kenji Tomita.
\newblock {Evolution of Irregularities in a Chaotic Early Universe}.
\newblock {\em Prog. Theor. Phys.}, 54:730, 1975.
\newblock \href {https://doi.org/10.1143/PTP.54.730} {\path{doi:10.1143/PTP.54.730}}.

\bibitem{Matarrese:1993zf}
Sabino Matarrese, Ornella Pantano, and Diego Saez.
\newblock {General relativistic dynamics of irrotational dust: Cosmological implications}.
\newblock {\em Phys. Rev. Lett.}, 72:320--323, 1994.
\newblock \href {https://arxiv.org/abs/astro-ph/9310036} {\path{arXiv:astro-ph/9310036}}, \href {https://doi.org/10.1103/PhysRevLett.72.320} {\path{doi:10.1103/PhysRevLett.72.320}}.

\bibitem{Acquaviva:2002ud}
Viviana Acquaviva, Nicola Bartolo, Sabino Matarrese, and Antonio Riotto.
\newblock {Second order cosmological perturbations from inflation}.
\newblock {\em Nucl. Phys. B}, 667:119--148, 2003.
\newblock \href {https://arxiv.org/abs/astro-ph/0209156} {\path{arXiv:astro-ph/0209156}}, \href {https://doi.org/10.1016/S0550-3213(03)00550-9} {\path{doi:10.1016/S0550-3213(03)00550-9}}.

\bibitem{Mollerach:2003nq}
Silvia Mollerach, Diego Harari, and Sabino Matarrese.
\newblock {CMB polarization from secondary vector and tensor modes}.
\newblock {\em Phys. Rev. D}, 69:063002, 2004.
\newblock \href {https://arxiv.org/abs/astro-ph/0310711} {\path{arXiv:astro-ph/0310711}}, \href {https://doi.org/10.1103/PhysRevD.69.063002} {\path{doi:10.1103/PhysRevD.69.063002}}.

\bibitem{Ananda:2006af}
Kishore~N. Ananda, Chris Clarkson, and David Wands.
\newblock {The Cosmological gravitational wave background from primordial density perturbations}.
\newblock {\em Phys. Rev. D}, 75:123518, 2007.
\newblock \href {https://arxiv.org/abs/gr-qc/0612013} {\path{arXiv:gr-qc/0612013}}, \href {https://doi.org/10.1103/PhysRevD.75.123518} {\path{doi:10.1103/PhysRevD.75.123518}}.

\bibitem{Baumann:2007zm}
Daniel Baumann, Paul~J. Steinhardt, Keitaro Takahashi, and Kiyotomo Ichiki.
\newblock {Gravitational Wave Spectrum Induced by Primordial Scalar Perturbations}.
\newblock {\em Phys. Rev. D}, 76:084019, 2007.
\newblock \href {https://arxiv.org/abs/hep-th/0703290} {\path{arXiv:hep-th/0703290}}, \href {https://doi.org/10.1103/PhysRevD.76.084019} {\path{doi:10.1103/PhysRevD.76.084019}}.

\bibitem{Mukhanov:2005sc}
V.~Mukhanov.
\newblock {\em {Physical Foundations of Cosmology}}.
\newblock Cambridge University Press, Oxford, 2005.
\newblock \href {https://doi.org/10.1017/CBO9780511790553} {\path{doi:10.1017/CBO9780511790553}}.

\bibitem{Abe:2020sqb}
Katsuya~T. Abe, Yuichiro Tada, and Ikumi Ueda.
\newblock {Induced gravitational waves as a cosmological probe of the sound speed during the QCD phase transition}.
\newblock {\em JCAP}, 06:048, 2021.
\newblock \href {https://arxiv.org/abs/2010.06193} {\path{arXiv:2010.06193}}, \href {https://doi.org/10.1088/1475-7516/2021/06/048} {\path{doi:10.1088/1475-7516/2021/06/048}}.

\bibitem{Espinosa:2018eve}
Jos\'e~Ram\'on Espinosa, Davide Racco, and Antonio Riotto.
\newblock {A Cosmological Signature of the SM Higgs Instability: Gravitational Waves}.
\newblock {\em JCAP}, 09:012, 2018.
\newblock \href {https://arxiv.org/abs/1804.07732} {\path{arXiv:1804.07732}}, \href {https://doi.org/10.1088/1475-7516/2018/09/012} {\path{doi:10.1088/1475-7516/2018/09/012}}.

\bibitem{Kohri:2018awv}
Kazunori Kohri and Takahiro Terada.
\newblock {Semianalytic calculation of gravitational wave spectrum nonlinearly induced from primordial curvature perturbations}.
\newblock {\em Phys. Rev. D}, 97(12):123532, 2018.
\newblock \href {https://arxiv.org/abs/1804.08577} {\path{arXiv:1804.08577}}, \href {https://doi.org/10.1103/PhysRevD.97.123532} {\path{doi:10.1103/PhysRevD.97.123532}}.

\bibitem{Carr:2020gox}
Bernard Carr, Kazunori Kohri, Yuuiti Sendouda, and Jun'ichi Yokoyama.
\newblock {Constraints on primordial black holes}.
\newblock {\em Rept. Prog. Phys.}, 84(11):116902, 2021.
\newblock \href {https://arxiv.org/abs/2002.12778} {\path{arXiv:2002.12778}}, \href {https://doi.org/10.1088/1361-6633/ac1e31} {\path{doi:10.1088/1361-6633/ac1e31}}.

\bibitem{Bavera:2021wmw}
Simone~S. Bavera, Gabriele Franciolini, Giulia Cusin, Antonio Riotto, Michael Zevin, and Tassos Fragos.
\newblock {Stochastic gravitational-wave background as a tool for investigating multi-channel astrophysical and primordial black-hole mergers}.
\newblock {\em Astron. Astrophys.}, 660:A26, 2022.
\newblock \href {https://arxiv.org/abs/2109.05836} {\path{arXiv:2109.05836}}, \href {https://doi.org/10.1051/0004-6361/202142208} {\path{doi:10.1051/0004-6361/202142208}}.

\bibitem{NANOGrav:2020bcs}
Zaven Arzoumanian et~al.
\newblock {The NANOGrav 12.5 yr Data Set: Search for an Isotropic Stochastic Gravitational-wave Background}.
\newblock {\em Astrophys. J. Lett.}, 905(2):L34, 2020.
\newblock \href {https://arxiv.org/abs/2009.04496} {\path{arXiv:2009.04496}}, \href {https://doi.org/10.3847/2041-8213/abd401} {\path{doi:10.3847/2041-8213/abd401}}.

\bibitem{NANOGrav:2021flc}
Zaven Arzoumanian et~al.
\newblock {Searching for Gravitational Waves from Cosmological Phase Transitions with the NANOGrav 12.5-Year Dataset}.
\newblock {\em Phys. Rev. Lett.}, 127(25):251302, 2021.
\newblock \href {https://arxiv.org/abs/2104.13930} {\path{arXiv:2104.13930}}, \href {https://doi.org/10.1103/PhysRevLett.127.251302} {\path{doi:10.1103/PhysRevLett.127.251302}}.

\bibitem{Xue:2021gyq}
Xiao Xue et~al.
\newblock {Constraining Cosmological Phase Transitions with the Parkes Pulsar Timing Array}.
\newblock {\em Phys. Rev. Lett.}, 127(25):251303, 2021.
\newblock \href {https://arxiv.org/abs/2110.03096} {\path{arXiv:2110.03096}}, \href {https://doi.org/10.1103/PhysRevLett.127.251303} {\path{doi:10.1103/PhysRevLett.127.251303}}.

\bibitem{Nakai:2020oit}
Yuichiro Nakai, Motoo Suzuki, Fuminobu Takahashi, and Masaki Yamada.
\newblock {Gravitational Waves and Dark Radiation from Dark Phase Transition: Connecting NANOGrav Pulsar Timing Data and Hubble Tension}.
\newblock {\em Phys. Lett. B}, 816:136238, 2021.
\newblock \href {https://arxiv.org/abs/2009.09754} {\path{arXiv:2009.09754}}, \href {https://doi.org/10.1016/j.physletb.2021.136238} {\path{doi:10.1016/j.physletb.2021.136238}}.

\bibitem{DiBari:2021dri}
Pasquale Di~Bari, Danny Marfatia, and Ye-Ling Zhou.
\newblock {Gravitational waves from first-order phase transitions in Majoron models of neutrino mass}.
\newblock {\em JHEP}, 10:193, 2021.
\newblock \href {https://arxiv.org/abs/2106.00025} {\path{arXiv:2106.00025}}, \href {https://doi.org/10.1007/JHEP10(2021)193} {\path{doi:10.1007/JHEP10(2021)193}}.

\bibitem{Sakharov:2021dim}
Alexander~S. Sakharov, Yury~N. Eroshenko, and Sergey~G. Rubin.
\newblock {Looking at the NANOGrav signal through the anthropic window of axionlike particles}.
\newblock {\em Phys. Rev. D}, 104(4):043005, 2021.
\newblock \href {https://arxiv.org/abs/2104.08750} {\path{arXiv:2104.08750}}, \href {https://doi.org/10.1103/PhysRevD.104.043005} {\path{doi:10.1103/PhysRevD.104.043005}}.

\bibitem{Li:2021qer}
Shou-Long Li, Lijing Shao, Puxun Wu, and Hongwei Yu.
\newblock {NANOGrav signal from first-order confinement-deconfinement phase transition in different QCD-matter scenarios}.
\newblock {\em Phys. Rev. D}, 104(4):043510, 2021.
\newblock \href {https://arxiv.org/abs/2101.08012} {\path{arXiv:2101.08012}}, \href {https://doi.org/10.1103/PhysRevD.104.043510} {\path{doi:10.1103/PhysRevD.104.043510}}.

\bibitem{Ashoorioon:2022raz}
Amjad Ashoorioon, Kazem Rezazadeh, and Abasalt Rostami.
\newblock {NANOGrav signal from the end of inflation and the LIGO mass and heavier primordial black holes}.
\newblock {\em Phys. Lett. B}, 835:137542, 2022.
\newblock \href {https://arxiv.org/abs/2202.01131} {\path{arXiv:2202.01131}}, \href {https://doi.org/10.1016/j.physletb.2022.137542} {\path{doi:10.1016/j.physletb.2022.137542}}.

\bibitem{Benetti:2021uea}
Micol Benetti, Leila~Lobato Graef, and Sunny Vagnozzi.
\newblock {Primordial gravitational waves from NANOGrav: A broken power-law approach}.
\newblock {\em Phys. Rev. D}, 105(4):043520, 2022.
\newblock \href {https://arxiv.org/abs/2111.04758} {\path{arXiv:2111.04758}}, \href {https://doi.org/10.1103/PhysRevD.105.043520} {\path{doi:10.1103/PhysRevD.105.043520}}.

\bibitem{Barir:2022kzo}
Joel Barir, Michael Geller, Chen Sun, and Tomer Volansky.
\newblock {Gravitational waves from incomplete inflationary phase transitions}.
\newblock {\em Phys. Rev. D}, 108(11):115016, 2023.
\newblock \href {https://arxiv.org/abs/2203.00693} {\path{arXiv:2203.00693}}, \href {https://doi.org/10.1103/PhysRevD.108.115016} {\path{doi:10.1103/PhysRevD.108.115016}}.

\bibitem{Hindmarsh:2022awe}
Mark Hindmarsh and Jun'ya Kume.
\newblock {Multi-messenger constraints on Abelian-Higgs cosmic string networks}.
\newblock {\em JCAP}, 04:045, 2023.
\newblock \href {https://arxiv.org/abs/2210.06178} {\path{arXiv:2210.06178}}, \href {https://doi.org/10.1088/1475-7516/2023/04/045} {\path{doi:10.1088/1475-7516/2023/04/045}}.

\bibitem{Gouttenoire:2023naa}
Yann Gouttenoire and Tomer Volansky.
\newblock {Primordial black holes from supercooled phase transitions}.
\newblock {\em Phys. Rev. D}, 110(4):043514, 2024.
\newblock \href {https://arxiv.org/abs/2305.04942} {\path{arXiv:2305.04942}}, \href {https://doi.org/10.1103/PhysRevD.110.043514} {\path{doi:10.1103/PhysRevD.110.043514}}.

\bibitem{Ellis:2020ena}
John Ellis and Marek Lewicki.
\newblock {Cosmic String Interpretation of NANOGrav Pulsar Timing Data}.
\newblock {\em Phys. Rev. Lett.}, 126(4):041304, 2021.
\newblock \href {https://arxiv.org/abs/2009.06555} {\path{arXiv:2009.06555}}, \href {https://doi.org/10.1103/PhysRevLett.126.041304} {\path{doi:10.1103/PhysRevLett.126.041304}}.

\bibitem{Datta:2020bht}
Satyabrata Datta, Ambar Ghosal, and Rome Samanta.
\newblock {Baryogenesis from ultralight primordial black holes and strong gravitational waves from cosmic strings}.
\newblock {\em JCAP}, 08:021, 2021.
\newblock \href {https://arxiv.org/abs/2012.14981} {\path{arXiv:2012.14981}}, \href {https://doi.org/10.1088/1475-7516/2021/08/021} {\path{doi:10.1088/1475-7516/2021/08/021}}.

\bibitem{Samanta:2020cdk}
Rome Samanta and Satyabrata Datta.
\newblock {Gravitational wave complementarity and impact of NANOGrav data on gravitational leptogenesis}.
\newblock {\em JHEP}, 05:211, 2021.
\newblock \href {https://arxiv.org/abs/2009.13452} {\path{arXiv:2009.13452}}, \href {https://doi.org/10.1007/JHEP05(2021)211} {\path{doi:10.1007/JHEP05(2021)211}}.

\bibitem{Buchmuller:2020lbh}
Wilfried Buchmuller, Valerie Domcke, and Kai Schmitz.
\newblock {From NANOGrav to LIGO with metastable cosmic strings}.
\newblock {\em Phys. Lett. B}, 811:135914, 2020.
\newblock \href {https://arxiv.org/abs/2009.10649} {\path{arXiv:2009.10649}}, \href {https://doi.org/10.1016/j.physletb.2020.135914} {\path{doi:10.1016/j.physletb.2020.135914}}.

\bibitem{Blasi:2020mfx}
Simone Blasi, Vedran Brdar, and Kai Schmitz.
\newblock {Has NANOGrav found first evidence for cosmic strings?}
\newblock {\em Phys. Rev. Lett.}, 126(4):041305, 2021.
\newblock \href {https://arxiv.org/abs/2009.06607} {\path{arXiv:2009.06607}}, \href {https://doi.org/10.1103/PhysRevLett.126.041305} {\path{doi:10.1103/PhysRevLett.126.041305}}.

\bibitem{Gorghetto:2021fsn}
Marco Gorghetto, Edward Hardy, and Horia Nicolaescu.
\newblock {Observing invisible axions with gravitational waves}.
\newblock {\em JCAP}, 06:034, 2021.
\newblock \href {https://arxiv.org/abs/2101.11007} {\path{arXiv:2101.11007}}, \href {https://doi.org/10.1088/1475-7516/2021/06/034} {\path{doi:10.1088/1475-7516/2021/06/034}}.

\bibitem{Buchmuller:2021mbb}
Wilfried Buchmuller, Valerie Domcke, and Kai Schmitz.
\newblock {Stochastic gravitational-wave background from metastable cosmic strings}.
\newblock {\em JCAP}, 12(12):006, 2021.
\newblock \href {https://arxiv.org/abs/2107.04578} {\path{arXiv:2107.04578}}, \href {https://doi.org/10.1088/1475-7516/2021/12/006} {\path{doi:10.1088/1475-7516/2021/12/006}}.

\bibitem{Blanco-Pillado:2021ygr}
Jose~J. Blanco-Pillado, Ken~D. Olum, and Jeremy~M. Wachter.
\newblock {Comparison of cosmic string and superstring models to NANOGrav 12.5-year results}.
\newblock {\em Phys. Rev. D}, 103(10):103512, 2021.
\newblock \href {https://arxiv.org/abs/2102.08194} {\path{arXiv:2102.08194}}, \href {https://doi.org/10.1103/PhysRevD.103.103512} {\path{doi:10.1103/PhysRevD.103.103512}}.

\bibitem{Ferreira:2022zzo}
Ricardo~Z. Ferreira, Alessio Notari, Oriol Pujolas, and Fabrizio Rompineve.
\newblock {Gravitational waves from domain walls in Pulsar Timing Array datasets}.
\newblock {\em JCAP}, 02:001, 2023.
\newblock \href {https://arxiv.org/abs/2204.04228} {\path{arXiv:2204.04228}}, \href {https://doi.org/10.1088/1475-7516/2023/02/001} {\path{doi:10.1088/1475-7516/2023/02/001}}.

\bibitem{An:2023idh}
Haipeng An and Chen Yang.
\newblock {Gravitational waves produced by domain walls during inflation}.
\newblock {\em Phys. Rev. D}, 109(12):123508, 2024.
\newblock \href {https://arxiv.org/abs/2304.02361} {\path{arXiv:2304.02361}}, \href {https://doi.org/10.1103/PhysRevD.109.123508} {\path{doi:10.1103/PhysRevD.109.123508}}.

\bibitem{Qiu:2023wbs}
Ze-Yu Qiu and Zhao-Huan Yu.
\newblock {Gravitational waves from cosmic strings associated with pseudo-Nambu-Goldstone dark matter*}.
\newblock {\em Chin. Phys. C}, 47(8):085104, 2023.
\newblock \href {https://arxiv.org/abs/2304.02506} {\path{arXiv:2304.02506}}, \href {https://doi.org/10.1088/1674-1137/acd9bf} {\path{doi:10.1088/1674-1137/acd9bf}}.

\bibitem{Zeng:2023jut}
Zhen-Min Zeng, Jing Liu, and Zong-Kuan Guo.
\newblock {Enhanced curvature perturbations from spherical domain walls nucleated during inflation}.
\newblock {\em Phys. Rev. D}, 108(6):063005, 2023.
\newblock \href {https://arxiv.org/abs/2301.07230} {\path{arXiv:2301.07230}}, \href {https://doi.org/10.1103/PhysRevD.108.063005} {\path{doi:10.1103/PhysRevD.108.063005}}.

\bibitem{King:2023cgv}
Stephen~F. King, Danny Marfatia, and Moinul~Hossain Rahat.
\newblock {Toward distinguishing Dirac from Majorana neutrino mass with gravitational waves}.
\newblock {\em Phys. Rev. D}, 109(3):035014, 2024.
\newblock \href {https://arxiv.org/abs/2306.05389} {\path{arXiv:2306.05389}}, \href {https://doi.org/10.1103/PhysRevD.109.035014} {\path{doi:10.1103/PhysRevD.109.035014}}.

\bibitem{Bhaumik:2020dor}
Nilanjandev Bhaumik and Rajeev~Kumar Jain.
\newblock {Small scale induced gravitational waves from primordial black holes, a~stringent lower mass bound, and the imprints of an early matter to~radiation transition}.
\newblock {\em Phys. Rev. D}, 104(2):023531, 2021.
\newblock \href {https://arxiv.org/abs/2009.10424} {\path{arXiv:2009.10424}}, \href {https://doi.org/10.1103/PhysRevD.104.023531} {\path{doi:10.1103/PhysRevD.104.023531}}.

\bibitem{Vagnozzi:2020gtf}
Sunny Vagnozzi.
\newblock {Implications of the NANOGrav results for inflation}.
\newblock {\em Mon. Not. Roy. Astron. Soc.}, 502(1):L11--L15, 2021.
\newblock \href {https://arxiv.org/abs/2009.13432} {\path{arXiv:2009.13432}}, \href {https://doi.org/10.1093/mnrasl/slaa203} {\path{doi:10.1093/mnrasl/slaa203}}.

\bibitem{Namba:2020kij}
Ryo Namba and Motoo Suzuki.
\newblock {Implications of Gravitational-wave Production from Dark Photon Resonance to Pulsar-timing Observations and Effective Number of Relativistic Species}.
\newblock {\em Phys. Rev. D}, 102:123527, 2020.
\newblock \href {https://arxiv.org/abs/2009.13909} {\path{arXiv:2009.13909}}, \href {https://doi.org/10.1103/PhysRevD.102.123527} {\path{doi:10.1103/PhysRevD.102.123527}}.

\bibitem{Sugiyama:2020roc}
Sunao Sugiyama, Volodymyr Takhistov, Edoardo Vitagliano, Alexander Kusenko, Misao Sasaki, and Masahiro Takada.
\newblock {Testing Stochastic Gravitational Wave Signals from Primordial Black Holes with Optical Telescopes}.
\newblock {\em Phys. Lett. B}, 814:136097, 2021.
\newblock \href {https://arxiv.org/abs/2010.02189} {\path{arXiv:2010.02189}}, \href {https://doi.org/10.1016/j.physletb.2021.136097} {\path{doi:10.1016/j.physletb.2021.136097}}.

\bibitem{Zhou:2020kkf}
Zihan Zhou, Jie Jiang, Yi-Fu Cai, Misao Sasaki, and Shi Pi.
\newblock {Primordial black holes and gravitational waves from resonant amplification during inflation}.
\newblock {\em Phys. Rev. D}, 102(10):103527, 2020.
\newblock \href {https://arxiv.org/abs/2010.03537} {\path{arXiv:2010.03537}}, \href {https://doi.org/10.1103/PhysRevD.102.103527} {\path{doi:10.1103/PhysRevD.102.103527}}.

\bibitem{Lin:2021vwc}
Jiong Lin, Shengqing Gao, Yungui Gong, Yizhou Lu, Zhongkai Wang, and Fengge Zhang.
\newblock {Primordial black holes and scalar induced gravitational waves from Higgs inflation with noncanonical kinetic term}.
\newblock {\em Phys. Rev. D}, 107(4):043517, 2023.
\newblock \href {https://arxiv.org/abs/2111.01362} {\path{arXiv:2111.01362}}, \href {https://doi.org/10.1103/PhysRevD.107.043517} {\path{doi:10.1103/PhysRevD.107.043517}}.

\bibitem{Rezazadeh:2021clf}
Kazem Rezazadeh, Zeinab Teimoori, Saeid Karimi, and Kayoomars Karami.
\newblock {Non-Gaussianity and secondary gravitational waves from primordial black holes production in $\alpha $-attractor inflation}.
\newblock {\em Eur. Phys. J. C}, 82(8):758, 2022.
\newblock \href {https://arxiv.org/abs/2110.01482} {\path{arXiv:2110.01482}}, \href {https://doi.org/10.1140/epjc/s10052-022-10735-w} {\path{doi:10.1140/epjc/s10052-022-10735-w}}.

\bibitem{Kawasaki:2021ycf}
Masahiro Kawasaki and Hiromasa Nakatsuka.
\newblock {Gravitational waves from type II axion-like curvaton model and its implication for NANOGrav result}.
\newblock {\em JCAP}, 05:023, 2021.
\newblock \href {https://arxiv.org/abs/2101.11244} {\path{arXiv:2101.11244}}, \href {https://doi.org/10.1088/1475-7516/2021/05/023} {\path{doi:10.1088/1475-7516/2021/05/023}}.

\bibitem{Ahmed:2021ucx}
Waqas Ahmed, M.~Junaid, and Umer Zubair.
\newblock {Primordial black holes and gravitational waves in hybrid inflation with chaotic potentials}.
\newblock {\em Nucl. Phys. B}, 984:115968, 2022.
\newblock \href {https://arxiv.org/abs/2109.14838} {\path{arXiv:2109.14838}}, \href {https://doi.org/10.1016/j.nuclphysb.2022.115968} {\path{doi:10.1016/j.nuclphysb.2022.115968}}.

\bibitem{Yi:2022ymw}
Zhu Yi and Qin Fei.
\newblock {Constraints on primordial curvature spectrum from primordial black holes and scalar-induced gravitational waves}.
\newblock {\em Eur. Phys. J. C}, 83(1):82, 2023.
\newblock \href {https://arxiv.org/abs/2210.03641} {\path{arXiv:2210.03641}}, \href {https://doi.org/10.1140/epjc/s10052-023-11233-3} {\path{doi:10.1140/epjc/s10052-023-11233-3}}.

\bibitem{Yi:2022anu}
Zhu Yi.
\newblock {Primordial black holes and scalar-induced gravitational waves from the generalized Brans-Dicke theory}.
\newblock {\em JCAP}, 03:048, 2023.
\newblock \href {https://arxiv.org/abs/2206.01039} {\path{arXiv:2206.01039}}, \href {https://doi.org/10.1088/1475-7516/2023/03/048} {\path{doi:10.1088/1475-7516/2023/03/048}}.

\bibitem{Zhao:2023xnh}
Ji-Xiang Zhao, Xiao-Hui Liu, and Nan Li.
\newblock {Primordial black holes and scalar-induced gravitational waves from the perturbations on the inflaton potential in peak theory}.
\newblock {\em Phys. Rev. D}, 107(4):043515, 2023.
\newblock \href {https://arxiv.org/abs/2302.06886} {\path{arXiv:2302.06886}}, \href {https://doi.org/10.1103/PhysRevD.107.043515} {\path{doi:10.1103/PhysRevD.107.043515}}.

\bibitem{Cai:2023uhc}
Yong Cai, Mian Zhu, and Yun-Song Piao.
\newblock {Primordial Black Holes from Null Energy Condition Violation during Inflation}.
\newblock {\em Phys. Rev. Lett.}, 133(2):021001, 2024.
\newblock \href {https://arxiv.org/abs/2305.10933} {\path{arXiv:2305.10933}}, \href {https://doi.org/10.1103/PhysRevLett.133.021001} {\path{doi:10.1103/PhysRevLett.133.021001}}.

\bibitem{Goncharov:2021oub}
Boris Goncharov et~al.
\newblock {On the Evidence for a Common-spectrum Process in the Search for the Nanohertz Gravitational-wave Background with the Parkes Pulsar Timing Array}.
\newblock {\em Astrophys. J. Lett.}, 917(2):L19, 2021.
\newblock \href {https://arxiv.org/abs/2107.12112} {\path{arXiv:2107.12112}}, \href {https://doi.org/10.3847/2041-8213/ac17f4} {\path{doi:10.3847/2041-8213/ac17f4}}.

\bibitem{Chen:2021rqp}
S.~Chen et~al.
\newblock {Common-red-signal analysis with 24-yr high-precision timing of the European Pulsar Timing Array: inferences in the stochastic gravitational-wave background search}.
\newblock {\em Mon. Not. Roy. Astron. Soc.}, 508(4):4970--4993, 2021.
\newblock \href {https://arxiv.org/abs/2110.13184} {\path{arXiv:2110.13184}}, \href {https://doi.org/10.1093/mnras/stab2833} {\path{doi:10.1093/mnras/stab2833}}.

\bibitem{Antoniadis:2022pcn}
J.~Antoniadis et~al.
\newblock {The International Pulsar Timing Array second data release: Search for an isotropic gravitational wave background}.
\newblock {\em Mon. Not. Roy. Astron. Soc.}, 510(4):4873--4887, 2022.
\newblock \href {https://arxiv.org/abs/2201.03980} {\path{arXiv:2201.03980}}, \href {https://doi.org/10.1093/mnras/stab3418} {\path{doi:10.1093/mnras/stab3418}}.

\bibitem{Hajkarim:2019nbx}
Fazlollah Hajkarim and J\"urgen Schaffner-Bielich.
\newblock {Thermal History of the Early Universe and Primordial Gravitational Waves from Induced Scalar Perturbations}.
\newblock {\em Phys. Rev. D}, 101(4):043522, 2020.
\newblock \href {https://arxiv.org/abs/1910.12357} {\path{arXiv:1910.12357}}, \href {https://doi.org/10.1103/PhysRevD.101.043522} {\path{doi:10.1103/PhysRevD.101.043522}}.

\bibitem{Franciolini:2023wjm}
Gabriele Franciolini, Davide Racco, and Fabrizio Rompineve.
\newblock {Footprints of the QCD Crossover on Cosmological Gravitational Waves at Pulsar Timing Arrays}.
\newblock {\em Phys. Rev. Lett.}, 132(8):081001, 2024.
\newblock \href {https://arxiv.org/abs/2306.17136} {\path{arXiv:2306.17136}}, \href {https://doi.org/10.1103/PhysRevLett.132.081001} {\path{doi:10.1103/PhysRevLett.132.081001}}.

\bibitem{Ferreira:1997hj}
Pedro~G. Ferreira and Michael Joyce.
\newblock {Cosmology with a primordial scaling field}.
\newblock {\em Phys. Rev. D}, 58:023503, 1998.
\newblock \href {https://arxiv.org/abs/astro-ph/9711102} {\path{arXiv:astro-ph/9711102}}, \href {https://doi.org/10.1103/PhysRevD.58.023503} {\path{doi:10.1103/PhysRevD.58.023503}}.

\bibitem{Pallis:2005bb}
C.~Pallis.
\newblock {Kination-dominated reheating and cold dark matter abundance}.
\newblock {\em Nucl. Phys. B}, 751:129--159, 2006.
\newblock \href {https://arxiv.org/abs/hep-ph/0510234} {\path{arXiv:hep-ph/0510234}}, \href {https://doi.org/10.1016/j.nuclphysb.2006.06.003} {\path{doi:10.1016/j.nuclphysb.2006.06.003}}.

\bibitem{Redmond:2018xty}
Kayla Redmond, Anthony Trezza, and Adrienne~L. Erickcek.
\newblock {Growth of Dark Matter Perturbations during Kination}.
\newblock {\em Phys. Rev. D}, 98(6):063504, 2018.
\newblock \href {https://arxiv.org/abs/1807.01327} {\path{arXiv:1807.01327}}, \href {https://doi.org/10.1103/PhysRevD.98.063504} {\path{doi:10.1103/PhysRevD.98.063504}}.

\bibitem{Co:2021lkc}
Raymond~T. Co, David Dunsky, Nicolas Fernandez, Akshay Ghalsasi, Lawrence~J. Hall, Keisuke Harigaya, and Jessie Shelton.
\newblock {Gravitational wave and CMB probes of axion kination}.
\newblock {\em JHEP}, 09:116, 2022.
\newblock \href {https://arxiv.org/abs/2108.09299} {\path{arXiv:2108.09299}}, \href {https://doi.org/10.1007/JHEP09(2022)116} {\path{doi:10.1007/JHEP09(2022)116}}.

\bibitem{Gouttenoire:2021jhk}
Yann Gouttenoire, Geraldine Servant, and Peera Simakachorn.
\newblock {Kination cosmology from scalar fields and gravitational-wave signatures}.
\newblock 11 2021.
\newblock \href {https://arxiv.org/abs/2111.01150} {\path{arXiv:2111.01150}}.

\bibitem{Chang:2021afa}
Chia-Feng Chang and Yanou Cui.
\newblock {Gravitational waves from global cosmic strings and cosmic archaeology}.
\newblock {\em JHEP}, 03:114, 2022.
\newblock \href {https://arxiv.org/abs/2106.09746} {\path{arXiv:2106.09746}}, \href {https://doi.org/10.1007/JHEP03(2022)114} {\path{doi:10.1007/JHEP03(2022)114}}.

\bibitem{Dalianis:2019asr}
Ioannis Dalianis and George Tringas.
\newblock {Primordial black hole remnants as dark matter produced in thermal, matter, and runaway-quintessence postinflationary scenarios}.
\newblock {\em Phys. Rev. D}, 100(8):083512, 2019.
\newblock \href {https://arxiv.org/abs/1905.01741} {\path{arXiv:1905.01741}}, \href {https://doi.org/10.1103/PhysRevD.100.083512} {\path{doi:10.1103/PhysRevD.100.083512}}.

\bibitem{Bhattacharya:2019bvk}
Sukannya Bhattacharya, Subhendra Mohanty, and Priyank Parashari.
\newblock {Primordial black holes and gravitational waves in nonstandard cosmologies}.
\newblock {\em Phys. Rev. D}, 102(4):043522, 2020.
\newblock \href {https://arxiv.org/abs/1912.01653} {\path{arXiv:1912.01653}}, \href {https://doi.org/10.1103/PhysRevD.102.043522} {\path{doi:10.1103/PhysRevD.102.043522}}.

\bibitem{Bhattacharya:2020lhc}
Sukannya Bhattacharya, Subhendra Mohanty, and Priyank Parashari.
\newblock {Implications of the NANOGrav result on primordial gravitational waves in nonstandard cosmologies}.
\newblock {\em Phys. Rev. D}, 103(6):063532, 2021.
\newblock \href {https://arxiv.org/abs/2010.05071} {\path{arXiv:2010.05071}}, \href {https://doi.org/10.1103/PhysRevD.103.063532} {\path{doi:10.1103/PhysRevD.103.063532}}.

\bibitem{Ireland:2023avg}
Aurora Ireland, Stefano Profumo, and Jordan Scharnhorst.
\newblock {Primordial gravitational waves from black hole evaporation in standard and nonstandard cosmologies}.
\newblock {\em Phys. Rev. D}, 107(10):104021, 2023.
\newblock \href {https://arxiv.org/abs/2302.10188} {\path{arXiv:2302.10188}}, \href {https://doi.org/10.1103/PhysRevD.107.104021} {\path{doi:10.1103/PhysRevD.107.104021}}.

\bibitem{Bhattacharya:2023ztw}
Sukannya Bhattacharya.
\newblock {Primordial Black Hole Formation in Non-Standard Post-Inflationary Epochs}.
\newblock {\em Galaxies}, 11(1):35, 2023.
\newblock \href {https://arxiv.org/abs/2302.12690} {\path{arXiv:2302.12690}}, \href {https://doi.org/10.3390/galaxies11010035} {\path{doi:10.3390/galaxies11010035}}.

\bibitem{Ghoshal:2023sfa}
Anish Ghoshal, Yann Gouttenoire, Lucien Heurtier, and Peera Simakachorn.
\newblock {Primordial black hole archaeology with gravitational waves from cosmic strings}.
\newblock {\em JHEP}, 08:196, 2023.
\newblock \href {https://arxiv.org/abs/2304.04793} {\path{arXiv:2304.04793}}, \href {https://doi.org/10.1007/JHEP08(2023)196} {\path{doi:10.1007/JHEP08(2023)196}}.

\bibitem{Yuan:2020iwf}
Chen Yuan and Qing-Guo Huang.
\newblock {Gravitational waves induced by the local-type non-Gaussian curvature perturbations}.
\newblock {\em Phys. Lett. B}, 821:136606, 2021.
\newblock \href {https://arxiv.org/abs/2007.10686} {\path{arXiv:2007.10686}}, \href {https://doi.org/10.1016/j.physletb.2021.136606} {\path{doi:10.1016/j.physletb.2021.136606}}.

\bibitem{Atal:2021jyo}
Vicente Atal and Guillem Dom\`enech.
\newblock {Probing non-Gaussianities with the high frequency tail of induced gravitational waves}.
\newblock {\em JCAP}, 06:001, 2021.
\newblock \href {https://arxiv.org/abs/2103.01056} {\path{arXiv:2103.01056}}, \href {https://doi.org/10.1088/1475-7516/2021/06/001} {\path{doi:10.1088/1475-7516/2021/06/001}}.

\bibitem{Chang:2022nzu}
Zhe Chang, Yu-Ting Kuang, Xukun Zhang, and Jing-Zhi Zhou.
\newblock {Primordial black holes and third order scalar induced gravitational waves*}.
\newblock {\em Chin. Phys. C}, 47(5):055104, 2023.
\newblock \href {https://arxiv.org/abs/2209.12404} {\path{arXiv:2209.12404}}, \href {https://doi.org/10.1088/1674-1137/acc649} {\path{doi:10.1088/1674-1137/acc649}}.

\bibitem{Li:2023qua}
Jun-Peng Li, Sai Wang, Zhi-Chao Zhao, and Kazunori Kohri.
\newblock {Primordial non-Gaussianity f $_{NL}$ and anisotropies in scalar-induced gravitational waves}.
\newblock {\em JCAP}, 10:056, 2023.
\newblock \href {https://arxiv.org/abs/2305.19950} {\path{arXiv:2305.19950}}, \href {https://doi.org/10.1088/1475-7516/2023/10/056} {\path{doi:10.1088/1475-7516/2023/10/056}}.

\bibitem{Bartolo:2007vp}
N.~Bartolo, S.~Matarrese, A.~Riotto, and A.~Vaihkonen.
\newblock {The Maximal Amount of Gravitational Waves in the Curvaton Scenario}.
\newblock {\em Phys. Rev. D}, 76:061302, 2007.
\newblock \href {https://arxiv.org/abs/0705.4240} {\path{arXiv:0705.4240}}, \href {https://doi.org/10.1103/PhysRevD.76.061302} {\path{doi:10.1103/PhysRevD.76.061302}}.

\bibitem{NANOGrav:2023hvm}
Adeela Afzal et~al.
\newblock {The NANOGrav 15 yr Data Set: Search for Signals from New Physics}.
\newblock {\em Astrophys. J. Lett.}, 951(1):L11, 2023.
\newblock \href {https://arxiv.org/abs/2306.16219} {\path{arXiv:2306.16219}}, \href {https://doi.org/10.3847/2041-8213/acdc91} {\path{doi:10.3847/2041-8213/acdc91}}.

\bibitem{Pagano:2015hma}
Luca Pagano, Laura Salvati, and Alessandro Melchiorri.
\newblock {New constraints on primordial gravitational waves from Planck 2015}.
\newblock {\em Phys. Lett. B}, 760:823--825, 2016.
\newblock \href {https://arxiv.org/abs/1508.02393} {\path{arXiv:1508.02393}}, \href {https://doi.org/10.1016/j.physletb.2016.07.078} {\path{doi:10.1016/j.physletb.2016.07.078}}.

\bibitem{Urrutia:2023mtk}
Juan Urrutia, Ville Vaskonen, and Hardi Veerm\"ae.
\newblock {Gravitational wave microlensing by dressed primordial black holes}.
\newblock {\em Phys. Rev. D}, 108(2):023507, 2023.
\newblock \href {https://arxiv.org/abs/2303.17601} {\path{arXiv:2303.17601}}, \href {https://doi.org/10.1103/PhysRevD.108.023507} {\path{doi:10.1103/PhysRevD.108.023507}}.

\bibitem{Young:2019osy}
Sam Young.
\newblock {The primordial black hole formation criterion re-examined: Parametrisation, timing and the choice of window function}.
\newblock {\em Int. J. Mod. Phys. D}, 29(02):2030002, 2019.
\newblock \href {https://arxiv.org/abs/1905.01230} {\path{arXiv:1905.01230}}, \href {https://doi.org/10.1142/S0218271820300025} {\path{doi:10.1142/S0218271820300025}}.

\bibitem{NANOGrav:2023hfp}
Gabriella Agazie et~al.
\newblock {The NANOGrav 15 yr Data Set: Constraints on Supermassive Black Hole Binaries from the Gravitational-wave Background}.
\newblock {\em Astrophys. J. Lett.}, 952(2):L37, 2023.
\newblock \href {https://arxiv.org/abs/2306.16220} {\path{arXiv:2306.16220}}, \href {https://doi.org/10.3847/2041-8213/ace18b} {\path{doi:10.3847/2041-8213/ace18b}}.

\bibitem{Ellis:2023dgf}
John Ellis, Malcolm Fairbairn, Gert H\"utsi, Juhan Raidal, Juan Urrutia, Ville Vaskonen, and Hardi Veerm\"ae.
\newblock {Gravitational waves from supermassive black hole binaries in light of the NANOGrav 15-year data}.
\newblock {\em Phys. Rev. D}, 109(2):L021302, 2024.
\newblock \href {https://arxiv.org/abs/2306.17021} {\path{arXiv:2306.17021}}, \href {https://doi.org/10.1103/PhysRevD.109.L021302} {\path{doi:10.1103/PhysRevD.109.L021302}}.

\bibitem{Bi:2023tib}
Yan-Chen Bi, Yu-Mei Wu, Zu-Cheng Chen, and Qing-Guo Huang.
\newblock {Implications for the supermassive black hole binaries from the NANOGrav 15-year data set}.
\newblock {\em Sci. China Phys. Mech. Astron.}, 66(12):120402, 2023.
\newblock \href {https://arxiv.org/abs/2307.00722} {\path{arXiv:2307.00722}}, \href {https://doi.org/10.1007/s11433-023-2252-4} {\path{doi:10.1007/s11433-023-2252-4}}.

\bibitem{Zhang:2023lzt}
Chao Zhang, Ning Dai, Qing Gao, Yungui Gong, Tong Jiang, and Xuchen Lu.
\newblock {Detecting new fundamental fields with pulsar timing arrays}.
\newblock {\em Phys. Rev. D}, 108(10):104069, 2023.
\newblock \href {https://arxiv.org/abs/2307.01093} {\path{arXiv:2307.01093}}, \href {https://doi.org/10.1103/PhysRevD.108.104069} {\path{doi:10.1103/PhysRevD.108.104069}}.

\bibitem{Ellis:2023tsl}
John Ellis, Marek Lewicki, Chunshan Lin, and Ville Vaskonen.
\newblock {Cosmic superstrings revisited in light of NANOGrav 15-year data}.
\newblock {\em Phys. Rev. D}, 108(10):103511, 2023.
\newblock \href {https://arxiv.org/abs/2306.17147} {\path{arXiv:2306.17147}}, \href {https://doi.org/10.1103/PhysRevD.108.103511} {\path{doi:10.1103/PhysRevD.108.103511}}.

\bibitem{Kitajima:2023vre}
Naoya Kitajima and Kazunori Nakayama.
\newblock {Nanohertz gravitational waves from cosmic strings and dark photon dark matter}.
\newblock {\em Phys. Lett. B}, 846:138213, 2023.
\newblock \href {https://arxiv.org/abs/2306.17390} {\path{arXiv:2306.17390}}, \href {https://doi.org/10.1016/j.physletb.2023.138213} {\path{doi:10.1016/j.physletb.2023.138213}}.

\bibitem{Wang:2023len}
Ziwei Wang, Lei Lei, Hao Jiao, Lei Feng, and Yi-Zhong Fan.
\newblock {The nanohertz stochastic gravitational wave background from cosmic string loops and the abundant high redshift massive galaxies}.
\newblock {\em Sci. China Phys. Mech. Astron.}, 66(12):120403, 2023.
\newblock \href {https://arxiv.org/abs/2306.17150} {\path{arXiv:2306.17150}}, \href {https://doi.org/10.1007/s11433-023-2262-0} {\path{doi:10.1007/s11433-023-2262-0}}.

\bibitem{Lazarides:2023ksx}
George Lazarides, Rinku Maji, and Qaisar Shafi.
\newblock {Superheavy quasistable strings and walls bounded by strings in the light of NANOGrav 15~year data}.
\newblock {\em Phys. Rev. D}, 108(9):095041, 2023.
\newblock \href {https://arxiv.org/abs/2306.17788} {\path{arXiv:2306.17788}}, \href {https://doi.org/10.1103/PhysRevD.108.095041} {\path{doi:10.1103/PhysRevD.108.095041}}.

\bibitem{Eichhorn:2023gat}
Astrid Eichhorn, Rafael~R. Lino~dos Santos, and Jo\~ao~Lucas Miqueleto.
\newblock {From quantum gravity to gravitational waves through cosmic strings}.
\newblock {\em Phys. Rev. D}, 109(2):026013, 2024.
\newblock \href {https://arxiv.org/abs/2306.17718} {\path{arXiv:2306.17718}}, \href {https://doi.org/10.1103/PhysRevD.109.026013} {\path{doi:10.1103/PhysRevD.109.026013}}.

\bibitem{Chowdhury:2023opo}
Debika Chowdhury, Gianmassimo Tasinato, and Ivonne Zavala.
\newblock {Dark energy, D-branes and pulsar timing arrays}.
\newblock {\em JCAP}, 11:090, 2023.
\newblock \href {https://arxiv.org/abs/2307.01188} {\path{arXiv:2307.01188}}, \href {https://doi.org/10.1088/1475-7516/2023/11/090} {\path{doi:10.1088/1475-7516/2023/11/090}}.

\bibitem{Servant:2023mwt}
G\'eraldine Servant and Peera Simakachorn.
\newblock {Constraining postinflationary axions with pulsar timing arrays}.
\newblock {\em Phys. Rev. D}, 108(12):123516, 2023.
\newblock \href {https://arxiv.org/abs/2307.03121} {\path{arXiv:2307.03121}}, \href {https://doi.org/10.1103/PhysRevD.108.123516} {\path{doi:10.1103/PhysRevD.108.123516}}.

\bibitem{Antusch:2023zjk}
Stefan Antusch, Kevin Hinze, Shaikh Saad, and Jonathan Steiner.
\newblock {Singling out SO(10) GUT models using recent PTA results}.
\newblock {\em Phys. Rev. D}, 108(9):095053, 2023.
\newblock \href {https://arxiv.org/abs/2307.04595} {\path{arXiv:2307.04595}}, \href {https://doi.org/10.1103/PhysRevD.108.095053} {\path{doi:10.1103/PhysRevD.108.095053}}.

\bibitem{Yamada:2023thl}
Masaki Yamada and Kazuya Yonekura.
\newblock {Dark baryon from pure Yang-Mills theory and its GW signature from cosmic strings}.
\newblock {\em JHEP}, 09:197, 2023.
\newblock \href {https://arxiv.org/abs/2307.06586} {\path{arXiv:2307.06586}}, \href {https://doi.org/10.1007/JHEP09(2023)197} {\path{doi:10.1007/JHEP09(2023)197}}.

\bibitem{Ge:2023rce}
Shuailiang Ge.
\newblock {Stochastic gravitational wave background: birth from string-wall death}.
\newblock {\em JCAP}, 06:064, 2024.
\newblock \href {https://arxiv.org/abs/2307.08185} {\path{arXiv:2307.08185}}, \href {https://doi.org/10.1088/1475-7516/2024/06/064} {\path{doi:10.1088/1475-7516/2024/06/064}}.

\bibitem{Basilakos:2023xof}
Spyros Basilakos, Dimitri~V. Nanopoulos, Theodoros Papanikolaou, Emmanuel~N. Saridakis, and Charalampos Tzerefos.
\newblock {Gravitational wave signatures of no-scale supergravity in NANOGrav and beyond}.
\newblock {\em Phys. Lett. B}, 850:138507, 2024.
\newblock \href {https://arxiv.org/abs/2307.08601} {\path{arXiv:2307.08601}}, \href {https://doi.org/10.1016/j.physletb.2024.138507} {\path{doi:10.1016/j.physletb.2024.138507}}.

\bibitem{Fujikura:2023lkn}
Kohei Fujikura, Sudhakantha Girmohanta, Yuichiro Nakai, and Motoo Suzuki.
\newblock {NANOGrav signal from a dark conformal phase transition}.
\newblock {\em Phys. Lett. B}, 846:138203, 2023.
\newblock \href {https://arxiv.org/abs/2306.17086} {\path{arXiv:2306.17086}}, \href {https://doi.org/10.1016/j.physletb.2023.138203} {\path{doi:10.1016/j.physletb.2023.138203}}.

\bibitem{Addazi:2023jvg}
Andrea Addazi, Yi-Fu Cai, Antonino Marciano, and Luca Visinelli.
\newblock {Have pulsar timing array methods detected a cosmological phase transition?}
\newblock {\em Phys. Rev. D}, 109(1):015028, 2024.
\newblock \href {https://arxiv.org/abs/2306.17205} {\path{arXiv:2306.17205}}, \href {https://doi.org/10.1103/PhysRevD.109.015028} {\path{doi:10.1103/PhysRevD.109.015028}}.

\bibitem{Bai:2023cqj}
Yang Bai, Ting-Kuo Chen, and Mrunal Korwar.
\newblock {QCD-collapsed domain walls: QCD phase transition and gravitational wave spectroscopy}.
\newblock {\em JHEP}, 12:194, 2023.
\newblock \href {https://arxiv.org/abs/2306.17160} {\path{arXiv:2306.17160}}, \href {https://doi.org/10.1007/JHEP12(2023)194} {\path{doi:10.1007/JHEP12(2023)194}}.

\bibitem{Megias:2023kiy}
Eugenio Megias, Germano Nardini, and Mariano Quiros.
\newblock {Pulsar timing array stochastic background from light Kaluza-Klein resonances}.
\newblock {\em Phys. Rev. D}, 108(9):095017, 2023.
\newblock \href {https://arxiv.org/abs/2306.17071} {\path{arXiv:2306.17071}}, \href {https://doi.org/10.1103/PhysRevD.108.095017} {\path{doi:10.1103/PhysRevD.108.095017}}.

\bibitem{Han:2023olf}
Chengcheng Han, Ke-Pan Xie, Jin~Min Yang, and Mengchao Zhang.
\newblock {Self-interacting dark matter implied by nano-Hertz gravitational waves}.
\newblock {\em Phys. Rev. D}, 109(11):115025, 2024.
\newblock \href {https://arxiv.org/abs/2306.16966} {\path{arXiv:2306.16966}}, \href {https://doi.org/10.1103/PhysRevD.109.115025} {\path{doi:10.1103/PhysRevD.109.115025}}.

\bibitem{Zu:2023olm}
Lei Zu, Chi Zhang, Yao-Yu Li, Yuchao Gu, Yue-Lin~Sming Tsai, and Yi-Zhong Fan.
\newblock {Mirror QCD phase transition as the origin of the nanohertz Stochastic Gravitational-Wave Background}.
\newblock {\em Sci. Bull.}, 69:741--746, 2024.
\newblock \href {https://arxiv.org/abs/2306.16769} {\path{arXiv:2306.16769}}, \href {https://doi.org/10.1016/j.scib.2024.01.037} {\path{doi:10.1016/j.scib.2024.01.037}}.

\bibitem{Ghosh:2023aum}
Tathagata Ghosh, Anish Ghoshal, Huai-Ke Guo, Fazlollah Hajkarim, Stephen~F. King, Kuver Sinha, Xin Wang, and Graham White.
\newblock {Did we hear the sound of the Universe boiling? Analysis using the full fluid velocity profiles and NANOGrav 15-year data}.
\newblock {\em JCAP}, 05:100, 2024.
\newblock \href {https://arxiv.org/abs/2307.02259} {\path{arXiv:2307.02259}}, \href {https://doi.org/10.1088/1475-7516/2024/05/100} {\path{doi:10.1088/1475-7516/2024/05/100}}.

\bibitem{Xiao:2023dbb}
Yang Xiao, Jin~Min Yang, and Yang Zhang.
\newblock {Implications of nano-Hertz gravitational waves on electroweak phase transition in the singlet dark matter model}.
\newblock {\em Sci. Bull.}, 68:3158--3164, 2023.
\newblock \href {https://arxiv.org/abs/2307.01072} {\path{arXiv:2307.01072}}, \href {https://doi.org/10.1016/j.scib.2023.11.025} {\path{doi:10.1016/j.scib.2023.11.025}}.

\bibitem{Li:2023bxy}
Shao-Ping Li and Ke-Pan Xie.
\newblock {Collider test of nano-Hertz gravitational waves from pulsar timing arrays}.
\newblock {\em Phys. Rev. D}, 108(5):055018, 2023.
\newblock \href {https://arxiv.org/abs/2307.01086} {\path{arXiv:2307.01086}}, \href {https://doi.org/10.1103/PhysRevD.108.055018} {\path{doi:10.1103/PhysRevD.108.055018}}.

\bibitem{DiBari:2023upq}
Pasquale Di~Bari and Moinul~Hossain Rahat.
\newblock {Split Majoron model confronts the NANOGrav signal and cosmological tensions}.
\newblock {\em Phys. Rev. D}, 110(5):055019, 2024.
\newblock \href {https://arxiv.org/abs/2307.03184} {\path{arXiv:2307.03184}}, \href {https://doi.org/10.1103/PhysRevD.110.055019} {\path{doi:10.1103/PhysRevD.110.055019}}.

\bibitem{Cruz:2023lnq}
Juan~S. Cruz, Florian Niedermann, and Martin~S. Sloth.
\newblock {NANOGrav meets Hot New Early Dark Energy and the origin of neutrino mass}.
\newblock {\em Phys. Lett. B}, 846:138202, 2023.
\newblock \href {https://arxiv.org/abs/2307.03091} {\path{arXiv:2307.03091}}, \href {https://doi.org/10.1016/j.physletb.2023.138202} {\path{doi:10.1016/j.physletb.2023.138202}}.

\bibitem{Gouttenoire:2023bqy}
Yann Gouttenoire.
\newblock {First-Order Phase Transition Interpretation of Pulsar Timing Array Signal Is Consistent with Solar-Mass Black Holes}.
\newblock {\em Phys. Rev. Lett.}, 131(17):171404, 2023.
\newblock \href {https://arxiv.org/abs/2307.04239} {\path{arXiv:2307.04239}}, \href {https://doi.org/10.1103/PhysRevLett.131.171404} {\path{doi:10.1103/PhysRevLett.131.171404}}.

\bibitem{Ahmadvand:2023lpp}
Moslem Ahmadvand, Ligong Bian, and Soroush Shakeri.
\newblock {Heavy QCD axion model in light of pulsar timing arrays}.
\newblock {\em Phys. Rev. D}, 108(11):115020, 2023.
\newblock \href {https://arxiv.org/abs/2307.12385} {\path{arXiv:2307.12385}}, \href {https://doi.org/10.1103/PhysRevD.108.115020} {\path{doi:10.1103/PhysRevD.108.115020}}.

\bibitem{An:2023jxf}
Haipeng An, Boye Su, Hanwen Tai, Lian-Tao Wang, and Chen Yang.
\newblock {Phase transition during inflation and the gravitational wave signal at pulsar timing arrays}.
\newblock {\em Phys. Rev. D}, 109(12):L121304, 2024.
\newblock \href {https://arxiv.org/abs/2308.00070} {\path{arXiv:2308.00070}}, \href {https://doi.org/10.1103/PhysRevD.109.L121304} {\path{doi:10.1103/PhysRevD.109.L121304}}.

\bibitem{Kitajima:2023cek}
Naoya Kitajima, Junseok Lee, Kai Murai, Fuminobu Takahashi, and Wen Yin.
\newblock {Gravitational waves from domain wall collapse, and application to nanohertz signals with QCD-coupled axions}.
\newblock {\em Phys. Lett. B}, 851:138586, 2024.
\newblock \href {https://arxiv.org/abs/2306.17146} {\path{arXiv:2306.17146}}, \href {https://doi.org/10.1016/j.physletb.2024.138586} {\path{doi:10.1016/j.physletb.2024.138586}}.

\bibitem{Guo:2023hyp}
Shu-Yuan Guo, Maxim Khlopov, Xuewen Liu, Lei Wu, Yongcheng Wu, and Bin Zhu.
\newblock {Footprints of axion-like particle in pulsar timing array data and James Webb Space Telescope observations}.
\newblock {\em Sci. China Phys. Mech. Astron.}, 67(11):111011, 2024.
\newblock \href {https://arxiv.org/abs/2306.17022} {\path{arXiv:2306.17022}}, \href {https://doi.org/10.1007/s11433-024-2445-1} {\path{doi:10.1007/s11433-024-2445-1}}.

\bibitem{Blasi:2023sej}
Simone Blasi, Alberto Mariotti, A\"aron Rase, and Alexander Sevrin.
\newblock {Axionic domain walls at Pulsar Timing Arrays: QCD bias and particle friction}.
\newblock {\em JHEP}, 11:169, 2023.
\newblock \href {https://arxiv.org/abs/2306.17830} {\path{arXiv:2306.17830}}, \href {https://doi.org/10.1007/JHEP11(2023)169} {\path{doi:10.1007/JHEP11(2023)169}}.

\bibitem{Gouttenoire:2023ftk}
Yann Gouttenoire and Edoardo Vitagliano.
\newblock {Domain wall interpretation of the PTA signal confronting black hole overproduction}.
\newblock {\em Phys. Rev. D}, 110(6):L061306, 2024.
\newblock \href {https://arxiv.org/abs/2306.17841} {\path{arXiv:2306.17841}}, \href {https://doi.org/10.1103/PhysRevD.110.L061306} {\path{doi:10.1103/PhysRevD.110.L061306}}.

\bibitem{Barman:2023fad}
Basabendu Barman, Debasish Borah, Suruj Jyoti~Das, and Indrajit Saha.
\newblock {Scale of Dirac leptogenesis and left-right symmetry in the light of recent PTA results}.
\newblock {\em JCAP}, 10:053, 2023.
\newblock \href {https://arxiv.org/abs/2307.00656} {\path{arXiv:2307.00656}}, \href {https://doi.org/10.1088/1475-7516/2023/10/053} {\path{doi:10.1088/1475-7516/2023/10/053}}.

\bibitem{Lu:2023mcz}
Bo-Qiang Lu, Cheng-Wei Chiang, and Tianjun Li.
\newblock {Clockwork axion footprint on nanohertz stochastic gravitational wave background}.
\newblock {\em Phys. Rev. D}, 109(10):L101304, 2024.
\newblock \href {https://arxiv.org/abs/2307.00746} {\path{arXiv:2307.00746}}, \href {https://doi.org/10.1103/PhysRevD.109.L101304} {\path{doi:10.1103/PhysRevD.109.L101304}}.

\bibitem{Babichev:2023pbf}
E.~Babichev, D.~Gorbunov, S.~Ramazanov, R.~Samanta, and A.~Vikman.
\newblock {NANOGrav spectral index \ensuremath{\gamma}=3 from melting domain walls}.
\newblock {\em Phys. Rev. D}, 108(12):123529, 2023.
\newblock \href {https://arxiv.org/abs/2307.04582} {\path{arXiv:2307.04582}}, \href {https://doi.org/10.1103/PhysRevD.108.123529} {\path{doi:10.1103/PhysRevD.108.123529}}.

\bibitem{Gelmini:2023kvo}
Graciela~B. Gelmini and Jonah Hyman.
\newblock {Catastrogenesis with unstable ALPs as the origin of the NANOGrav 15 yr gravitational wave signal}.
\newblock {\em Phys. Lett. B}, 848:138356, 2024.
\newblock \href {https://arxiv.org/abs/2307.07665} {\path{arXiv:2307.07665}}, \href {https://doi.org/10.1016/j.physletb.2023.138356} {\path{doi:10.1016/j.physletb.2023.138356}}.

\bibitem{Zhang:2023nrs}
Zhao Zhang, Chengfeng Cai, Yu-Hang Su, Shiyu Wang, Zhao-Huan Yu, and Hong-Hao Zhang.
\newblock {Nano-Hertz gravitational waves from collapsing domain walls associated with freeze-in dark matter in light of pulsar timing array observations}.
\newblock {\em Phys. Rev. D}, 108(9):095037, 2023.
\newblock \href {https://arxiv.org/abs/2307.11495} {\path{arXiv:2307.11495}}, \href {https://doi.org/10.1103/PhysRevD.108.095037} {\path{doi:10.1103/PhysRevD.108.095037}}.

\bibitem{Vagnozzi:2023lwo}
Sunny Vagnozzi.
\newblock {Inflationary interpretation of the stochastic gravitational wave background signal detected by pulsar timing array experiments}.
\newblock {\em JHEAp}, 39:81--98, 2023.
\newblock \href {https://arxiv.org/abs/2306.16912} {\path{arXiv:2306.16912}}, \href {https://doi.org/10.1016/j.jheap.2023.07.001} {\path{doi:10.1016/j.jheap.2023.07.001}}.

\bibitem{Inomata:2023zup}
Keisuke Inomata, Kazunori Kohri, and Takahiro Terada.
\newblock {Detected stochastic gravitational waves and subsolar-mass primordial black holes}.
\newblock {\em Phys. Rev. D}, 109(6):063506, 2024.
\newblock \href {https://arxiv.org/abs/2306.17834} {\path{arXiv:2306.17834}}, \href {https://doi.org/10.1103/PhysRevD.109.063506} {\path{doi:10.1103/PhysRevD.109.063506}}.

\bibitem{Cai:2023dls}
Yi-Fu Cai, Xin-Chen He, Xiao-Han Ma, Sheng-Feng Yan, and Guan-Wen Yuan.
\newblock {Limits on scalar-induced gravitational waves from the stochastic background by pulsar timing array observations}.
\newblock {\em Sci. Bull.}, 68:2929--2935, 2023.
\newblock \href {https://arxiv.org/abs/2306.17822} {\path{arXiv:2306.17822}}, \href {https://doi.org/10.1016/j.scib.2023.10.027} {\path{doi:10.1016/j.scib.2023.10.027}}.

\bibitem{Wang:2023ost}
Sai Wang, Zhi-Chao Zhao, Jun-Peng Li, and Qing-Hua Zhu.
\newblock {Implications of pulsar timing array data for scalar-induced gravitational waves and primordial black holes: Primordial non-Gaussianity fNL considered}.
\newblock {\em Phys. Rev. Res.}, 6(1):L012060, 2024.
\newblock \href {https://arxiv.org/abs/2307.00572} {\path{arXiv:2307.00572}}, \href {https://doi.org/10.1103/PhysRevResearch.6.L012060} {\path{doi:10.1103/PhysRevResearch.6.L012060}}.

\bibitem{Ebadi:2023xhq}
Reza Ebadi, Soubhik Kumar, Amara McCune, Hanwen Tai, and Lian-Tao Wang.
\newblock {Gravitational waves from stochastic scalar fluctuations}.
\newblock {\em Phys. Rev. D}, 109(8):083519, 2024.
\newblock \href {https://arxiv.org/abs/2307.01248} {\path{arXiv:2307.01248}}, \href {https://doi.org/10.1103/PhysRevD.109.083519} {\path{doi:10.1103/PhysRevD.109.083519}}.

\bibitem{Gouttenoire:2023nzr}
Yann Gouttenoire, Sokratis Trifinopoulos, Georgios Valogiannis, and Miguel Vanvlasselaer.
\newblock {Scrutinizing the primordial black hole interpretation of PTA gravitational waves and JWST early galaxies}.
\newblock {\em Phys. Rev. D}, 109(12):123002, 2024.
\newblock \href {https://arxiv.org/abs/2307.01457} {\path{arXiv:2307.01457}}, \href {https://doi.org/10.1103/PhysRevD.109.123002} {\path{doi:10.1103/PhysRevD.109.123002}}.

\bibitem{Liu:2023ymk}
Lang Liu, Zu-Cheng Chen, and Qing-Guo Huang.
\newblock {Implications for the non-Gaussianity of curvature perturbation from pulsar timing arrays}.
\newblock {\em Phys. Rev. D}, 109(6):L061301, 2024.
\newblock \href {https://arxiv.org/abs/2307.01102} {\path{arXiv:2307.01102}}, \href {https://doi.org/10.1103/PhysRevD.109.L061301} {\path{doi:10.1103/PhysRevD.109.L061301}}.

\bibitem{Abe:2023yrw}
Katsuya~T. Abe and Yuichiro Tada.
\newblock {Translating nano-Hertz gravitational wave background into primordial perturbations taking account of the cosmological QCD phase transition}.
\newblock {\em Phys. Rev. D}, 108(10):L101304, 2023.
\newblock \href {https://arxiv.org/abs/2307.01653} {\path{arXiv:2307.01653}}, \href {https://doi.org/10.1103/PhysRevD.108.L101304} {\path{doi:10.1103/PhysRevD.108.L101304}}.

\bibitem{Unal:2023srk}
Caner Unal, Alexandros Papageorgiou, and Ippei Obata.
\newblock {Axion-gauge dynamics during inflation as the origin of pulsar timing array signals and primordial black holes}.
\newblock {\em Phys. Lett. B}, 856:138873, 2024.
\newblock \href {https://arxiv.org/abs/2307.02322} {\path{arXiv:2307.02322}}, \href {https://doi.org/10.1016/j.physletb.2024.138873} {\path{doi:10.1016/j.physletb.2024.138873}}.

\bibitem{Yi:2023mbm}
Zhu Yi, Qing Gao, Yungui Gong, Yue Wang, and Fengge Zhang.
\newblock {Scalar induced gravitational waves in light of Pulsar Timing Array data}.
\newblock {\em Sci. China Phys. Mech. Astron.}, 66(12):120404, 2023.
\newblock \href {https://arxiv.org/abs/2307.02467} {\path{arXiv:2307.02467}}, \href {https://doi.org/10.1007/s11433-023-2266-1} {\path{doi:10.1007/s11433-023-2266-1}}.

\bibitem{Firouzjahi:2023lzg}
Hassan Firouzjahi and Alireza Talebian.
\newblock {Induced gravitational waves from ultra slow-roll inflation and pulsar timing arrays observations}.
\newblock {\em JCAP}, 10:032, 2023.
\newblock \href {https://arxiv.org/abs/2307.03164} {\path{arXiv:2307.03164}}, \href {https://doi.org/10.1088/1475-7516/2023/10/032} {\path{doi:10.1088/1475-7516/2023/10/032}}.

\bibitem{Salvio:2023ynn}
Alberto Salvio.
\newblock {Supercooling in radiative symmetry breaking: theory extensions, gravitational wave detection and primordial black holes}.
\newblock {\em JCAP}, 12:046, 2023.
\newblock \href {https://arxiv.org/abs/2307.04694} {\path{arXiv:2307.04694}}, \href {https://doi.org/10.1088/1475-7516/2023/12/046} {\path{doi:10.1088/1475-7516/2023/12/046}}.

\bibitem{You:2023rmn}
Zhi-Qiang You, Zhu Yi, and You Wu.
\newblock {Constraints on primordial curvature power spectrum with pulsar timing arrays}.
\newblock {\em JCAP}, 11:065, 2023.
\newblock \href {https://arxiv.org/abs/2307.04419} {\path{arXiv:2307.04419}}, \href {https://doi.org/10.1088/1475-7516/2023/11/065} {\path{doi:10.1088/1475-7516/2023/11/065}}.

\bibitem{Bari:2023rcw}
Pritha Bari, Nicola Bartolo, Guillem Dom\`enech, and Sabino Matarrese.
\newblock {Gravitational waves induced by scalar-tensor mixing}.
\newblock {\em Phys. Rev. D}, 109(2):023509, 2024.
\newblock \href {https://arxiv.org/abs/2307.05404} {\path{arXiv:2307.05404}}, \href {https://doi.org/10.1103/PhysRevD.109.023509} {\path{doi:10.1103/PhysRevD.109.023509}}.

\bibitem{Ye:2023xyr}
Gen Ye and Alessandra Silvestri.
\newblock {Can the Gravitational Wave Background Feel Wiggles in Spacetime?}
\newblock {\em Astrophys. J. Lett.}, 963(1):L15, 2024.
\newblock \href {https://arxiv.org/abs/2307.05455} {\path{arXiv:2307.05455}}, \href {https://doi.org/10.3847/2041-8213/ad2851} {\path{doi:10.3847/2041-8213/ad2851}}.

\bibitem{HosseiniMansoori:2023mqh}
Seyed~Ali Hosseini~Mansoori, Fereshteh Felegray, Alireza Talebian, and Mohammad Sami.
\newblock {PBHs and GWs from \ensuremath{\mathbb{T}}$^{2}$-inflation and NANOGrav 15-year data}.
\newblock {\em JCAP}, 08:067, 2023.
\newblock \href {https://arxiv.org/abs/2307.06757} {\path{arXiv:2307.06757}}, \href {https://doi.org/10.1088/1475-7516/2023/08/067} {\path{doi:10.1088/1475-7516/2023/08/067}}.

\bibitem{Cheung:2023ihl}
Kingman Cheung, C.~J. Ouseph, and Po-Yan Tseng.
\newblock {NANOGrav and other PTA signals and PBH from the modified Higgs inflation}.
\newblock {\em Eur. Phys. J. C}, 84(9):906, 2024.
\newblock \href {https://arxiv.org/abs/2307.08046} {\path{arXiv:2307.08046}}, \href {https://doi.org/10.1140/epjc/s10052-024-13268-6} {\path{doi:10.1140/epjc/s10052-024-13268-6}}.

\bibitem{Balaji:2023ehk}
Shyam Balaji, Guillem Dom\`enech, and Gabriele Franciolini.
\newblock {Scalar-induced gravitational wave interpretation of PTA data: the role of scalar fluctuation propagation speed}.
\newblock {\em JCAP}, 10:041, 2023.
\newblock \href {https://arxiv.org/abs/2307.08552} {\path{arXiv:2307.08552}}, \href {https://doi.org/10.1088/1475-7516/2023/10/041} {\path{doi:10.1088/1475-7516/2023/10/041}}.

\bibitem{Jin:2023wri}
Jia-Heng Jin, Zu-Cheng Chen, Zhu Yi, Zhi-Qiang You, Lang Liu, and You Wu.
\newblock {Confronting sound speed resonance with pulsar timing arrays}.
\newblock {\em JCAP}, 09:016, 2023.
\newblock \href {https://arxiv.org/abs/2307.08687} {\path{arXiv:2307.08687}}, \href {https://doi.org/10.1088/1475-7516/2023/09/016} {\path{doi:10.1088/1475-7516/2023/09/016}}.

\bibitem{Das:2023nmm}
Barnali Das, Nur Jaman, and M.~Sami.
\newblock {Gravitational wave background from quintessential inflation and NANOGrav data}.
\newblock {\em Phys. Rev. D}, 108(10):103510, 2023.
\newblock \href {https://arxiv.org/abs/2307.12913} {\path{arXiv:2307.12913}}, \href {https://doi.org/10.1103/PhysRevD.108.103510} {\path{doi:10.1103/PhysRevD.108.103510}}.

\bibitem{Ben-Dayan:2023lwd}
Ido Ben-Dayan, Utkarsh Kumar, Udaykrishna Thattarampilly, and Amresh Verma.
\newblock {Probing the early Universe cosmology with NANOGrav: Possibilities and limitations}.
\newblock {\em Phys. Rev. D}, 108(10):103507, 2023.
\newblock \href {https://arxiv.org/abs/2307.15123} {\path{arXiv:2307.15123}}, \href {https://doi.org/10.1103/PhysRevD.108.103507} {\path{doi:10.1103/PhysRevD.108.103507}}.

\bibitem{Jiang:2023gfe}
Jun-Qian Jiang, Yong Cai, Gen Ye, and Yun-Song Piao.
\newblock {Broken blue-tilted inflationary gravitational waves: a joint analysis of NANOGrav 15-year and BICEP/Keck 2018 data}.
\newblock {\em JCAP}, 05:004, 2024.
\newblock \href {https://arxiv.org/abs/2307.15547} {\path{arXiv:2307.15547}}, \href {https://doi.org/10.1088/1475-7516/2024/05/004} {\path{doi:10.1088/1475-7516/2024/05/004}}.

\bibitem{Liu:2023pau}
Lang Liu, Zu-Cheng Chen, and Qing-Guo Huang.
\newblock {Probing the equation of state of the early Universe with pulsar timing arrays}.
\newblock {\em JCAP}, 11:071, 2023.
\newblock \href {https://arxiv.org/abs/2307.14911} {\path{arXiv:2307.14911}}, \href {https://doi.org/10.1088/1475-7516/2023/11/071} {\path{doi:10.1088/1475-7516/2023/11/071}}.

\bibitem{Yi:2023tdk}
Zhu Yi, Zhi-Qiang You, and You Wu.
\newblock {Model-independent reconstruction of the primordial curvature power spectrum from PTA data}.
\newblock {\em JCAP}, 01:066, 2024.
\newblock \href {https://arxiv.org/abs/2308.05632} {\path{arXiv:2308.05632}}, \href {https://doi.org/10.1088/1475-7516/2024/01/066} {\path{doi:10.1088/1475-7516/2024/01/066}}.

\bibitem{Bhaumik:2023wmw}
Nilanjandev Bhaumik, Rajeev~Kumar Jain, and Marek Lewicki.
\newblock {Ultralow mass primordial black holes in the early Universe can explain the pulsar timing array signal}.
\newblock {\em Phys. Rev. D}, 108(12):123532, 2023.
\newblock \href {https://arxiv.org/abs/2308.07912} {\path{arXiv:2308.07912}}, \href {https://doi.org/10.1103/PhysRevD.108.123532} {\path{doi:10.1103/PhysRevD.108.123532}}.

\bibitem{Yuan:2023ofl}
Chen Yuan, De-Shuang Meng, and Qing-Guo Huang.
\newblock {Full analysis of the scalar-induced gravitational waves for the curvature perturbation with local-type non-Gaussianities}.
\newblock {\em JCAP}, 12:036, 2023.
\newblock \href {https://arxiv.org/abs/2308.07155} {\path{arXiv:2308.07155}}, \href {https://doi.org/10.1088/1475-7516/2023/12/036} {\path{doi:10.1088/1475-7516/2023/12/036}}.

\bibitem{Gorji:2023sil}
Mohammad~Ali Gorji, Misao Sasaki, and Teruaki Suyama.
\newblock {Extra-tensor-induced origin for the PTA signal: No primordial black hole production}.
\newblock {\em Phys. Lett. B}, 846:138214, 2023.
\newblock \href {https://arxiv.org/abs/2307.13109} {\path{arXiv:2307.13109}}, \href {https://doi.org/10.1016/j.physletb.2023.138214} {\path{doi:10.1016/j.physletb.2023.138214}}.

\bibitem{Figueroa:2023zhu}
Daniel~G. Figueroa, Mauro Pieroni, Angelo Ricciardone, and Peera Simakachorn.
\newblock {Cosmological Background Interpretation of Pulsar Timing Array Data}.
\newblock {\em Phys. Rev. Lett.}, 132(17):171002, 2024.
\newblock \href {https://arxiv.org/abs/2307.02399} {\path{arXiv:2307.02399}}, \href {https://doi.org/10.1103/PhysRevLett.132.171002} {\path{doi:10.1103/PhysRevLett.132.171002}}.

\bibitem{Geller:2023shn}
Michael Geller, Subhajit Ghosh, Sida Lu, and Yuhsin Tsai.
\newblock {Challenges in interpreting the NANOGrav 15-year dataset as early Universe gravitational waves produced by an ALP induced instability}.
\newblock {\em Phys. Rev. D}, 109(6):063537, 2024.
\newblock \href {https://arxiv.org/abs/2307.03724} {\path{arXiv:2307.03724}}, \href {https://doi.org/10.1103/PhysRevD.109.063537} {\path{doi:10.1103/PhysRevD.109.063537}}.

\bibitem{Li:2023yaj}
Yao-Yu Li, Chi Zhang, Ziwei Wang, Ming-Yang Cui, Yue-Lin~Sming Tsai, Qiang Yuan, and Yi-Zhong Fan.
\newblock {Primordial magnetic field as a common solution of nanohertz gravitational waves and the Hubble tension}.
\newblock {\em Phys. Rev. D}, 109(4):043538, 2024.
\newblock \href {https://arxiv.org/abs/2306.17124} {\path{arXiv:2306.17124}}, \href {https://doi.org/10.1103/PhysRevD.109.043538} {\path{doi:10.1103/PhysRevD.109.043538}}.

\bibitem{Lambiase:2023pxd}
Gaetano Lambiase, Leonardo Mastrototaro, and Luca Visinelli.
\newblock {Astrophysical neutrino oscillations after pulsar timing array analyses}.
\newblock {\em Phys. Rev. D}, 108(12):123028, 2023.
\newblock \href {https://arxiv.org/abs/2306.16977} {\path{arXiv:2306.16977}}, \href {https://doi.org/10.1103/PhysRevD.108.123028} {\path{doi:10.1103/PhysRevD.108.123028}}.

\bibitem{Borah:2023sbc}
Debasish Borah, Suruj Jyoti~Das, and Rome Samanta.
\newblock {Imprint of inflationary gravitational waves and WIMP dark matter in pulsar timing array data}.
\newblock {\em JCAP}, 03:031, 2024.
\newblock \href {https://arxiv.org/abs/2307.00537} {\path{arXiv:2307.00537}}, \href {https://doi.org/10.1088/1475-7516/2024/03/031} {\path{doi:10.1088/1475-7516/2024/03/031}}.

\bibitem{Datta:2023vbs}
Satyabrata Datta and Rome Samanta.
\newblock {Fingerprints of GeV scale right-handed neutrinos on inflationary gravitational waves and PTA data}.
\newblock {\em Phys. Rev. D}, 108(9):L091706, 2023.
\newblock \href {https://arxiv.org/abs/2307.00646} {\path{arXiv:2307.00646}}, \href {https://doi.org/10.1103/PhysRevD.108.L091706} {\path{doi:10.1103/PhysRevD.108.L091706}}.

\bibitem{Murai:2023gkv}
Kai Murai and Wen Yin.
\newblock {A novel probe of supersymmetry in light of nanohertz gravitational waves}.
\newblock {\em JHEP}, 10:062, 2023.
\newblock \href {https://arxiv.org/abs/2307.00628} {\path{arXiv:2307.00628}}, \href {https://doi.org/10.1007/JHEP10(2023)062} {\path{doi:10.1007/JHEP10(2023)062}}.

\bibitem{Niu:2023bsr}
Xuce Niu and Moinul~Hossain Rahat.
\newblock {NANOGrav signal from axion inflation}.
\newblock {\em Phys. Rev. D}, 108(11):115023, 2023.
\newblock \href {https://arxiv.org/abs/2307.01192} {\path{arXiv:2307.01192}}, \href {https://doi.org/10.1103/PhysRevD.108.115023} {\path{doi:10.1103/PhysRevD.108.115023}}.

\bibitem{Choudhury:2023kam}
Sayantan Choudhury.
\newblock {Single field inflation in the light of Pulsar Timing Array Data: quintessential interpretation of blue tilted tensor spectrum through Non-Bunch Davies initial condition}.
\newblock {\em Eur. Phys. J. C}, 84(3):278, 2024.
\newblock \href {https://arxiv.org/abs/2307.03249} {\path{arXiv:2307.03249}}, \href {https://doi.org/10.1140/epjc/s10052-024-12625-9} {\path{doi:10.1140/epjc/s10052-024-12625-9}}.

\bibitem{Cannizzaro:2023mgc}
Enrico Cannizzaro, Gabriele Franciolini, and Paolo Pani.
\newblock {Novel tests of gravity using nano-Hertz stochastic gravitational-wave background signals}.
\newblock {\em JCAP}, 04:056, 2024.
\newblock \href {https://arxiv.org/abs/2307.11665} {\path{arXiv:2307.11665}}, \href {https://doi.org/10.1088/1475-7516/2024/04/056} {\path{doi:10.1088/1475-7516/2024/04/056}}.

\bibitem{Zhu:2023lbf}
Mian Zhu, Gen Ye, and Yong Cai.
\newblock {Pulsar timing array observations as possible hints for nonsingular cosmology}.
\newblock {\em Eur. Phys. J. C}, 83(9):816, 2023.
\newblock \href {https://arxiv.org/abs/2307.16211} {\path{arXiv:2307.16211}}, \href {https://doi.org/10.1140/epjc/s10052-023-11963-4} {\path{doi:10.1140/epjc/s10052-023-11963-4}}.

\bibitem{Aghaie:2023lan}
Mohammad Aghaie, Giovanni Armando, Alessandro Dondarini, and Paolo Panci.
\newblock {Bounds on ultralight dark matter from NANOGrav}.
\newblock {\em Phys. Rev. D}, 109(10):103030, 2024.
\newblock \href {https://arxiv.org/abs/2308.04590} {\path{arXiv:2308.04590}}, \href {https://doi.org/10.1103/PhysRevD.109.103030} {\path{doi:10.1103/PhysRevD.109.103030}}.

\bibitem{He:2023ado}
Song He, Li~Li, Sai Wang, and Shao-Jiang Wang.
\newblock {Constraints on holographic QCD phase transitions from PTA observations}.
\newblock {\em Sci. China Phys. Mech. Astron.}, 68(1):210411, 2025.
\newblock \href {https://arxiv.org/abs/2308.07257} {\path{arXiv:2308.07257}}, \href {https://doi.org/10.1007/s11433-024-2468-x} {\path{doi:10.1007/s11433-024-2468-x}}.

\bibitem{Bian:2023dnv}
Ligong Bian, Shuailiang Ge, Jing Shu, Bo~Wang, Xing-Yu Yang, and Junchao Zong.
\newblock {Gravitational wave sources for pulsar timing arrays}.
\newblock {\em Phys. Rev. D}, 109(10):L101301, 2024.
\newblock \href {https://arxiv.org/abs/2307.02376} {\path{arXiv:2307.02376}}, \href {https://doi.org/10.1103/PhysRevD.109.L101301} {\path{doi:10.1103/PhysRevD.109.L101301}}.

\bibitem{Wu:2023hsa}
Yu-Mei Wu, Zu-Cheng Chen, and Qing-Guo Huang.
\newblock {Cosmological interpretation for the stochastic signal in pulsar timing arrays}.
\newblock {\em Sci. China Phys. Mech. Astron.}, 67(4):240412, 2024.
\newblock \href {https://arxiv.org/abs/2307.03141} {\path{arXiv:2307.03141}}, \href {https://doi.org/10.1007/s11433-023-2298-7} {\path{doi:10.1007/s11433-023-2298-7}}.

\bibitem{Press:1973iz}
William~H. Press and Paul Schechter.
\newblock {Formation of galaxies and clusters of galaxies by selfsimilar gravitational condensation}.
\newblock {\em Astrophys. J.}, 187:425--438, 1974.
\newblock \href {https://doi.org/10.1086/152650} {\path{doi:10.1086/152650}}.

\bibitem{Bond:1990iw}
J.~R. Bond, S.~Cole, G.~Efstathiou, and Nick Kaiser.
\newblock {Excursion set mass functions for hierarchical Gaussian fluctuations}.
\newblock {\em Astrophys. J.}, 379:440, 1991.
\newblock \href {https://doi.org/10.1086/170520} {\path{doi:10.1086/170520}}.

\bibitem{Lacey:1993iv}
Cedric~G. Lacey and Shaun Cole.
\newblock {Merger rates in hierarchical models of galaxy formation}.
\newblock {\em Mon. Not. Roy. Astron. Soc.}, 262:627--649, 1993.

\bibitem{Kormendy:2013dxa}
John Kormendy and Luis~C. Ho.
\newblock {Coevolution (Or Not) of Supermassive Black Holes and Host Galaxies}.
\newblock {\em Ann. Rev. Astron. Astrophys.}, 51:511--653, 2013.
\newblock \href {https://arxiv.org/abs/1304.7762} {\path{arXiv:1304.7762}}, \href {https://doi.org/10.1146/annurev-astro-082708-101811} {\path{doi:10.1146/annurev-astro-082708-101811}}.

\bibitem{Girelli:2020goz}
Giacomo Girelli, Lucia Pozzetti, Micol Bolzonella, Carlo Giocoli, Federico Marulli, and Marco Baldi.
\newblock {The stellar-to-halo mass relation over the past 12 Gyr: I. Standard $\Lambda$CDM model}.
\newblock {\em Astron. Astrophys.}, 634:A135, 2020.
\newblock \href {https://arxiv.org/abs/2001.02230} {\path{arXiv:2001.02230}}, \href {https://doi.org/10.1051/0004-6361/201936329} {\path{doi:10.1051/0004-6361/201936329}}.

\bibitem{Ellis:2023owy}
John Ellis, Malcolm Fairbairn, Gert H\"utsi, Martti Raidal, Juan Urrutia, Ville Vaskonen, and Hardi Veerm\"ae.
\newblock {Prospects for Future Binary Black Hole GW Studies in Light of PTA Measurements}.
\newblock {\em Astron. Astrophys.}, 676:A38, 2023.
\newblock \href {https://arxiv.org/abs/2301.13854} {\path{arXiv:2301.13854}}, \href {https://doi.org/10.1051/0004-6361/202346268} {\path{doi:10.1051/0004-6361/202346268}}.

\bibitem{Begelman:1980vb}
M.~C. Begelman, R.~D. Blandford, and M.~J. Rees.
\newblock {Massive black hole binaries in active galactic nuclei}.
\newblock {\em Nature}, 287:307--309, 1980.
\newblock \href {https://doi.org/10.1038/287307a0} {\path{doi:10.1038/287307a0}}.

\bibitem{Phinney:2001di}
E.~S. Phinney.
\newblock {A Practical theorem on gravitational wave backgrounds}.
\newblock 7 2001.
\newblock \href {https://arxiv.org/abs/astro-ph/0108028} {\path{arXiv:astro-ph/0108028}}.

\bibitem{lamb2023rapid}
William~G Lamb, Stephen~R Taylor, and Rutger van Haasteren.
\newblock Rapid refitting techniques for bayesian spectral characterization of the gravitational wave background using pulsar timing arrays.
\newblock {\em Physical Review D}, 108(10):103019, 2023.

\bibitem{Enoki:2006kj}
Motohiro Enoki and Masahiro Nagashima.
\newblock {The Effect of Orbital Eccentricity on Gravitational Wave Background Radiation from Cosmological Binaries}.
\newblock {\em Prog. Theor. Phys.}, 117:241, 2007.
\newblock \href {https://arxiv.org/abs/astro-ph/0609377} {\path{arXiv:astro-ph/0609377}}, \href {https://doi.org/10.1143/PTP.117.241} {\path{doi:10.1143/PTP.117.241}}.

\bibitem{LIGOScientific:2023lpe}
A.~G. Abac et~al.
\newblock {Search for Eccentric Black Hole Coalescences during the Third Observing Run of LIGO and Virgo}.
\newblock {\em Astrophys. J.}, 973(2):132, 2024.
\newblock \href {https://arxiv.org/abs/2308.03822} {\path{arXiv:2308.03822}}, \href {https://doi.org/10.3847/1538-4357/ad65ce} {\path{doi:10.3847/1538-4357/ad65ce}}.

\bibitem{Kelley:2017lek}
Luke~Zoltan Kelley, Laura Blecha, Lars Hernquist, Alberto Sesana, and Stephen~R. Taylor.
\newblock {The Gravitational Wave Background from Massive Black Hole Binaries in Illustris: spectral features and time to detection with pulsar timing arrays}.
\newblock {\em Mon. Not. Roy. Astron. Soc.}, 471(4):4508--4526, 2017.
\newblock \href {https://arxiv.org/abs/1702.02180} {\path{arXiv:1702.02180}}, \href {https://doi.org/10.1093/mnras/stx1638} {\path{doi:10.1093/mnras/stx1638}}.

\bibitem{2022arXiv220802794N}
Rohan~P. {Naidu}, Pascal~A. {Oesch}, David~J. {Setton}, Jorryt {Matthee}, Charlie {Conroy}, Benjamin~D. {Johnson}, John~R. {Weaver}, Rychard~J. {Bouwens}, Gabriel~B. {Brammer}, Pratika {Dayal}, Garth~D. {Illingworth}, Laia {Barrufet}, Sirio {Belli}, Rachel {Bezanson}, Sownak {Bose}, Kasper~E. {Heintz}, Joel {Leja}, Ecaterina {Leonova}, Rui {Marques-Chaves}, Mauro {Stefanon}, Sune {Toft}, Arjen {van der Wel}, Pieter {van Dokkum}, Andrea {Weibel}, and Katherine~E. {Whitaker}.
\newblock {Schrodinger's Galaxy Candidate: Puzzlingly Luminous at $z\approx17$, or Dusty/Quenched at $z\approx5$?}
\newblock {\em arXiv e-prints}, August 2022.

\bibitem{2022ApJ...940L..55F}
Steven~L Finkelstein, Micaela~B Bagley, Pablo~Arrabal Haro, Mark Dickinson, Henry~C Ferguson, Jeyhan~S Kartaltepe, Casey Papovich, Denis Burgarella, Dale~D Kocevski, Kartheik~G Iyer, et~al.
\newblock A long time ago in a galaxy far, far away: A candidate z~ 12 galaxy in early jwst ceers imaging.
\newblock {\em The Astrophysical journal letters}, 940(2):L55, 2022.
\newblock \href {https://arxiv.org/abs/2207.12474} {\path{arXiv:2207.12474}}.

\bibitem{2023Natur.616..266L}
Ivo {Labb{\'e}}, Pieter {van Dokkum}, Erica {Nelson}, Rachel {Bezanson}, Katherine~A. {Suess}, Joel {Leja}, Gabriel {Brammer}, Katherine {Whitaker}, Elijah {Mathews}, Mauro {Stefanon}, and Bingjie {Wang}.
\newblock {A population of red candidate massive galaxies 600 Myr after the Big Bang}.
\newblock 616(7956):266--269, April 2023.
\newblock \href {https://arxiv.org/abs/2207.12446} {\path{arXiv:2207.12446}}, \href {https://doi.org/10.1038/s41586-023-05786-2} {\path{doi:10.1038/s41586-023-05786-2}}.

\bibitem{2023ApJS..265....5H}
Yuichi {Harikane}, Masami {Ouchi}, Masamune {Oguri}, Yoshiaki {Ono}, Kimihiko {Nakajima}, Yuki {Isobe}, Hiroya {Umeda}, Ken {Mawatari}, and Yechi {Zhang}.
\newblock {A Comprehensive Study of Galaxies at z 9-16 Found in the Early JWST Data: Ultraviolet Luminosity Functions and Cosmic Star Formation History at the Pre-reionization Epoch}.
\newblock 265(1):5, March 2023.
\newblock \href {https://arxiv.org/abs/2208.01612} {\path{arXiv:2208.01612}}, \href {https://doi.org/10.3847/1538-4365/acaaa9} {\path{doi:10.3847/1538-4365/acaaa9}}.

\bibitem{Liu:2022bvr}
Boyuan Liu and Volker Bromm.
\newblock {Accelerating Early Massive Galaxy Formation with Primordial Black Holes}.
\newblock {\em Astrophys. J. Lett.}, 937(2):L30, 2022.
\newblock \href {https://arxiv.org/abs/2208.13178} {\path{arXiv:2208.13178}}, \href {https://doi.org/10.3847/2041-8213/ac927f} {\path{doi:10.3847/2041-8213/ac927f}}.

\bibitem{Menci:2022wia}
N.~Menci, M.~Castellano, P.~Santini, E.~Merlin, A.~Fontana, and F.~Shankar.
\newblock {High-redshift Galaxies from Early JWST Observations: Constraints on Dark Energy Models}.
\newblock {\em Astrophys. J. Lett.}, 938(1):L5, 2022.
\newblock \href {https://arxiv.org/abs/2208.11471} {\path{arXiv:2208.11471}}, \href {https://doi.org/10.3847/2041-8213/ac96e9} {\path{doi:10.3847/2041-8213/ac96e9}}.

\bibitem{Biagetti:2022ode}
Matteo Biagetti, Gabriele Franciolini, and Antonio Riotto.
\newblock {High-redshift JWST Observations and Primordial Non-Gaussianity}.
\newblock {\em Astrophys. J.}, 944(2):113, 2023.
\newblock \href {https://arxiv.org/abs/2210.04812} {\path{arXiv:2210.04812}}, \href {https://doi.org/10.3847/1538-4357/acb5ea} {\path{doi:10.3847/1538-4357/acb5ea}}.

\bibitem{Hutsi:2022fzw}
Gert H\"utsi, Martti Raidal, Juan Urrutia, Ville Vaskonen, and Hardi Veerm\"ae.
\newblock {Did JWST observe imprints of axion miniclusters or primordial black holes?}
\newblock {\em Phys. Rev. D}, 107(4):043502, 2023.
\newblock \href {https://arxiv.org/abs/2211.02651} {\path{arXiv:2211.02651}}, \href {https://doi.org/10.1103/PhysRevD.107.043502} {\path{doi:10.1103/PhysRevD.107.043502}}.

\bibitem{Parashari:2023cui}
Priyank Parashari and Ranjan Laha.
\newblock {Primordial power spectrum in light of JWST observations of high redshift galaxies}.
\newblock {\em Mon. Not. Roy. Astron. Soc.}, 526(1):L63--L69, 2023.
\newblock \href {https://arxiv.org/abs/2305.00999} {\path{arXiv:2305.00999}}, \href {https://doi.org/10.1093/mnrasl/slad107} {\path{doi:10.1093/mnrasl/slad107}}.

\bibitem{Hassan:2023asd}
Sultan Hassan et~al.
\newblock {JWST Constraints on the UV Luminosity Density at Cosmic Dawn: Implications for 21 cm Cosmology}.
\newblock {\em Astrophys. J. Lett.}, 958(1):L3, 2023.
\newblock \href {https://arxiv.org/abs/2305.02703} {\path{arXiv:2305.02703}}, \href {https://doi.org/10.3847/2041-8213/ad0239} {\path{doi:10.3847/2041-8213/ad0239}}.

\bibitem{Ellis:2024wdh}
John Ellis, Malcolm Fairbairn, Gert H\"utsi, Juan Urrutia, Ville Vaskonen, and Hardi Veerm\"ae.
\newblock {Consistency of JWST Black Hole Observations with NANOGrav Gravitational Wave Measurements}.
\newblock {\em Astron. Astrophys.}, 691:A270, 2024.
\newblock \href {https://arxiv.org/abs/2403.19650} {\path{arXiv:2403.19650}}, \href {https://doi.org/10.1051/0004-6361/202450846} {\path{doi:10.1051/0004-6361/202450846}}.

\bibitem{NANOGrav:2023pdq}
Gabriella Agazie et~al.
\newblock {The NANOGrav 15 yr Data Set: Bayesian Limits on Gravitational Waves from Individual Supermassive Black Hole Binaries}.
\newblock {\em Astrophys. J. Lett.}, 951(2):L50, 2023.
\newblock \href {https://arxiv.org/abs/2306.16222} {\path{arXiv:2306.16222}}, \href {https://doi.org/10.3847/2041-8213/ace18a} {\path{doi:10.3847/2041-8213/ace18a}}.

\bibitem{EPTA:2023gyr}
J.~Antoniadis et~al.
\newblock {The second data release from the European Pulsar Timing Array - V. Search for continuous gravitational wave signals}.
\newblock {\em Astron. Astrophys.}, 690:A118, 2024.
\newblock \href {https://arxiv.org/abs/2306.16226} {\path{arXiv:2306.16226}}, \href {https://doi.org/10.1051/0004-6361/202348568} {\path{doi:10.1051/0004-6361/202348568}}.

\bibitem{Sato-Polito:2021efu}
Gabriela Sato-Polito and Marc Kamionkowski.
\newblock {Pulsar-timing measurement of the circular polarization of the stochastic gravitational-wave background}.
\newblock {\em Phys. Rev. D}, 106(2):023004, 2022.
\newblock \href {https://arxiv.org/abs/2111.05867} {\path{arXiv:2111.05867}}, \href {https://doi.org/10.1103/PhysRevD.106.023004} {\path{doi:10.1103/PhysRevD.106.023004}}.

\bibitem{Sato-Polito:2023spo}
Gabriela Sato-Polito and Marc Kamionkowski.
\newblock {Exploring the spectrum of stochastic gravitational-wave anisotropies with pulsar timing arrays}.
\newblock {\em Phys. Rev. D}, 109(12):123544, 2024.
\newblock \href {https://arxiv.org/abs/2305.05690} {\path{arXiv:2305.05690}}, \href {https://doi.org/10.1103/PhysRevD.109.123544} {\path{doi:10.1103/PhysRevD.109.123544}}.

\bibitem{Caprini:2009fx}
Chiara Caprini, Ruth Durrer, Thomas Konstandin, and Geraldine Servant.
\newblock {General Properties of the Gravitational Wave Spectrum from Phase Transitions}.
\newblock {\em Phys. Rev. D}, 79:083519, 2009.
\newblock \href {https://arxiv.org/abs/0901.1661} {\path{arXiv:0901.1661}}, \href {https://doi.org/10.1103/PhysRevD.79.083519} {\path{doi:10.1103/PhysRevD.79.083519}}.

\bibitem{Domenech:2020kqm}
Guillem Dom\`enech, Shi Pi, and Misao Sasaki.
\newblock {Induced gravitational waves as a probe of thermal history of the universe}.
\newblock {\em JCAP}, 08:017, 2020.
\newblock \href {https://arxiv.org/abs/2005.12314} {\path{arXiv:2005.12314}}, \href {https://doi.org/10.1088/1475-7516/2020/08/017} {\path{doi:10.1088/1475-7516/2020/08/017}}.

\bibitem{Hindmarsh:1994re}
M.~B. Hindmarsh and T.~W.~B. Kibble.
\newblock {Cosmic strings}.
\newblock {\em Rept. Prog. Phys.}, 58:477--562, 1995.
\newblock \href {https://arxiv.org/abs/hep-ph/9411342} {\path{arXiv:hep-ph/9411342}}, \href {https://doi.org/10.1088/0034-4885/58/5/001} {\path{doi:10.1088/0034-4885/58/5/001}}.

\bibitem{Jeannerot:2003qv}
Rachel Jeannerot, Jonathan Rocher, and Mairi Sakellariadou.
\newblock {How generic is cosmic string formation in SUSY GUTs}.
\newblock {\em Phys. Rev. D}, 68:103514, 2003.
\newblock \href {https://arxiv.org/abs/hep-ph/0308134} {\path{arXiv:hep-ph/0308134}}, \href {https://doi.org/10.1103/PhysRevD.68.103514} {\path{doi:10.1103/PhysRevD.68.103514}}.

\bibitem{King:2020hyd}
Stephen~F. King, Silvia Pascoli, Jessica Turner, and Ye-Ling Zhou.
\newblock {Gravitational Waves and Proton Decay: Complementary Windows into Grand Unified Theories}.
\newblock {\em Phys. Rev. Lett.}, 126(2):021802, 2021.
\newblock \href {https://arxiv.org/abs/2005.13549} {\path{arXiv:2005.13549}}, \href {https://doi.org/10.1103/PhysRevLett.126.021802} {\path{doi:10.1103/PhysRevLett.126.021802}}.

\bibitem{Dvali:2003zj}
Gia Dvali and Alexander Vilenkin.
\newblock {Formation and evolution of cosmic D strings}.
\newblock {\em JCAP}, 03:010, 2004.
\newblock \href {https://arxiv.org/abs/hep-th/0312007} {\path{arXiv:hep-th/0312007}}, \href {https://doi.org/10.1088/1475-7516/2004/03/010} {\path{doi:10.1088/1475-7516/2004/03/010}}.

\bibitem{Copeland:2003bj}
Edmund~J. Copeland, Robert~C. Myers, and Joseph Polchinski.
\newblock {Cosmic F and D strings}.
\newblock {\em JHEP}, 06:013, 2004.
\newblock \href {https://arxiv.org/abs/hep-th/0312067} {\path{arXiv:hep-th/0312067}}, \href {https://doi.org/10.1088/1126-6708/2004/06/013} {\path{doi:10.1088/1126-6708/2004/06/013}}.

\bibitem{Yamada:2022aax}
Masaki Yamada and Kazuya Yonekura.
\newblock {Cosmic F- and D-strings from pure Yang\textendash{}Mills theory}.
\newblock {\em Phys. Lett. B}, 838:137724, 2023.
\newblock \href {https://arxiv.org/abs/2204.13125} {\path{arXiv:2204.13125}}, \href {https://doi.org/10.1016/j.physletb.2023.137724} {\path{doi:10.1016/j.physletb.2023.137724}}.

\bibitem{Yamada:2022imq}
Masaki Yamada and Kazuya Yonekura.
\newblock {Cosmic strings from pure Yang\textendash{}Mills theory}.
\newblock {\em Phys. Rev. D}, 106(12):123515, 2022.
\newblock \href {https://arxiv.org/abs/2204.13123} {\path{arXiv:2204.13123}}, \href {https://doi.org/10.1103/PhysRevD.106.123515} {\path{doi:10.1103/PhysRevD.106.123515}}.

\bibitem{Sakellariadou:2004wq}
Mairi Sakellariadou.
\newblock {A Note on the evolution of cosmic string/superstring networks}.
\newblock {\em JCAP}, 04:003, 2005.
\newblock \href {https://arxiv.org/abs/hep-th/0410234} {\path{arXiv:hep-th/0410234}}, \href {https://doi.org/10.1088/1475-7516/2005/04/003} {\path{doi:10.1088/1475-7516/2005/04/003}}.

\bibitem{Blanco-Pillado:2017rnf}
Jose~J. Blanco-Pillado, Ken~D. Olum, and Xavier Siemens.
\newblock {New limits on cosmic strings from gravitational wave observation}.
\newblock {\em Phys. Lett. B}, 778:392--396, 2018.
\newblock \href {https://arxiv.org/abs/1709.02434} {\path{arXiv:1709.02434}}, \href {https://doi.org/10.1016/j.physletb.2018.01.050} {\path{doi:10.1016/j.physletb.2018.01.050}}.

\bibitem{Martins:1995tg}
C.~J. A.~P. Martins and E.~P.~S. Shellard.
\newblock {String evolution with friction}.
\newblock {\em Phys. Rev. D}, 53:575--579, 1996.
\newblock \href {https://arxiv.org/abs/hep-ph/9507335} {\path{arXiv:hep-ph/9507335}}, \href {https://doi.org/10.1103/PhysRevD.53.R575} {\path{doi:10.1103/PhysRevD.53.R575}}.

\bibitem{Martins:1996jp}
C.~J. A.~P. Martins and E.~P.~S. Shellard.
\newblock {Quantitative string evolution}.
\newblock {\em Phys. Rev. D}, 54:2535--2556, 1996.
\newblock \href {https://arxiv.org/abs/hep-ph/9602271} {\path{arXiv:hep-ph/9602271}}, \href {https://doi.org/10.1103/PhysRevD.54.2535} {\path{doi:10.1103/PhysRevD.54.2535}}.

\bibitem{Martins:2000cs}
C.~J. A.~P. Martins and E.~P.~S. Shellard.
\newblock {Extending the velocity dependent one scale string evolution model}.
\newblock {\em Phys. Rev. D}, 65:043514, 2002.
\newblock \href {https://arxiv.org/abs/hep-ph/0003298} {\path{arXiv:hep-ph/0003298}}, \href {https://doi.org/10.1103/PhysRevD.65.043514} {\path{doi:10.1103/PhysRevD.65.043514}}.

\bibitem{Avelino:2012qy}
P.~P. Avelino and L.~Sousa.
\newblock {Scaling laws for weakly interacting cosmic (super)string and p-brane networks}.
\newblock {\em Phys. Rev. D}, 85:083525, 2012.
\newblock \href {https://arxiv.org/abs/1202.6298} {\path{arXiv:1202.6298}}, \href {https://doi.org/10.1103/PhysRevD.85.083525} {\path{doi:10.1103/PhysRevD.85.083525}}.

\bibitem{Sousa:2013aaa}
L.~Sousa and P.~P. Avelino.
\newblock {Stochastic Gravitational Wave Background generated by Cosmic String Networks: Velocity-Dependent One-Scale model versus Scale-Invariant Evolution}.
\newblock {\em Phys. Rev. D}, 88(2):023516, 2013.
\newblock \href {https://arxiv.org/abs/1304.2445} {\path{arXiv:1304.2445}}, \href {https://doi.org/10.1103/PhysRevD.88.023516} {\path{doi:10.1103/PhysRevD.88.023516}}.

\bibitem{Blanco-Pillado:2011egf}
Jose~J. Blanco-Pillado, Ken~D. Olum, and Benjamin Shlaer.
\newblock {Large parallel cosmic string simulations: New results on loop production}.
\newblock {\em Phys. Rev. D}, 83:083514, 2011.
\newblock \href {https://arxiv.org/abs/1101.5173} {\path{arXiv:1101.5173}}, \href {https://doi.org/10.1103/PhysRevD.83.083514} {\path{doi:10.1103/PhysRevD.83.083514}}.

\bibitem{Blanco-Pillado:2013qja}
Jose~J. Blanco-Pillado, Ken~D. Olum, and Benjamin Shlaer.
\newblock {The number of cosmic string loops}.
\newblock {\em Phys. Rev. D}, 89(2):023512, 2014.
\newblock \href {https://arxiv.org/abs/1309.6637} {\path{arXiv:1309.6637}}, \href {https://doi.org/10.1103/PhysRevD.89.023512} {\path{doi:10.1103/PhysRevD.89.023512}}.

\bibitem{Blanco-Pillado:2015ana}
Jose~J. Blanco-Pillado, Ken~D. Olum, and Benjamin Shlaer.
\newblock {Cosmic string loop shapes}.
\newblock {\em Phys. Rev. D}, 92(6):063528, 2015.
\newblock \href {https://arxiv.org/abs/1508.02693} {\path{arXiv:1508.02693}}, \href {https://doi.org/10.1103/PhysRevD.92.063528} {\path{doi:10.1103/PhysRevD.92.063528}}.

\bibitem{Blanco-Pillado:2017oxo}
Jose~J. Blanco-Pillado and Ken~D. Olum.
\newblock {Stochastic gravitational wave background from smoothed cosmic string loops}.
\newblock {\em Phys. Rev. D}, 96(10):104046, 2017.
\newblock \href {https://arxiv.org/abs/1709.02693} {\path{arXiv:1709.02693}}, \href {https://doi.org/10.1103/PhysRevD.96.104046} {\path{doi:10.1103/PhysRevD.96.104046}}.

\bibitem{Blanco-Pillado:2019vcs}
Jose~J. Blanco-Pillado, Ken~D. Olum, and Jeremy~M. Wachter.
\newblock {Energy-conservation constraints on cosmic string loop production and distribution functions}.
\newblock {\em Phys. Rev. D}, 100(12):123526, 2019.
\newblock \href {https://arxiv.org/abs/1907.09373} {\path{arXiv:1907.09373}}, \href {https://doi.org/10.1103/PhysRevD.100.123526} {\path{doi:10.1103/PhysRevD.100.123526}}.

\bibitem{Blanco-Pillado:2019tbi}
Jose~J. Blanco-Pillado and Ken~D. Olum.
\newblock {Direct determination of cosmic string loop density from simulations}.
\newblock {\em Phys. Rev. D}, 101(10):103018, 2020.
\newblock \href {https://arxiv.org/abs/1912.10017} {\path{arXiv:1912.10017}}, \href {https://doi.org/10.1103/PhysRevD.101.103018} {\path{doi:10.1103/PhysRevD.101.103018}}.

\bibitem{Auclair:2019wcv}
Pierre Auclair et~al.
\newblock {Probing the gravitational wave background from cosmic strings with LISA}.
\newblock {\em JCAP}, 04:034, 2020.
\newblock \href {https://arxiv.org/abs/1909.00819} {\path{arXiv:1909.00819}}, \href {https://doi.org/10.1088/1475-7516/2020/04/034} {\path{doi:10.1088/1475-7516/2020/04/034}}.

\bibitem{Burden:1985md}
C.~J. Burden.
\newblock {Gravitational Radiation From a Particular Class of Cosmic Strings}.
\newblock {\em Phys. Lett. B}, 164:277--281, 1985.
\newblock \href {https://doi.org/10.1016/0370-2693(85)90326-0} {\path{doi:10.1016/0370-2693(85)90326-0}}.

\bibitem{Cui:2019kkd}
Yanou Cui, Marek Lewicki, and David~E. Morrissey.
\newblock {Gravitational Wave Bursts as Harbingers of Cosmic Strings Diluted by Inflation}.
\newblock {\em Phys. Rev. Lett.}, 125(21):211302, 2020.
\newblock \href {https://arxiv.org/abs/1912.08832} {\path{arXiv:1912.08832}}, \href {https://doi.org/10.1103/PhysRevLett.125.211302} {\path{doi:10.1103/PhysRevLett.125.211302}}.

\bibitem{Blasi:2020wpy}
Simone Blasi, Vedran Brdar, and Kai Schmitz.
\newblock {Fingerprint of low-scale leptogenesis in the primordial gravitational-wave spectrum}.
\newblock {\em Phys. Rev. Res.}, 2(4):043321, 2020.
\newblock \href {https://arxiv.org/abs/2004.02889} {\path{arXiv:2004.02889}}, \href {https://doi.org/10.1103/PhysRevResearch.2.043321} {\path{doi:10.1103/PhysRevResearch.2.043321}}.

\bibitem{Gouttenoire:2019kij}
Yann Gouttenoire, G\'eraldine Servant, and Peera Simakachorn.
\newblock {Beyond the Standard Models with Cosmic Strings}.
\newblock {\em JCAP}, 07:032, 2020.
\newblock \href {https://arxiv.org/abs/1912.02569} {\path{arXiv:1912.02569}}, \href {https://doi.org/10.1088/1475-7516/2020/07/032} {\path{doi:10.1088/1475-7516/2020/07/032}}.

\bibitem{Witten:1984rs}
Edward Witten.
\newblock {Cosmic Separation of Phases}.
\newblock {\em Phys. Rev. D}, 30:272--285, 1984.
\newblock \href {https://doi.org/10.1103/PhysRevD.30.272} {\path{doi:10.1103/PhysRevD.30.272}}.

\bibitem{Hogan:1986qda}
C.~J. Hogan.
\newblock {Gravitational radiation from cosmological phase transitions}.
\newblock {\em Mon. Not. Roy. Astron. Soc.}, 218:629--636, 1986.

\bibitem{Lewicki:2022pdb}
Marek Lewicki and Ville Vaskonen.
\newblock {Gravitational waves from bubble collisions and fluid motion in strongly supercooled phase transitions}.
\newblock {\em Eur. Phys. J. C}, 83(2):109, 2023.
\newblock \href {https://arxiv.org/abs/2208.11697} {\path{arXiv:2208.11697}}, \href {https://doi.org/10.1140/epjc/s10052-023-11241-3} {\path{doi:10.1140/epjc/s10052-023-11241-3}}.

\bibitem{Kamionkowski:1993fg}
Marc Kamionkowski, Arthur Kosowsky, and Michael~S. Turner.
\newblock {Gravitational radiation from first order phase transitions}.
\newblock {\em Phys. Rev. D}, 49:2837--2851, 1994.
\newblock \href {https://arxiv.org/abs/astro-ph/9310044} {\path{arXiv:astro-ph/9310044}}, \href {https://doi.org/10.1103/PhysRevD.49.2837} {\path{doi:10.1103/PhysRevD.49.2837}}.

\bibitem{Hindmarsh:2015qta}
Mark Hindmarsh, Stephan~J. Huber, Kari Rummukainen, and David~J. Weir.
\newblock {Numerical simulations of acoustically generated gravitational waves at a first order phase transition}.
\newblock {\em Phys. Rev. D}, 92(12):123009, 2015.
\newblock \href {https://arxiv.org/abs/1504.03291} {\path{arXiv:1504.03291}}, \href {https://doi.org/10.1103/PhysRevD.92.123009} {\path{doi:10.1103/PhysRevD.92.123009}}.

\bibitem{Hindmarsh:2016lnk}
Mark Hindmarsh.
\newblock {Sound shell model for acoustic gravitational wave production at a first-order phase transition in the early Universe}.
\newblock {\em Phys. Rev. Lett.}, 120(7):071301, 2018.
\newblock \href {https://arxiv.org/abs/1608.04735} {\path{arXiv:1608.04735}}, \href {https://doi.org/10.1103/PhysRevLett.120.071301} {\path{doi:10.1103/PhysRevLett.120.071301}}.

\bibitem{Hindmarsh:2017gnf}
Mark Hindmarsh, Stephan~J. Huber, Kari Rummukainen, and David~J. Weir.
\newblock {Shape of the acoustic gravitational wave power spectrum from a first order phase transition}.
\newblock {\em Phys. Rev. D}, 96(10):103520, 2017.
\newblock [Erratum: Phys.Rev.D 101, 089902 (2020)].
\newblock \href {https://arxiv.org/abs/1704.05871} {\path{arXiv:1704.05871}}, \href {https://doi.org/10.1103/PhysRevD.96.103520} {\path{doi:10.1103/PhysRevD.96.103520}}.

\bibitem{Ellis:2018mja}
John Ellis, Marek Lewicki, and Jos\'e~Miguel No.
\newblock {On the Maximal Strength of a First-Order Electroweak Phase Transition and its Gravitational Wave Signal}.
\newblock {\em JCAP}, 04:003, 2019.
\newblock \href {https://arxiv.org/abs/1809.08242} {\path{arXiv:1809.08242}}, \href {https://doi.org/10.1088/1475-7516/2019/04/003} {\path{doi:10.1088/1475-7516/2019/04/003}}.

\bibitem{Cutting:2019zws}
Daniel Cutting, Mark Hindmarsh, and David~J. Weir.
\newblock {Vorticity, kinetic energy, and suppressed gravitational wave production in strong first order phase transitions}.
\newblock {\em Phys. Rev. Lett.}, 125(2):021302, 2020.
\newblock \href {https://arxiv.org/abs/1906.00480} {\path{arXiv:1906.00480}}, \href {https://doi.org/10.1103/PhysRevLett.125.021302} {\path{doi:10.1103/PhysRevLett.125.021302}}.

\bibitem{Hindmarsh:2019phv}
Mark Hindmarsh and Mulham Hijazi.
\newblock {Gravitational waves from first order cosmological phase transitions in the Sound Shell Model}.
\newblock {\em JCAP}, 12:062, 2019.
\newblock \href {https://arxiv.org/abs/1909.10040} {\path{arXiv:1909.10040}}, \href {https://doi.org/10.1088/1475-7516/2019/12/062} {\path{doi:10.1088/1475-7516/2019/12/062}}.

\bibitem{Ellis:2020awk}
John Ellis, Marek Lewicki, and Jos\'e~Miguel No.
\newblock {Gravitational waves from first-order cosmological phase transitions: lifetime of the sound wave source}.
\newblock {\em JCAP}, 07:050, 2020.
\newblock \href {https://arxiv.org/abs/2003.07360} {\path{arXiv:2003.07360}}, \href {https://doi.org/10.1088/1475-7516/2020/07/050} {\path{doi:10.1088/1475-7516/2020/07/050}}.

\bibitem{Kosowsky:1992vn}
Arthur Kosowsky and Michael~S. Turner.
\newblock {Gravitational radiation from colliding vacuum bubbles: envelope approximation to many bubble collisions}.
\newblock {\em Phys. Rev. D}, 47:4372--4391, 1993.
\newblock \href {https://arxiv.org/abs/astro-ph/9211004} {\path{arXiv:astro-ph/9211004}}, \href {https://doi.org/10.1103/PhysRevD.47.4372} {\path{doi:10.1103/PhysRevD.47.4372}}.

\bibitem{Cutting:2018tjt}
Daniel Cutting, Mark Hindmarsh, and David~J. Weir.
\newblock {Gravitational waves from vacuum first-order phase transitions: from the envelope to the lattice}.
\newblock {\em Phys. Rev. D}, 97(12):123513, 2018.
\newblock \href {https://arxiv.org/abs/1802.05712} {\path{arXiv:1802.05712}}, \href {https://doi.org/10.1103/PhysRevD.97.123513} {\path{doi:10.1103/PhysRevD.97.123513}}.

\bibitem{Ellis:2019oqb}
John Ellis, Marek Lewicki, Jos\'e~Miguel No, and Ville Vaskonen.
\newblock {Gravitational wave energy budget in strongly supercooled phase transitions}.
\newblock {\em JCAP}, 06:024, 2019.
\newblock \href {https://arxiv.org/abs/1903.09642} {\path{arXiv:1903.09642}}, \href {https://doi.org/10.1088/1475-7516/2019/06/024} {\path{doi:10.1088/1475-7516/2019/06/024}}.

\bibitem{Lewicki:2019gmv}
Marek Lewicki and Ville Vaskonen.
\newblock {On bubble collisions in strongly supercooled phase transitions}.
\newblock {\em Phys. Dark Univ.}, 30:100672, 2020.
\newblock \href {https://arxiv.org/abs/1912.00997} {\path{arXiv:1912.00997}}, \href {https://doi.org/10.1016/j.dark.2020.100672} {\path{doi:10.1016/j.dark.2020.100672}}.

\bibitem{Cutting:2020nla}
Daniel Cutting, Elba~Granados Escartin, Mark Hindmarsh, and David~J. Weir.
\newblock {Gravitational waves from vacuum first order phase transitions II: from thin to thick walls}.
\newblock {\em Phys. Rev. D}, 103(2):023531, 2021.
\newblock \href {https://arxiv.org/abs/2005.13537} {\path{arXiv:2005.13537}}, \href {https://doi.org/10.1103/PhysRevD.103.023531} {\path{doi:10.1103/PhysRevD.103.023531}}.

\bibitem{Lewicki:2020jiv}
Marek Lewicki and Ville Vaskonen.
\newblock {Gravitational wave spectra from strongly supercooled phase transitions}.
\newblock {\em Eur. Phys. J. C}, 80(11):1003, 2020.
\newblock \href {https://arxiv.org/abs/2007.04967} {\path{arXiv:2007.04967}}, \href {https://doi.org/10.1140/epjc/s10052-020-08589-1} {\path{doi:10.1140/epjc/s10052-020-08589-1}}.

\bibitem{Giese:2020znk}
Felix Giese, Thomas Konstandin, Kai Schmitz, and Jorinde van~de Vis.
\newblock {Model-independent energy budget for LISA}.
\newblock {\em JCAP}, 01:072, 2021.
\newblock \href {https://arxiv.org/abs/2010.09744} {\path{arXiv:2010.09744}}, \href {https://doi.org/10.1088/1475-7516/2021/01/072} {\path{doi:10.1088/1475-7516/2021/01/072}}.

\bibitem{Ellis:2020nnr}
John Ellis, Marek Lewicki, and Ville Vaskonen.
\newblock {Updated predictions for gravitational waves produced in a strongly supercooled phase transition}.
\newblock {\em JCAP}, 11:020, 2020.
\newblock \href {https://arxiv.org/abs/2007.15586} {\path{arXiv:2007.15586}}, \href {https://doi.org/10.1088/1475-7516/2020/11/020} {\path{doi:10.1088/1475-7516/2020/11/020}}.

\bibitem{Lewicki:2020azd}
M.~Lewicki and V.~Vaskonen.
\newblock Gravitational waves from colliding vacuum bubbles in gauge theories.
\newblock {\em Eur. Phys. J. C}, 81(5):1--10, 2021.
\newblock \href {https://doi.org/10.1140/epjc/s10052-021-09232-3} {\path{doi:10.1140/epjc/s10052-021-09232-3}}.

\bibitem{Liu:2021svg}
Jing Liu, Ligong Bian, Rong-Gen Cai, Zong-Kuan Guo, and Shao-Jiang Wang.
\newblock {Primordial black hole production during first-order phase transitions}.
\newblock {\em Phys. Rev. D}, 105(2):L021303, 2022.
\newblock \href {https://arxiv.org/abs/2106.05637} {\path{arXiv:2106.05637}}, \href {https://doi.org/10.1103/PhysRevD.105.L021303} {\path{doi:10.1103/PhysRevD.105.L021303}}.

\bibitem{Lewicki:2023ioy}
Marek Lewicki, Piotr Toczek, and Ville Vaskonen.
\newblock {Primordial black holes from strong first-order phase transitions}.
\newblock {\em JHEP}, 09:092, 2023.
\newblock \href {https://arxiv.org/abs/2305.04924} {\path{arXiv:2305.04924}}, \href {https://doi.org/10.1007/JHEP09(2023)092} {\path{doi:10.1007/JHEP09(2023)092}}.

\bibitem{Gorton:2022fyb}
Matthew Gorton and Anne~M. Green.
\newblock {Effect of clustering on primordial black hole microlensing constraints}.
\newblock {\em JCAP}, 08(08):035, 2022.
\newblock \href {https://arxiv.org/abs/2203.04209} {\path{arXiv:2203.04209}}, \href {https://doi.org/10.1088/1475-7516/2022/08/035} {\path{doi:10.1088/1475-7516/2022/08/035}}.

\bibitem{Petac:2022rio}
Mihael Peta\v{c}, Julien Lavalle, and Karsten Jedamzik.
\newblock {Microlensing constraints on clustered primordial black holes}.
\newblock {\em Phys. Rev. D}, 105(8):083520, 2022.
\newblock \href {https://arxiv.org/abs/2201.02521} {\path{arXiv:2201.02521}}, \href {https://doi.org/10.1103/PhysRevD.105.083520} {\path{doi:10.1103/PhysRevD.105.083520}}.

\bibitem{Ricotti:2007au}
Massimo Ricotti, Jeremiah~P. Ostriker, and Katherine~J. Mack.
\newblock {Effect of Primordial Black Holes on the Cosmic Microwave Background and Cosmological Parameter Estimates}.
\newblock {\em Astrophys. J.}, 680:829, 2008.
\newblock \href {https://arxiv.org/abs/0709.0524} {\path{arXiv:0709.0524}}, \href {https://doi.org/10.1086/587831} {\path{doi:10.1086/587831}}.

\bibitem{Horowitz:2016lib}
Benjamin Horowitz.
\newblock {Revisiting Primordial Black Holes Constraints from Ionization History}.
\newblock 12 2016.
\newblock \href {https://arxiv.org/abs/1612.07264} {\path{arXiv:1612.07264}}.

\bibitem{Clesse:2020ghq}
Sebastien Clesse and Juan Garcia-Bellido.
\newblock {GW190425, GW190521 and GW190814: Three candidate mergers of primordial black holes from the QCD epoch}.
\newblock {\em Phys. Dark Univ.}, 38:101111, 2022.
\newblock \href {https://arxiv.org/abs/2007.06481} {\path{arXiv:2007.06481}}, \href {https://doi.org/10.1016/j.dark.2022.101111} {\path{doi:10.1016/j.dark.2022.101111}}.

\bibitem{Allahverdi:2020bys}
Rouzbeh Allahverdi et~al.
\newblock {The First Three Seconds: a Review of Possible Expansion Histories of the Early Universe}.
\newblock 6 2020.
\newblock \href {https://arxiv.org/abs/2006.16182} {\path{arXiv:2006.16182}}, \href {https://doi.org/10.21105/astro.2006.16182} {\path{doi:10.21105/astro.2006.16182}}.

\bibitem{Vilenkin:1984ib}
Alexander Vilenkin.
\newblock {Cosmic Strings and Domain Walls}.
\newblock {\em Phys. Rept.}, 121:263--315, 1985.
\newblock \href {https://doi.org/10.1016/0370-1573(85)90033-X} {\path{doi:10.1016/0370-1573(85)90033-X}}.

\bibitem{Hiramatsu:2013qaa}
Takashi Hiramatsu, Masahiro Kawasaki, and Ken'ichi Saikawa.
\newblock {On the estimation of gravitational wave spectrum from cosmic domain walls}.
\newblock {\em JCAP}, 02:031, 2014.
\newblock \href {https://arxiv.org/abs/1309.5001} {\path{arXiv:1309.5001}}, \href {https://doi.org/10.1088/1475-7516/2014/02/031} {\path{doi:10.1088/1475-7516/2014/02/031}}.

\bibitem{Kibble:1976sj}
T.~W.~B. Kibble.
\newblock {Topology of Cosmic Domains and Strings}.
\newblock {\em J. Phys. A}, 9:1387--1398, 1976.
\newblock \href {https://doi.org/10.1088/0305-4470/9/8/029} {\path{doi:10.1088/0305-4470/9/8/029}}.

\bibitem{Gelmini:1988sf}
Graciela~B. Gelmini, Marcelo Gleiser, and Edward~W. Kolb.
\newblock {Cosmology of Biased Discrete Symmetry Breaking}.
\newblock {\em Phys. Rev. D}, 39:1558, 1989.
\newblock \href {https://doi.org/10.1103/PhysRevD.39.1558} {\path{doi:10.1103/PhysRevD.39.1558}}.

\bibitem{Coulson:1995nv}
D.~Coulson, Z.~Lalak, and Burt~A. Ovrut.
\newblock {Biased domain walls}.
\newblock {\em Phys. Rev. D}, 53:4237--4246, 1996.
\newblock \href {https://doi.org/10.1103/PhysRevD.53.4237} {\path{doi:10.1103/PhysRevD.53.4237}}.

\bibitem{Larsson:1996sp}
Sebastian~E. Larsson, Subir Sarkar, and Peter~L. White.
\newblock {Evading the cosmological domain wall problem}.
\newblock {\em Phys. Rev. D}, 55:5129--5135, 1997.
\newblock \href {https://arxiv.org/abs/hep-ph/9608319} {\path{arXiv:hep-ph/9608319}}, \href {https://doi.org/10.1103/PhysRevD.55.5129} {\path{doi:10.1103/PhysRevD.55.5129}}.

\bibitem{Preskill:1991kd}
John Preskill, Sandip~P. Trivedi, Frank Wilczek, and Mark~B. Wise.
\newblock {Cosmology and broken discrete symmetry}.
\newblock {\em Nucl. Phys. B}, 363:207--220, 1991.
\newblock \href {https://doi.org/10.1016/0550-3213(91)90241-O} {\path{doi:10.1016/0550-3213(91)90241-O}}.

\bibitem{Fields:2019pfx}
Brian~D. Fields, Keith~A. Olive, Tsung-Han Yeh, and Charles Young.
\newblock {Big-Bang Nucleosynthesis after Planck}.
\newblock {\em JCAP}, 03:010, 2020.
\newblock [Erratum: JCAP 11, E02 (2020)].
\newblock \href {https://arxiv.org/abs/1912.01132} {\path{arXiv:1912.01132}}, \href {https://doi.org/10.1088/1475-7516/2020/03/010} {\path{doi:10.1088/1475-7516/2020/03/010}}.

\bibitem{Ramberg:2022irf}
Nicklas Ramberg, Wolfram Ratzinger, and Pedro Schwaller.
\newblock {One \ensuremath{\mu} to rule them all: CMB spectral distortions can probe domain walls, cosmic strings and low scale phase transitions}.
\newblock {\em JCAP}, 02:039, 2023.
\newblock \href {https://arxiv.org/abs/2209.14313} {\path{arXiv:2209.14313}}, \href {https://doi.org/10.1088/1475-7516/2023/02/039} {\path{doi:10.1088/1475-7516/2023/02/039}}.

\bibitem{Hasegawa:2019jsa}
Takuya Hasegawa, Nagisa Hiroshima, Kazunori Kohri, Rasmus S.~L. Hansen, Thomas Tram, and Steen Hannestad.
\newblock {MeV-scale reheating temperature and thermalization of oscillating neutrinos by radiative and hadronic decays of massive particles}.
\newblock {\em JCAP}, 12:012, 2019.
\newblock \href {https://arxiv.org/abs/1908.10189} {\path{arXiv:1908.10189}}, \href {https://doi.org/10.1088/1475-7516/2019/12/012} {\path{doi:10.1088/1475-7516/2019/12/012}}.

\bibitem{deSalas:2015glj}
P.~F. de~Salas, M.~Lattanzi, G.~Mangano, G.~Miele, S.~Pastor, and O.~Pisanti.
\newblock {Bounds on very low reheating scenarios after Planck}.
\newblock {\em Phys. Rev. D}, 92(12):123534, 2015.
\newblock \href {https://arxiv.org/abs/1511.00672} {\path{arXiv:1511.00672}}, \href {https://doi.org/10.1103/PhysRevD.92.123534} {\path{doi:10.1103/PhysRevD.92.123534}}.

\bibitem{Ferrer:2018uiu}
Francesc Ferrer, Eduard Masso, Giuliano Panico, Oriol Pujolas, and Fabrizio Rompineve.
\newblock {Primordial Black Holes from the QCD axion}.
\newblock {\em Phys. Rev. Lett.}, 122(10):101301, 2019.
\newblock \href {https://arxiv.org/abs/1807.01707} {\path{arXiv:1807.01707}}, \href {https://doi.org/10.1103/PhysRevLett.122.101301} {\path{doi:10.1103/PhysRevLett.122.101301}}.

\bibitem{Gelmini:2023ngs}
Graciela~B. Gelmini, Jonah Hyman, Anna Simpson, and Edoardo Vitagliano.
\newblock {Primordial black hole dark matter from catastrogenesis with unstable pseudo-Goldstone bosons}.
\newblock {\em JCAP}, 06:055, 2023.
\newblock \href {https://arxiv.org/abs/2303.14107} {\path{arXiv:2303.14107}}, \href {https://doi.org/10.1088/1475-7516/2023/06/055} {\path{doi:10.1088/1475-7516/2023/06/055}}.

\bibitem{Yuan:2021qgz}
Chen Yuan and Qing-Guo Huang.
\newblock {A topic review on probing primordial black hole dark matter with scalar induced gravitational waves}.
\newblock {\em iScience}, 24:102860, 2021.
\newblock \href {https://arxiv.org/abs/2103.04739} {\path{arXiv:2103.04739}}, \href {https://doi.org/10.1016/j.isci.2021.102860} {\path{doi:10.1016/j.isci.2021.102860}}.

\bibitem{Inomata:2019yww}
Keisuke Inomata and Takahiro Terada.
\newblock {Gauge Independence of Induced Gravitational Waves}.
\newblock {\em Phys. Rev. D}, 101(2):023523, 2020.
\newblock \href {https://arxiv.org/abs/1912.00785} {\path{arXiv:1912.00785}}, \href {https://doi.org/10.1103/PhysRevD.101.023523} {\path{doi:10.1103/PhysRevD.101.023523}}.

\bibitem{DeLuca:2019ufz}
V.~De~Luca, G.~Franciolini, A.~Kehagias, and A.~Riotto.
\newblock {On the Gauge Invariance of Cosmological Gravitational Waves}.
\newblock {\em JCAP}, 03:014, 2020.
\newblock \href {https://arxiv.org/abs/1911.09689} {\path{arXiv:1911.09689}}, \href {https://doi.org/10.1088/1475-7516/2020/03/014} {\path{doi:10.1088/1475-7516/2020/03/014}}.

\bibitem{Yuan:2019fwv}
Chen Yuan, Zu-Cheng Chen, and Qing-Guo Huang.
\newblock {Scalar induced gravitational waves in different gauges}.
\newblock {\em Phys. Rev. D}, 101(6):063018, 2020.
\newblock \href {https://arxiv.org/abs/1912.00885} {\path{arXiv:1912.00885}}, \href {https://doi.org/10.1103/PhysRevD.101.063018} {\path{doi:10.1103/PhysRevD.101.063018}}.

\bibitem{Domenech:2020xin}
Guillem Dom\`enech and Misao Sasaki.
\newblock {Approximate gauge independence of the induced gravitational wave spectrum}.
\newblock {\em Phys. Rev. D}, 103(6):063531, 2021.
\newblock \href {https://arxiv.org/abs/2012.14016} {\path{arXiv:2012.14016}}, \href {https://doi.org/10.1103/PhysRevD.103.063531} {\path{doi:10.1103/PhysRevD.103.063531}}.

\bibitem{Grishchuk:1974ny}
L.~P. Grishchuk.
\newblock {Amplification of gravitational waves in an istropic universe}.
\newblock {\em Zh. Eksp. Teor. Fiz.}, 67:825--838, 1974.

\bibitem{Starobinsky:1979ty}
Alexei~A. Starobinsky.
\newblock {Spectrum of relict gravitational radiation and the early state of the universe}.
\newblock {\em JETP Lett.}, 30:682--685, 1979.

\bibitem{Kuroyanagi:2014nba}
Sachiko Kuroyanagi, Tomo Takahashi, and Shuichiro Yokoyama.
\newblock {Blue-tilted Tensor Spectrum and Thermal History of the Universe}.
\newblock {\em JCAP}, 02:003, 2015.
\newblock \href {https://arxiv.org/abs/1407.4785} {\path{arXiv:1407.4785}}, \href {https://doi.org/10.1088/1475-7516/2015/02/003} {\path{doi:10.1088/1475-7516/2015/02/003}}.

\bibitem{Kuroyanagi:2020sfw}
Sachiko Kuroyanagi, Tomo Takahashi, and Shuichiro Yokoyama.
\newblock {Blue-tilted inflationary tensor spectrum and reheating in the light of NANOGrav results}.
\newblock {\em JCAP}, 01:071, 2021.
\newblock \href {https://arxiv.org/abs/2011.03323} {\path{arXiv:2011.03323}}, \href {https://doi.org/10.1088/1475-7516/2021/01/071} {\path{doi:10.1088/1475-7516/2021/01/071}}.

\bibitem{Smith:2006nka}
Tristan~L. Smith, Elena Pierpaoli, and Marc Kamionkowski.
\newblock {A new cosmic microwave background constraint to primordial gravitational waves}.
\newblock {\em Phys. Rev. Lett.}, 97:021301, 2006.
\newblock \href {https://arxiv.org/abs/astro-ph/0603144} {\path{arXiv:astro-ph/0603144}}, \href {https://doi.org/10.1103/PhysRevLett.97.021301} {\path{doi:10.1103/PhysRevLett.97.021301}}.

\bibitem{Boyle:2007zx}
Latham~A. Boyle and Alessandra Buonanno.
\newblock {Relating gravitational wave constraints from primordial nucleosynthesis, pulsar timing, laser interferometers, and the CMB: Implications for the early Universe}.
\newblock {\em Phys. Rev. D}, 78:043531, 2008.
\newblock \href {https://arxiv.org/abs/0708.2279} {\path{arXiv:0708.2279}}, \href {https://doi.org/10.1103/PhysRevD.78.043531} {\path{doi:10.1103/PhysRevD.78.043531}}.

\bibitem{Caprini:2018mtu}
Chiara Caprini and Daniel~G. Figueroa.
\newblock {Cosmological Backgrounds of Gravitational Waves}.
\newblock {\em Class. Quant. Grav.}, 35(16):163001, 2018.
\newblock \href {https://arxiv.org/abs/1801.04268} {\path{arXiv:1801.04268}}, \href {https://doi.org/10.1088/1361-6382/aac608} {\path{doi:10.1088/1361-6382/aac608}}.

\bibitem{Maggiore:1999vm}
Michele Maggiore.
\newblock {Gravitational wave experiments and early universe cosmology}.
\newblock {\em Phys. Rept.}, 331:283--367, 2000.
\newblock \href {https://arxiv.org/abs/gr-qc/9909001} {\path{arXiv:gr-qc/9909001}}, \href {https://doi.org/10.1016/S0370-1573(99)00102-7} {\path{doi:10.1016/S0370-1573(99)00102-7}}.

\bibitem{Machado:2018nqk}
Camila~S. Machado, Wolfram Ratzinger, Pedro Schwaller, and Ben~A. Stefanek.
\newblock {Audible Axions}.
\newblock {\em JHEP}, 01:053, 2019.
\newblock \href {https://arxiv.org/abs/1811.01950} {\path{arXiv:1811.01950}}, \href {https://doi.org/10.1007/JHEP01(2019)053} {\path{doi:10.1007/JHEP01(2019)053}}.

\bibitem{Machado:2019xuc}
Camila~S. Machado, Wolfram Ratzinger, Pedro Schwaller, and Ben~A. Stefanek.
\newblock {Gravitational wave probes of axionlike particles}.
\newblock {\em Phys. Rev. D}, 102(7):075033, 2020.
\newblock \href {https://arxiv.org/abs/1912.01007} {\path{arXiv:1912.01007}}, \href {https://doi.org/10.1103/PhysRevD.102.075033} {\path{doi:10.1103/PhysRevD.102.075033}}.

\bibitem{Co:2021rhi}
Raymond~T. Co, Keisuke Harigaya, and Aaron Pierce.
\newblock {Gravitational waves and dark photon dark matter from axion rotations}.
\newblock {\em JHEP}, 12:099, 2021.
\newblock \href {https://arxiv.org/abs/2104.02077} {\path{arXiv:2104.02077}}, \href {https://doi.org/10.1007/JHEP12(2021)099} {\path{doi:10.1007/JHEP12(2021)099}}.

\bibitem{Fonseca:2019ypl}
Nayara Fonseca, Enrico Morgante, Ryosuke Sato, and G\'eraldine Servant.
\newblock {Axion fragmentation}.
\newblock {\em JHEP}, 04:010, 2020.
\newblock \href {https://arxiv.org/abs/1911.08472} {\path{arXiv:1911.08472}}, \href {https://doi.org/10.1007/JHEP04(2020)010} {\path{doi:10.1007/JHEP04(2020)010}}.

\bibitem{Chatrchyan:2020pzh}
Aleksandr Chatrchyan and Joerg Jaeckel.
\newblock {Gravitational waves from the fragmentation of axion-like particle dark matter}.
\newblock {\em JCAP}, 02:003, 2021.
\newblock \href {https://arxiv.org/abs/2004.07844} {\path{arXiv:2004.07844}}, \href {https://doi.org/10.1088/1475-7516/2021/02/003} {\path{doi:10.1088/1475-7516/2021/02/003}}.

\bibitem{Ratzinger:2020oct}
Wolfram Ratzinger, Pedro Schwaller, and Ben~A. Stefanek.
\newblock {Gravitational Waves from an Axion-Dark Photon System: A Lattice Study}.
\newblock {\em SciPost Phys.}, 11:001, 2021.
\newblock \href {https://arxiv.org/abs/2012.11584} {\path{arXiv:2012.11584}}, \href {https://doi.org/10.21468/SciPostPhys.11.1.001} {\path{doi:10.21468/SciPostPhys.11.1.001}}.

\bibitem{Eroncel:2022vjg}
Cem Er\"oncel, Ryosuke Sato, Geraldine Servant, and Philip S\o{}rensen.
\newblock {ALP dark matter from kinetic fragmentation: opening up the parameter window}.
\newblock {\em JCAP}, 10:053, 2022.
\newblock \href {https://arxiv.org/abs/2206.14259} {\path{arXiv:2206.14259}}, \href {https://doi.org/10.1088/1475-7516/2022/10/053} {\path{doi:10.1088/1475-7516/2022/10/053}}.

\bibitem{DiLuzio:2020wdo}
Luca Di~Luzio, Maurizio Giannotti, Enrico Nardi, and Luca Visinelli.
\newblock {The landscape of QCD axion models}.
\newblock {\em Phys. Rept.}, 870:1--117, 2020.
\newblock \href {https://arxiv.org/abs/2003.01100} {\path{arXiv:2003.01100}}, \href {https://doi.org/10.1016/j.physrep.2020.06.002} {\path{doi:10.1016/j.physrep.2020.06.002}}.

\bibitem{Marsh:2015xka}
David J.~E. Marsh.
\newblock {Axion Cosmology}.
\newblock {\em Phys. Rept.}, 643:1--79, 2016.
\newblock \href {https://arxiv.org/abs/1510.07633} {\path{arXiv:1510.07633}}, \href {https://doi.org/10.1016/j.physrep.2016.06.005} {\path{doi:10.1016/j.physrep.2016.06.005}}.

\bibitem{Agrawal:2017eqm}
Prateek Agrawal, Gustavo Marques-Tavares, and Wei Xue.
\newblock {Opening up the QCD axion window}.
\newblock {\em JHEP}, 03:049, 2018.
\newblock \href {https://arxiv.org/abs/1708.05008} {\path{arXiv:1708.05008}}, \href {https://doi.org/10.1007/JHEP03(2018)049} {\path{doi:10.1007/JHEP03(2018)049}}.

\bibitem{Zhang:2021mks}
Jun Zhang, Zhenwei Lyu, Junwu Huang, Matthew~C. Johnson, Laura Sagunski, Mairi Sakellariadou, and Huan Yang.
\newblock {First Constraints on Nuclear Coupling of Axionlike Particles from the Binary Neutron Star Gravitational Wave Event GW170817}.
\newblock {\em Phys. Rev. Lett.}, 127(16):161101, 2021.
\newblock \href {https://arxiv.org/abs/2105.13963} {\path{arXiv:2105.13963}}, \href {https://doi.org/10.1103/PhysRevLett.127.161101} {\path{doi:10.1103/PhysRevLett.127.161101}}.

\bibitem{Baryakhtar:2020gao}
Masha Baryakhtar, Marios Galanis, Robert Lasenby, and Olivier Simon.
\newblock {Black hole superradiance of self-interacting scalar fields}.
\newblock {\em Phys. Rev. D}, 103(9):095019, 2021.
\newblock \href {https://arxiv.org/abs/2011.11646} {\path{arXiv:2011.11646}}, \href {https://doi.org/10.1103/PhysRevD.103.095019} {\path{doi:10.1103/PhysRevD.103.095019}}.

\bibitem{Kitajima:2017peg}
Naoya Kitajima, Toyokazu Sekiguchi, and Fuminobu Takahashi.
\newblock {Cosmological abundance of the QCD axion coupled to hidden photons}.
\newblock {\em Phys. Lett. B}, 781:684--687, 2018.
\newblock \href {https://arxiv.org/abs/1711.06590} {\path{arXiv:1711.06590}}, \href {https://doi.org/10.1016/j.physletb.2018.04.024} {\path{doi:10.1016/j.physletb.2018.04.024}}.

\bibitem{Wang:2022sxn}
Yijun Wang, Kris Pardo, Tzu-Ching Chang, and Olivier Dor\'e.
\newblock {Constraining the stochastic gravitational wave background with photometric surveys}.
\newblock {\em Phys. Rev. D}, 106(8):084006, 2022.
\newblock \href {https://arxiv.org/abs/2205.07962} {\path{arXiv:2205.07962}}, \href {https://doi.org/10.1103/PhysRevD.106.084006} {\path{doi:10.1103/PhysRevD.106.084006}}.

\bibitem{Sathyaprakash:2012jk}
B.~Sathyaprakash et~al.
\newblock {Scientific Objectives of Einstein Telescope}.
\newblock {\em Class. Quant. Grav.}, 29:124013, 2012.
\newblock [Erratum: Class.Quant.Grav. 30, 079501 (2013)].
\newblock \href {https://arxiv.org/abs/1206.0331} {\path{arXiv:1206.0331}}, \href {https://doi.org/10.1088/0264-9381/29/12/124013} {\path{doi:10.1088/0264-9381/29/12/124013}}.

\bibitem{Planck:2018jri}
Y.~Akrami et~al.
\newblock {Planck 2018 results. X. Constraints on inflation}.
\newblock {\em Astron. Astrophys.}, 641:A10, 2020.
\newblock \href {https://arxiv.org/abs/1807.06211} {\path{arXiv:1807.06211}}, \href {https://doi.org/10.1051/0004-6361/201833887} {\path{doi:10.1051/0004-6361/201833887}}.

\bibitem{Kristiano:2022maq}
Jason Kristiano and Jun'ichi Yokoyama.
\newblock {Constraining Primordial Black Hole Formation from Single-Field Inflation}.
\newblock {\em Phys. Rev. Lett.}, 132(22):221003, 2024.
\newblock \href {https://arxiv.org/abs/2211.03395} {\path{arXiv:2211.03395}}, \href {https://doi.org/10.1103/PhysRevLett.132.221003} {\path{doi:10.1103/PhysRevLett.132.221003}}.

\bibitem{Maldacena:2002vr}
Juan~Martin Maldacena.
\newblock {Non-Gaussian features of primordial fluctuations in single field inflationary models}.
\newblock {\em JHEP}, 05:013, 2003.
\newblock \href {https://arxiv.org/abs/astro-ph/0210603} {\path{arXiv:astro-ph/0210603}}, \href {https://doi.org/10.1088/1126-6708/2003/05/013} {\path{doi:10.1088/1126-6708/2003/05/013}}.

\bibitem{Senatore:2009cf}
Leonardo Senatore and Matias Zaldarriaga.
\newblock {On Loops in Inflation}.
\newblock {\em JHEP}, 12:008, 2010.
\newblock \href {https://arxiv.org/abs/0912.2734} {\path{arXiv:0912.2734}}, \href {https://doi.org/10.1007/JHEP12(2010)008} {\path{doi:10.1007/JHEP12(2010)008}}.

\bibitem{BICEP:2021xfz}
P.~A.~R. Ade et~al.
\newblock {Improved Constraints on Primordial Gravitational Waves using Planck, WMAP, and BICEP/Keck Observations through the 2018 Observing Season}.
\newblock {\em Phys. Rev. Lett.}, 127(15):151301, 2021.
\newblock \href {https://arxiv.org/abs/2110.00483} {\path{arXiv:2110.00483}}, \href {https://doi.org/10.1103/PhysRevLett.127.151301} {\path{doi:10.1103/PhysRevLett.127.151301}}.

\bibitem{Assassi:2012et}
Valentin Assassi, Daniel Baumann, and Daniel Green.
\newblock {Symmetries and Loops in Inflation}.
\newblock {\em JHEP}, 02:151, 2013.
\newblock \href {https://arxiv.org/abs/1210.7792} {\path{arXiv:1210.7792}}, \href {https://doi.org/10.1007/JHEP02(2013)151} {\path{doi:10.1007/JHEP02(2013)151}}.

\bibitem{Riotto:2023hoz}
Antonio Riotto.
\newblock {The Primordial Black Hole Formation from Single-Field Inflation is Not Ruled Out}.
\newblock 1 2023.
\newblock \href {https://arxiv.org/abs/2301.00599} {\path{arXiv:2301.00599}}.

\bibitem{Kristiano:2023scm}
Jason Kristiano and Jun'ichi Yokoyama.
\newblock {Note on the bispectrum and one-loop corrections in single-field inflation with primordial black hole formation}.
\newblock {\em Phys. Rev. D}, 109(10):103541, 2024.
\newblock \href {https://arxiv.org/abs/2303.00341} {\path{arXiv:2303.00341}}, \href {https://doi.org/10.1103/PhysRevD.109.103541} {\path{doi:10.1103/PhysRevD.109.103541}}.

\bibitem{Riotto:2023gpm}
A.~Riotto.
\newblock {The Primordial Black Hole Formation from Single-Field Inflation is Still Not Ruled Out}.
\newblock 3 2023.
\newblock \href {https://arxiv.org/abs/2303.01727} {\path{arXiv:2303.01727}}.

\bibitem{Firouzjahi:2023aum}
Hassan Firouzjahi.
\newblock {One-loop corrections in power spectrum in single field inflation}.
\newblock {\em JCAP}, 10:006, 2023.
\newblock \href {https://arxiv.org/abs/2303.12025} {\path{arXiv:2303.12025}}, \href {https://doi.org/10.1088/1475-7516/2023/10/006} {\path{doi:10.1088/1475-7516/2023/10/006}}.

\bibitem{Firouzjahi:2023ahg}
Hassan Firouzjahi and Antonio Riotto.
\newblock {Primordial Black Holes and loops in single-field inflation}.
\newblock {\em JCAP}, 02:021, 2024.
\newblock \href {https://arxiv.org/abs/2304.07801} {\path{arXiv:2304.07801}}, \href {https://doi.org/10.1088/1475-7516/2024/02/021} {\path{doi:10.1088/1475-7516/2024/02/021}}.

\bibitem{Choudhury:2023vuj}
Sayantan Choudhury, Mayukh~R. Gangopadhyay, and M.~Sami.
\newblock {No-go for the formation of heavy mass Primordial Black Holes in Single Field Inflation}.
\newblock {\em Eur. Phys. J. C}, 84(9):884, 2024.
\newblock \href {https://arxiv.org/abs/2301.10000} {\path{arXiv:2301.10000}}, \href {https://doi.org/10.1140/epjc/s10052-024-13218-2} {\path{doi:10.1140/epjc/s10052-024-13218-2}}.

\bibitem{Choudhury:2023jlt}
Sayantan Choudhury, Sudhakar Panda, and M.~Sami.
\newblock {PBH formation in EFT of single field inflation with sharp transition}.
\newblock {\em Phys. Lett. B}, 845:138123, 2023.
\newblock \href {https://arxiv.org/abs/2302.05655} {\path{arXiv:2302.05655}}, \href {https://doi.org/10.1016/j.physletb.2023.138123} {\path{doi:10.1016/j.physletb.2023.138123}}.

\bibitem{Choudhury:2023rks}
Sayantan Choudhury, Sudhakar Panda, and M.~Sami.
\newblock {Quantum loop effects on the power spectrum and constraints on primordial black holes}.
\newblock {\em JCAP}, 11:066, 2023.
\newblock \href {https://arxiv.org/abs/2303.06066} {\path{arXiv:2303.06066}}, \href {https://doi.org/10.1088/1475-7516/2023/11/066} {\path{doi:10.1088/1475-7516/2023/11/066}}.

\bibitem{Choudhury:2023hvf}
Sayantan Choudhury, Sudhakar Panda, and M.~Sami.
\newblock {Galileon inflation evades the no-go for PBH formation in the single-field framework}.
\newblock {\em JCAP}, 08:078, 2023.
\newblock \href {https://arxiv.org/abs/2304.04065} {\path{arXiv:2304.04065}}, \href {https://doi.org/10.1088/1475-7516/2023/08/078} {\path{doi:10.1088/1475-7516/2023/08/078}}.

\bibitem{Calzetta:1986ey}
E.~Calzetta and B.~L. Hu.
\newblock {Closed Time Path Functional Formalism in Curved Space-Time: Application to Cosmological Back Reaction Problems}.
\newblock {\em Phys. Rev. D}, 35:495, 1987.
\newblock \href {https://doi.org/10.1103/PhysRevD.35.495} {\path{doi:10.1103/PhysRevD.35.495}}.

\bibitem{Jordan:1986ug}
R.~D. Jordan.
\newblock {Effective Field Equations for Expectation Values}.
\newblock {\em Phys. Rev. D}, 33:444--454, 1986.
\newblock \href {https://doi.org/10.1103/PhysRevD.33.444} {\path{doi:10.1103/PhysRevD.33.444}}.

\bibitem{Adshead:2009cb}
Peter Adshead, Richard Easther, and Eugene~A. Lim.
\newblock {The 'in-in' Formalism and Cosmological Perturbations}.
\newblock {\em Phys. Rev. D}, 80:083521, 2009.
\newblock \href {https://arxiv.org/abs/0904.4207} {\path{arXiv:0904.4207}}, \href {https://doi.org/10.1103/PhysRevD.80.083521} {\path{doi:10.1103/PhysRevD.80.083521}}.

\bibitem{Sloth:2006nu}
Martin~S. Sloth.
\newblock {On the one loop corrections to inflation. II. The Consistency relation}.
\newblock {\em Nucl. Phys. B}, 775:78--94, 2007.
\newblock \href {https://arxiv.org/abs/hep-th/0612138} {\path{arXiv:hep-th/0612138}}, \href {https://doi.org/10.1016/j.nuclphysb.2007.04.012} {\path{doi:10.1016/j.nuclphysb.2007.04.012}}.

\bibitem{Inomata:2022yte}
Keisuke Inomata, Matteo Braglia, and Xingang Chen.
\newblock {Questions on calculation of primordial power spectrum with large spikes: the resonance model case}.
\newblock {\em JCAP}, 04:011, 2023.
\newblock \href {https://arxiv.org/abs/2211.02586} {\path{arXiv:2211.02586}}, \href {https://doi.org/10.1088/1475-7516/2023/04/011} {\path{doi:10.1088/1475-7516/2023/04/011}}.

\bibitem{Jarnhus:2007ia}
Philip~R. Jarnhus and Martin~S. Sloth.
\newblock {de Sitter limit of inflation and nonlinear perturbation theory}.
\newblock {\em JCAP}, 02:013, 2008.
\newblock \href {https://arxiv.org/abs/0709.2708} {\path{arXiv:0709.2708}}, \href {https://doi.org/10.1088/1475-7516/2008/02/013} {\path{doi:10.1088/1475-7516/2008/02/013}}.

\bibitem{Kristiano:2024ngc}
Jason Kristiano and Jun'ichi Yokoyama.
\newblock {Generating large primordial fluctuations in single-field inflation for PBH formation}.
\newblock 5 2024.
\newblock \href {https://arxiv.org/abs/2405.12149} {\path{arXiv:2405.12149}}.

\bibitem{Kristiano:2024vst}
Jason Kristiano and Jun'ichi Yokoyama.
\newblock {Comparing sharp and smooth transitions of the second slow-roll parameter in single-field inflation}.
\newblock {\em JCAP}, 10:036, 2024.
\newblock \href {https://arxiv.org/abs/2405.12145} {\path{arXiv:2405.12145}}, \href {https://doi.org/10.1088/1475-7516/2024/10/036} {\path{doi:10.1088/1475-7516/2024/10/036}}.

\bibitem{Adshead:2008gk}
Peter Adshead, Richard Easther, and Eugene~A. Lim.
\newblock {Cosmology With Many Light Scalar Fields: Stochastic Inflation and Loop Corrections}.
\newblock {\em Phys. Rev. D}, 79:063504, 2009.
\newblock \href {https://arxiv.org/abs/0809.4008} {\path{arXiv:0809.4008}}, \href {https://doi.org/10.1103/PhysRevD.79.063504} {\path{doi:10.1103/PhysRevD.79.063504}}.

\bibitem{Carr:2020xqk}
Bernard Carr and Florian Kuhnel.
\newblock {Primordial Black Holes as Dark Matter: Recent Developments}.
\newblock {\em Ann. Rev. Nucl. Part. Sci.}, 70:355--394, 2020.
\newblock \href {https://arxiv.org/abs/2006.02838} {\path{arXiv:2006.02838}}, \href {https://doi.org/10.1146/annurev-nucl-050520-125911} {\path{doi:10.1146/annurev-nucl-050520-125911}}.

\bibitem{Namjoo:2012aa}
Mohammad~Hossein Namjoo, Hassan Firouzjahi, and Misao Sasaki.
\newblock {Violation of non-Gaussianity consistency relation in a single field inflationary model}.
\newblock {\em EPL}, 101(3):39001, 2013.
\newblock \href {https://arxiv.org/abs/1210.3692} {\path{arXiv:1210.3692}}, \href {https://doi.org/10.1209/0295-5075/101/39001} {\path{doi:10.1209/0295-5075/101/39001}}.

\bibitem{Chen:2013eea}
Xingang Chen, Hassan Firouzjahi, Eiichiro Komatsu, Mohammad~Hossein Namjoo, and Misao Sasaki.
\newblock {In-in and $\delta N$ calculations of the bispectrum from non-attractor single-field inflation}.
\newblock {\em JCAP}, 12:039, 2013.
\newblock \href {https://arxiv.org/abs/1308.5341} {\path{arXiv:1308.5341}}, \href {https://doi.org/10.1088/1475-7516/2013/12/039} {\path{doi:10.1088/1475-7516/2013/12/039}}.

\bibitem{Cai:2018dkf}
Yi-Fu Cai, Xingang Chen, Mohammad~Hossein Namjoo, Misao Sasaki, Dong-Gang Wang, and Ziwei Wang.
\newblock {Revisiting non-Gaussianity from non-attractor inflation models}.
\newblock {\em JCAP}, 05:012, 2018.
\newblock \href {https://arxiv.org/abs/1712.09998} {\path{arXiv:1712.09998}}, \href {https://doi.org/10.1088/1475-7516/2018/05/012} {\path{doi:10.1088/1475-7516/2018/05/012}}.

\bibitem{Pi:2022ysn}
Shi Pi and Misao Sasaki.
\newblock {Logarithmic Duality of the Curvature Perturbation}.
\newblock {\em Phys. Rev. Lett.}, 131(1):011002, 2023.
\newblock \href {https://arxiv.org/abs/2211.13932} {\path{arXiv:2211.13932}}, \href {https://doi.org/10.1103/PhysRevLett.131.011002} {\path{doi:10.1103/PhysRevLett.131.011002}}.

\bibitem{Passaglia:2018ixg}
Samuel Passaglia, Wayne Hu, and Hayato Motohashi.
\newblock {Primordial black holes and local non-Gaussianity in canonical inflation}.
\newblock {\em Phys. Rev. D}, 99(4):043536, 2019.
\newblock \href {https://arxiv.org/abs/1812.08243} {\path{arXiv:1812.08243}}, \href {https://doi.org/10.1103/PhysRevD.99.043536} {\path{doi:10.1103/PhysRevD.99.043536}}.

\bibitem{Figueroa:2020jkf}
Daniel~G. Figueroa, Sami Raatikainen, Syksy Rasanen, and Eemeli Tomberg.
\newblock {Non-Gaussian Tail of the Curvature Perturbation in Stochastic Ultraslow-Roll Inflation: Implications for Primordial Black Hole Production}.
\newblock {\em Phys. Rev. Lett.}, 127(10):101302, 2021.
\newblock \href {https://arxiv.org/abs/2012.06551} {\path{arXiv:2012.06551}}, \href {https://doi.org/10.1103/PhysRevLett.127.101302} {\path{doi:10.1103/PhysRevLett.127.101302}}.

\bibitem{Figueroa:2021zah}
Daniel~G. Figueroa, Sami Raatikainen, Syksy Rasanen, and Eemeli Tomberg.
\newblock {Implications of stochastic effects for primordial black hole production in ultra-slow-roll inflation}.
\newblock {\em JCAP}, 05(05):027, 2022.
\newblock \href {https://arxiv.org/abs/2111.07437} {\path{arXiv:2111.07437}}, \href {https://doi.org/10.1088/1475-7516/2022/05/027} {\path{doi:10.1088/1475-7516/2022/05/027}}.

\bibitem{Green:2004wb}
Anne~M. Green, Andrew~R. Liddle, Karim~A. Malik, and Misao Sasaki.
\newblock {A New calculation of the mass fraction of primordial black holes}.
\newblock {\em Phys. Rev. D}, 70:041502, 2004.
\newblock \href {https://arxiv.org/abs/astro-ph/0403181} {\path{arXiv:astro-ph/0403181}}, \href {https://doi.org/10.1103/PhysRevD.70.041502} {\path{doi:10.1103/PhysRevD.70.041502}}.

\bibitem{Motohashi:2023syh}
Hayato Motohashi and Yuichiro Tada.
\newblock {Squeezed bispectrum and one-loop corrections in transient constant-roll inflation}.
\newblock {\em JCAP}, 08:069, 2023.
\newblock \href {https://arxiv.org/abs/2303.16035} {\path{arXiv:2303.16035}}, \href {https://doi.org/10.1088/1475-7516/2023/08/069} {\path{doi:10.1088/1475-7516/2023/08/069}}.

\bibitem{Leach:2002ar}
Samuel~M. Leach, Andrew~R. Liddle, Jerome Martin, and Dominik~J Schwarz.
\newblock {Cosmological parameter estimation and the inflationary cosmology}.
\newblock {\em Phys. Rev. D}, 66:023515, 2002.
\newblock \href {https://arxiv.org/abs/astro-ph/0202094} {\path{arXiv:astro-ph/0202094}}, \href {https://doi.org/10.1103/PhysRevD.66.023515} {\path{doi:10.1103/PhysRevD.66.023515}}.

\bibitem{Pimentel:2012tw}
Guilherme~L. Pimentel, Leonardo Senatore, and Matias Zaldarriaga.
\newblock {On Loops in Inflation III: Time Independence of zeta in Single Clock Inflation}.
\newblock {\em JHEP}, 07:166, 2012.
\newblock \href {https://arxiv.org/abs/1203.6651} {\path{arXiv:1203.6651}}, \href {https://doi.org/10.1007/JHEP07(2012)166} {\path{doi:10.1007/JHEP07(2012)166}}.

\bibitem{Jackson:2023obv}
Joseph H.~P. Jackson, Hooshyar Assadullahi, Andrew~D. Gow, Kazuya Koyama, Vincent Vennin, and David Wands.
\newblock {The separate-universe approach and sudden transitions during inflation}.
\newblock {\em JCAP}, 05:053, 2024.
\newblock \href {https://arxiv.org/abs/2311.03281} {\path{arXiv:2311.03281}}, \href {https://doi.org/10.1088/1475-7516/2024/05/053} {\path{doi:10.1088/1475-7516/2024/05/053}}.

\bibitem{Domenech:2023dxx}
Guillem Dom\`enech, Gerson Vargas, and Te\'ofilo Vargas.
\newblock {An exact model for enhancing/suppressing primordial fluctuations}.
\newblock {\em JCAP}, 03:002, 2024.
\newblock \href {https://arxiv.org/abs/2309.05750} {\path{arXiv:2309.05750}}, \href {https://doi.org/10.1088/1475-7516/2024/03/002} {\path{doi:10.1088/1475-7516/2024/03/002}}.

\bibitem{Pajer:2013ana}
Enrico Pajer, Fabian Schmidt, and Matias Zaldarriaga.
\newblock {The Observed Squeezed Limit of Cosmological Three-Point Functions}.
\newblock {\em Phys. Rev. D}, 88(8):083502, 2013.
\newblock \href {https://arxiv.org/abs/1305.0824} {\path{arXiv:1305.0824}}, \href {https://doi.org/10.1103/PhysRevD.88.083502} {\path{doi:10.1103/PhysRevD.88.083502}}.

\bibitem{Creminelli:2004yq}
Paolo Creminelli and Matias Zaldarriaga.
\newblock {Single field consistency relation for the 3-point function}.
\newblock {\em JCAP}, 10:006, 2004.
\newblock \href {https://arxiv.org/abs/astro-ph/0407059} {\path{arXiv:astro-ph/0407059}}, \href {https://doi.org/10.1088/1475-7516/2004/10/006} {\path{doi:10.1088/1475-7516/2004/10/006}}.

\bibitem{ParticleDataGroup:2022pth}
R.~L. Workman et~al.
\newblock {Review of Particle Physics}.
\newblock {\em PTEP}, 2022:083C01, 2022.
\newblock \href {https://doi.org/10.1093/ptep/ptac097} {\path{doi:10.1093/ptep/ptac097}}.

\bibitem{Bassett:2005xm}
Bruce~A. Bassett, Shinji Tsujikawa, and David Wands.
\newblock {Inflation dynamics and reheating}.
\newblock {\em Rev. Mod. Phys.}, 78:537--589, 2006.
\newblock \href {https://arxiv.org/abs/astro-ph/0507632} {\path{arXiv:astro-ph/0507632}}, \href {https://doi.org/10.1103/RevModPhys.78.537} {\path{doi:10.1103/RevModPhys.78.537}}.

\bibitem{Allahverdi:2010xz}
Rouzbeh Allahverdi, Robert Brandenberger, Francis-Yan Cyr-Racine, and Anupam Mazumdar.
\newblock {Reheating in Inflationary Cosmology: Theory and Applications}.
\newblock {\em Ann. Rev. Nucl. Part. Sci.}, 60:27--51, 2010.
\newblock \href {https://arxiv.org/abs/1001.2600} {\path{arXiv:1001.2600}}, \href {https://doi.org/10.1146/annurev.nucl.012809.104511} {\path{doi:10.1146/annurev.nucl.012809.104511}}.

\bibitem{Amin:2014eta}
Mustafa~A. Amin, Mark~P. Hertzberg, David~I. Kaiser, and Johanna Karouby.
\newblock {Nonperturbative Dynamics Of Reheating After Inflation: A Review}.
\newblock {\em Int. J. Mod. Phys. D}, 24:1530003, 2014.
\newblock \href {https://arxiv.org/abs/1410.3808} {\path{arXiv:1410.3808}}, \href {https://doi.org/10.1142/S0218271815300037} {\path{doi:10.1142/S0218271815300037}}.

\bibitem{Starobinsky:1980te}
Alexei~A. Starobinsky.
\newblock {A New Type of Isotropic Cosmological Models Without Singularity}.
\newblock {\em Phys. Lett. B}, 91:99--102, 1980.
\newblock \href {https://doi.org/10.1016/0370-2693(80)90670-X} {\path{doi:10.1016/0370-2693(80)90670-X}}.

\bibitem{Mishra:2019pzq}
Swagat~S. Mishra and Varun Sahni.
\newblock {Primordial Black Holes from a tiny bump/dip in the Inflaton potential}.
\newblock {\em JCAP}, 04:007, 2020.
\newblock \href {https://arxiv.org/abs/1911.00057} {\path{arXiv:1911.00057}}, \href {https://doi.org/10.1088/1475-7516/2020/04/007} {\path{doi:10.1088/1475-7516/2020/04/007}}.

\bibitem{Ozsoy:2021pws}
Ogan \"Ozsoy and Gianmassimo Tasinato.
\newblock {Consistency conditions and primordial black holes in single field inflation}.
\newblock {\em Phys. Rev. D}, 105(2):023524, 2022.
\newblock \href {https://arxiv.org/abs/2111.02432} {\path{arXiv:2111.02432}}, \href {https://doi.org/10.1103/PhysRevD.105.023524} {\path{doi:10.1103/PhysRevD.105.023524}}.

\bibitem{Balaji:2022zur}
Shyam Balaji, H.~V. Ragavendra, Shiv~K. Sethi, Joseph Silk, and L.~Sriramkumar.
\newblock {Observing Nulling of Primordial Correlations via the 21-cm Signal}.
\newblock {\em Phys. Rev. Lett.}, 129(26):261301, 2022.
\newblock \href {https://arxiv.org/abs/2206.06386} {\path{arXiv:2206.06386}}, \href {https://doi.org/10.1103/PhysRevLett.129.261301} {\path{doi:10.1103/PhysRevLett.129.261301}}.

\bibitem{Tasinato:2023ukp}
Gianmassimo Tasinato.
\newblock {Large |\ensuremath{\eta}| approach to single field inflation}.
\newblock {\em Phys. Rev. D}, 108(4):043526, 2023.
\newblock \href {https://arxiv.org/abs/2305.11568} {\path{arXiv:2305.11568}}, \href {https://doi.org/10.1103/PhysRevD.108.043526} {\path{doi:10.1103/PhysRevD.108.043526}}.

\bibitem{Fumagalli:2023hpa}
Jacopo Fumagalli.
\newblock {Absence of one-loop effects on large scales from small scales in non-slow-roll dynamics}.
\newblock 5 2023.
\newblock \href {https://arxiv.org/abs/2305.19263} {\path{arXiv:2305.19263}}.

\bibitem{Firouzjahi:2024psd}
Hassan Firouzjahi.
\newblock {Loop corrections in the bispectrum in ultraslow-roll inflation with PBHs formation}.
\newblock {\em Phys. Rev. D}, 110(4):043519, 2024.
\newblock \href {https://arxiv.org/abs/2403.03841} {\path{arXiv:2403.03841}}, \href {https://doi.org/10.1103/PhysRevD.110.043519} {\path{doi:10.1103/PhysRevD.110.043519}}.

\bibitem{Iacconi:2023ggt}
Laura Iacconi, David Mulryne, and David Seery.
\newblock {Loop corrections in the separate universe picture}.
\newblock {\em JCAP}, 06:062, 2024.
\newblock \href {https://arxiv.org/abs/2312.12424} {\path{arXiv:2312.12424}}, \href {https://doi.org/10.1088/1475-7516/2024/06/062} {\path{doi:10.1088/1475-7516/2024/06/062}}.

\bibitem{Davies:2023hhn}
Matthew~W. Davies, Laura Iacconi, and David~J. Mulryne.
\newblock {Numerical 1-loop correction from a potential yielding ultra-slow-roll dynamics}.
\newblock {\em JCAP}, 04:050, 2024.
\newblock \href {https://arxiv.org/abs/2312.05694} {\path{arXiv:2312.05694}}, \href {https://doi.org/10.1088/1475-7516/2024/04/050} {\path{doi:10.1088/1475-7516/2024/04/050}}.

\bibitem{Tasinato:2023ioq}
Gianmassimo Tasinato.
\newblock {Non-Gaussianities and the large |\ensuremath{\eta}| approach to inflation}.
\newblock {\em Phys. Rev. D}, 109(6):063510, 2024.
\newblock \href {https://arxiv.org/abs/2312.03498} {\path{arXiv:2312.03498}}, \href {https://doi.org/10.1103/PhysRevD.109.063510} {\path{doi:10.1103/PhysRevD.109.063510}}.

\bibitem{Firouzjahi:2023bkt}
Hassan Firouzjahi.
\newblock {Revisiting loop corrections in single field ultraslow-roll inflation}.
\newblock {\em Phys. Rev. D}, 109(4):043514, 2024.
\newblock \href {https://arxiv.org/abs/2311.04080} {\path{arXiv:2311.04080}}, \href {https://doi.org/10.1103/PhysRevD.109.043514} {\path{doi:10.1103/PhysRevD.109.043514}}.

\bibitem{Tada:2023rgp}
Yuichiro Tada, Takahiro Terada, and Junsei Tokuda.
\newblock {Cancellation of quantum corrections on the soft curvature perturbations}.
\newblock {\em JHEP}, 01:105, 2024.
\newblock \href {https://arxiv.org/abs/2308.04732} {\path{arXiv:2308.04732}}, \href {https://doi.org/10.1007/JHEP01(2024)105} {\path{doi:10.1007/JHEP01(2024)105}}.

\bibitem{Inomata:2024lud}
Keisuke Inomata.
\newblock {Superhorizon Curvature Perturbations Are Protected against One-Loop Corrections}.
\newblock {\em Phys. Rev. Lett.}, 133(14):141001, 2024.
\newblock \href {https://arxiv.org/abs/2403.04682} {\path{arXiv:2403.04682}}, \href {https://doi.org/10.1103/PhysRevLett.133.141001} {\path{doi:10.1103/PhysRevLett.133.141001}}.

\bibitem{Ballesteros:2024zdp}
Guillermo Ballesteros and Jes\'us~Gamb\'\i{}n Egea.
\newblock {One-loop power spectrum in ultra slow-roll inflation and implications for primordial black hole dark matter}.
\newblock {\em JCAP}, 07:052, 2024.
\newblock \href {https://arxiv.org/abs/2404.07196} {\path{arXiv:2404.07196}}, \href {https://doi.org/10.1088/1475-7516/2024/07/052} {\path{doi:10.1088/1475-7516/2024/07/052}}.

\bibitem{Dine:1995kz}
Michael Dine, Lisa Randall, and Scott~D. Thomas.
\newblock {Baryogenesis from flat directions of the supersymmetric standard model}.
\newblock {\em Nucl. Phys. B}, 458:291--326, 1996.
\newblock \href {https://arxiv.org/abs/hep-ph/9507453} {\path{arXiv:hep-ph/9507453}}, \href {https://doi.org/10.1016/0550-3213(95)00538-2} {\path{doi:10.1016/0550-3213(95)00538-2}}.

\bibitem{Blinov:2019rhb}
Nikita Blinov, Matthew~J Dolan, Patrick Draper, and Jonathan Kozaczuk.
\newblock {Dark matter targets for axionlike particle searches}.
\newblock {\em Phys. Rev. D}, 100(1):015049, 2019.
\newblock \href {https://arxiv.org/abs/1905.06952} {\path{arXiv:1905.06952}}, \href {https://doi.org/10.1103/PhysRevD.100.015049} {\path{doi:10.1103/PhysRevD.100.015049}}.

\bibitem{Firouzjahi:2012iz}
Hassan Firouzjahi, Anne Green, Karim Malik, and Moslem Zarei.
\newblock {Effect of curvaton decay on the primordial power spectrum}.
\newblock {\em Phys. Rev. D}, 87(10):103502, 2013.
\newblock \href {https://arxiv.org/abs/1209.2652} {\path{arXiv:1209.2652}}, \href {https://doi.org/10.1103/PhysRevD.87.103502} {\path{doi:10.1103/PhysRevD.87.103502}}.

\bibitem{Kodama:1984ziu}
Hideo Kodama and Misao Sasaki.
\newblock {Cosmological Perturbation Theory}.
\newblock {\em Prog. Theor. Phys. Suppl.}, 78:1--166, 1984.
\newblock \href {https://doi.org/10.1143/PTPS.78.1} {\path{doi:10.1143/PTPS.78.1}}.

\bibitem{Jedamzik:1998hc}
Karsten Jedamzik.
\newblock {Could MACHOS be primordial black holes formed during the QCD epoch?}
\newblock {\em Phys. Rept.}, 307:155--162, 1998.
\newblock \href {https://arxiv.org/abs/astro-ph/9805147} {\path{arXiv:astro-ph/9805147}}, \href {https://doi.org/10.1016/S0370-1573(98)00067-2} {\path{doi:10.1016/S0370-1573(98)00067-2}}.

\bibitem{Escriva:2022bwe}
Albert Escriv\`a, Eleni Bagui, and Sebastien Clesse.
\newblock {Simulations of PBH formation at the QCD epoch and comparison with the GWTC-3 catalog}.
\newblock {\em JCAP}, 05:004, 2023.
\newblock \href {https://arxiv.org/abs/2209.06196} {\path{arXiv:2209.06196}}, \href {https://doi.org/10.1088/1475-7516/2023/05/004} {\path{doi:10.1088/1475-7516/2023/05/004}}.

\bibitem{Gow:2020bzo}
Andrew~D. Gow, Christian~T. Byrnes, Philippa~S. Cole, and Sam Young.
\newblock {The power spectrum on small scales: Robust constraints and comparing PBH methodologies}.
\newblock {\em JCAP}, 02:002, 2021.
\newblock \href {https://arxiv.org/abs/2008.03289} {\path{arXiv:2008.03289}}, \href {https://doi.org/10.1088/1475-7516/2021/02/002} {\path{doi:10.1088/1475-7516/2021/02/002}}.

\bibitem{Pi:2020otn}
Shi Pi and Misao Sasaki.
\newblock {Gravitational Waves Induced by Scalar Perturbations with a Lognormal Peak}.
\newblock {\em JCAP}, 09:037, 2020.
\newblock \href {https://arxiv.org/abs/2005.12306} {\path{arXiv:2005.12306}}, \href {https://doi.org/10.1088/1475-7516/2020/09/037} {\path{doi:10.1088/1475-7516/2020/09/037}}.

\bibitem{Yuan:2019wwo}
Chen Yuan, Zu-Cheng Chen, and Qing-Guo Huang.
\newblock {Log-dependent slope of scalar induced gravitational waves in the infrared regions}.
\newblock {\em Phys. Rev. D}, 101(4):043019, 2020.
\newblock \href {https://arxiv.org/abs/1910.09099} {\path{arXiv:1910.09099}}, \href {https://doi.org/10.1103/PhysRevD.101.043019} {\path{doi:10.1103/PhysRevD.101.043019}}.

\bibitem{Wang:2023sij}
Sai Wang, Zhi-Chao Zhao, and Qing-Hua Zhu.
\newblock {Constraints on scalar-induced gravitational waves up to third order from a joint analysis of BBN, CMB, and PTA data}.
\newblock {\em Phys. Rev. Res.}, 6(1):013207, 2024.
\newblock \href {https://arxiv.org/abs/2307.03095} {\path{arXiv:2307.03095}}, \href {https://doi.org/10.1103/PhysRevResearch.6.013207} {\path{doi:10.1103/PhysRevResearch.6.013207}}.

\bibitem{Linde:1986fe}
Andrei~D. Linde.
\newblock { Eternally existing selfreproducing inflationary universe}.
\newblock {\em Phys. Scripta T}, 15:169, 1987.
\newblock \href {https://doi.org/10.1088/0031-8949/1987/T15/024} {\path{doi:10.1088/0031-8949/1987/T15/024}}.

\bibitem{Goncharov:1987ir}
A.~S. Goncharov, Andrei~D. Linde, and Viatcheslav~F. Mukhanov.
\newblock {The Global Structure of the Inflationary Universe}.
\newblock {\em Int. J. Mod. Phys. A}, 2:561--591, 1987.
\newblock \href {https://doi.org/10.1142/S0217751X87000211} {\path{doi:10.1142/S0217751X87000211}}.

\bibitem{Drees:2022aea}
Manuel Drees and Yong Xu.
\newblock {Large field polynomial inflation: parameter space, predictions and (double) eternal nature}.
\newblock {\em JCAP}, 12:005, 2022.
\newblock \href {https://arxiv.org/abs/2209.07545} {\path{arXiv:2209.07545}}, \href {https://doi.org/10.1088/1475-7516/2022/12/005} {\path{doi:10.1088/1475-7516/2022/12/005}}.

\bibitem{Azhar:2018lzd}
Feraz Azhar and Abraham Loeb.
\newblock {Gauging Fine-Tuning}.
\newblock {\em Phys. Rev. D}, 98(10):103018, 2018.
\newblock \href {https://arxiv.org/abs/1809.06220} {\path{arXiv:1809.06220}}, \href {https://doi.org/10.1103/PhysRevD.98.103018} {\path{doi:10.1103/PhysRevD.98.103018}}.

\bibitem{Kasuya:2009up}
Shinta Kasuya and Masahiro Kawasaki.
\newblock {Axion isocurvature fluctuations with extremely blue spectrum}.
\newblock {\em Phys. Rev. D}, 80:023516, 2009.
\newblock \href {https://arxiv.org/abs/0904.3800} {\path{arXiv:0904.3800}}, \href {https://doi.org/10.1103/PhysRevD.80.023516} {\path{doi:10.1103/PhysRevD.80.023516}}.

\bibitem{Hellerman:2005yi}
Simeon Hellerman and Johannes Walcher.
\newblock {Dark matter and the anthropic principle}.
\newblock {\em Phys. Rev. D}, 72:123520, 2005.
\newblock \href {https://arxiv.org/abs/hep-th/0508161} {\path{arXiv:hep-th/0508161}}, \href {https://doi.org/10.1103/PhysRevD.72.123520} {\path{doi:10.1103/PhysRevD.72.123520}}.

\end{thebibliography}

\cleardoublepage
\appendix 
\chapter{Appendix of Chapter 3}
\section{On the convergence of the power-series expansion}\label{app:Radius}
Consider the power series expansion of eq.\,(\ref{eq:MasterX})
\begin{align}
\sum_{n = 1}^{\infty} c_n(r_{\rm dec}) \zeta_{\rm G}^{n} = \log\big[X(r_{\rm dec},\zeta_{\rm G})\big]\,.
\label{eq:PoweSeriesExpansion}
\end{align}
The above equality is valid only within the radius of convergence of the series expansion.
First, consider for simplicity the case with $r_{\rm dec}=1$. The coefficients $c_n(1)$ take the form
\begin{align}
c_n(1) = \frac{(-1)^{n+1}}{n(2/3)^{n-1}}\,,
\end{align}
and the radius of convergence is given by
\begin{align}
R = \lim_{n\to \infty}\left|\frac{c_n(1)}{c_{n+1}(1)}\right| = \frac{2}{3}\,,
\end{align}
so that we have
\begin{align}
\sum_{n = 1}^{\infty} c_n(1) \zeta_{\rm G}^{n} = 
 \log\big[X(1,\zeta_{\rm G})\big] = \frac{2}{3}\log\bigg(
1+ \frac{3}{2}\zeta_{\rm G}
\bigg)\,,~~~~~{\rm for}\,\,\,-\frac{2}{3} < \zeta_{\rm G} < +\frac{2}{3}\,.
\end{align}
This result makes sense since the function  $\log(1+ 3\zeta_{\rm G}/2)$ has a singularity when $\zeta_{\rm G} = -2/3$ and the series expansion centered in $\zeta_{\rm G} = 0$ is limited by its presence. 

Consider now the case $r_{\rm dec}\neq 1$. 
For definiteness, we consider $r_{\rm dec} = 0.5$. 
In this case, we find that the radius of convergence is given by $R \simeq 0.173$. 
Unfortunately, 
we do not have analytic expressions for the coefficients 
 $c_n(r_{\rm dec})$ with $r_{\rm dec} \neq 1$ so the aforementioned result must be derived and understood without relying on standard methods.
 
First of all, let us visualize the situation with the help of a graph.
\begin{figure}[!h!]
\begin{center}
$$\includegraphics[width=.45\textwidth]{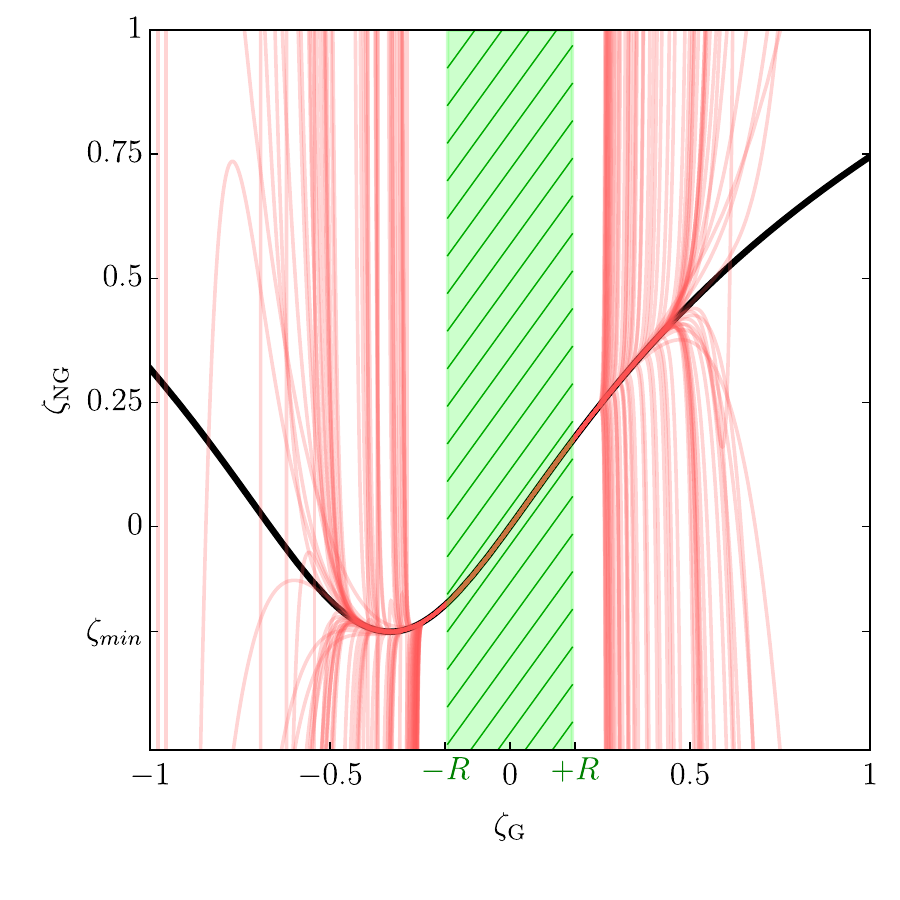}
\qquad\qquad\includegraphics[width=.45\textwidth]{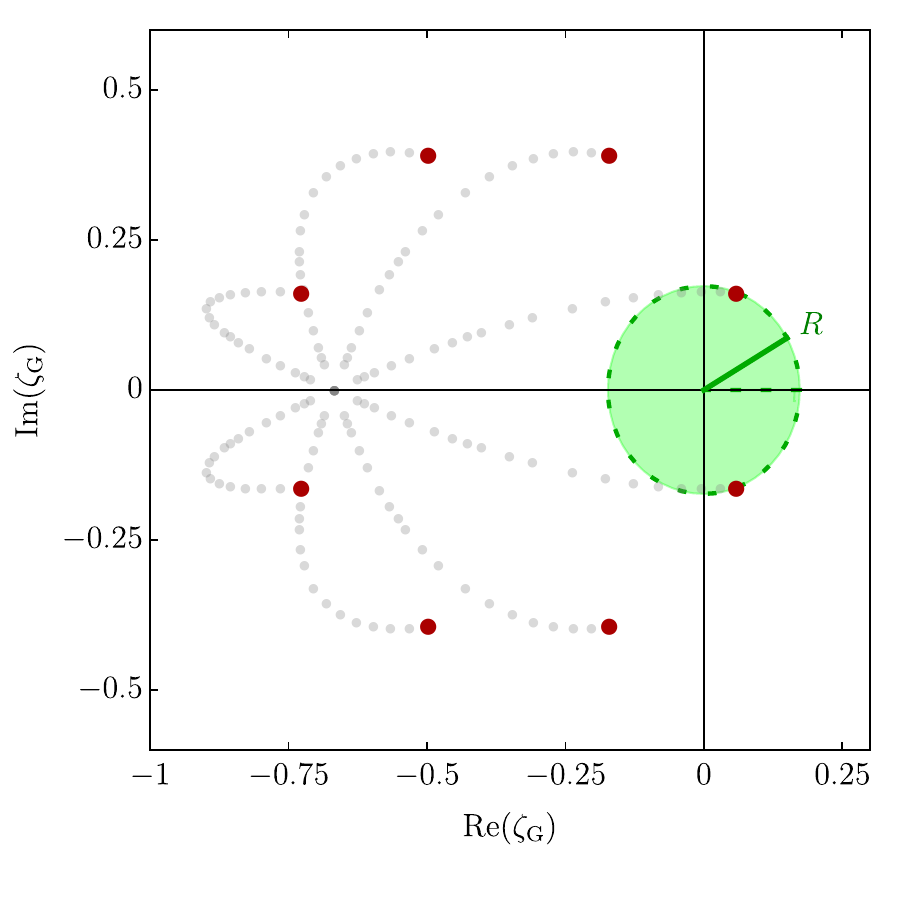}$$
\caption{
\textbf{\textit{	Left panel. }}Comparison between the function $\zeta =  \log\big[X(r_{\rm dec},\zeta_{\rm G})\big]$ with $r_{\rm dec}=0.5$ (solid black line) 
and its power series expansion $\sum_{n = 1}^{N} c_n(r_{\rm dec}) \zeta_{\rm G}^{n}$ for increasing 
values of $N$ (light red lines). The region 
shaded in green is limited by the condition $-R< \zeta_{\rm G} < +R$ with radius of convergence 
$R \simeq 0.173$. 
	\textbf{\textit{	Right panel. }}The red dots mark the position of the branch points of the function 
$P(r_{\rm dec},\zeta_{\rm G})$ in eq.\,(\ref{eq:EqRip}) with $r_{\rm dec}=0.5$. To visualize these points, 
the independent variable $\zeta_{\rm G}$ is promoted to be complex. 
The green region is the disk of convergence with radius $R \simeq 0.173$; the latter measures the 
distance from the center of the expansion (that is the origin of the complex plane in this case) of the closest branch point. 
The trajectories marked by light gray dots indicate how the branch points move as we increase $r_{\rm dec}$ from $r_{\rm dec}=0.5$ to $r_{\rm dec}=1$.
 }\label{fig:Radius}  
\end{center}
\end{figure} 
 In the left panel of fig.\,\ref{fig:Radius} we compare, as function of $\zeta_{\rm G}$, the full 
 expression $\log\big[X(r_{\rm dec},\zeta_{\rm G})\big]$ (solid black line) with its power-series expansion 
 $\sum_{n = 1}^{N} c_n(r_{\rm dec}) \zeta_{\rm G}^{n}$ truncated at increasing values of $N$ (light red lines); 
 the region shaded in green is limited by $-R<\zeta_{\rm G}<+R$ with $R \simeq 0.173$. 
 As clear from this plot, within the radius of convergence the equality in 
 eq.\,(\ref{eq:PoweSeriesExpansion}) holds true but as soon as we consider values of 
 $\zeta_{\rm G}$ outside the green band the power-series badly diverges from the exact result. 
\\This plot provides a numerical confirmation that indeed $R \simeq 0.173$. 
\\However, it is natural to ask what is the origin of this number and how we computed it. 
 At first sight, this is indeed puzzling. 
 In the case $r_{\rm dec} = 1$, as previously discussed, the meaning of the radius of convergence was clear given the 
 singularity of the function $\log(
1+ 3\zeta_{\rm G}/2)$: the radius of convergence was just the distance from the origin of the singularity 
 (including which analyticity would be lost).
 
Moreover, if we take $r_{\rm dec} = 0.5$ the function 
$\log\big[X(r_{\rm dec},\zeta_{\rm G})\big]$ does not present any problem. This is indeed evident 
from the very same plot in the left panel of fig.\,\ref{fig:Radius} that we just discussed: 
the black curve is a perfectly 
smooth curve. What is the obstruction that sets $R \simeq 0.173$? 
To answer this question, let us discuss a simple example. 
Consider the geometric series
\begin{align}
\sum_{k=0}^{\infty}(-x)^{2k} = \frac{1}{1+x^2}\,,~~~~~{\rm for}\,\,\,-1<x<1\,.
\end{align}
The radius of convergence is $R=1$. However, if we plot the function $f(x) = 1/(1+x^2)$ we notice 
that this is a perfectly smooth curve that peaks at the origin and dies off at $\pm \infty$ without any singularity 
on the real axis.
Notice that this situation is qualitatively very similar to the one we discussed before.
What limits $R=1$? The (well-known) answer is that, because of analytical continuation, the function $f(x) = 1/(1+x^2)$ actually knows about its singularities in the complex plane located at $x = \pm i$. The radius of convergence, therefore, is still measuring the distance from the origin of the closest singularity with the difference that now the singularity in question is displaced from the real axis.  

Something analogous happens in our case. The radius of convergence $R \simeq 0.173$ measures the distance from the center of the closest branch-point singularity of the square root that enters in the function
 \begin{align}
P(r_{\rm dec},\zeta_{\rm G}) \equiv 
\frac{(2r_{\rm dec} + 3\zeta_{\rm G})^{4}}{16 r_{\rm dec}^2} + \sqrt{
(1-r_{\rm dec})^3(3+r_{\rm dec}) + \frac{(2r_{\rm dec} + 3\zeta_{\rm G})^8}{256 r_{\rm dec}^4}
}\,.\label{eq:EqRip}
\end{align}
The argument of the square root is a polynomial of degree 8 in $\zeta_{\rm G}$ with branch points located as in the right panel of fig.\,\ref{fig:Radius} 
(the red dots, with $\zeta_{\rm G}$ promoted to be a complex variable) 
and the singularity closest to $\zeta_{\rm G} = 0$ sets the radius of convergence, as shown in the plot (notice that, since the polynomial 
has real coefficients, its 
roots are pairs of complex conjugate numbers). 

\begin{figure}[t]
\begin{center}
$$\includegraphics[width=.45\textwidth]{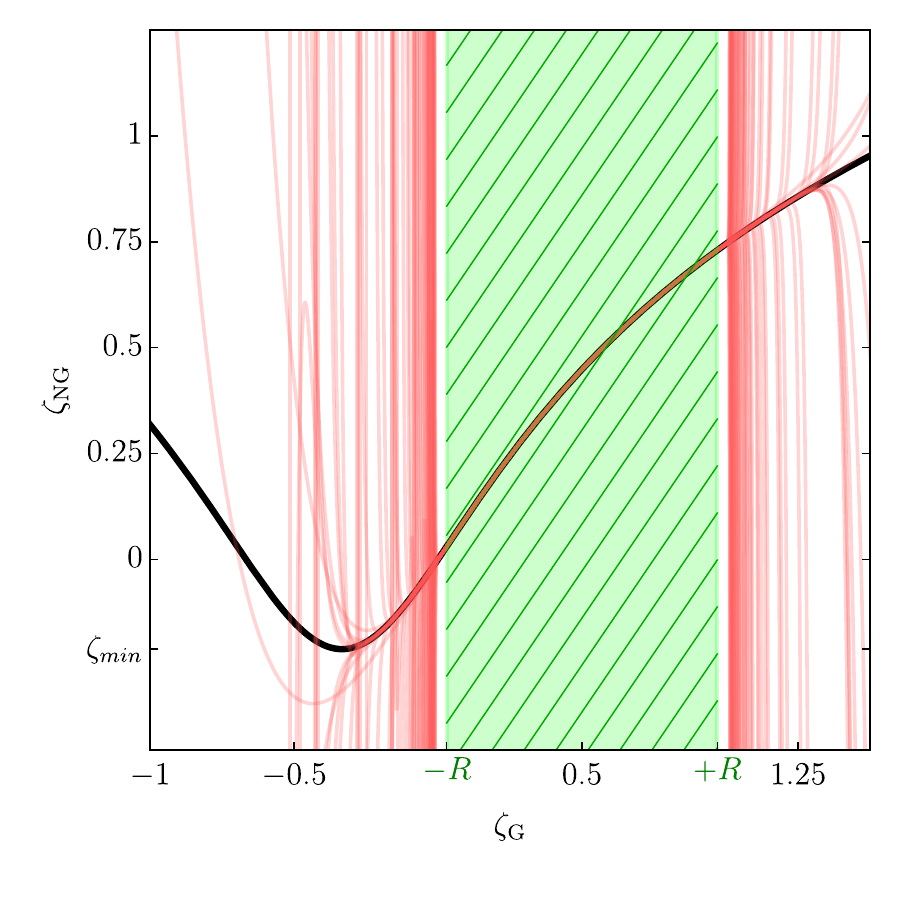}
\qquad\qquad\includegraphics[width=.45\textwidth]{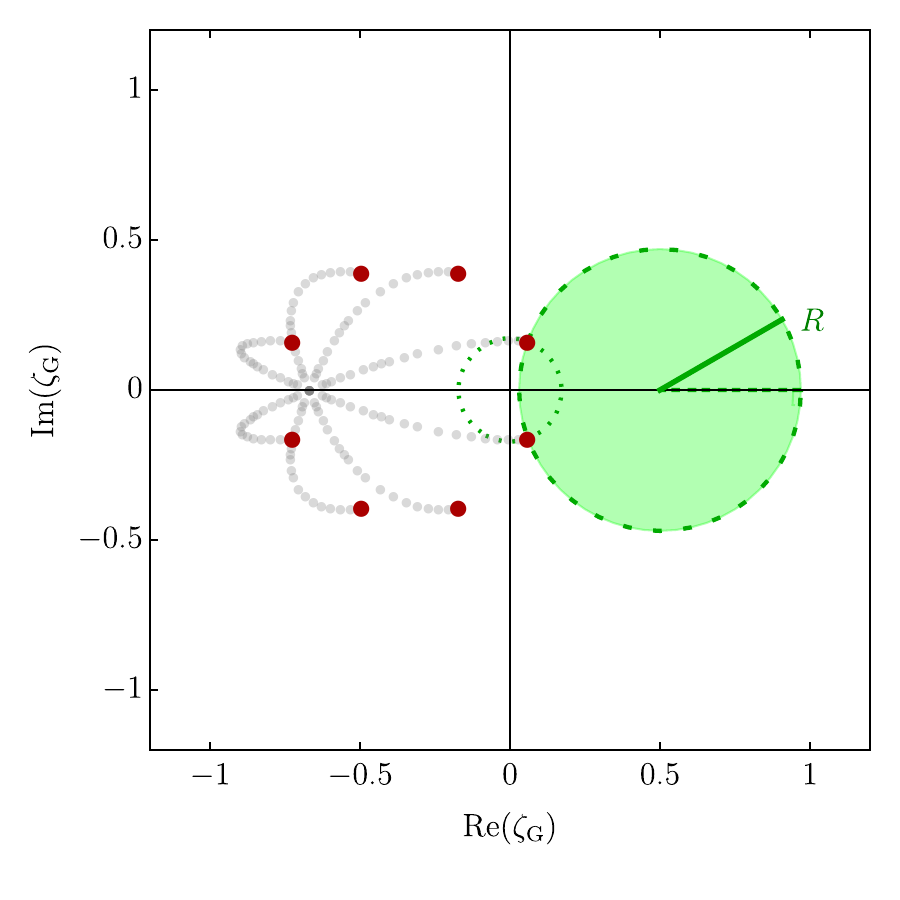}$$
\caption{ 
Same as in fig.\,\ref{fig:Radius} but with the center of the power series expansion that is now 
set by $\zeta_{\rm G} = 0.5$ (see text for details).
 }\label{fig:RadiusOff}  
\end{center}
\end{figure} 
For completeness, we also show in light gray how these branch point singularities move as we increase the 
value of $r_{\rm dec}$ from $r_{\rm dec}=0.5$ towards $r_{\rm dec}=1$; as expected from our previous result, 
as $r_{\rm dec}\to 1$ the singularities eventually  collapse on the real singularity 
located at $\zeta_{\rm G} = -2/3$.  

As a side comment, notice that it is possible to tune the region of convergence if one takes a power-series expansion centered around a point that is not the origin.
This is shown in fig.\,\ref{fig:RadiusOff}. 
For illustration, we expand around the point $\zeta_{\rm G} = 0.5$. 
In the left panel of fig.\,\ref{fig:RadiusOff} we see that the region of convergence 
is way bigger than it was before, and we find $R\simeq 0.47$. 
In light of our previous discussion, this is now clear. 
If we consider the right panel of fig.\,\ref{fig:RadiusOff} we see that now the distance 
between the center of the expansion and the closest branch point singularity increased (since we moved the former 
towards the right while the latter obviously remained fixed) thus giving a wider region of convergence.
\section{PBH abundance from peak theory}
It is well known that the Press-Schechter approach does not agree with the theory of peaks (see, e.g., Refs.~\cite{Green:2004wb, Young:2014ana, DeLuca:2019qsy}).
A technical drawback of peaks theory is that it is not clear how to include NGs in the computation of the abundance in this approach using a generic functional form for the curvature perturbation field. Hence, for comparison, we limit our analysis for the peak theory approach to the case of negligible primordial NGs\footnote{The impact of local-type NG on PBH abundance in peak theory is expected to be reduced,  especially for higher-order terms \cite{Young:2022phe} and for peaked power spectra. Indeed,  the compaction is volume-averaged over the scale of the perturbation and depends on the curvature perturbation at its edge, rather than its centre.}.

In this approach, the starting point is the number density of peaks with a height in the interval $(\nu,\nu+\td \nu)$. When $\nu \gg 1$, it is given by~\cite{Bardeen:1985tr}
\be
    \frac{\td \mathcal{N}}{\td \nu}
    =\frac{1}{4 \pi^2 r_m^3}\left(\frac{\sigma_{cc}}{\sigma_{c}}\right)^3 \nu^3 \exp \left(-\frac{\nu^2}{2}\right)\,,
\ee
where we introduced the rescaled peak height $\nu \equiv \mathcal{C}_1 / \sigma_c$ and the first rescaled moment of the distribution~\cite{Balaji:2023ehk}
\be
    \sigma_{cc}^2 = \frac{4 \Phi^2}{9} \int_0^{\infty} \frac{\td k}{k}\left(k r_m\right)^6 W^2\left(k, r_m\right)P^{T}_\zeta\,.
\ee
The number of peaks within a Hubble volume is therefore
\be
    \frac{\td N_k}{\td \nu} 
    = \frac{4\pi}{3} r_m^3 \mathcal{N}
\ee
and, since $N_k \ll 1$, the probability of finding at least one peak within a range $(\nu,\nu+\td \nu)$ within a Hubble volume is $P_k \approx \td N_k$.
The mass fraction of PBHs is then~\cite{Yoo:2018kvb,Yoo:2019pma,Gow:2020bzo,Franciolini:2022tfm}
\be
    P_k(\mathcal{C}) 
    = \frac{1}{3 \pi}\left(\frac{\sigma_{cc}}{\sigma_{c}}\right)^3 \nu^3 \exp \left(-\frac{\nu^2}{2}\right)
\ee
and the domain of integration agrees with Eq.~\ref{eq:RegionD}. Thus, peaks theory predicts an additional $\nu^3$ factor to the collapse probability in the Gaussian case.

Recasting the quantities in terms of $M_k$, we can express the PBH abundance as
\begin{align}\label{eq:df_PBH_peak}
    \frac{\td f_{\rm PBH}}{\td \ln M_{\rm PBH}} 
    = &\frac{1}{\Omega_{\rm DM}} \int_{M_{\rm H}^{\rm min}}
    \frac{d M_{\rm H}}{M_{\rm H}} \left(\frac{M_{\rm eq}}{M_{\rm H}}\right)^{1/2}\left(\frac{M_{\rm PBH}}{\mathcal{K} M_{\rm H}}\right)^{\frac{1+\gamma}{\gamma}}
    \nonumber\\
    & \times \frac{{\cal K}}{\gamma} \frac{\left(2 \Phi \left(1-\sqrt{\Lambda}\right)\right)^3}{3\pi \sigma_c^4\Lambda^{1/2}} \left(\frac{\sigma_{cc}}{\sigma_c}\right)^3 
     \exp\left[{\frac{-2\Phi^2}{\sigma_c^2}\left(1-\sqrt{\Lambda}\right)^2}\right]\,.
\end{align}

\chapter{Appendix of Chapter 5}
\section{Asymptotics of the GW power spectrum}\label{supp:1}

\begin{figure}[h]
  \centering
  \includegraphics[width=0.95\textwidth]{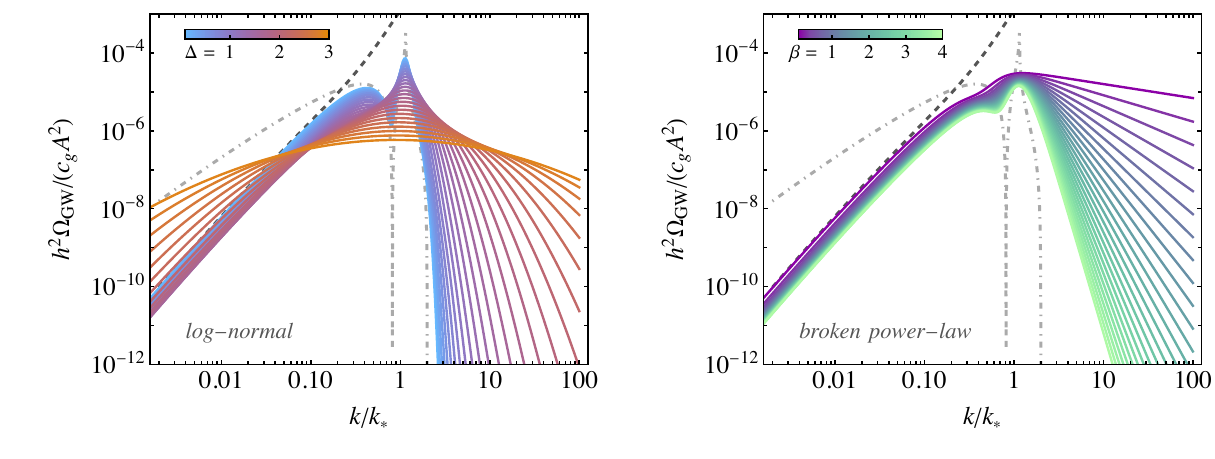}
  \caption{Examples of SIGW spectra induced by LN \emph{(left panel)} and BPL \emph{(right panel)} scalar curvature power spectra with $\alpha = 4$ and $\gamma = 1$ in a range of parameters. The dashed line shows the asymptotic tail~\eqref{eq:GW_asympt} for $\beta = 0.1$ and $\Delta = 0.1$. The dot-dashed line shows the SIGW spectrum for a monochromatic curvature power spectrum.
  }
  \label{fig:SIGW_examples}
\end{figure}

We report here some analytic expressions for the low $k$ asymptotics of GW spectra generated by an enhanced feature in the curvature fluctuations with BPL and LN shapes, i.e. Eqs.~\eqref{eq:PPL} and \eqref{eq:PLN} respectively (see also Refs.~\cite{Pi:2020otn,Yuan:2019wwo}). 
The corresponding SIGW spectra are shown in Fig.~\ref{fig:SIGW_examples}. Sufficiently far away from the peak, we can drop the $s$ dependence inside $P_\zeta$ and expand the transfer functions in the computation of $\Omega_{\rm Gw}$
\bea\label{eq:GW_appr}
    \Omega_{\rm GW} (k)
    \stackrel{k\ll k_*}{\sim}  \frac{4 c_g \Omega_r}{5} k^3 \int_0^{\infty} \frac{\td q}{q^4} \,P(q)^2 \left[ \pi^2 + \ln^2\left( \frac{3 e^2}{4} \frac{k^2}{q^2}\right) \right]
    \propto k^3(1 + \tilde A \ln^2(k/\tilde k)), 
\eea
where $\tilde A$ and $\tilde k = \mathcal{O}(k_{*})$ are parameters that depend on the shape of the curvature power spectrum and $c_g \equiv g_*/g_*^0 \,\left(g_{*s}/g_{*s}^0\right)^{-4/3}$. For the power spectra \eqref{eq:PPL} and \eqref{eq:PLN}, the asymptotics can be obtained explicitly
\bea\label{eq:GW_asympt}
    \Omega^{\rm PL}_{\rm GW} (k)/k^3  
    &\sim \frac{4 c_g \Omega_r}{5 } 
    \frac{\gamma (\alpha/\beta + 1)^{2 \gamma}}{\alpha +\beta}
    \left(\frac{\beta }{\alpha }\right)^{-x_-} 
    \frac{\Gamma \left(x_+\right) \Gamma \left(x_-\right)}{\Gamma(x_- + x_+)}
    \bigg[\pi^2 
    + \frac{\gamma ^2}{(\alpha+\beta)^2}\left(\psi'\left(x_-\right)+\psi'\left(x_+\right)\right) \\
    & +  
    \left(\log \left(3k^2/4\right) + 2 + \frac{2\gamma}{\alpha +\beta } \left( \psi\left(x_+\right) - \psi\left(x_-\right) + \log \left(\alpha/\beta\right)\right) \right)^2\bigg] 
    \\
    \Omega^{\rm PL}_{\rm GW} (k)/k^3
    &\sim  \frac{2 c_g \Omega_r}{5\sqrt{\pi} \Delta}  e^{\frac{9 \Delta^2}{4}} 
    \left[\pi^2 + 2 \sigma^2 + \left(\log \left(3k^2/4\right)+2+3 \sigma ^2\right)^2\right]
\eea
where $x_- \equiv \gamma (2\alpha-3)/(\alpha+\beta)$, $x_+ \equiv \gamma (2 \beta+3)/(\alpha+\beta)$ and $\psi$ denotes the polygamma function. This tail fits the numerically evaluated SIGW tails well, as can be seen in Fig.~\ref{fig:SIGW_examples}.

At the $k \gg k_{*}$ tail, the SIGW spectrum tracks roughly $\mathcal{P}^2_{\zeta}(k)$ when it is sufficiently wide, that is, wider than the SIGW spectrum corresponding to the monochromatic curvature spectrum.

\begin{figure}[t]
  \centering
  \includegraphics[width=0.49\textwidth]{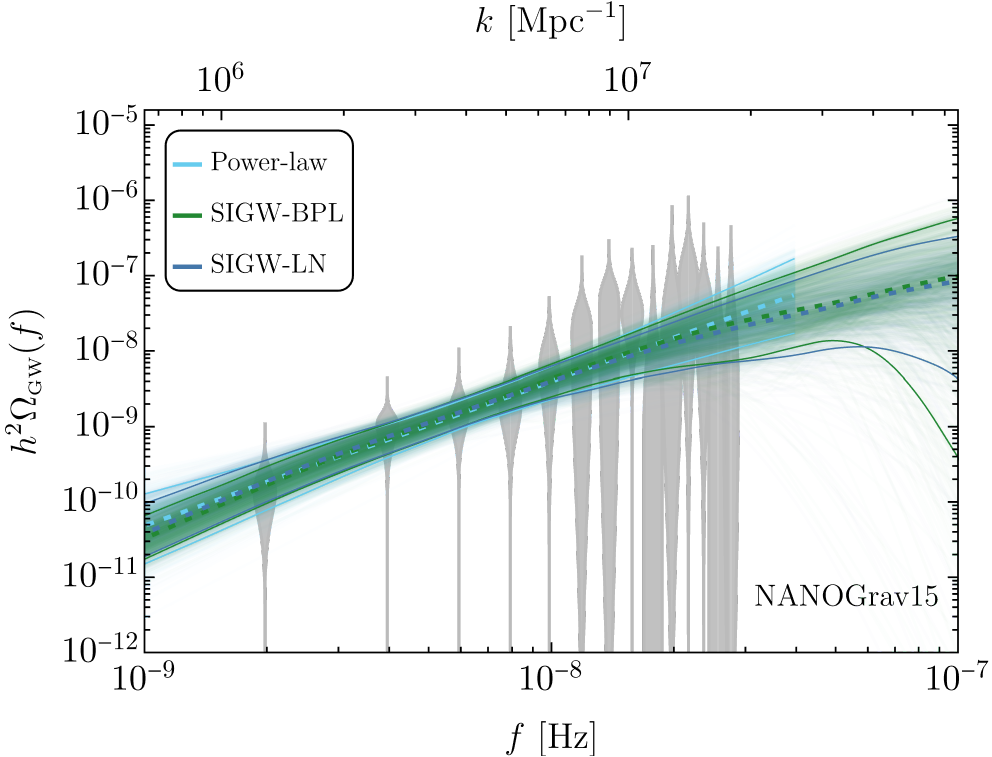}
  \includegraphics[width=0.49\textwidth]{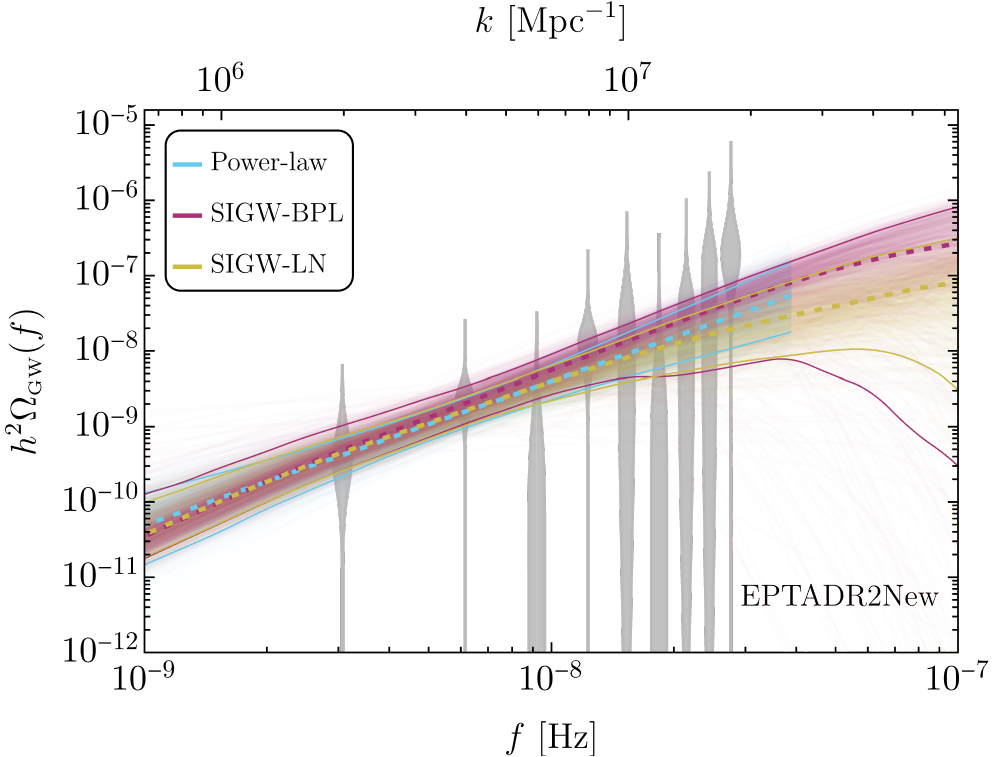}
  \caption{ 
SGWB spectra for the SIGW from BPL and LN models fitted to the NANOGrav15 data ({\it left panel}) and to the  EPTA data ({\it right panel}). 
For reference, we also show the SGWB power-law model used in~\cite{NANOGrav:2023gor, NANOGrav:2023hde}. 
The colored bands indicate the 90\%  C.I., while the gray posteriors show the NANOGrav15~\cite{NANOGrav:2023gor, NANOGrav:2023hde} and EPTA~\cite{EPTA:2023fyk, EPTA:2023sfo} data.}
  \label{fig:fits}
\end{figure}

In Fig.~\ref{fig:fits}, we show the best fit SGWB spectra for both the BPL and LN models compared to the NANOGrav15 (left panel) and EPTA (right panel) datasets.

\section{Dependency of the PBH formation parameters on the curvature power spectral shape}

The density contrast field $\delta$ averaged over a spherical region of areal radius $R_m\equiv a(t_H)r_me^{\zeta(r_m)}$,  and the compaction function $\mathcal{C}(r)$, defined as twice the local mass excess over the areal radius and evaluated at the scale $r_m$ at which it is maximized, are related by
\begin{align}\label{eq:Dicotomy}
\delta = \frac{1}{V_b(r_m,t_H)}
\int_{S^2_{R_m}}d\vec{x}\,
\delta\rho(\vec{x},t_H)
= \frac{\delta M(r_m,t_H)}{M_b(r_m,t_H)} 
=
\frac{2\left[M(r,t) - M_b(r,t)\right]}{R(r,t)} 
= \mathcal{C}\,.
\end{align}
As discussed in Ref.~\cite{Musco:2018rwt} and Refs.~therein, the gravitational collapse that triggers the formation of a PBH takes place
when the maximum of the compaction function $\mathcal{C}(r_m)$ is larger than a certain threshold value. Using Eq.\,(\ref{eq:Dicotomy}), we can relate threshold values in the compaction $\mathcal{C}_{\rm th}$ 
to the threshold for the density contrast $\delta_{\rm th}$.

In full generality, the threshold $\mathcal{C}_{\rm th}$ (or $\delta_{\rm th}$) depends on the shape of the collapsing overdensities, which is controlled by the curvature power spectrum~\cite{Germani:2018jgr, Musco:2018rwt, Musco:2020jjb}.
In this work, we follow Ref.~\cite{Musco:2020jjb}, where a prescription of how to compute the collapse parameter as a function of the curvature spectrum is derived. 
For completeness, we briefly report here the main steps to compute the threshold $\delta_{\rm th}$
and the shape parameter $\alpha_{\rm s}$. First of all, the maximum of the compaction function is located the the radius $r_m$, which can be found
solving numerically the integral equation
\begin{equation}
\int \frac{d k}{ k}\left[\left(k^2 {r}_m^2-1\right) 
\frac{\sin \left(k {r}_m\right)}{k {r}_m}+\cos \left(k {r}_m\right)\right]  P^{T}_\zeta(k) =0.
\end{equation}
Consequently, the shape parameter $\alpha_{\rm s}$ is obtained using 
\begin{equation}
F(\alpha_{\rm s})[1+F(\alpha_{\rm s})] \alpha_{\rm s}
=
-\frac{1}{2}\left[1+{r}_m \frac{\int d k k \cos \left(k {r}_m\right) P^{T}_\zeta(k)}{\int d k \sin \left(k {r}_m\right) P^{T}_\zeta(k)}\right],
\end{equation}
 where we introduced
 $  F(\alpha_{\rm s})
 =
 \left \{
 1-\frac{2}{5} e^{-1 / \alpha_{\rm s}} 
 \alpha_{\rm s}^{1-5 / 2 \alpha_{\rm s}}
 / 
 \left [
 \Gamma\left({5}/{2 \alpha_{\rm s}}\right)
 -\Gamma\left({5}/{2 \alpha_{\rm s}}, {1}/{\alpha_{\rm s}}
 \right)
 \right]
 \right \}^{1/2}
 $ to shorten the notation. 
Finally, once we determined shape parameter $\alpha_{\rm s}$, we can compute the threshold $\delta_{\rm th}$ using the relation \cite{Escriva:2019phb}
 \begin{equation}
\delta_{\rm th} \simeq \frac{4}{15} e^{-1 / \alpha_{\rm s}} \frac{\alpha_{\rm s}^{1-5 / 2 \alpha_{\rm s}}}{\Gamma\left(\frac{5}{2 \alpha_{\rm s}}\right)-\Gamma\left(\frac{5}{2 \alpha_{\rm s}}, \frac{1}{\alpha_{\rm s}}\right)}.
\end{equation}

\begin{figure}[h]
\begin{center}
$$\includegraphics[width=.48\textwidth]{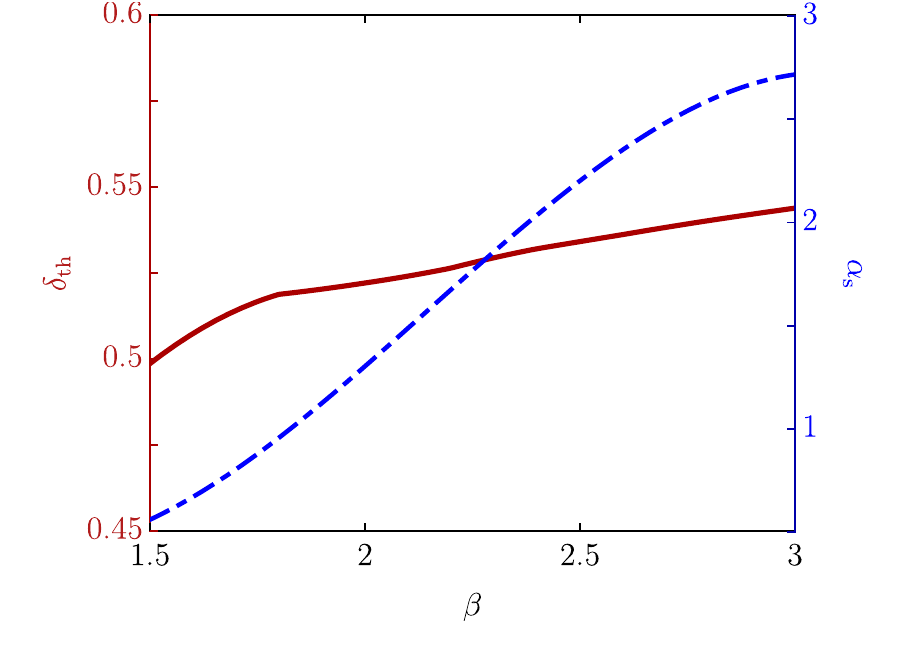}
\qquad\includegraphics[width=.48\textwidth]{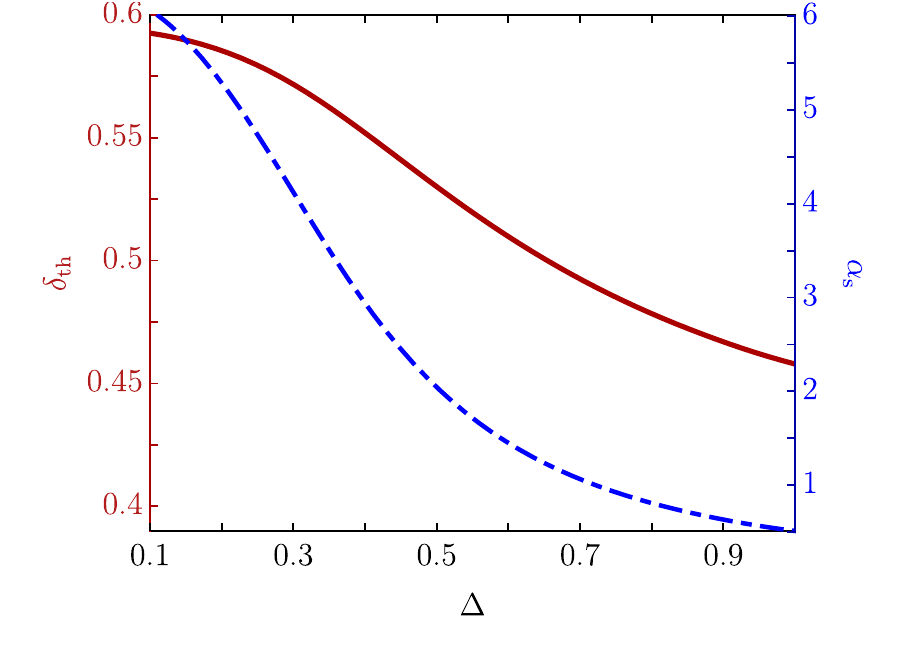}$$
\caption{ Threshold $\delta_{\rm th}$ and NG shape parameter $\alpha_{\rm s}$ for different shape of the power spectrum for a BPL \emph{(left panel)}, where we fix $\alpha=4$ and $\gamma=1$, and a log-normal \emph{(right panel)}.
 }\label{fig:ThShape}  
\end{center}
\end{figure}

In Fig.~\ref{fig:ThShape} we show how the threshold $\delta_{\rm th}$ and the shape parameter $\alpha_{\rm s}$ change respect the shape of the power spectrum for a BPL (right panel) and a LN (left panel).
It is worth emphasizing a general trend. As the spectrum becomes narrower, the threshold for collapse rises up to ${\cal O}({20})\%$ larger values. 
This is because, for broader spectra, more modes participate in the evolution of the collapsing overdensity, leading to a reduction of pressure gradients that facilitates the PBH formation, see Ref.~\cite{Musco:2020jjb} for more details.
It is important to note that the prescription outlined in Ref.~\cite{Musco:2020jjb} to compute the threshold for PBH collapse only accounts for NGs arising from the non-linear relation between the density contrast and the curvature perturbations. In principle, also primordial NGs beyond the quadratic approximation should be taken into account when computing the threshold value.
Following Refs.~\cite{Kehagias:2019eil,Escriva:2022pnz}, it appears that the effect on the threshold is small and at most of the order of a few percent. Interestingly, it follows the same impact NGs have on the statistics, i.e. negative (positive) NGs would tend to increase (decrease) the required amplitude of spectra. Therefore, neglecting this effect we are conservative. We left the inclusion of this effect for future work.

Due to a softening of the equation of state, the formation of PBH becomes more efficient during the QCD transition \cite{Jedamzik:1998hc, Byrnes:2018clq, Franciolini:2022tfm, Escriva:2022bwe, Musco:2023dak}. 
Consequently, the formation parameters change due to the softening of the equation of state. 
In our analysis presented in the main text, we computed $\alpha_{\rm s}$ for each power spectrum,
and used the results obtained in Ref.\cite{Musco:2023dak}, where  the formation parameters $\gamma(M_H), \mathcal{K}(M_H),\Phi(M_H)$ and $\mathcal{\delta}_{\rm th}(M_H)$ (or equivalently $\mathcal{C}_{\rm th}(M_H)(M_H)$) are obtained a function of $\alpha_s$.

\section{Primordial non-Gaussianities}\label{App:Curvaton}

\begin{figure}[h]
  \centering
  \includegraphics[width=0.325\textwidth]{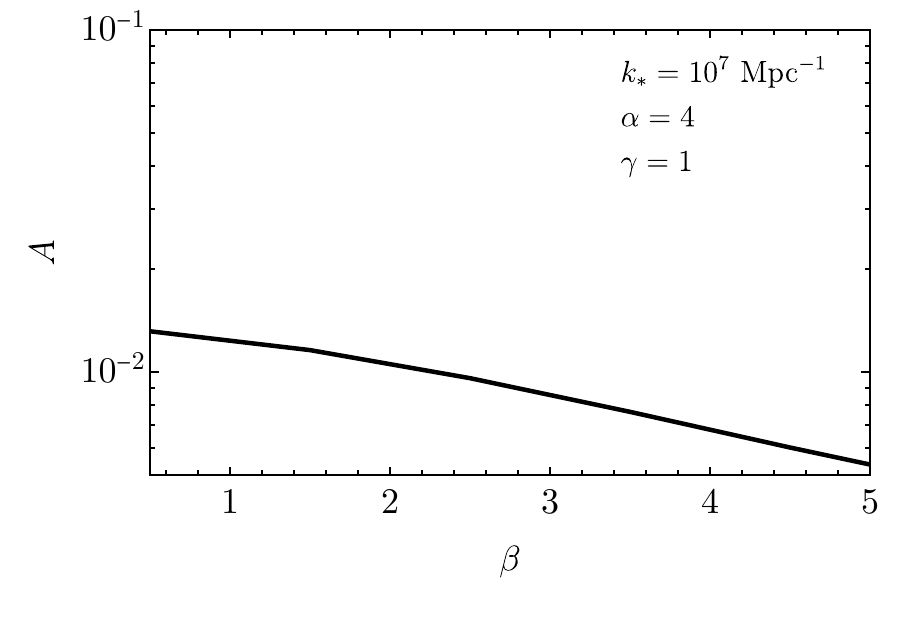}
  \includegraphics[width=0.325\textwidth]{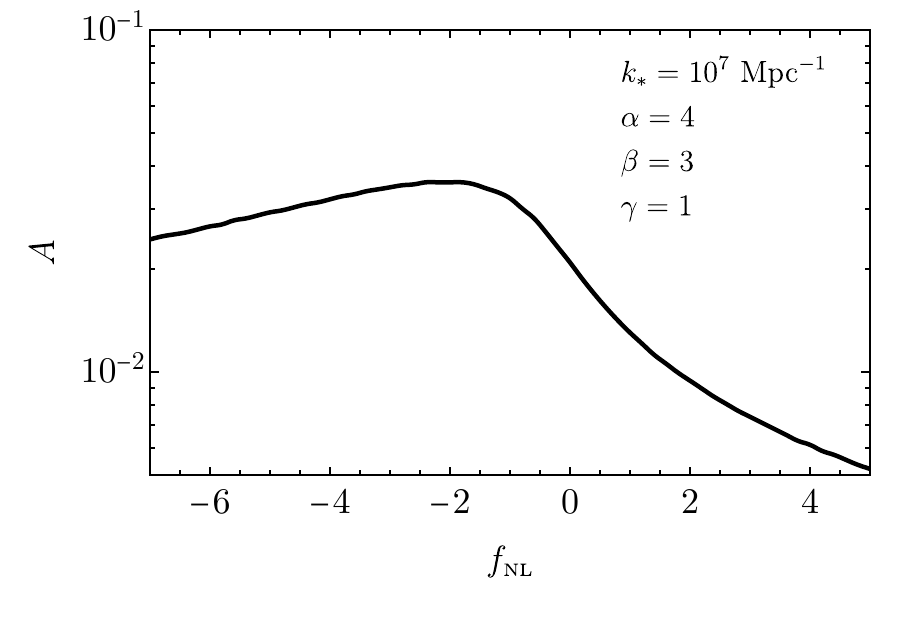}
  \includegraphics[width=0.325\textwidth]{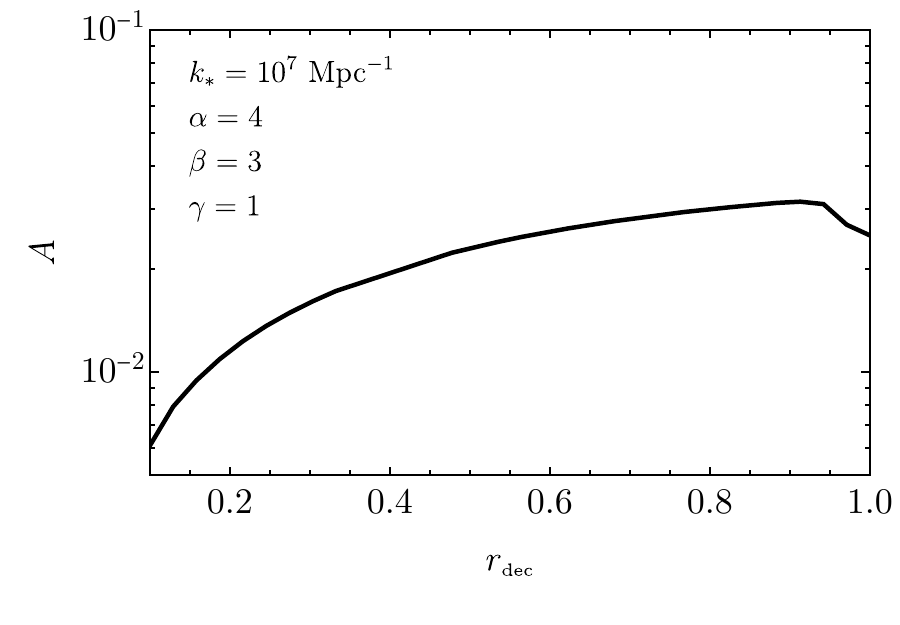}
  \caption{\emph{Left panel}: Amplitude values for a BPL curvature spectrum (Eq.~\eqref{eq:PPL}), fixing $\alpha=4$ and $\gamma=1$, in order to get $f_{\rm PBH}=1$ for quasi-inflection-point models with different $\beta$ values. 
  In order to isolate the effect of NGs, in this plot only we fix $\delta_{\rm th}=0.54$, as used in the other panels. 
  \emph{Middle panel}: Amplitudes required for $f_{\rm PBH}=1$ in the case of a BPL (Eq.~\eqref{eq:PPL}),  $\alpha=4$, $\beta=3$ and $\gamma=1$, in a model that only presents quadratic NGs (see  Eq.~\eqref{eq:FirstExpansion}) with different values of $f_{\rm NL}$. 
  \emph{Right panel}: The same as the middle panel but for the curvaton model (Eq.~\eqref{eq:MasterXX}) as a function of $r_{\rm dec}$.}
  \label{fig:AppFNL}
\end{figure}

In this section, we discuss a few relevant details about the NG models considered in this work. 
We summarize the behavior of the amplitude required to produce $f_{\rm PBH }=1$ depending on the various types of NGs discussed in this paper in Fig.~\ref{fig:AppFNL}, assuming a reference BPL power spectrum.

\subsubsection{Quasi-inflection point} 
As described already in the main text, when presenting results inspired by the quasi-inflection point, the NG parameter $\beta$ is inherently determined by the UV slope of the power spectrum. As we can see from the left panel in Fig.~\ref{fig:AppFNL}, when one shrinks the shape of the power spectrum and keeps the threshold for collapse constant, we find that the amplitude should decrease to obtain $f_{\rm PBH} = 1$. 
This is because NGs become larger with increasing $\beta$ in such models. 

One can also easily see that, by expanding Eq.~\eqref{eq:zeta_IP} at second order, 
$f_{\rm NL} = 5 \beta/12 $. 
Using the reference value $\beta = 3$ shown in Fig.~\ref{fig:abundance}, one finds $f_{\rm NL} = 5/4$. As a consequence, we expect the NG correction to the SGWB to be below ${\cal O}(10^{-2})$, and no relevant correction from higher order terms in $\Omega_{\rm GW}$ is expected.

\subsubsection{Quadratic non-Gaussianities} 

\begin{figure}[h]
\begin{center}
\includegraphics[width=.325\textwidth]{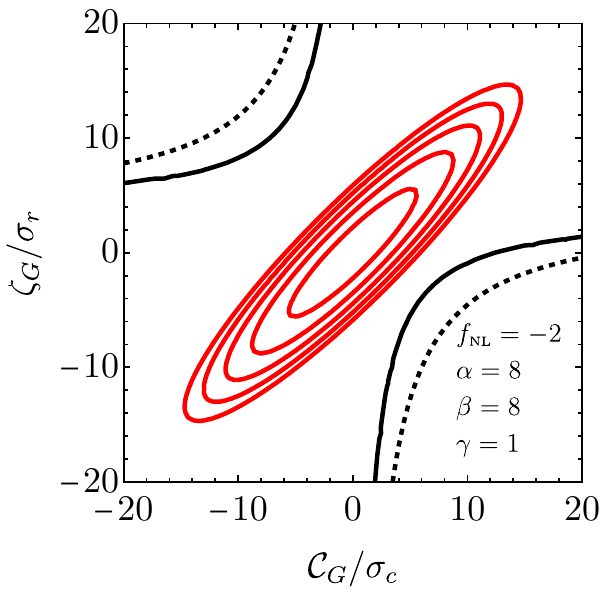}
\includegraphics[width=.325\textwidth]{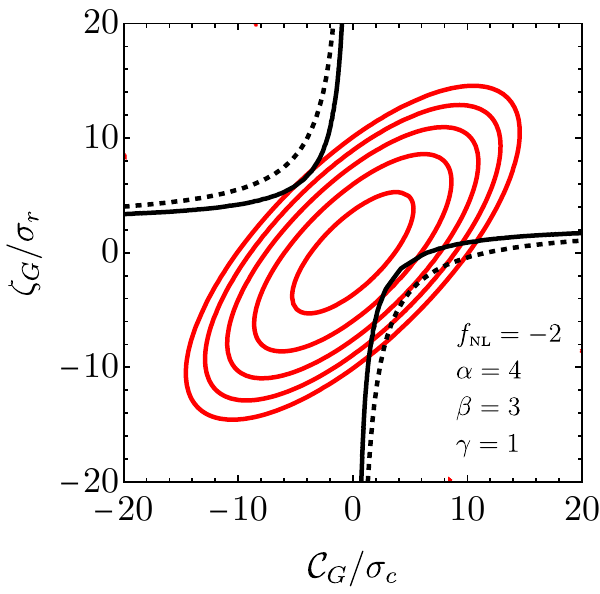}
\includegraphics[width=.325\textwidth]{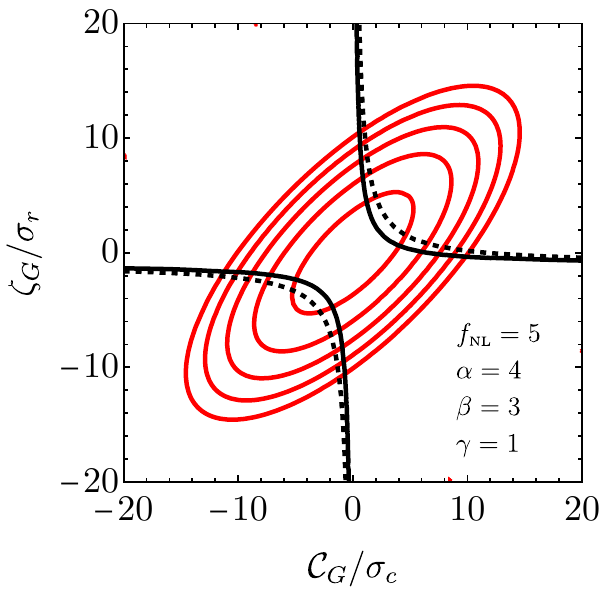}
\caption{ Two dimensional PDF as a function of $({\cal C}_{\rm G}, \zeta_{\rm G})$ compared to the over-threshold condition ${\cal C}>{\cal C}_{\rm th}$.
In all panels, we considered the BPL power spectrum with an amplitude $A = 0.05$. The red lines indicates the contour lines corresponding to $\log_{10}(P_{\rm G}) = {-45,-35,-25,-15,-5}$. The collapse of type-I PBHs takes place between the black solid and dashed lines (see more details in Ref.~\cite{Ferrante:2022mui}). 
\textit{Left panel:} Example of a very narrow power spectrum with $\alpha = \beta = 8$. The abundance is suppressed in the presence of negative $f_{\rm NL} $ by the strong correlation between ${\cal C}_{\rm G}$ and $\zeta_{\rm G}$ obtained for narrow spectra.
\textit{Center panel:}
Example of negative non-Gaussianity and representative BPL spectrum. The PBH formation is sourced by regions of small $\zeta_{\rm G}$ and positive ${\cal C}_{\rm  G}$ or both negative ${\cal C}_{\rm G}$ and $\zeta_{\rm G}$.
\textit{Right panel:}
Example with positive $f_{\rm NL}$, showing the region producing PBHs populates the correlated quadrants of the plot, at odds with that is found in the other panels. 
 }\label{fig:quadfnl}  
\end{center}
\end{figure}

Generically, one might mistakenly assume that by pushing the coefficient $f_{\rm NL}$ towards larger negative values, the required value of the power spectral amplitude associated with $f_{\rm PBH} \simeq 1$ would subsequently rise. 
As we show in the middle plot of Fig.~\ref{fig:AppFNL}, however, a maximum amplitude is reached for $f_{\rm NL} \simeq -2$.
The reason for the appearance of such a peak and the subsequent decrease of A for $f_{\rm NL}<-2$ can be understood as follows. 
When $f_{\rm NL}$ is negative, one can still push the compaction function beyond the threshold, provided  $\zeta_{\rm G}$ and ${\cal C}_{\rm G}$ had opposite signs or one has small $\zeta_{\rm G}$ and positive ${\cal C}_{\rm G}$. 
In Fig.~\ref{fig:quadfnl}, we show the probability distribution for both Gaussian parameters $\zeta_{\rm G}$ and ${\cal C}_{\rm G}$ (red contours), compared to the overthreshold condition (between the black lines).
For realistic spectra, one always finds sufficient support of the PDF in the anti-correlated direction, and thus obtain a sizeable PBH abundance. Only in the limit of a very narrow power spectrum, one finds $\gamma_{cr}$ converging towards unity, which means a perfect correlation between $\zeta_{\rm G}$ and ${\cal C}_{\rm G}$, that leads to very small overlap with the parameter space producing overthreshold perturbation of the compaction function. 
This explains the appearance of a sharp rise of $A$ in the results Ref.~\cite{Young:2022phe} (see their Fig. 2), which, is however, expected only for extremely narrow spectra.

For this ansatz and with $f_{\rm NL}=-2$, we find that the amplitude of the power spectrum saturating the abundance of PBH is around $A \simeq 10^{-1.4}$, so we should expect a correction to the SIGW spectrum of the order $A(3/5f_{\rm NL})^2\simeq{\cal O}(0.05)$.
Still, this is subdominant compared to the leading order term used in this paper.

\subsubsection{Curvaton models} 
When presenting results inspired by the curvaton model, we will focus on primordial NG (derived analytically within the sudden-decay approximation~\cite{Sasaki:2006kq})
\be\label{eq:MasterXX}
    \zeta = \log\big[X(r_{\rm dec},\zeta_{\rm G})\big]\,,
\ee
with
\begin{subequations}
\begin{align}\label{eq:XFunction2}
    X &\equiv \frac{\sqrt{K}\left(1 + \sqrt{A K^{-\frac32}-1}\right)}{(3+r_{\rm dec})^{\frac13}}, 
    \\
    K & \equiv \frac{1}{2}\left((3+r_{\rm dec})^{\frac13}(r_{\rm dec}-1)P^{-\frac13} + P^{\frac13}\right), 
    \\
    P &\equiv A^2 + \sqrt{A^4 + (3+r_{\rm dec})(1-r_{\rm dec})^3}\,,
    \\
    A &\equiv \left(1 + \frac{3\zeta_{\rm G}}{2r_{\rm dec}} \right)^{2}r_{\rm dec}\,.
\end{align}
\end{subequations}
The parameter $r_{\rm dec}$ is the weighted fraction of the curvaton energy density  $\rho_{\phi}$ to the total energy density at the time of curvaton decay, defined by
\be
    r_{\rm dec} \equiv 
    \left.\frac{3 \rho_{\phi}}{3 \rho_{\phi} + 4 \rho_{\gamma}}\right|_{\rm curvaton\,\,decay}\,,
\ee
where $\rho_{\gamma}$ is the energy density stored in radiation after reheating. 
Thus, $r_{\rm dec}$ depends on the physical assumptions about the physics of the curvaton within a given model.

For comparison, the coefficients in the series expansion
$
    \zeta = \zeta_{\rm G} + (3/5)f_{\rm NL} \zeta_{\rm G}^2 + (3/5)^2g_{\rm NL} \zeta_{\rm G}^3 + \ldots
$
are given by
\be
    f_{\rm NL} = \frac{5}{3}\left(\frac{3}{4 r_{\rm dec}} - 1 - \frac{r_{\rm dec}}{2}\right), \qquad
    g_{\rm NL} = \frac{25}{54}\left(-\frac{9}{r_{\rm dec}} + \frac{1}{2} + 10r_{\rm dec} + 3r_{\rm dec}^2 \right)\,.
\ee
At $r_{\rm dec} \to 1$, the fluctuations arise from the curvaton field only, as it completely dominates the energy density budget at the time of decay. This gives
$
    \zeta = (2/3)\ln\left[1 + (3/2) \zeta_{\rm G}\right],
$
and $f_{\rm NL} = -5/4$, $g_{\rm NL} = 25/12$. Note that this mimics NG in inflection point models~\eqref{eq:zeta_IP} with an unphysical $\beta = -3$.
For the benchmark case we used in the main text, $r_{\rm dec}=0.9$, one finds $f_{\rm NL} = -1$, $g_{\rm NL} = 0.9$.
Also, we determine that the order of magnitude of the NG correction to the SIGW should be of order $A(3/5f_{\rm NL})^2\simeq {\cal O}(0.01)$, and we do not expect any relevant correction from higher order terms in  the computation of the $\Omega_{\rm GW}$.
\section{Second-order induced gravitational waves with non-Gaussianities}\label{app:NGSIGW}

We summarize here the formulas used to compute the spectrum of SIGWs when non-Gaussianities are included, referring the reader to Refs.~\cite{Unal:2018yaa, Cai:2018dig, Yuan:2020iwf, Adshead:2021hnm, Abe:2022xur, Chang:2022nzu, Garcia-Saenz:2022tzu, Li:2023qua} for more details. 
We assume that the curvature perturbation can be written in terms of its Gaussian component, as 
$
\zeta = \zeta_{\rm G} + F_{\rm NL}\zeta_{\rm G}^2 .
$
The non-linear parameter is often normalised as $F_{\rm NL } = (3/5) f_{\rm NL} $.

The higher-order corrections to the GW power spectrum can then be written as
\begin{equation}\label{eq:Omegabar-total}
   \Omega_{\rm GW}(T) 
    =\Omega_{\rm GW}^{(0)} (T)
    +\Omega_{\rm GW}^{(1)} (T)
    +\Omega_{\rm GW} ^{(2)} (T) \ ,
\end{equation}
where each contribution scales as
${\cal P}_h (k,\eta) ^{(n)} (\eta,k) \propto A^2 (A F_{\rm NL}^2)^n$ and the first order is given in Eq.~\eqref{eq:P_h_ts}. 
The energy-density fraction spectrum of SIGW at the current epoch is then given by Eq.~\eqref{eq:OmegaGWtoday}.

\begin{figure}[ht!]
    \begin{center}
    \includegraphics[width=0.8\columnwidth]{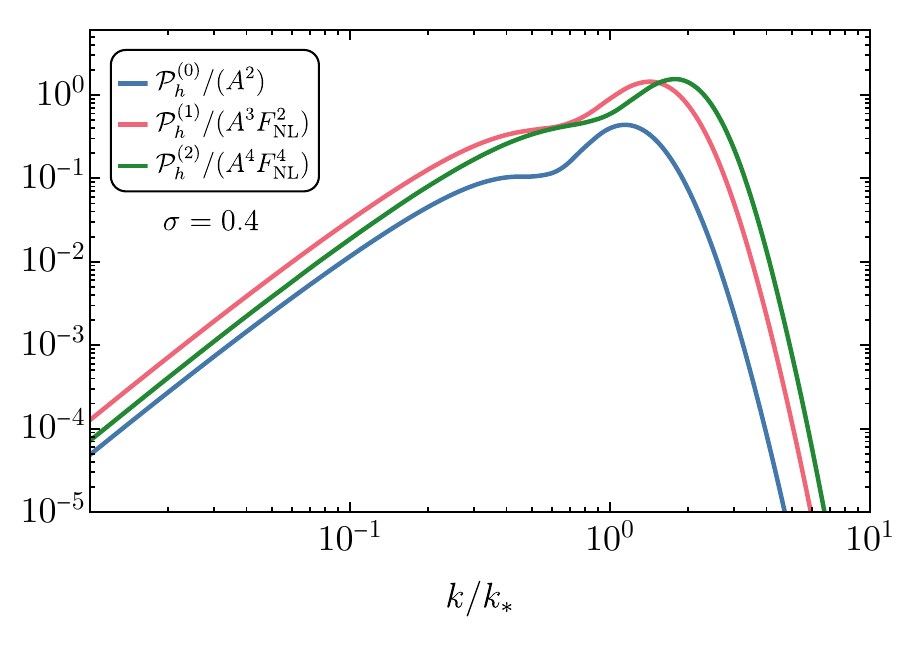}
    \end{center}
    \vspace{-1cm}
    \caption{Contributions to the SIGW spectrum from the leading orders in perturbation theory, where we have factored out the relevant combination of the amplitude $A$ and non-Gaussianity $F_{\rm NL}$.
    This plot assumes a log-normal curvature power spectrum centered at $k_*$ that is characterized by a width $\sigma = 0.4$.}
    \label{fig:SIGW-fNL_spectra}
\end{figure}

In accordance with Refs.~\cite{Adshead:2021hnm, Li:2023qua, Wang:2023sij, Ragavendra:2021qdu, Atal:2021jyo}, the higher-order terms in \eqref{eq:Omegabar-total} can be expressed as
\begin{align}
\Omega_{\rm GW}^{(1)} (T)
=& \frac{2F_{\rm NL}^2}{3} 
\prod_{i=1}^2 
\biggl[\int_1^\infty \td t_i \int_{-1}^1 \td s_i\, \frac{16}{(t_i^2-s_i^2)^2}\biggr] 
\Bigg\{ \frac{1}{2}
\overline{J^2 (s_1,t_1)} 
{\cal P}_\zeta  (v_1 v_2 k) 
{\cal P}_\zeta  (u_1 k) 
{\cal P}_\zeta  (v_1 u_2 k)
\nonumber
\\
+& \int_0^{2\pi} \frac{\td \varphi_{12}}{2\pi}\, \cos 2\varphi_{12} \overline{J (s_1,t_1) J (s_2,t_2)}
\frac{{\cal P}_\zeta (v_2 k)}{v_2^3}
\frac{{\cal P}_\zeta  (w_{12} k)}{w_{12}^3} 
\bigg[
\frac{{\cal P}_\zeta (u_2 k)}{u_2^3} +
\frac{{\cal P}_\zeta  (u_1 k)}{u_1^3} 
\bigg] \Bigg\}
\label{eq:Omega-C}
\end{align}
and
\begin{align}
\label{eq:Omega-N}  
\Omega_{\rm GW}^{(2)} (T)
&= \frac{F_{\rm NL}^4}{6} 
\prod_{i=1}^3 
\biggl[\int_1^\infty \td t_i \int_{-1}^1 \td s_i\, \frac{16}{(t_i^2-s_i^2)^2}\biggr] 
\Bigg\{\frac{1}{2}\overline{J^2 (s_1,t_1)} 
{\cal P}_\zeta (v_1 v_2 k) 
{\cal P}_\zeta (v_1 u_2 k) 
{\cal P}_\zeta (u_1 v_3 k) 
{\cal P}_\zeta (u_1 u_3 k)
\\
&
+ \int_0^{2\pi} \frac{\td \varphi_{12}}{2\pi}\frac{\td \varphi_{23}}{2\pi}
\cos (2\varphi_{12} )
\overline{J (s_1,t_1) J (s_2,t_2)}
\frac{{\cal P}_\zeta(u_3 k)}{u_3^3} 
\frac{{\cal P}_\zeta (w_{13} k)}{w_{13}^3}
\frac{{\cal P}_\zeta (w_{23} k)}{w_{23}^3}
\bigg[
    \frac{{\cal P}_\zeta (v_3 k)}{v_3^3} 
+   \frac{{\cal P}_\zeta (w_{123} k)}{w_{123}^3}
\bigg]\Bigg\} \ , 
\nonumber
\end{align}
where we defined $s_i=u_i-v_i$, $t_i=u_i+v_i$, and 
\begin{subequations}
\begin{eqnarray}
    y_{ij}&=&\frac{\cos\varphi_{ij}}{4}\sqrt{(t_i^2-1)(t_j^2-1)(1-s_i^2)(1-s_j^2)}+\frac{1}{4}(1-s_it_i)(1-s_jt_j)\ , \\
    w_{ij}&=&\sqrt{v_i^2+v_j^2-y_{ij}}\, , 
    \qquad 
    w_{123}=\sqrt{v_1^2+v_2^2+v_3^2+y_{12}-y_{13}-y_{23}}\ .
\end{eqnarray}
\end{subequations}
The time-averaged integrated transfer functions $J(u,v)$ were derived in Refs.~\cite{Espinosa:2018eve,Kohri:2018awv,Atal:2021jyo,Adshead:2021hnm,Li:2023qua}, and are
\begin{align}\label{eq:J-ave-12}
& \overline{ J (s_i,t_i)J (s_j,t_j) }
=
\frac{9}{8}\frac{\left(t_i^2-1\right) \left(t_j^2-1\right) \left(1-s_i^2\right) \left(1-s_j^2\right) \left(t_i^2+s_i^2-6\right) \left(t_j^2+s_j^2-6\right) }{\left(t_i^2-s_i^2\right)^3 \left(t_j^2-s_j^2\right)^3}
\nonumber\\
&
\Bigg[
\left( 
    \left(t_i^2+s_i^2-6\right) 
    \ln \left| \frac{t_i^2-3}{3-s_i^2}\right| 
-   2\left(t_i^2-s_i^2\right)
\right) \left(
    \left(t_j^2+s_j^2-6\right) 
    \ln \left| \frac{t_j^2-3}{3-s_j^2}\right|
-   2\left(t_j^2-s_j^2\right)
\right)
\nonumber\\
&
+   \pi^2\Theta\left(t_i-\sqrt{3}\right) \Theta\left(t_j-\sqrt{3}\right) \left(t_i^2+s_i^2-6\right) \left(t_j^2+s_j^2-6\right)
\Bigg]\ .
\end{align}
Assuming a log-normal curvature power spectrum of the form \eqref{eq:PLN}, as done in Section~\ref{sec:SIGWth}, we can compute the different contributions to the SGWB, factoring out the overall scaling with the spectral amplitude $A$ and the non-linear parameter $F_{\rm NL}$, as shown in Fig.~\ref{fig:SIGW-fNL_spectra}.

The presence of higher-order corrections to the SIGW spectrum could be inferred from a modulation of the peak close to $k\approx k_*$, if the combination $A F_{\rm NL}^2$ is sufficiently large. 
On the other hand, the shape of the low-frequency tail at $k \ll k_*$ is maintained, i.e., 
\begin{equation}
    \Omega_{\rm GW} (k \ll k_*) \, {\propto}\,  k^3(1 + \tilde A \ln^2(k/\tilde k))\, ,
\end{equation}
where $\tilde  A$ and $\tilde k = \mathcal{O}(k_{*})$ are parameters that depend mildly on the shape of the curvature power spectrum.

\begin{figure}[h!]
    \includegraphics[width=0.4\textwidth]{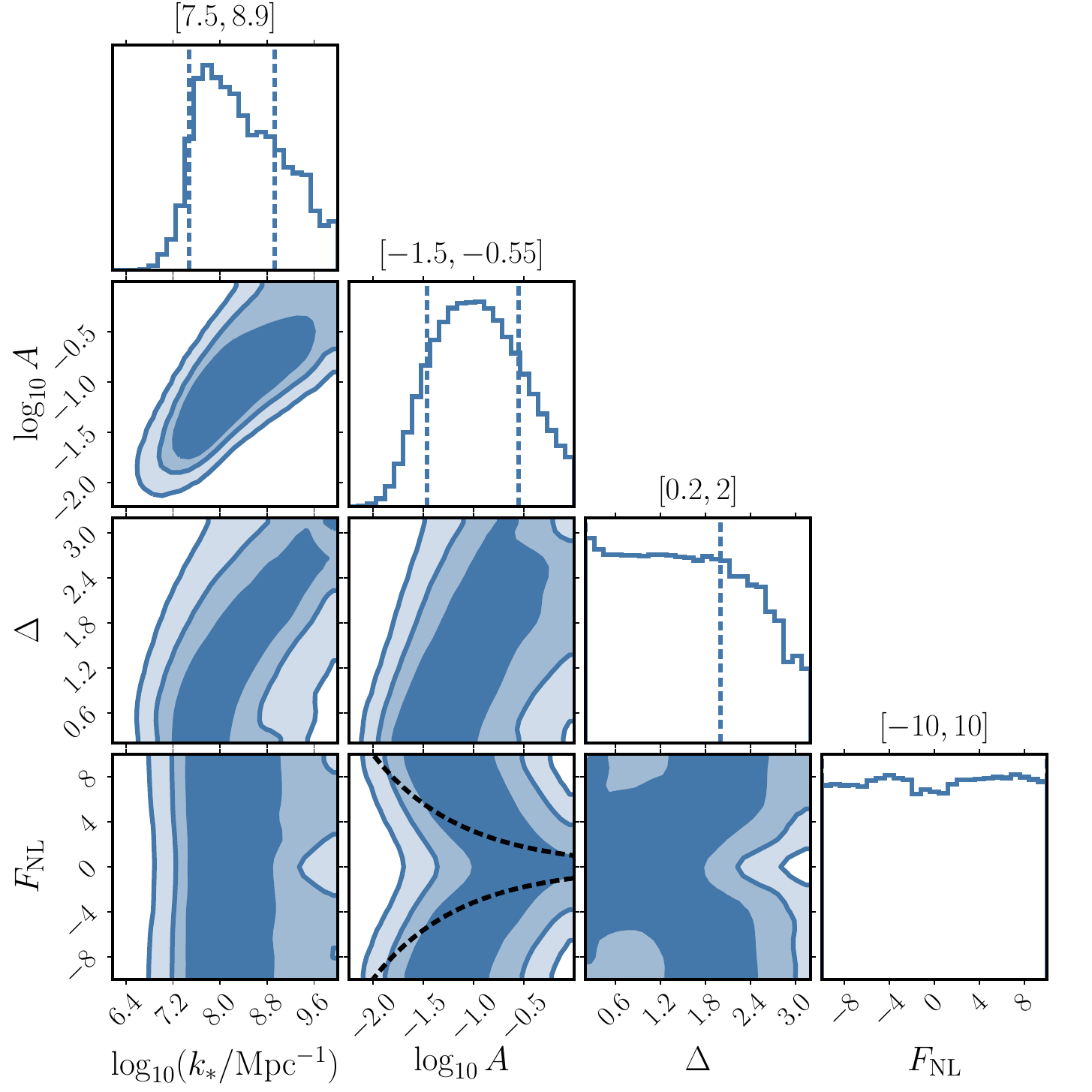}
\includegraphics[width=0.67\textwidth]{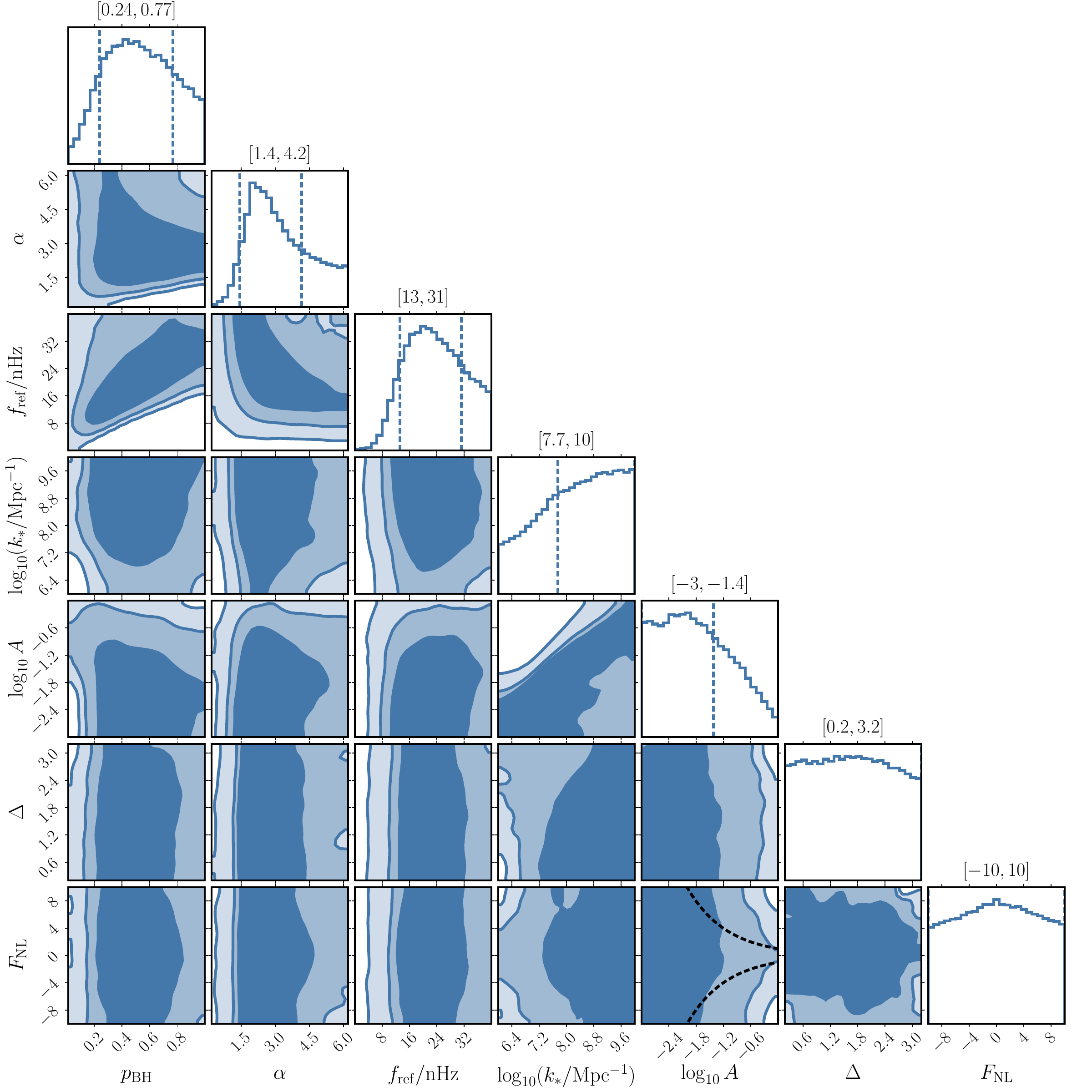}
    \caption{
Same as Fig.~\ref{fig:SIGWonly_posteriors}
but including the possibility of non-Gaussian curvature perturbations with a non-linear parameter $F_{\rm NL}$.
    }
    \label{fig:SIGW_posteriors-fNL}
\end{figure}

As the PTA observations could only be compatible with the infra-red tail of a SIGW, due to the preference for a positive tilt of the spectrum, we conclude that higher-order corrections do not affect significantly the results presented in the main text, while, in full generality, current GW observations are not able to break the degeneracy in the $(A,F_{\rm NL}^2)$ plane. 
As a by-product of our analysis, at odds with the claim in Refs.~\cite{Liu:2023ymk,Figueroa:2023zhu}, we conclude that current data are not able to constrain the presence of non-Gaussianities at PTA scales, beyond the reach of large-scale structure and CMB probes. This is because large values of $F_{\rm NL}$ remain possible, provided the curvature power spectral amplitude is sufficiently small. 
This is confirmed by the posterior distributions shown in Fig.~\ref{fig:SIGW_posteriors-fNL}, where we report results for non-Gaussian SIGW-only fits (left panel) and for non-Gaussian SIGWs together with SMBH binaries (right panel). 
In both panels, the posterior distribution for $F_{\rm NL}$ is flat over the entire prior range.
\chapter{Appendix of Chapter 6}
\section{Dynamics of curvature modes, some essential results}\label{app:TimeDer}

The main purpose of this appendix is to understand, both numerically and analytically, the behavior of the 
time derivative $d\zeta_k/dN$. 

\begin{figure}[h]
\begin{center}
$$\includegraphics[width=.495\textwidth]{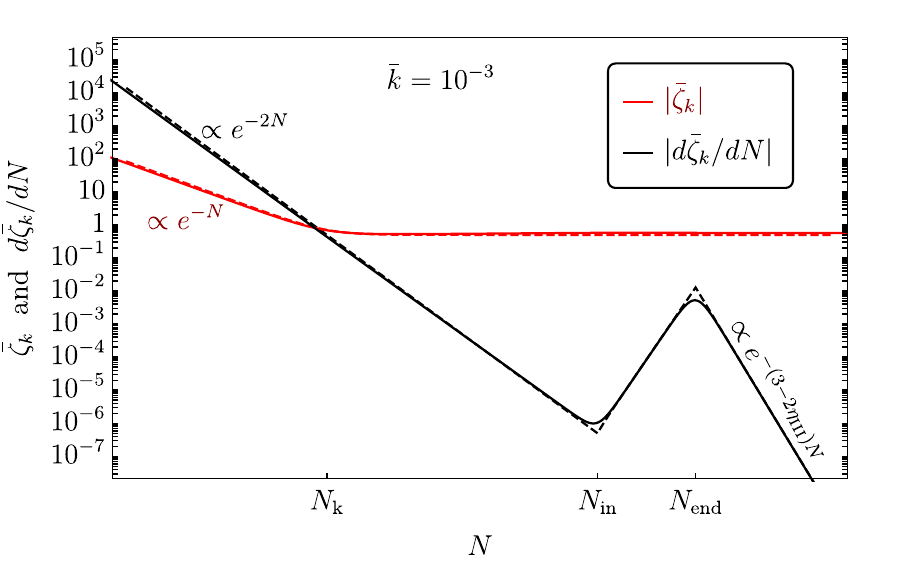}~
\includegraphics[width=.495\textwidth]{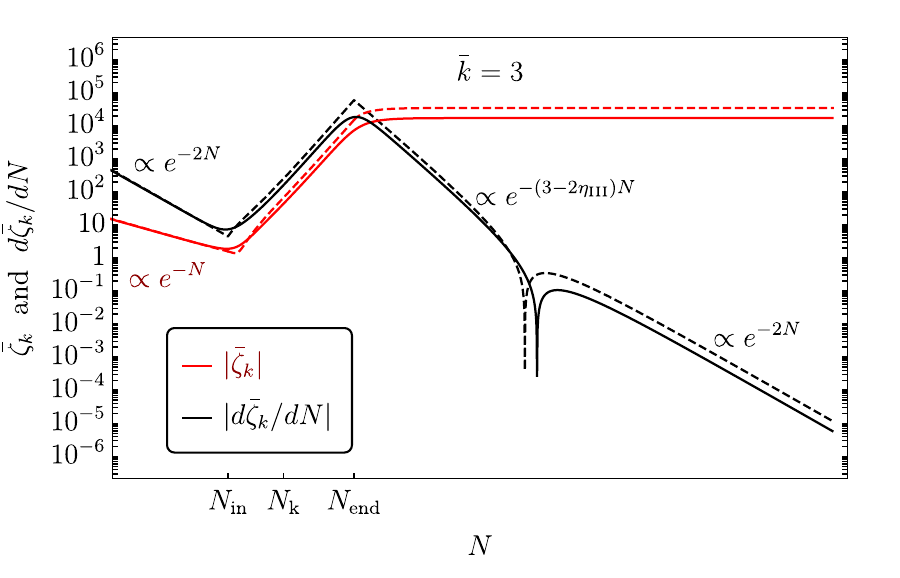}$$\vspace{-0.5cm}
\caption{\em Comparison of the time evolution of 
$|\bar{\zeta}_k|$ and $|d\bar{\zeta}_k/dN|$ computed numerically (solid lines) and with the analytical approximation (dashed lines) within the reverse enginnering approach. We take $\bar{k} = 10^{-3}$ (left panel) and $\bar{k} = 1$ (right panel). 
To draw this figure we consider the benchmark values $\eta_{\textrm{\rm II}} = 3.5$, $\eta_{\textrm{\rm III}} = 0$, $\Delta N_{\textrm{\rm USR}} = 2.5$ and $\delta N = 0.3$.
}\label{fig:DynDer}  
\end{center}
\end{figure}

We rewrite the M-S equation in the form
\begin{align}
\frac{d^2\zeta_k}{dN^2} + (3+\epsilon - 2\eta)\frac{d\zeta_k}{dN} + \frac{k^2}{(aH)^2} \zeta_k = 0\,.
\end{align}
Assuming $\epsilon \approx 0$, constant $\eta$ and constant $H$, this equation admits the solution 
\begin{align}
\zeta_k(N) \propto e^{-\left(\frac{3}{2}-\eta\right)N}
\left[
c_1\,J_{\frac{3}{2}-\eta}\left(
\bar{k}e^{N_{\textrm{in}}-N}
\right)
\Gamma\left(
\frac{5}{2}-\eta
\right) + 
c_2\,J_{-\frac{3}{2}+\eta}\left(\bar{k}e^{N_{\textrm{in}}-N}\right)
\Gamma\left(
-\frac{1}{2}+\eta
\right)
\right]\,,\label{eq:AnalRk}
\end{align}
where $J_{\alpha}(x)$ are Bessel functions of the first kind and $\Gamma(x)$ is the Euler gamma function. 
Consequently, we find
\begin{align}
\frac{d\zeta_k}{dN}(N) \propto e^{-\left(\frac{5}{2}-\eta\right)N}\left[
-c_1\,J_{\frac{1}{2}-\eta}\left(
\bar{k}e^{N_{\textrm{in}}-N}
\right)
\Gamma\left(
\frac{5}{2}-\eta
\right) + 
c_2\,J_{-\frac{1}{2}+\eta}\left(\bar{k}e^{N_{\textrm{in}}-N}\right)
\Gamma\left(
-\frac{1}{2}+\eta
\right)
\right]\,.\label{eq:AnaldRk}
\end{align}
This approximation is applicable 
for $N< N_{\textrm{in}}$ with $\eta = 0$,
for $N_{\textrm{in}} < N < N_{\textrm{end}}$ with $\eta = \eta_{\textrm{II}}$ and for $N > N_{\textrm{end}}$ with $\eta = \eta_{\textrm{III}}$.
We have the following asymptotic behaviors  
\begin{align}
J_{\alpha}(x) \sim 
\left\{
\begin{array}{ccc}
 1/\sqrt{x} & \textrm{for}  & x\gg 1  \\
x^{\alpha}  & \textrm{for}  & x\ll 1
\end{array}
\right.~~~~~
\textrm{where}~~~~~~x\equiv \bar{k}e^{N_{\textrm{in}} - N} = 
e^{N_k - N}\,.\label{eq:Asy}
\end{align}
Consequently, we highlight the following scalings.
\begin{itemize}
\item[$\circ$]
On sub-horizon scales, we find
\begin{align}
\textrm{sub-horizon scales,\,\,}N\ll N_k
~~~~~~
\zeta_k(N) \sim e^{-(1-\eta)N}
~~~\textrm{and}~~~
\frac{d\zeta_k}{dN}(N) \sim e^{-(2-\eta)N}\,.
\end{align}
The above scaling 
implies, for instance, that before the USR phase (that is, for $N<N_{\textrm{in}}$ with $\eta =0$) sub-horizon modes decay according to $\zeta_k\sim  e^{-N}$ and 
$d\zeta_k/dN \sim e^{-2N}$. 
\item[$\circ$]
On super-horizon scales, we find
\begin{align}
\textrm{super-horizon scales,\,\,}N\gg N_k
~~
\zeta_k(N) \sim c_1\, e^{-(3-2\eta)N} + c_2~\textrm{and}~
\frac{d\zeta_k}{dN}(N) \sim -c_1\, e^{-(3-2\eta)N} + c_2\,e^{-2N}\,.\label{eq:SuXder}
\end{align}
Consider a mode that is super-horizon after the end of the USR phase (that is, for $N>N_{\textrm{end}}$ with $\eta =\eta_{\textrm{III}} < 0$). 
Eq.\,(\ref{eq:SuXder}) tells us that $d\zeta_k/dN$ is given by the superposition of two functions:  the first one decays faster, as $e^{-(3-2\eta_{\textrm{III}})N}$, while the second one decays slower, as $e^{-2N}$. 
On the contrary, $\zeta_k$ quickly settles to a constant value.
\item[$\circ$] Consider the evolution during the USR phase. We have 
$\eta = \eta_{\textrm{II}} > 3/2$ 
and $N_{\textrm{in}} < N < N_{\textrm{end}}$. 
We have two possibilities that are relevant to our analysis. 
\begin{enumerate}
\item If the mode is way outside the horizon at the beginning of the USR phase, it stays constant even though its derivative exponentially grows because of the term $\sim e^{-(3-2\eta_{\textrm{II}})N}$.
\item Consider a mode that crosses the Hubble horizon during the USR phase. 
The curvature perturbation (and its derivative) grows because of 
the factor $e^{-(3/2-\eta_{\textrm{II}})N}$. However, it is not immediate to find the exact scaling in time because in this case none of the approximations in eq.\,(\ref{eq:Asy}) can be applied.
\end{enumerate}
\end{itemize}
All the above features, even though obtained in the context of the over-simplified framework given by eq.\,(\ref{eq:AnalRk}) and 
eq.\,(\ref{eq:AnaldRk}), are valid in general.  
In fig.\,\ref{fig:DynDer}, we 
plot $|\zeta_k|$ and 
$|d\zeta_k/dN|$. We checked that all the relevant scaling properties discussed above are indeed verified.
It is possible to derive some useful analytical approximations.

First of all, we consider the Wronskian condition
\begin{align}
 i\left[
 u_k^{\prime}(\tau)
 u_k^*(\tau) - 
 u_k^{\prime\,*}(\tau)
 u_k(\tau)
 \right] = 1\,,
\end{align}
which we rewrite as
\begin{align}
i(aH)\left[
\frac{du_k}{dN}(N)u_k^*(N) - 
\frac{du_k^*}{dN}(N)u_k(N)
\right] = 1\,.
\end{align}
As far  as $du_k/dN$ is concerned, we find
\begin{align}
  \frac{du_k}{dN} = 
  a\sqrt{2\epsilon}(1+\epsilon-\eta)\zeta_k + a\sqrt{2\epsilon}\frac{d\zeta_k}{dN}\,,
\end{align}
so that the Wronskian condition reads
\begin{align}
 \textrm{Im}\bigg[
 \zeta_k(N)
 \frac{d\zeta_k^*}{dN}(N)
 \bigg] = \frac{H^2}{4\epsilon_{\textrm{ref}}\bar{\epsilon}(N)(aH)^3}\,.
\end{align}
If we introduce the field $\bar{\zeta}_k$ as in eq.\,(\ref{eq:BarField}), we find 
\begin{align}
 W(N)\equiv  \textrm{Im}\bigg[
 \bar{\zeta}_k(N)
 \frac{d\bar{\zeta}_k^*}{dN}(N)
 \bigg] = 
 -\textrm{Im}\bigg[
 \bar{\zeta}_k^*(N)
 \frac{d\bar{\zeta}_k}{dN}(N)
 \bigg]
 = \frac{\bar{k}^3 }{4\bar{\epsilon}(N)}e^{3(N_{\textrm{in}}- N)}\,,
\end{align}
with $\epsilon(N)$ given by eq.\,(\ref{eq:DynEps}) for generic $\delta N$. 
In the limit $\delta N\to 0$ and 
at time $N = N_{\textrm{end}}$, we find
\begin{align}
\lim_{\delta N \to 0}W(N_{\textrm{end}}) = 
\frac{\bar{k}^3}{4}e^{(2\eta_{\textrm{II}}-3)(N_{\textrm{end}}-
N_{\textrm{in}})} = 
\frac{\bar{k}^3}{4}
\left(\frac{k_{\textrm{end}}}{k_{\textrm{in}}}\right)^{2\eta_{\textrm{II}}-3}
= \frac{k^3}{4}
\left(\frac{k_{\textrm{end}}^{2\eta_{\textrm{II}}-3}}{
k_{\textrm{in}}^{2\eta_{\textrm{II}}}
}\right)
\,.\label{eq:Wrowro}
\end{align}
If we further take $\eta_{\textrm{II}} = 3$, the above equation is compatible with ref.\,\cite{Kristiano:2022maq}.

We now consider the limit $\delta N\to 0$ and the case
$\eta_{\textrm{II}} = 3$. 
In this case, it is possible to compute  the function $\bar{\zeta}_q(N)$ by solving analytically the M-S equations in both the SR (for $N\leqslant N_{\textrm{in}}$) and USR (for $N_{\textrm{in}} \leqslant N \leqslant N_{\textrm{end}}$) regime and then matching the solutions at $N_{\textrm{in}}$, as done in ref.\,\cite{Kristiano:2022maq} (see also refs.\,\cite{Byrnes:2018txb,Ballesteros:2020qam}).  We find ($x\equiv e^{\Delta N_{\textrm{USR}}}$)
\begin{align}
|\bar{\zeta}_q(N_{\rm end})|^2  = 
&\frac{x^6}{8\bar{q}^6}
\left[
9 + 18 \bar{q}^2 + 9 \bar{q}^4 + 2 \bar{q}^6 + 3(-3 + 7 \bar{q}^4)\cos\left(2\bar{q} - \frac{2\bar{q}}{x}\right) 
 -6 \bar{q} (3 + 4 \bar{q}^2 - \bar{q}^4)\sin\left(2\bar{q} - \frac{2\bar{q}}{x}\right)
\right]+ \nn\\
&\frac{x^5}{8\bar{q}^6}
\left[
12 \bar{q}^2 (-3 - 4 \bar{q}^2 + \bar{q}^4)
\cos\left(2\bar{q} - \frac{2\bar{q}}{x}\right) 
 -6 \bar{q} (-3 + 7 \bar{q}^4)
 \sin\left(2\bar{q} - \frac{2\bar{q}}{x}\right)
\right]+
\nn\\
&\frac{x^4}{8\bar{q}^6}\left[
\bar{q}^2 (9 + 18 \bar{q}^2 + 9 \bar{q}^4 + 2 \bar{q}^6)
+ \bar{q}^2(9  -21 \bar{q}^4)
\cos\left(2\bar{q} - \frac{2\bar{q}}{x}\right) 
 -6 \bar{q}^3 (-3 - 4 \bar{q}^2 + \bar{q}^4)
 \sin\left(2\bar{q} - \frac{2\bar{q}}{x}\right)
\right]\,,
\end{align}
which enters into the computation of eq.\,(\ref{eq:ModeInte}).

\section{Starobinsky model in slow-roll approximation}\label{app:Staro}

In this appendix, we present the semi-analytical expressions corresponding to the main quantities that characterize the Starobinsky inflationary model in the slow-roll approximation. 
The model is based on the scalar potential
\begin{align}
V_{\textrm{Staro}}(\phi) = 
\frac{3M^2\MPl^2}{4}
\left[1 - 
\exp\left(
-\sqrt{\frac{2}{3}}
\frac{\phi}{\MPl}
\right)
\right]^2\,.
\end{align}
For ease of reading, we introduce the function
\begin{align}
f_x \equiv -\frac{1}{3}
(3+2\sqrt{3})e^{-1-2/\sqrt{3} - 4x/3}\,.\label{eq:ShortHandfx}
\end{align}
We indicate with $\Delta N_{\star}$ the number of $e$-folds between the horizon crossing of the CMB pivot scale $k_{\star} \equiv 0.05$ Mpc$^{-1}$ and the end of inflation.
The field value at the end of inflation is
\begin{align}
\phi_{\textrm{end}} = 
\sqrt{\frac{3}{2}}
\log\left(1+\frac{2}{\sqrt{3}}\right)\MPl\,.
\end{align}
The field value at the CMB pivot  scale is
\begin{equation}
\phi_{\textrm{CMB}} = 
-\frac{1}{\sqrt{6}}\big[
3+2\sqrt{3}+4
\Delta N_{\star} 
+ 
\log(-135+78\sqrt{3}) + 3W_{-1}(f_{\Delta N_{\star}})\big]
\MPl\,,
\end{equation}
where $W_{-1}(z)$ is the branch with 
$k=-1$ of the Lambert W function $W_k(z)$. 
The scalar spectral index at the CMB pivot scale reads
\begin{align}
n_s = 1-\frac{16}{3[1+W_{-1}(f_{\Delta N_{\star}})]^2} 
+ \frac{8}{
3[1+W_{-1}(f_{\Delta N_{\star}})]
}\,,
\end{align}
while for the tensor-to-scalar ratio we find
\begin{align}
r = 
\frac{64}{
3[1+W_{-1}(f_{\Delta N_{\star}})]^2
}\,.
\end{align}
The amplitude of the scalar power spectrum at the CMB pivot scale is
\begin{align}
A_s = \frac{3M^2[1+W_{-1}(f_{\Delta N_{\star}})]^4}{128\pi^2 \MPl^2 
W_{-1}(f_{\Delta N_{\star}})^2}\,.
\end{align}
Finally, the square of the Hubble rate at the end of inflation is 
\begin{align}
H(N_{\textrm{end}})^2\equiv H_{\textrm{end}}^2 =  
3\left(\frac{7}{2}-2\sqrt{3}\right)M^2\,,
\end{align}
while its value at the time of horizon crossing for the CMB pivot scale  $k_{\star}$ is
\begin{align}
H(N_{\star})^2\equiv
H_{\star}^2 = \frac{M^2}{4}
\left[
1+\frac{1}{W_{-1}(f_{\Delta N_{\star}})}
\right]^2\,.
\end{align}
The value of the Hubble rate at the $e$-fold time $N_k$ at which the generic comoving wavenumber $k$ crosses the inverse comoving Hubble radius (that is, the time $N_k$ defined by the condition $k=a(N_k)H(N_k)$) can be obtained from the scaling
\begin{align}
\frac{k}{k_{\star}}=\frac{
a(N_k)H(N_k)
}{
a(N_{\star})H(N_{\star})
}\to
H(N_k)\equiv H_k = 
\left(\frac{k}{k_{\star}}\right)e^{N_* - N_k}H_{\star}\,.
\end{align}
\section{Validity of the EFT approach and fine tuning of the model}\label{app:EFT}
In order to discuss some remarks concerning our model from the perspective of the effective field theory (EFT) approach and on the fine tuning, we recast the parameters value in Tab.I using the potential form in eq.\ref{eq:Pot1}. The best parametrization values read
\\
\begin{center}
\begin{tabular}{||c||c|c|c|c|c||}
\hline Parameter & $\bar{c}_2$ & $\bar{c}_3$ & $\bar{c}_5$ & $\bar{c}_6$ & $c_4 M^4 / g^2$ \\
\hline Value & 2.053 & -2.360 & -0.164 & 0.009 & $5\times 10^{-12}$ \\
\hline \hline
\end{tabular}
\end{center}
TABLE II: Numerical values (with three digits precision) for the benchmark realization of our model in terms of a polynomial potential as in eq.\ref{eq:Pot1}.
\subsection{Remarks on the EFT approach}
Expliciting the dimensional factor in the field $x=\phi/\bar{M}_{\rm Pl}$, when we impose the constraint from the amplitude $A_S$ of the scalar power spectrum at CMB scales, we immediately find the value 
\begin{equation}\label{eq:Ratio}
    \left(c_4/g^2\right) \left(M/\bar{M}_{\rm Pl}\right)^4= 5\times10^{-12}.
\end{equation}
Requiring that the dimensionless quantities in our model are of order $\mathcal{O}(1)$, in such a way to consider our theory natural in the Wilsonian sense, we can interpret $\bar{g}=\left(M^2/g\bar{M}^2_{\rm Pl}\right)$ as a coupling and the smallness of the ratio in eq.\ref{eq:Ratio} has interesting consequences. Indeed if we impose $c_4=\mathcal{O}(1)$, we get $\bar{g}\simeq10^{-6}$, indicating that the ultraviolet completion from which our scalar EFT arises is a weakly coupled theory. Moreover, considering the parameters $\bar{c}_{i}$, we find that their values alternate in sign. This is expected, as they must conspire to yield the stationary inflection point. Furthermore, $\bar{c_2}$ and $\bar{c}_{3}$ are both of order $O(1)$. Different is the situation for $\bar{c}_5$ and $\bar{c}_6$. Their smallness seems to contradict the idea of having coefficients of order unity as in a Wilson EFT. In principle, they represent coefficients of higher-dimensional operators, and then, they could potentially be generated at the loop level. In this sense, the higher order coefficients are naturally accompanied by an additional loop suppression factor.

Moreover in order to have the right enhancement of the curvature power spectrum, the inflaton should start from planckian values and a small modification of the initial conditions can spoil the inflationary parameters values at the CMB scales. Indeed in general the probability of having such precise values for the initial condition is very small. One can try to justify these small probability with some anthropic reasoning or requiring eternal inflation that can happen in models with flat potential near the initial saddle point (see refs.\cite{Linde:1986fe,Goncharov:1987ir,Drees:2022aea}).In the context of the multiverse picture, the small initial probability should be multiplied by  the number of vacua in the landscape. The latter could well be a gargantuan number, and it may turn what seems to be an extremely unlucky event (from the point of view of one single universe) into a plausible property of the landscape.

Even though the discussion presented here is simplistic, given that our model is based on simplifications, such as assuming a single coupling and mass scale, it remains of considerable interest how inflationary observables manage to provide some insights into the fundamental properties of the underlying theoretical framework.
\subsection{Fine tuning of the model}
It is well known that models for PBH production are sensitive to a huge fine tuning of the model parameters. In this appendix we try to address, following ref.\,\cite{Cole:2023wyx}, how large the fine tuning is in this model. We consider shifts in the four parameters of the polynomial potential, i.e $\bar{c}_2$, $\bar{c}_3$, $\bar{c}_5$ and $\bar{c}_6$. We compute the largest shifts, at order of magnitude, which do not make the power spectrum grow larger than unity. We find that for $\bar{c}_2$, $\bar{c}_3$, $\bar{c}_5$ the tuning is as at $\mathcal{O}(10^{-7})$ while the most tuned parameter is $\bar{c}_6$ with a required fine tuning of $\mathcal{O}(10^{-8})$.
\\It is possible to connect the fine-tuning on these parameters to physical observables $\mathcal{A}$, such as the abundance and inflationary quantities of the CMB, by introducing, as already done in refs.~\cite{Azhar:2018lzd,Cole:2023wyx}, the following dimensionless quantity 
\begin{equation}
    \epsilon_{\mathcal{A}}=\frac{\mathrm{d} \log \mathcal{A}}{\mathrm{d} \log p}.
\end{equation}
For the model under consideration, we computed $\epsilon$ for several observables by varying the "most fine-tuned" parameter $\bar{c}_6$. We consider the fiducial value $\bar{c}_6=0.009$, and we shifted its value by a small shift $\delta$. The results are shown in the following table:
\begin{center}
\begin{tabular}{||c||c|c|c|c||}
\hline $\bar{c}_6$  & $\epsilon_{P_{\rm peak}}$ & $\epsilon_{f_{\rm PBH}}$ & $\epsilon_{n_s}$ & $\epsilon_{ r}$  \\
\hline $0.009$ & $-1 \times 10^{8}$ & $-3 \times 10^{9}$ & $-6 \times 10^{4}$ & $5\times 10^{5} $ \\
\hline \hline
\end{tabular}
\end{center}
TABLE III: Numerical values of the fine-tuning parameter $\epsilon$ for several observables by varying the fiducial value of $\bar{c}_6$ in the polynomial potential of Eq.~\ref{eq:Pot1}.

\begin{figure}[!t]
	\centering
\includegraphics[width=0.6\textwidth]{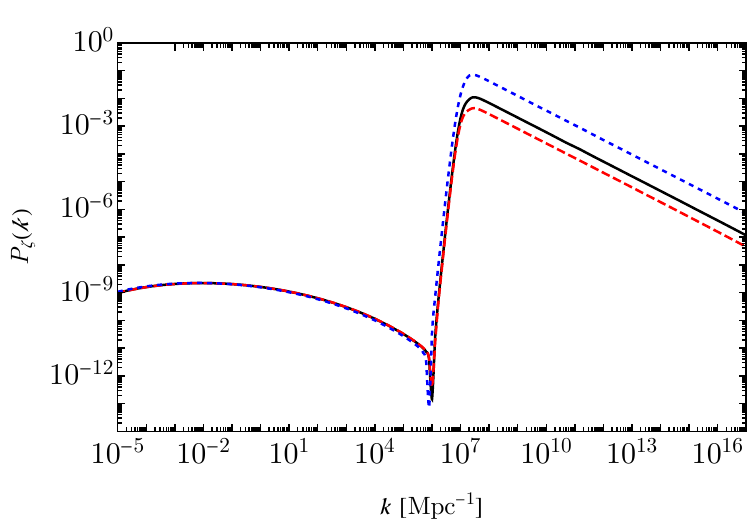}
	\caption{Power spectra for the potential in eq.\ref{eq:Pot1}. Fiducial values correspond to the black line, dashed lines correspond increasing/decreasing (red/blue) the parameter $\bar{c}_6$ by a factor $10^{-8}$. We tune the initial conditions in order to have the peak at same scale $k$.}
\label{Fig:Fine}
\end{figure}

In ref.~\cite{Cole:2023wyx}, a similar analysis was performed for polynomial potentials up to dimension $5$ with the inclusion of a dimension $1$ operator. In their analysis, they found that the cubic $\bar{c}_3$ and the fifth operator $\bar{c}_5$ were the most sensitive parameters, with $\epsilon_{P_{\rm peak}}\simeq -2 \times 10^{8}$ and $\epsilon_{f_{\rm PBH}} \simeq -5 \times 10^{9}$ respectively. From our analysis one can wrongly think that the CMB parameters are not so sensitive respect the abundance to the fine tuning of the parameter. This is not completely true. Indeed in terms of $\epsilon$ the changes are symmetric if the shift is positive or negative. This is not true if we want to fix the position of the main peak. Indeed the position of the main peak is shifted changing only one parameter. We show in Fig.\,\ref{Fig:Fine} the power spectrum and how it changes with the small fine tuning on the parameter $\bar{c}_6$, tuning the initial condition $\phi_0$ in order to have the peak at same scale $k_{\rm peak}$. Now we can observe that the enhancement of the power spectrum is not symmetric for a given perturbation of the fiducial value if we fix the position of the main peak. This has a huge impact on the CMB inflationary observables. Indeed modifying the amplitude of the peak at a fixed scale $k$ has in general the effect of change the CMB predictions, and in some cases to not generate anymore a successful inflation. For instance in the case we reported in Fig.\ref{Fig:Fine} the scalar spectral index decrease of a factor of circa $0.01$ when we increase the amplitude of the peak while it remains nearly constant when we decrease it.
Then, if we look at the CMB parameters, fixing the position of the peak, using the parameter $\epsilon$ as done in ref.~\cite{Cole:2023wyx} to determine the fine-tuning of a model is not a suitable choice. The searching for a more suitable set of parameters for the study of the fine tuning of models is left as future work.

\section{Details of the scalar potential}\label{app:Para}

\begin{align}
&\alpha_2  =\frac{4 x_0^3 [x_1^2 (\beta_2 - 2 \beta_3) - 2 x_0 x_1 (\beta_3 + \beta_5) +
    x_0^2 (-2 \beta_5 + \beta_6)]}{
x_0^2 + 4 x_0 x_1 + x_1^2
    }\,,\\
&-\alpha_3  = \frac{4 x_0^2 [-x_1^2 (\beta_2 - 4 \beta_3) + 
   4 x_0 x_1 (\beta_3 + 2 \beta_5) + x_0^2 (8 \beta_5 - 5 \beta_6)]}{
   x_0^2 + 4 x_0 x_1 + x_1^2
   }\,, \\
&\alpha_5  =\frac{4 x_1^3 [x_0^2 (\beta_2 - 2 \beta_3) - 2 x_0 x_1 (\beta_3 + \beta_5) +
    x_1^2 (-2 \beta_5 + \beta_6)]}{
x_0^2 + 4 x_0 x_1 + x_1^2    
    }\,,\\
&\alpha_6  = \frac{4 x_1^2 [x_0^2 (\beta_2 - 4 \beta_3) - 
   4 x_0 x_1 (\beta_3 + 2 \beta_5) + x_1^2 (-8 \beta_5 + 5 \beta_6)]}{
 x_0^2 + 4 x_0 x_1 + x_1^2  
   }\,.
\end{align}
\chapter{Appendix of Chapter 7}
\section{Fine tuning and initial conditions in the curvaton model}\label{app:Ini}
In order for the model to produce the needed enhancement of the angular perturbations, $\varphi$ has to start its rolling from some value $\varphi_* = \mathcal{O}(\bar{M}_{\rm Pl})$. Ref.\,\cite{Kasuya:2009up} justifies the Planckian initial condition by arguing the existence of some
 pre-inflationary phase during which the field $\varphi$ gets a negative Hubble-induced mass term. 
Instead of invoking some custom-made pre-inflationary physics, let us try to understand if this initial condition can be considered as a natural outcome of inflationary dynamics. 

\begin{figure}[h]
\begin{center}
$$\includegraphics[width=.49\textwidth]{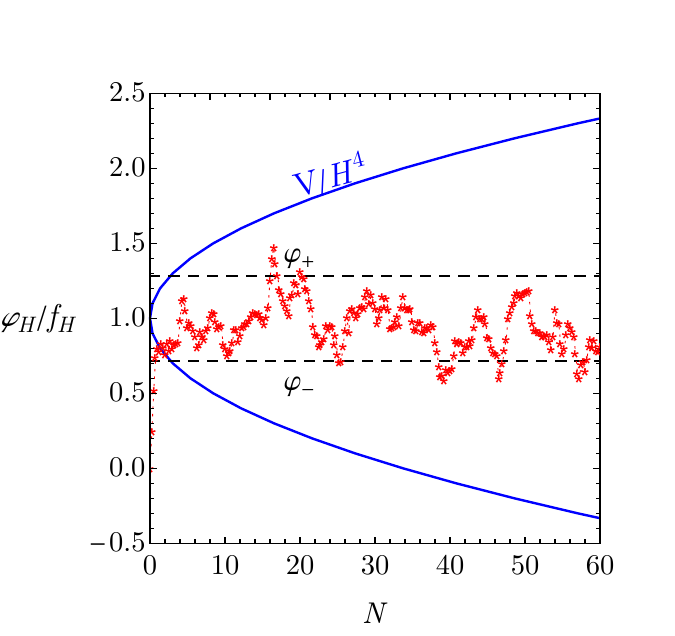}
\qquad\includegraphics[width=.49\textwidth]{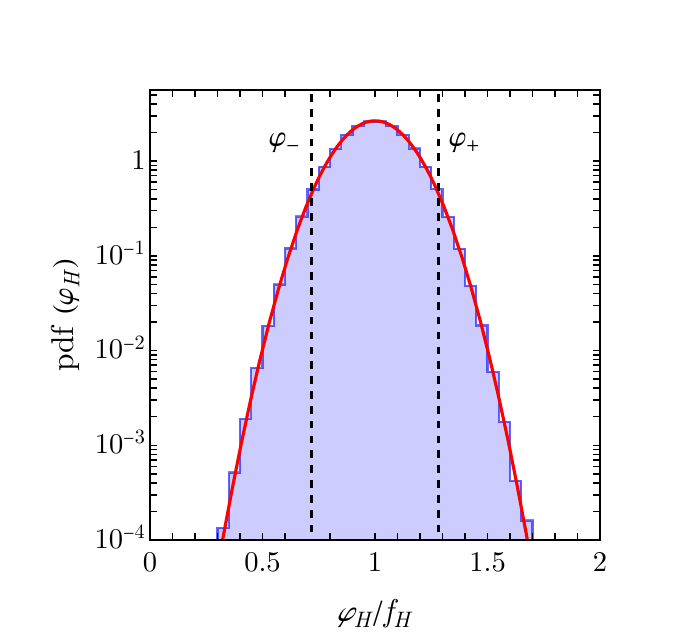}$$
\caption{ Left panel: Red dots represent the stochastic dynamics of the field $\varphi$ in terms of the number of $e$-folds N while the blue line is the quadratic potential in equation (\ref{eq:EffectiveV}).
Right panel: The histogram represents the distribution of the different numerical solutions of equation (\ref{eq:QuantumKicks}) while the red line is the Gaussian function in equation (\ref{eq:PDFGauss}).} \label{fig:Stoca} 
\end{center}
\end{figure}

During inflation, the stochastic dynamics of the field $\varphi$ in the de Sitter background is described by the equation
\begin{align}\label{eq:QuantumKicks}
\frac{d^2\varphi_H}{dN^2} + 3\frac{d\varphi_H}{dN} + c(\varphi_H -f_H) = 
\eta_H\,,
~~~~~~~~
\langle \eta_H(N)\eta_H(N^{\prime})\rangle = \frac{9}{4\pi^2}\delta(N - N^{\prime})\,,
\end{align}
where the left-hand side of eq.\,(\ref{eq:QuantumKicks}) describes the evolution of the long-wavelength modes 
while the right-hand side represents the quantum noise sourced by the short-wavelength ones. 
Notice that this stochastic picture is applicable in the range of values $0 < c < 9/4$ that we consider in the curvaton model under examination.
A realization of the numerical solution of the above stochastic differential equation is shown in the left panel of fig.\,\ref{fig:Stoca}.
One can compute several times the solution of the above stochastic equation but it is very likely (in a way the we shall quantify in a moment) that the outcome will be always similar: the motion of the field $\varphi_H$ remains confined close to the minimum of the potential.  
More in detail, we can define the limiting values
\begin{align}
\varphi_{\pm} = f_H \pm \frac{3}{2\pi c}\,,
\end{align}
which correspond to the two dashed lines in fig.\,\ref{fig:Stoca}. 
The random motion of the field $\varphi_H$ does not overcome these two values. 
This is simple to understand.
When $|\varphi_H| > \varphi_{\pm}$, the classical displacement (per unit Hubble time) becomes larger than the amplitude of the quantum jump and the field $\varphi_H$ is pushed towards the minimum of the potential.
We can actually do better and compute the probability to find, after $N$ $e$-fold of inflation and at some position in space, some specific value of the field $\varphi_H$. 
This probability can be computed numerically by solving many times eq.\,(\ref{eq:QuantumKicks}) and extracting from the resulting statistical sample the corresponding  
PDF or by solving the Fokker-Planck equation. The two procedure agree, and we find that, after few $e$-folds, the PDF is well described by the Gaussian distribution
\begin{align}\label{eq:PDFGauss}
{\it pdf}(\varphi_H) = \frac{1}{\sqrt{2\pi}\sigma}\exp\left[
-\frac{(\varphi_H - f_H)^2}{2\sigma^2}
\right]\,,~~~~~~~{\rm with\,variance}~~~\sigma = \frac{1}{2\pi}\sqrt{\frac{3}{2c}}\,.
\end{align}
Clearly, the probability to find $\varphi_H$ at Planckian values is an exponentially small number and one should admit a certain degree of fine-tuning in the initial conditions of the model.

One can try to justify the small probability in eq.\,(\ref{eq:PDFGauss}) with some anthropic reasoning. 
In the context of the multiverse picture, one should multiply the small probability in eq.\,(\ref{eq:PDFGauss}) times the number of vacua in the landscape.
The latter could well be a gargantuan number, and it may turn what seems to be an extremely unlucky event (from the point of view of one single universe) 
into a plausible property of the landscape.

This perspective is acceptable as long as we link the properties of the model to some 
anthropic observable like  the ratio of dark matter to baryon matter\,\cite{Hellerman:2005yi}.

\section{Perturbations in a Friedmann-Robertson-Walker universe}\label{app:Pert}

In this appendix we review the formalism used in our analysis for the description of scalar
 perturbations in a Friedmann-Robertson-Walker universe with multiple interacting fluids. 
 We refer the reader to ref.\,\cite{Kodama:1984ziu} for a more comprehensive discussion.

The perturbed line element is
 \begin{align}\label{eq:MetricPerturbation}
ds^2 = 
&
-[1+2A(t,\vec{x})]dt^2 + 2a(t)\partial_{i}B(t,\vec{x})dx^i dt
\nonumber \\
& + 
  a(t)^2\big\{
 [1-2\psi(t,\vec{x})]\delta_{ij} + 2\partial_{ij}E(t,\vec{x})
  \big\}dx^i dx^j\,.
 \end{align} 
 Derivatives with respect to the cosmic time $t$ are indicated with a dot, $\dot{}\equiv d/dt$.
 The intrinsic curvature of a spatial hypersurface, $R_3$, is given by $R_3 = (4/a^2)\nabla^2\psi$.\footnote{
We write $ds^2 = g_{\mu\nu}dx^{\mu}dx^{\nu} = [g^{(0)}_{\mu\nu}(t) + 
\delta g_{\mu\nu}(t,\vec{x})]
dx^{\mu}dx^{\nu}$ with $g^{(0)}_{\mu\nu}(t) = {\rm diag}(-1,a(t)^2,a(t)^2,a(t)^2)$. 
Notice that in eq.\,(\ref{eq:MetricPerturbation}) the matrix $\partial_{ij}E(t,\vec{x})$ is {\it not} traceless. 
Alternatively, one can define the perturbed metric as
 \begin{align}\label{eq:MetricPerturbation2}
ds^2 = -[1+2A(t,\vec{x})]dt^2 + 2a(t)\partial_{i}B(t,\vec{x})dx^i dt + 
  a(t)^2\big\{
 [1-2D(t,\vec{x})]\delta_{ij} + 2\bar{\partial}_{ij}E(t,\vec{x})
  \big\}dx^i dx^j\,,
 \end{align} 
with $\bar{\partial}_{ij} = \partial_{ij} - (1/3)\delta_{ij}\nabla^2 E(t,\vec{x})$. 
The matrix $\bar{\partial}_{ij}E(t,\vec{x})$ is now, by construction, traceless. 
Of course, $D(t,\vec{x})$ in eq.\,(\ref{eq:MetricPerturbation2}) differs from $\psi(t,\vec{x})$ in eq.\,(\ref{eq:MetricPerturbation}). 
In particular, using eq.\,(\ref{eq:MetricPerturbation2}) the intrinsic
curvature of a spatial hypersurface is now given by $\bar{R}_3 = (4/a^2)\nabla^2(D + \nabla^2E/3)$. 
Consequently, the spatially-flat gauge is defined by $\psi = 0$ if one takes eq.\,(\ref{eq:MetricPerturbation}) but 
by $D  = -\nabla^2E/3$ if one takes eq.\,(\ref{eq:MetricPerturbation2}).
} 
We split the Einstein field equations $G_{\mu\nu}\equiv R_{\mu\nu} - \frac{1}{2}g_{\mu\nu}R = 8\pi G_N T_{\mu\nu}$ 
into a zero-order  system of equations 
$G_{\mu\nu}^{(0)} = 8\pi G_N T_{\mu\nu}^{(0)}$
describing the background dynamics (discussed in section\,\ref{sec:AfterInfla}) 
plus linear perturbations $\delta G_{\mu\nu} = 8\pi G_N \delta T_{\mu\nu}$ 
(relevant for the computations carried out in section\,\ref{eq:Per}). 
The reduced Planck mass $\bar{M}_{\rm Pl}$ is related to the Newton gravitational constant $G_N$ by 
$\bar{M}_{\rm Pl}^2 = 1/8\pi G_N$.

The stress-energy tensor of a fluid with energy density $\rho$, isotropic pressure $P$ and four-velocity $u^{\mu}$ 
is given by $T^{\mu}_{\,\,\nu} = (\rho + P)u^{\mu}u_{\nu} + P\delta^{\mu}_{\,\,\nu}$.\footnote{
We set to zero the anisotropic stress tensor. 
Scalar fields and perfect fluids cannot support anisotropic stress.} 
The total four velocity is subject to the constraint $u^{\mu}u_{\mu} = -1$. 
At the linear order in the perturbations, we have $u_{\mu} = (-(1+A), a(\partial_i v + \partial_i B))$ where 
$v$ is the total scalar velocity potential.
Consequently, we find
$T^{\mu}_{\,\,\nu} = 
{\rm diag}(-\rho,P,P,P) + \delta T^{\mu}_{\,\,\nu}$
with 
\begin{align}\label{eq:EnergyMomentumTensor}
\delta T^{0}_{\,\,0} = -\delta\rho\,,
~~~~~
\delta T^{0}_{\,\,i} = (\rho + P)a(\partial_i B + \partial_i v)\,,
~~~~~
\delta T^{i}_{\,\,0} = - \frac{(\rho + P)}{a}\partial_i v\,,
~~~~~
\delta T^{i}_{\,\,j} = \delta^{i}_{j}\,\delta P\,. 
\end{align}
The $^0_{\,\,0}$ and $^0_{\,\,i}$ components of the perturbed Einstein field equations -- that are
 $\delta G^0_{\,\,0} = 
8\pi G_N \delta T^0_{\,\,0}$ and 
$\delta G^0_{\,\,i} = 
8\pi G_N \delta T^0_{\,\,i}$, respectively -- 
 read
\begin{align}
3H(\dot{\psi} + HA) - \frac{\nabla^2}{a^2}\big[
\psi + H(\underbrace{a^2 \dot{E} - aB}_{\equiv\,\chi})
\big] 
 + 4\pi G_N \delta\rho & = 0\,,\label{eq:Funda1}\\
\dot{\psi} + HA + 4\pi G_N(\rho + P)\underbrace{a(B + v)}_{\equiv\,V} & = 0\,,\label{eq:Funda2}
\end{align}
where we defined the scalar shear $\chi \equiv a^2 \dot{E} - aB$ and the total covariant velocity perturbation 
$V\equiv a(B+v)$. 
The off-diagonal spatial components of the perturbed Einstein field equations give 
the time-evolution of the scalar shear 
\begin{align}\label{eq:EvoChi}
\dot{\chi} + H\chi - A + \psi = 0\,.
\end{align}
Finally, the spatial trace of the perturbed Einstein field equations, combined with eq.\,(\ref{eq:EvoChi}), gives 
\begin{align}
\ddot{\psi} + 3H\dot{\psi} + H\dot{A} + (3H^2 + 2\dot{H})A - 4\pi G_N \delta P = 0\,.\label{eq:Funda3} 
\end{align}

Through the Bianchi identities, the Einstein field equations 
imply the local conservation of {\it total} energy and momentum, that is $\nabla_{\mu}T^{\mu\nu} = 0$. 
In the multiple fluid case the total energy-momentum tensor is the sum of the energy-momentum tensors of the
individual fluids, $T^{\mu\nu} = \sum_{\alpha}T^{\mu\nu}_{\alpha}$. 
For each fluid, we write the local energy-momentum transfer 4-vector as 
$\nabla_{\mu}T^{\mu\nu}_{\alpha} = Q^{\nu}_{\alpha}$ with the constraint $\sum_{\alpha}Q^{\nu}_{\alpha} = 0$.
At the zero-order in the perturbations, 
the time component $\nu = 0$ of $\nabla_{\mu}T^{\mu\nu} = 0$ gives 
$\dot{\rho} + 3H(\rho + P) = 0$; at the linear order in the perturbations, on the contrary, we find
\begin{align}\label{eq:EnCon}
\dot{\delta\rho} + 3H(\delta\rho + \delta P) - 3\dot{\psi}(\rho + P) + 
\frac{\nabla^2}{a^2}\big[(\rho + P)(\chi + V)\big] = 0\,.
\end{align}
For a single fluid identified by the label $\alpha$,  
$\nabla_{\mu}T^{\mu 0}_{\alpha} = Q^{0}_{\alpha}$ gives the zero-order result
$\dot{\rho}_{\alpha}+3H(\rho_{\alpha} + P_{\alpha}) = Q_{\alpha}$ while its perturbed version reads
\begin{align}\label{eq:IndividualEnCon}
\dot{\delta\rho}_{\alpha} + 3H(\delta\rho_{\alpha} + \delta P_{\alpha}) - 3\dot{\psi}(\rho_{\alpha} + P_{\alpha}) + 
\frac{\nabla^2}{a^2}\big[(\rho_{\alpha} + P_{\alpha})(\chi + V_{\alpha})\big] - AQ_{\alpha} - \delta Q_{\alpha}= 0\,,
\end{align}
where we defined the covariant velocity perturbation of the $\alpha$-fluid as 
$V_{\alpha} \equiv a(v_{\alpha} + B)$ where $v_{\alpha}$ 
is the  scalar velocity potential for the $\alpha$-fluid. 
The total fluid perturbations are related to the individual fluid quantities by 
\begin{align}
\delta\rho \equiv \sum_{\alpha}\delta\rho_{\alpha}\,,~~~~~~~~~~
\delta P \equiv \sum_{\alpha}\delta P_{\alpha}\,,~~~~~~~~~~
V = \sum_{\alpha}\frac{\rho_{\alpha} + P_{\alpha}}{\rho + P}\,V_{\alpha}\,,
\end{align}
through which one can get eq.\,(\ref{eq:EnCon}) by summing 
eqs.\,(\ref{eq:IndividualEnCon}). In eq.\,(\ref{eq:IndividualEnCon}) $Q_{\alpha}$ is the energy transfer to the 
$\alpha$-fluid and $\delta Q_{\alpha}$ its perturbation. 
Momentum conservation, $\nabla_{\mu}T^{\mu i}_{\alpha} = Q^{i}_{\alpha}$, gives for the $\alpha$-fluid 
the equation 
\begin{align}\label{eq:IndividualMomCon}
\dot{V}_{\alpha} + \bigg[
\frac{Q_{\alpha}}{(\rho_{\alpha} + P_{\alpha})}(1+c_{\alpha}^2) - 3Hc_{\alpha}^2 
\bigg]V_{\alpha} + A + 
\frac{1}{\rho_{\alpha} + P_{\alpha}}\big(
\delta P_{\alpha} - Q_{\alpha}V
\big) = 0\,,
\end{align}
where $c_{\alpha}^2 \equiv \dot{P}_{\alpha}/\dot{\rho}_{\alpha}$ is the adiabatic sound speed of the $\alpha$-fluid. 
We consider in the above equation the case of zero momentum transfer among the fluids. 
The total momentum conservation equation $\nabla_{\mu}T^{\mu i} = 0$ is given by
\begin{align}\label{eq:MomCon}
\dot{V} - 3Hc_s^2 V + A + \frac{1}{\rho + P}\,\delta P = 0\,,
\end{align}
where $c_s^2 \equiv \dot{P}/\dot{\rho}$ is the total adiabatic speed of sound which can be written as a weighted sum of the adiabatic sound speeds of the
individual fluids 
\begin{align}
c_s^2 = \sum_{\alpha}\frac{\dot{\rho}_{\alpha}}{\dot{\rho}}c_{\alpha}^2\,.
\end{align}
In summary, the relevant equations are eqs.\,(\ref{eq:Funda1},\,\ref{eq:Funda2},\,\ref{eq:EvoChi}\,,\ref{eq:Funda3}) 
with the energy and momentum conservation in eqs.\,(\ref{eq:EnCon},\,\ref{eq:IndividualEnCon}) 
and eqs.\,(\ref{eq:IndividualMomCon},\,\ref{eq:MomCon}).
 
We consider the description of the dynamics of the scalar perturbations in terms of gauge-invariant quantities.

We define {\it i)} the total curvature perturbation on
uniform density hypersurfaces
\begin{align}
\zeta \equiv -\psi - H\frac{\delta\rho}{\dot{\rho}} = \sum_{\alpha}\frac{\dot{\rho}_{\alpha}}{\dot{\rho}}\,\zeta_{\alpha}\,,
~~~~~~~{\rm with}~~~~~~\zeta_{\alpha} \equiv -\psi - H\frac{\delta\rho_{\alpha}}{\dot{\rho}_{\alpha}}\,,
\label{eq:ZetaDefinititon}
\end{align}
{\it ii)} the total comoving curvature perturbation 
\begin{align}
\mathcal{R} \equiv \psi - HV = 
\sum_{\alpha}\frac{\rho_{\alpha} + P_{\alpha}}{\rho + P}\,\mathcal{R}_{\alpha}\,,
~~~~~~~{\rm with}~~~~~~
\mathcal{R}_{\alpha} \equiv \psi - HV_{\alpha}\,,\label{eq:DDefiR}
\end{align}
and {\it iii)} the curvature perturbation on
uniform shear hypersurfaces
\begin{align}
\Psi \equiv \psi + H\chi\,,~~~~~~~{\rm with}~~~~~~\chi \equiv a^2 \dot{E} - aB\,.
\end{align}
From now on, we move to consider the dynamics in Fourier space. 
This implies the formal substitution $\nabla^2 \to -k^2$, where $k$ is the comoving wavenumber (with $k\equiv |\vec{k}|$). 
Furthermore, each perturbed quantity should be now understood as a specific Fourier mode with comoving wavenumber $k$. 
Notice that the system formed by eq.\,(\ref{eq:Funda1}) and eq.\,(\ref{eq:Funda2}), together with the 
equations governing the background dynamics, gives the relation
\begin{align}\label{eq:ZetaR}
3\dot{H}(\zeta + \mathcal{R}) = \frac{k^2}{a^2}\,\Psi\,,~~~~
{\rm or\,equivalently\,}~~~-\frac{3}{2}\big(1+P/\rho\big)(\zeta + \mathcal{R}) = \frac{k^2}{(a H)^2}\,\Psi\,,
\end{align}
which shows that on super-Hubble scale, where $k^2/(aH)^2 \ll 1$, we have $-\zeta \simeq  \mathcal{R}$. 

The evolution equation for $\zeta_{\alpha}$ can be obtained by taking the time derivative of its definition 
in eq.\,(\ref{eq:ZetaDefinititon}) and using eq.\,(\ref{eq:IndividualEnCon}) for $\dot{\delta\rho}_{\alpha}$ 
(in conjunction
with eq.\,(\ref{eq:Funda1}) and the background equations). We find
\begin{align}\label{eq:GaugeInv1}
\dot{\zeta}_{\alpha} =  
- 
\frac{\dot{H}Q_{\alpha}}{H\dot{\rho}_{\alpha}}\big(
\zeta - \zeta_{\alpha}
\big) 
 + 
 \frac{k^2}{3a^2 H}\bigg[
 \Psi - \bigg(
 1 - \frac{Q_{\alpha}}{\dot{\rho}_{\alpha}}
 \bigg)\mathcal{R}_{\alpha}
 \bigg]
 \nonumber 
 \\
 +
 \frac{3H^2}{\dot{\rho}_{\alpha}}\underbrace{\big(
\delta P_{\alpha} - c_{\alpha}^2 \delta\rho_{\alpha}
\big)}_{\equiv\,\delta P_{{\rm intr,}\alpha}}
- \frac{H}{\dot{\rho}_{\alpha}}\underbrace{\bigg(
\delta Q_{\alpha} - \frac{\dot{Q}_{\alpha}\delta\rho_{\alpha}}{\dot{\rho}_{\alpha}}
\bigg)}_{\equiv\,\delta Q_{{\rm intr,}\alpha}}
 \,,
\end{align}
where the combination $\delta P_{\alpha} - c_{\alpha}^2 \delta\rho_{\alpha}$ defines 
the so-called intrinsic non-adiabatic pressure perturbation of the $\alpha$-fluid $P_{{\rm intr,}\alpha}$.
For a barotropic fluid, that is a fluid with equation of state $P_{\alpha} = P_{\alpha}(\rho_{\alpha})$, 
the intrinsic non-adiabatic pressure perturbation vanishes since in this case 
we simply have $\delta P_{\alpha} = (\dot{P}_{\alpha}/\dot{\rho}_{\alpha})\delta \rho_{\alpha}$.

The combination 
$\delta Q_{\alpha} - \dot{Q}_{\alpha}\delta\rho_{\alpha}/\dot{\rho}_{\alpha}$ defines the so-called
intrinsic non-adiabatic energy transfer perturbations of the $\alpha$-fluid $\delta Q_{{\rm intr,}\alpha}$. 
Notice that $\delta Q_{{\rm intr,}\alpha}$ vanishes if 
the energy transfer $Q_{\alpha}$ is a function of the density $\rho_{\alpha}$ so that 
  $\delta Q_{\alpha} = (\dot{Q}_{\alpha}/\dot{\rho}_{\alpha})\delta \rho_{\alpha}$.

From the definition in eq.\,(\ref{eq:DDefiR}) 
we write $\dot{\mathcal{R}}_{\alpha} = \dot{\psi} - \dot{H}V_{\alpha} - H\dot{V}_{\alpha}$. 
The evolution equation for  $\mathcal{R}_{\alpha}$, therefore, can be obtained from the momentum conservation in 
eq.\,(\ref{eq:IndividualMomCon}). Using the background equations and eq.\,(\ref{eq:Funda1}), we find
\begin{align}\label{eq:GaugeInv2}
\dot{\mathcal{R}}_{\alpha} = 
(\mathcal{R} - \mathcal{R}_{\alpha})\bigg(
\frac{Q_{\alpha}}{\rho_{\alpha} + P_{\alpha}} - \frac{\dot{H}}{H}
\bigg)  - \frac{c_{\alpha}^2 \dot{\rho}_{\alpha}}{\rho_{\alpha} + P_{\alpha}}
\big(\zeta_{\alpha} + \mathcal{R}_{\alpha}\big) + \frac{H}{\rho_{\alpha} + P_{\alpha}}
\big(
\delta P_{\alpha} - c_{\alpha}^2 \delta\rho_{\alpha}
\big)\,.
\end{align}
Similarly, eq.\,(\ref{eq:EnCon}) gives the evolution equation for the total curvature perturbation $\zeta$, and we find
\begin{align}\label{eq:GaugeInv3}
\dot{\zeta} = -\frac{H}{\rho + P}\underbrace{\big(
\delta P - c_s^2 \delta \rho
\big)}_{=\,\delta P_{\rm nad}} + \frac{k^2}{3a^2 H}\big(\Psi - \mathcal{R}\big)\,,
\end{align}
where on the right-hand side we used the definition of the non-adiabatic pressure perturbation 
$\delta P \equiv \delta P_{\rm nad} + c_s^2 \delta \rho$. 
In the presence of more than one fluid, the total non-adiabatic pressure perturbation $\delta P_{\rm nad}$ 
consists of two parts, $\delta P_{\rm nad} \equiv \delta P_{\rm intr} + \delta P_{\rm rel}$. 
The first part is due to the intrinsic entropy perturbation of each fluid, 
$\delta P_{\rm intr} = \sum_{\alpha}\delta P_{{\rm intr,}\alpha}$ with $\delta P_{{\rm intr,}\alpha}$
as defined in eq.\,(\ref{eq:GaugeInv1}); 
the second part of the non-adiabatic pressure perturbation, $\delta P_{\rm rel}$, 
is due to the relative entropy perturbation $\mathcal{S}_{\alpha\beta}\equiv 3(\zeta_{\alpha} - \zeta_{\beta})$
between different fluids
\begin{align}
\delta P_{\rm rel} = -\frac{1}{6H\dot{\rho}}\sum_{\alpha,\beta}\dot{\rho}_{\alpha}\dot{\rho}_{\beta}
(c_{\alpha}^2 - c_{\beta}^2)\mathcal{S}_{\alpha\beta} = 
 -\frac{1}{2H\dot{\rho}}\sum_{\alpha,\beta}\dot{\rho}_{\alpha}\dot{\rho}_{\beta}
(c_{\alpha}^2 - c_{\beta}^2)(\zeta_{\alpha} - \zeta_{\beta})\,.
\end{align}

From eq.\,(\ref{eq:MomCon}), on the other hand, we get
\begin{align}\label{eq:GaugeInv4}
\dot{\mathcal{R}} = 
\bigg(
\frac{\dot{H}}{H} + 3Hc_s^2 
\bigg)\big(\zeta + \mathcal{R}\big) + \frac{H}{\rho + P}\big(
\delta P - c_{\alpha}^2 \delta\rho
\big) - \frac{k^2}{3a^2 H}\,\Psi\,.
\end{align}
\subsection{Perturbations dynamics in the Axion-curvaton model}

We now interpret eqs.\,(\ref{eq:GaugeInv1},\,\ref{eq:GaugeInv2},\,\ref{eq:GaugeInv3},\,\ref{eq:GaugeInv4}) 
in light of the curvaton model studied in the main body of this paper.  
We have two fluid species, namely the curvaton and the radiation field, that are identified, respectively, with the labels $\alpha = \phi,\gamma$. The curvaton field decays into radiation with a decay rate $\Gamma_{\phi}$, which we take to be a constant.  

Consider the dynamics during phase I. We set $\Gamma_{\phi} = 0$ so that we do not have energy transfer between the scalar field and radiation. Furthermore we neglect the curvature perturbation and comoving curvature perturbation of the radiation. Hence the eqs.\,\ref{eq:ZetaDefinititon} and eqs.\,\ref{eq:DDefiR} simply read as
\begin{align}
&\zeta_{\phi} \equiv -\psi - H\frac{\delta\rho_{\phi}}{\dot{\rho}_{\phi}} = \frac{2 \ \delta\theta_0}{3 \ \theta_0}\,,
~~~~~~{\rm and}~~~~~~\zeta \equiv  \sum_{\alpha}\frac{\dot{\rho}_{\alpha}}{\dot{\rho}}\,\zeta_{\alpha}= \frac{\dot{\rho}_{\phi}}{\dot{\rho}_{\phi}+\dot{\rho}_{\gamma}}\zeta_{\phi}\,,\label{eq:Zetaaxion}
\\
&\mathcal{R}_{\phi} \equiv \psi - HV_{\phi}=  \frac{ \delta\theta_0}{ \theta_0}\frac{H \ \theta}{\delta\dot{\theta}}\,,
~~~~~~{\rm and}~~~~~~
\mathcal{R} \equiv \sum_{\alpha}\frac{\rho_{\alpha} + P_{\alpha}}{\rho + P}\,\mathcal{R}_{\alpha}=\frac{\rho_{\phi} + P_{\phi}}{\rho_{\phi} +(4/3)\rho_{\gamma} +P_{\phi}}\mathcal{R}_{\phi}\, ,\label{eq:Raxion}
\end{align}
where in this case the time-dependent quantities are evaluated solving the system given by eqs.\,(\ref{eq:DynBGSim1}-\ref{eq:DynBGSim3}).
We can use these equations as the initial conditions for the system that describe the evolution of the perturbations during phase II and III.
For the sake of clarity, we introduced explicitly the subscript $_k$ to remark that perturbations are Fourier modes with fixed comoving wavenumber $k$.

Now we consider the case of phase II+III, as defined in section\,\ref{sec:AfterInfla}.
The scalar field $\phi$ verifies the Klein-Gordon equation of motion $\ddot{\phi} + 3H\dot{\phi} + \mathcal{V}^{\prime}(\phi) = 0$ 
and its energy density and pressure are given by
\begin{align}\label{eq:ScalarFluid}
\rho_{\phi} = \frac{1}{2}\dot{\phi}^2 + \mathcal{V}(\phi)\,,~~~~~P_{\phi} = \frac{1}{2}\dot{\phi}^2 - \mathcal{V}(\phi)\,,
\end{align}
with adiabatic speed of sound
\begin{align}
c_{\phi}^2 = \frac{\dot{P}_{\phi}}{\dot{\rho}_{\phi}} = 1 + \frac{2\mathcal{V}^{\prime}(\phi)}{3H\dot{\phi}}\,. 
\end{align}
The energy transfer from the curvaton field to radiation is described by 
$Q_{\phi} = -\Gamma_{\phi}\rho_{\phi}$ (and, consequently, 
$Q_{\gamma} = \Gamma_{\phi}\rho_{\phi}$).
Radiation is a perfect fluid with $P_{\gamma} = \rho_{\gamma}/3$.
The perturbations in the energy transfer are described by 
$\delta Q_{\phi} = -\Gamma_{\phi}\delta\rho_{\phi}$ and 
$\delta Q_{\gamma} = \Gamma_{\phi}\delta\rho_{\phi}$ (where, as stated before, we are assuming that $\delta\Gamma_{\phi} = 0$). 
Consequently, as discussed below eq.\,(\ref{eq:GaugeInv1}), we have 
\begin{align}
\delta Q_{{\rm intr,}\phi} = 
\delta Q_{\phi} - \frac{\dot{Q}_{\phi}\delta\rho_{\phi}}{\dot{\rho}_{\phi}} = 0\,.
\end{align}
During this phase, we have $P_{\phi} = 0$, and the scalar field behaves as a pressure-less fluid. 
Consequently, we have $\delta P_{{\rm intr,}\phi} = 0$. 
Eq.\,(\ref{eq:GaugeInv1}), therefore, simplifies to 
\begin{align}
\dot{\zeta}_{\phi} =  - 
\frac{\dot{H}Q_{\phi}}{H\dot{\rho}_{\phi}}\big(
\zeta - \zeta_{\phi}
\big)
 + 
 \frac{k^2}{3a^2 H}\bigg[
 \Psi - \bigg(
 1 - \frac{Q_{\phi}}{\dot{\rho}_{\phi}}
 \bigg)\mathcal{R}_{\phi}
 \bigg]\,.
\end{align}
Using the background dynamics, and introducing the $e$-fold time as time variable, 
we recast the previous equation in the form
\begin{align}\label{eq:Final1}
\left.\frac{d\zeta_{\phi,k}}{dN}\right|_{\rm phase\,II+III} = 
\frac{(3+\Omega_{\gamma})\Gamma_{\phi}}{2(3H + \Gamma_{\phi})}\,(\zeta_k - \zeta_{\phi,k}) 
+ \frac{k^2}{3(aH)^2}\,\Psi_k - \frac{k^2}{(aH)^2}\frac{H}{(3H + \Gamma_{\phi})}\,\mathcal{R}_{\phi,k},
\end{align}
where the notation $\left.\right|_{\rm phase\,II+III}$ remarks the fact that the corresponding evolution equation 
is strictly valid during the oscillating and decaying phase. 
To close the system, we need the evolution of $\zeta$, $\mathcal{R}$ and $\mathcal{R}_{\phi}$ (given that $\Psi$ is related to $\zeta$ and $\mathcal{R}$ via eq.\,(\ref{eq:ZetaR})).
Since radiation has a well-defined equation of state, we also have $\delta P_{{\rm intr,}\gamma} = 0$. 
Consequently, we find
\begin{align}\label{eq:Pnad}
\delta P_{\rm nad} \equiv \delta P_{\rm intr} + \delta P_{\rm rel} = 
\delta P_{{\rm intr,}\gamma} + \delta P_{{\rm intr,}\phi} + \delta P_{\rm rel}
= \delta P_{\rm rel} 
= \frac{\dot{\rho}_{\gamma}\dot{\rho}_{\phi}}{3H\dot{\rho}}(\zeta_{\phi} - \zeta_{\gamma}) 
= 
\frac{\dot{\rho}_{\phi}}{3H}(\zeta_{\phi} - \zeta)\,.
\end{align}
where in the last step we used eq.\,(\ref{eq:ZetaDefinititon}).  Eq.\,(\ref{eq:GaugeInv3}) gives
\begin{align}\label{eq:Final2}
\left.\frac{d\zeta_k}{dN}\right|_{\rm phase\,II+III} = \frac{(3H + \Gamma_{\phi})\Omega_{\phi}}{(3+\Omega_{\gamma})H}\,
(\zeta_{\phi,k} - \zeta_k) + \frac{k^2}{3(aH)^2}\,(\Psi_k - \mathcal{R}_k).
\end{align}
The evolution of $\mathcal{R}_k$ is governed by 
\begin{align}\label{eq:Final3}
\left.\frac{d\mathcal{R}_k}{dN}\right|_{\rm phase\,II+III} = 
\bigg[
\frac{1}{H}\frac{dH}{dN} + \frac{
4H\Omega_{\gamma} - \Omega_{\phi}\Gamma_{\phi}
}{
H(3+\Omega_{\gamma})
}
\bigg]\mathcal{R}_k + 
\bigg(
\frac{1}{H}\frac{dH}{dN} + 1
\bigg)\zeta_k 
- \frac{(3H + \Gamma_{\phi})\Omega_{\phi}}{H(3+\Omega_{\gamma})}\,\zeta_{\phi,k} - 
\frac{k^2}{3(aH)^2}\,\Psi_k,
\end{align}
while we find for $\mathcal{R}_{\phi,k}$
\begin{align}\label{eq:Final4}
\left.\frac{d\mathcal{R}_{\phi,k}}{dN}\right|_{\rm phase\,II+III} = -\bigg(
\frac{\Gamma_{\phi}}{H} + \frac{1}{H}\frac{dH}{dN}
\bigg)(\mathcal{R}_k - \mathcal{R}_{\phi,k}).
\end{align}
The system formed by eqs.\,(\ref{eq:Final1},\,\ref{eq:Final2},\,\ref{eq:Final3},\,\ref{eq:Final4}) 
is subject to the relation
\begin{align}
\frac{3}{H}\frac{dH}{dN}(\zeta_k + \mathcal{R}_k) = \frac{k^2}{(aH)^2}\,\Psi_k
\end{align}
We solve numerically the evolution described by eqs.\,(\ref{eq:Final1},\,\ref{eq:Final2},\,\ref{eq:Final3},\,\ref{eq:Final4})  
, that is valid during phase II and phase III ($P_{\phi} = 0$), in order to get the correct value of the curvature perturbation $\zeta_k$ for the power spectrum defined in eq.\,(\ref{eq:FinaalPS}).

\cleardoublepage
\pagenumbering{gobble}
\thispagestyle{plain}			
\chapter*{Acknowledgements}
\subsection*{Some professional\footnote{Not so much by the way} acknowledgments}
A PhD journey is an intense and often challenging path, filled with obstacles and lessons. Yet, as I reflect on the past few years, I see them as some of the most beautiful and fulfilling of my life. For this reason, as I reach the end of this journey, I feel the need and deep desire to thank the people who made it all possible and meaningful.

First, I would like to express my gratitude to Prof. Alfredo Urbano. His guidance taught me the value of perseverance and hard work, and showed me that research, at times, can also be a playful endeavor.

At the start of my PhD, as I was learning to take my first steps in the world of research, I had the privilege of meeting my own little Chiron, Gabriele Franciolini. Thank you for everything you taught me and for being a role model, both professionally and personally. (Even if you don’t appreciate my memes, you’ll always be an inspiration to me!)

Among the Roman hills, I had the good fortune of meeting extraordinary people who made each day memorable with a mix of equations, discussions, and a few\footnote{many} Negroni. My heartfelt thanks to Giacomo, Danilo, Antonio, Fede, Lorenzo, and Filippo for your friendship and unwavering support.

In the middle of my second year, I left the warm embrace of Rome for the cold forests of Estonia, where I found exceptional collaborators. A special thanks to Hardi, who reminded me, during a particularly difficult time in my life, why I love what I do. I will always be indebted to you! My experience in Estonia was made unforgettable thanks to Ville, Juan, Kristjan, Christian, and Nico, who welcomed me as one of their own and made me feel at home even while far away.

In my final year, life blessed me with a second family. This gift would not have been possible without Toni, who welcomed me to Geneva as if I were his own son. Beyond the scientific contributions, I am profoundly grateful for your kindness, generosity, and the support you gave me during difficult times. Alongside your extraordinary presence, you gave me a group of brothers: il Sindaco Davide, il Demone Gabri, il Cravattaro Andrea, il Dio Greco Nicco, l'Americano Francesco, and the Coaches Stefy and Patrick. I cannot imagine how I would have made it through without all of you. I hope to have you by my side, today and always.

Finally, a special thanks goes to those who have always been there for me, especially during the hardest times, when the problems were far greater than just academic ones: Davide, Stefano, Ludo, Cri, and Carmine.
\newpage
\subsection*{Some unprofessional neapolitan acknowledgments}
Riprendendo le parole del meraviglioso artista napoletano Renato Carosone 
\begin{align}
    &\textit{"Comme te po' capi' chi te vo' ben} \nn \\
&\textit{Si tu le parle miezo americano"} \nn
\end{align}
per ringraziare le persone che mi sono sempre state accanto, necessariamente devo farlo nella lingua con cui ho sempre comunicato con loro: Italiano-Napoletano. Per fare ciò volevo riprendere le parole di un altro artista napoletano, Geolier, che con i suoi testi e la sua musica è stato per me un compagno di viaggio durante questi anni.
\begin{align}
    &\textit{"So aumentat e crep ngop e marciapied} \nn \\
&\textit{Pur mamma a fat e crep ngop a facc} \nn \\
&\textit{E m chied} \nn \\
&\textit{Si mammà c sta quant arriv in alt} \nn \\
&\textit{Giust p na soddisfazion personal"} \nn
\end{align}
A mia madre, la persona più forte che conosco. A lei che è la dimostrazione che sono gli uomini ad aver bisogno di una donna e mai il contrario. Carme' ha frnut e chiov!
\begin{align}
    &\textit{"A piccril co palloc cugliev semp na frnes} \nn \\
&\textit{Mo ng sta chiu o pallon ne a signor nta fnest"} \nn 
\end{align}
A mia Nonna. Non solo le lasagne ma la vita non ha più lo stesso sapore da quando non ci sei più. Mi manchi.
\begin{align}
    &\textit{"Ag rat t l'ammor mij a stu suol} \nn \\
&\textit{A lun ma salvaguardat} \nn \\
&\textit{Annascost ro sol"}\nn
\end{align}
Alle mie sorelle, Oriana e Antonella, per avermi sempre protetto nonostante la distanza.
\begin{align}
    &\textit{"Na major m ricet tu ija lua o dialet a miezz} \nn \\
&\textit{Com facc} \nn \\
&\textit{A chesta gent chi cio spieg}\nn \\
&\textit{Maij stat personag so rimast scugnzziel"}\nn
\end{align}
Al mio essere sempre me stesso. Che la vita possa non cambiarmi mai!
\vspace{1.5cm}
\hfill
\\Iovino Antonio Junior, Winter 2024

\newpage				
\thispagestyle{empty}
\mbox{}

\end{document}